%% file: main.tex
\documentclass[final,3p,ptimes,onecolumn,letterpaper]{elsarticle}

\usepackage{graphicx,color}

\usepackage{amssymb}
\usepackage{amsfonts}
\usepackage{amsmath}
\usepackage{epsfig}
\usepackage{enumitem}

\usepackage{bm}
\usepackage{xstring}
\usepackage{xspace}
\usepackage{etoolbox}
\xspaceaddexceptions{]\}}
\usepackage{fix-cm}
\usepackage{framed}
\usepackage{float}
\usepackage{tabularx}
\usepackage[caption = false]{subfig}

\usepackage{footnote}
\usepackage{threeparttable}

\usepackage{multirow}
\usepackage{floatpag}
\floatpagestyle{plain}
\usepackage{bigfoot}
\usepackage[normalem]{ulem}

\usepackage{minitoc}
\mtcsettitle{secttoc}{}
\usepackage{hyperref}

\input{defs.tex}

\journal{Physics Reports}

\numberwithin{equation}{section}

\begin{document}

\begin{frontmatter}

\title{\Huge{Large-Scale Galaxy Bias}}

\author[ITT,UG]{Vincent~Desjacques}
\ead{dvince@physics.technion.ac.il}

\author[PSU]{Donghui~Jeong}
\ead{djeong@psu.edu}

\author[MPA]{Fabian~Schmidt}
\ead{fabians@mpa-garching.mpg.de}

\address[ITT]{Physics Department, Technion, 3200003 Haifa, Israel}

\address[UG]{D\'epartement de Physique Th\'eorique and Center for Astroparticle Physics, Universit\'e de Gen\`eve, 24 quai Ernest Ansermet, CH-1221 Gen\`eve 4, Switzerland}

\address[PSU]{
Department of Astronomy and Astrophysics, and Institute for Gravitation and the Cosmos,
The Pennsylvania State University, University Park, PA 16802, USA
}

\address[MPA]{
Max-Planck-Institut f\"ur Astrophysik, Karl-Schwarzschild-Stra\ss e~1, 85748 Garching, Germany
}

\date{\today}

\begin{abstract}
 This review presents a comprehensive overview of galaxy bias, that is, the statistical relation between the distribution of galaxies and matter. We focus on large scales where cosmic density fields are quasi-linear. On these scales, the clustering of galaxies can be described by a perturbative bias expansion, and the complicated physics of galaxy formation is absorbed by a finite set of coefficients of the expansion, called \emph{bias parameters}. The review begins with a detailed derivation of this very important result, which forms the basis of the rigorous perturbative description of galaxy clustering, under the assumptions of General Relativity and Gaussian, adiabatic initial conditions. Key components of the bias expansion are all leading local gravitational observables, which include the matter density but also tidal fields and their time derivatives. We hence expand the definition of \emph{local bias} to encompass all these contributions. This derivation is followed by a presentation of the peak-background split in its general form, which elucidates the physical meaning of the bias parameters, and a detailed description of the connection between bias parameters and galaxy statistics. We then review the excursion-set formalism and peak theory which provide predictions for the values of the bias parameters. In the remainder of the review, we consider the generalizations of galaxy bias required in the presence of various types of cosmological physics that go beyond pressureless matter with adiabatic, Gaussian initial conditions: primordial non-Gaussianity, massive neutrinos, baryon-CDM isocurvature perturbations, dark energy, and modified gravity. Finally, we discuss how the description of galaxy bias in the galaxies' rest frame is related to clustering statistics measured from the observed angular positions and redshifts in actual galaxy catalogs.
\end{abstract}


\end{frontmatter}

\clearpage

\technote{* Sections that are more technical in nature and not essential to the understanding of later sections are marked with an asterisk.}

\dosecttoc
\tableofcontents

\clearpage

\input{intro.tex}

\input{evolution.tex}

\input{PBS.tex}

\input{measurements.tex}

\input{exset.tex}

\input{peaks.tex}

\input{NG.tex}

\input{beyondCDM.tex}

\input{observations.tex}

\input{summary.tex}

\clearpage
\appendix

\input{App_statistics.tex}

\clearpage
\input{App_SPT.tex}

\clearpage
\input{App_biastrans.tex}

\clearpage
\input{App_halofinder.tex}

\clearpage
\bibliographystyle{arxiv_physrev}
\bibliography{references}

\end{document}

%% file: defs.tex
\newcommand{\refeq}[1]{Eq.~(\ref{eq:#1})}          
\newcommand{\refeqs}[2]{Eqs.~(\ref{eq:#1})--(\ref{eq:#2})}          
          
\newcommand{\refEq}[1]{Eq.~(\ref{eq:#1})}          
\newcommand{\reffig}[1]{Fig.~\ref{fig:#1}}          
\newcommand{\reffigs}[2]{Figs.~\ref{fig:#1}--\ref{fig:#2}}          
\newcommand{\reftab}[1]{Tab.~\ref{tab:#1}}
\newcommand{\refsec}[1]{Sec.~\ref{sec:#1}}
\newcommand{\refsection}[1]{Section~\ref{sec:#1}}
\newcommand{\refsecs}[2]{Sec.~\ref{sec:#1}--\ref{sec:#2}}
\newcommand{\refapp}[1]{\ref{app:#1}}

\def\hmpc{\,h^{-1}\,{\rm Mpc}}
\def\hmsun{\,h^{-1}\,M_{\odot}}
\def\hmmpc{\,h\,{\rm Mpc}^{-1}}
\newcommand{\Mpch}{\,h^{-1}\,{\rm Mpc}}
\newcommand{\nsMpch}{h^{-1}\,{\rm Mpc}}
\newcommand{\iMpch}{\,h\,{\rm Mpc}^{-1}}
\newcommand{\Msunh}{\,h^{-1}\,M_{\odot}}

\newcommand{\be}{\begin{equation}}
\newcommand{\ee}{\end{equation}}
\newcommand{\bea}{\begin{eqnarray}}
\newcommand{\eea}{\end{eqnarray}}
\newcommand{\vs}{\nonumber\\} 
\def\ba#1\ea{\begin{align}#1\end{align}}
\def\la{\langle}
\def\ra{\rangle}
\def\dd{d}
\newcommand{\<}{\langle}
\renewcommand{\>}{\rangle}

\newcommand{\figsource}[1]{From~\cite{#1}.}

\newcommand{\perm}[1]{ \expandafter\ifstrempty\expandafter{#1} {\mbox{perm.}} {\mbox{$#1$ perm.}} }

\let\Var\relax
\DeclareMathOperator{\Var}{Var}
\newcommand{\Varof}[1]{\Var\left[#1\right]}

\def\Plin{P_\text{L}}

\def\Peps{P_\eps^{\{0\}}}
\def\PepsL{P_{\eps^L}^{\{0\}}}
\def\Plapleps{P_\eps^{\{2\}}}
\def\Beps{B_\eps^{\{0\}}}
\def\BepsL{B_{\eps^L}^{\{0\}}}
\def\Pepsepsd{P_{\eps\eps_\d}^{\{0\}}}
\def\PepsepsdL{P_{\eps^L\eps_\d^L}^{\{0\}}}

\newcommand{\LIMD}{LIMD\xspace}

\def\otd{\text{td}}

\def\Dstar{\frac{D_*}{D}}
\def\Dstarinline{D_*/D}

\newcommand{\avnh}{\overline{n}_h}

\newcommand{\avng}{\overline{n}_g}

\newcommand{\dc}{\delta_\text{cr}}

\def\intscale{\text{env}}

\newcommand{\rhob}{\overline{\rho}_m}
\newcommand{\Om}{\Omega_m}

\newcommand{\Omn}{\Omega_{m0}}

\newcommand{\OLn}{\Omega_{\Lambda0}}

\newcommand{\rhop}{\overline{\varrho}_m}
\newcommand{\rhopb}{\overline{\varrho}_b}

\def\ngp{\text{n}_g}

\def\avngp{\overline{\text{n}}_g}
\def\avnhp{\overline{\text{n}}_h}
\def\ngptilde{\tilde{\text{n}}_g}

\def\lapl{\nabla^2}
\def\laplsq{\nabla^4}
\def\ilapl{\nabla^{-2}} % inverse Laplace

\DeclareMathOperator{\tr}{tr}

\def\trans{\top}

\renewcommand{\Re}{\operatorname{Re}}
\renewcommand{\Im}{\operatorname{Im}}

\renewcommand{\ln}{\operatorname{ln}}
\renewcommand{\det}{\operatorname{det}}

\DeclareMathOperator{\erf}{erf}
\DeclareMathOperator{\erfc}{erfc}

\def\convD{\frac{\text{D}}{{\text{D}}\tau}}
\def\convDinline{{\text{D}}/{\text{D}}\tau}
\def\convDD{\frac{\text{D}}{{\text{D}}\ln D}}

\newcommand\technote[1]{%
  \begingroup
  \renewcommand\thefootnote{}\footnote{#1}%
  \addtocounter{footnote}{-1}%
  \endgroup
}

\newcommand{\bfem}[1]{\textbf{\emph{#1}}}

\newcommand{\vcc}{\mathrm{C}}
\newcommand{\vii}{\mathrm{I}}
\renewcommand{\O}{\mathcal{O}}
\newcommand{\U}{\mathcal{U}}
\newcommand{\Del}{\mathcal{D}}
\newcommand{\T}{\mathcal{T}}
\newcommand{\Q}{\mathcal{Q}}
\newcommand{\Mm}{\mathcal{M}}
\newcommand{\cH}{\mathcal{H}}
\newcommand{\G}{\mathcal{G}}

\renewcommand{\v}[1]{\bm{#1}}

\newcommand{\vk}{\bm{k}}
\newcommand{\vp}{\bm{p}}
\newcommand{\vq}{\bm{q}}
\newcommand{\vr}{\bm{r}}
\newcommand{\vu}{\bm{u}}
\newcommand{\vv}{\tilde{\bm{v}}}
\newcommand{\vw}{\bm{w}}
\newcommand{\vx}{\bm{x}}
\newcommand{\vy}{\bm{y}}

\newcommand{\bfx}{\bm{x}}
\newcommand{\bfr}{\bm{r}}
\newcommand{\bfk}{\bm{k}}

\newcommand{\D}{\Delta}
\newcommand{\s}{\sigma}
\renewcommand{\d}{\delta}
\renewcommand{\a}{\alpha}

\newcommand{\eps}{\varepsilon}

\def\L{\Lambda}
\def\xfl{\v{x}_\text{fl}}
\def\knl{k_\textnormal{\textsc{nl}}}
\def\Rnl{R_\textnormal{\textsc{nl}}}
\newcommand{\vkhat}{\hat{\v{k}}}
\newcommand{\vn}{\boldsymbol{\nabla}}
\def\LO{\textnormal{\textsc{lo}}}
\def\NLO{\textnormal{\textsc{nlo}}}

\newcommand{\trho}{\tilde{\rho}_m}
\newcommand{\trhop}{\tilde{\varrho}_m}
\newcommand{\tOmn}{\tilde{\Omega}_{m0}}
\newcommand{\tOKn}{\tilde{\Omega}_{K0}}
\newcommand{\tOLn}{\tilde{\Omega}_{\Lambda0}}
\def\tt{\tilde t_{0}}

\def\aup{\nearrow}
\def\adown{\searrow}
\def\anone{\rightarrow}

\def\dell{\delta_{\ell}}

\newcommand{\cQF}{\mathcal{Q}}
\newcommand{\rhat}{\hat{r}}
\newcommand{\vrhat}{\hat{\v{r}}}

\newcommand{\fnl}{f_\textnormal{\textsc{nl}}}
\newcommand{\taunl}{\tau_\textnormal{\textsc{nl}}}
\newcommand{\gnl}{g_\textnormal{\textsc{nl}}}
\newcommand{\Kernl}{K_\textnormal{\textsc{nl}}}
\newcommand{\DeltabNG}{\Delta b_1}
\newcommand{\Fs}[1]{\mathcal{F}^{(#1)}} % shape factor

\newcommand{\chit}{\tilde{\chi}}
\newcommand{\zt}{\tilde{z}}
\def\Mag{\mathcal{M}}
\newcommand{\nhat}{\hat{n}}
\newcommand{\vnhat}{\hat{\v{n}}}

\newcommand{\vpsi}{\bm{s}}   % displacement vector
\newcommand{\grad}{\boldsymbol{\nabla}} % gradient vector
\newcommand{\veta}{{\boldsymbol\eta}}
\newcommand{\vomg}{{\boldsymbol\Omega}}
\newcommand{\vlpk}{\bm{s}_\text{pk}}
\newcommand{\vxpk}{\bm{x}_\text{pk}}
\newcommand{\vvpk}{\bm{v}_\text{pk}}
\newcommand{\vqpk}{\bm{q}_\text{pk}}
\newcommand{\npk}{n_\text{pk}}
\newcommand{\fpk}{f_\text{pk}}
\newcommand{\bnpk}{\overline{n}_\text{pk}}
\newcommand{\bnesp}{\overline{n}_\text{ESP}}
\newcommand{\dpk}{\delta_\text{pk}^L}
\newcommand{\svpk}{\sigma_{v,\text{pk}}}
\newcommand{\nesp}{n_\text{ESP}}
\newcommand{\fesp}{f_\text{ESP}}
\newcommand{\xpk}{\xi_\text{pk}^L}
\newcommand{\ppk}{P_\text{pk}^L}
\newcommand{\xpkE}{\xi_\text{pk}}

\newcommand{\xpd}{\xi_{\text{pk},\delta}^L}

\makeatletter
\newlength{\apb@width}
\newcommand{\autoparbox}[2][c]{\settowidth{\apb@width}{#2}\parbox[#1]{\apb@width}{#2}}
\newcommand{\includegraphicsbox}[2][]{\autoparbox{\includegraphics[#1]{#2}}}
\makeatother

\def\Ctabzero{0}
\def\Ctabzero{}

%% file: intro.tex
\section{Introduction}
\label{sec:intro}

\secttoc

The observed distribution of galaxies, quasars, and clusters of galaxies---the 
large-scale structure of the Universe, \reffig{pie}---is one of the 
foundations of our knowledge about the history of the Universe.  
These tracers can be observed out to cosmological distances, and can thus be 
used to survey significant fractions of the observable Universe.  
If we understand how the distribution of tracers is related to the underlying distribution of matter, we can access a wealth of information on the composition of the Universe, properties of dark matter, dark energy and gravity, as well as the nature of the process that produced the initial seeds of structure.  The relation between luminous tracers and matter, which is known as \emph{bias}, thus forms a key ingredient in the interpretation of the observed large-scale structure.

\subsection{Historical review} 
\label{sec:history}

Perhaps the first example of cosmological conclusions drawn from sky surveys
is \cite{ryle/clarke:1961}, who showed that 
the observed flux distribution of radio sources is inconsistent with a static homogeneous Universe.  
Beyond this qualitative conclusion, it is difficult to extract information from the 1-point function of 
galaxies.  Thus, most cosmological inferences have been based on the next-order statistic of the galaxy density field, the two-point 
correlation function and its Fourier transform, the power spectrum \cite{peebles:1973} (a basic introduction to the description of statistical fields is provided in \refapp{stat}).  
Two-point statistics were first measured with significant signal-to-noise ratio in early 
surveys of galaxies and clusters of galaxies
\cite{hauser/peebles:1973,peebles/hauser:1974,peebles:1975,groth/peebles:1977,
seldner/peebles:1978,kirshner/oemler/schechter:1978,bahcall/soneira:1983,klypin/kopilov:1983,davis/peebles:1983}.  
Already in these first measurements it became clear that the correlation function of galaxies and clusters 
is not the same, which implies that they cannot both be unbiased 
tracers of the matter density fluctuations. Consider
a simple ansatz which locally relates the 
density contrast of galaxies or clusters of galaxies to that of matter 
at a fixed time:
\be
\d_g(\vx) \equiv \frac{n_g(\vx)}{\avng} - 1
= b_1\, \d(\vx) = b_1 \left( \frac{\rho_m(\vx)}{\rhob}-1\right)\,,
\label{eq:bias1}
\ee
where all quantities are evaluated at the same fixed time, which we leave implicit here, $\avng$ is the 
mean \emph{comoving} (see \refsec{notation}) number density of galaxies, while
$\rhob$ is the comoving background matter density, and $b_1$ is a 
parameter that we call \emph{bias}.  Then, the two-point function of galaxies (or clusters of galaxies)
is enhanced by a factor of $(b_1)^2$ over the matter two-point function.  
If we allow for clusters to have a larger bias parameter than galaxies, 
their different observed correlation functions can be explained.  

A relation of the form \refeq{bias1} with $b_1 \neq 1$ implies that the galaxy 
density $n_g(\vx)$ is not linearly proportional to $\rho_m(\vx)$,
as otherwise their fractional perturbations would be equal.  Instead,
the galaxy density has to be a nonlinear function of the matter density.  
In his seminal paper \cite{kaiser:1984}, Kaiser laid out a physical picture for such 
a nonlinear function, by positing that clusters form at the locations of rare, high-density excursions of the matter density 
field (we will describe this ansatz in  detail in \refsec{localbias}).  
This argument was subsequently refined by Bardeen, Bond, Kaiser, and Szalay (BBKS \cite{bardeen/etal:1986}), who derived the statistics of peaks in 
Gaussian random fields (\refsec{peaks}). Further, Refs.~\cite{kaiser:1984,bardeen/etal:1986} and, subsequently, 
Refs.~\cite{cole/kaiser:1989,mo/white:1996,sheth/tormen:1999} formulated the ``peak-background split,'' which establishes a connection between the bias parameters and the mean abundance of tracers (\refsec{PBS}).
Importantly, in the meantime Ref.~\cite{fry/gaztanaga:1993} showed that bias can be made much more general than the specific examples considered in \cite{kaiser:1984,bardeen/etal:1986}.  

The advent of larger galaxy surveys, in particular the CfA \cite{CfAsurvey:I} and APM 
\cite{maddox/efstathiou/etal:1990} 
surveys, allowed for the first robust cosmological inferences from galaxy clustering 
\cite{CfAsurvey:II,davis/peebles:1983,baumgart/fry:1991,park/gott/costa:1992}, for which bias is a crucial ingredient.  In particular,  
Ref.~\cite{davis/etal:1985} showed that the two-point correlation function 
of galaxies measured in the CfA and APM surveys was impossible to 
reconcile with the predictions of a cold dark matter (CDM) dominated cosmology (with matter density parameter 
$\Om \leq 1$) unless a bias parameter $b_1$ is introduced following \cite{kaiser:1984}.  
Moreover, the results of these surveys played an important role in establishing the now familiar standard model of cosmology, the spatially flat $\Lambda$CDM model.  To this day, this model consistently describes both the large-scale galaxy power spectrum and the cosmic microwave background (CMB) \cite{tegmark/etal:2004,BAO/2dF,wigglez_final:2012,wmap9:2013,vipers/RSD,planck:2015-overview,BOSSdr12:2016,DES1yr:cosmology}, and any significant departures from this model are by now tightly constrained.  We briefly recap this history here.  

\begin{figure}[t!]
\centering
\includegraphics[width=0.8\textwidth]{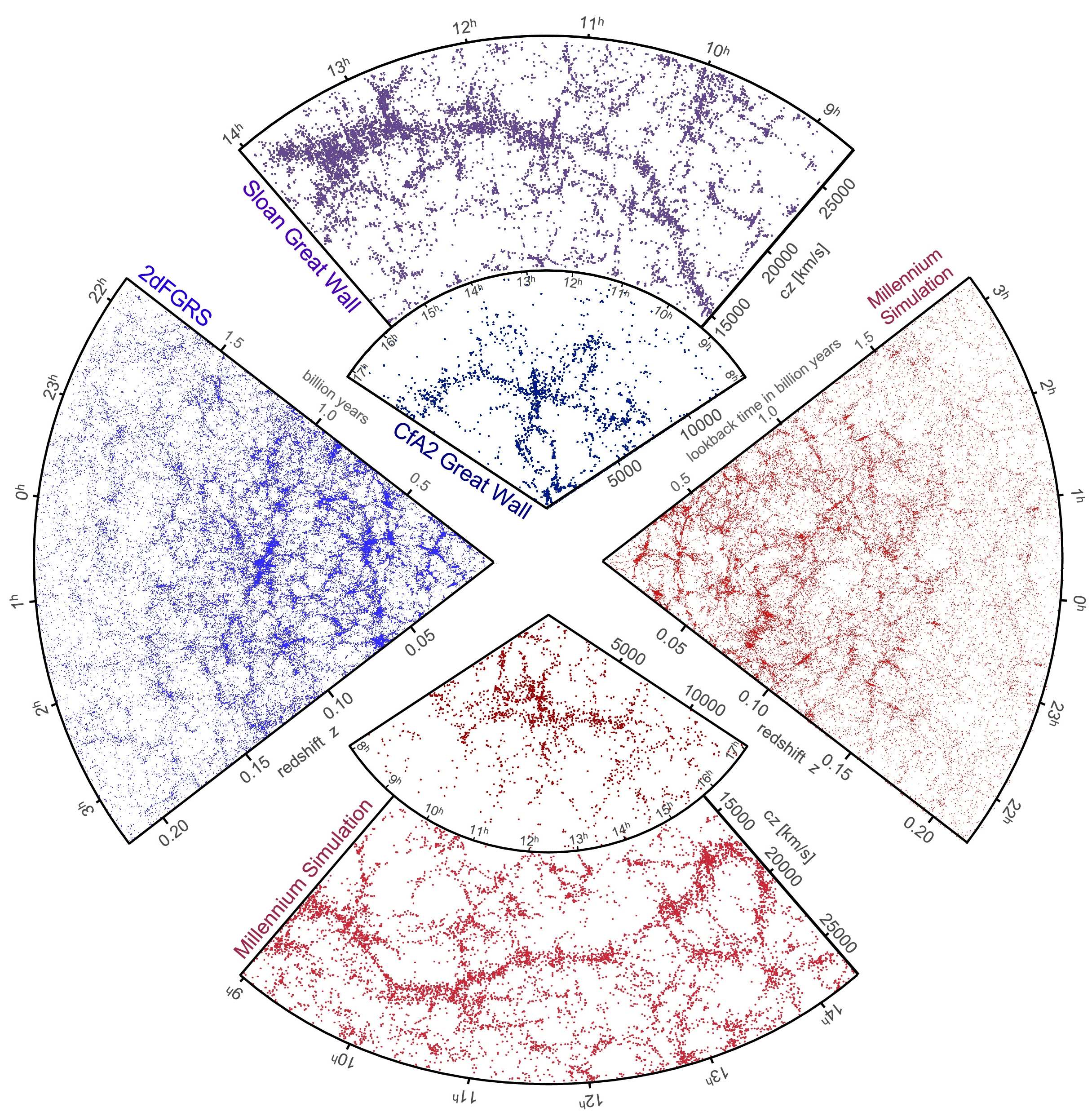}
\caption{Two-dimensional slice projections (pie diagram) of the measured locations of galaxies in the CfA2, 2dF, and SDSS galaxy redshift surveys (top left half).  The lower right half shows the location of galaxies which were assigned to dark matter halos in the \emph{Millennium} gravity-only N-body simulation using a semi-analytical prescription.  It is apparent that the simulation, which assumes a flat $\Lambda$CDM cosmology, qualitatively reproduces the observed large-scale structure of the Universe very well.  
\figsource{millennium:2006}
\label{fig:pie}}
\end{figure}

In the early 1990s, the flat, matter-dominated Einstein-de Sitter (EdS) Universe \cite{EdS}, was favored on theoretical grounds. Ref.~\cite{efstathiou/bond/white:1992} pointed out that this scenario made inconsistent predictions for the relative amplitude between 
large-scale and smaller scale clustering of the APM galaxies, given the 
constraints on the primordial amplitude of perturbations from CMB temperature 
fluctuations from COBE \cite{COBE}.  
Essentially, since the shape of the two-point function of tracers, even those that are biased according to \refeq{bias1}, is the same as that of matter on large scales, the former could be used to rule out the shape of the matter correlation function predicted by the EdS scenario.  
Moreover, this cosmological model did not correctly describe 
galaxy velocity statistics \cite{davis/etal:1992}.  As noted in \cite{efstathiou/sutherland/maddox:1990}, the 
introduction of a cosmological constant, with a magnitude that corresponds to roughly 80 percent of the present-day 
total mean energy density, could resolve the discrepancy (see also the earlier discussion in \cite{peebles:1984}).  
This eventually led to the establishment of the standard flat $\Lambda$CDM cosmology, whose confrontation with observations has been enormously successful \cite{tegmark/etal:2004,BAO/2dF,wigglez_final:2012,wmap9:2013,vipers/RSD,planck:2015-overview,BOSSdr12:2016,DES1yr:cosmology,DES1yr:clustering}. 
We refer the reader to \cite{percival:2007} for a review of recent cosmological constraints from galaxy clustering.  

Another milestone was reached in the detection of the baryon acoustic oscillation (BAO) feature \cite{eisenstein/hu} in the two-point correlation
function of the SDSS \cite{BAO/SDSS} and the power spectrum of the 2dF \cite{BAO/2dF} data sets (see \reffig{pie} for representations of these surveys).    
This feature can be used as a robust standard ruler to measure the 
expansion history of the Universe \cite{seo/eisenstein:2003,
seo/eisenstein:2007}.
In addition, redshift-space distortions \cite{Jackson:1972} allow for a measurement of the growth rate of structure  \cite{sargent/turner:1977,kaiser:1987}.  
These probes are by now part of the core science of a number of ongoing\footnote{Dark Energy Survey (DES) \cite{DES:2016}, eBOSS \cite{eBOSS}, HETDEX \cite{HETDEX}, HSC \cite{HSC}, KiDS \cite{KiDS}}
and next-generation experiments.\footnote{DESI \cite{DESI}, Euclid \cite{amendola/etal:2013}, 4MOST \cite{4MOST}, LSST \cite{LSST}, PFS \cite{PFS}, WFIRST \cite{WFIRST}}  

Despite these successes, the exploitation of the cosmological information in large-scale galaxy surveys is still in an early stage:    
(i) Beyond the two-point function, many studies suggest that higher $n$-point functions contain significant amounts of information 
as well \cite{cooray/hu:2001,sefusatti/scoccimarro:2005,sefusatti/etal:2006,sefusatti/komatsu:2007} (\refsec{measurements}).  
(ii) The potential of galaxy clustering to test General Relativity on 
scales of 30--150~Mpc has only recently been realized \cite{zhang/etal:2007,reyes/etal:2010,blake/etal:2016,pullen/etal:2016} (\refsec{modgrav}).  (iii) Departures from perfect
Gaussian initial conditions can leave distinct imprints in the large-scale clustering of biased tracers
\cite{scoccimarro/etal:2004,dalal/etal:2008}.  This leads to the fascinating (and surprising) prospect that galaxy 
clustering can provide insights on early-Universe physics that are complementary to those from the CMB (\refsec{NG}).

A common theme of all these, and many other possible applications of galaxy statistics is that they critically rely on a robust physical description of the
relation between galaxies and matter: \bfem{galaxy bias}.  

We emphasize that
galaxies are not the only tracers of large-scale structure.  Important other
examples include groups and clusters of galaxies, voids, quasars, the Lyman-$\alpha$
forest, line intensity mapping of H$\alpha$, CO, the 21cm hydrogen hyperfine structure
transition, and others, as well as diffuse backgrounds such as the cosmic infrared background.  Keeping this in mind,
in this review we will use the term ``galaxies'' as a convenient stand-in for general
tracers of the large-scale structure;  aspects that are specific to different
kinds of tracers are discussed in \refsec{observations}.  

\subsection{From initial conditions to observed galaxies: the role of bias} 
\label{sec:intro:role}

In order to be able to extract cosmological information from the observed clustering of galaxies, we need a reliable model, or, better yet, theory for the statistics of galaxies, given the properties of the very small ($\delta\sim 10^{-4}$) initial perturbations in the early Universe, and the background cosmology.  The different ingredients necessary in this endeavor are summarized in the flowchart in \reffig{flowchart}.  

First, we require a theoretical description of the distribution of matter itself, which is governed by the growth under gravitational collapse starting from the initial perturbations (see \cite{einasto:1992,coles:2001} for reviews).  
Unlike the CMB, which is accurately described by linear (first-order) perturbation theory, the perturbations in the matter density of the Universe at redshifts $z \lesssim 10$ are nonlinear.  Crucially, the degree of nonlinearity depends on the scales considered.  
One can broadly divide large-scale structure into two regimes: large, \emph{quasi-linear scales}, where perturbation theory (see below) converges to the correct result if carried out to sufficiently high order;  and the smaller \emph{nonlinear scales}, which cannot be described by perturbation theory.  The description of LSS on nonlinear scales has to rely on numerical simulations and simplified heuristic models.  In this review, we focus on quasi-linear scales.  While this neglects the cosmological information that can be extracted from LSS on small scales, the key advantage of restricting to large scales, as we will see, is that we are able to obtain a rigorous theory of galaxy clustering.  
In order to describe the distribution of matter on quasi-linear scales, nonlinear cosmological perturbation theory (PT) approaches have been developed (\cite{peebles:1980}; see \cite{bernardeau/etal:2001} for a review).  PT provides a robust theoretical foundation for the description of the quasi-linear matter density and tidal field, and has made significant progress in the past decade with the development of several new technical approaches.\footnote{Renormalized perturbation theory \citep{crocce/scoccimarro:2006,bernardeau/etal:2008,montesano/etal:2010}, renormalization group approach \citep{mcdonald:2007},  closure theory \citep{taruya/hiramatsu:2008}, Lagrangian perturbation theory \citep{buchert:1992,matsubara:2008,zheligovsky/frisch}, TimeRG theory \citep{matarrese/pietroni:2007,pietroni:2008}, effective field theory \citep{baumann/etal:2012,carrasco/etal:2012,carrasco/etal:2014,hertzberg:2014,porto/etal:2014,senatore:2015,senatore/zaldarriaga:2014,baldauf/etal:2015b}, time-sliced perturbation theory \cite{blas/etal:2015}, among others.}  
In this review, we will only rely on fairly basic results of perturbation theory.  These are summarized in \refapp{SPT}.  

Now, galaxy surveys of course do not measure the matter density field itself, but rather the distribution of galaxies or other tracers, that is, of highly nonlinear objects which are the result of a complex formation process.  \emph{Bias} describes, in a statistical sense, the relation of the distribution of these objects to that of matter.  Clearly, this is a very complicated relation in general: galaxy formation takes place over long periods of time and in interaction with the formation of structure in the matter distribution, and is currently far from being understood in detail (see \cite{mo/vandenbosch/white:2010} for an overview).  
Given a perturbation-theory based description of the large-scale
matter density and tidal fields on quasi-linear scales, the goal of the theory of galaxy bias is 
to write the local number density of galaxies $n_g$ as the most general
function of the properties of the large-scale environment that is allowed by general covariance under coordinate transformations.  
Key theoretical advances have been made in the understanding of bias over the past decade, paralleling those for the matter density field mentioned above.  

Remarkably, on quasi-linear scales, all the complexities of galaxy formation can be absorbed into a finite number of parameters (at each order in perturbations and at fixed time), the \emph{bias parameters}.  This is a nontrivial result, and relies on the fact that on large scales, structure formation is completely determined by the action of gravity.  In fact, one can show that, at linear order in perturbation theory (and assuming Gaussian, adiabatic initial conditions), the ansatz \refeq{bias1} is correct and complete, up to an additional additive noise term.  
More generally, the statistics of galaxies at a given
order in perturbation theory are determined by a well-defined set of bias parameters which can be constrained using these statistics.  
In this way, we effectively marginalize over the unknown details of the
galaxy formation process, while robustly extracting cosmological information
from galaxy surveys.  
This review provides a comprehensive overview of this approach, and connects it
to the other aspects of the theory of galaxy clustering on quasi-linear scales, as summarized in \reffig{flowchart}.  

\begin{framed}
\noindent 
To summarize, the \emph{perturbative theory of galaxy clustering}, valid on quasi-linear scales, is based on two key ingredients:  (i) a perturbation theory prediction for the matter density and tidal field;  (ii) a complete parametrization of galaxy bias at each order in perturbation theory.  
\end{framed}

Despite the complexities of galaxy formation mentioned above, there is a robust, well-established fact that we can build on: galaxies reside in massive, gravitationally bound structures called \emph{halos}. 
As dark matter makes up approximately 80\% percent of all matter, the potential
well of gravitationally bound structures is dominated by dark matter.   
Consequently, the halos that host galaxies are dark-matter dominated.  
The connection between galaxies and halos is well established numerically and 
observationally, for example through stacked weak gravitational lensing \cite{guzik/seljak:2002,brainerd/blandford/smail:1996}.  
The lower half of \reffig{pie} shows an example of how halos identified in a gravity-only simulation can be populated with galaxies to realistically reproduce their observed distribution.  Thus, the understanding of the large-scale
clustering of dark matter halos is a physically well-motivated intermediate step toward the same for the clustering of galaxies themselves.  Moreover, the formation,
structure, and clustering of halos can be studied reliably and in detail
via gravity-only N-body simulations.  Hence, several sections of this review 
will deal with numerical results on, and physical models of, halo clustering
(in particular, \refsecs{measurements}{peaks}).

\subsection{Notation and terminology}
\label{sec:notation}

Throughout the bulk of this review, we work in conformal-Newtonian gauge
and comoving coordinates, and restrict to scalar perturbations. Moreover,
we assume a spatially flat background, as supported by observations, although
this assumption is of no relevance to the topic of bias. Then,
the perturbed Friedmann-Robertson-Walker (FRW) metric can be written as
\be
ds^2 = a^2(\tau)\left[- (1+2\Phi) d\tau^2
+ (1-2\Psi) \d_{ij} dx^i dx^j \right]\,,
\label{eq:metriccN}
\ee
where in General Relativity $\Phi=\Psi$ in the absence of anisotropic stress.  
The matter density contrast $\d$ is, in general, not a local observable.  
Thus, it should not really appear by itself in a bias expansion [such as in \refeq{bias1}].  
On scales much smaller than the comoving horizon, $\cH \equiv aH$, however, 
this issue is irrelevant for all popular gauge choices.  Strictly speaking, the matter density perturbation
$\d$ should be understood as being defined in synchronous-comoving gauge throughout this review \cite{baldauf/etal:2011,gaugePk}.  We will discuss these issues in \refsec{GR}.  

We will use the reference cosmology defined in \reftab{ref_cosmology} for all numerical results, unless otherwise indicated.  Note that numerical results and figures
taken from the published literature are based on different cosmological parameters.  Throughout the review, the matter ($\rho_m$), galaxy ($n_g$), and halo densities ($n_h$) are defined as \emph{comoving densities}, i.e.~as physical densities multiplied by $a^3$; for example, $\rho_m = a^3 \varrho_m$, where $\varrho_m$ is the physical matter density.  Further, we define the time-dependent density parameter for any component $X$ as $\Omega_X(\tau) \equiv \overline{\varrho}_X(\tau) / \varrho_{\rm crit}(\tau)$,
where $\varrho_{\rm crit}(\tau) = 3 H^2(\tau)/8\pi G$.  Their values today are denoted as $\Omega_{X0}$.  

Let us now generalize the simple linear bias in \refeq{bias1}, by writing
\be
\d_g(\vx,\tau) = \sum_O b_O(\tau) O(\vx,\tau) \,.
\label{eq:bias2}
\ee
Here, $O$ are operators, or statistical fields, which describe properties
of the galaxies' environment on which their density can depend.  Each
operator is multiplied by a corresponding bias parameter $b_O$, which, at fixed time,
is merely a number.  
\refeq{bias1} already provides one example, with $O=\delta$ and $b_O = b_1$.  
This is easily generalized to a whole set of operators $O=\delta^N$, $N \geq 1$.  
The corresponding bias parameters have historically been known
as ``local bias parameters'' $b_N$ (see \refsec{localbias}), defined through
\be
b_N \equiv N!\, b_{\d^N}\,.
\label{eq:bNdef}
\ee 
However, as we describe below, we will broaden the definition of \emph{local bias} to a more general, physically motivated class.  Whenever we need to single out the restricted set of bias parameters in \refeq{bNdef}, we will refer to the expansion in a power series in $\d$ solely as \emph{local-in-matter-density (\LIMD)} bias.

Note that we will not distinguish in notation between bias parameters of halos and those of galaxies.  This is because almost all of the bias parameters we encounter in the review are equally relevant for galaxies and halos.  Where results only apply to halos, we will clearly indicate this.

\begin{table*}
\centering
\begin{tabular}{l|l|l}
\hline
\hline
Baryon density parameter & $\Omega_{b0}h^2$ & 0.022307 \\
CDM density parameter & $\Omega_{c0}h^2$ & 0.11865 \\
Neutrino density parameter & $\Omega_{\nu0}h^2$ & 0.000638 \vs
Cosmological constant parameter & $\Omega_{\Lambda0}$ & 0.69179 \\
Hubble constant today & $H_0 = h \: 100\, {\rm km}\, {\rm s}^{-1}\, {\rm Mpc}^{-1}$ & $67.78\,{\rm km}\,{\rm s}^{-1}\,{\rm Mpc}^{-1}$ \\
Scalar spectral index & $n_s$ & 0.9672 \\
Scalar power spectrum amplitude & $\mathcal{A}_s$ & $2.147\times 10^{-9}$ \\
\hline
Matter power spectrum normalization at $t_0$ & $\sigma_8 = \s(R=8\Mpch,z=0)$ & 0.8166 \\[1pt]
\hline
\hline
\end{tabular}
\caption{Parameters of the flat $\Lambda$CDM reference cosmology used for numerical results (maximum likelihood values for ``{\sf base\_plikHM\_TTTEEE\_lowTEB\_lensing\_post\_BAO\_H080p6\_JLA}'' from Planck 2015 \cite{planck:2015-overview,planck:2015-parameter}).  Here, the density parameters are defined as $\Omega_{X0} = \overline{\varrho}_X(t_0) / \varrho_{\rm crit}(t_0)$, where $\rho_{\rm crit}$ is the critical density, and $t_0$ denotes today's epoch. $\sigma_8$ is a derived parameter.}
\label{tab:ref_cosmology}
\end{table*}

\begin{table*}
\centering
\begin{threeparttable}[b]
\begin{tabular}{l|l}
\hline
\hline
Probability density function (PDF) & $p(x)$ \\
Spatial derivatives, Laplace operator & $\partial_i \equiv \partial/\partial x^i$,  $\lapl \equiv \d^{ij} \partial_i\partial_j$, \\
 & \  \  $\Del_{ij} \equiv \frac{\partial_i\partial_j}{\lapl} - \frac13 \d_{ij}$ \\
Fourier transform & $f(\vk) \equiv \int d^3 \vx\, f(\vx) e^{-i\vk\cdot\vx}$ \\
Momentum integral & $\int_{\vk} \equiv \int \frac{d^3 \vk}{(2\pi)^3}$ \\
Sum notation & $\vk_{1\cdots n} \equiv \vk_1 + \cdots + \vk_n$ \\
Connected $n$-point function\tnote{1} & $\< O(\vk_1) \cdots O(\vk_n)\>_c$ \\
Dirac delta distribution\tnote{2} 
 & $\d_D(\vx) = \int_{\vp} e^{i\vp\cdot\vx}$  \\
$n$-point correlator without & $\< O(\vk_1) \cdots O(\vk_n)\>'\,,$ where \\
\  \  momentum conservation & $\  \  \< O(\vk_1) \cdots O(\vk_n)\> = \< O(\vk_1) \cdots O(\vk_n)\>' \,(2\pi)^3 \d_D(\vk_{1\cdots n})$ \\
Kronecker symbol & $\d_{ij},\,\d_{NM}$ \\
Heaviside step function & $\Theta_H(x) =1$ for $x>0$ and 0 otherwise \\
Complementary error function &
$\erfc(x) = 1 - \erf(x) = \frac{2}{\sqrt{\pi}}\int_x^{\infty}\!\! du\,  e^{-u^2}$ \\[3pt]
\hline
Laguerre polynomials & $L_n(x)$ \\
Legendre polynomials & $\mathcal{L}_l(x)$ \\
Probabilists' Hermite polynomials & $H_N(x)$ \\
\hline
\hline
\end{tabular}
{\footnotesize
\begin{tablenotes}
\item[1] See \refapp{stat}.
\item[2] Note that this implies $(2\pi)^3\int_{\vk}\delta_D(\vk-\vk_0)f(\vk)=f(\vk_0)$.
\end{tablenotes}
}
\caption{List of mathematical symbols and notations.}
\label{tab:math}
\end{threeparttable}
\end{table*}
The subsequent sections will derive exactly which operators appear in the
proper, complete bias expansion.  In addition to the term \emph{\LIMD} defined above, we assign certain names to
different categories of operators that will appear in the following, which
we state here mainly for readers with experience in the subject.  The definition
of these categories can be skipped on a first reading,
as it jumps somewhat ahead of the proceedings.  Specifically, we distinguish
three categories of bias terms:
\begin{itemize}
\item \emph{Local bias:} this includes all operators $O$ that involve exactly two spatial derivatives\footnote{Note that the operator $\partial_i\partial_j/\lapl$ counts as zero net spatial derivatives.} for each instance of the gravitational potential $\Phi$.  
This includes the \LIMD terms, i.e. powers of the density $\d^N$, as $\d$ is related to $\lapl\Phi$ through the Poisson equation (\reftab{notation}).  It also includes powers of the tidal field, and time derivatives of the density and tidal field.  The physical reasoning behind this category is that these operators constitute the leading \emph{local gravitational observables} of long-wavelength spacetime perturbations.  
Note that $\Phi$ itself, or first derivatives $\partial_i\Phi$, are not locally observable and hence not included in this category, as required by the equivalence principle.  
\item \emph{Higher-derivative bias:} this includes operators that involve more than two derivatives acting on a single instance of $\Phi$;  for example, 
$\lapl\d$ or $(\partial_i\partial_k\partial_l\Phi)^2$.  These are clearly
also local gravitational observables. However, they are sub-leading in the limit of very long-wavelength density perturbations.  Moreover, each additional
spatial derivative has to be multiplied by a spatial scale $R_*$ in order to render the contribution to $\d_g$ dimensionless [see \refeq{bias2}].  $R_*$
corresponds to the characteristic spatial scale of the formation of the galaxies considered. 
\item \emph{Nonlocal bias:} this class, finally, includes operators with fewer than two, or a fractional number of derivatives of the potential $\Phi$.  
These terms cannot be induced by gravitational evolution or local physical processes, as they are 
forbidden by the equivalence principle, and so they must be imprinted in the
initial conditions (with a single minor exception described in \refsec{baryons}).  The most important case is that of primordial non-Gaussianity, considered in \refsec{NG}.
\end{itemize}
The term ``scale-dependent bias'' has been used frequently in the literature
to denote a nontrivial function of wavenumber $k$ in the Fourier-space
relation between $\d_g(\vk)$ and $\d(\vk)$, or other operators.  Unfortunately,
this term can apply equally to the classes of higher-derivative bias
and nonlocal bias, which, as we argued above, have distinct physical origin.  
We will only use the term ``scale-dependent bias'' in the context
of primordial non-Gaussianity (\refsec{NG}), specifically to denote the leading 
nonlocal term appearing in that case.

For reference, mathematical symbols and conventions used throughout this review
are summarized in \reftab{math}, abbreviations are listed in \reftab{abbrev}, and a reference list of physical variables is given in \reftab{notation}.

\begin{table*}[b]
\centering
\begin{tabular}{l|l}
\hline
\hline
BAO  & Baryon acoustic oscillation \\
CMB  & Cosmic microwave background \\
EdS  & Einstein-de Sitter (flat, matter-dominated Universe) \\
EFT  & Effective field theory \\
ESP  & Excursion-set peaks \\
FoF  & Friends-of-friends algorithm (\refapp{halofinder}) \\
$\L$CDM & $\Lambda$ cold dark matter \\
LIMD & Local in matter density (previously commonly known as ``local bias'') \\
LPT  & Lagrangian perturbation theory\\
LO   & Leading order (tree level)\\
NLO  & Next-to-leading order (1-loop)\\
PBS  & Peak-background split \\
PNG  & Primordial non-Gaussianity \\
PT   & Perturbation theory \\
RSD  & Redshift-space distortions \\
SO   & Spherical overdensity (\refapp{halofinder}) \\
SPT  & Eulerian standard perturbation theory \\
\hline
\hline
\end{tabular}
\caption{List of abbreviations used in the text.}
\label{tab:abbrev}
\end{table*}

\begin{table*}
\thisfloatpagestyle{empty} % to not have page number on page with big table
\centering
\begin{threeparttable}[b]
\begin{tabular}{l|l|l}
\hline
\hline
Quantity & Symbol & Defining relation\\
\hline
Conformal time & $\tau$ & $d\tau \equiv a^{-1} dt$ \\
Eulerian comoving coordinate & $\vx$ & \refeq{metriccN} \\
Time derivative & $\dot f$  &  $\dot f\equiv df/dt$ \\
Hubble rate & $H$ & $H \equiv \dot a / a$ \\
Conformal Hubble rate & $\cH$ & $\cH \equiv a^{-1} da/d\tau = aH$ \\
Mean comoving (physical) matter density &  $\rhob$~($\rhop$) & 
$\rhob(\tau) \equiv a^3(\tau) \rhop(\tau)$ \\
Mean comoving halo number density %per log. mass interval
& $\avnh$ & $\avnh(M,z) \equiv  \partial^4\overline{N}_h /(\partial^3 \vx\, \partial \ln M)$ \\
\hline %%%%%%%%%%%%%%%%%%%%%%%%%%%%%%%%%%%%%%%%%%%%%%%%%%%%%%%%%%%%%%%%%
Linear matter growth factor
& $D(\tau)$ & \refeq{Deom} \\
Logarithmic growth rate & $f(\tau)$ & $f \equiv d\ln D/d\ln a$ \\
Gravitational potential & $\Phi$ & \refeq{metriccN} \\
Primordial Bardeen potential & $\phi$ & $\Phi(k)|_{\rm mat.~dom.} = T(k) \phi(k)$ \\
Primordial curvature pert. in comoving gauge & $\mathcal{R}$ & $\mathcal{R} = (5/3) \phi$ in matter domination\\
\hline %%%%%%%%%%%%%%%%%%%%%%%%%%%%%%%%%%%%%%%%%%%%%%%%%%%%%%%%%%%%%%%%%
Lagrangian comoving coordinate & $\vq$ & $\vq = \lim_{\tau\to 0} \xfl(\vq,\tau)$ \\
Comoving coordinate of fluid trajectory & $\xfl(\vq,\tau)$ & $\xfl(\vq,\tau) \equiv \vq + \v{s}(\vq,\tau)$ \\
Lagrangian displacement & $\v{s}(\vq,\tau)$ & \refeq{seom} \\
Peculiar fluid velocity & $\v{v}$ & $\v{v} \equiv a\,\dot{\vx} = d\vx/d\tau$ \\
\hline %%%%%%%%%%%%%%%%%%%%%%%%%%%%%%%%%%%%%%%%%%%%%%%%%%%%%%%%%%%%%%%%%
Matter density contrast\tnote{1} & $\d$ & \refeq{bias1};~~$\d(\vx,\tau) = 2/(3\Om \cH^2) \lapl \Phi(\vx,\tau)$ \\
Density contrast of galaxies (general tracer) & $\d_g$ & \refeq{bias1} \\
Halo density contrast & $\d_h$ & $\d_h(\vx,\tau) \equiv n_h(\vx,\tau)/\avnh(\tau) - 1$ \\
Lagrangian halo density contrast\tnote{2} & $\d_h^L(\vq,\tau_0)$ & \reftab{EulLagr} on p.~\pageref{tab:EulLagr} \\
Tidal field & $K_{ij}$ & $K_{ij} \equiv (\partial_i\partial_j/\lapl-\d_{ij}/3)\d$ \\
Linearly extrapolated initial density field & $\d^{(1)}$  & $\d^{(1)}(\vk, \tau) \equiv \Mm(k,\tau) \phi(\vk)$~~~[\refeq{deltaphi}] \\
Operator constructed out of density field & $O$ & e.g., $O(\vx,\tau) = [\d(\vx,\tau)]^2$ \\
Smoothed field & $O_R$  & $O_R(\vx, \tau) \equiv \int d^3\vy\,O(\vx+\vy) W_R(\vy)$ \\
Operator at $n$-th order in perturbation theory & $O^{(n)}$ & e.g., $O^{(2)}(\vx,\tau) = [\d^{(1)}(\vx,\tau)]^2$ \\
\hline %%%%%%%%%%%%%%%%%%%%%%%%%%%%%%%%%%%%%%%%%%%%%%%%%%%%%%%%%%%%%%%%%
Linear matter power spectrum & $\Plin(k,\tau)$ & $\Plin(k,\tau) \equiv \< \d^{(1)}(\vk,\tau)\d^{(1)}(\vk',\tau)\>'$  \\
Variance of linear density field on scale $R$ & $\sigma^2(R)$ 
& $ \s^2(R) \equiv \int_{\vk} \Plin(k) W_R^2(k)$ \\
Generalized spectral moment\tnote{3} & $\s_n^2(R)$ & $\s_n^2(R) \equiv \int_{\vk} k^{2n} \Plin(k) W_R^2(k)$~~~[\refeq{mspec}] \\
Critical density (collapse threshold) & $\dc \simeq 1.686$ & \refeq{deltacdef} \\
Peak significance & $\nu_c$ & $\nu_c \equiv \dc / \sigma(R)$ \\
Multiplicity function & $\nu_c f(\nu_c)$ & \refeq{vfv} \\
\hline %%%%%%%%%%%%%%%%%%%%%%%%%%%%%%%%%%%%%%%%%%%%%%%%%%%%%%%%%%%%%%%%%
Bias parameter\tnote{4} \  with respect to operator $O$ & $b_O$ & $\d_h(\vx,\tau) = \sum_O b_O(\tau) [O](\vx,\tau)$ 
\\
$N$-th order \LIMD bias parameter & $b_N$ & $b_N \equiv N!\, b_{\d^N} $~~[\refeq{bNdef}] \\
Lagrangian bias parameter & $b_O^L$ & $\d_h^L(\vq,\tau_0) = \sum_O b_O^L(\tau_0) [O^L](\vq,\tau_0)$ \\
\hline %%%%%%%%%%%%%%%%%%%%%%%%%%%%%%%%%%%%%%%%%%%%%%%%%%%%%%%%%%%%%%%%%
Filter function\tnote{5} \ on scale $R$ & $W_R(x),\,W_R(k)$ & See \refapp{stat:Fourier}. \\
\hline %%%%%%%%%%%%%%%%%%%%%%%%%%%%%%%%%%%%%%%%%%%%%%%%%%%%%%%%%%%%%%%%%
Lagrangian radius of halos & $R(M)$ & $R(M) \equiv (3 M /4\pi \rhob)^{1/3}$~~[\refeq{MofR}]\\
Large smoothing scale & $R_\ell$ & $R_\ell \gg R(M)$ \\
Operator smoothed on large scale & $O_\ell$ & $O_\ell(\vx) \equiv \int d^3 \vy\, O(\vx+\vy) W_{R_\ell}(\vy)$ \\
\hline
\hline
\end{tabular}
{\footnotesize
\begin{tablenotes}
\item[1] In synchronous-comoving gauge, see \refsec{notation}.
\item[2] For halos identified at time $\tau_0$; $\d_h$ satisfies the continuity equation by definition (see \refsec{evol1}).
\item[3] Here we allow for $n \in \mathbb{R}$.
\item[4] This is the physical, renormalized bias, see \refsec{renorm}.
\item[5] Filter functions are normalized such that $\int d^3\vx\,W_R(x) = 1$ and $\lim_{k\to0} W_R(k) = 1$.
\end{tablenotes}
}
\caption{List of symbols and notation used throughout the review.}
\label{tab:notation}
\end{threeparttable}
\end{table*}

\subsection{List of new results}
\label{sec:newres}

Beyond summarizing the current state of the field, this review also includes
numerous new results, mainly related to connections between different aspects
of bias that had previously been overlooked. We briefly list these new results
here, for the benefit of expert readers. 

\begin{itemize}
  % 2
\item A complete enumeration of higher-derivative bias up to second order (\refsec{higherderiv}), and of stochastic contributions to the galaxy density and velocity (\refsec{stoch}).
\item A derivation of the connection between peculiar acceleration of galaxies and the leading velocity bias and higher-derivative bias (\refsec{velbias}), elucidating the reason for the different time dependences of velocity bias obtained in the literature.
  % 3
\item A proof that the Eulerian peak-background split bias parameters are exactly the large-scale renormalized bias parameters (\refsec{PBSrenorm}).
  % 4
\item A previously-overlooked stochastic contribution to the halo-matter cross-power spectrum at next-to-leading (1-loop) order.
\item A rigorous derivation of the connection between bias parameters inferred from moments and ``scatter-plot'' methods with those measured through $n$-point functions (\refsecs{bmom}{bscatter}).  The former measure different bias parameters (which we call \emph{moments biases}), a fact which has not been recognized previously.
  % 5
\item Excursion-set bias parameters obtained using a numerical solution of the exact Langevin equation for a tophat filter (\refsec{exset_summary}). 
  % 6
  % 7
\item Forecasts for constraints on primordial non-Gaussianity from the power spectrum and bispectrum of galaxies for current and upcoming surveys (\refsec{fnl_fisher}), including, for the first time, all relevant bias parameters and stochastic terms.
  % 8
  % 9
  % 10
  % A
  % B
  % C
  % D
\end{itemize}

\begin{figure}[t]
\centering
\includegraphics[width=0.9\textwidth]{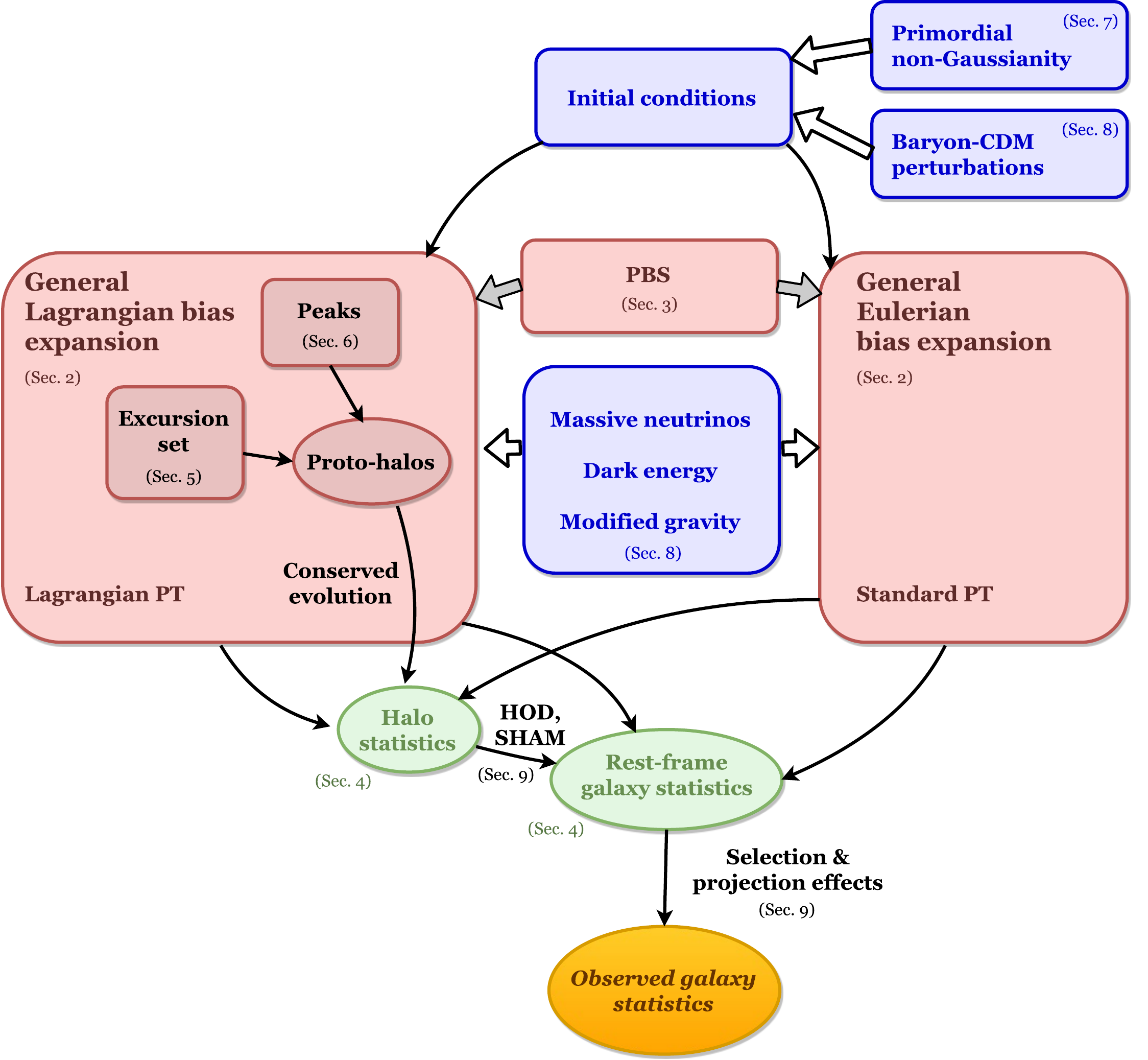}
\caption{
Schematic outline of the theoretical prediction of observed galaxy statistics.  Given the statistics of the initial conditions, perturbative bias expansions predict the rest-frame galaxy density as well as that of dark matter halos.  This expansion can either be done using Lagrangian (left) or Eulerian frames (right).   Crucially, the general bias expansion in either frame is \emph{mathematically equivalent}. The bias expansion is closely connected to the perturbation theory of the matter density field (Lagrangian [LPT] and standard Eulerian [SPT] perturbation theory, respectively).  The peak-background split (PBS) informs the bias expansion by relating the bias parameters to responses of the mean tracer abundance.  
The peak and excursion-set approaches are a special case of the Lagrangian bias expansion, and predict the proto-halo density, which is connected to the statistics of halos at low redshift through conserved evolution.  The statistics of halos in turn can be related to those of galaxies through halo occupation distribution (HOD) and subhalo abundance matching (SHAM) approaches. Finally, the connection between rest-frame and observed galaxy statistics involves selection and projection effects (such as redshift-space distortions).  
Cosmological physics enters in the initial conditions through primordial non-Gaussianity and isocurvature perturbations between baryons and CDM.  It also affects the evolution of structure, and consequently the bias expansion, through the effects of massive neutrinos, dark energy, and modified gravity.  
\label{fig:flowchart}}
\end{figure}

\clearpage
\subsection{Guide for the reader}
\label{sec:outline}

In the following, we describe the outline of the review.  We recommend
that readers begin with \refsec{evolution}, 
and continue to \refsec{PBS} and \ref{sec:measurements}.  \refsec{exset} and following
are, for the most part, independent of each other and can be read in
arbitrary order.  The connection between the different topics is illustrated in the flowchart, \reffig{flowchart} on the previous page.  
Below is a brief outline of the contents of each section:

\begin{itemize}[leftmargin=48pt,itemsep=0pt]
\item[\refsection{evolution}:] This section provides a pedagogical introduction to the general perturbative bias expansion.  We begin with a simple example, bias that is local in the matter density (\LIMD) in Lagrangian space, and then turn to a more realistic study of the gravitational evolution of
proto-halos---i.e. the Lagrangian patches which collapse to form virialized halos.  These examples set the stage for our derivation of the general perturbative bias expansion.
\item[\refsection{PBS}:] This section presents physical arguments, which have become known under the term \emph{peak-background split}, that can be used to derive the actual
values of the bias parameters.  We begin with general physical arguments valid for
any tracer, and specialize to dark matter halos of a given mass afterwards.
\item[\refsection{measurements}:] We present various methods of measuring bias parameters, for example
auto-correlations and cross-correlations with matter, and ``scatter plot''
methods.  In each case, we show rigorously which bias parameters are measured and derive the leading theoretical uncertainties.  
While we do not make any assumptions about the tracers considered, we do not include redshift-space distortions and other observational complications here (see \refsection{observations}), and consequently focus on the application to simulations.  
We also review measurements of the bias parameters of halos and galaxies, including assembly bias. 
\item[\refsection{exset}:] This section presents the excursion-set approach to calculating the abundance and clustering of dark matter halos.  We discuss in detail the so-called cloud-in-cloud problem, choice of filter, and barrier shape, as well as assembly bias.  
\item[\refsection{peaks}:] The second physical model of halos, peaks in the Lagrangian density field, is described in this section.  We also discuss the recent framework for merging excursion sets and peaks (ESP).  
\item[\refsection{NG}:] While \refsecs{evolution}{peaks} assume Gaussian initial conditions, which is an excellent first-order approximation, we discuss in detail the impact of non-Gaussian initial conditions on galaxy bias here.  We derive both the general bias expansion (extension of \refsec{evolution}) and the prediction in the excursion-set and peak approaches (extensions of \refsecs{exset}{peaks}).  We also review measurements on simulations with non-Gaussian initial conditions, and present idealized forecasted constraints on primordial non-Gaussianity from galaxy surveys.
\item[\refsection{beyondCDM}:] In this section, we relax the assumption of adiabatic perturbations in a single pressureless (CDM+baryon) fluid made in the previous sections.  That is, we consider the impact of massive neutrinos and relative (isocurvature) perturbations between baryons and CDM induced by pre-recombination plasma oscillations, as well as Compton drag after reionization.  Finally, we consider the impact  on bias of dark energy perturbations and modifications to General Relativity.
\item[\refsection{observations}:] Strictly speaking, galaxy bias provides a relation between the galaxy (or halo) density and local gravitational observables in the 
galaxy rest frame.  This section describes the selection and projection effects that enter when relating the rest-frame galaxy statistics to observations;  this includes redshift-space distortions as well as so-called relativistic and light-cone effects.  We also briefly describe empirical models connecting halos and galaxies.  
\end{itemize}
We conclude with an outlook in \refsec{summary}.  The appendices contain:
\begin{itemize}[leftmargin=66pt,itemsep=0pt]
\item[\refapp{stat}:] an overview of statistical field theory.
\item[\refapp{SPT}:] an introduction to perturbation theory in large-scale structure.
\item[\refapp{biastrans}:] relations between different conventions for the bias parameters.
\item[\refapp{halofinder}:] a brief overview of halo finding algorithms.
\end{itemize}

%% file: evolution.tex
\clearpage
\section{From local-in-matter-density bias to the general perturbative bias expansion}
\label{sec:evolution}

\secttoc

This section provides a detailed introduction into the general 
perturbative description of galaxy bias.  The final result of this section
is fully general, and applies to any large-scale structure tracer.  
In particular, although bias has been studied 
extensively for dark matter halos identified in N-body simulations,
the general bias expansion is \emph{not restricted to halos}. 
Bias is a complex problem, which goes significantly beyond the simple well-known \LIMD
relation $\d_g = b_1 \d + (b_2/2) \d^2 + \cdots$.  For this reason, this section provides a detailed,
step-by-step treatment of the problem.  Readers mostly
interested in a summary of the relevant equations will find precisely that
in \refsec{evol:summary}.  

The ultimate goal of bias is to describe the observed
statistics of galaxies, such as the galaxy two-point correlation function 
$\xi_g(r)$, 
over a certain range of scales, in terms of a finite number of terms
(various correlation functions of matter and space-time perturbations)
and associated bias parameters.  These bias parameters can be understood
as coefficients of operators $O(\vx,\tau)$ in an expansion of the galaxy 
number density perturbation of the general form
\be
\d_g(\vx,\tau) = \sum_O b_O(\tau) O(\vx,\tau)\,.
\label{eq:biasrel1}
\ee
Once certain physical assumptions about the background
cosmology and the nature of the initial conditions are made, galaxy statistics 
then contain sufficient information to constrain parameters of the 
cosmological model even after the free bias parameters have been marginalized
over.  The goal of the general perturbative bias expansion is to determine
which operators have to be included in the sum of \refeq{biasrel1} in order
to describe galaxy clustering down to a certain minimum scale.  The relative
importance of operators can be ranked by their order in cosmological
perturbation theory (see \refapp{SPT} for a brief overview, and 
\cite{bernardeau/etal:2001} for a comprehensive review).  On scales where
perturbation theory is valid, higher-order terms are successively 
suppressed so that the expansion converges (see \reffig{xi_thr} for an
illustration in the context of a simple toy model);  as stated in
\refsec{intro}, in this review we restrict ourselves to these scales throughout.  
Thus, \emph{the general 
perturbative bias expansion consists of an enumeration of all operators
that are relevant at a given order in perturbation theory} (and given
order in spatial derivatives, as we will see).    

The derivation of this general \emph{local} bias expansion, which contains
the operators that are relevant in the large-scale limit, 
is the topic of \refsecs{localbias}{general}.  
We proceed in a pedagogical fashion, beginning with the simplest example, namely a 
special case of local-in-matter density (\LIMD) bias in Lagrangian space,
motivated by the spherical collapse approximation to
halo formation (\refsec{localbias}).  
We then turn to a more realistic study of the gravitational evolution of
conserved tracers, using two complementary
approaches, in \refsecs{dynamics}{evol2}.  These yield the
general local bias expansion up to cubic order, and set the stage for our
derivation of the general bias expansion at all orders, in \refsec{general}.  
Readers interested only in the full bias expansion can jump to
\refsec{general} directly.  

At this point, we stress again that
we define a \emph{local operator} to mean any leading local gravitational observable,
which includes the matter density and tidal field, as well as further
operators we encounter for the first time in \refsec{evol2}.  On the other
hand, we refer to the restricted expansion which only contains powers of the
matter density perturbation $\d$ as \emph{local-in-matter density, or \LIMD}
bias, a case frequently known in the previous literature under the name
``local bias.''  Note that the \LIMD expansions in Eulerian and Lagrangian frames are not equivalent (\refsec{evol1}).

In addition to the perturbative order, any physical biased tracer 
introduces a spatial scale which controls the importance of so-called
\emph{higher-derivative operators} in the expansion \refeq{biasrel1}.  
This spatial scale, which we will denote as $R_*$ in 
the following, quantifies the \emph{size of the spatial region involved in the process of galaxy formation.}  
That is, the abundance of galaxies depends on the detailed matter distribution
(as well as the other local gravitational observables such as tidal fields) within a region of size $R_*$.  This leads
to an expansion in spatial derivatives, i.e. powers of $R_* \partial/\partial x^i$.  Such terms are thus known as higher-derivative operators, and
are the topic of \refsec{higherderiv}.  These higher-derivative operators
also take into account all non-gravitational physics influencing the galaxy
formation process, such as gas heating and cooling as well as radiative
and kinetic feedback processes.  In this context, we also discuss
\emph{bias of galaxy velocities} (\refsec{velbias});  the latter affect
the observed galaxy density field through redshift-space distortions 
(\refsec{RSD}) and are thus a key ingredient in the modeling of the 
observed galaxy statistics.

There is one more ingredient we need to consider.  The relation between the galaxy density field and the operators $O$ written in \refeq{biasrel1} 
is deterministic.  In reality, whether a galaxy forms at a given location
depends on the initial conditions on very small scales, whose random
phases are not determined by the large-scale perturbations included in
the bias expansion.  This randomness, or \emph{stochasticity}, has
to be included separately in the bias expansion, leading to
\be
\d_g(\vx,\tau) = \mbox{[\refeq{biasrel1}]} + \eps(\vx,\tau) + \sum_O \eps_O(\vx,\tau) O(\vx,\tau)\,,
\ee
where the fields $\eps,\,\eps_O$ are uncorrelated with the large-scale
perturbations described by the operators $O$
and uncorrelated among themselves on large scales.  Their contribution
to galaxy statistics on large scales can then again be described by a finite set of
parameters.  These terms are discussed in \refsec{stoch}.  

Finally, the last two sections of this section deal with the 
embedding of the general bias expansion within relativistic 
perturbation theory (\refsec{GR}) and renormalization
(\refsec{renorm}).   Specifically, \refsec{renorm} discusses the rigorous
machinery beneath perturbative galaxy bias, and provides
the mathematical proof of the completeness of the bias expansion 
argued on physical grounds in \refsec{general}.  While essential from a 
theoretical perspective, it can be skipped by readers mostly interested in 
the observational and measurement aspects of bias.  The section concludes
with the above mentioned summary in \refsec{evol:summary}.  

\subsection{A toy model: \LIMD in Lagrangian space}
\label{sec:localbias}

\begin{figure}[t]
\centering
\includegraphics[trim=0cm 9cm 0cm 6cm,clip,width=0.5\textwidth]{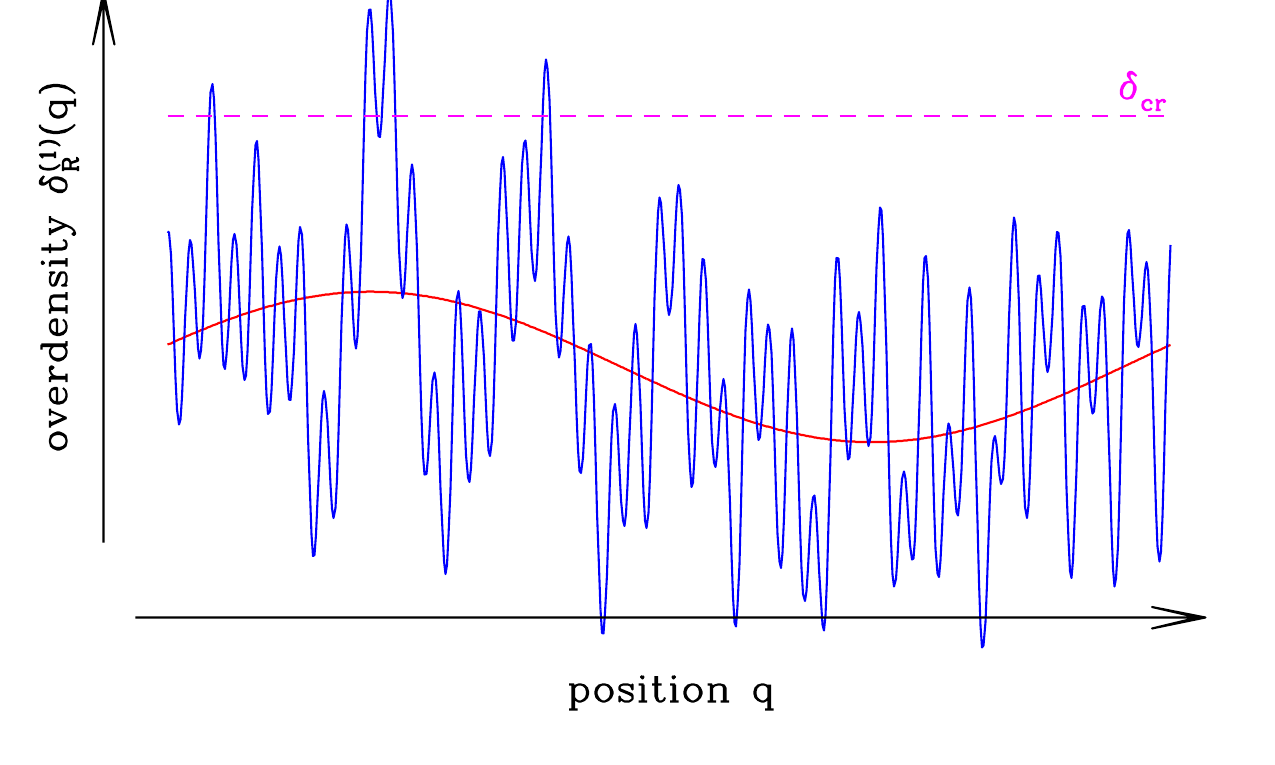}
\caption{
Illustration of the toy model of \refsec{localbias}.  The solid blue line shows the smoothed density field $\d_R^{(1)}(\vq)$, while the red line
indicates a long-wavelength perturbation.  The dashed horizontal line
marks the threshold overdensity $\dc$.
\label{fig:pbsplit}}
\end{figure}

Let us begin our discussion of galaxy bias with a simplified example.  
We assume that dark matter halos, within which observed galaxies reside,
simply correspond to overdense regions (above a threshold) in 
Lagrangian space, that is, in the initial matter density
field extrapolated to a desired reference time using the linear growth.  
We denote this field, often referred to as linear density field,
by $\d^{(1)}$, and assume it to be Gaussian 
(see \refsec{NG} for the non-Gaussian case).  
This toy model was first studied quantitatively in \cite{kaiser:1984}.  
In order to trace halos identified at low redshift back to the initial conditions, we imagine
following the trajectory of the constituent particles 
of a given halo back to the initial time $\tau=0$.  The region occupied by
these particles is referred to as the \emph{proto-halo}.  
Since the initial density field is 
arbitrarily close to uniform, the proto-halo volume of a halo of mass $M$ is $M/\rhob$, from which we
can define the \emph{Lagrangian radius} of the halo via
$R(M) = (3M/4\pi\rhob)^{1/3}$.  Hence, in order to define ``regions above threshold''
which eventually collapse to form halos, we filter the initial (linear) density
field on the scale $R(M)$, denoting this as $\d^{(1)}_R$ (\reffig{pbsplit}).  The shape of the filter
is not relevant for this discussion; we list popular filters in 
\refapp{stat:Fourier}.  
The comoving Lagrangian number density of proto-halos is then defined as
\be
n_{\rm thr}^L(\vq) \equiv \Theta_H(\d_R^{(1)}(\vq) - \dc)\,,
\label{eq:nhthr}
\ee
where $\dc$ is a fixed density threshold and $\Theta_H$ denotes the
Heaviside step function.
Note that the ``density'' $n_{\rm thr}^{L}$ defined here corresponds, up to normalizing factors which we neglect here, to the mass-weighted cumulative number density of halos above mass $M$ (see for example \cite{long} and \refsec{exset}).  
In the sketch \reffig{pbsplit}, the proto-halo number density is unity
whenever the blue solid line crosses the threshold indicated by the 
horizontal line.  As this toy model describes ``thresholded regions,'' 
it is often referred to as ``thresholding.''

The statistics of the Gaussian field $\d_R^{(1)}$ are 
completely  described by the two-point correlation function (\refapp{stat})
\be
\xi_{{\rm L},R}(r) = \< \d_R^{(1)}(\vq) \d_R^{(1)}(\vq+\v{r})\>\,,
\ee
where $\xi_{{\rm L},R}(r)$ is the filtered version of the linear matter correlation function $\xi_{\rm L}(r) = \lim_{R\to 0} \xi_{{\rm L},R}(r)$ [the Fourier transform of the linear power spectrum $\Plin(k)$, \refeq{xilin}], and $\xi_{{\rm L},R}(0) = \s^2(R)$ is the variance of the filtered density field.  
The mean ``number density'' of proto-halos is obtained by taking the expectation value
of \refeq{nhthr},
\be
\< n_{\rm thr}^L(\vq) \>
= \frac{1}{\sqrt{2\pi}} \int_{\nu_c}^\infty e^{-\nu^2/2} d\nu
= \frac12 \erfc[\nu_c/\sqrt{2}] \equiv p_1\,, \quad 
\nu_c \equiv \frac{\dc}{\s(R)}\,.
\label{eq:p1thr}
\ee
This shows that, for a high threshold $\nu_c > 1$, proto-halos become exponentially 
rare.  Note that any normalizing factors that we neglect here do not affect the calculation of the bias, which we turn to next.

The Lagrangian two-point function of equal-mass proto-halos at separation $r$ is given by the ratio
of the pair probability $p_2(\vq,\vq+\v{r})$ of finding two proto-halos at positions $\vq$ and $\vq+\v{r}$, divided
by the 1-point probability squared $(p_1)^2$ \cite{peebles:1980,kaiser:1984}:
\ba
1 + \xi_h^L(r) = \frac{p_2(\vq,\vq+\v{r})}{p_1^2} = \sqrt{\frac2\pi} \left[\erfc(\nu_c/\sqrt{2})\right]^{-2} \int_{\nu_c}^\infty e^{-\nu^2/2} 
\erfc\left[ \frac{\nu_c - \nu \hat\xi(r)}{\sqrt{2 \{1 - \hat\xi^2(r)\}}}
\right] d\nu\,.
\label{eq:xihthr1}
\ea
where $\hat\xi(r) \equiv \xi_{{\rm L},R}(r)/\s^2(R)$.  This relation follows
directly from integrating over the Gaussian likelihood of the
density field $\d_R^{(1)}$.  
Again, $\xi_h^L(r)$ is the two-point
function of proto-halos in the initial conditions, as emphasized by the superscript
$L$, extrapolated to $z=0$ using
linear theory.  If $\xi_{{\rm L},R}(r)$ is sufficiently small,
we can expand \refeq{xihthr1} in a series,
\be
\xi_h^L(r) = \sum_{N=1}^\infty \frac1{N!} (b_N^L)^2 \left[\xi_{{\rm L},R}(r)\right]^N\,,
\label{eq:xihthr}
\ee
where the $r$-independent coefficients $b_N^L$, the \emph{bias parameters}, 
are given by \cite{kaiser:1984,szalay:1988,PBSpaper}
\be
b_N^L = \sqrt{\frac2\pi} \left[
\erfc\left(\frac{\nu_c}{\sqrt{2}}\right)\right]^{-1}
\frac{e^{-\nu_c^2/2}}{\sigma^N(R)}
H_{N-1}(\nu_c)
\stackrel{\nu_c\gg 1}{=} \frac{\nu_c^N}{\s^N(R)} + \O\left(\nu_c^{N-1}\right)\,.
\label{eq:bPBSthr}
\ee
The superscript $L$ on the bias parameters indicates that they refer to
the Lagrangian density field.  Note that $b_1^L$ is positive definite, which is not true for general tracers.  Moreover, in the limit of low mass $R\to 0$, $\sigma(R)\to \infty$ and $b_1^L$ 
approaches zero; this is because the simple thresholding procedure 
\refeq{nhthr} does not correctly describe low-mass objects
due to the ``cloud-in-cloud'' problem.  
We will return to this in \refsec{exset}.

\begin{figure}[t]
\centering
\includegraphics[width=\textwidth]{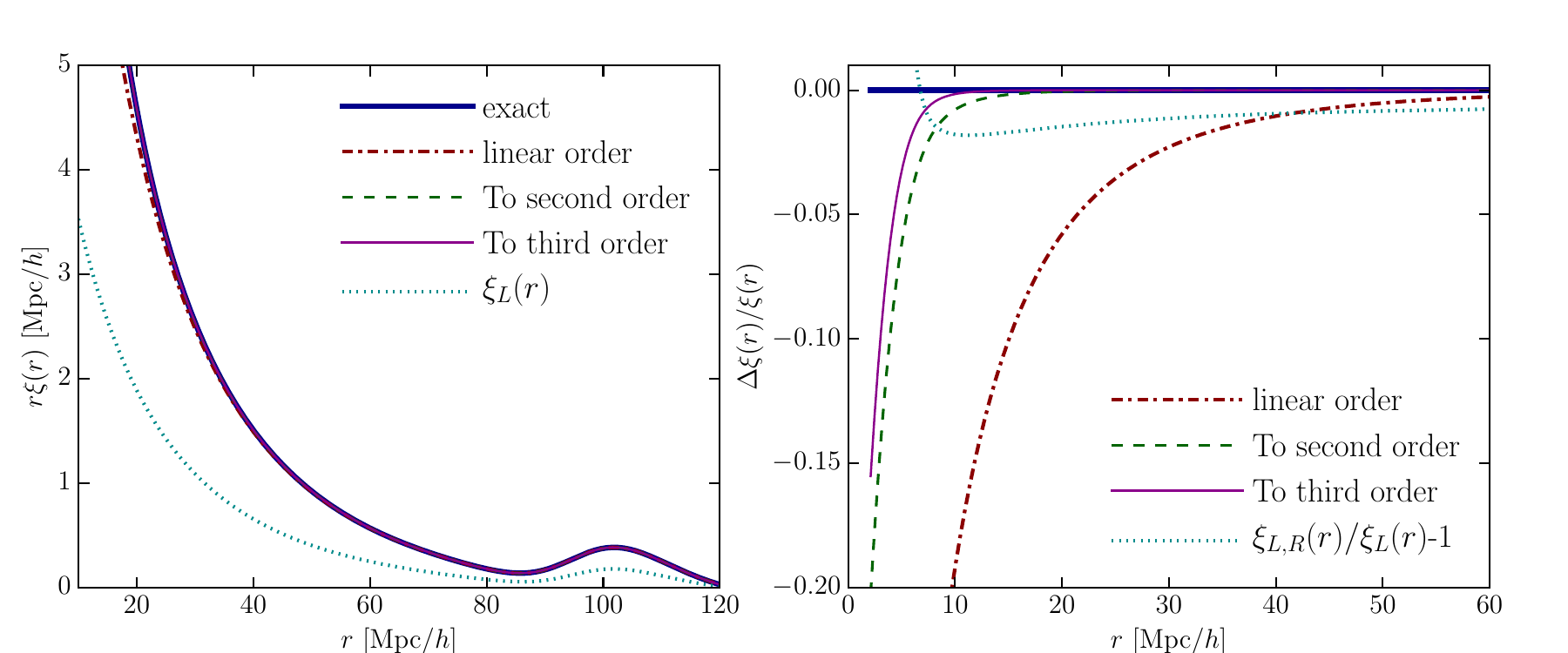}
\caption{
Correlation function in Lagrangian space of thresholded regions in the initial
density field extrapolated to $z=0$.  The smoothing scale $R = 4.21 \Mpch$ (mass scale $2.5\cdot 10^{13}\Msunh$) is chosen to correspond to $b_1^L = 1.5$ via \refeq{bPBSthr}.  
\textit{Left panel:} We show the exact result \refeq{xihthr1}, multiplied by
$r$ to better show the large-scale behavior, as well
as the series expansion \refeq{xihthr} truncated at different orders.  
For comparison, the cyan dotted line shows the linear, unfiltered matter 
correlation function.  The bump at $r \approx 100 \Mpch$ is the BAO feature.
\textit{Right panel:}  Relative deviation of the truncated series expansion
from the exact result.  The dotted line shows the relative deviation of the
unfiltered linear contribution $(b_1^L)^2 \xi_{\rm L}$ from the filtered
contribution $(b_1^L)^2 \xi_{{\rm L},R}$.
\label{fig:xi_thr}}
\end{figure}

Since in the real Universe, matter density fluctuations are small on large scales, we have $\xi_{{\rm L}}(r) \to 0$ as $r \to \infty$.  This means that, as long as we are interested in the proto-halo correlation
function on large scales, it is sufficient to keep only the first few terms
in the expansion \refeq{xihthr}.  In particular, the leading term is
\be
\xi_h^L(r) = (b_1^L)^2 \xi_{{\rm L},R}(r)\,,
\ee
which is what one obtains from the simple bias relation \refeq{bias1}.  
Thus, the proto-halo correlation function is directly proportional to the matter correlation function on large scales, i.e. both have the same $r$-dependence.  
However, $\xi_h^L$ is enhanced by the factor $(b_1^L)^2$, which becomes large for rare, high-mass proto-halos [$\nu_c \gg 1$; \refeq{bPBSthr}].  
The picture in \reffig{pbsplit} delivers an intuitive explanation for why
this happens.  When adding a large-scale density perturbation (red solid line)
to the matter density field, the abundance of rare regions above threshold
responds much more sensitively than the matter density itself, which, as we
will see in \refsec{PBS}, is an alternative, exactly equivalent definition
of bias.  This larger response in turn leads to enhanced clustering on large scales
\cite{kaiser:1984,bardeen/etal:1986}. The latter effect, shown in \reffig{xi_thr}, was the original motivation of \cite{kaiser:1984}, who showed that the correlation function of rare objects such as massive galaxy clusters is enhanced relative to that of the underlying matter on \emph{all} scales.  

\reffig{xi_thr} shows that the series expansion \refeq{xihthr} converges rapidly to the exact result on scales $\gtrsim 10 \Mpch$.  
This approximately corresponds to the scale where
the matter correlation function becomes of order 1.  Thus,
separations $r \gtrsim 10 \Mpch$ are amenable to a perturbative description
in the context of this toy model.  
On the other hand, on smaller scales,
higher-order terms are no longer smaller than lower-order terms.  This means 
that a general perturbative expansion is not guaranteed to converge to the
correct result.  While for the toy model considered here we have an exact
result, this is not the case for real galaxies, so that the restriction
to perturbative scales is the only way to guarantee a theoretical
error that is under rigorous control.  Note that the scale where
the expansion \refeq{xihthr} breaks down is related to the \emph{nonlinear scale} $\Rnl$ 
(see \refapp{SPT}; $\Rnl \sim 10-20 \Mpch$ at redshift zero),  
at which the perturbative description of the nonlinear matter density field 
itself breaks down.  
This scale becomes smaller at higher redshifts;  hence, the range of
scales accessible to perturbation theory is larger at high redshifts
\cite{jeong/komatsu:2006}.  

The expression for the proto-halo two-point correlation function, \refeq{xihthr}, 
is not specific to the ansatz \refeq{nhthr} we started from.  
Indeed, if we write $\d_h(\vq) = n_{\rm thr}^L(\vq)/\<n_{\rm thr}^L\> - 1$ as a formal series
expansion in $\d_R^{(1)}$,
\be
\d_h^L(\vq) = b_1^L \d_R^{(1)}(\vq) + \frac12 b_2^L \left([\d_R^{(1)}(\vq)]^2 - \s^2(R) \right) + \cdots\,,
\label{eq:dhlocal}
\ee
then \refeq{xihthr} is obtained directly when discarding all terms that involve
zero-lag correlators (terms containing factors of 
$\< [\d_R^{(1)}(\vq)]^n\>$; see \refapp{stat}).  One can similarly
derive all higher $N$-point functions and cumulants of $\d_h(\vq)$ \cite{bernardeau:1996,PBSpaper}.  
Historically, \refeq{dhlocal} is known as the \emph{local bias expansion} \cite{fry/gaztanaga:1993}, since $\d_h(\vq)$ is
a local function of the filtered matter density field.  As we have discussed
in the beginning of \refsec{evolution}, \refeq{dhlocal} only contains a subset
of the local bias terms according to the definition used in this review.  
To be specific, we refer to \refeq{dhlocal} as \emph{local-in-matter-density (\LIMD) bias}.  Even though not complete, as we will see, the \LIMD expansion is sufficiently general
to describe an ansatz of the type \refeq{nhthr} for any function $n_{\rm thr}^L(\vq) = F\left(\d_R^{(1)}[\vq]\right)$ \cite{fry:1986,coles:1993}.  Furthermore, while \refeq{xihthr} is a prediction specific to Gaussian density fields, the \LIMD ansatz allows statements about the statistics of
$n_{\rm thr}^L$ to be made even for general density statistics \cite{coles:1993}, and
for density fields whose $N$-point functions obey hierarchical scaling laws
\cite{scherrer/weinberg:1998,coles/melott/munshi:1999}.  

This fact, combined with the fact that the series expansion \refeq{xihthr}
converges to the true result on large scales (\reffig{xi_thr}), shows the power of the
perturbative bias expansion:  it can capture a very general class of 
tracers (in the case of \refeq{dhlocal}, a tracer whose abundance is described by an arbitrary local function of the initial matter density field) via a small
set of free parameters, while guaranteed to converge to the correct result on sufficiently
large scales.  

Moreover, the proto-halo number density does not have to be an \emph{exactly} local
function of $\d_R^{(1)}$ in order for the \LIMD expansion \refeq{dhlocal} to be effective, as already noticed by \cite{fry/gaztanaga:1993}:  we only need to require that the nonlocality is restricted to scales much smaller than the scale $r$ at which the correlation function is calculated. 
This is illustrated in the right panel
of \reffig{xi_thr}, which shows the ratio of the correlation function of the filtered matter density field, $\xi_{{\rm L},R}(r)$, to that of the unfiltered field $\xi_{\rm L}(r)$.  Clearly, the smoothed density field is nonlocally related to the un-smoothed density field, however this manifests itself in the correlation function only on scales smaller than a few times the 
smoothing radius.
This exemplifies that, as shown in \cite{coles:1993,
coles/melott/munshi:1999,narayanan/berlind/weinberg:2000}, in order to change 
the shape of the galaxy correlation function $\xi_h^L(r)$ relative to 
the matter correlation function $\xi_{{\rm L}}(r)$ on \emph{large} scales,
say on hundreds of Mpc, 
galaxy formation has to be highly nonlocal, i.e. nonlocal on a scale of order a hundred Mpc.  
The only exception to this rule, which however does
not invalidate the point, is that there can be an effect on large scales 
if there are sharp features in the matter correlation function, for example
the BAO feature at $r \approx 100 \Mpch$.  As we will see in \refsec{higherderiv},
this effect can be taken into account to high accuracy with only one additional
bias parameter.

Notice that we did not include a zeroth order bias $b_0^L$, i.e. a constant offset, 
in the bias relation \refeq{dhlocal} because of the requirement $\<\d_h \>=0$.  This is only true
for a \emph{deterministic bias relation} as assumed in \refeq{dhlocal}. 
In reality, both the presence of random small-scale fluctuations 
(which affect the abundance of halos and galaxies, but are not correlated with the 
long-wavelength fluctuations)
and the fact that halos are a discrete sample of the underlying density field
demand that we allow for \emph{stochastic fields} in the bias expansion.  
At lowest order, this adds a stochastic variable $\eps(\vq)$ with vanishing mean to \refeq{dhlocal}.
We begin including stochastic contributions from the next section.  
  
So far, we have provided a description of proto-halo statistics in Lagrangian space.  
Naturally, we need to ask how to translate these results to the 
observationally relevant statistics of the evolved halo field at lower 
redshifts (Eulerian space). 
Let us again consider a simplified setup, namely a large spherical 
region of radius $R_\ell$; specifically, following \cite{mo/white:1996}, 
we assume that $R_\ell$ is much 
larger than the typical separation between the halos considered. 
We allow this region to have an over- or underdensity $\d_\ell(\tau)$, but assume
the Universe to be unperturbed otherwise on large scales.  Then, we
expand the fractional overdensity $\d_{h,\ell}$ of halos within this region
with respect to the global mean at a given redshift as
\begin{equation}
\delta_{h,\ell}(\tau) = \frac{\bar{N}_\ell}{{\avnh} V_\ell(\tau)}-1 =
b_1^E(\tau) \delta_\ell(\tau) + \frac12 b_2^E(\tau) 
\delta_\ell^2 + \cdots\;,
\label{eq:dhsc}
\end{equation}
where $\bar{N}_\ell$ is the number of virialized halos in the 
region, and ${\avnh}$ is the global mean abundance 
(recall that all densities are comoving and thus unaffected by the 
dilution due to the expansion of the Universe), 
while $V_\ell(\tau)$ is the comoving volume of the spherical region.  
Analogously to \refeq{dhlocal}, and following \cite{mo/white:1996}, \refeq{dhsc} defines the Eulerian bias parameters $b_N^E$ on the right-hand side.  
Note that these bias parameters formally become independent of the size of the region only in the limit $R_\ell\to\infty$. For simplicity however, we shall omit this dependence here. This subtlety will be addressed in \refsec{bmom}.

We now want to relate the Eulerian bias parameters to the Lagrangian 
bias parameters $b_N^L$, by using the number conservation of halos.
Since by definition we refer to proto-halos in the initial conditions which are the exact progenitors of halos identified at time $\tau$, their number is conserved.  
The evolution of the matter overdensity $\d_\ell$ or equivalently $V_\ell(\tau)$ 
on the other hand follows 
spherical collapse \citep{gunn/gott:1972} (see \refsec{sph_collapse}), which
is uniquely determined from the initial overdensity $\d_\ell^{(1)}$ of the region.  
The spherical collapse evolution is valid up to shell crossing, i.e. as
long as $R_\ell > 0$; up to that point, the mass in 
each shell is conserved.  This implies that the Lagrangian
density perturbation of halos is given by $1+\d_{h,\ell}^L = \bar{N}_\ell/({\avnh} V_\ell^L)$ where $V_\ell^L$ is the Lagrangian volume of the region, which
is related to the Eulerian volume $V_\ell$ by 
\be
V_\ell^L = (1+\dell) V_\ell\,.
\label{eq:Vell}
\ee
Therefore, we obtain the following relation between $\d_{h,\ell}(\tau)$ and $\d_{h,\ell}^L$:
\begin{equation}
\label{eq:scbias}
1+\delta_{h,\ell}(\tau) =\bigl[1+\delta_{h,\ell}^L\bigr] \left[1+\delta_\ell(\tau)\right]
= 1 + \bigl(1+ b_1^L\bigr)\delta_\ell + \O(\d_\ell^2)\,, 
\end{equation}
where on the right-hand side we have expanded the result in powers of $\d_\ell$
using \refeq{dhlocal}, filtered on the scale $R_\ell$.  Comparison with \refeq{dhsc} then shows that $b_1^E=1+b_1^L$.  Halos which have vanishing bias ($b_1^L=0$) in the initial conditions, so that $\delta_h^L=0$ and they are uniformly distributed at the initial epoch, will hence remain unbiased relative to the matter at any later time $\tau$.\footnote{Here, we have implicitly assumed that proto-halos comove with the matter fluid.  We will discuss this in more detail in \refsec{dynamics} and \refsec{velbias}.} 

The calculation of the Eulerian bias parameters can be extended to all orders within this picture, since the mapping between $\delta_\ell(\tau)$ and $\delta_\ell^{(1)}$ can be calculated for any given cosmology by integrating the spherical collapse equations.  Specifically, one 
obtains the series expansion \cite{bernardeau:1992,mo/jing/white:1997,fosalba/gaztanaga:1998}
\ba
\label{eq:scseries}
\delta_\ell(\tau) =\:& \sum_{k=1}^\infty a_k \left[\delta_\ell^{(1)}(\tau)\right]^k \,;
&  a_1 =\:& 1\,, \quad a_2 = \frac{17}{21}\,, \quad a_3 = \frac{341}{567}\,,
\quad a_4 = \frac{55805}{130977}\,, \quad\cdots \vs
\delta_\ell^{(1)}(\tau) =\:& \sum_{k=1}^\infty a^{\rm I}_k \left[\delta_\ell(\tau)\right]^k \,;
 & a^{\rm I}_1 =\:& 1\,, \quad a^{\rm I}_2 = -\frac{17}{21}\,, \quad a^{\rm I}_3 = \frac{2815}{3969}\,,
\quad a^{\rm I}_4 = -\frac{590725}{916839}\,, \quad\cdots \,. 
\ea
Here, the coefficients are only strictly valid for Einstein-de Sitter (EdS), where $\d_\ell^{(1)}(\tau) \propto a = 1/(1+z)$.  However,
they are also highly accurate for other cosmologies \cite{wagner/etal:2015}, as long as we remain
sufficiently far from the collapse of the spherical region of size $R_\ell$.  
As pointed out in \cite{matsubara:2011,wagner/etal:2015}, the
coefficients $a^{\rm I}_3, a^{\rm I}_4$ in the inverse relation were given incorrectly
in \cite{mo/jing/white:1997}.   
Substitution of \refeq{scseries} into \refeq{scbias} then yields a unique 
prediction for the \emph{evolved} halo correlation function,
\be
\xi_h(r,\tau) = \sum_{N=1}^\infty \frac1{N!} [b_N^E(\tau)]^2 \left[\xi_{R}(r,\tau)\right]^N\,,
\label{eq:xihthrE}
\ee
with Eulerian bias parameters $b_N^E$ given by
\begin{align}
\label{eq:scbiasEL}
b_1^E(\tau) =\:& 1 + b_1^L(\tau) \\
b_2^E(\tau) =\:& 2\left(1 + a^{\rm I}_2\right) b^L_1(\tau) + b^L_2(\tau) \vs
b_3^E(\tau) =\:& 6\left(a^{\rm I}_2 + a^{\rm I}_3\right) b^L_1(\tau) + 3\left(1 +2 a^{\rm I}_2\right)b^L_2(\tau) 
+ b^L_3(\tau) \,.
\nonumber
\end{align}
These important relations were first derived by \cite{mo/white:1996,mo/jing/white:1997}.

We have thus achieved our goal of deriving the bias parameters describing 
statistics of halos in the evolved density field given a Lagrangian \LIMD
bias expansion \refeq{dhlocal}.  
Clearly, in the context of a spherically symmetric density perturbation, a deterministic \LIMD bias relation in 
the initial conditions is transformed into a similar relation in the evolved distribution (Eulerian \LIMD).   
We will see in the next section that this is a consequence of the spherical
symmetry assumed in this derivation:  in general, gravitational evolution 
starting from a \LIMD bias expansion generates additional terms in the Eulerian
bias relation which are beyond \LIMD (yet still local).  In essence, we have neglected tidal fields which play a similarly important role as density perturbations.  
The next subsections will look more closely at the gravitational evolution
of proto-halos or -galaxies.

\begin{figure}[t]
\centering
\includegraphics[width=0.6\textwidth]{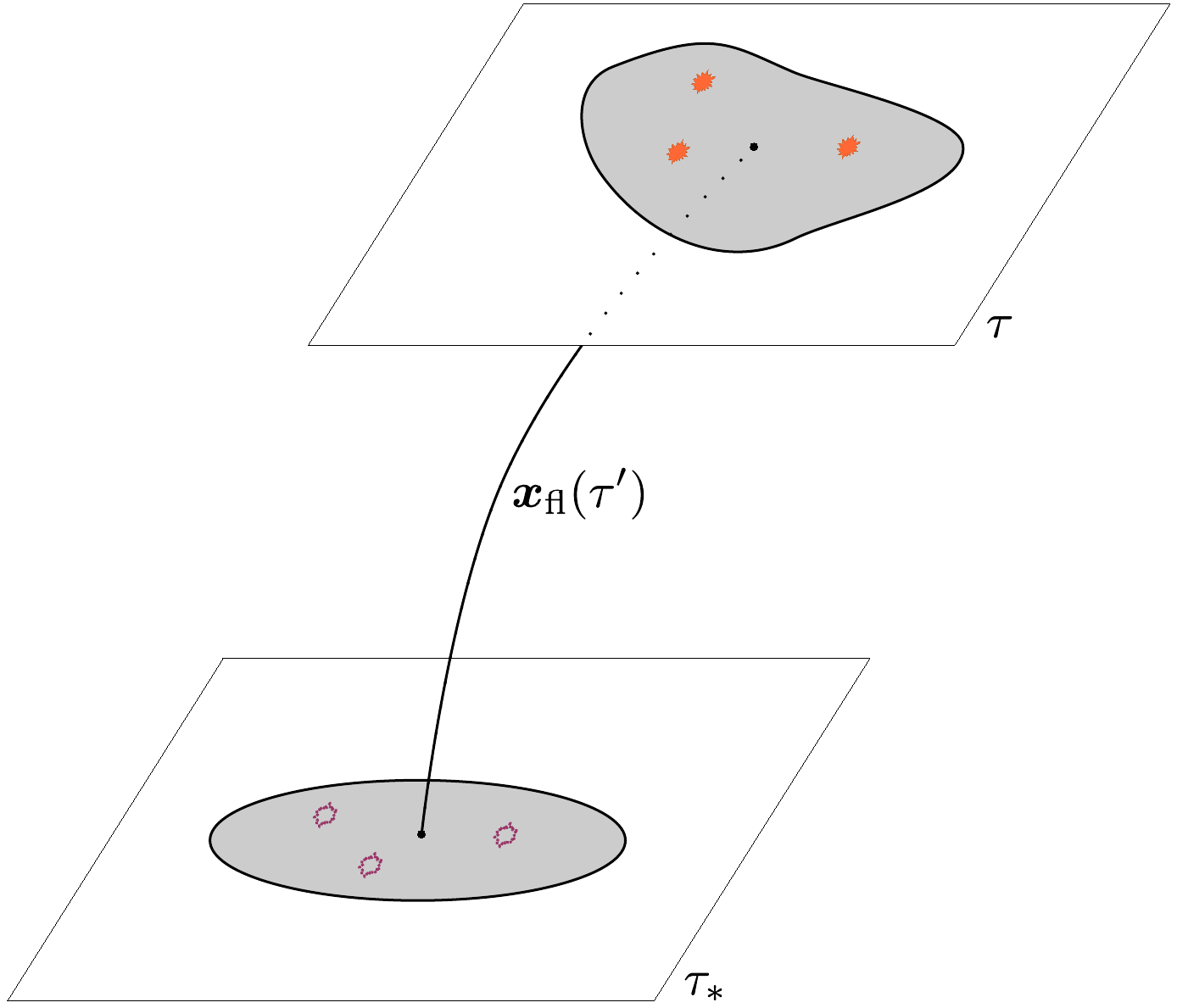}
\caption{Sketch of the setup considered in \refsecs{dynamics}{evol2}.  
Galaxies form instantaneously at $\tau=\tau_*$, where they are described by an 
initial bias relation (lower slice), and are comoving with the matter.  
After this, the evolution is governed by number conservation and the
comoving assumption up until time $\tau$ (upper slice), where 
they are assumed to be observed.   The grey region denotes a Lagrangian
volume encompassing three galaxies which gets deformed by nonlinear
gravitational evolution.  Since galaxies comove with matter, their
density is similarly affected.  
\label{fig:sketch_evol}}
\end{figure}

\subsection{Gravitational evolution: general considerations}
\label{sec:dynamics}

We now investigate in more detail the interplay of gravitational evolution
and bias.  Specifically, we consider the toy example of a sample of galaxies
that formed instantaneously at a fixed time $\tau_*$, that are comoving with matter, and whose number is 
conserved afterwards (\reffig{sketch_evol}).
The discussion here is not specific to dark matter halos, 
and is applicable to the statistics of any conserved tracer of the large-scale 
structure.  Hence, we will use the term ``galaxies'' instead of ``halos''
in the following.  
Given a bias relation for $\d_g^*$ at time $\tau_*$, involving the 
matter density and other locally observable quantities, the goal of this 
section is to 
derive how this bias relation evolves under gravity while conserving the
number of tracers to some later time $\tau > \tau_*$.  
We will then be able to express the bias parameters $b_n$ at time $\tau$ in
terms of the bias parameters $b_n^*$ at the ``formation time'' $\tau_*$, 
generalizing \refeq{scbiasEL} which is specific to a spherical region
and \LIMD bias in Lagrangian space.  

With this ansatz, we essentially describe a galaxy sample undergoing ``passive evolution.''  
However, beyond being a mere toy model, this ansatz can be considered a \emph{Green's function} for galaxies in the sense that a realistic galaxy sample (consisting of a broad range of formation times) can be described as a superposition
of many samples with instantaneous formation at various epochs $\tau_*$.  Thus, if our bias
expansion correctly describes a sample formed at an arbitrary time $\tau_*$
and evolved to time $\tau$, then this bias expansion is sufficiently general
to describe any galaxy sample provided all bias parameters are allowed to be
free.

Similarly to the matter density field itself, there is no closed solution
for the nonlinear evolution of a conserved galaxy density field.  Instead,
we adopt a perturbation theory approach which successfully describes the 
evolution on sufficiently large scales.  Unlike the discussion in the
previous section, we do not assume spherical symmetry;  indeed, the lack
of this symmetry is the reason why we have to add additional local terms to our
bias expansion beyond the \LIMD case, \refeq{dhsc}.  
Moreover, we will focus on scales much
larger than the Lagrangian radius of halos throughout, and hence drop the
subscript $R$ denoting the filtering in the following.  
In \refsec{localbias}, we found this to be accurate empirically, however
this will be justified rigorously in \refsec{higherderiv}.  

The equation of motion governing the evolution of a conserved tracer is 
the continuity equation \citep{nusser/davis:1994,fry:1996,tegmark/peebles:1998,catelan/etal:1998}:
\be
\convD \d_g = -\theta (1+\delta_g)\,,
\label{eq:contg}
\ee
where 
\be
\convD \equiv \frac{\partial}{\partial\tau} + v^i \frac{\partial}{\partial x^i}
\label{eq:DDtau}
\ee
is the convective (or Lagrangian) time derivative, 
$v^i$ is the peculiar velocity of the cosmic matter fluid, and 
$\theta = \partial_{x,i}v^i$ is the velocity divergence.  

In \refeq{contg}, we have implicitly assumed that 
there is \emph{no velocity bias} of the galaxies with respect to matter, 
that is, galaxies comove with the matter fluid. 
A detailed discussion of velocity bias, and the proof
that it is absent on large scales, is the
topic of \refsec{velbias}, but we elaborate briefly here.  
Consider a sufficiently large patch surrounding a given galaxy, say several
times the Lagrangian radius of the galaxy's host halo (e.g. the grey region in \reffig{sketch_evol}).  We then
determine the center-of-mass velocity of this patch.  Any peculiar
velocity of the galaxy with respect to the center-of-mass has to be
due either to the detailed matter distribution within the patch, or
non-gravitational forces such as momentum transfer due to baryonic
feedback processes or radiation pressure.  That is, for a sufficiently
large patch, the galaxy's relative velocity with respect to the matter is only
determined by physics within the patch, which is equivalent to stating
that on large scales galaxies comove with matter.  In the absence
of non-gravitational forces, this is already required by the 
equivalence principle:  test bodies, regardless of their nature---be they a galaxy, black hole, or dark matter particle---fall at the same rate in an external (large-scale) gravitational field.

\begin{table*}[t]
\centering
\begin{threeparttable}[b]
\begin{tabular}{l|l|l}
\hline
\hline
Quantity & Eulerian expression & Lagrangian expression\\
\hline
Lagrangian displacement & \multicolumn{2}{c}{$\v{s}(\vq,\tau)$} \\
Deformation tensor & \multicolumn{2}{c}{$M_{i}^{\  j}(\vq,\tau) \equiv \partial_{q,i} s^j(\vq,\tau)$} \\
\hline
Spatial coordinate & $\vx$ & $\vq$ \\
Spatial derivative & $\partial_{x,i}$ & $\partial_{q,i} = (\d_i^{\  j} + M_i^{\  j}) \partial_{x,j}$ \\
Fluid trajectory & $\xfl(\vq,\tau) \equiv \vq + \v{s}(\vq,\tau)$ & $\vq$ \\
Fluid velocity & $\v{v} = d\xfl/d\tau = \partial_\tau \v{s}(\vq,\tau)$ & 0 \\
Convective derivative & $\convDinline = \partial_\tau + v^i \partial_{x,i}$
& $\convDinline = \partial_\tau$ \\
\hline
Matter overdensity & $\d(\vx,\tau)  = |\v{1}+\v{M}(\vq,\tau)|^{-1}-1$ & $0$ \\[3pt]
Conserved tracer overdensity\tnote{1} &
$\begin{aligned} \d_g(\xfl[\vq,\tau],\tau)  =\:& \left(1+\d(\xfl[\vq,\tau],\tau)\right) \\
& \times \left(1+\d_g^L(\vq)\right) - 1 \end{aligned}$ 
& $\displaystyle \d_g^L(\vq)
= \frac{1 + \d_g[\xfl(\vq,\tau),\tau]}{1 + \d[\xfl(\vq,\tau),\tau]} - 1$
\\[9pt]
\hline
\end{tabular}
{\footnotesize
\begin{tablenotes}
\item[1] This only holds for a tracer whose number is conserved from time $\tau=0$ to time $\tau$ (e.g., proto-halos of halos identified at time $\tau$).
\end{tablenotes}
}
\caption{Summary of relations between Eulerian and Lagrangian space.}
\label{tab:EulLagr}
\end{threeparttable}
\end{table*}

In order to evolve $\d_g$, we also need the perturbative solution of $\d$ and $v^i$.  
In the context of standard perturbation theory (SPT; see \refapp{SPT} for a brief overview), where dark matter 
is treated as a pressureless ideal fluid, the fluid velocity $\v{v}$ is 
curl-free and can thus be written as $v^i = (\partial^i/\lapl) \theta$.  
Then, the density perturbation $\d$ and velocity divergence $\theta$ 
obey the continuity and Euler equations,
\ba
\label{eq:Cont}
\convD \d  =\:& - \theta (1+\delta),  \\
\convD \theta  =\:& - \cH \theta - (\partial_x^i v^j)^2 - \frac32 \Om \cH^2 \d  \,.
\label{eq:Euler}
\ea
There are essentially two equivalent approaches to deriving the mapping from
$\d_g(\tau_*)$ to $\d_g(\tau)$.  First, one can directly integrate
\refeq{contg} along the fluid trajectory, making use of the known
perturbative solutions of the SPT equations.  Second, one can iteratively
solve the set of equations \refeqs{contg}{Euler} to obtain a joint
perturbative solution for $\d_g$ and $\d$.  We will outline
each approach in turn in the following sections, as they illustrate 
different aspects of gravitational evolution and bias.  

Apart from the matter density field $\d$, in the following we will also
encounter the tidal field, which we define through the scaled dimensionless
quantity $K_{ij}$,
\be\label{eq:Del}
K_{ij} \equiv \Del_{ij} \delta
= \frac{2}{3\Om\cH^2}\partial_i\partial_j\Phi - \frac13 \delta_{ij} \d
\,; \quad
\Del_{ij} \equiv \left(\frac{\partial_i\partial_j}{\lapl} - \frac13 \d_{ij} \right)\,,
\ee
where the second equality follows from the Poisson equation [\refeq{Poisson}].  
Equivalently, in Fourier space
\be
K_{ij}(\vk) = \left[\frac{k_i k_j}{k^2} - \frac13 \d_{ij}\right] \d(\vk)\,.
\label{eq:KijF}
\ee
Furthermore, we will denote quantities at $n$-th order in perturbation
theory with a superscript $(n)$.  For example, the second-order density
field is denoted as $\d^{(2)}$.

\subsection{Evolution from the continuity equation}
\label{sec:evol1}

We now proceed to solve the continuity equation \refeq{contg} for our
conserved galaxies as follows \cite{nusser/davis:1994,fry:1996,catelan/etal:1998,chan/scoccimarro/sheth:2012}.  
We first divide \refeq{contg}
and \refeq{Cont} by $1+\d_g$ and $1+\d$ respectively.  This yields
\be
\frac1{1+\d_g}\convD \d_g  = - \theta = \frac1{1+\d}\convD \d\,.
\label{eq:ngcons}
\ee
For convenience, we now write $\d_g$ and $\d$ in terms of Lagrangian coordinates
$\vx\to\vq$, in which case the convective
time derivatives become partial derivatives with respect to $\tau$;  
see \reftab{EulLagr}, where relations between various quantities in the Eulerian and Lagrangian frames are summarized.  
Then, the integral along the fluid trajectory becomes trivial, and
we obtain
\be
\ln [1 + \d_g(\xfl(\tau),\tau)] = \ln [1 + \d(\xfl(\tau),\tau)] 
+ \ln \left[\frac{1 + \d_g(\vx_*, \tau_*)}{1+\d(\vx_*,\tau_*)}\right]
\label{eq:dgcons}
\ee
for $\tau \geq \tau_*$.  We have fixed the integration constant by introducing the galaxy overdensity $\d_g(\vx_*,\tau_*)$ on the formation time slice.  Here, $\xfl(\tau)$ denotes the Eulerian coordinate of the fluid trajectory corresponding to a fixed Lagrangian position $\vq = \xfl(\tau=0)$, 
and $\vx_* \equiv \xfl(\tau_*)$ denotes the position on the formation time slice.  By introducing
the displacement $\v{s}$ through $\xfl(\tau) = \vq + \v{s}(\vq,\tau)$,
we can write the equation of motion governing the fluid trajectory as 
\be
\left(\frac{\partial^2}{\partial\tau^2} + \cH \frac{\partial}{\partial\tau}
\right) \v{s}(\vq,\tau) = -\vn \Phi\Big(\vq+\v{s}(\vq,\tau),\tau\Big)\,,
\label{eq:seom}
\ee
with initial condition $\v{s}(\vq,\tau=0)=0$.  This equation was first derived by \cite{dmitriev/zeldovich:1964}. 
At linear order, we can neglect the appearance of $\v{s}$ in the argument of
$\vn\Phi(\xfl)$, and this yields \refeq{lindisp} in \refapp{SPT}.  

Thus, if we know the bias relation at $\tau_*$, \refeq{dgcons}
supplies us with the bias relation at all later times.  This has 
been derived in \cite{matsubara:2008,chan/scoccimarro/sheth:2012} (see also \cite{carlson/reid/white:2013}).  It is important to note that conserved evolution
relates $\d_g(\vx,\tau)$ and $\d_g(\vx_*,\tau_*)$ at two different times
along \emph{the same fluid trajectory} (solid line in \reffig{sketch_evol}).  
Mathematically, this is due to the convective derivatives in \refeq{ngcons};  
physically, it states that galaxies and matter fall at the same rate in the
large-scale gravitational field, and hence co-evolve along the same
trajectories in the absence of an initial velocity bias.

We can make \refeq{dgcons} even more clear by rewriting it as
\be
1 + \d_g\Big|_\tau = \frac{1 + \d|_\tau}{1+\d|_{\tau_*}} (1+ \d_g|_{\tau_*})\,,
\label{eq:dgconsn}
\ee
where a vertical bar $|_\tau$ denotes a quantity evaluated at $\tau$ 
on a fixed fluid trajectory.  This simply states that the density ratio
of two conserved, initially comoving fluids remains constant under gravitational
evolution, as required by the equivalence principle;  we will encounter this
again in \refsec{baryons} when considering the different initial conditions for baryons and CDM set in the early Universe.  
Moreover, letting $\tau_*\to 0$, so that $\d|_{\tau_*} \to 0$,
we recover the well-known relation between Eulerian and Lagrangian densities
of a conserved tracer (see \reftab{EulLagr}), derived for the special case
of a spherical perturbation in \refeq{scbias}.

In the following, we will solve \refeq{dgconsn} to second order in perturbations, that is, up to quadratic terms in $\d^{(1)}$ and $K_{ij}^{(1)}$.  Denoting
second-order terms with a superscript $(2)$, we easily obtain
\ba
1 + \d_g^{(1)}(\vx,\tau) + \d_g^{(2)}(\vx,\tau) =\:& 1 + \d^{(1)} - \d^{(1)}_* + \d_{g*}^{(1)} \vs
& + \d^{(2)} - \d_*^{(2)} + \d_{g*}^{(2)} + [\d_*^{(1)}]^2 - \d^{(1)} \d_*^{(1)}
+ \left[\d^{(1)}  - \d_*^{(1)} \right] \d_{g*}^{(1)}\,,
\label{eq:dgconsexp}
\ea
where $f_* \equiv f(\vx_*,\tau_*)$, while quantities without a subscript $*$
are evaluated at $(\vx,\tau)$.  Here, we have separated linear and second
order terms into the first and second line.  Note however that the 
distinction between $\vx_* = \xfl(\tau_*)$ and $\vx$ is itself first order
in perturbations.  This is simply because the fluid trajectory in an
unperturbed FRW spacetime is $\xfl =$~const.  
Specifically, using the definition of the Lagrangian
displacement [\reftab{EulLagr}, see \refeq{lindisp} for the linear-order expression], we have at linear order
\be
\vx_*(\vx,\tau) = \vx + \v{s}(\vq, \tau_*) - \v{s}(\vq, \tau) = \vx 
+ \left(\Dstar -1 \right) \v{s}_{(1)}(\vx,\tau) + \cdots \,,
\ee
where $\Dstarinline \equiv D(\tau_*) / D(\tau)$, and $D(\tau)$ is the linear growth
factor defined in \refeq{Deom}.  Thus, we can neglect the distinction
between $\vx$ and $\vx_*$ in the argument of the terms in the second line of
\refeq{dgconsexp}, while those in the first line need to be expanded.  
For example, we obtain
\be
\d^{(1)}(\vx,\tau) - \d^{(1)}(\vx_*,\tau_*)
= \left(1-\Dstar\right) \d^{(1)} - \left(\Dstar-1\right) \Dstar s_{(1)}^i \partial_i \d^{(1)}\,,
\ee
where on the right-hand side all quantities are evaluated at $(\vx,\tau)$.  

Finally, we need a relation for the galaxy density $\d_{g*} = \d_g(\vx_*,\tau_*)$.   We write
\ba
\d_g^{(1+2)}(\vx_*,\tau_*) =\:& b_1^* [\d^{(1)} + \d^{(2)}](\vx_*, \tau_*) 
+ \eps^*(\vx_*,\tau_*) 
+ \frac12 b_2^* \left[\left(\d^{(1)}\right)_*\right]^2  + b_{K^2}^* 
\left[\left(K^{(1)}_{ij}\right)_*\right]^2 
\,,
\label{eq:dgstar}
\ea
where all quantities are evaluated at $\vx_*, \tau_*$.  
We now have allowed for a dependence of $\d_g$ on the tidal field squared
$(K_{ij})^2$, since this is a local observable and of the same order in perturbations
as $\d^2$, and thus is expected to be of similar relevance as the term
$\propto b_2^*$.  Note that because the tidal field is traceless ($\tr [K_{ij}]=0$), the tidal field cannot enter the bias expansion at linear order.
We have also included the leading correction to a
deterministic bias relation, namely a stochastic contribution $\eps^*$ to the
galaxy density field which we consider to be first order.  By definition, $\eps^*$ is assumed to be uncorrelated with the
matter variables.  We did not include any further stochastic contribution at second
order, a point we will return to below.  
Note that $b_1^*$ multiplies both $\d^{(1)}$ and $\d^{(2)}$, since $b_1^*$ is a physical bias parameter that has to be independent of the perturbative order we are working in (it corresponds to the
response of the mean density of galaxies to a change in the background
matter density, as we will discuss in \refsec{PBS}).  

Now we can simply collect the linear and second-order contributions to the
galaxy overdensity at $(\vx,\tau)$ from \refeq{dgconsexp}, to obtain
\ba
\d_g^{(1)}(\vx,\tau) =\:& \left(1 + \Dstar [b_1^* - 1] \right) \d^{(1)}(\vx,\tau) + \eps^*\vs
\d_g^{(2)}(\vx,\tau) =\:& \left\{1 + [b_1^* -1] \left(\Dstar\right)^2 \right\} \d^{(2)}
+ \left\{\Dstar[b_1^*-1] - \left(\Dstar\right)^2[b_1^*-1] + \frac12 b_2^* \left(\Dstar\right)^2 
\right\} [\d^{(1)}]^2
\vs
& + b_{K^2}^* \left(\Dstar\right)^2 [K_{ij}^{(1)}]^2
+\left(\Dstar - 1\right) \Dstar [b_1^* - 1] s_{(1)}^i \partial_i \d^{(1)}
- \left(\Dstar-1\right) \eps^* \d^{(1)} 
\vs
& + \left(\Dstar-1\right) s^i_{(1)}\partial_i \eps^*\,,
\ea
where on the right-hand side all quantities are evaluated at $(\vx,\tau)$,
except for $\eps^*$ which is evaluated at $(\vx,\tau_*)$.  
This corresponds to Eq.~(53) of
\cite{chan/scoccimarro/sheth:2012} (who do not include stochasticity however).\footnote{Note that $\v{\Psi}$ in Ref.~\cite{chan/scoccimarro/sheth:2012}
should stand for the displacement from the formation time to the present, which we write as $(\Dstarinline-1)\v{s}_{(1)}(\vx,\tau)$.}   
We define
\be
b_1^E = 1 + \Dstar [b_1^* - 1]
\label{eq:b1E}
\ee
as the \emph{linear Eulerian bias} (at time $\tau>\tau_*$). This relation first appears in \cite{nusser/davis:1994}, who pointed out that the bias of
galaxies must approach unity as time goes by. We will come back to this shortly.
Note that if we let $\Dstarinline \to 0$ while keeping $b_1^L \equiv (\Dstarinline) b_1^*$ finite,
corresponding to a formation at $\tau=0$ with subsequent conserved evolution, 
we obtain the relation
$b_1^E = 1 + b_1^L$ as in the spherical collapse evolution (\refsec{localbias}).  
We will return to this case below.  

Following our discussion above, the coefficient of the second-order density $\d^{(2)}$ in $\d_g$ has
to be $b_1^E$ as well.  Thus, we separate out $b_1^E \d^{(2)}$ in $\d_g^{(2)}$, and insert the expression for $\delta^{(2)}$ in terms of the 
linear-order $\d^{(1)}$, $K_{ij}^{(1)}$, and displacement term [\refeq{d2}], 
into the remainder.  Reordering terms, we finally obtain the expression for 
the galaxy density contrast to second order:
\ba
\d_g^{(1+2)} =\:& b_1^E \left[\d^{(1)}+\d^{(2)}\right] + \eps^* 
+ \frac12 b_2^E [\d^{(1)}]^2 
+ b_{K^2}^E [K_{ij}^{(1)}]^2 
\vs
& - \left(\Dstar-1\right) \eps^* \d^{(1)} + \left(\Dstar-1\right) s^i_{(1)}\partial_i \eps^*
\,,
\label{eq:dg2nd_inst2}
\ea
where all quantities (again, except for $\eps^*$) are evaluated at $(\vx,\tau)$ and the 
\emph{second-order Eulerian bias parameters} are given by \cite{sheth/chan/scoccimarro:2012}  
\ba
b_2^E(\tau) =\:& b_2^* \left(\Dstar\right)^2 + \frac{8}{21}\left(1-\Dstar\right)[b_1^E-1]
\vs
b_{K^2}^E(\tau) =\:& b_{K^2}^* \left(\Dstar\right)^2 -\frac27 \left(1-\Dstar\right)[b_1^E-1]\,. 
\label{eq:b2E}
\ea
A relation of the type \refeq{dg2nd_inst2} has been derived in \cite{fry:1996}, who, crucially, did not separate out the contribution
$b_1^E \d^{(2)}$ at second order however.  
Let us discuss this interesting result.
\begin{itemize}
\item The displacement term $s^i\partial_i \d$ has canceled out of the
deterministic bias relation. 
This is in fact expected, since the appearance of a nonzero displacement term 
would mean the galaxy has moved away from the fluid trajectory. 
Since we have only considered gravity here 
and neglected all non-gravitational momentum transfer effects, such a displacement cannot happen 
by way of the Equivalence Principle. The displacement $\v{s}$ is still explicit in the stochastic term at second order, since the
stochastic field is defined on the formation time slice.  
However, since $\eps^*$ is completely described by 
a spatially independent 1-point probability distribution function (\refsec{stoch}), 
the displacement term does not contribute to the galaxy clustering statistics.  
This does change however in the case of non-Gaussianity in the initial conditions, as we will see in \refsec{bNG}.  
\item For fixed formation time $\tau_*$, $\Dstarinline=D(\tau_*)/D(\tau)$ monotonically decreases towards later 
times. Thus, the galaxy density field becomes progressively less biased with respect to matter as $\tau\to\infty$.
\item At second order, evolution induces a term $\propto \eps^* \d^{(1)}$.  
If we imagine a realistic galaxy sample that includes galaxies with various formation times, following the ``Green's function'' approach
mentioned above,
$\d_g^{(2)}$ contains a superposition of individual contributions  $\eps^*\d^{(1)}$ from various formation times $\tau_*$.
Instead of attempting to model the distribution of formation times of all galaxies in the sample,
these contributions can be absorbed by introducing a second stochastic field $\eps_\d$ in addition to $\eps$, which enters the
bias expansion as $\eps_\d\, \d$.  In general, there will be some nonzero
(but not perfect) cross-correlation between the fields $\eps$ and $\eps_\d$.  
We return to this in \refsec{stoch}.  
\end{itemize}

Finally, we briefly consider the cross-correlation coefficient $r$
between galaxies and matter.  This is at linear order given by
\cite{tegmark/peebles:1998,taruya/koyama/soda:1999}
\be
r_{gm} \equiv \frac{\<\d_g \d\>}{\sqrt{\<\d_g \d_g\> \<\d \d\>}}
= \left[1 + \left(\Dstar\right)^2 \frac{\<(\eps^*)^2\>}{(b_1^E)^2 \<(\d^*)^2\> }\right]^{-1/2}\,.
\label{eq:crosscorrlinear}
\ee
At this order, $r_{gm}$ differs from unity solely due to the stochastic 
term $\eps^*$.  Note that if we wanted to derive the leading nonlinear 
correction to \refeq{crosscorrlinear}, we would need to include terms up to third order in perturbation theory.  We defer this until \refsec{npt1loop}.   
\refeq{crosscorrlinear} is equally valid in real and in Fourier space.  
While in Fourier space, $r_{gm}$ is generally less than 1 on all scales, 
in real space it is equal 
to 1 if $|\bfx_2-\bfx_1|$ is sufficiently large, since the correlation function 
of $\eps^*$ vanishes at large separations (see \refsec{stoch} for a more precise discussion).  
For fixed $\tau_*$ and $\eps^*$, the cross-correlation coefficient 
$r_{gm}(\tau)$ asymptotes toward 1 as $\tau\to\infty$, similar to $b_1^E$.  
In other words, $(b_1,\<\eps^2\>)=(1,0)$ is a fixed point: if galaxies are unbiased and perfectly correlated
with matter at some time $\tau_0$, this remains true for all $\tau>\tau_0$.  \\

\bfem{Bias expansions at initial and final time:}  Let us now go back to the gravitational evolution of a bias relation given
in the initial conditions considered in \refsec{localbias}, i.e. of Lagrangian bias (see also \cite{matarrese/etal:97}).  
This limit is obtained
by letting $\tau_*\to 0$ while keeping $(\Dstarinline) b_1^* \equiv b_1^L$, $\eps^* \equiv \eps^L$, $(\Dstarinline)^2 b_2^* \equiv b_2^L$, $(\Dstarinline)^2 b_{K^2}^* \equiv b_{K^2}^L$, fixed.  This is
the approach taken with the predictions from the excursion-set
(\refsec{exset}) and peak approaches (\refsec{peaks}).  
We then obtain the following expression at time $\tau$:
\ba
\d_g^{(1+2)}(\vx,\tau) =\:& (1+b_1^L) \left[\d^{(1)}+\d^{(2)}\right] + (1+\d^{(1)}) \eps^L 
+ \frac12 \left\{\frac{8}{21} b_1^L +  b_2^L \right\} [\d^{(1)}]^2 
+ \left[-\frac27 b_1^L + b_{K^2}^L \right] [K_{ij}^{(1)}]^2
\,,
\label{eq:dg2nd_L}
\ea
where we have dropped the stochastic displacement
term ${\bf s}_{(1)}\cdot\nabla\eps^L$
as it does not contribute to observables (see above).  To illustrate
the significance of \refeq{dg2nd_L}, we can contrast it with an example
with no evolution, obtained by setting $\tau_*=\tau$, which contains exactly
the terms included in the initial bias relation \refeq{dgstar}:
\ba
\d_g^{(1+2)}(\vx,\tau) =\:& b_1^E \left[\d^{(1)} + \d^{(2)} \right] + \eps 
+ \frac12 b_2^E  [\d^{(1)}]^2 
+ b_{K^2}^E [K_{ij}^{(1)}]^2 
\,.
\label{eq:dg2nd_E}
\ea
Now, letting $b_{K^2}^L = 0$ in \refeq{dg2nd_L}, we obtain the limit of
\emph{\LIMD Lagrangian bias} (cf. \refsec{localbias}).  At finite
time, it leads to a bias with respect to the tidal field squared of 
$b_{K^2}^E = -2/7 b_1^L = -2/7 (b_1^E-1)$.  Thus, \emph{a \LIMD expansion in the
initial conditions is inconsistent with a \LIMD expansion
at the evolved time $\tau$} [\refeq{dg2nd_E} with $b_{K^2}^E=0$], \emph{and vice versa}, unless bias is trivial, $b_1^* = 1$ and $b_n^* = 0$ for $n>1$.  
This was first derived by \cite{catelan/etal:1998,catelan/porciani/kamionkowski:2000}, who 
pointed out that Eulerian \LIMD and Lagrangian \LIMD bias lead to different 
three-point functions.  Refs.~\cite{matsubara:2008,chan/scoccimarro/sheth:2012,wang/szalay:2012}
also discuss the relation between Eulerian \LIMD and Lagrangian \LIMD bias. 
One might therefore consider a measurement of a Eulerian tidal bias of 
$b_{K^2}^E = -2/7 (b_1^E-1)$ as a confirmation of Lagrangian \LIMD bias.   
However, this interpretation assumes a
formation time of $\tau_*\to 0$ with subsequent conserved evolution.  
In fact, the same relation would be measured for any conserved tracer with 
finite formation time $\tau_*$ and nonzero $b_{K^2}^*$, as long as
\be
b_{K^2}^* = - \frac27 [b_1^*-1]\,.
\ee
Indeed, there is no reason why the formation of halos or galaxies should in general be independent of the local tidal field \cite{catelan/etal:1998,heavens/matarrese/verde:1998,smith/etal:2007,mcdonald/roy:2009,chan/scoccimarro/sheth:2012,baldauf/etal:2012}.  

Finally, comparing \refeq{dg2nd_L} and \refeq{dg2nd_E}, 
we also see that a stochastic term $\propto \eps\, \d^{(1)}$ appears in the former, but is absent in \refeq{dg2nd_E}.  
Hence, bias expansions at initial time and evolved time are only equivalent if we
also allow for a second-order stochastic term $\eps_\delta \delta$ at formation time.  Essentially, while $\eps$ corresponds to stochastic fluctuations
in the galaxy density, $\eps_\delta$ corresponds to stochastic fluctuations in 
the linear bias. 
This effect is an integral part of the general bias expansion starting at 
second order.\\

At this point, it is worth pointing out that the time evolution of the bias parameters of a realistic, observationally selected galaxy sample is difficult to predict.  As mentioned at
the beginning of \refsec{dynamics}, a realistic sample consists
of galaxies that have formed at various different times. Consider such a sample,
where the normalized distribution of formation times is denoted as $p(\tau_*)$.  Then, the bias parameter $b_O(\tau)$ corresponding to an operator $O$ at time of measurement $\tau$ is given by
\be
b_O(\tau) = \int b_O^E(\tau| \{ b_{O'}^*\} ,\tau_*) p(\tau_*) d\tau_*\,,
\label{eq:bOsuperposition}
\ee
where $b_O^E(\tau|\{b_{O'}^*\},\tau_*)$ is the Eulerian bias at time $\tau$ 
given a set of bias parameter $\{ b_{O'}^* \}$ at formation time $\tau_*$,
which at second order is given exactly by the relations in \refeq{b1E} and
\refeq{b2E};  we can write the stochastic fields $\eps,\,\eps_O$ in a similar
way.  Thus, without detailed knowledge
of the formation-time distribution, constraints on the bias parameters at one redshift will 
not lead to a prediction of the bias parameters at another redshift.  
Moreover, the distribution of formation times itself at a given point 
could well depend on the large-scale density and tidal field, which would also 
affect the time evolution of the bias parameters.

We stress, however, that as long as we keep all relevant terms at a given order, 
the presence of a distribution of formation times does not lead to new bias parameters.  
Thus, in most applications, as long as we consider the bias parameters to be 
effectively free functions of time, we arrive at the correct description 
regardless of the formation history of galaxies.  Moreover, the calculation 
above shows that the bias parameters are expected to evolve slowly, 
namely on the same Hubble time scale as the growth of matter perturbations themselves: $d\ln b_O/d\tau \sim \cH$. 

\subsection{Evolution from a joint perturbative solution}
\label{sec:evol2}

An alternative to the approach described in the previous section
is to directly solve the full set consisting of \refeq{contg}
and \refeqs{Cont}{Euler} for $\d_g,\,\d$, and $\theta$
\cite{elia/etal:2011,chan/scoccimarro/sheth:2012,MSZ} (see also \cite{somogyi/smith:2010}).  
At linear order in PT, one obtains second-order ordinary differential equations (ODE) for $\d$ and $\d_g$, which can be combined to yield an ODE for $b_1(\tau)$ \cite{basilakos/plionis:2001}.  
The standard approach to solving these equations at nonlinear order is to
work in Fourier space, in which case \refeqs{Cont}{Euler} become
\refeqs{contF}{eulerF} in \refapp{SPT}.  
In the context of the evolution of bias, this approach has the disadvantage
that it mixes local physical effects (which are important for bias) with 
advection terms from the fluid flow such as $s^i\partial_i \d,\,s^i\partial_i \d_g$ (which are identical between matter and galaxies and thus are of no relevance for bias).  

One option to circumvent this issue is to use Lagrangian perturbation 
theory (LPT) \cite{buchert:1992,buchert:1994,bouchet/etal:1995,matsubara:2008,carlson/reid/white:2013}.  The fundamental quantity of LPT is the deformation tensor
$M_{ij} \equiv \partial_{q,i} s_j$.  The evolved matter density is given by
\be
1+\d(\vx,\tau) = |\v{1} + \v{M}|_{\vq,\tau}^{-1}\,,
\ee
where we have assumed the single-stream regime (in the multi-stream regime, one should sum over all solutions of $\vx = \vq + \v{s}(\vq,\tau)$).  
Now, to describe biased tracers, we have to correspondingly introduce a galaxy deformation tensor $\v{M}_g$.  Note that the displacement $\v{s}$ itself is still the same for matter and galaxies, as long as velocity bias can be neglected
(see \refsec{velbias}).

One then immediately obtains the Lagrangian version of \refeq{dgcons},
\be
- \ln [1 + \d_g(\vx[\vq,\tau], \tau)] =
\ln |\v{1} + \v{M}_g|_{\vq,\tau} = \ln |\v{1} + \v{M}|_{\vq,\tau}
+ \ln \left[ \frac{|\v{1} + \v{M}_g|}{|\v{1} + \v{M}|}
\right]_{\vq,\tau_*}\,.
\ee
So far, this is very similar to the derivation of \refsec{evol1}, with
the advantage that there is no need to deal with the displacement terms as
all terms are explicitly evaluated at a fixed Lagrangian coordinate.  
On the other hand, the initial bias relation at $\tau_*$ now relates
$\v{M}_g$ to $\v{M}$:
\be
|\v{1} + \v{M}_g(\vq,\tau_*)|^{-1} = 
b_{\tr M}^* \tr \v{M}
+ b_{\tr(M^2)}^* \tr[\v{M} \v{M}] + b_{(\tr M)^2}^* (\tr \v{M})^2  
+ \eps^*
+ \eps_{\tr M}^* \tr \v{M}
+ \cdots\,,
\label{eq:biasLPT}
\ee
where all terms on the right-hand side are evaluated at $(\vq,\tau_*)$.  
This is equivalent to the bias expansion in \refeq{dgstar}, in the sense that,
at second order in perturbations, we can convert each term in \refeq{biasLPT}
into a linear combination of the terms in \refeq{dgstar}.  However, this
bias relation does not allow us to easily read off the standard bias
parameters ($b_1$, $b_2$ and $b_{K^2}$ to second order), since their 
contributions are spread over all terms in \refeq{biasLPT}, and terms at all higher orders as well.  

Another approach to solving \refeq{contg} and 
\refeqs{Cont}{Euler} proceeds by integrating the equations along the fluid flow.  In this approach, which was introduced by \cite{MSZ} and dubbed ``convective SPT,'' the advection terms never appear explicitly, and only terms relevant for bias are present.  On the other hand, we still deal explicitly with the Eulerian density and tidal field, allowing us to connect to the results of the previous section as well as subsequent sections.   We briefly present this approach here, with details given in \refapp{convSPT}.  We stress that regardless of the approach taken, the final result has to be the same, and we follow the convective SPT approach only because the resulting expressions offer a clear physical interpretation.

We continue to assume the absence of velocity bias; that is, galaxies are comoving with the matter.  
The system which we started from, \refeq{contg} and \refeqs{Cont}{Euler}, can be written in compact form as
\ba
\convD \v{\Psi} =\:& - \v{\sigma}\cdot\v{\Psi} + \v{S} 
\label{eq:Psigeom}
\ea
where we have defined the vector $\v{\Psi} = ( \d_g,\,\d,\,\theta )$, 
$\v{\sigma}$ is a matrix that solely depends on the FRW background,
and $\v{S}$ is a source term which is at least second order in
perturbation theory [see \refeq{Psigdef}].  

The system \refeq{Psigeom} can be solved order by order in a straightforward
manner, as detailed in \refapp{convSPT}.  Care needs to be taken in 
deriving the source term in order to allow for an integration along 
the fluid trajectory.  
In the following, we will present the solution up to third order.  
Earlier results up to this order can be found in 
\cite{ohta/kayo/taruya:2004,chan/scoccimarro/sheth:2012,saito/etal:14,Biagetti:2014pha}.  
To begin, we need to provide an expression for the galaxy density
at the initial ``formation'' time $\tau_*$.   Assuming instantaneous
formation as before, we include all terms composed of the density
and tidal field, and include all relevant stochastic
terms, up to third order:
\ba
\d_g^* =\:& \sum_{n=1}^3 \frac{b_n^*}{n!} [\d^*]^n 
+ b_{K^2}^* \tr \left[ (K_{ij}^*)^2 \right]
+ b_{K^3}^* \tr \left[ (K_{ij}^*)^3 \right]
+ b_{\d K^2} \d^* \tr \left[ (K_{ij}^*)^2 \right] \vs
& + \eps^* + \eps^*_{\d} \d^* + \eps^*_{\d^2} [\d^*]^2
+ \eps^*_{K^2} \tr \left[ (K_{ij}^*)^2 \right] 
\,,
\label{eq:dgIC}
\ea
where here and throughout, a superscript $*$ indicates that a quantity is evaluated at
$\vx_* \equiv \vx_{\rm fl}(\tau_*)$ and $\tau_*$.  The stochastic fields 
$\eps^*,\,\eps^*_X$ ($X=\d$, $\d^2$, $K^2$) are
assumed to be first-order random fields.  Note that a term of
the form $\eps_\d \d$ has already appeared through second-order evolution
in the previous section, indicating that it should be included at second
order.  Similar reasoning applies to the new third order stochastic
terms.

Up to second order, we recover the results of \refsec{evol1}. The solution is given by [\refeq{Psig2}] 
\ba
\d_g^{(1+2)}(\vx,\tau) = b_1^E(\tau) \left[\d^{(1)} + \d^{(2)}\right] + \frac12 b_2^E(\tau) (\d^{(1)})^2 + b_{K^2}^E(\tau) (K^{(1)}_{ij})^2 +  \eps^E_{\d}(\tau) \d^{(1)} \,,
\ea
where on the right-hand side, all quantities are evaluated at the
\emph{Lagrangian}
position $\vq$ corresponding to $(\vx,\tau)$, and the Eulerian bias parameters are
\ba
b_1^E(\tau) =\:& 1 + \Dstar [b_1^* - 1] \vs
b_2^E(\tau) =\:& b_2^* \left(\Dstar\right)^2 + \frac8{21} (b_1^*-1) \Dstar \left(1-\Dstar\right) \vs
b_{K^2}^E(\tau) =\:& b_{K^2}^* \left(\Dstar\right)^2 - \frac27 (b_1^*-1) \Dstar \left(1-\Dstar\right)\,,
\label{eq:bE2}
\ea
while the Eulerian stochasticity is
\ba
\eps^E_\d(\tau) = \eps^*_\d \Dstar - \left(\Dstar-1\right) \eps^*
\,,
\label{eq:epsE2}
\ea
as derived independently in \refsec{evol1}.  
Note that $\eps^E_\d$ has one power of $\Dstarinline$ less than the other
second-order Eulerian quantities, since the fields $\eps^*_X$ are defined at $\tau_*$.  
As mentioned above, $\d_g(\vx,\tau)$ is written here as a local function of the
matter fields and stochastic variables evaluated at $\vq = \vx(\tau=0)$, i.e. Eulerian quantities evaluated at a fixed Lagrangian position. In order to transform to a Eulerian position $\vx$, one adds the second-order displacement term to $\d_g$ (as well as analogously to $\d$ and $\theta$) through
\be
- s_{(1)}^i(\vx,\tau) \partial_i \d_g(\vx,\tau)\,.
\ee
Going to third order, we obtain
\ba
\d_{g}^{(3)}(\vx,\tau) =\:& b_1^E \d^{(3)} + b_2^E \d^{(1)} \d^{(2)} + 2 b_{K^2}^E K^{(1)}_{ij} K^{(2)}_{ij} + \frac16 b_3^E (\d^{(1)})^3 + b_{K^3}^E (K^{(1)}_{ij})^3 + b_{\d K^2} \d^{(1)} (K^{(1)}_{ij})^2 
\vs
& + b_{\otd}^E O_{\otd}^{(3)} + \eps_\d^E \d^{(2)} + \eps_{\d^2}^E (\d^{(1)})^2 
+ \eps_{K^2}^E (K^{(1)}_{ij})^2 
\,,
\label{eq:dg3f}
\ea
where 
\ba
O^{(3)}_{\otd} \equiv \frac{8}{21} 
K^{(1)}_{ij} \Del^{ij} \left[(\d^{(1)})^2 
-\frac32 (K^{(1)})^2 \right]
\,.
\label{eq:Otddef}
\ea
Here and throughout, we use the short-hand notation
\be
K^2 \equiv (K_{ij})^2 \equiv \tr[ K\, K ] = K_{ij}K_{ji}
\quad\mbox{and}\quad
K^3 \equiv (K_{ij})^3\equiv \tr[ K\,K\, K ] = K_{ij}K_{jk}K_{ki}\,.
\ee
Again, in order to obtain the density at a fixed order in standard Eulerian perturbation
theory, we need to displace $\d_g$ from a fixed Lagrangian position $\vq$ to the Eulerian position, by expanding in the argument $\xfl[\vq,\tau]$.  This is given in \refeq{Psigtotal}.  As this coordinate shift is somewhat tangential to the topic of bias,
we do not repeat the results here.

The third-order Eulerian bias parameters are given by
\ba
b_3^E(\tau) =\:& b_3^* \left(\Dstar\right)^3 + \left[
(b_1^*-1) \frac4{1323} \left(199 - 35 \Dstar\right) 
+ \frac{13}7 b_2^* \Dstar 
\right]
\Dstar \left(\Dstar-1\right)
\vs 
b_{\d K^2}^E(\tau) 
=\:& b_{\d K^2}^* \left(\Dstar\right)^3 + \left[
- (b_1^*-1) \frac1{147} \left(33+7 \Dstar\right) 
+ \left(\frac{2}7 b_2^* + b_{K^2}^*\right) \Dstar
\right]
\Dstar \left(\Dstar-1\right)
\vs
b_{K^3}^E(\tau) 
=\:& b_{K^3}^* \left(\Dstar\right)^3 + \left[
(b_1^*-1) \frac2{63} \left(-11 + 7 \Dstar\right) 
+ 2 b_{K^2}^* \Dstar 
\right]
\Dstar \left(\Dstar-1\right)
\vs
b_{\otd}^E(\tau) =\:& \left[
(b_1^*-1) \frac16\left(\Dstar - \frac{23}7\right) 
+  \frac52 b_{K^2}^* \Dstar 
\right]
\Dstar \left(\Dstar-1\right)
\,.
\label{eq:biasEthirdorder}
\ea
We see that, as expected, all $b_O^E$ converge to $b_O^*$ for $\Dstarinline \to 1$ (when $\tau_*\to\tau$), with the exception of $O_{\otd}^{(3)}$ since we did not allow for it at the formation time [\refeq{dgIC}].  Also, they vanish for $\Dstarinline \to 0$, unless $b_O^*$ diverge in the limit
$\tau_*\to 0$, which is usually assumed when writing a Lagrangian bias relation.  Specifically, we can easily obtain the prediction for $b_O^E$ assuming 
Lagrangian \LIMD bias, by letting $\tau_* \to 0$ and all cubic $b_O^* = 0$ apart from $b_3^*(\Dstarinline)^3 = b_3^L$.  This yields
\ba
b_3^E(\tau) =\:& b_3^L - \frac{796}{1323} b_1^L - \frac{13}7 b_2^L
\vs 
b_{\d K^2}^E(\tau) =\:& \frac{11}{49} b_1^L - \frac27 b_2^L
\vs
b_{K^3}^E(\tau) =\:& \frac{22}{63} b_1^L 
\vs
b_{\otd}^E(\tau) =\:& \frac{23}{42} b_1^L \,.
\label{eq:biasEthirdorderLL}
\ea
The first line matches \refeq{scbiasEL} exactly, since 
it is the only term that remains for a spherically symmetric 
perturbation, whose evolution is governed exactly by the spherical collapse
solution.

Finally, the Eulerian stochastic terms at third order become
\ba
\eps_{\d^2}^E(\tau) =\:& \eps^*_{\d^2} \left(\Dstar\right)^2 - \frac{4}{21} (\eps_\d^* - \eps^*)
\Dstar \left(\Dstar-1\right)
\vs
\eps_{K^2}^E(\tau) =\:& \eps^*_{K^2} \left(\Dstar\right)^2 + \frac27 (\eps_\d^* - \eps^*)
\Dstar \left(\Dstar-1\right)\,.
\label{eq:epsE3}
\ea
In analogy to the second-order case, we see that the third-order
stochastic terms $\eps_{\d^2},\,\eps_{K^2}$ are induced by gravitational
evolution, even if they are absent initially.  Thus, following the discussion at the end of \refsec{evol1}, we should allow for these terms in the general bias expansion.  

Let us now turn to the operator $O_{\otd}^{(3)}$ which has appeared in the third-order evolved bias relation \refeq{dg3f}. While the operator $O_{\otd}^{(3)}$ is third order in perturbations, it cannot be expressed locally in terms of the linear density and 
tidal field, as is clear from \refeq{Otddef}. Therefore, including this operator goes beyond the initial 
bias relation \refeq{dgIC}: that is, starting from third order in 
perturbation theory, \emph{a bias expansion involving
only the local density and tidal field is not sufficient.}   
Interestingly, the operator $O_\otd^{(3)}$ can be expressed in a variety of
equivalent ways at third order, as derived in \refapp{biastrans}: 
\begin{itemize} \label{other:Otdrel}
\item $O_\otd^{(3)} = 2 K^{ij}\Del_{ij} [\d+ (f\cH)^{-1} \theta] + F(\d, K_{lm})$,
where 
$f\equiv d\ln D/d\ln a$ is the logarithmic linear growth rate, and 
$F(\d, K_{lm})$ denotes cubic local combinations of density and tidal field as enumerated in \refeq{dgIC} [see \refeqs{Gamma3def}{Gamma3third} for the precise relation].  Thus, $O_\otd^{(3)}$ is related to the local difference of the tidal field, $\propto \Del_{ij} \d$, and velocity shear, $\propto -\Del_{ij} \theta$.  This type of operator was first considered in \cite{mcdonald/roy:2009}.  
\item $O_\otd^{(3)} = (4/5) K^{ij} \left([\cH f]^{-1} \convDinline - 1\right) K_{ij} + G(\d, K_{lm})$ [\refeq{OtdPi2} with \refeq{piij2def}].  We see that $O_\otd^{(3)}$ is proportional to the \emph{convective time derivative} of the tidal field (at second order) along the fluid flow.
\item $O_\otd^{(3)} = (8/3) [M^{ij} - (1/3) \d^{ij} \tr M] \left([\cH f]^{-1} \partial_\tau - 1\right) M_{ij}$ [\refeq{OtdLagr}].  This is the Lagrangian expression of $O_\otd^{(3)}$, showing that this operator is related to the time derivative of the Lagrangian deformation tensor (recall that in Lagrangian coordinates, convective time derivatives reduce to partial derivatives; \reftab{EulLagr}).
\end{itemize}
All these equivalent formulations show that the operator $O_\otd^{(3)}$ is 
a local observable.  An observer comoving with the galaxy could measure
it, for example, by measuring the proper time derivative of the local
tidal field.  Indeed, we would expect any quantity that emerges from
the gravitational evolution of a conserved tracer to be a local 
gravitational observable.  
Thus, given that $O_\otd^{(3)}$ has exactly two spatial derivatives for each
power of the potential, counting $\Del_{ij}$ as zero net derivatives, it is justified to include it in our category of \emph{local bias operators}.  

Let us briefly pause to consider how our treatment, which has gone
to third order in perturbations, might continue to higher orders.  
In particular, what terms beyond simple combinations of $\d$ and $K_{ij}$
would one expect at fourth and higher order?  Our first encounter of
such a term, $O_{\otd}^{(3)}$ in \refeq{Otddef}, 
shows that at third order, only a certain combination of $\Del^{ij}(\d^2)$ and
$\Del^{ij}(K^2)$ appears, not each one of them individually.  One might
then wonder if, following \cite{mcdonald/roy:2009}, it is sufficient to include the velocity shear $(\partial_i\partial_j/\lapl)\theta$ in addition to $\d,\,K_{ij}$.  
As we will see in the next section, this is not the case.  At fourth
order, the velocity shear is no longer sufficient.

Before continuing to the general bias expansion, it is worth considering how
these results change when allowing for an expansion history beyond EdS.  Specifically, we continue to assume the validity of GR, but allow for a nonzero curvature, and cosmological constant $\Lambda$ or dark energy component (where we neglect the effect of dark energy perturbations; we will discuss further generalizations in \refsec{neutrinos} and \refsec{modgrav}).  
As shown in \refapp{convSPT:b}, the equations of motion maintain the same structure as in EdS.  We also show there that no new type of bias operator appears up to including third order for a general expansion history, although the time evolution of bias parameters for conserved tracers [\refeq{biasEthirdorder}] is modified.  Further, the departure from EdS of the equations of motion is completely
quantified by the quantity $\Om(a)/f^2(a)-1$.  Since for $\Lambda$CDM, as well as most viable dark energy expansion histories, we have $f \approx \Om^{0.55}$ \cite{linder:2005}, this quantity is approximately
\be
\Om(a)/f^2(a)-1  \approx -0.1 \,\ln\, \Om\,.
\ee
For $\Omega_{m0} \geq 0.3$, this quantity remains less than 0.13 at all redshifts.  
This explains why the EdS approximation in perturbation theory calculations, with $a(\tau)$ replaced with $D(\tau)$ in the final result, is numerically accurate.

\subsection{General perturbative bias expansion}
\label{sec:general}

Building upon the specific second- and third-order results we have presented 
above, the goal of this section is to derive a general framework for the 
perturbative bias expansion.
That is, we aim to derive a  set of bias parameters that is 
\emph{sufficient to describe the statistics of any large-scale structure 
tracer}, within the realm of perturbation theory.  
In the following, we will continue to use the term ``galaxy'', although one
should keep in mind that the results are more generally applicable to other
tracers.  In fact, the general bias expansion can also be applied to the local matter power spectrum \cite{barreira/schmidt:2017}.

In \refsecs{evol1}{evol2}, we have seen specific bias relations involving 
the density and tidal field, as well as an additional operator $O_\otd^{(3)}$, 
and their stochastic counterparts such as $\eps_\d$, $\eps_{K^2}$.
While the previous derivations assumed a conserved tracer, this is easily
generalized by employing the functions $b_O^E(\tau;\{b_{O'}^*\},\tau_*)$
as Green's functions, as discussed at the end of \refsec{evol1}.  
We now abandon the assumption of number conservation entirely, and
allow our galaxies to form and merge arbitrarily.  
In general, a deterministic bias relation for an arbitrary galaxy sample can be written as
\be
\d_g(\vx,\tau) = \sum_O b_O(\tau) O(\vx,\tau)\,,
\label{eq:biasrel}
\ee
where $\d_g = n_g/\avng - 1$ is the density contrast
of the galaxies, $b_O$ are bias parameters, and $O$ are operators constructed
out of the matter density field, gravitational potential, and in general other
perturbations.  The statistics of the galaxies at a fixed time are then given 
by the statistics of the matter density field,
potential, and so on at the same time, and a finite set of numbers $b_O$.  
Note that this relation is only useful if we have a finite set of operators and hence a limited number of bias parameters which can be marginalized over while retaining cosmological information.  
Refs.~\cite{matsubara:2011,matsubara:2014} have considered generalizations 
of bias parameters to bias functions, $b_N \to b_N(\vx_1,\cdots, \vx_N)$.  
However, in order to retain predictive power, we need to condense
these functionals into a finite set of terms.  This is precisely the
goal of this section.

In the context of cosmological perturbation theory,
we thus need to identify which operators 
should be included in the bias relation up to any given order in 
perturbations.  
We will refer to such a set of operators as \emph{basis}.  Of course,
such a basis is not unique, since any linearly independent combination of 
the operators in one basis represents another basis.  One significant
example is choosing operators given in the Lagrangian frame, for example
in terms of the distortion tensor $\v{M}$ (\reftab{EulLagr}) (\emph{Lagrangian bias}), or in terms of quantities at the final, observation time (\emph{Eulerian bias}), such as \refeq{dg3f}.  We will give explicit examples of both.

In the previous section, we have seen that, when starting from a bias
expansion that includes powers of the density field and tidal field, i.e.
$O \in \{ \d, \,\d^2,\, (K_{ij})^2,\, \dots\}$ at one time $\tau_*$,
then conserved evolution under gravity introduces additional bias operators,  
in particular $O_{\otd}^{(3)}$ in \refeq{dg3f}, at a later time $\tau$.  
In this sense, a bias expansion restricted to powers of the density and tidal
field is incomplete, since
it does not allow us to describe, for example, a galaxy that has formed 
at some earlier time (or even in the initial conditions) and then passively evolved following number conservation;  see also the introduction of Ref.~\cite{matsubara:2011} which nicely describes this problem.  

As a starting point, we will work under the following assumptions in the remaining subsections:  
\begin{itemize}
\item Gravitation is described by General Relativity (GR).
\item The impact of massive neutrinos and dark energy perturbations can be neglected.
\item The initial conditions are Gaussian and adiabatic, and any isocurvature perturbations induced by physics in the early Universe can be neglected.
\end{itemize}
We will relax the first two assumptions in \refsec{beyondCDM}, and the last assumption in \refsec{NG} as well as \refsec{beyondCDM} (for isocurvature perturbations between baryons and CDM).  The last point implies that we only consider the adiabatic growing mode.  In a standard $\Lambda$CDM cosmology, this is accurate at the percent-level (see \refsec{baryons}).

\subsubsection{Spacetime picture of bias and evolution}
\label{sec:framework}

\begin{figure}[t]
\centering
\includegraphics[width=0.6\textwidth]{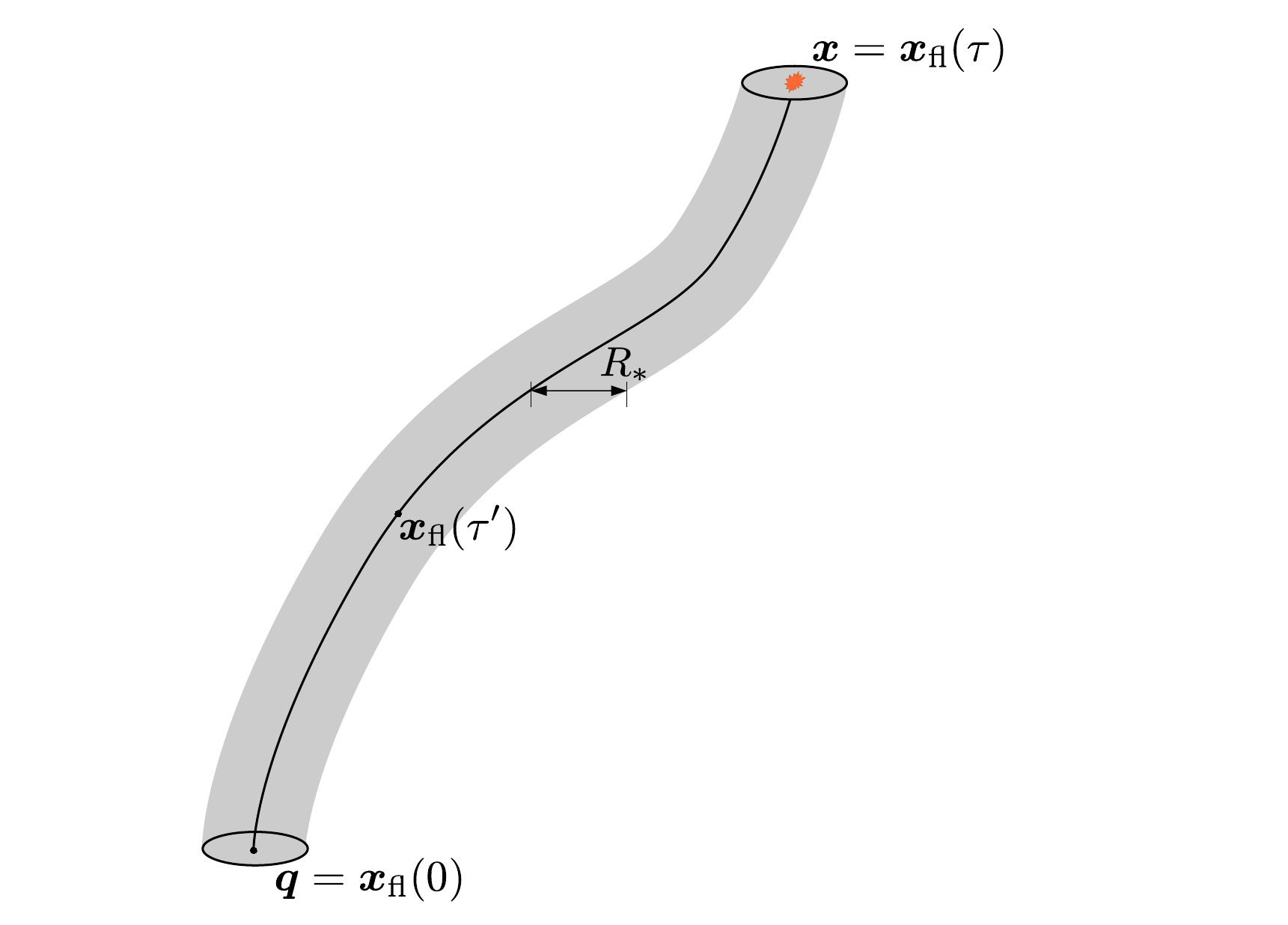}
\caption{
Sketch of the spacetime region involved in the formation of tracers such as halos or galaxies.  Time is running vertically.  The solid line denotes the fluid trajectory $\xfl(\tau')$ from a Lagrangian position $\vq = \xfl(\tau=0)$ to a Eulerian position $\vx = \xfl(\tau)$ at time $\tau$.  The shaded region with a comoving spatial extent of order $R_*$ denotes the region from which the matter within the galaxy and its host halo originates, or the region of influence feedback processes---whichever is larger.  
\label{fig:CFCsketch}}
\end{figure}

The key physical feature of the formation of tracers such as halos and 
galaxies is that it happens over \emph{long time scales} 
\cite{mo/vandenbosch/white:2010}, while the formation process 
is limited to relatively \emph{small spatial scales} \cite{vandaalen/etal:2014}.
With long time scales here we mean that the formation takes place over 
an appreciable fraction of the age of the Universe.  
On the other hand, the matter that forms dark matter halos comes 
from within a region of a few Mpc in comoving size.  In other words, the 
spacetime region that encompasses the formation and evolution
of galaxies is of the ``spaghetti'' shape sketched in \reffig{CFCsketch}.  This is
related to the well-known fact that the nonlinear scale, at which the
fractional density perturbations become order unity, is much smaller than the
Hubble horizon.  This latter fact forms the basis of all perturbation-theory
approaches to large-scale structure, and it is of similarly
crucial importance in the derivation of bias relations.  

Suppose that the abundance of galaxies at position $\vx$ only depends on the
distribution of matter in a finite region around $\vx$, of
characteristic comoving length scale $R_*$.  We will call $R_*$ the ``nonlocality scale''
which is understood to be defined on a certain time slice 
(for example, the final or initial times). We will further discuss its 
physical significance in \refsec{higherderiv}.  For dark matter halos, $R_*$ is expected to be of order the Lagrangian radius.   
Now consider the case where we look at statistics of galaxies or halos on 
scales much larger than $R_*$.  Then, we can approximate the 
bias relation as effectively \emph{local in space}, thus reducing
the bias expansion from a functional expansion to an ordinary Taylor 
expansion as in \refeq{biasrel} (see \refsec{higherderiv} for the functional 
expansion).  In  effective field theory (EFT) language (\cite{goldberger/rothstein,baumann/etal:2012,hertzberg:2014,senatore:2015}; see \cite{porto:2016} for a review), 
the spatially local approximation provides the low-energy effective description of the full, complicated dynamics of the formation of galaxies (\refsec{renorm}).  

When considering galaxy formation as effectively local, the only quantities that
are relevant for the formation of galaxies are then the \emph{density and the
tidal field} $\partial_i\partial_j \Phi(\xfl(\tau'),\tau')$ \emph{along the 
trajectory of a Lagrangian patch enclosing the galaxy} 
(\cite{senatore:2015,MSZ};  Ref.~\cite{wang/szalay:2012} only considered the matter density along the fluid trajectory).   
One way to prove this statement is to invoke the equivalence
principle, which states that in a free-falling frame, such as that comoving
with the trajectory $\xfl(\tau)$, the leading locally observable gravitational effect is given by second derivatives of the metric tensor.  Moreover, 
essentially all tracers of the LSS are non-relativistic.  Then, the only
relevant component of the metric tensor is the time-time-component.  
On sub-horizon scales, this is in turn equivalent to the
tensor $\partial_i\partial_j\Phi$, where $\Phi$ is the gravitational potential
defined in \refeq{metriccN}.  This tensor can further be decomposed into
the trace $\lapl\Phi$ which is directly related to the density perturbation
$\d$ through the Poisson equation; and the trace-free part $K_{ij}$ [\refeq{KijF}], which quantifies the tidal field proper.  
An alternative, more rigorous derivation of the same result is given by the
Conformal Fermi Coordinate (CFC) approach \cite{CFCorig,CFCpaper1}, which
clarifies the meaning of the density perturbation and $K_{ij}$ in the
relativistic context.  We will return to this in \refsec{GR}.  

This reasoning provides the physical justification for our definition of \emph{local bias}
(\refsec{notation}) as encompassing all terms in the general bias expansion that are constructed (without any further spatial derivatives) out of the density
and tidal field along the fluid trajectory:  these are precisely the leading local gravitational observables for a comoving observer.  In conformal-Newtonian gauge, these terms 
are characterized by exactly two spatial derivatives acting on each power of the potential $\Phi$.  
Note that we do not need to assume a conserved, passively evolving galaxy sample here.  
Any gravitational interactions such as mergers \cite{simon:2005} 
do not, on sufficiently large scales, depend on any property apart from the local density and tidal field.  A galaxy sample that preferentially resides in halos formed from recent major mergers might have a larger nonlocality scale $R_*$ than that of typical halos of the same mass.  Nevertheless, it will be a finite scale, and presumably still of order the Lagrangian radius of these halos as argued above.  

In our reasoning we did however
implicitly assume that the small-scale initial conditions, i.e. those
of much smaller scale than the large-scale correlations we are interested in,
are statistically uncorrelated over large scales.  This is the case for
Gaussian initial conditions, which we assume in this section.

Now, let us formalize our reasoning.  The dependence on $\d(\xfl(\tau'),\tau')$ and $K_{ij}(\xfl(\tau'),\tau')$ can be written as multiple time integrals over the fluid trajectory.  For example, in the simplest case, for a given operator $O$ constructed out of $\d$ and $K_{ij}$, we can formally expand the time integral as \cite{senatore:2015}
\ba
\d_g(\vx,\tau) \supset\:& \int^\tau d\tau' \, f_O(\tau,\tau') O(\xfl(\tau'),\tau') \vs
=\:& \left[ \int^\tau d\tau' \, f_O(\tau,\tau')\right] O(\vx,\tau)
+ \left[\int^\tau d\tau' \, (\tau'-\tau) f_O(\tau,\tau')\right] \convD O(\vx,\tau) + \cdots\,,
\label{eq:dg_taylor}
\ea
where $\convDinline$ is the convective derivative along the fluid flow [\refeq{DDtau}].  
In order to provide a basis of operators at a fixed time,
as demanded by \refeq{biasrel}, we thus have to include time derivatives along the fluid flow,
such as $\text{D}(\partial_i\partial_j\Phi)/\text{D}\tau$, in the basis of operators.  
Including time derivatives, of arbitrary order, of powers of the density field and tidal field 
then provides a complete basis of operators for the local bias expansion
(that is, at lowest order in \emph{spatial} derivatives).  However, this is not very satisfying:
since the formation of galaxies happens on long time scales,
the higher-order terms not written in \refeq{dg_taylor} are not necessarily 
smaller than the ones that we include;  in fact, this seems to suggest that we 
need infinitely many operators in our bias expansion. 
Fortunately however, at fixed order in perturbation theory, only a finite 
number of time derivatives are linearly independent, and thus the basis can be 
completed with a finite number of operators. The physical reason is that 
the time evolution of the large-scale, quasi-linear perturbations 
is predicted at any given order in perturbation theory.  Moreover,
given our assumptions, they 
evolve at exactly the same rate on linear scales, since the linear growth 
factor is scale-independent.  Thus, the departure from the linear growth
rate is higher order in perturbations.
We will now show precisely how to use this fact to obtain a finite set of operators.

\subsubsection{Lagrangian basis of operators}
\label{sec:basisL}

To begin with, let us work on the initial time slice, in Lagrangian
coordinates $\vq = \vx_{\rm fl}(\tau=0)$.  This simplifies the treatment,
since in Lagrangian coordinates convective time derivatives $\convDinline$
[\refeq{DDtau}] reduce to simple time derivatives $\partial/\partial\tau$.

Consider a Lagrangian operator $O_L(\vq,\tau)$ 
constructed out of $d_O$ factors of $\partial_{q,i}\partial_{q,j}\Phi^{(1)}(\vq)$ (or, equivalently, the distortion tensor $M_{ij}$).  
In perturbation theory, it can be written as
\be
O_L(\vq,\tau) = D^{d_O}(\tau) O_L^{(d_O)}(\vq,\tau_0)
+ D^{d_O+1}(\tau) O_L^{(d_O+1)}(\vq,\tau_0) + \cdots\,,
\label{eq:Oexp}
\ee
where $D(\tau)$ is the linear growth factor, normalized to some reference time 
$\tau_0$.  By construction, $d_O$ is the perturbative order of the leading contribution to $O_L$.  
The operators $O_L^{(n)}(\vq,\tau_0)$, $n= d_O,\,d_O+1, \dots$, correspond to the contributions to $O_L$ at $n$-th order in perturbation theory, evaluated at the reference time $\tau_0$.  
Here, we have assumed for simplicity
that the $n$-th order growth factor is given by the linear growth factor
to the $n$-th power.  This is only strictly valid for an EdS
(flat matter-dominated) Universe where $D(\tau) = a(\tau)$, although also generally very accurate
for other cosmologies such as $\Lambda$CDM.  We will discuss this at the end of \refsec{basisE}.  

Given the general relation \refeq{Oexp}, allowing for time derivatives of $O_L$ in the bias expansion [see \refeq{dg_taylor}] is,
at $n$-th order in perturbation theory, equivalent to including the contributions 
$O_L^{(m)}$ ($m\leq n$) at \emph{each order} individually.  
This is because, at this order in perturbation theory, 
the time derivatives of any operator $O_L$
are given by linear combinations of the $O_L^{(m)}$ ($m\le n$).  
Consider 
for example a second-order Lagrangian operator, which in perturbation theory
can be written as, following \refeq{Oexp}, $O_L(\vq,\tau) = D^2(\tau) O_L^{(2)}(\vq,\tau_0) + D^3(\tau) O_L^{(3)}(\vq,\tau_0) + \cdots$.  
Then, in third-order perturbation theory, the $n$-th convective time derivative of $O_L$ is given by
\be
\left(\convD\right)^n O_L(\vq,\tau)\Big|^{(3)} = \frac{\partial^n}{\partial\tau^n} O_L(\vq,\tau)\Big|^{(3)} = \left(\frac{d^n}{d\tau^n} D^2(\tau)\right) O_L^{(2)}(\vq,\tau_0)
+ \left(\frac{d^n}{d\tau^n} D^3(\tau)\right) O_L^{(3)}(\vq,\tau_0)\,.
\label{eq:dOexp}
\ee
At any given fixed time, this is just a linear combination of $O_L^{(2)}(\vq,\tau_0)$
and $O_L^{(3)}(\vq,\tau_0)$.  This continues to hold correspondingly 
at any fixed, higher order in perturbation theory, and becomes even more obvious 
when transforming the time coordinate from $\tau$ to $\ln D(\tau)$: then, the right-hand side of \refeq{dOexp} simply becomes $2^n D^2(\tau) O^{(2)} + 3^n D^3(\tau) O^{(3)}$.  

Note that, even when starting with an operator $O_L$ that is a local combination of $\partial_{q,i}\partial_{q,j}\Phi$, the higher-order terms
$O_L^{(n)}$ generated by time evolution are in general not expressible
as local combinations of $\partial_{q,i}\partial_{q,j}\Phi$.  Instead,
terms involving $\partial^i\partial^j/\lapl$ acting on powers of $\partial_l\partial_m\Phi$ appear, just as we have seen with $O_\otd^{(3)}$ in \refeq{Otddef}.  
Fundamentally, this is a consequence of the fact that gravity acts over
long distances, so that the gravitational evolution of the tidal field
cannot be approximated as local \cite{kofman/pogosyan:1995,bertschinger:1995}.  
In particular, the invariant definition of the tidal field is a certain projection of the
Weyl tensor \cite{ip/schmidt:2016}, which corresponds to the specific part of the
Riemann tensor that is not locally related to the stress-energy tensor via
the Einstein equations.  

Crucially however, while we assume that galaxy formation is local 
(we relax this assumption in \refsec{higherderiv}), 
we do \emph{not} have to assume that gravity is local.  
Indeed, one can straightforwardly derive
the evolution of the tidal field in perturbation theory, and take that
into account in the bias expansion, namely through the terms $O_L^{(m)}$ described above.  One finds that the time derivatives
of the tidal field only contain a small subset of all possible operators 
constructed out of $\partial^i\partial^j/\lapl$ acting on powers of 
$\partial_l\partial_m\Phi$.     
Only these specific operators should be included in the bias expansion, 
because only these terms correspond to local observables, essentially time
derivatives of the tidal field along the fluid flow.    

Let us now construct an explicit Lagrangian basis of bias operators.  It is convenient to write these in terms of the Lagrangian distortion tensor introduced 
in \refsec{evol2},
\be
M_{ij} \equiv \frac{\partial s_j}{\partial q^i}\, .
\ee
Note that at linear order, $M_{ij}^{(1)}$ is directly proportional to
$\partial_{q,i}\partial_{q,j} \Phi^{(1)}$. 
Knowing that we can always recast the time derivatives as a sum of higher-order operators, we simply have to take
all scalar contractions of the contributions $M^{(n)}_{ij}$ at 
each perturbative order;  up to quadratic order, this was already written
in \refeq{biasLPT}.  However, we do not need to include
$\tr[M^{(n)}] \equiv \d^{ij} M^{(n)}_{ij}$ with $n>1$, as these terms
can always be expressed in terms of lower-order operators through
the equations of motion for $M_{ij}$ (see \cite{zheligovsky/frisch,matsubara:2015} for the explicit expression of the latter).  The basis up to fourth order then is~\cite{MSZ}\footnote{We have added two operators at fourth order that were missing in \cite{MSZ}; specifically, the first two terms in the very last line of \refeq{LagrBasis}.}
\bea
{\rm 1^{st}} \ && \ \tr[M^{(1)}]  \nonumber \\[3pt] 
{\rm 2^{nd}} \ && \ \tr[(M^{(1)})^2]\,,\  (\tr[M^{(1)}])^2 \nonumber\\[3pt] 
{\rm 3^{rd}} \ && \ \tr[(M^{(1)})^3 ]\,,\ \tr[(M^{(1)})^2 ]  \tr[M^{(1)}],\ (\tr[M^{(1)}])^3\,,\ \tr[M^{(1)} M^{(2)}] \label{eq:LagrBasis}  \\[3pt] 
{\rm 4^{th}} \ && \ \tr[(M^{(1)})^4 ]\,,\ \tr[(M^{(1)})^3 ]  \tr[M^{(1)}]\,,\  \left(\tr[(M^{(1)})^2 ]\right)^2\,,\ (\tr[M^{(1)}])^4\,,\nonumber\\
&& \ \tr[M^{(1)}] \tr[M^{(1)} M^{[2]}]\,,\  \tr[M^{(1)} M^{(1)} M^{(2)}]\,,
\ \tr[M^{(1)} M^{(3)}]\,, \   \tr[M^{(2)} M^{(2)}]\, . 
\nonumber
\eea
At fourth order, we have used the fact that, as a $3\times 3$ symmetric matrix, $M_{ij}^{(1)}$ is characterized by three independent rotational invariants, allowing us to eliminate $\tr[ (M^{(1)})^2] (\tr[ M^{(1)}])^2$. Starting at third order, $M_{ij}$ is no longer symmetric; that is, the displacement vector also has a curl component. However, at each order in perturbations, the antisymmetric part of $M_{ij}$ can be re-expressed in terms of combinations of lower-order contributions to the symmetric part, $M_{(ij)} \equiv (M_{ij}+M_{ji})/2$, via the equations of motion \cite{zheligovsky/frisch,matsubara:2015}. Hence, it is sufficient to write the bias expansion purely in terms of the contributions to the symmetric part, $M_{(ij)}^{(n)}$.

All of the operators in \refeq{LagrBasis} are easily evaluated in Lagrangian perturbation theory.  
The first instance of a convective time derivative appears at third order
in the bias expansion through the operator $\tr[M^{(1)} M^{(2)}]$, which is
precisely related to the operator $O_\otd^{(3)}$ introduced in \refsec{evol2} (see p.~\pageref{other:Otdrel}).  Note however that, while $M^{(1)}_{ij}$ and $M^{(2)}_{ij}$ can be rephrased in terms of second derivatives of the gravitational and velocity potentials, this is no longer true for $M^{(3)}_{ij}$ which appears at fourth order. Fortunately, with this construction it is now obvious how to extend the complete set of bias operators to higher orders.  
Thus, \refeq{LagrBasis} is directly applicable to calculate statistics of biased
tracers in LPT.  The disadvantage, as has already been mentioned after 
\refeq{biasLPT}, is that \refeq{LagrBasis} is not very convenient to connect to 
well-known perturbative bias expansions.  This is because the matter 
density contrast $\d$ and the tensor $\v{M}$ are nonlinearly related through
$\d(\vq) = |\v{1}+\v{M}|^{-1}-1$.  Thus, the well-known linear bias term $b_1 \d(\vq)$ contributes to many terms in the list \refeq{LagrBasis}.  
In the next section, we will derive an equivalent
basis that is closer to standard perturbative bias expansions.

\subsubsection{Eulerian basis of operators}
\label{sec:basisE}

A Eulerian basis can similarly be constructed out of 
$\partial_{x,i}\partial_{x,j}\Phi(\vx,\tau)$ and its convective time 
derivatives.  Here, we follow Ref.~\cite{MSZ} who have defined\footnote{Note that the prefactor was absorbed into the definition of $\Phi$ there.}
\be
\Pi^{[1]}_{ij}(\vx,\tau) = \frac{2}{3\Om\cH^2} \partial_{x,i}\partial_{x,j}\Phi(\vx,\tau)
=
K_{ij}(\vx,\tau) + \frac13 \delta_{ij}\d(\vx,\tau)\,,
\label{eq:hatPi}
\ee
which in the notation of \refsec{evol1} and \refsec{evol2} contains 
$\d = \tr \Pi^{[1]}$ and $K_{ij}$ as the trace-free part of 
$\Pi^{[1]}_{ij}$.  
Note that the superscript $[1]$, to be distinguished from $(1)$, refers to the
fact that $\Pi^{[1]}$ \emph{starts} at first order in perturbation theory,
but contains higher-order terms as well.
We then define the higher-order tensors $\Pi^{[n]}$ recursively 
by convective time derivatives:
\be
\Pi^{[n]}_{ij} = \frac{1}{(n-1)!} \left[(\cH f)^{-1}\convD \Pi^{[n-1]}_{ij} - (n-1) \Pi^{[n-1]}_{ij}\right]\,.
\label{eq:Pindef}
\ee
Note that $\Pi^{[n]}_{ij}$ is symmetric for any $n$. 
By construction, the lowest-order contribution to $\Pi^{[n]}$ in perturbation theory is at $n$-th order.  This helps us keep track of all the relevant terms at any given order.  However, unlike in the Lagrangian case, 
here the $\Pi^{[n]}$ are not simply the $n$-th order perturbative 
contributions to $\Pi^{[1]}$.  
This is because in the Eulerian case, convective time derivatives
[\refeq{DDtau}] are not equal to ordinary derivatives $\partial/\partial\tau$. 
The reason why a basis constructed out of $\Pi^{[n]}_{ij}$ is 
complete, however, is exactly analogous to the Lagrangian case discussed 
above: at $n$-th order in perturbation theory, $\Pi^{[1]}_{ij}$, which contains
the matter density and tidal field up to $n$-th order, only has $n$ different
time dependences $D(\tau), D^2(\tau), \cdots D^n(\tau)$.  
Therefore, any higher convective time derivative $(\convDinline)^m$ with $m > n$ can be expressed in terms of the first, 
second, $\cdots,\,n$-th time derivatives, 
when neglecting terms higher than $n$-th order in perturbation theory.  

The quantity $\tr[\Pi^{[n]}]$ is a linear combination of convective time derivatives of the Eulerian density perturbation $\d$.  At any given order, these can be written as combinations of lower-order operators, by way of the Eulerian fluid equations (for example, $\tr\left[\Pi^{(2)}\right]= (17/21)\d^2+(2/7)(K_{ij})^2$ in second-order PT; see \refapp{biastrans}), 
and can thus be excluded from the basis
for $n>1$.  This is in analogy with $\tr[M^{(n)}]$ in the Lagrangian basis.  
Correspondingly, the bias coefficients of these terms can be seen as
integrals over the kernels $f_{\tr \Pi^{[1]}}(\tau,\tau')$, $f_{(\tr\Pi^{[1]})^2}(\tau,\tau',\tau'')$ and so on, introduced in \refeq{dg_taylor} above, against progressively higher powers of the growth factor \cite{senatore:2015}.  

Up to fourth order, we therefore have, in exact formal analogy to \refeq{LagrBasis},
\bea
{\rm 1^{st}} \ && \ \tr[\Pi^{[1]}] \label{eq:EulBasis} \\[3pt] 
{\rm 2^{nd}} \ && \ \tr[(\Pi^{[1]})^2]\,,\  (\tr[\Pi^{[1]}])^2 \nonumber\\[3pt] 
{\rm 3^{rd}} \ && \ \tr[(\Pi^{[1]})^3 ]\,,\ \tr[(\Pi^{[1]})^2 ]  \tr[\Pi^{[1]}]\,,\ (\tr[\Pi^{[1]}])^3\,,\ \tr[\Pi^{[1]} \Pi^{[2]}] \nonumber \\[3pt] 
{\rm 4^{th}} \ && \ \tr[(\Pi^{[1]})^4 ]\,,\ \tr[(\Pi^{[1]})^3 ]  \tr[\Pi^{[1]}]\,,\  \left(\tr[(\Pi^{[1]})^2 ]\right)^2\,,\ (\tr[\Pi^{[1]}])^4\, ,\nonumber\\
&& \  \tr[\Pi^{[1]}] \tr[\Pi^{[1]} \Pi^{[2]}]\,,\  \tr[\Pi^{[1]} \Pi^{[1]} \Pi^{[2]} ]\,,
\  \tr[\Pi^{[1]} \Pi^{[3]}]\,,\  \tr[\Pi^{[2]} \Pi^{[2]}]\,. \nonumber
\eea
This basis offers the advantage of having a close connection to the standard
Eulerian bias expansion.  For example, the coefficient of
the term $(\tr[\Pi^{[1]}])^n$ is precisely $b_{\delta^n}=b_n/n!$, since $\tr[\Pi^{[1]}(\vx,\tau)] = \d(\vx,\tau)$ at all orders.  
The term $\tr[(\Pi^{[1]})^2] = (K_{ij})^2 + \d^2/3$ on the other hand contains
the tidal field squared.  Of course, as in the Lagrangian case, explicit time derivatives appear in the bias expansion at third order through the operator 
$\tr[\Pi^{[1]}\Pi^{[2]}]$, which again is directly related to the operator 
$O_\otd^{(3)}$ (see p.~\pageref{other:Otdrel} and \refapp{biastrans}). As mentioned after \refeq{LagrBasis}, gravitational and velocity shear are no longer sufficient in the bias expansion starting at fourth order, as evidenced by the appearance of $\Pi^{[3]}_{ij}$.\\

We now discuss the key approximation made in constructing the convenient bases
\refeq{LagrBasis} and \refeq{EulBasis}, namely that all operator contributions at
a given perturbative order $n$ have the same time dependence, 
$[D(\tau)]^n$.  This is only strictly true in an EdS Universe, 
while in $\Lambda$CDM and quintessence cosmologies
new time dependences appear at each new order.  For example, 
second-order operators can have a time dependence given by $[D(\tau)]^2$ 
or by $D_2(\tau)$, 
where $D_2(\tau) \propto \int D^2 d\ln D$ is the second-order growth factor.  
This means that the operators in the bases described above are in general not 
sufficient anymore.  However, we show in \refapp{convSPT:b} that 
the first instance of a new term in the bias expansion appears only at fourth order.  Specifically, the $\tr[\Pi^{[1]} \Pi^{[3]}]$ term in \refeq{EulBasis}
splits into two terms which, however, have to have very similar bias coefficients if 
the nonlinear growth factors approximately obey $D_n(\tau) \simeq [D(\tau)]^n$;
the departures from this relation are at the percent level for a standard $\Lambda$CDM cosmology. 
This means that the additional
operators added to complete the operator bases described here will be
$(i)$ fourth and higher order; $(ii)$ suppressed by a numerical prefactor
$\lesssim 0.1$ relative to the terms included in \refeq{EulBasis}. 
They will thus be irrelevant in most practical applications.  
Note that, since we only work to third order in perturbation theory there,
all results given in \refsec{measurements} hold in $\Lambda$CDM and 
quintessence cosmologies.  

\subsection{Higher-derivative bias}
\label{sec:higherderiv}

In the treatment of bias so far, we have approximated
the formation of halos and galaxies as perfectly local in a spatial sense.   
After reordering the time derivatives along the fluid trajectory, 
we have written the bias expansion for $\delta_g(\vx,\tau)$ in terms of  
operators evaluated at the same location: $O(\vx,\tau)$ in the Eulerian basis, 
or $O_L(\vq,\tau)$ in the Lagrangian basis.  
However, we know that the formation of halos and galaxies involves the 
collapse of matter from a finite region in space, and thus, the
\emph{local bias} expansion derived above cannot be completely correct
on all scales.  In this section, we study the limitation of the spatially-local approximation and derive the set of additional operators to include in the expansion \refeq{biasrel}.  We refer to these operators as \emph{higher-derivative operators}.  
They naturally arise in peak theory or the 
excursion-set approach \cite{bardeen/etal:1986,desjacques/crocce/etal:2010,musso/sheth:2012}
(see \refsec{exset} and \refsec{peaks} for a detailed discussion).  

In order to incorporate the deviation from perfect locality of galaxy
formation, we should replace the local operators $O(\vx,\tau)$ 
appearing in \refeq{biasrel} with \emph{functionals} 
\cite{matsubara:1999,coles/erdogdu:2007}.  
For example, the linear-order operator in the Eulerian basis, $O=\d$, 
now becomes
\ba
b_\d(\tau) \d(\vx,\tau) \to\:& \int d^3 \vy\, F_\d(\vy,\tau) \d(\vx+\vy,\tau) \,,
\label{eq:convL}
\ea
where $F_\d(\vy,\tau)$ is a kernel that is in general time dependent.  
Here, we have used the homogeneity of the Universe, or the absence of preferred 
locations, which dictates that $F_\delta$ is independent of $\vx$.  
We can now perform a formal series expansion of $\d$ around 
$\vx$, leading to
\ba
b_\d(\tau) \d(\vx,\tau) \to\:&
\left[\int d^3 \vy\, F_\d(\vy,\tau) \right] \d(\vx,\tau)
+ \left[\frac16  \int d^3 \vy\, 
|\vy|^2\, 
F_\d(\vy,\tau)\right]  \lapl_x \d(\vx,\tau)
+ \cdots \vs
=\:& b_\d(\tau) \d(\vx,\tau) + b_{\lapl\d}(\tau)\lapl_x \d(\vx,\tau) + \cdots\,,
\label{eq:hddelta}
\ea
where statistical isotropy demands the absence of any preferred directions with which the derivative operators 
could be contracted.  

Hence, there is no term linear in spatial derivatives, and the leading higher-derivative term involves the Laplacian of $\d(\vx,\tau)$.  We then 
identify the standard linear bias $b_\d = b_1$ with the ``total mass'' of the kernel, while the second integral, the ``moment of inertia'' 
of the kernel, defines a new higher-derivative bias parameter $b_{\lapl\d}$.  
This bias parameter has dimension $[\rm{length}]^2$;  the characteristic scale $R_*$ that sets the magnitude of 
$|b_{\lapl\d}| \sim R_*^2$ is the scale of the spatial support of the kernel $F_\d(\vy,\tau)$, which we identify 
with the nonlocality scale of the tracer $R_*$.  
Note that $\lapl\d$ is a local observable for a comoving observer and, following our discussion in \refsec{framework}, 
should thus be included in the general bias expansion in any case. Such a term naturally appears for example in the peak approach 
(\refsec{peaks}) without any explicit nonlocality, induced by the constraint of a negative curvature of the smoothed density field.  
The formal derivation from a kernel as in \refeq{hddelta} provides another physical interpretation for the 
bias parameter $b_{\nabla^2\d}$ and its relation to the spatial scale $R_*$.

Let us briefly discuss how baryonic effects are also captured by 
higher-derivative terms.   While this statement holds for general perturbations of the stress tensor of matter, we consider the effect of gas 
pressure perturbations $\delta p$ here for simplicity.  These are sourced by density perturbations, so that at linear order
\be
\delta p^{(1)} = c_s^2 \rhopb \delta^{(1)}\,,
\label{eq:deltap}
\ee
where $c_s$ is the sound speed, and $\rhopb$ is the mean physical baryon density. The relevant quantity for dynamics is the gradient $\vn \d p$ of the pressure. Thus, if 
we allow for the galaxy density to depend on $\vn\delta p$, we have to add the 
leading scalar quantities that can be constructed out of $\vn\delta p$ to the bias expansion, leading to 
\be
\delta_g \supset 
b_{\lapl \delta p} \lapl \delta p 
+ b_{(\vn\delta p)^2} (\vn\delta p)^2 
\propto b_{\lapl \delta p} \lapl\d + b_{(\vn\delta p)^2} (\vn\d)^2\,.
\label{eq:hddeltap}
\ee
We see that these terms are higher-derivative and, moreover, the only term linear in perturbations is exactly of the same form as in \refeq{hddelta}.  We will discuss nonlinear higher-derivative terms, such as $(\vn\d)^2$, below.  
On the other hand, in contrast to gas, radiation can travel large distances and hence, in principle, lead to a dependence of the local galaxy density on the matter distribution within a very large region; i.e. it can significantly increase the scale $R_*$.  
We will discuss realistic estimates of these effects below. 
For now, let us continue with the general discussion based on \refeq{hddelta}.  

In \refeq{convL}, the kernel $F_\d$ describes how the formation of galaxies 
depends on the precise distribution of matter in the vicinity of $\vx$.  
Therefore, it is sensible to connect the second moment of $F_\d$ with 
the nonlocality scale $R_*$ introduced in \refsec{framework}.  
If $F_\d$ is given by a typical filtering kernel, e.g. a Gaussian or tophat [cf. \refeqs{sharp-k_filter}{top-hat_filter}], 
then we expect that $b_{\lapl\d} > 0$:  if we are at the location of a peak in the density field, then the smoothed
density will be smaller than the un-smoothed one.  On the other hand, 
$\lapl \d < 0$ at such a location, and so a positive $b_{\lapl\d}$ leads to the
expected behavior, i.e. that peaks are damped by smoothing.  
Consider for example our toy model of \refsec{localbias}.  At linear order, the proto-halo density perturbation can be written as $\d_{h,L}(\vq) = b_1^L \d_R^{(1)}(\vq)$.  A formal expansion of the smoothed density field $\d_R^{(1)}$ around the un-smoothed linear density $\d^{(1)}$ yields
\be
\d_{h,L}(\vq,\tau) = b_1^L \d^{(1)}(\vq,\tau) + b_{\lapl\d}^L \lapl_q \d^{(1)}(\vq,\tau) + \cdots\,, \quad\mbox{where}\quad
b_{\lapl\d}^L = \frac1{10} b_1^L R^2 \,.
\label{eq:bhderiv_thr}
\ee
Here we have assumed a tophat filtering kernel to be specific.  
Note that $b_{\lapl\d}^L$ does not necessarily have to be positive;  peaks of the density field provide an illuminating example for the case $b_{\lapl\d}^L<0$ (see \refsec{peaks}). 
Continuing the derivative expansion, the next higher correction in \refeq{bhderiv_thr} scales
as $R^4 \laplsq \d$.

In Fourier space, the term proportional to $b_{\lapl\d}$ corresponds to a ``scale-dependent bias'' $-b_{\lapl\d} k^2 \d \propto (R_* k)^2 \d$, with higher corrections scaling as $(R_* k)^{2n} \d$.  However, let us emphasize again that 
this is an expansion in powers of $k^2$, rather than a general function $f(k)$,
which is how the term ``scale-dependent bias'' is sometimes interpreted.  
To make this distinction clear, we will use the term higher-derivative bias
throughout.    

At linear order, which we have assumed in \refeq{hddelta} and \refeq{bhderiv_thr}, the distinction between higher-derivative terms in Eulerian and Lagrangian frames, i.e.~$\lapl_x\d$ vs.~$\lapl_q\d$, is irrelevant.  This is no longer the case at nonlinear order in perturbations. Moreover, the relation between Lagrangian and Eulerian higher-derivative biases is complicated by \emph{velocity bias}, unlike the case in the local bias expansion, as we will see in the next section.
Crucially, the difference between derivatives with respect to $\vq$ and those with respect to $\vx$, which involves the distortion tensor $\v{M}$ [\reftab{EulLagr}], can be absorbed by nonlinear higher-derivative terms, which then render bias expansions in Eulerian and Lagrangian frames equivalent again.  We will return to this below.

\begin{figure}[t]
\centering
\includegraphics[width=\textwidth]{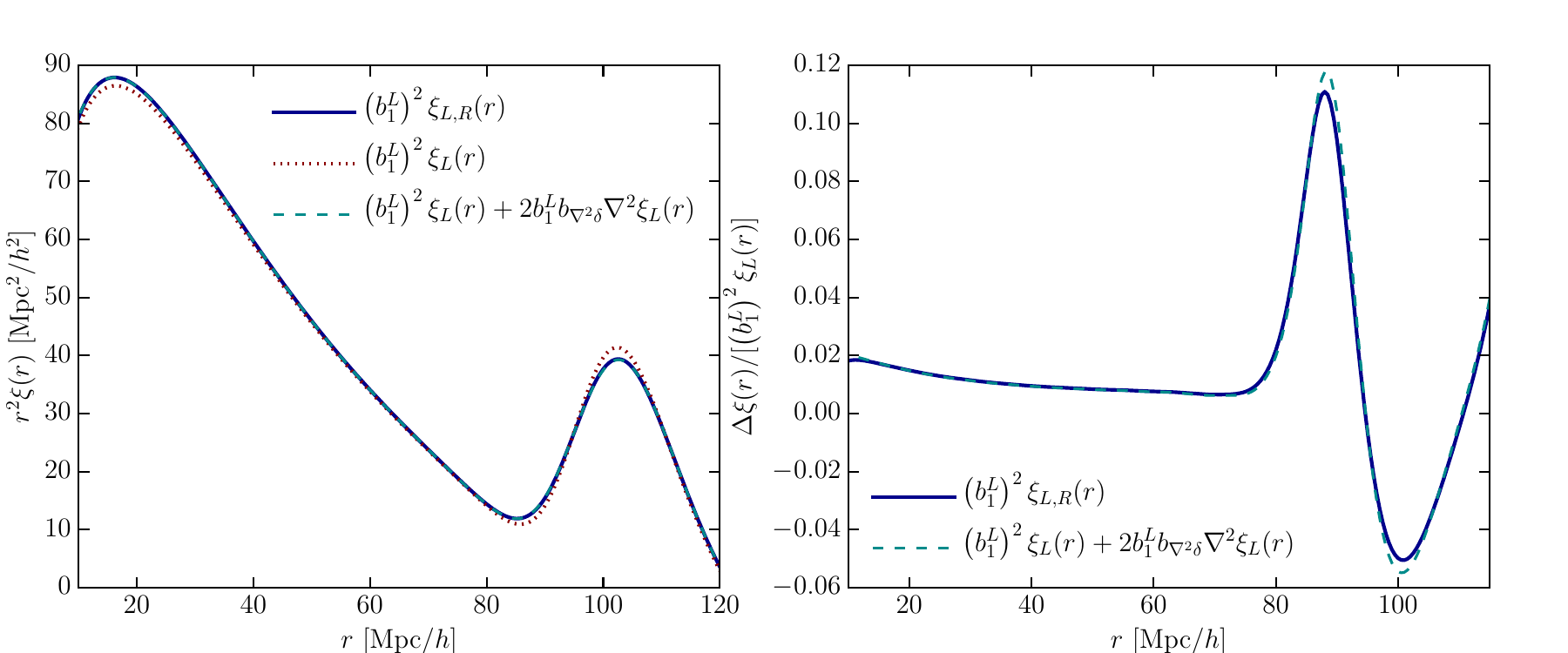}
\caption{
Illustration of the effect of higher-derivative bias in the context
of the thresholding toy model of \refsec{localbias}.  We show two-point correlation functions multiplied by $r^2$ in order to better illustrate the effect on the BAO feature.  
The values for $R = 4.21 \Mpch$ (mass scale $2.5\cdot 10^{13}\Msunh$), $b_1^L = 1.5$ and $z=0$ are the same as in \reffig{xi_thr}.  
\textit{Left panel:} Linear-order contribution $(b_1^L)^2 \xi_{{\rm L},R}(r)$ to \refeq{xihthr} (solid), and the linear \LIMD bias prediction $(b_1^L)^2 \xi_{\rm L}(r)$ (dotted); note that the former contains the filtering kernel while the latter
does not.  Some differences are seen on small scales and especially around the
BAO feature.  When adding the leading higher-derivative bias [\refeq{bhderiv_thr}] to the \LIMD prediction (dashed), the 
perturbative bias expansion matches the smoothed two-point correlation 
function very well.  
\textit{Right panel:}  Relative deviation of the results shown in the left panel from the linear \LIMD prediction.
\label{fig:xi_thr_hderiv}}
\end{figure}

In terms of the two-point correlation function of galaxies, the leading 
contribution is of the form
\be
\xi_{gg}(r)\Big|_\text{higher deriv.} = 2 b_1 b_{\lapl\d}\, \nabla_r^2 \xi(r)\,,
\label{eq:nonlocterm}
\ee
where $\xi(r)$ is the matter two-point correlation function.  
Note that this term can become observationally relevant
not just if the correlation scale $r$ is of order $R_*$, but on the much 
larger scale of the BAO feature
($r\sim 100\Mpch$) because of the narrow width of this feature 
\citep{desjacques:2008,desjacques/crocce/etal:2010}.  This is
illustrated,  in the context of the
thresholding toy model of \refsec{localbias}, in \reffig{xi_thr_hderiv} (left panel), which shows the
effect of smoothing (on the scale $R\simeq 4\Mpch$) on the two-point correlation 
function.  The smoothing damps the BAO feature slightly.  We also
show the effect of including the leading higher-derivative contribution
through \refeq{nonlocterm} and \refeq{bhderiv_thr}.  The right panel of \reffig{xi_thr_hderiv} clearly shows that the leading higher-derivative
term captures the bulk of the smoothing effect.  Thus, by 
including one additional bias parameter, we incorporate the
leading effect of the finite size of the thresholded regions, 
and improve the precision of the bias expansion significantly.  
At this point, it is worth
emphasizing that the higher-derivative terms 
primarily modify the amplitude of the BAO feature.
If the BAO scale is determined by marginalizing over the broad-band shape of 
the galaxy two-point correlation 
function, as is usually done in BAO studies, then 
the estimated scale is insensitive to $b_{\lapl\d}$.  On the other hand, if our 
goal is to extract the full information from the galaxy two-point correlation
function, this term has to be included (\refsec{bnpt}).

As we have seen, higher-derivative biases introduce an additional spatial scale, $R_*$,
into the perturbative bias expansion (unlike the \emph{local} bias expansion 
at leading order in derivatives,
which only involves the same Hubble time scale that 
governs the evolution of matter itself).  
The significance of this new scale is that it provides a fundamental cutoff
for the perturbative approach to galaxy clustering:  when the scale $r$ on which
we measure correlations approaches $R_*$, all higher-derivative terms become
relevant, and the perturbative description loses all predictive power.  
The same effect happens in Fourier space for $k \sim R_*^{-1}$.  Thus,
even if we were able to predict the properties of the matter density field
perfectly, the nonlocality of the formation of galaxies sets a fundamental
limit on the scales over which we can describe the statistics of galaxies
perturbatively.  
In practice, even for halos, it is still unclear whether higher-order local bias terms 
or higher-derivative operators provide the actual cutoff of the perturbative
theory.  For galaxies, the cutoff will most likely
depend strongly on the specific galaxy sample considered.  In the 
following, we discuss approximate estimates of and constraints on $R_*$
for different tracers:
\begin{itemize}
\item \bfem{Dark matter halos:} Since halo formation in N-body simulations is governed exclusively by gravity, one expects $R_*$ to be 
comparable to the Lagrangian radius $R(M)$ of halos.  This is because
the matter within a given halo originates from a region of size $R(M)$.  
This is also borne out by nonlinear models of halo formation such as the excursion set
(\refsec{exset}) and peaks of the Lagrangian density field (\refsec{peaks}), 
and indeed the toy model of \refsec{localbias},
where $R(M)$ is the filter scale used to define the significance
$\nu_c = \dc / \sigma(R)$  [just as we found in \refeq{bhderiv_thr}].  
As discussed above however, the precise value and indeed \emph{sign} of $b_{\lapl\d}$ depend on the details of the model considered.  
\item \bfem{Galaxies:}  If the properties of galaxies in a given
sample are completely determined by those of their host halos, 
as assumed in the halo occupation distribution and abundance matching
approaches (\refsec{HOD}), then the scale $R_*$ for these galaxies is
given by that of the host halos, i.e. $R(M)$.  
On the other hand, if the local rate of
galaxy formation is significantly modified by the radiation field (e.g., the
flux of ionizing UV radiation), then $R_*$ could be as large as the
absorption length of this radiation \cite{babul/white:1991,bower/etal:1993,schmidt/beutler:2017},
which can be several hundred Mpc.  Thermal heating of the intergalactic
medium by high-energy cosmic-ray cascades is another possibility
for long-range influences \cite{broderick/etal:2012,lamberts/etal:2015}, as
these cosmic rays have large mean-free paths as well.  
These effects can in turn strongly modify the shape of the galaxy two-point 
function and thus affect the measured position of the BAO feature \cite{pritchard/furlanetto/kamionkowski:2007,coles/erdogdu:2007}.  
Similar conclusions hold for large-scale outflows, for example, or strong jets launched by active galactic nuclei, although the scale of these phenomena
is expected to be at most tens of Mpc \cite{vandaalen/etal:2014} and thus 
significantly smaller than the mean free path of UV photons.
\item \bfem{Line emission from diffuse gas:}  This is an interesting case,
which includes the Lyman-$\alpha$ forest as well as intensity mapping.  Ignoring the effects of any large-scale fluctuations in the ambient radiation field, 
the nonlocality scale of the gas is of order the Jeans scale, $R_* \sim 1/k_J$,
which is very small ($1/k_J \sim 0.1 \iMpch$ \cite{schaye:2001,meiksin:2009}) for the relatively cold gas observed using these channels.  Combined with the fact that these tracers are observed
at fairly high redshifts, where the nonlinear scale is small, this suggests
that line emission from the intergalactic medium can be modeled accurately
to very small scales.  Unfortunately, the line emission depends on the
ionization state of the medium, which in turn is controlled by the 
ambient radiation field \cite{meiksin/white:2004,croft:2004}.  As mentioned above, the mean free path of
ionizing radiation in the intergalactic medium is very large, so that
in fact $R_*$ for these tracers is \emph{not} small.  Ref.~\cite{pontzen:2014}
provides a very clear illustration of this effect 
on large-scale statistics of the Lyman-$\alpha$ forest.  
The information loss by being restricted to very large scales above $R_*$
can however be reduced by explicitly modeling this effect through measured 
cross-correlations with sources of ionizing radiation such as galaxies
and quasars \cite{pontzen/etal:2014}.  \\
\end{itemize}

Finally, we turn to higher-derivative corrections to nonlinear operators in the bias expansion.  While technically more complicated, this follows in strict analogy to the linear case leading to \refeq{hddelta};  the remainder of the section is not essential for the subsequent developments and can be skipped on a first reading.  Let us work again in Eulerian coordinates, noting that exactly the same reasoning goes through in the Lagrangian basis in terms of derivatives with respect to $\vq$.  As described in \refsec{basisE}, each operator in the local basis can be written as
\be
O(\vx,\tau) = \Pi^{[i_1]}(\vx,\tau) \cdots \Pi^{[i_n]}(\vx,\tau)\,,
\ee 
where $\Pi_{ij}^{[n]}$ is defined in \refeqs{hatPi}{Pindef}, and we have suppressed tensor indices in this expression since they are irrelevant for the following arguments.  
Going beyond locality then means that we should introduce a kernel $F_O(\vy_1,\vy_2,\cdots,\vy_n;\tau)$ to replace terms in the local bias expansion so that,
again suppressing tensor indices,
\ba
b_O(\tau) O(\vx,\tau) \to\:& \int d^3 \vy_1\cdots d^3 \vy_n\,
F_O(\vy_1,\cdots,\vy_n;\tau) \Pi^{[i_1]}(\vx+\vy_1,\tau) \cdots \Pi^{[i_n]}(\vx+\vy_n,\tau)\,.
\label{eq:convNL}
\ea
Note that the operators $\Pi^{[n]}_{ij}$ are local observables, and the formation of 
halos can depend on the detailed distribution of the $\Pi^{[n]}_{ij}$ within the scale $R_*$.  Thus, they also form the building blocks for the general higher-derivative expansion.  It is then straightforward to perform the same Taylor expansion around $\vx$ as in \refeq{hddelta}, for each factor $\Pi^{[i_j]}$, resulting at leading order in terms of the type
\ba
& \Pi^{[i_1]}(\vx,\tau) \cdots \left[\lapl_x \Pi^{[i_j]}(\vx,\tau)\right] \cdots \Pi^{[i_n]}(\vx,\tau)
\quad\mbox{and}\quad \vs
& \Pi^{[i_1]}(\vx,\tau) \cdots \left[\partial_{x,k}\Pi^{[i_j]}(\vx,\tau)\right] \cdots
\left[\partial_x^l\Pi^{[i_k]}(\vx,\tau)\right]
\cdots \Pi^{[i_n]}(\vx,\tau)\,.
\ea
Note that the indices $k$ and $l$ in the second line can be
contracted among themselves or with the tensor indices of the 
$\Pi^{[n]}_{ij}$.  
\emph{Thus, in order to obtain the complete set of higher-derivative operators, we
have to allow for all contractions of derivatives on each
$\Pi$ factor in the operator basis \refeq{EulBasis}, including contractions
with the tensor indices of the $\Pi$ themselves.}  
The analogous construction works for the Lagrangian basis \refeq{LagrBasis} in terms of $\partial_q$ acting on $M_{ij}^{[n]}$.  
Including the complete set of higher-derivative terms is also necessary and sufficient to ensure that the Lagrangian 
and Eulerian bias expansions are equivalent at higher order in derivatives.  

Explicitly, the leading higher-derivative terms $\O(R_*^2)$ in the Eulerian basis are, up to second order in perturbations, given by
\bea
\O(R_*^2):\  {\rm 1^{st}} \ && \  \lapl\, \tr[\Pi] \label{eq:hdEulBasis} \\[3pt] 
{\rm 2^{nd}} \ && \ \tr[(\lapl \Pi) \Pi]\,,\  
\tr[(\partial_i \Pi) (\partial^i\Pi)]\,,\  
(\lapl \tr[\Pi])  \tr[\Pi]\,,\  
(\partial_i \tr[\Pi]) (\partial^i \tr[\Pi])
\,, \vs
&& \  \partial_k \Pi_{ij} \partial^i \Pi^{kj}\,,\  
\Pi^{kl} \partial_k \partial_l \tr[\Pi]\,,\  
\Pi^{kl} \partial_k \partial_i \Pi^i_{\  l}\,,
\partial^i\Pi_{ij} \partial^j \tr[\Pi]\,,\  \partial^i\Pi_{ij} \partial_k \Pi^{jk}\,,
\nonumber
\eea
where we have denoted $\Pi \equiv \Pi^{[1]}$ for clarity. Note that 
the second line includes the $(\nabla\delta)^2$ term
already discussed in the context of pressure perturbations, \refeq{hddeltap}.  
Clearly, the number of higher-derivative operators and corresponding bias
coefficients increases rapidly toward higher order in perturbation theory.  
However, as outlined above, all the contributions quoted here are suppressed
by a factor of $(R_* k)^2$ on large scales relative to the leading
terms in \refeq{EulBasis}.

\subsection{Velocity bias}
\label{sec:velbias}

Starting from \refeq{contg}, we have assumed that galaxies and matter comove 
along the same
fluid trajectories.  We now show why this is consistent within the local bias expansion.  
Galaxies in general experience different peculiar forces than the matter fluid.  This is both due to the fact that galaxies are strongly influenced by baryonic 
physics, and that galaxy formation happens within a spatial region of finite size $R_*$ (\refsec{higherderiv}),
so that their center-of-mass acceleration is some weighted mean
of the local gravitational acceleration field within this region.  
Let us denote the peculiar
acceleration of galaxies with respect to the matter fluid as $\v{a}_g$.   
The Euler equation for the galaxy velocity field $\v{v}_g$ is then
\be
\frac{\partial}{\partial \tau} \v{v}_g + \cH \v{v}_g + (\v{v}_g\cdot\nabla)\v{v}_g
= -\nabla\Phi+\v{a}_g\,.
\label{eq:Eulerg}
\ee
Note that, in the effective field theory of the matter fluid (EFT \cite{baumann/etal:2012}, \refapp{EFT}), a peculiar acceleration $\v{a}_m$ also appears 
in the Euler equation for the matter fluid velocity $\v{v}$.  There, $a_m^i=-\partial_j\tau^{ij}/\rho_m$ is sourced by the effective stress tensor of matter $\tau^{ij}$ that
is induced by integrating out the small-scale non-perturbative modes (see \refapp{EFT}).  
Strictly speaking, we define $\v{a}_g$ here as the difference between the
large-scale acceleration field of galaxies and $\v{a}_m$ (denoted as $\v{f}_g-\v{f}$ in \cite{MSZ}).  
Subtracting the Euler equation from \refeq{Eulerg} and defining $\v{v}_{\rm rel} \equiv \v{v}_g - \v{v}$, one can then easily show that \cite{MSZ}
\ba
\convD\v{v}_{\rm rel} + \cH \v{v}_{\rm rel} + (\v{v}_{\rm rel}\cdot\nabla)\v{v}_g =\:& \v{a}_g \,,  \label{eq:vrel}
\ea
where $\convDinline$ is throughout defined with respect to the matter velocity $\v{v}$. 
We see that $\v{a}_g$ is the source of galaxy velocity bias $\v{v}_g-\v{v}$.  
If we assume a conserved tracer in the spirit of \refsec{evol2}, the continuity equation becomes
\ba
\convD(\delta_g-\delta) =\:& -\theta(\delta_g-\delta)-\nabla\cdot [(1+\delta_g)\v{v}_{\rm rel}]\,.
\label{eq:conthd}
\ea
Our general results on velocity bias derived below are independent of the conserved-tracer assumption, however.  

Since an observer in a given galaxy can in principle measure the relative acceleration between the galaxy's center of mass and the matter fluid, 
$\v{a}_g$ is a locally observable quantity.  As such, its effective large-scale expansion can again only involve gravitational observables, that is, $\partial_i\partial_j\Phi$ and its derivatives.  
Expanding $\v{a}_g$ in powers of $\Phi$, isotropy implies that each term involving $n$ $\Phi$ fields must have at least $2n+1$ derivatives.  This is simple to see: any local observable composed of $n$ powers of $(\partial_i\partial_j\Phi)$, say, has an even number of indices.  In order to construct a vector $a_g^k$, we have to add an additional derivative $\partial^k$, leading to $2n+1$ derivatives in total.  The single leading term ($n=1$) is given by $a_g^i \propto \partial^i \lapl\Phi$.    Physically,
the peculiar acceleration can be sourced both by (statistical) differences
in the local gravitational potential gradient for galaxies and matter,
induced for example by smoothing or the peak constraint, and by non-gravitational effects such as the pressure forces discussed around \refeqs{deltap}{hddeltap}.  Indeed, we have seen in the previous section that both effects lead to higher-derivative terms [\refeq{hddelta} and \refeq{hddeltap}, respectively].  

We have thus proven that galaxy velocity bias is a \emph{higher-derivative} effect,
which justifies why we were able to consistently set $\v{v}_g = \v{v}$ in \refsecs{dynamics}{evol2}. The general velocity bias expansion then consists of all 3-vectors that can be constructed from the local gravitational observables and their spatial derivatives, following the same procedure as described in the previous section. At leading order, there is only a single contribution, which we will consider below.

Note that a velocity bias that is not suppressed by derivatives has been studied in
the literature \cite{elia/etal:2011,chan/scoccimarro/sheth:2012}.  However, as
argued here, such a velocity bias violates the equivalence principle.  
On the other hand, if the galaxy density depends on multiple fluids, such as baryons and CDM, a \emph{relative}
velocity perturbation between these fluids is a local observable even without any derivatives.  
If present in the initial conditions on large scales, such a relative velocity between baryons and CDM also needs to be taken into account in the bias expansion, as described in \refsec{baryons}. 

By construction, our expansion only includes the leading terms in the
large-scale limit.  On small scales, a velocity bias of sub-halos within
massive dark matter halos is well established in simulations 
\cite{carlberg:1994,colin/klypin/kravtsov:2000,jennings/baugh/hatt:2015}.  
These fully nonlinear effects are beyond the reach of the perturbative
treatment described here.  
Further, while the galaxy velocity is unbiased at the level of the
local bias expansion, the galaxy momentum density $\v{j}_g = (1+\d_g) \v{v}_g$ is not,
since it is weighted by the galaxy density.  All measurements of velocities
of halos and galaxies in simulations and observations naturally yield
the momentum density and must be carefully re-weighted to obtain an accurate
estimate of $\v{v}_g$ \cite{pueblas/scoccimarro:09,zheng/zhang/jing:2015}.  
Ref.~\cite{zheng/zhang/jing:2015} empirically verified that halo velocity 
bias on scales $k \leq 0.1 \iMpch$ is less than 2\%.  

Let us now consider the leading higher-derivative velocity bias at linear order,
and its effect on the evolution of bias.  In order to solve \refeq{vrel}
at linear order, we need an expression for the peculiar acceleration $\v{a}_g$ 
(again, strictly speaking it is the effective relative acceleration between galaxies
and matter on large scales).  As discussed above, 
$\v{a}_g$ has at least three derivatives acting on $\Phi$.  At linear
order and leading order in derivatives, $\v{a}_g$ thus has to be proportional to $\vn\d$, i.e.
\be
\v{a}_g(\xfl(\tau),\tau) = A_g(\tau) \vn\d(\xfl(\tau),\tau)\,.
\label{eq:a_gi}
\ee
Other possible choices, such as $\lapl \v{v}$, are equivalent to $\vn\d$ at linear order.  
The dimensionless proportionality constant $A_g(\tau)$, essentially a bias parameter quantifying the 
effective acceleration due to small-scale perturbations, is in general a free function of time.  $A_g$ is expected to be of the same order
as the leading higher-derivative biases (\refsec{higherderiv}), i.e. $A_g = \O(R_*^2\cH^2)$.  Note
that since $R_*$ is a spatial scale and there are no preferred directions
in the galaxy's rest frame, the lowest nontrivial dependence on $R_*$ has
to be of this form (cf. the low-$k$ expansion of a generic spherically
symmetric convolution kernel $W_R(k) = c_1 + c_2 k^2 R^2 + \cdots$).  

We can then immediately integrate the linearized version of \refeq{vrel} to obtain
\ba
\v{v}_g(\vx,\tau) =\:& \v{v}(\vx,\tau) + \frac1{a(\tau)} \int_0^\tau d\tau'\, a(\tau') \v{a}_g (\vx,\tau')
= \v{v}(\vx,\tau) + \beta_{\vn\d}(\tau) \vn \d(\vx,\tau) \,,\quad\mbox{where}
\vs
\beta_{\vn\d}(\tau) \equiv\:& \frac1{a(\tau)D(\tau)} \int_0^\tau d\tau'\, 
a(\tau') D(\tau') A_g(\tau')
\,.
\label{eq:bthetarel}
\ea
Thus, we can trade the acceleration bias $A_g(\tau)$ for a
bias in the galaxy velocity $\beta_{\vn\d}(\tau)$.  We will reserve the notation $\beta_O(\tau)$ for velocity-bias parameters. 
It is instructive to consider two simple limiting cases for the time
dependence of $\v{a}_g(\tau)$.  
If $\v{a}_g(\tau)=$~const, implying $A_g(\tau) \propto D^{-1}(\tau)$, and assuming an
EdS Universe for simplicity, \refeq{bthetarel} immediately
yields $\beta_{\vn\d}(\tau) \lapl\d(\vx,\tau) \propto \tau$, i.e.
the velocity-bias term has the same time dependence as $\v{v}$ in the standard growing mode itself.  In this case, one can write
\be
\v{v}_g(\vx,\tau) \stackrel{\v{a}_g = \rm const}{=} 
\left[1 + \beta_{\lapl\v{v}} \lapl \right] \v{v}(\vx,\tau)\,,
\ee
where $\beta_{\lapl\v{v}}$ is constant.  This time evolution was first proposed in the context of the peak model 
\cite{desjacques/crocce/etal:2010,baldauf/desjacques/seljak:2015} (\refsec{velocitybias}).  

If on the other hand $\v{a}_g(\tau) \propto \d_D(\tau-\tau_*)$, corresponding to an 
instantaneous boost of galaxy velocities relative to matter at time $\tau_*$, then
$(\v{v}_g-\v{v})(\vx,\tau) \propto a^{-1}(\tau) \vn \d(\xfl(\tau_*),\tau_*) \propto a^{-1}(\tau)$ for $\tau > \tau_*$ [equivalently, $\beta_{\vn\d} \propto (a D)^{-1}$, \refeq{bthetarel}].  
This corresponds to the decaying relative-velocity mode 
of a system of two fluids coupled by gravity \cite{chan/scoccimarro/sheth:2012}, and can be understood as the usual 
decay of peculiar velocities (when not sourced) in an expanding Universe.  Thus, the different results on the
evolution of large-scale velocity bias obtained in the literature are a consequence
of different assumptions on the time evolution of the relative acceleration $\v{a}_g$.  

Finally, we consider the setup studied in \refsecs{dynamics}{evol2}.  That is, we prescribe a bias relation at a ``formation time'' $\tau_*$, and assume conserved evolution of the tracers afterwards.  We can then integrate the continuity equation \refeq{conthd} to obtain the galaxy density $\d_g$.  At linear order and including the leading higher-derivative term, we write the galaxy density at an initial time $\tau_*$ as
\be
\d_g(\vx,\tau_*) = b_1^* \d(\vx,\tau_*) + b_{\lapl\d}^* \lapl \d(\vx,\tau_*)\,.
\ee
At some later time $\tau$, $\d_g$ is then given by
\ba
\label{eq:velbG}
\d_g(\vx,\tau) 
=\:& b_1^E(\tau) \d(\vx,\tau) + b_{\lapl\d}^E(\tau) \lapl \d(\vx,\tau)
\vs
\mbox{where}\quad b_{\lapl\d}^E(\tau) =\:& 
b_{\lapl\d}^* \Dstar 
- \int_{\tau_*}^\tau d\tau' \beta_{\vn\d}(\tau') \frac{D(\tau')}{D(\tau)}\,.
\ea
We have used $\Dstarinline \equiv D(\tau_*)/D(\tau)$ and $b_1^E(\tau) = 1 + (b_1^*-1)(\Dstarinline)$
as defined in \refsec{evol1}. 
This generalizes the result of \cite{desjacques/crocce/etal:2010} to a time-dependent
$\beta_{\vn\d}$. Following our arguments after \refeq{a_gi}, the two contributions 
to $b_{\lapl\d}^E$ are expected to be of the same order $\sim R_*^2$, noting that 
$A_g \sim R_*^2\cH^2$, while $\beta_{\vn\d} \sim R_*^2 \cH$.  
We see that the evolution of the higher-derivative bias in the density depends on 
the time evolution of the velocity bias.  
Unlike the case of the local bias expansion, where the evolution of
bias parameters for a conserved tracer at a given order is uniquely determined by the knowledge
of the bias parameters \emph{at fixed time}, at higher order in derivatives 
we require knowledge of the velocity bias $\beta_{\vn\d}$ along
the \emph{entire trajectory}.  This is because higher-derivative terms
allow for non-gravitational effects to play a role, whose evolution is not necessarily
tied to the gravitational evolution of matter.  We stress that it is
still possible to completely describe galaxy bias at a fixed time,
on sufficiently large scales, in terms of a finite number of bias parameters.  

Going back to the 
two limiting cases for $A_g$ considered above, we find that, for
$\v{a}_g =$~const, the velocity bias contribution to $b_{\lapl\d}^E(\tau)$
is also constant (in EdS), so that for $\tau\gg \tau_*$
this eventually becomes the dominant contribution, as is the case in the 
peak model (\cite{desjacques/crocce/etal:2010}; see \refsec{velocitybias}).  
On the other hand, for an initial boost, $A_g \propto \d_D(\tau-\tau_*)$,  
this term decays as $\tau^{-3}$ in EdS (this corresponds to the adiabatic
decaying mode of the density field).  In this case, the initial higher-derivative bias contribution, which scales as $b_{\lapl\d}^* \Dstarinline \propto \tau^{-2}$, dominates over the velocity bias contribution at late times.  Of course, all of these results are only valid under the assumption of conserved evolution.

\subsection{Stochasticity}
\label{sec:stoch}

The bias expansion described so far captures the impact of long-wavelength 
perturbations on the galaxy density.  That is, we have ignored
the influence of small-scale perturbations on the formation of galaxies,
which is stochastic, as the small-scale initial conditions are not correlated
over long distances (this is a consequence of the assumed Gaussianity of the initial
conditions).  In order to take this into
account, we have to introduce stochastic fields in the bias relation, as already done in
\refsecs{evol1}{evol2}.  This is related to a phenomenon known in the literature as \emph{stochastic bias}  \cite{dekel/lahav:1999,taruya/soda:1999,matsubara:1999}.  
Further, in \refsecs{evol1}{evol2}
we have seen that stochasticity on one time slice couples to gravitational
evolution, and thereby introduces further stochastic terms at higher order
such as the term $\eps_\delta \delta$ in \refeq{dg2nd_inst2}.  

The fully general set of stochastic contributions then consists of
all terms of the deterministic bias expansion $O$, with independent stochastic
parameter $\eps_O$.  Crucially, since the $\eps_O$ are uncorrelated with
large-scale perturbations, they are 
\emph{completely described by their joint 1-point PDF} on large scales, or equivalently their real-space
moments $\<\eps_O \eps_{O'}\>$, $\<\eps_O \eps_{O'} \eps_{O''}\>$, and so on.  
Consider as an example the galaxy two-point function, which we will
describe in detail in \refsec{bnpt}.  The leading stochastic contribution
is
\ba
& \< \d_g(\vx_1) \d_g(\vx_2) \>\Big|_\text{leading stoch.} 
= \< \eps(\vx_1) \eps(\vx_2) \> = \Peps\, \d_D(\vx_1-\vx_2)\,,
\label{eq:Pggstoch}
\ea
where
\be
\Peps \equiv \lim_{k \to 0} \< \eps(\vk) \eps(\vk') \>'\,.
\ee
is the power spectrum of the field $\eps$ in the large-scale limit.  
Here and throughout, a prime on an expectation value denotes that
the momentum-conserving Dirac delta function is to be dropped.  
Note that in real space, the stochastic contributions are localized
at zero lag (but see below).  

Next, we consider the galaxy three-point function, whose leading stochastic contributions are 
\ba
& \< \d_g(\vx_1) \d_g(\vx_2) \d_g(\vx_3) \>\Big|_\text{leading stoch.} = \< \eps(\vx_1) \eps(\vx_2) \eps(\vx_3) \> + \left\{\< \eps(\vx_1) (\eps_\d \d)(\vx_2) \d(\vx_3) \> + \perm{5} \right\}\vs
& \qquad = \Beps\, \d_D(\vx_1-\vx_2) \d_D(\vx_2-\vx_3) + 
\left\{2 \Pepsepsd\,  \d_D(\vx_1-\vx_2) \< \d(\vx_2) \d(\vx_3) \> + \perm{3} \right\}\,,
\label{eq:Bgggstoch}
\ea
where 
\ba
\Beps \equiv \lim_{k,k' \to 0} \< \eps(\vk) \eps(\vk') \eps(\vk'') \>'
\quad\mbox{and}\quad
\Pepsepsd \equiv \lim_{k\to 0} \< \eps(\vk) \eps_\d(\vk')\>'\,.
\label{eq:Bepsdef}
\ea
We see that, when expanding correlators using
Wick's theorem, only correlators of the $\eps_O$ among themselves remain.  
This is a consequence of the fact that the $\eps_O$ are uncorrelated
with the long-wavelength perturbations, such that $\< \eps_O \d\>=0$.  

Since the coefficients $\eps_O$ themselves are first-order 
perturbations, the complete set of stochastic terms up to third order becomes,
working in the Eulerian local basis and thus neglecting higher-derivative operators,
\bea
{\rm 1^{st}} \ && \ \eps \label{eq:stochBasis} \\[3pt] 
{\rm 2^{nd}} \ && \  \eps_\d \tr[\Pi^{[1]}] \nonumber\\[3pt] 
{\rm 3^{rd}} \ && \  \eps_{\tr[\Pi^2]} \tr[(\Pi^{[1]})^2]\,,\  \eps_{(\tr \Pi)^2} (\tr[\Pi^{[1]}])^2 \,, \nonumber
\eea
with an analogous expansion in the Lagrangian case.  Note that these
terms have already been included in the third-order derivation of
\refsec{evol2}.  Going back to \refeq{epsE2} and \refeq{epsE3}, we see that
gravitational evolution mixes the various stochastic terms.  This shows
that we also have to allow for covariance (cross-correlation) between different stochastic
fields $\eps_O$ and $\eps_{O'}$, as written in \refeq{Bgggstoch}.  
The higher-derivative terms will similarly have stochastic counterparts, for example $\eps_{\lapl\d} \lapl\d$.   

Finally, we also need to take into account that
the galaxy density contrast is not expected to depend on the small-scale perturbations in 
an exactly local sense, but rather depends on their spatial distribution within 
a finite region of order $R_*$ 
(much like its dependence on the long-wavelength perturbations).  
This implies the necessity of including 
higher derivatives of the stochastic moments.  In Fourier space, this translates into a series expansion in $k^2$, that is,
\be
\<\eps_O (\vk) \eps_{O'}(\vk') \>' = P^{\{0\}}_{\eps_O\eps_{O'}} + P^{\{2\}}_{\eps_O\eps_{O'}}\, k^2
+ P^{\{4\}}_{\eps_O\eps_{O'}}\, k^4 + \cdots\,,
\label{eq:epsOgeneral}
\ee
where the $P^{\{n\}}_{\eps_O\eps_{O'}}$ are only functions of time.  Moreover, following
the discussion above, we expect the higher-derivative moments to scale approximately as $|P^{\{n\}}_{\eps_O\eps_{O'}}| \sim R_*^n P^{\{0\}}_{\eps_O\eps_{O'}}$. 
Note however, that the stochastic correlators arise through the coupling of two or more small-scale (non-perturbative) modes to a large-scale (perturbative) mode. The boundary between these two regimes is the nonlinear scale $\knl$ [\refeq{knldef}]. Thus, stochastic contributions with a different scaling which involves $\knl$ instead of $R_*$ are also possible. We return to observational constraints on the stochastic correlators in \refsec{meas:stoch}.

In the context of field theory, these terms are referred to as 
``contact terms'', as they formally correspond, in real space, to taking $n$ powers of the Laplacian on a Dirac-delta two-point correlation function.  
However, care must be taken when deriving the corresponding result in real space.  \refeq{epsOgeneral}
cannot be simply Fourier transformed, since it is a low-$k$ expansion
and the Fourier transform relies on the contribution of all $k$ modes.  
Physically, we
expect the correlations of the stochastic terms to be localized to 
scales of order $R_*$ or less, i.e.
\be
\< \eps_O(\vx) \eps_{O'}(\vx+\vr) \> \stackrel{r \gg R_*}{\longrightarrow} 0\,.
\label{eq:epsOreal}
\ee
On scales much larger than $R_*$, they can then be approximated as 
effective delta functions, $\< \eps_O(\vx) \eps_{O'}(\vx+\vr)\> = P^{\{0\}}_{\eps_O\eps_{O'}} \d_D(\vr)$, equivalent to keeping the leading term in \refeq{epsOgeneral}.  
As discussed above, the real-space counterparts to the sub-leading terms in
\refeq{epsOgeneral} are formal derivatives of $\d_D(\vr)$.  This reflects the fact that on scales $r \sim R_*$, any perturbative description has to 
break down, similar to the breakdown of the expansion \refeq{epsOgeneral}
for $k \sim 1/R_*$.

In order to illustrate these considerations, let us adopt a toy model
for the power spectrum of $\eps(\vk)$ that is presumed valid on all scales:
\be
P_{\eps}(k) \equiv \< \eps(\vk) \eps(\vk') \>' = \frac1{\avng} \left[1 - c\, e^{-k^2 R_*^2}\right]\,,
\ee
where $\avng$ is the mean galaxy number density and $c \in [0,1]$ is a constant.  
For $k \gg 1/R_*$, this approaches the Poisson shot noise $1/\avng$, while 
in the low-$k$ limit, this asymptotes to a smaller value $\Peps = (1-c)/\avng$.  
Qualitatively, this behavior matches the expectation from halo exclusion,
where $c$ corresponds to the Lagrangian volume fraction occupied by the
halos considered \cite{smith/etal:2007,hamaus/seljak/etal:2010,baldauf/seljak/etal:2013,chan/etal:2014}.  
We can Fourier transform this power spectrum to obtain the corresponding two-point correlation function, yielding
\be
\xi_{\eps}(r) = \frac{1}{\avng}\left[\d_D(\vr) - c(8\pi^{3/2} R_*^3) e^{-r^2/4R_*^2}\right]\,.
\ee
Clearly, this obeys \refeq{epsOreal}.  
Note that the apparently unphysical delta function will always yield finite results in practice, 
since the correlation function cannot be measured at strictly zero separation, and is instead 
integrated over a finite volume.  If the volume that is integrated over is much larger than $R_*^3$, then the variance of $\eps$ becomes independent of scale, just as expected for an effective white-noise distribution.  
Thus, while the stochastic terms contribute to Fourier-space statistics of
galaxies on all scales, in real space they only contribute at
small separations $r \sim R_*$, and to statistics that involve zero-lag correlators.  

Finally, there are also stochastic contributions to galaxy velocities.
As in \refsec{velbias}, we consider the effective relative velocity between
galaxies and matter.  In addition to the leading-order deterministic
velocity bias $\beta_{\lapl \v{v}}$, we now include a stochastic contribution:
\be
\left[v_g^i - v^i\right](\vx,\tau) = \beta_{\lapl \v{v}}(\tau) \lapl v^i(\vx,\tau) + \eps_v^i(\vx,\tau)\,,
\ee
where the vector $\eps_v^i$ is characterized by the same properties as the stochastic
contributions to the galaxy density: it is uncorrelated with the large-scale operators, and completely described by its zero-lag moments.  Compared to the leading stochastic field $\eps$ in the galaxy density, an additional constraint is imposed on these moments by the equivalence principle.  For the same reasons as described for the deterministic velocity bias $\beta_{\lapl\v{v}}$ in \refsec{velbias}, the stochastic contribution to $\v{v}_g-\v{v}$ has to be constructed from local observables, such as the density and tidal field; this holds both for gravitational and non-gravitational sources of $\v{v}_g-\v{v}$.  Thus, on large scales the
relative velocity can always be written as a total derivative of a nonlinear
function of the density and tidal field: $[v_g^i - v^i](\vx) = \partial^i F(\delta(\vx), K_{lm}(\vx))$. This fact implies that the power spectrum of the stochastic relative velocity between galaxies and matter has the following scaling on large scales:
\be
\lim_{k\to 0} \< \eps_v^i(\vk) \eps_v^j(\vk') \>' = k^i k^j P^{\{2\}}_{\eps_v}\,,
\label{eq:Pepsv}
\ee
where $P^{\{2\}}_{\eps_v}$ is a positive constant.  That is, there is no white-noise term
$P^{\{0\}}_{\eps_v}$ unlike for $\eps$.  Similarly, we have, for the
cross-correlation with $\eps$, $\lim_{k\to 0} \< \eps_v^j(\vk) \eps(\vk') \>' = i k^j P^{\{1\}}_{\eps\eps_v}$.  The stochastic velocity bias plays an important role for galaxy statistics in redshift space \cite{perko/etal:2016}. 

Above, we have argued that
$P^{\{2\}}_{\eps} \sim R_*^2 P^{\{0\}}_{\eps} \sim R_*^2/\avng$, if stochasticity
in the galaxy number density is close to Poisson.  We now give
a rough order-of-magnitude estimate for the stochastic amplitudes of the velocity. First, for $P^{\{1\}}_{\eps\eps_v}$, one can follow the reasoning described after \refeq{velbG}, to roughly estimate that this term should not be larger than of order $R_*^2 \cH P^{\{0\}}_{\eps\eps}$, and, similarly, $P^{\{2\}}_{\eps_v}\lesssim R_*^4 \cH^2 P^{\{0\}}_{\eps\eps}$.  We can provide a sharper estimate by adopting a concrete physical model. Let us assume
that the stochastic velocity contribution of a given galaxy sample is due to
the virial velocities within the host halos of mass $M$ and Eulerian radius $R_E$ of these galaxies.  The variance of virial velocities is $\s_{v,\rm vir}^2 \approx GM/R_E$.  Matching this variance to \refeq{Pepsv} yields
\ba
\s_{v,\rm vir}^2 = \int_{\vk} P_{\eps_v}(k) \sim \frac1{2\pi^2} \int_0^{1/R_E} k^4 dk\: P^{\{2\}}_{\eps_v} %\sim \frac{GM}{R_E}
\quad\Rightarrow\quad
P^{\{2\}}_{\eps_v} \sim \frac{GM}{R_E}\, R_E^5\,.
\ea
Here, we have cut off the integral at $k \sim 1/R_E$, and
neglected all prefactors since this is a very rough estimate.
Since $\Phi \sim GM/R_E$ is approximately mass-independent, we find that
$P^{\{2\}}_{\eps_v}$ scales as $R_E^5$, or $M^{5/3}$.  This is the same
scaling as dynamical mass estimates based on velocity statistics
(e.g., \cite{schmidt:2010}).  

Including the stochastic terms listed here completes the parametrization of bias 
under the assumptions listed in the beginning of \refsec{general}:  
General Relativity, a non-relativistic matter fluid, and adiabatic Gaussian initial conditions.  The general bias expansion is summarized in \refsec{evol:summary}.

\subsection{Galaxy bias in the relativistic context}
\label{sec:GR}

So far in this section, we have restricted our discussion of galaxy bias 
to sub-horizon scales $k\gg aH$, where most of the information in large-scale 
structure resides.  This is because most of the measured modes are small-scale modes, while a survey with comoving volume of order $H_0^{-3}$ will only measure a handful of modes with $k \sim H_0$.  However, future large-scale galaxy surveys will have sufficiently large volume to probe such scales.  
We now consider how the bias can be extended to describe the galaxy density contrast on arbitrarily large scales.  

It is important to note that the leading corrections to the sub-horizon 
treatment of bias, which describes the rest-frame galaxy density, scale as $(aH/k)^2$, and are thus very small unless we 
consider very large scales.  On those large
scales, linear perturbation theory is an excellent approximation.  
Hence, for most practical purposes, it is sufficient to consider 
relativistic effects at linear order.  
In fact, the only requirement necessary to make the bias expansion consistent in the
relativistic context at linear order
is that the time slicing (constant-time hypersurface)
chosen to perform the bias expansion must correspond to a constant proper 
time of comoving observers, which is realized by working in the 
synchronous-comoving gauge \cite{wands/slosar:2009,baldauf/etal:2011,gaugePk} 
(Ref.~\cite{yoo:2014,bertacca/etal:2015} showed how to extend this to second order in 
relativistic perturbations).  The reason is that, apart from the local
density, the galaxy density depends on the local age of the Universe,
or proper time along the fluid trajectory up to the time of observation.  
In other words, at fixed proper time, galaxies are at a fixed evolutionary stage, in which case the bias expansion in terms of the galaxy density is sufficient at linear order.

The conformal Fermi coordinates (CFC) introduced in \cite{CFCorig,CFCpaper1} 
provide a useful way of defining bias operators in the relativistic context. 
The CFC are the natural coordinates a cosmologist living in a given galaxy 
would use.  Briefly, the construction of CFC starts from a time-like geodesic,
namely the fluid trajectory $x_{\rm fl}^\mu(t_F)$ parametrized by the proper time $t_F$,
with tangent vector $U^\mu\equiv \partial x_{\rm fl}^\mu(t_F)/\partial t_F$.
This frame can
be constructed locally for any spacetime, as long as a geodesic congruence
exists within a neighborhood of the geodesic considered, and is thus
not restricted to the assumption of a perturbed FRW spacetime.  
However, in the actual Universe, which has perturbations on all scales,
we perform the construction on a coarse-grained metric which only has
contributions from Fourier modes below some cutoff $k < \L$.  
Then, the velocity 4-divergence 
$\vartheta \equiv \nabla_\mu U^\mu$ along the fluid defines the local Hubble rate $H_F(t_F) \equiv \vartheta/3$.  The Hubble rate can be integrated to yield the local scale factor $a_F(t_F)$ (up to an arbitrary overall normalization).  One can then construct a coordinate system with time coordinate $\tau_F$ defined through $a_F d\tau_F = dt_F$ so that the metric becomes
\be
g_{\mu\nu}^F(x_F) = a_F^2(\tau_F) \left[ \eta_{\mu\nu} + A_{\mu\nu,kl}(\tau_F) x_F^k x_F^l + \O(\vx_F^3) \right]\,.
\label{eq:CFCmetric}
\ee
That is, \emph{along the entire trajectory} $x^\mu_{\rm fl}(\tau)$, all large-scale cosmological perturbations ($k < \L$) are absorbed into the local
scale factor $a_F$, and the corrections $\propto \vx_F^2$, $A_{\mu\nu,kl}$.  
Note that $A_{00,kl}$ is trace-free with respect to $kl$, $A_{00,kl}\d^{kl} = 0$.  

For nonrelativistic tracers such as galaxies, the spatial components of the
metric are dynamically irrelevant, and the CFC spacetime is characterized
completely by the local expansion rate $H_F(\tau_F)$ and the purely tidal
perturbation to the time-time component of the metric, 
$K_F^{ij} \equiv A_{00,ij}$.  These are the same 6 degrees of freedom as found in the quasi-Newtonian description described in the previous sections.  
Thus, a fully relativistic basis of bias operators is given by all scalar combinations
of $H_F$ and $K_F^{ij}$, as well as their derivatives with respect to
$\tau_F$ and $x_F^i$.  The latter are suppressed by the spatial scale $R_*$
as before.

To illustrate this, let us consider an adiabatic perturbation at linear order.  
As discussed above, this is the most relevant case.  As shown in \cite{CFCpaper2}, 
the isotropic part of the perturbation is absorbed into $a_F$, which
then obeys the Friedmann equation,
\be
H_F^2 = \frac{8\pi G}{3} \varrho_F + \frac{\text{K}_F}{a_F^2}\,,
\ee
in terms of the \emph{local physical} (not comoving) CFC-frame matter density $\varrho_F$ and curvature $\text{K}_F =$~const.  
That is, the local matter density together with the 
initial conditions (curvature) completely specifies the isotropic part
of the spacetime along the entire fluid trajectory.  
The CFC approach thus provides one proof of the \emph{separate universe} conjecture \cite{lemaitre:1933,barrow/saich:1993,cole:1997,baldauf/etal:2011,sherwin/zaldarriaga:2012}.  
Moreover, $\text{K}_F$ is directly related to the matter density perturbation $\d_{\rm sc}$ in synchronous-comoving gauge by \cite{baldauf/etal:2011,CFCpaper2}
\be
\frac{\text{K}_F}{H_0^2} = \frac53 \Omega_{m0} \frac{\delta_{\rm sc}(\tau)}{D_\text{md}(\tau)}\,,
\label{eq:KFsc2}
\ee
where $D_\text{md}(\tau)$ is the linear growth factor normalized to $a(\tau)$ during matter domination (for $a(\tau) \ll 1$). $\delta_{\rm sc}$ is in turn proportional to $\lapl\Phi$ in conformal-Newtonian gauge.  
This shows that, at linear order, the physical variable with respect to which galaxies are biased is $\d_{\rm sc}$ or equivalently $\lapl\Phi$ 
\cite{challinor/lewis:2011,baldauf/etal:2011,gaugePk}.  
Note that the separate-universe picture holds fully nonlinearly for 
isotropic perturbations.  For such configurations, the complete bias
expansion thus consists of powers of $\text{K}_F$ or equivalently $\varrho_F$
(and spatial derivatives thereof).  This provides another interpretation for
the fact that we do not have to include convective time derivatives of $\d$ in the
bias expansion.  
On the other hand, the anisotropic part is encoded in the trace-free perturbations $A_{00,kl} = K_F^{kl}$ and $A_{ij,kl} = \delta_{ij} K_F^{kl}$, while $A_{0i,kl}=0$.  

Beyond linear order, there are nontrivial relativistic effects.  First, relativistic effects due to the motion of galaxies enter at order $(v/c)^2 \sim 10^{-5}(v/[1000\,\text{km}\, \text{s}^{-1}])^2$.  Second, the galaxy density couples to vector and tensor metric perturbations, which in turn are produced by large-scale structure at nonlinear order \cite{matarrese/mollerach/bruni:1997,baumann/etal:2007,boubekeur/etal:2008} (and primordially, in case of tensor modes);  one can show \cite{ip/schmidt:2016} that in the CFC frame, the nonlinearly generated vector and tensor modes only start to contribute at third order in perturbations.\footnote{In case of galaxy shapes, \emph{primordial} tensor modes actually lead to effects at linear order which can become relevant on large scales \cite{GWshear,schmidt/pajer/zaldarriaga}.}
These ``genuine'' relativistic effects in large-scale structure, while interesting, are most likely too small to be of observational relevance for current and upcoming surveys.

Apart from such genuine relativistic effects, the results of this section give a rigorous description of galaxy clustering in relativistic perturbation theory \emph{in a specific frame}, the CFC.  In order to connect to observations, we then have to transform the galaxy density in CFC to observed coordinates.  
This is straightforward to do, given that the 
galaxy number density transforms as the $0$-component of a four-vector, namely the galaxy current vector $j_g^\mu$.   
The coordinate transformation is determined by tracing photon geodesics from 
the galaxy's position to the observer.  We shall describe this in 
\refsec{projection}.

\subsection{Renormalization: bare vs. physical bias parameters*}
\label{sec:renorm}

\technote{* This section is of a more technical nature and is not essential for the remainder of the review.  However, we encourage readers to go through the non-technical introduction, \refsec{renorm:intro}.}

\subsubsection{Motivation}
\label{sec:renorm:intro}

So far in \refsec{evolution}, we have enumerated which bias terms to include 
to express the galaxy density in an expansion of the form
\be
\d_g(\vx,\tau) = \sum_{O} \left[ b_O(\tau) + \eps_O(\vx,\tau) \right] O(\vx,\tau) +  \eps(\vx,\tau)\,.
\ee
However, there is a subtlety when relating this bias expansion to the statistics of galaxies.  
To understand this issue, let us disregard nonlinear gravitational
evolution for the moment, and go back to the case of \LIMD bias in Lagrangian
space considered in \refsec{localbias}.  We start from the \LIMD bias expansion \refeq{dhlocal}, now written as
\be
\d_g^L(\vq) = c_1 \d_R(\vq) + \frac12 c_2 \left(\d_R^2(\vq) - \s^2(R) \right) + \frac16 c_3 \d_R^3(\vq) + \cdots\,,
\label{eq:dhlocal2}
\ee
where we have dropped the superscript $(1)$ on the matter density $\d_R$ for clarity, and denoted the Lagrangian bias parameters with $c_i$, for a reason that will become clear shortly.  Now, given that $\d_R = \d_R^{(1)}$ is Gaussian, we can easily work out the two-point function of $\d_g^L$ in Lagrangian space, given \refeq{dhlocal2}:
\be
\xi_g^L(r) = \left[c_1^2 + c_1 c_3 \s^2(R) \right] \xi_{{\rm L},R}(r)
+ \frac12 c_2^2 [\xi_{{\rm L},R}(r)]^2 + \cdots\,.
\label{eq:xigbare}
\ee
We immediately see a difference to the expansion \refeq{xihthr} that we obtained from
an explicit calculation of the pair probability in \refsec{localbias}:  
while in \refeq{xihthr}, the coefficient of the leading term $\xi_{{\rm L},R}(r)$
is $(b_1^L)^2$, here we have a coefficient
$c_1 (c_1 + c_3 \s^2(R))$.  Moreover, if we were to continue the expansion
in \refeq{dhlocal2} to higher orders, we would obtain additional contributions
to $\xi_g^L(r)$ that are proportional to 
$c_1c_{2n+1}\s^{2n}(R) \xi_{{\rm L},R}(r)$ and thus
contribute at leading order on large scales.  This is clearly in contradiction
with the spirit of the perturbative bias expansion, which is based on the principle
that the contributions of higher-order bias terms are suppressed on
large scales.  Further, if we choose a small smoothing scale $R$ such that
$\s(R)$ is not much less than 1, then these higher-order contributions 
change the amplitude of $\xi_g^L(r)$ by order one.  This contradicts 
\refsec{higherderiv}, where we have argued that any smoothing scale involved should be irrelevant on large scales.  

On the other hand, \refeq{xihthr} shows exactly the desired behavior, with
higher-order bias terms (as well as the effects of smoothing) being suppressed on large scales.  The solution
to the undesirable situation posed by \refeq{xigbare} is that the 
coefficients $c_n$ in the bias expansion \refeq{dhlocal2} are not physical, but ``bare''
bias parameters.  Instead, the physical bias parameters are introduced
as coefficients of \emph{renormalized operators}.  In the present case,
\be
c_1 \d_R \to b^L_1 [\d_R] \equiv b^L_1 \d_R \,;\quad
c_3 \d_R^3 \to b^L_3 [\d_R^3] \equiv b^L_3 (\d_R^3 - 3 \s^2(R) \d_R)\,.
\ee
Inserting these relations into \refeq{dhlocal2}, we see that this removes the undesired
contribution to $\xi_g^L(r)$, and we recover 
$\xi_g^L(r) = (b_1^L)^2 \xi_{{\rm L},R}(r)$ in the large-scale limit.  
An equivalent approach is to define $b^L_1 \equiv c_1 + c_3 \s^2(R)/2$, as
done in \cite{mcdonald:2007}.  Here, the scale $R$ should really be seen
as an arbitrary smoothing scale (denoted as $\L^{-1}$ below) 
whose value becomes irrelevant in the end.  
Thus, renormalization is an essential part of the connection between bias
expansion and galaxy statistics.  

In the case of a Gaussian density field, the renormalized \LIMD operators are simply given at all orders by Hermite polynomials \cite{szalay:1988},
\be
[\d_N] = \s^N(R) H_N\left(\frac{\d_R}{\s(R)}\right)\,.
\label{eq:dNHermite}
\ee
Suitable orthogonal polynomials can be defined for other operators, such as 
$(\vn\delta)^2$ \cite{lazeyras/musso/desjacques:2015}. 
This approach is discussed in \refsec{PBSpeaks} in the context of peak theory, but it is applicable to any Lagrangian bias scheme.  
However, this method cannot be directly applied to the evolved, Eulerian density field as the latter is highly non-Gaussian on 
small scales.  Nevertheless, even in this case, renormalized operators can be systematically derived order by order in perturbation theory.  
We stress again that all physical statements about the values of bias parameters, for example a large bias for rare 
objects, refer to the \emph{physical, renormalized bias parameters}, i.e. the coefficients of the renormalized operators.

The aim of this section is to describe in detail the renormalization procedure that connects the bias expansion derived so far for the \emph{evolved} galaxy density field to the observable statistics of galaxies on large scales.  In the course of this derivation, we make the physical arguments of the
previous sections more formal and rigorous, and connect to the
EFT language (\cite{goldberger/rothstein,baumann/etal:2012,senatore:2015}; see \cite{porto:2016} for a review).  
Renormalization in the context
of galaxy bias was first considered by \cite{mcdonald:2007}, and later
expanded on in \cite{PBSpaper,assassi/etal,senatore:2015,MSZ}.

\subsubsection{Equivalence principle and local gravitational observables}
\label{sec:renorm:boost}

In essence, the perturbative bias expansion attempts to connect the proper rest-frame density of galaxies at 
some time $\tau$ and position $\vx$ to the initial conditions 
(for example, produced by inflation) at time $\tau=0$.  In full generality,
the proper comoving density $n_g$ can be written as
\be
n_g(\vx,\tau) = F_g[\Phi^{(1)}(\vy)](\vx,\tau)\,,
\label{eq:ngformal}
\ee
where $F_g$ is a nonlinear functional of the initial potential perturbations $\Phi^{(1)}(\vy) = \Phi(\vy,\tau=0)$.  
The assumption of statistical homogeneity of the Universe
dictates that the functional be invariant under a constant spatial
coordinate shift $\vx \to \vx + \bm{\xi}$.  This implies that $F_g$ only
involves the combination $\vy-\vx$.  
More generally, $n_g$, being the 0-component of the current vector $j_g^\mu$, 
has specific transformation properties under general coordinate (or gauge) transformations.  In particular, it transforms as a 3-scalar under coordinate rotations on a fixed time slice, a property we have already used from the beginning (e.g., when introducing the term $(K_{ij})^2$ in \refsec{evol1}).  

A particular \emph{generalized Galilean} coordinate transformation, which corresponds to a time-dependent but spatially uniform coordinate shift, is of special 
relevance to large-scale structure \cite{rosen:1972,weinberg:2003,creminelli/etal:2013}.  
In General Relativity, the equivalence principle holds,
which implies that we can remove a pure-gradient metric perturbation by going to the free-falling frame of comoving observers, who move on trajectories $\vx_{\rm fl}(\tau')$ with $\tau'\in [0,\tau]$. 
Specifically, for the conformal-Newtonian gauge metric \refeq{metriccN},
a pure-gradient potential perturbation,
\be
\Phi(\vx,\tau) = \Psi(\vx,\tau) = \v{A}(\tau) \cdot\vx\,,
\ee
can be removed by performing a spatially constant but \emph{time-dependent} spatial translation,
\be
\vx \to \vx + \v{\xi}(\tau); \quad
\v{v} \to \v{v} + \partial_\tau{\v{\xi}}(\tau)\,.
\label{eq:boost}
\ee
Under this translation, the $00$-component of the metric transforms as 
\be
\Phi \to \Phi - (\partial_\tau^2{\v{\xi}} + \cH \partial_\tau{\v{\xi}})\cdot \vx\,.
\ee
Thus, if we demand that $\v{\xi}(\tau)$ solves
\be
\frac{\partial}{\partial\tau^2} \v{\xi} + \cH \frac{\partial}{\partial\tau}\v{\xi} = \vn \Phi(\tau) = \v{A}(\tau)\,,
\label{eq:xieom}
\ee
then $\vn\Phi$ vanishes in the transformed coordinates at all times.  
We recognize \refeq{xieom} as the equation for (minus) the Lagrangian displacement $\v{\xi}=-\v{s}$, \refeq{seom}.  That is, $\vx + \v{\xi}(\tau)$ corresponds to the Lagrangian coordinate of a fluid trajectory in the presence of a long-wavelength (pure-gradient) mode.  An observer comoving with this trajectory will experience no gravitational force, as $\vn\Phi=0$.  Clearly however, we can only remove first spatial derivatives of the potential, while second spatial derivatives lead to locally observable gravitational effects.  
This is the physical content
of the well-known \emph{consistency relations} in large-scale structure
\cite{kehagias/riotto:2013,goldberger/hui/nicolis:2013,creminelli/etal:2013,peloso/pietroni:2013,peloso/pietroni:2014,valageas:2014}, 
which phrase the requirement that the local physics must
be invariant under time-dependent translations as conditions relating
certain limits of $N$-point functions to $(N-1)$-point functions.  
Following our discussion, the consistency relations hold
for biased tracers as well \cite{kehagias/etal:2014,horn/hui/xiao:2014}, on scales much larger than $R_*$.  

Working in the fluid coordinates, we can then write $F_g$ fully generally as
\be
n_g(\vx,\tau) = F_g\Big[\partial_i\partial_j\Phi(\vx_{\rm fl}'(\tau'),\tau')\Big](\vx,\tau)\,,
\label{eq:ngformal2}
\ee
where $F_g$ now only involves the difference $\vx'_{\rm fl}-\vx_{\rm fl}$ at various times $\tau'$ (with $0 \leq \tau' \leq \tau$).  We could have equivalently written \refeq{ngformal2} in terms of $\partial_{q,i}\partial_{q,j}\Phi(\vq,\tau)$.  
This nonlinear functional of $\partial_i\partial_j\Phi$ is significantly more restrictive than the full functional $F_g[\Phi]$.  Moreover, since $\partial_i\partial_j\Phi$ corresponds to the leading locally observable effect of the gravitational field, this way of writing the functional isolates the actual physical impact of long-wavelength perturbations on the galaxy density.  On the other hand, $F_g[\Phi^{(1)}(\vy)]$ contains a large number of unphysical gauge modes.  
The price we have payed for this reduction in degrees of freedom is that we need to include the dependence on $\partial_i\partial_j\Phi$ along the entire fluid trajectory.  Physically, this makes sense, since, as argued in \refsec{framework}, the formation of halos and galaxies is not an instantaneous process and thus depends on the local environment throughout cosmic time.  We also see that the dependence is now on the full potential $\Phi$ rather than the initial potential $\Phi^{(1)}$, a point to which we will return below.   

The requirement that $n_g$ be a spatial scalar is then simply achieved by
only allowing terms where all indices in the factors of $\partial_i\partial_j\Phi$ are contracted.  
We stress again that any dependence on other local observables, such as the matter density or velocity shear, and time derivatives thereof, are implicitly included in the arguments listed in \refeq{ngformal2}, as these can themselves be expressed as integrals over $\partial_i\partial_j\Phi(\xfl')$ via the equations of motion.   
One might wonder whether the relation \refeq{ngformal2} depends on when the spatial derivatives are taken, e.g. in Eulerian frame ($\partial_{x,i}\partial_{x,j}$) or Lagrangian frame ($\partial_{q,i}\partial_{q,j}$).   However, the different spatial derivatives can be transformed into each other (see \reftab{EulLagr}), where the transformation itself just depends on $\partial_i\partial_j\Phi$ along the fluid trajectory.  Hence, the expansion is independent of the time slice chosen for the spatial derivatives.  

Further, following our physical arguments in \refsec{framework}, there is a spatial scale $R_*$ 
such that the functional kernel in $F_g$ becomes negligible if $|\vx_{\rm fl}'-\vx_{\rm fl}| \gg R_*$.  In general, $R_*$ is a function of time $\tau'$; in this case, what we refer to as $R_*$ should always be seen as ${\rm max}_{\tau'} [ R_*(\tau') ]$.  

We now expand the functional in \refeq{ngformal2} in time.  For this
we consider an operator $O(\vx_{\rm fl}'(\tau'), \tau')$ which is
composed of any nonlinear combination of $\partial_i\partial_j\Phi(\xfl'(\tau'),\tau')$
(we will deal with the functional dependence on $\vx_{\rm fl}'$ below).  
The time dependence then formally is [as in \refeq{dg_taylor}]
\be
\int d\tau'\,F_O(\tau;\tau') O(\vx_{\rm fl}'(\tau'),\tau')
= \sum_{n=0}^\infty \frac1{n!}\left[\int d\tau' (\tau'-\tau)^n F_O(\tau;\tau')\right]
\left(\convD\right)^n O(\vx_{\rm fl}'(\tau),\tau)\,,
\label{eq:functau}
\ee
where we have expanded $O$ in terms of convective time derivatives around $\tau'=\tau$, and $\xfl'(\tau)$ denotes a fluid trajectory in the vicinity of $\xfl(\tau)$.    Of course, we could have equivalently expanded around $\tau=0$.  
The factors in brackets are coefficients which only depend on $\tau$.  
With this, the functional \refeq{ngformal2} can be written \emph{on a single time slice}, by including convective derivatives of all combinations of
$\partial_i\partial_j\Phi$:
\be
n_g(\vx,\tau) = F_g\Big[\partial_i\partial_j\Phi(\vx',\tau),
\text{D}\partial_i\partial_j\Phi(\vx',\tau)/\text{D}\tau, \dots ;\,\tau
\Big](\vx)\,,
\label{eq:ngformalE}
\ee
which is understood to also include nonlinear mixed terms such as
$(\partial_k\partial_l\Phi) \text{D}(\partial^k\partial^l \Phi)/\text{D}\tau$ [including
these terms captures functionals involving products of operators such as
$\int d\tau'\int d\tau'' F(\tau;\tau',\tau'') O_1(\tau')O_2(\tau'')$]. 
\refeq{ngformalE} is still a functional in terms of the spatial dependence $\vx'$, but is local in time.  Again, any reference time $\tilde\tau \neq \tau$ could be chosen for
the expansion on the right-hand side, as long as $\vx$ is replaced with $\vx_{\rm fl}(\tilde\tau)$.  

\subsubsection{Coarse graining and bare bias expansion}

In the context of the perturbative bias expansion, our goal is to isolate
the dependence of $n_g(\vx,\tau)$ on large-scale perturbations, with the
ultimate goal of deriving statistics of $n_g$ for some large scale $r$, 
or Fourier wavenumber $k \propto 1/r$.  The theory itself, once applied to measurements, will tell us what ``large'' here means precisely.

\begin{figure}[t]
\centering
\includegraphics[trim=0cm 0cm 0cm 0cm,clip,width=0.7\textwidth]{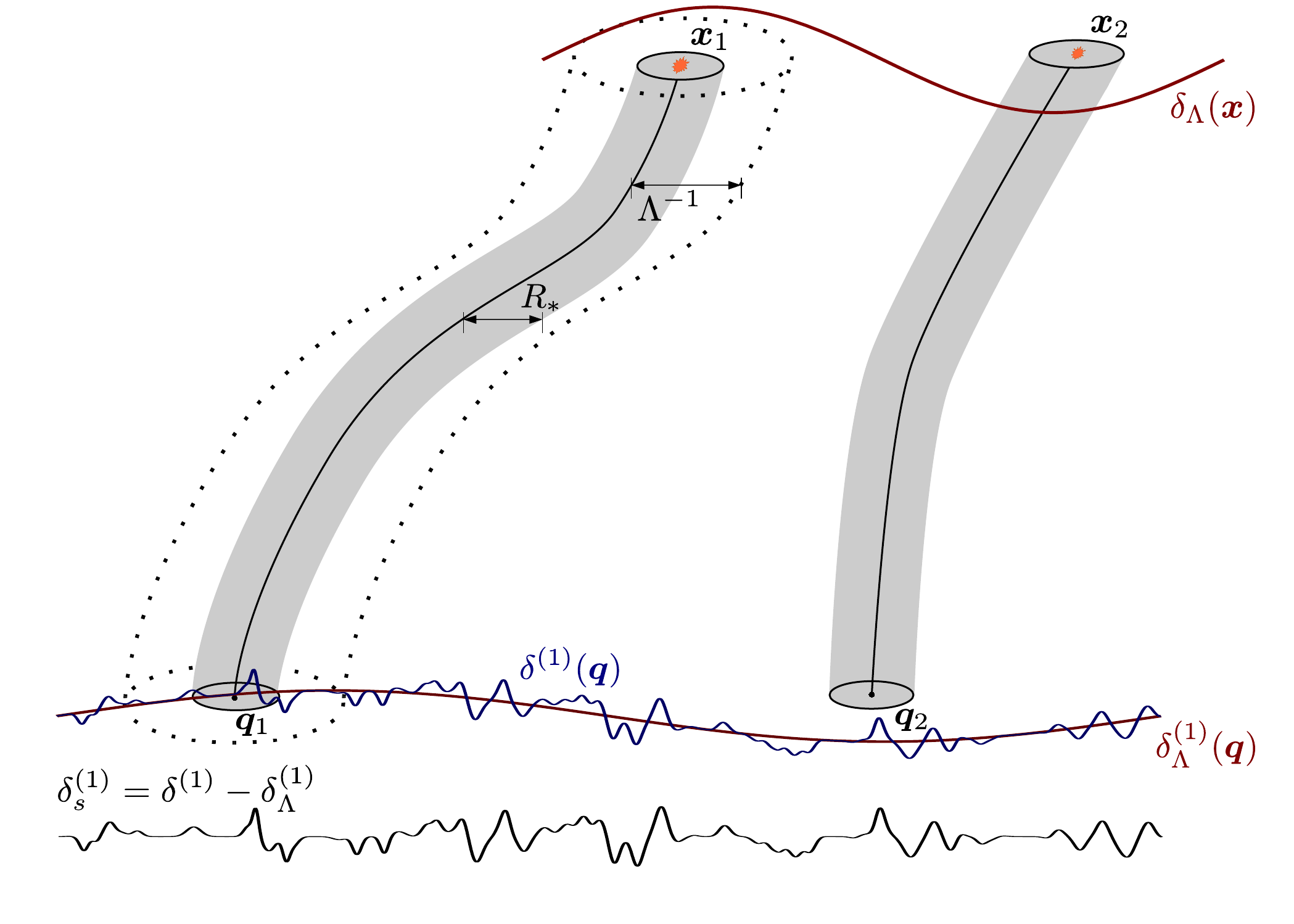}
\caption{Sketch of the setup and quantities used in the derivation of bias renormalization. Time runs vertically as in \reffig{CFCsketch}.  
The observed galaxies reside at Eulerian positions $\vx_1,\,\vx_2$ and
each form within a spacetime region centered around the fluid trajectories
to each point as in \reffig{CFCsketch}.  The top of the plot corresponds to
the epoch of observation, while the bottom denotes the initial condition
(Lagrangian positions $\vq_1,\,\vq_2$).  The blue line near the bottom 
shows the total initial (linear) density field.  We coarse-grain the galaxy and
matter fields on the scale $\L^{-1}$ (dotted region), resulting in a smoothed
large-scale density field $\d_\L$ shown by red solid lines.  The black line
at the bottom shows the small-scale density field in the initial conditions, 
which is statistically the same everywhere.  
For clarity, we only represent the density field, although in reality
the galaxy density is a function of all components of $\partial_i\partial_j\Phi$ as well as its convective time derivatives.  
\label{fig:sketch}}
\end{figure}

Let us coarse-grain, or filter, \refeq{ngformalE} on a spatial scale 
$\L^{-1}$ which is much smaller than $r$, or, equivalently, $\L \gg k$ for all
wavenumbers $k$ of interest (see \reffig{sketch}).  Since this is a crucial
step, we carefully describe it here.  
We let $W_\L(|\vx|)$ denote the filtering kernel, assumed to
be isotropic and normalized to unity (which is a natural assumption 
since any anisotropy would correspond to introducing preferred directions).  
We denote any filtered quantity with a subscript $\L$, e.g. 
$\Phi_{\L}$.
In EFT language, $\L$ provides our UV 
(ultraviolet; high-energy, or small-scale) cutoff, 
introduced to render loop integrals finite.  We can think of $\d_{\L}(\vx) \propto \lapl \Phi_{\L}$ as the average density perturbation within a region $\U$ of size $\Lambda^{-1}$ centered on $\vx$ (dotted region in \reffig{sketch}).  
Complementary to $\Phi_{\L}$, we define the small-scale potential perturbations as 
\be
\Phi_{s}(\vx,\tau) \equiv \Phi(\vx,\tau) - \Phi_{\L}(\vx,\tau)\,.
\label{eq:Phisdef}
\ee
Note that the cutoff $\L$ is merely a computational tool whose value we are free to determine;  
any observable, including statistics of galaxies as well as the physical bias
parameters, have to be independent of the value of $\Lambda$. 
This can in fact be used as a sanity check for any predictions for observables.  
Thus, $\L^{-1}$ must be distinguished from the
\emph{physical} scale $R_*$ which controls the amplitude of the higher
derivative bias parameters discussed in \refsec{higherderiv}, and is, for
halos, typically of order the Lagrangian radius.  
In fact, demanding that
galaxy statistics be independent of $\L$ forces us to introduce precisely these higher
derivative terms with physical renormalized coefficients, as we will see
in \refsec{rderiv}.  However, going step by step,
we will again start by neglecting higher-derivative terms.  In this case,
we implicitly assume that $\L^{-1}$ is larger than $R_*$.  

Note that we could have equally chosen to perform
the coarse-graining in terms of the initial potential perturbations $\Phi^{(1)}$.  
The nonlinear gravitational potential in Eulerian coordinates, $\Phi(\vx,\tau)$, and $\Phi^{(1)}(\vq)$ are nonlinearly related.  Their relation, however, can be
expanded in terms of the same operators as we will include in the bias
expansion. Since coarse-graining (convolution) and multiplication
do not commute, there are differences in the result obtained by 
coarse-graining in $\Phi^{(1)}$;  however, these differences are 
at higher order in spatial derivatives, and will be absorbed by higher
derivative operators (\refsec{rderiv}).

If we make $\Lambda$ sufficiently small 
(that is, the smoothing scale sufficiently large), then 
the functional dependence on $\partial_i\partial_j\Phi$ in \refeq{ngformalE} 
separates into an \emph{ordinary} 
[as in the local expansion in \refeq{biasrel}] dependence on the variable 
$\partial_i\partial_j\Phi_{\L}(\vx)$, while the functional dependence is restricted to $\partial_i\partial_j\Phi_{s}(\vx')$:
\be
n_{g,\L}(\vx) = F_{g,\L}\left[\partial_i\partial_j\Phi_{\L}(\vx),\,
\text{D}\partial_i\partial_j\Phi_{\L}(\vx)/\text{D}\tau,\,\dots\,
;\   \partial_i\partial_j\Phi_{s}(\vx'),\, \text{D}\partial_i\partial_j\Phi_{s}(\vx')/\text{D}\tau,\,\dots\right]\,,
\label{eq:Fp2}
\ee
where here and in the following we suppress the time arguments $\tau$ for 
clarity, since they are the same on both sides of \refeq{Fp2}.  We now
prove this statement.  First, \refeq{Fp2} holds if and only if
$\Lambda^{-1}$ is much larger than $R_*$, the spatial scale over which the functional in \refeq{ngformalE} has significant support.  
We now expand \refeq{ngformalE} into
a series of linear, quadratic, and higher-order functionals, in analogy with the expansion of the functional in time \refeq{functau}.  
By inserting $\Phi = \Phi_\L + \Phi_s$ [\refeq{Phisdef}], any term without time derivatives in this expansion can be written as
\ba
\Biggl[\prod_{i=1}^n \int d^3 \vx_i\Biggl] \, F^{(n)}_{i_1j_1\cdots i_n j_n}(\vx_1,\cdots,\vx_n)
\partial_{i_1}\partial_{j_1} \Phi(\xfl+\vx_1) & \cdots 
\partial_{i_n}\partial_{j_n} \Phi(\xfl+\vx_n) 
\label{eq:funcx}\\
\simeq \sum_{k=0}^n \frac{n!}{(n-k)!} \partial_{i_1}\partial_{j_1} \Phi_{\L}(\xfl)
\cdots \partial_{i_k}\partial_{j_k} \Phi_{\L}(\xfl) 
\Biggl\{
\Biggl[\prod_{i=1}^n &\int d^3\vx_i\Biggl]\: F^{(n)}_{i_1j_1\cdots i_n j_n} (\vx_1,\cdots \vx_n) 
\vs
 & \times 
\partial_{i_{k+1}}\partial_{j_{k+1}}\Phi_{s}(\xfl+\vx_{k+1})
\cdots \partial_{i_{n}}\partial_{j_{n}}\Phi_{s}(\xfl+\vx_{n})
\Biggl\}
\,,
\nonumber
\ea
where we have approximated $\Phi_{\L}(\xfl[\tau])$ as constant over the 
region over which the kernel $F^{(n)}$ is nonzero, as implied by the condition  
$\Lambda^{-1} \gg R_*$.  This allowed us to pull the factors of $\Phi_\L(\xfl+\vx_i)$ outside the integral.  In \refsec{rderiv} we will show that going beyond this approximation leads to the higher-derivative terms discussed in \refsec{higherderiv}.  
Further, we have assumed without loss of generality that the kernels  $F^{(n)}_{i_1j_1\cdots i_nj_n}$ are symmetrized over the indices $i_kj_k$ and the associated spatial positions $\vx_k$.  
The same reasoning also goes through for terms with time derivatives.  
If we think of the spatial derivatives as being with respect to 
$\vq$,
i.e. Lagrangian derivatives, then $\convDinline$ commutes with $\partial_i$.  
For Eulerian spatial derivatives, this is not the case, however the term $v^i\partial_i$ in the convective derivative cancels via the time dependence that appears in the spatial argument $\xfl(\tau)$ of all instances of the potential $\Phi$.  

The terms in curly brackets in \refeq{funcx} correspond to weighted integrals
over the small-scale fluctuations.  
It is then clear that we can write \refeq{Fp2} as
\be
n_{g,\Lambda}(\vx) = \sum_{1,\  O_\Lambda} 
c_{O,\Lambda}\left[\partial_i\partial_j\Phi_s,\,\text{D}\partial_i\partial_j\Phi_s/\text{D}\tau,\,\dots\right] O_{\Lambda}(\vx)\,,
\label{eq:Fp3}
\ee
where the coefficients $c_{O,\Lambda}[\partial_i\partial_j\Phi_s, \cdots]$ are still functionals of the small-scale modes
$\partial_i\partial_j\Phi_s$.
The sum runs over 1 and all scalar operators constructed out of
$\partial_i\partial_j\Phi_\L$ and its time derivatives, i.e.~\emph{precisely the set described in \refsec{framework}}, 
with a zeroth-order constant related to the mean density of the galaxies.  
However, here the operators are constructed
out of the coarse-grained quantity $\partial_i\partial_j\Phi_\L$.  
We have made this explicit through the subscript $\L$.  

Now, by construction, the short modes have no support at low $k$ in Fourier
space.  Specifically, $\Phi_s(\vk) = [1 - W_\L(k)] \Phi(\vk) \stackrel{k\to 0}{\longrightarrow} \O(k^2\L^2) \Phi(\vk)$.  The small-scale modes in the initial
conditions $\Phi^{(1)}_s$
are then independent of long-wavelength modes $\Phi^{(1)}_\L$ for Gaussian initial conditions (see lower panel of \reffig{sketch}).  At finite time, they do 
depend on long-wavelength modes through the gravitational influence of the
latter.  
However, the gravitational effects on the evolution of the small-scale
modes are also precisely captured by $\partial_i\partial_j\Phi_\L$ and its convective time derivatives, i.e. the operators $O_\L$ that appear in the sum in \refeq{Fp3}.  This can be shown for example by considering the Lagrangian evolution 
equations of the matter density field (Sec.~4 of \cite{MSZ}).  
Then, at any given time, we can expand the evolved 
small-scale statistics and their time derivatives
in terms of the long wavelength operators $O_\Lambda$ 
and the initial small-scale modes $\Phi^{(1)}_s$ in \refeq{Fp3}.  Finally, we reorder the sum to write
the coefficients $c_{O,\Lambda}$ as functionals of the small-scale modes 
in the \emph{initial} conditions:
\be
c_{O,\Lambda}\left[\partial_i\partial_j\Phi_s,\,\text{D}\partial_i\partial_j\Phi_s/\text{D}\tau,\,\dots\right]
\to c_{O,\Lambda}[\partial_i\partial_j\Phi^{(1)}_s]\,.
\ee
Note that we can only do this reordering consistently if the basis contains 
all operators that appear in the solution for the gravitational evolution of the 
small-scale modes $\Phi_s$ up to a given order in perturbation theory.  
Only in this case is the operator basis closed under renormalization, which
we will discuss below.  Since, on large scales, the small-scale initial conditions $\Phi_s^{(1)}$ are uncorrelated with each other as well as the coarse-grained operators, the dependence of $c_{O,\Lambda}$ on $\partial_i\partial_j\Phi^{(1)}_s$ can be completely described by stochastic fields $\eps_{\Lambda,O}$; that is,
fields that are entirely characterized by their one-point PDF.  This is of course only valid on scales much larger than $\Lambda^{-1}$.  We then obtain the expression
\be
n_{g,\Lambda}(\vx) = n_{0,\L} \left[1 + \eps_{\L}(\vx) + \sum_{O} \Big\{ c_{O,\Lambda} 
+\eps_{O,\Lambda}(\vx) \Big\} O_{\Lambda}(\vx) \right]\,,
\label{eq:Fp4}
\ee
where the sum runs over the operators described in \refsec{general}, and $\eps_{O,\Lambda}$ are stochastic fields with zero expectation value
which are uncorrelated
with any other coarse-grained fields (in particular the $O_{\Lambda}$),
and only have zero-lag correlations among themselves.  
We have pulled out an overall factor $n_{0,\L}$ in \refeq{Fp4}, so that the quantities in brackets are dimensionless.  Note that the term $c_{1,\L}[\cdots]$ in \refeq{Fp3} now corresponds to $\eps_\L$.

\subsubsection{Renormalizing the bias parameters}
\label{sec:renormalization}

We see that the bias expansion written in \refeq{Fp4} contains exactly the operators 
of the general local bias expansion described in \refsec{general}.  However, now all terms in \refeq{Fp4} 
depend on the coarse-graining scale $\L$, which is an arbitrary cutoff
introduced in the calculation.  Nevertheless, if we have included
all necessary operators in the bias expansion up to a given order in perturbation
theory, we know that we should be able to describe the statistics of
$n_g$ up to that order without making reference to an unphysical smoothing
scale $\L$.  The goal of renormalization is to reorder the sum in \refeq{Fp4}
into a sum over $\L$-independent operators multiplied by $\L$-independent 
parameters:
\be
n_g(\vx) = \avng \left[1 + [\eps] + \sum_{O} \Big\{ b_O + [\eps_O] \Big\} [O] \right]\,,
\label{eq:Fprenorm}
\ee
where $[O]$ are renormalized operators which we will discuss below and 
$b_O,\,[\eps_O]$ are the physical, renormalized bias coefficients and stochastic fields.  Clearly, \refeq{Fprenorm} is not unique, as different linear combinations of operators $[O]$ will yield an equivalent expression with different bias parameters.  However, any complete renormalized bias expansion of the form \refeq{Fprenorm} can be related unambiguously to any other (see \refapp{biastrans}).  In the following, we will focus on the operators $[O]$ and biases $b_O$, and will not further consider the stochastic contributions;  their renormalization can be derived analogously.    

The reason why we need to renormalize the bias coefficients is that,
when calculating the large-scale statistics of $n_g$, we obtain
results that are strongly dependent on the artificial coarse-graining
scale $\L$ if we use the ``bare'' bias expansion \refeq{Fp4}.  This
already becomes apparent when taking the ensemble average of \refeq{Fp4},
which should yield the observed mean density of galaxies
$\< n_{g,\Lambda} \> = \avng$.  We obtain
\be
\avng = \< n_{g,\Lambda} \> = n_{0,\L} \left[ 1 + \sum_O  c_{O,\L} \< O_\L \> \right]
= n_{0,\L} \left[ 1 + c_{\d^2,\L} \<\d_\L^2\> + c_{K^2,\L} \< (K^{ij}_\L)^2 \> + \cdots \right]\,,
\label{eq:Fpbar}
\ee
where we have used the fact that $\<\eps_\L\> = 0 = \<\eps_{O,\L}\>$, and
have written the leading terms at second order. These terms
thus renormalize the ``bare'' mean number density $n_{0,\L}$, which is
physically meaningless due to its $\L$-dependence. We can infer from this relation, and the corresponding ensemble average of \refeq{Fprenorm}, our first \emph{renormalization condition} for the operators, namely that $\< [O] \> = 0$. Consider $O = \d^2$ as an example. We can define
\be
[\d^2] = \d_\L^2 - \s^2(\L) + \cdots\,,\quad\mbox{where}\quad
\s^2(\L) \equiv \left\< \left(\d_\L^{(1)}\right)^2\right\>\,,
\label{eq:d2renorm1}
\ee
and the ellipsis stands for higher-order contributions. Clearly, $\< [\d^2] \> = 0$ at lowest order, as desired. This trivial renormalization can clearly be performed for all other operators as well.

Next, let us consider the galaxy-matter cross-power spectrum, which is the simplest way to estimate the linear bias $b_\d$ (\refsec{bnpt}). 
For demonstration, we include both the leading term and the contribution $\propto \d_\L^2$ in the bare bias expansion:
\be
\< \d_g(\vk) \d(\vk') \> = c_{\d,\L} \< \d_\L(\vk) \d(\vk') \>
+ c_{\d^2,\L} \< (\d_\L^2)(\vk) \d(\vk') \>
+ \cdots\,,
\label{eq:bare2pt}
\ee
where we have inserted \refeq{Fp4} and the dots denote remaining second-order as well as higher-order terms.  
Our goal is to identify the bias $b_\d$ on very large scales $k \to 0$
through  
\be
\< \d_g(\vk) \d(\vk') \> \stackrel{k\to 0}{=}
b_\d \< \d^{(1)}(\vk) \d^{(1)}(\vk') \>\,.
\label{eq:renorm2pt}
\ee
Since, for fixed $\L$, both $\d_\L$ and $\d$ asymptote to the linear density 
field $\d^{(1)}$ in the large-scale limit, we see that the first term in \refeq{bare2pt} already shows the desired behavior. That is, at this order there is 
no need to renormalize $O = \d$, and we have
$[\d] = \d$ (this does not imply however that $b_\d = c_{\d,\L}$, as we
will see below; only in the limit $\L \to 0$ (infinite smoothing scale)
does the bare bias parameter asymptote to the renormalized one).

This is not the end of the story however.
Let us consider the second contribution in \refeq{bare2pt}. Note that $(\d_\L^2)(\vk)$ involves a convolution in Fourier space. That is, small-scale (high-wavenumber) modes contribute to this term, no matter how small $k$ is. Given the assumed Gaussian initial conditions, the leading contribution to $\< (\d_\L^2) \d \>$ appears at second order in perturbation theory, and is obtained by replacing each instance of $\d_\L$ with $\d_\L^{(2)}$ (the third possible contribution vanishes). This yields
\ba
\< (\d_\L^2)(\vk) \d(\vk') \>'\Big|_\LO =\:& 2 \left\< \left(\d_\L^{(1)}\d_\L^{(2)}\right)(\vk) \d^{(1)}(\vk') \right \>' \vs
=\:& 2\int_{\vp_1}\int_{\vp_2}\int_{\vp_3}\!\!\! (2\pi)^3 \d_D(\vk-\vp_{123}) W_\L(p_1) W_\L(|\vp_{23}|) F_2(\vp_2,\vp_3) \vs
& \hspace*{2cm} \times \< \d^{(1)}(\vp_1) \d^{(1)}(\vp_2) \d^{(1)}(\vp_3) \d^{(1)}(\vk') \>' \vs
=\:& \left[ 4 \int_{\vp} W_\L(p) W_\L(|\vp+\vk|) F_2(\vp,\vk) \Plin(p)\right] \Plin(k)
\vs
\stackrel{p \gg k}{=}\:& \left[4 \frac{17}{21} \int_{\vp} [W_\L(p)]^2 \Plin(p)\right] \Plin(k) = \frac{68}{21} \s^2(\L)\,\Plin(k)\,,
\label{eq:d2dbare}
\ea
where $F_2$ is the second-order perturbation theory kernel [\refeq{F2}].
Here, a subscript ``LO'' denotes the leading-order expression of the correlator, that is, at the lowest non-vanishing order in perturbation theory.
In the last line, we have assumed the large-scale limit $k \to 0$. Since the dominant contribution to the momentum integral comes from modes $p\sim \L \gg k$, it is appropriate to assume $p \gg k$. In this limit, the angle-average of $F_2$ yields $17/21$. 
Thus, the second-order bare bias operator $\d_\L^2$ leads to a contribution that is of the exact same shape on large scales as the linear bias term, namely $\propto \Plin(k)$, but depends on the variance $\s^2(\L)$ of the density field at the cutoff scale. This is clearly unsatisfactory, and very reminiscent of the issue we encountered with the cubic bias term in \refsec{renorm:intro}. Fortunately, there is a simple remedy: we add $\d_\L$ as counter-term to $[\d^2]$, extending \refeq{d2renorm1} to
\be
[\d^2] = \d_\L^2 - \s^2(\L)\left[1 + \frac{68}{21} \d_\L\right] + \cdots\,.
\label{eq:d2renorm2}
\ee
The same exercise can then be performed for the other bare bias operators, ensuring that \refeq{renorm2pt} actually holds.

However, the large-scale limit of the two-point function \refeq{renorm2pt} is not the only constraint. Consider for example the galaxy-matter-matter three-point function, which on large scales allows for a measurement of the two (physical, renormalized) second-order biases $b_{\d^2},\,b_{K^2}$, as will be described in detail in \refsec{bnpt}. Specifically, the desired large-scale limit of the three-point function is
\ba
\< \d_g(\vk) \d(\vk_1)\d(\vk_2) \> \stackrel{k_1,k_2 \to 0}{=}\:&
b_\d \< \d(\vk) \d(\vk_1)\d(\vk_2) \>_\LO
+  b_{\d^2} \< (\d^2)(\vk) \d(\vk_1)\d(\vk_2) \>_\LO \vs
& +  b_{K^2} \< (K_{ij})^2(\vk) \d(\vk_1)\d(\vk_2) \>_\LO\,,
\label{eq:renorm3pt2}
\ea
where, again, a subscript ``LO'' denotes the leading-order expression
in perturbation theory. 
Note in particular the absence of the coarse-graining scale $\L$ in this expression; technically, this is because the right-hand side does not contain any loop integrals.

In the bare bias expansion, \refeq{renorm3pt2} involves $c_{\d,\L} \< \d_\L(\vk) \d(\vk_1)\d(\vk_2)\>$. As in the case of the two-point function, this already asymptotes to the leading-order expression in \refeq{renorm3pt2} in the large-scale limit.  Thus, we again do not need to renormalize the operator $\d$.  This in fact continues at all orders in perturbation theory (but only at the level of the local bias expansion excluding higher-derivative operators), and is related to the fact that we did not need to include $\tr \Pi^{[n]}$ in the bias expansion in \refsec{general}.  
On the other hand, now the expression \refeq{d2renorm2} is not sufficient to renormalize $\d_\L^2$, as it yields a cutoff-dependent result at next-to-leading (NLO, or 1-loop) order, which is fourth order in perturbations, when inserted into \refeq{renorm3pt2}. In this case, we can cure the unphysical behavior by adding $(\d_\L)^2$ and $(K_\L^{ij})^2$ as counter-terms. Specifically, Ref.~\cite{assassi/etal} derived
\be
[\d^2] = \d_\L^2 - \s^2(\L) \left[1 + \frac{68}{21} \d_{\L} + \frac{24032}{6615} \d_{\L}^2
+ \frac{254}{2205} (K_{ij,\L})^2 \right]\,.
\ee
Of course, going to higher order in perturbations requires the addition of more counter-terms, such as $\s^2(\L) \d_\L^3$, as well as corrections to lower-order counter-terms such as $\s^4(\L) \d_\L$, each multiplied by a coefficient determined through the renormalization conditions. Similarly, $(K_\L^{ij})^2$ also needs to be renormalized. Crucially however, we do not need these relations to derive the large-scale galaxy statistics: at leading order, we can always be assured that the bias parameters that appear are the physical ones, thanks to renormalization. When including next-to-leading contributions, i.e. loop terms, we can simply drop those terms that are canceled by counter-terms to lower-order operators. They are easily identified by having the same scale dependence as the contribution from the lower-order operator, but with a cutoff-dependent amplitude (cf. \refeq{d2dbare}; this will be illustrated in \refsec{npt1loop}).\\

Finally, after these concrete examples, we can turn to systematically defining the conditions which the renormalized operators need to satisfy. A necessary condition for a properly renormalized operator $O$ in the bias expansion is that \emph{all $N$-point auto- and cross-correlations} of the galaxy density asymptote to their leading-order expressions on large scales. This is crucial, as it ensures that a single set of physical bias parameters describes all statistics of a given galaxy sample. Achieving this result might seem like a daunting task; however, many of the auto- and cross-correlations are 
redundant. Since, in our perturbative bias expansion, all operators
are constructed out of $\partial_i\partial_j\Phi^{(1)}$, or equivalently
in Fourier space $(k_ik_j/k^2) \d^{(1)}(\vk)$, it is sufficient to 
enforce the following \emph{renormalization conditions} for each operator in the bias expansion \cite{assassi/etal}:
\be
\lim_{k_i \to 0} \< [O](\vk) \d^{(1)}(\vk_1) \cdots \d^{(1)}(\vk_n)\>
= \< O(\vk) \d^{(1)}(\vk_1) \cdots \d^{(1)}(\vk_n) \>_\LO\,,
\label{eq:renormcond}
\ee
where $n=0,1,2,...$.  
The $n=0$ condition simply reads $\< [O](\vx) \> = 0$, which is what we found after \refeq{Fpbar}.  At any order
in perturbation theory, only a finite set of these conditions adds new
constraints.  Note that the constraints \refeq{renormcond} are 
sufficient to ensure that all statistics where $[O]$ contributes
at leading order, for example, the three-point function in case of $O = \d^2$,
approach the leading-order result on large scales.  This is because \refeq{renormcond} can be taken as building block to construct all other auto- and cross-correlations between any operators in the renormalized set. When $[O]$ appears in 
loop integrals, it can still lead to significant contributions 
at low $k$ that depend on small-scale modes.  However, these contributions can be expanded in powers of $k^2$ (they are analytic), and hence are absorbed by counter-terms to the renormalized stochastic fields $[\eps],\,[\eps_O]$, which show precisely the same large-scale behavior [cf. \refeq{epsOgeneral}].  We will encounter an example of this kind in \refsec{npt1loop} for $O=\d^2$.

In general, we can write the renormalized operators as a sum
over bare operators with $\L$-dependent coefficients,
\be
[O] = \sum_{O'} Z_{OO'}(\L) O'\,.
\label{eq:Orenorm}
\ee
The relation between the bare bias parameters $c_{O,\L}$ and their
renormalized counterparts $b_O$ is then given by matching
coefficients in the respective expansions, \refeq{Fp4} and \refeq{Fprenorm} (see also \refapp{biastrans}):
\be
b_O = \sum_{O'} (Z^{-1})_{O'O}(\L) c_{O',\L}\,,
\label{eq:bOrenorm}
\ee  
which involves the transpose of the inverse of the coefficient matrix $Z_{OO'}$.  That is,
in order to obtain an expression for the renormalized bias $b_O$, we
need to know which renormalized operators $[O']$ contain $O$
as counter-term.  For this reason, even though
$\d$ is not renormalized, so that $Z_{\d O'} = 0$ for any $O'\neq\d$,
the same is not true for $b_\d$, since $\d$ appears as a counter-term
for many other operators (including $\d^2$, as we have seen).  
Instead, we have \cite{assassi/etal}\footnote{Note the slightly different normalization of bias coefficients there, and that $\mathcal{G}_2 = (K_{ij})^2 - (2/3)\d^2$ in our notation (see \refapp{biastrans}).} 
\be
b_\d 
= c_{\d,\L} + \s^2(\L) \left[ \frac{68}{21} \left(c_{\d^2,\L} + \frac23 c_{K^2,\L} \right) + 3 c_{\d^3,\L} 
+ \frac23 c_{\d K^2,\L} + \frac{32}{63} c_{{\otd},\L}
\right]\,,
\label{eq:b1c1}
\ee
where $c_{\otd,\L}$ is the bare bias coefficient multiplying $O_\otd^{(3)}$.  
We will re-derive this expression from \refeqs{Fp4}{Fprenorm} 
using the peak-background split approach in \refsec{PBSrenorm}.  

The renormalized operators and bias parameters defined through 
\refeq{renormcond} and \refeq{bOrenorm} describe the $n$-point functions of galaxies in the
large-scale limit, and are thus particularly relevant to actual observations
(e.g. from galaxy redshift surveys).  It is however just
as well possible to  construct renormalized bias operators corresponding to other
observables.  Consider the joint moments of the smoothed halo density field
and matter density field on a \emph{physical} scale $R_\ell$, $\< [\d_{h,\ell}(\vx)]^n [\d_\ell(\vx)]^m \>$.  
As we will describe in detail in \refsecs{bmom}{bscatter}, these moments can be 
used to derive bias parameters, for example through the so-called hierarchical ansatz, or the ``scatter plot'' method (note that these moments are not
sufficient to unambiguously measure all bias coefficients, and mostly
restricted to the \LIMD biases $b_{\d^n}$).  
The renormalized bias parameters inferred using these measurements are
\emph{not} the same as those introduced above.  This is because the moments
are defined with respect to a specific scale $R_\ell$, and filtering kernel $W_{R_\ell}(k)$.  
The biases $b_O^{\rm m}(R_\ell)$ inferred from moments are renormalized bias
parameters in analogy to the $b_O(R_\ell)$, i.e. they are coefficients of renormalized
operators $[O]^{\rm m}_\ell$.  However, the latter are defined with different
renormalization conditions, specifically
\be
\< [O]^{\rm m}_\ell \> = 0 \quad\mbox{and}\quad
\< [O]^{\rm m}_\ell(\vx) [\d^{(1)}_{\ell}(\vx)]^n \>
= \< O[\d_{\ell},\cdots \d_{\ell}](\vx) [\d^{(1)}_{\ell}(\vx)]^n \>_\LO\,,
\label{eq:renormcondM}
\ee
where $O[\d_{\ell},\cdots \d_{\ell}]$ denotes the
operator $O$ constructed out of smoothed density fields; that is,
in case of $O = \d^n$, it stands for $O[\d_{\ell},\cdots \d_{\ell}](\vx) = (\d_{\ell}(\vx))^n$.  On the other hand, on the left-hand side, $O$ is smoothed \emph{after} taking the
nonlinear functional of the density field.  
It is straightforward to verify that the counter-terms for 
$[O]^{\rm m}_\ell$ scale in the same way with the cutoff as those for $[O]$;  
in other words, $[O]^{\rm m}_\ell$ and $[O]$ only differ by finite
$R_\ell$-dependent corrections.  The same correspondingly holds for 
$b_O^{\rm m}(R_\ell)$ and $b_O$.  The relation between the two sets
of bias parameters can be derived at any given order in perturbation
theory (\refsec{bmom}).  Note that the finite difference between $b_O^{\rm m}(R_\ell)$ and $b_O$ 
does not necessarily vanish as $R_\ell \to \infty$, as one might think, because
the finite terms involve different integrals over the linear power
spectrum.  These depend on the local spectral index of the power spectrum 
which, in turn, induces a logarithmic $R_\ell$ dependence as we will see in \refsec{bmom}. 

\subsubsection{Higher-derivative operators}
\label{sec:rderiv}

The considerations so far were limited to the operators in the local basis.  
This can be traced back to \refeq{funcx}, where we
approximated $\partial_i\partial_j\Phi_\L$ as approximately constant
over the scales over which the convolution kernel is nonzero.  
This corresponds to the lowest-order term in a Taylor series,
\be
\partial_i\partial_j\Phi_\L(\vx_1) = 
\partial_i\partial_j\Phi_\L(\vx)
+ \partial_i\partial_j\partial_k\Phi_\L(\vx) (\vx_1-\vx)^k
+ \frac12 \partial_i\partial_j\partial_k\partial_l\Phi_\L(\vx) (\vx_1-\vx)^k (\vx_1-\vx)^l + \cdots\,,
\ee
where we have written $\vx \equiv \xfl(\tau)$ for clarity. 
We can now include these sub-leading terms in \refeq{funcx}.  The leading
correction to \refeq{funcx} is 
\ba
 \prod_{i=1}^n \int d^3 \vx_i \, F^{(n)}_{i_1j_1\cdots i_n j_n}(\vx_1,\cdots,\vx_n)
\partial_{i_1}\partial_{j_1} \Phi(\vx_1) & \cdots 
\partial_{i_n}\partial_{j_n} \Phi(\vx_n) = 
\mbox{[right-hand side of \refeq{funcx}]} \vs
 + \sum_{k=0}^n \frac{n!}{(n-k)!} \partial_l\partial_m\partial_{i_1}\partial_{j_1} \Phi_{\L}(\vx)
\cdots \partial_{i_k}\partial_{j_k} \Phi_{\L}(\vx) 
\prod_{i=1}^n &\int d^3\vx_i\: (\vx_1-\vx)^l (\vx_1-\vx)^m\, F^{(n)}_{i_1j_1\cdots i_n j_n} (\vx_1,\cdots \vx_n) 
\vs
 & \times 
\partial_{i_{k+1}}\partial_{j_{k+1}}\Phi_{s}(\vx_{k+1})
\cdots \partial_{i_{n}}\partial_{j_{n}}\Phi_{s}(\vx_{n}) \vs
+ \cdots\,.\hspace{6.75cm} &
\label{eq:funcderiv}
\ea
Here, we have written only one term with both derivatives acting on
$\partial_{i_1}\partial_{j_1}\Phi_\L(\vx_1)$; the ellipsis in the last line
stands for other terms  which involve $\partial_l$ and $\partial_m$ each 
acting on one (i.e. not necessarily the same) of the 
other instances of $\partial_{i_k}\partial_{j_k}\Phi_\L$ and contracted with
one of the $i_k, j_k$ or with each other.  This is because neither the kernel $F^{(n)}$ nor the small-scale modes $\Phi_s$ have any preferred directions (apart from those induced by the long-wavelength modes themselves). For the same reason, the leading higher-derivative correction involves two powers of separations $(\vx_i-\vx)(\vx_j-\vx)$.  

We see that the higher-derivative terms have the same structure as the lowest-order term, \refeq{funcx}, with two differences: first, there are two more derivatives acting on the long-wavelength modes; second, the convolution over the small-scale modes
now involves a modified kernel,
\be
 (\vx_i - \vx)^l (\vx_j - \vx)^m F^{(n)}_{i_1j_1\cdots i_n j_n}\,.
\ee
This modified kernel, given our assumptions about the scales over which 
$F^{(n)}$ is nonzero, scales as $R_*^2$ times the kernel appearing
in \refeq{funcx}.  This then leads us to bare bias parameters 
that correspondingly scale as $R_*^2$ which multiply precisely the
higher-derivative operators described in detail in \refsec{higherderiv}.  
We can then let the smoothing scale $\L\to 0$, leading to renormalized
higher-derivative biases that scale as $R_*^2$.  
This reasoning continues correspondingly to higher orders.  

\clearpage
\subsection{Summary}
\label{sec:evol:summary}

In this section, we have described the general, perturbative bias 
expansion of arbitrary large-scale structure tracers.  Clearly, while well-defined and systematic, this 
expansion is nontrivial and consists of a complex set of terms.  For
this reason, we provide a brief summary here before moving on to
the following sections.  
The general bias expansion can be broken down into three ingredients:
\begin{itemize}
\item  \bfem{The deterministic local expansion, i.e.~at leading order in derivatives (\refsec{general}):} this series of terms of the form $b_O\, O$, where $O$ is an operator and $b_O$ is its associated bias parameter, includes as operators powers of the density and tidal field, as well as convective time derivatives of the tidal field, as summarized in \refeq{EulBasis} (in Eulerian space) and \refeq{LagrBasis} (in Lagrangian space).  Each operator $O$ has exactly two spatial derivatives acting on each occurrence of the gravitational potential $\Phi$ (where we count the differential operator $\partial_i\partial_j/\lapl$ as zero net derivatives).
\item \bfem{Stochastic contributions (\refsec{stoch}):}  in addition to the leading stochastic field $\eps$, there is an additional stochastic field $\eps_O$ associated with each operator $O$ in the bias expansion.  This can be interpreted as ``scatter'' in the deterministic bias parameter $b_O$.
\item \bfem{Higher-derivative terms (\refsec{higherderiv}):}  For each operator $O$ in the local
bias expansion, there are higher spatial derivative terms such as
$b_{\lapl O} \lapl O$ (and others; for the precise list of terms, see \refsec{higherderiv}).  Physically, these terms take into account the fact that galaxy
formation is not perfectly local.  The bias coefficients of these terms
have units of length to some power, for example ${\rm Mpc}^2$ in the case
of $b_{\lapl O}$.  The length scale that sets the value of these coefficients
is the physical ``nonlocality scale'' $R_*$ of galaxy formation;  on scales
of order $R_*$, these terms are un-suppressed, and any perturbative description
of galaxy bias necessarily breaks down.  Note that
the stochastic fields also have associated higher-derivative contributions,
which effectively capture the fact that the stochastic fields are expected
to be correlated over the scale $R_*$ (see \refsec{stoch} for
an example).
\end{itemize}
A further important result pertains to the relation between the galaxy
velocity field and the matter velocity (\bfem{velocity bias}, \refsec{velbias}):  velocity bias is guaranteed to be a higher-derivative effect. That is, we can write at lowest order in perturbations and derivatives,
\be
\v{v}_g = \v{v} + \beta_{\lapl \v{v}} \lapl \v{v}
+ \v{\eps}_v(\vx,\tau)\,,\label{eq:vgsummary}
\ee
where $\beta_{\lapl\v{v}} \sim R_*^2$ is related to the nonlocality scale of
galaxy formation (other possible terms such as $\propto \vn \d$ are equivalent
to $\lapl\v{v}$ at linear order), and $\v{\eps}_v$ is a stochastic velocity
contribution whose large-scale power spectrum scales as $k^2$ (\refsec{stoch}).

The sections referenced above provide all the ingredients necessary to
write down the general bias expansion at any desired order.  
To be specific, we now summarize the complete bias expansion of a general
galaxy sample up to third order:
\ba
\d_{g} =\:& b_1 [\d] + b_{\lapl\d} [\lapl\d] + [\eps] \vs
& + \frac12 b_2 \big[\d^2\big] + b_{K^2} \big[(K_{ij})^2\big] + \big[\eps_\d \d\big]
\vs
& + \frac16 b_3 \big[\d^3\big] + b_{\d K^2} \big[\d (K_{ij})^2\big] + b_{K^3} \big[(K_{ij})^3\big]
+ b_{\otd} \big[O_{\otd}^{(3)}\big]
+ \big[\eps_{\d^2} \d^2\big] + \big[\eps_{K^2} (K_{ij})^2\big] \vs
& + O(\d^4) + \O\big[R_*^2 \lapl(\d^2),\, R_*^4 \laplsq \d\big]\,.
\label{eq:dgsummary}
\ea
The brackets denote renormalized operators, as defined in \refsec{renormalization}.  
As indicated in the last line, the terms neglected here are either fourth
order in perturbation theory, or involve higher derivatives of nonlinear
operators, or four powers of spatial derivatives (all of these include stochastic contributions as well).  
Note that the number of higher-derivative terms to be kept depends on the
scale $R_*$ (recall that $b_{\lapl\d}\propto R_*^2$).  Here, we have assumed that $R_*$ is of order of the scale
where the matter density field becomes nonlinear, in which case it
is sufficient to keep only the leading higher-derivative term,
$b_{\lapl\d} \lapl\d$ (see \refsec{npt1loop}), when going to third order in perturbations. In this case, \refeq{vgsummary}
also provides the complete relation for galaxy velocities at this order.  
In practice, when analyzing
an actual galaxy sample, as many higher-derivative terms should be included
as the data are able to constrain (see the discussion in \cite{mcdonald/roy:2009}).  

In \refsec{measurements}, we will describe how \refeq{dgsummary} makes
predictions for galaxy (or halo) statistics.  Depending on the precise
statistic chosen, usually only a subset of the terms in the general bias expansion
contribute at a given order.  Before that however, in the next section (\refsec{PBS}), we will discuss
the physical interpretation of several of the bias parameters in 
\refeq{dgsummary} as responses of the galaxy number density to long-wavelength 
perturbations, an argument historically known as the ``peak-background split.''

%% file: PBS.tex
\clearpage
\section{Peak-background split: rigorous formulation and approximations}
\label{sec:PBS}

\secttoc

In the previous section, we have derived what bias parameters need to be
included at any given order in the perturbative bias expansion
to describe a general tracer of the large-scale structure.
We now turn to physical arguments that can be used to derive the actual
values of these bias parameters, as well as hierarchies and relations
between them.  We will begin with general physical arguments valid for
general tracers such as galaxies, before making simplifying assumptions which
mainly apply to dark matter halos.  For this, we use the \emph{peak-background
split (PBS)} approach in a more general sense than frequently understood;
this will lead to a well-defined physical interpretation of the
bias parameters that are measured from the correlation functions of
galaxies (i.e., the \emph{renormalized} bias parameters).

The decomposition of the density field into the sum of a low-amplitude
signal $\delta_\ell$ (background) with a large coherence length and a high-amplitude, noisy component $\delta_s=\delta-\delta_\ell$ (peak) with a small coherence length
was first introduced by \cite{kaiser:1984}, though the term
``peak-background split'' was
coined by \cite{bardeen/etal:1986}. The PBS simply states that
{\it a long-wavelength density perturbation acts like a local modification of
the background density for the purposes of the formation of halos and
galaxies}, since it can be considered constant over the spatial scale
within which tracers form
(that is, the scale $R_*$ introduced in \refsec{general}).
This separation of scales between the long-wavelength perturbation---on the scales on which we measure correlations---and the small-scale
perturbations---that are directly involved in the formation of tracers---is
precisely the physical argument made for the general bias expansion
in \refsec{framework}.

While the PBS argument was originally introduced for long-wavelength density
perturbations, and thus only allowed for a derivation of the \LIMD parameters
$b_N$
(e.g. \cite{kaiser:1984,cole/kaiser:1989,mo/white:1996,mo/jing/white:1997,sheth/tormen:1999}),
similar arguments can be made for other bias terms:
\begin{itemize}
\item those with respect to the tidal field (see \refsec{gen_barrier});
\item higher-derivative operators such as $\lapl\d$ or $(\grad\delta)^2$ (see \refsec{PBSpeaks}); and
\item bias contributions induced by primordial non-Gaussianity (see \refsec{NG}).
\end{itemize}
For the remainder of this section, we will focus on the bias with respect to
powers of the density $b_N$.

After the initial proposition of \cite{kaiser:1984}, the physical argument regarding
the separation of long- and short-wavelength modes has come to be
associated with theoretical approaches such as excursion set (\refsec{exset})
and peaks (\refsec{peaks}), that is, analytically tractable toy models of
large-scale structure. However, as we describe in \refsecs{bphys}{sepuni}, the PBS
argument is not specific to these models and is in fact \emph{exact}.
That is, when defined properly, the PBS predicts the exact bias parameters for
halos and galaxies.  One example of how
the exact PBS can be implemented is given in \refsec{sepuni};
the results of implementing this approach with N-body simulations
are presented later in \refsec{bsepuni}.
In \refsec{buniv}, we specialize to the case of tracers which follow
a so-called universal mass function.  This yields the well-known expressions
for the bias parameters commonly referred to as ``peak-background split
biases.''

\subsection{Bias parameters as responses}
\label{sec:bphys}

In \refsec{evolution}, we arrived at a bias expansion of
the galaxy number density perturbation of the form
\be
\d_g(\vx,\tau) = \sum_O b_O(\tau) [O](\vx,\tau)\,,
\label{eq:biasexpPBS}
\ee
where we have neglected stochastic terms, as they will not play a role
in this section.  As discussed in \refsec{renorm}, the physical,
measurable biases are really given as the coefficients of the
\emph{renormalized} operators $[O]$.  We will only deal with the renormalized bias parameters $b_O$ in \refsecs{bphys}{buniv}, although an interesting link to
the bare bias parameters and renormalization will be presented in \refsec{PBSrenorm}.

Now, we would like to ask: what is the physical interpretation of the bias coefficients
appearing in \refeq{biasexpPBS}?
Let us focus on the coefficients $b_N \equiv N!\,b_{\d^N}$ of the operator $\d^N$
[this is $\left(\tr \Pi^{[1]}\right)^N$ in the Eulerian basis
\refeq{EulBasis}].  Consider
a large region (much larger than the nonlocality scale $R_*$ of the galaxy
sample considered) characterized, at a given arbitrary time, by a mean fractional overdensity $\Delta$.
According to the peak-background split argument, the expectation value
of the \emph{physical} galaxy density $\ngp$ in such a region is given by the average abundance of tracers in a
fictitious FRW spacetime with modified physical background density
\be
\trhop = (1+\Delta)\,\rhop \,,
\label{eq:rhobprime}
\ee
where $\rhop$ is the fiducial background density.
We shall further elaborate on exactly what this ``fictitious spacetime'' is
in \refsec{sepuni}.  \refeq{rhobprime} is equivalent to adding
a uniform component $\rhop\Delta$ to the matter density.
Then, at a given point within this region where the density is
$\varrho_m(\vx) = \rhop[1+\d(\vx)]$ (here $\d$ is not necessarily small), the matter density is perturbed to
\be
\varrho_m(\vx)  \to \varrho_m(\vx) + \rhop\D  =
\rhop ( 1 + \d(\vx) + \Delta)\,,\quad\mbox{or}\quad
\d(\vx) \to \d(\vx) + \Delta\,.
\label{eq:rhobp}
\ee
Thus, we \emph{shift} all density perturbations by an amount $\Delta$.
Note that we add a fixed amount of \emph{uniform} matter density everywhere;
we do \emph{not} rescale the local matter density $\varrho_m$
by $1+\Delta$, which would also amplify the fluctuations $\d$.

We now introduce the strict definition of \emph{peak-background split bias parameters} $b_N$ ($N \geq 1$) as
the derivative of the mean physical number density of galaxies $\avngp$ with 
respect to $\Delta$:
\be
b_N \equiv \frac{1}{\avngp|_{\Delta=0}} \frac{\partial^N \avngp|_\Delta}{\partial \Delta^N}\Big|_{\Delta=0}.
\label{eq:bN}
\ee
Using \refeq{rhobprime}, we can also write this as
\be
b_N
= \frac{\rhop^N}{\avngp} \frac{\partial^N \avngp}{\partial \trhop^N}\Big|_{\rhop}\,,
\ee
where the derivatives are evaluated at the fiducial value of $\rhop$.
These relations show that, if we can predict the abundance of galaxies as a
function of the modified background density $\trhop$, at fixed time $t$, we can predict the \LIMD bias parameters $b_N$.  In the next section, we will
discuss precisely this approach of measuring biases, in particular for dark matter halos.
In the course of that calculation, we will also clarify the physical meaning of the transformation \refeq{rhobp}.

\subsection{Exact implementation of the PBS: separate-universe approach}
\label{sec:sepuni}

To begin, let us define a bit more carefully what we mean by the transformation
\refeqs{rhobprime}{rhobp}.  A uniform adiabatic density perturbation on an
FRW background is equivalent to a different (curved) FRW background, as first shown by
\cite{lemaitre:1933} (we continue to assume Gaussian initial conditions in this section).  This idea, often referred to as ``separate-universe'' approach, has been used in many
calculations and N-body simulations since then, first starting at linear order in the perturbation $\Delta$
\cite{barrow/saich:1993,cole:1997,mcdonald:2003,goldberg/vogeley:2004,sirko:2005,martino/sheth:2009,gnedin/kravtsov/rudd:2011,
baldauf/etal:2011,sherwin/zaldarriaga:2012,li/hu/takada:2014}, and more recently
generalized to fully nonlinear order \cite{wagner/etal:2014,CFCpaper2}.
The following brief description follows the one laid out in \cite{wagner/etal:2014}. We assume a $\Lambda$CDM cosmology throughout.

Consider a long-wavelength overdensity $\Delta(t)$ as in \refeq{rhobp}, distinguished there from the small-scale
perturbations $\d$.  Throughout, we will allow $\Delta$ to be fully nonlinear; the separate-universe approach is
not restricted to linear order in the density perturbation, as emphasized in \cite{CFCpaper2}.
Moreover, $t$ stands for the proper time of comoving observers throughout.
We then proceed to absorb the overdensity $\Delta$ into the physical background matter density of a modified cosmology
$\trhop(t)$ as
\ba
\rhop(t) \left[ 1 +\Delta(t)\right] = \trhop(t) \,. \label{eq:absorb}
\ea
Thus, instead of embedding the region with overdensity $\Delta$ in a
fiducial background Universe, one considers it as a separate universe with an altered cosmology.
We will now derive the parameters of this modified
cosmology in relation to $\Delta$, specifically the \emph{linearly extrapolated} present-day overdensity
\be
\Delta_0^{(1)} \equiv \Delta(t_i) \frac{D(t_0)}{D(t_i)}\,,
\ee
where $D$ is the linear growth function of the fiducial cosmology, $t_0$ is the present time, and $t_i$ denotes
an early time at which $\Delta(t_i)$ is still small. Again, we will not assume that $\Delta_0^{(1)}$ is small, although in
most practical applications one will choose $|\Delta_0^{(1)}| \lesssim 1$.

Expressed in terms of the standard cosmological parameters, i.e.~$\rhop(a=1)= \rhob = \Omn \frac{3H_0^2}{8\pi G}$ and
$H_0=100\,h\,\rm{km\, s}^{-1}\rm{Mpc}^{-1}$, \refeq{absorb} becomes
\ba
\frac{\Omn h^2}{a^3(t)} \left[ 1 +\Delta(t)\right] = \frac{\tOmn \tilde h^2}{\tilde a^3 (t)} \,,
\label{eq:omh2t}
\ea
where we used a tilde to denote quantities in the modified cosmology.
For the fiducial cosmology, we adopt the standard convention for the scale factor $a(t_0)=1$.
In contrast, for the modified cosmology, it is convenient to choose the normalization such that $\lim_{t\to 0} \tilde a(t)/a(t) = 1$. These conventions lead to (as $\Delta(t\to0)=0$)
\be
\Omn h^2=\tOmn \tilde h^2\,,\quad
\OLn h^2=\tOLn \tilde h^2\,.
\label{eq:omh2}
\ee
Note that this relation implies that the comoving matter densities defined using the respective scale factor are the same: $\trho \equiv \tilde a^3 \trhop = \rhob$.  
Introducing $\delta_a(t)$ through $\tilde a(t)=[1+\delta_a(t)] a(t)$, we find from \refeq{absorb}
\be
1 + \Delta(t) = [1 + \d_a(t)]^{-3}\,.
\label{eq:dadrho}
\ee
This is just a statement of mass conservation.  
The Friedmann equation for $a(t)$ is, assuming a flat fiducial cosmology
for simplicity, given by
\be
H^2(t) = \left(\frac{\dot a}a\right)^2
= \frac{8\pi G}3 \rhop(t) + \frac13 \Lambda \,,
\label{eq:Fr1}
\ee
where $\Lambda$ denotes the cosmological constant.
The same equation, but including curvature $\tilde{\text{K}}$ and modifying the
matter density, holds for $\tilde a(t)$:
\be
\tilde H^2(t) = \left(\frac{\dot{\tilde a}}{\tilde a}\right)^2 =
\frac{8\pi G}3 \trhop(t) + \frac13\Lambda
 - \frac{\tilde{\text{K}}}{\tilde a^2(t)}\,.
\label{eq:Fr1t}
\ee
One can then combine \refeqs{Fr1}{Fr1t} with the corresponding second
Friedmann equations to obtain an evolution equation for $\Delta$:
\be
\ddot\Delta + 2 H \dot\Delta - \frac43 \frac{\dot\Delta^2}{1+\Delta}
= 4\pi G \rhop\, (1+\Delta) \Delta\,.
\label{eq:drhoNL}
\ee
When linearizing this equation in $\Delta$, one recovers the equation for
the linear growth factor (given in \refeq{Deom} using $\tau$ rather than $t$ as  time variable). Beyond linear order, 
\refeq{drhoNL} is exactly the equation for the interior density of a spherical
tophat perturbation in a $\Lambda$CDM Universe (see \refsec{sph_collapse} and App.~A of \cite{HPMhalopaper}).
Taking the difference of \refeq{Fr1} and \refeq{Fr1t} yields a relation for the curvature $\tilde{\text{K}}$.  One can verify that $\tilde{\text{K}}$
is conserved \cite{wagner/etal:2014}, which is a necessary condition for \refeq{Fr1t} to
describe a physical FRW solution \cite{CFCpaper2}.

One can also generalize the separate-universe picture to include a dark energy component instead of a cosmological constant.  In that case, dark energy perturbations also need to be taken into account, and the conservation of curvature only holds outside the sound horizon of the dark energy component.  Recently, Ref.~\cite{hu/etal:2016} generalized the separate-universe approach to effectively include pressure and anisotropic stress perturbations, although this only holds strictly when following the evolution of non-relativistic fluids such as baryons and CDM.  We return to this issue in \refsec{modgrav}, and assume for the remainder of \refsec{PBS} that we are outside the sound horizon of all fluid components.
In this case, we can evaluate the curvature at an early time $t_i$, when the perturbation $\d_a$ is infinitesimal and the Universe is in matter domination.
We then have $H^2=H_0^2 \Omn a^{-3}$, $\dot\d_a=H\d_a$, and $\d_a=-\Delta/3$, with which one can derive
\be
\frac{\tilde{\text{K}}}{H_0^2} = \frac53 \frac{\Omn}{a(t_i)} \Delta(t_i)\,.
\ee
Alternatively, using the linear growth
factor $D_\text{md}$ normalized such that $D_\text{md}(t_i)=a(t_i)$, we can write
\be
\frac{\tilde{\text{K}}}{H_0^2} = \frac53 \frac {\Omn}{D_\text{md}(t_0)}  \Delta_0^{(1)}\,.
\label{eq:deltaKL}
\ee
We have thus recovered the relation between curvature and matter density in synchronous-comoving gauge given in \refsec{GR} [\refeq{KFsc2}]. 
Now let us derive the parameters of the modified cosmology. They are defined through the Friedmann equation at time $\tt$ where  $\tilde a(\tt)=1$.
Defining the fractional perturbation to the Hubble parameter $\d_H$ through $\tilde H(\tt) = H_0 [ 1 + \d_H]$
and using \refeq{omh2}, we obtain
\be
\tOmn = \Omn [ 1 + \d_H]^{-2}; \quad
\tOLn = \OLn [ 1 + \d_H]^{-2}\,.
\ee
Finally, in order to derive $\delta_H$ we can make use of the Friedmann equation at $\tt$, which yields, for a flat fiducial cosmology,
\be
\tOKn = -\frac{\tilde{\text{K}}}{\tilde H_0^2} 
=
1-\tOmn -\tOLn
= 1 - (1+\d_H)^{-2}\,.
\ee
Since $\tilde{\text{K}}$ is given by \refeq{deltaKL}, we
can use this relation to solve for $\d_H$:
\be
\d_H = \left(1-\frac{\tilde{\text{K}}}{H_0^2} \right)^{1/2} - 1\,.
\ee
Thus, $\d_H > 0$ if $\tilde{\text{K}} < 0$, i.e. in case of an underdense region ($\Delta_0^{(1)}<0$). 
There is no solution if $\tilde{\text{K}}/H_0^2 \geq 1$, or equivalently
$\Delta_0^{(1)} \geq 3 D_\text{md}(t_0) / (5 \Omn)$.  This is because for such a large
positive curvature, the Universe reaches turnaround at or before $\tilde a = 1$.
This is not a physical problem, it is merely not possible
to parametrize such a cosmology in the standard convention.
For practical applications, smaller values of $\Delta_0^{(1)}$ are in any case sufficient.

With these relations, it is straightforward to perform N-body simulations
which implement a uniform density perturbation as in \refeqs{rhobprime}{rhobp}.  The power spectrum used to generate the initial conditions is unmodified in shape and amplitude apart from the modification to the linear growth factor in the background $\tilde a(t)$.  One remaining subtlety is that
we want to output the data at a fixed physical time $t_{\rm out}$, which in
N-body codes is usually specified by the scale factor as $a(t_{\rm out})=a_{\rm out}$.  Therefore we need to determine the corresponding
scale factor in the modified cosmology as $\tilde a(t_{\rm out})=a_{\rm out}\left[1+\delta_a(t_{\rm out})\right]$, which can be done easily numerically for any given value of $\Delta_0^{(1)}$.

It is now straightforward to give an operational procedure to derive
the exact PBS biases \refeq{bN} for any tracer whose formation can be
simulated: we run simulations, of sufficient volume to
contain a statistical sample of the tracers of interest, up to a
given fixed proper time $t$, with different parameters following the 
prescription above, which implement various values of $\Delta(t)$. 
This allows for a measurement of $\avngp(t,\Delta(t))$. 
The Eulerian \LIMD biases for the tracers are then given as derivatives of the physical density of tracers $\avngp$ with respect to $\Delta$:
\be
b_N^E(t) = N!\, b_{\d^N}^E(t)
= \left.\frac1{\avngp(t,0)} \frac{\partial^N \avngp(t,\Delta(t))}{\partial[\Delta(t)]^N}\right|_{\Delta=0}\,,
\label{eq:bNgal}
\ee
which corresponds to the rigorous definition of \refeq{bN}.
Note that a nonlinear implementation of the separate-universe rescaling is 
essential if one wants to derive the higher-order biases starting with $b_2$;
that is, the implementation has to correctly take into account the nonlinear evolution
of the long-wavelength mode.

We now specialize to the most commonly considered case of dark matter halos
at fixed mass. We define $\avnh(M,t)$ as
the mean \emph{comoving} number density of halos per logarithmic mass interval.
Since the comoving density is related to the physical density through
$\avnh = a^3 \avnhp$, and correspondingly
$\avnh(\Delta) = \tilde a^3 \avnhp(\Delta)$ in the modified cosmology, the quantity $\avnh(\Delta)/\avnh(0)-1$ is the fractional \emph{Lagrangian} overdensity of halos induced by a long-wavelength density perturbation $\Delta$ [see \refeq{dadrho}].
Then, the Lagrangian \LIMD biases $b_N^L$ of halos are given as derivatives of $\avnh(M,t)$ with respect to $\Delta_0^{(1)}$, i.e.
\be
b_N^L(M,t) = N!\, b_{\d^N}^L(M,t)
= \left.\frac1{\avnh(M,t,0)} \frac{\partial^N \avnh(M,t,\Delta(t))}{\partial[\Delta_0^{(1)}]^N}\right|_{\Delta_0^{(1)}=0}\,.
\label{eq:bNhaloL}
\ee
Since we are considering a pure density perturbation, i.e. a spherically symmetric long-wavelength perturbation which does not induce a proper tidal field ($K_{ij}=0$), the mapping between $b_N^L$ and the Eulerian \LIMD biases $b_N^E$ is exactly given by the mapping described in \refsec{localbias}, in particular \refeq{scbiasEL}.
Note that it is essential that halos are identified in the separate-universe simulations with respect to a fixed physical density criterion in the fiducial cosmology;  in particular, the overdensity or linking length passed on to the halo finder (see \refapp{halofinder}) needs to be adjusted accordingly.
We will discuss results of these measurements in \refsec{bsepuni}.

\subsection{PBS biases for universal mass functions}
\label{sec:buniv}

After having defined the rigorous peak-background split biases, we now present
a well-known case where \refeq{bNhaloL} can be evaluated
analytically. Motivated by the excursion-set argument (which we will describe
in \refsec{exset}), the halo mass function, the mean comoving number density of dark matter halos per logarithmic mass interval, is often parametrized in the form
\ba
\avnh(M) \equiv\:& \frac{\partial^2 \bar{N}_h}{\partial V \partial \ln M} =
\frac{\rhob}{M}\nu_c(M)\:f[\nu_c(M)]\: J(M)
\label{eq:univ1}\\
\nu_c(M) \equiv\:& \frac{\dc}{\s(M)}; \quad
J(M) \equiv \left\lvert\frac{d\ln \s(M)}{d\ln  M}\right\lvert = \frac{d\ln\nu_c(M)}{d\ln M} \;,
\label{eq:Js}
\ea
where $\s(M) \equiv \s[R(M)]$ is the variance of the linear matter density field smoothed
on the Lagrangian radius $R$ which is related to the mass $M$ through $M = (4\pi/3) \rhob R^3(M)$, and
$\dc$ is the linearly extrapolated initial spherical overdensity that collapses at the time at which \refeq{univ1} is evaluated (see \refsec{sph_collapse}).
The scale-independent collapse threshold $\dc$ can be promoted to a mass-dependent \emph{barrier} $B(M)$.
We discuss different choices of barrier in \refsec{gen_barrier}.
Furthermore, the so-called multiplicity function $f(\nu_c)$ generally is an arbitrary function of $\nu_c$.
Finally, the Jacobian $J$ is included in order to convert from an interval
in the variance $\s(M)$ to a mass interval.

\refeq{univ1} is referred to as ``universal mass function'' and was first introduced by \cite{press/schechter:1974}.
More generally, an EdS Universe with a pure power-law matter power spectrum $\Plin(k) \propto k^n$ obeys a scaling symmetry
(e.g., \cite{pajer/zaldarriaga:2013}).  Then, the halo mass function as a function of $M$ and $z$ can be exactly written in the form \refeq{univ1},
since $\nu_c$ is invariant under the scaling symmetry and all moments of the matter density are directly proportional to $\sigma(M)$
\cite{sheth/tormen:1999}.
The fact that the linear matter power spectrum in the standard $\Lambda$CDM cosmology can be fairly well approximated as a power law over a range of scales
(thus approximately obeys the scaling symmetry) is one of the main reasons why \refeq{univ1} gives fairly accurate results in $\Lambda$CDM. In fact, the approximate symmetry can be used to rescale simulations from one cosmology to another \cite{angulo/white:2010}.
For precision estimates of the halo bias however, the departure of $\Lambda$CDM from a power-law EdS Universe must be taken into account.

In order to derive the bias parameters via \refeq{bNhaloL}, we need to know how $\avnh$ changes under a change in the background density of the Universe
[\refeq{rhobprime}].  Since the comoving density $\trho \equiv \tilde a^3 \trhop
$ is unchanged [see the discussion after \refeq{omh2t}], the only nontrivial contribution to the \LIMD bias comes from the response of the barrier to a change in the background density.\footnote{Note that since we are studying an infinite-wavelength perturbation, it does not contribute to the variance $\s(M)$; the effect of a finite wavelength of the perturbation is captured by higher-derivative biases.}
Ref.~\cite{mo/white:1996} argued that, under a long-wavelength density perturbation $\Delta_0^{(1)}$, the collapse threshold shifts to
\be
\dc \to \dc - \Delta_0^{(1)}(t_0) \;.
\ee
We now derive why and when this holds in the context of the exact separate-universe picture described in \refsec{sepuni}.
The threshold $\dc$ is defined as the linearly extrapolated initial density 
contrast of a region collapsing at a given proper time $t_0$.
Since General Relativity is scale-free, this threshold is independent of the size and enclosed mass of the perturbation, unless there
is an additional scale in the matter sector, for example the sound horizon of one of the fluid components, which we will ignore here.
In an EdS Universe, a spherical perturbation with a present-day linear fractional overdensity $\dc \approx 1.686$ collapses at $a=1$ (this value is derived in \refsec{sph_collapse}).
The same reasoning holds for more general expansion histories, where $\dc$ assumes other values (e.g., \cite{eke/etal:1996}). 
Since the evolution of a spherical perturbation is independent of the external Universe (by Birkhoff's theorem)
a perturbation of the same physical density 
$\varrho_\text{cr}\equiv \rhop(1+\dc)$ will collapse at the same proper time in a Universe with modified background density following \refeq{rhobprime}.
We now derive what this implies for the significance $\nu_c = \dc/\s(M) = (\varrho_\text{cr}-\rhop) / \d\varrho_{\rm RMS}$, which quantifies how rare 
fluctuations above a physical density threshold $\varrho_\text{cr}$ are in the \emph{linearly extrapolated initial density field}, given its root-mean-square fluctuation amplitude $\d\varrho_{\rm RMS} = \s(M) \rhop$.  Following our discussion in \refsec{sepuni}, we add a uniform matter density component $\Delta_0^{(1)}\,\rhop$ to the linear density field.  Then, the critical overdensity changes to
\be
\varrho_\text{cr} - \trhop = (1 + \dc)\rhop - (1 + \Delta_0^{(1)})\rhop = (\dc - \Delta_0^{(1)}) \rhop\,.
\ee
Thus, the significance is modified to
\be
\tilde\nu_c = \frac{\varrho_\text{cr}-\trhop}{\s(M)\rhop} = \frac{\dc - \Delta_0^{(1)}}{\s(M)}\,.
\label{eq:tildenuc}
\ee
For a mass function of the form \refeq{univ1}, changing the background density is thus equivalent to changing $\dc \to \dc - \Delta_0^{(1)}$.
\refeq{bNhaloL} applied to \refeq{univ1} thus immediately yields
\be
b_N^L(M) = \frac{(-1)^N}{[\sigma(M)]^N} \frac{1}{\nu_cf(\nu_c)}
\frac{d^N [\nu_c f(\nu_c)]}{d \nu_c^N}\,. 
\label{eq:bPBSuniv}
\ee
This is the widely known expression for the peak-background split Lagrangian bias parameters, which really
are a special case of \refeq{bNhaloL}.
Furthermore, \refeq{bPBSuniv} also holds if $f(\nu_c)$ is given by an integral over other variables
(such as, for example, the peak curvature), as long as these other variables are independent of $\dc$
\cite{desjacques/crocce/etal:2010,desjacques:2013,PBSpaper}.
Specifically, the result remains valid if $f(\nu_c)$ depends on ratios of spectral moments, e.g. $\s_1(M)/\s(M)$, since these are not changed by the long-wavelength density perturbation. 
\refeq{bPBSuniv} has been applied extensively to compute the Lagrangian and Eulerian \LIMD bias
parameters from analytic prescriptions of the multiplicity function;
see \cite{cole/kaiser:1989,mo/white:1996,sheth/tormen:1999} to only cite a few.
We will review some of these prescriptions in \refsec{exset} and \refsec{peaks}.

\begin{figure}[!ht]
\centering
\includegraphics[width=0.45\textwidth]{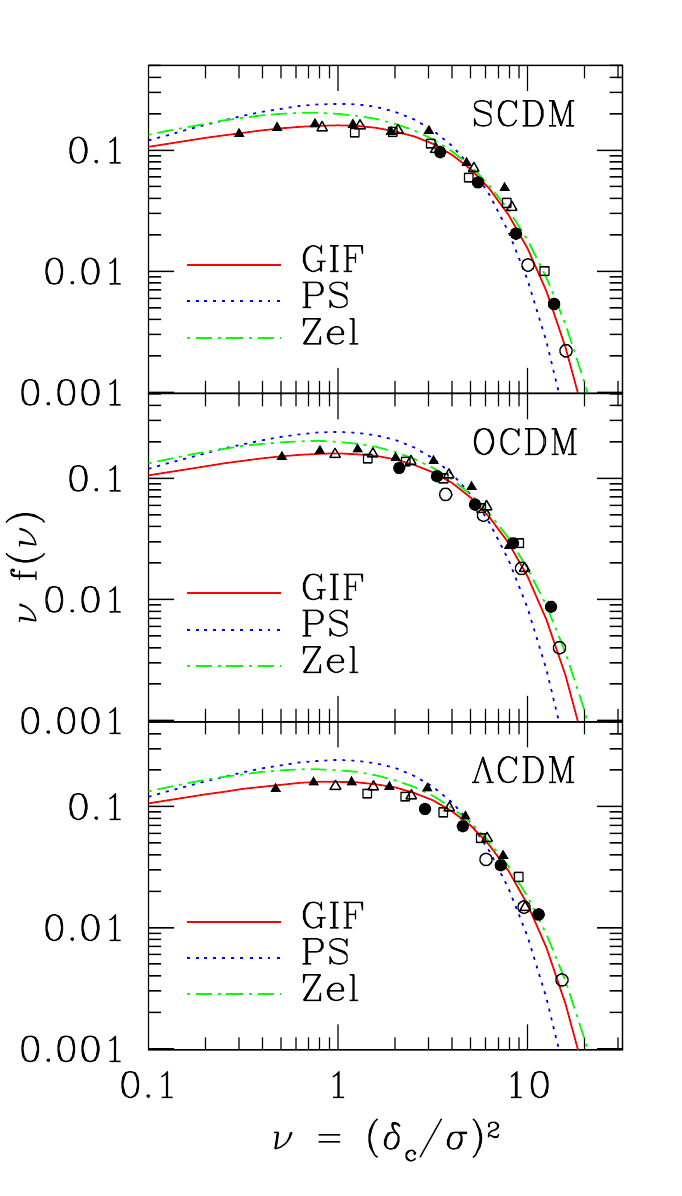}
\includegraphics[width=0.45\textwidth]{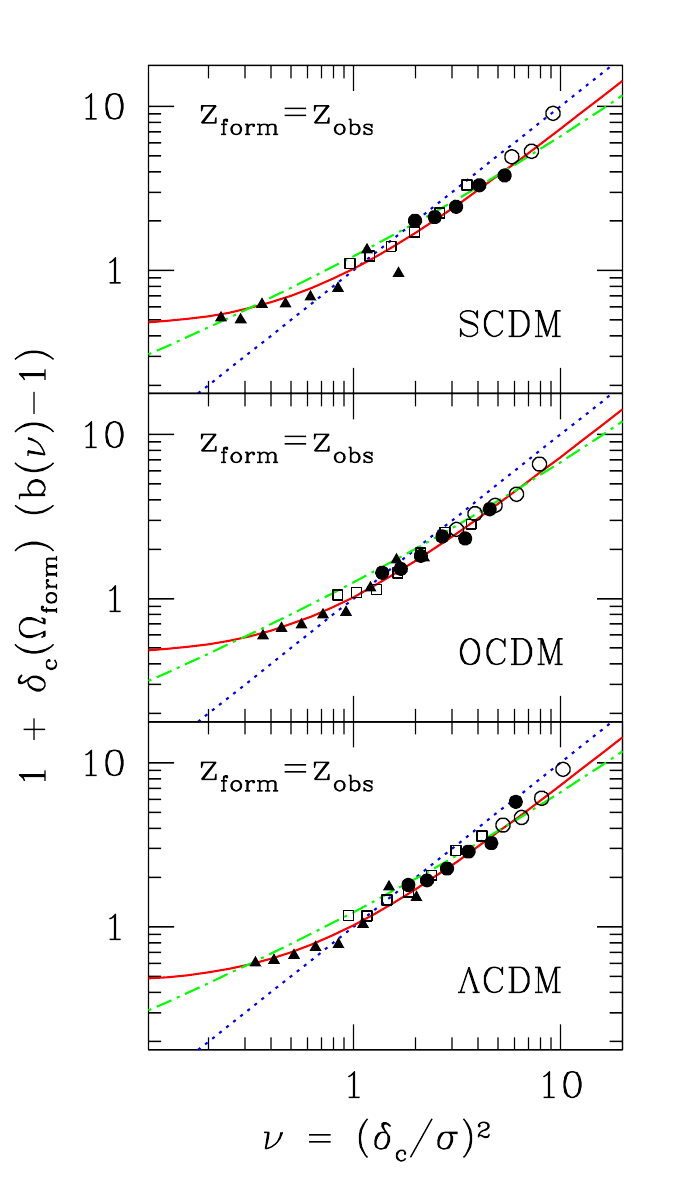}
\caption{The multiplicity function $f(\nu) \equiv \nu_c f(\nu_c)$ \textit{(left panels)} and the scaled linear Lagrangian \LIMD bias $b(\nu)-1 \equiv b_1^L$ \textit{(right panels)} of halos identified using an FoF halo finder at the output redshifts $z_\text{obs}=0$ (filled triangle), 1 (empty square), 2 (filled circle) and 4
(empty circle).  Note that $\nu \equiv (\nu_c)^2$ in our notation.
The halo bias was measured using the halo power spectrum (\refsec{bnpt}). $b_1^L$ has been rescaled as indicated by the axis label to show that, for a universal mass function, the dependence of the PBS bias parameters on cosmology arises only through
$\nu_c$. The solid (red) curves (labeled ``GIF'') represent the relation predicted by the Sheth-Tormen mass function [\refeq{nufnuST} and \refeq{bST}, respectively], whereas the
dotted (blue) curves (labeled ``PS'') show the relations which follow from the Press-Schechter mass function [\refeq{nufnuPS} and \refeq{bEPS}, respectively]. Finally, the (green)
dot-dashed line (labeled ``Zel'') is the mass function in the Zel'dovich approximation \cite{lee/shandarin:1998}.
\figsource{sheth/tormen:1999}
\label{fig:shethtormen}}
\end{figure}

Perhaps the most widely known calculation along these lines is the derivation
of the \LIMD bias parameters from the Press-Schechter (PS, \cite{press/schechter:1974}) and Sheth-Tormen (ST, \cite{sheth/tormen:1999}) mass functions. The Press-Schechter multiplicity function, which can be derived analytically starting from simple physical assumptions (see \refsec{PS}), is given by
\be
\nu_c f_\text{PS}(\nu_c) = \sqrt{\frac{2}{\pi}} \,\nu_c e^{-\nu_c^2/2}\,.
\label{eq:nufnuPS}
\ee
Applying \refeq{bPBSuniv} to this, we immediately obtain for the linear and second-order Lagrangian \LIMD bias parameters 
\ba
(b_1^L)_\text{PS} =\:& \frac{\nu_c^2-1}{\dc} \vs
(b_2^L)_\text{PS} =\:& \frac{\nu_c^2}{\dc^2} \left(\nu_c^2 - 3\right)\,.
\label{eq:bLPS}
\ea
We will re-derive these well-known results using the excursion-set formalism in \refsec{sharpk_bias}. We can now map these to Eulerian biases through \refeq{scbiasEL}, to obtain
\ba
(b_1^E)_\text{PS} =\:& 1 + \frac{\nu_c^2-1}{\dc} \vs
(b_2^E)_\text{PS} =\:& \frac8{21} \left(\frac{\nu_c^2-1}{\dc} \right) + \frac{\nu_c^2}{\dc^2} \left(\nu_c^2 - 3\right)\,.
\label{eq:bEPS}
\ea
Now, the agreement between the Press-Schechter mass function 
and the abundance of halos in N-body simulations is far from perfect,
as is apparent in the left panels of \reffig{shethtormen}. 
The same is also true for the linear bias parameter $b_1^E$ of simulated halos when compared with \refeq{bEPS} (right panels in \reffig{shethtormen}).
Following the suggestion of \cite{sheth/lemson:1999} that this discrepancy
primarily arises because of the limitations of the Press-Schechter model,
and not the peak-background-split argument,
Sheth and Tormen \cite{sheth/tormen:1999} derived an empirical expression for the multiplicity function which subsequently was shown to be of a functional form expected from a version of the ellipsoidal collapse model
(\cite{sheth/mo/tormen:2001,sheth/tormen:2002}; see \refsec{gen_barrier}),
\be
\nu_c f_\text{ST}(\nu_c) =
\sqrt{\frac{2}{\pi}}
\left[1+ \frac1{2^{p}\sqrt{\pi}}  \Gamma\left(\frac12-p\right) \right]^{-1}
\left[1+(q \nu_c^2)^{-p}\right] \sqrt{q}\,\nu_c e^{-q\nu_c^2/2}\;,
\label{eq:nufnuST}
\ee
where $p=0.3$ and $q=0.707$ were fitted to the measured halo mass function. 
Therefore, these parameters
generally depend on the precise definition of halos employed in the simulation analysis;
we describe the commonly used halo finding algorithms in \refapp{halofinder}.
The PS mass function corresponds to the special case $p=0$, $q=1$ in \refeq{nufnuST}. 
Note that \refeq{nufnuST} is normalized to satisfy the constraint
$\int M \avnh d\ln M = \rhob$ [as is \refeq{nufnuPS}]. 
The Eulerian PBS bias parameters $b_1^E$ and $b_2^E$
inferred from the multiplicity function \refeq{nufnuST} on
applying the same procedure as for the PS mass function, \refeq{bPBSuniv} and \refeq{scbiasEL}, are obtained as
\begin{align}
\label{eq:bST}
(b_1^E)_\text{ST} &= 1 + \frac{q\nu_c^2-1}{\dc}+\frac{2p/\dc}{1+(q\nu_c^2)^p} \\
(b_2^E)_\text{ST} &= \frac{8}{21}\left(\frac{q\nu_c^2-1}{\dc}+\frac{2p/\dc}{1+(q\nu^2)^p}\right)
+ \frac{q\nu_c^2}{\dc^2}\left(q\nu_c^2-3\right) \\
&\quad
+ \frac{2p}{\dc^2} \left(-1+2p +2q\nu_c^2\right)\frac{1}{1+(q\nu_c^2)^p}
\nonumber \;.
\end{align}
Ref.~\cite{sheth/tormen:1999} found that \refeqs{nufnuST}{bST} provide a much better fit to the mass function and
linear bias of halos than the Press-Schechter prediction.
\reffig{shethtormen} shows the multiplicity function $\nu_c f(\nu_c)$ (left panels) and scaled linear Lagrangian bias $b_1^L$ (right panels) for halos identified at redshift $z_\text{obs}=0$, 1, 2 and 4. Clearly, the Sheth-Tormen mass function is a better fit to the data at all masses and redshifts, and so is the PBS bias prediction based on this mass function. As emphasized by \cite{sheth/tormen:1999}, this demonstrates that knowledge of the mass function also means knowledge of clustering, through the peak-background split.

The accuracy of \refeq{bPBSuniv} was further quantified in \cite{manera/etal:2010,hoffmann/bel/gaztanaga:2015},
who empirically calibrated the multiplicity function of the halos identified in their simulations.
They found that the bias parameters derived from this multiplicity function via \refeq{bPBSuniv} are accurate at the 10--20\% level.  Following our discussion in this section, this limited accuracy is due to the limitations of the universal mass function prescription [\refeq{univ1}], and not due to the peak-background split argument itself, which is exact when defined rigorously.

\refeq{bPBSuniv} is based on a mass-independent critical density, or \emph{barrier}, $\dc$.
One can generalize this to a general barrier $B(M)$, so that $\nu_c \equiv B(M)/\s(M)$, which is intended to capture other physical effects,
such as ellipsoidal collapse and the scatter due to the specific realization of small-scale initial
conditions (see \refsec{gen_barrier}).
We can now make the \emph{assumption} that \refeq{tildenuc} analogously holds for a general barrier, yielding
\be
\tilde\nu_c = \frac{B(M) - \Delta_0^{(1)}}{\s(M)}\,.
\ee
Then, \refeq{bPBSuniv} remains valid even for a mass-dependent barrier.
Let us thus summarize the two assumptions made in \refeq{univ1} and \refeq{bPBSuniv}: $(i)$ the abundance of halos
depends on the amplitude of the matter power spectrum only through a single moment $\sigma(M)$; $(ii)$ the barrier
$B(M)$ is unchanged by the presence of a long-wavelength density perturbation, which allows us to derive the bias parameters from the mass function.
Birkhoff's theorem ensures that the second assumption is true for a spherically symmetric setup.  However, real halos are not spherically symmetric, and tidal fields cannot be ignored, rendering the second assumption an approximation.

Note that, for a general barrier, we have two free functions at our disposal: $f(\nu_c)$ and $B(M)$.  These can, for
example, be
estimated from the mass function $\avnh(M)$ and the linear bias $b_1(M)$.
In that case, measurements of $b_1(M)$ can be matched exactly, and the PBS biases only become
predictive for $b_N$ with $N=2,3,\cdots$.  We will discuss the comparison with N-body simulations in detail in \refsec{bsepuni}.

\subsection{Renormalized biases and the PBS*}
\label{sec:PBSrenorm}

\technote{* This section is of a more technical nature and is not essential for the remainder of the review.}

We now make the connection between the PBS bias parameters defined as response
to the transformation \refeq{rhobp} and the renormalized bias parameters
defined in \refsec{renorm}.
We can formally obtain the average number density of galaxies $\avng = \< n_g\>$
in a Universe with modified background density $\trhop = (1 + \Delta) \rhop $, starting from the expansion in terms of the coarse-grained
fields $\d_\L, K_{ij,\L}, ...$, \refeq{Fpbar}.  For this, we will assume that
the coarse-graining scale $\L^{-1}$ is sufficiently large so that NLO (1-loop)
perturbation theory applies.  This assumption is being made merely in order
to be able to connect to the perturbative relation between bare and renormalized
bias parameters.  Expanding up to cubic order and keeping terms that
depend on $\Delta$, \refeq{Fpbar} yields
\ba
\< n_g \>|_\Delta
= n_{0,\L} \bigg[& 1 + c_{\d,\L} \<\d_\L\>_\Delta + c_{\d^2,\L} \<\d_\L^2\>_\Delta + c_{K^2,\L} \< (K_{ij,\L})^2 \>_\Delta +  c_{\d^3,\L} \< \d_\L^3\>_\Delta +  c_{\d K^2,\L} \< \d_\L (K_{ij,\L})^2 \>_{\Delta} \vs
& + c_{\otd,\L} \< O_{\otd,\L}\>_\D + \cdots \bigg]\,.
\label{eq:FpbarD}
\ea
Note that in the presence of $\Delta$, the expectation value of the
overdensity $\left<\d_\L\right>_\Delta$ no longer vanishes.  Here, both $n_{0,\L}$ and the $c_{O,\L}$ refer to the
quantities defined in the Universe with background density
$\rhob$, i.e. $\Delta=0$.  Let us expand this expression to linear order in $\Delta$ around $\Delta=0$.  The expectation values on the right-hand side are easily calculated at cubic order in perturbation theory.  For the linear term, we trivially have
$\<\d_\L\>_\Delta = \Delta$.
In the cubic terms, we can also directly insert $\d_\L \to \d_\L + \Delta$ to obtain
\ba
\< \d_\L^3\>_\Delta =\:& \< \d_\L^3\> + 3 \< \d_\L^2 \> \Delta+ \O(\Delta^2) \vs
\< \d_\L (K_{ij,\L})^2 \>_\Delta =\:& \< \d_{\L} (K_{ij,\L})^2 \> + \< (K_{ij,\L})^2 \> \Delta + \O(\Delta^2) \vs
\< O_{\otd,\L}\>_\Delta =\:& \frac{16}{21} \< (K_{ij,\L})^2\> \Delta + \O(\Delta^2)
\,,
\ea
where expectation values without subscripts denote those in the fiducial
cosmology ($\Delta=0$), and $\<(K_{ij,\L})^2\>=(2/3)\< \d_\L^2\>$. The last line can be obtained from the definition
of $O_\otd$, \refeq{Otddef}, and noting that $\Del_{ij} [ (\d_\L)^2 - 3/2 (K_{ij,\L})^2]$ becomes, upon replacing $\d_\L(\vx)$ with $\d_\L(\vx) + \D$,
$\Del_{ij} [ (\d_\L)^2 - 3/2 (K_{ij,\L})^2] + 2 (K_{ij,\L}+ \d_\L \d_{ij}/3) \Delta$.

In order to derive the vacuum expectation value of the quadratic operators,
$\< \d_\L^2\>_\D,\,\<(K_{ij,\L})^2\>_\D$,
we need to insert the change in the amplitude of the small-scale power spectrum in the presence of a long-wavelength mode.
This can be calculated directly in second-order perturbation theory, either using \refeq{d2}, taking the angle-averaged squeezed limit of the
leading-order bispectrum, or calculating the growth factor in a modified background cosmology (\refsec{sepuni}) \cite{bernardeau/etal:2002}.
Any of these calculations leads to
\ba
\< \d_\L^2 \>_\Delta =\:& \<\d_\L^2\>\left(1 + \frac{68}{21} \Delta\right) + \O (\Delta^2)
\quad\mbox{and}\quad
\< K_{ij,\L}^2 \>_\Delta = \< K_{ij,\L}^2\>\left(1 + \frac{68}{21} \Delta\right) + \O (\Delta^2)
\,.
\ea
Putting everything together, we obtain
\be
\< n_g \>|_\Delta
= \< n_g\> \left[ 1 + c_{\d,\L} \Delta + \Delta\, \s^2(\L) \left\{ \frac{68}{21} \left(c_{\d^2,\L}  + \frac23 c_{K^2,\L}\right)
+ 3 c_{\d^3,\L} + \frac23 c_{\d K^2,\L}
+ \frac{32}{63} c_{\otd,\L}
\right\} + \O(\Delta^2)\right]\,,
\label{eq:Fpbar1}
\ee
where $\s^2(\L) \equiv \< \d_\L^2\>$ and we have absorbed the
$\Delta$-independent terms into $\< n_g \>$ (consistently at the order we
are working in).

Following \refeq{bN}, we now introduce the strict definition of
the PBS bias parameters $b_N$ ($N \geq 1$) as the derivative of $\<n_g\>$ with respect to $\Delta$.
Evaluating the derivative on \refeq{Fpbar1}, we immediately see that
$b_1$ defined through \refeq{bN} becomes
\be
b_1 = c_{\d,\L} + \s^2(\L)\left\{ \frac{68}{21} \left(c_{\d^2,\L}  + \frac23 c_{K^2,\L}\right)
+ 3 c_{\d^3,\L} + \frac23 c_{\d K^2,\L}
+ \frac{32}{63} c_{\otd,\L}
\right\}\,.
\ee
This exactly coincides with the renormalized
bias parameter given in the previous section, \refeq{b1c1}.  Thus,
the peak-background split bias parameters understood in the sense of \refeq{bN}
are \emph{exactly the renormalized bias coefficients} which describe large-scale correlations of tracers \cite{matsubara:2008,PBSpaper}.
This proof can be straightforwardly generalized to $b_2,\,b_3,\cdots$,
which then necessitates higher-order perturbative solutions for $\d, K_{ij}$,
and so on.  When neglecting gravitational evolution and performing the
renormalization purely in Lagrangian space, Ref.~\cite{PBSpaper} has shown that
this result holds to all orders and for all bias parameters $b_N$.
Moreover, the same renormalized biases describe both tracer auto- and cross-correlations as well as higher $N$-point functions such as the bispectrum,
as will be discussed in detail in the following \refsec{measurements}.
This is ensured by the renormalization conditions applied in \refsec{renorm}.

It is worth emphasizing again the difference between the PBS bias parameters $b_N$ and the
bare $c_{\d^N,\L}$:  the $b_N$ quantify the response of
the cosmic mean abundance of tracers to a change in the background density
of the Universe; specifically, they do not make any reference to the scale $\L$.  The $c_{\d^N,\L}$ on the other hand quantify the average
response of the abundance of tracers within a region of size $\L^{-1}$ to a change in the
average density $\d_\L$ within that region, evaluated at $\d_\L=0$; they thus
necessarily depend on the cutoff, i.e. the filtering kernel $W_\L$ and scale $\L^{-1}$.
On the other hand, after renormalization we only need a prediction for
$\<n_g\>$ as function of the background density $\rhop$ to calculate
the \LIMD bias parameters $b_N$. Note also that the $b_N$ are closely related to the resummed bias propagators defined in \cite{matsubara:2011} [see Eqs.~(83)--(84) there], while the bare bias parameters $c_n$ correspond to the bare propagators [Eqs.~(1)--(2) in that paper].

%% file: measurements.tex
\clearpage
\section{Measuring galaxy and halo bias}
\label{sec:measurements}

\secttoc

Having described the general framework of the bias expansion (\refsec{evolution}), 
and the physical interpretation of bias parameters offered by the peak-background split argument 
(\refsec{PBS}), we now turn to the connection with clustering statistics and actual measurements of the bias parameters.  
We will discuss various observables which involve bias parameters and can be used to measure them, including
the auto-correlation functions of galaxies and cross-correlations between 
galaxies and matter, as well as moments-based and ``scatter plot'' methods.  
In each case, we define the range of scales over which the perturbative
bias expansion holds, and derive how the parameters inferred in each method
are precisely related to the bias parameters defined in \refsec{evolution}.  
Since this is an important point, let us emphasize again: on the scales where
perturbation theory applies, and barring any systematics in the measurement, 
\emph{all of the various methods
to estimate bias recover the unique, large-scale bias parameters $b_O$
defined in \refsec{evolution}.}  This holds, of course, up to linear combinations
of bias parameters at a given order, which are a matter of choice of basis;  
we give relations between various popular choices in \refapp{biastrans}.  
In this sense, the bias parameters are \emph{physical quantities rather than observable-dependent fitting parameters.}  
This is not the case for phenomenological fitting relations between, for 
example, the galaxy power spectrum and the matter power spectrum for a 
fiducial cosmological model (such as, for instance, the ``$Q$ model'' of 
\cite{BAO/2dF}).  We will not discuss these fitting functions here.

The final goal of the bias expansion is, of course, to predict observed galaxy
statistics given a set of bias parameters as well as a cosmological model
with its associated predictions for the statistics of matter.  
In this section, we will simplify the treatment by ignoring complications 
that, while important for actual observed galaxy statistics, are not directly 
related to bias, the most important being \emph{redshift-space distortions}.  
These, along with other issues affecting observed galaxy statistics, will be discussed 
in \refsec{observations}.  
The idealized relations given below are thus strictly valid only for tracers extracted from numerical 
simulations---such as dark matter halos---for which we have access to the 
real-space clustering.  
Correspondingly, we will denote the tracer overdensity as $\d_h$ in this section, and 
mostly review measurements of halo bias here.  These also provide the context for the 
models of halo bias discussed in the following sections \refsecs{exset}{peaks}.   
We emphasize again that \emph{all results are applicable also to galaxy statistics} 
once the effects described in \refsec{observations} are added in.  

Furthermore, we continue to restrict ourselves to the case of Gaussian initial conditions 
in this section.  Results for galaxy clustering with primordial non-Gaussianity are 
discussed in \refsec{NG}. We will focus on bias parameters up to third order, 
as derived in \refsec{evol2} and summarized in \refsec{evol:summary}. This is essentially the current state-of-the-art in the published literature on galaxy and halo bias. The lowest-order biases are phenomenologically the 
most important ones. While linear- and second-order terms are sufficient to describe the galaxy 
power spectrum and bispectrum (three-point function) at leading order, respectively, cubic 
order terms contribute to the next-to-leading (1-loop) correction to the galaxy power spectrum.

At this order, the galaxy as well as halo density contrast can be expressed as
\ba
\d_h =\:& b_1 \d + b_{\lapl\d} \lapl\d + \eps \vs
& + \frac12 b_2 \d^2 + b_{K^2} (K_{ij})^2 + \eps_\d \d 
\vs
& + \frac16 b_3 \d^3 + b_{\d K^2} \d (K_{ij})^2 + b_{K^3} (K_{ij})^3 
+ b_{\rm td} O_{\rm td}^{(3)} 
+ \eps_{\d^2} \d^2 + \eps_{K^2} (K_{ij})^2 \vs
& + \O(\d^4) + \O[R_*^2 \lapl(\d^2),\, R_*^4 \laplsq \d]\;,
\label{eq:deltag2}
\ea
where $\eps$ and $\eps_O$ are stochastic fields that are uncorrelated with $\d,\,K_{ij}$ 
(see \refsec{stoch}), and we have listed, in the last line, examples of higher-order terms 
that are neglected in this expression.  Although not made explicit here, the operators appearing in \refeq{deltag2} 
are strictly the \emph{renormalized} operators, and the $b_O$ are consequently renormalized 
bias parameters. Indeed, as we are dealing with measurements as well as physical models of 
bias in the following, all instances of the bias parameters appearing in the remainder of 
the review are renormalized (with the exception of \refsec{bNG}, where, for the purposes of 
derivation, the bare bias parameters appear).  

Note that all observables, fields, and bias parameters depend on the time $\tau$ of observation. For clarity, however, we will drop the time argument throughout this section. 
We divide the various procedures to measure bias parameters into four 
categories: 
\begin{enumerate}
\item \emph{$n$-point correlation functions:} two- and three-point correlation functions, 
or power spectrum and bispectrum in Fourier space (\refapp{stat});
\item \emph{Moments:} one-point statistics of the halo and matter density smoothed on a large scale $R_\ell$;
\item \emph{Scatter plots} of the halo density as a function of the local matter density;
\item \emph{Responses:} an implementation of the separate-universe approach 
(or exact PBS) described in \refsec{sepuni}.  
\end{enumerate}
We will successively describe each approach in 
\refsec{bnpt}--\ref{sec:bsepuni}. We then present a brief overview of
published results for bias parameters (and stochastic amplitudes) obtained using these various methods
for dark matter halos in simulations, and observed galaxies, in \refsec{meas:meas}. Finally, in \refsec{assembly} we will 
discuss the phenomenon of
 \emph{assembly bias}, which is specific to dark matter halos.  
This refers to the fact that the bias parameters of halos depend on other halo properties than just their mass.

\subsection{$n$-point correlation functions}
\label{sec:bnpt}

We begin with the measurement of bias parameters from halo $n$-point 
correlation functions. 
This approach is most closely related to the discussion 
of \refsec{evolution}. The lowest-order statistics that allow us to 
unambiguously measure the first- and second-order bias parameters are, respectively, the two- 
and three-point functions. We consider both halo auto-correlations as well as 
cross-correlations with matter.  Once projection effects are included,
auto-correlations are readily measurable for galaxies in real survey data.  
In addition, measuring stacked weak gravitational lensing 
around galaxies (\emph{galaxy-galaxy lensing}) yields 
the galaxy-matter cross-correlation function projected along the line of sight 
\cite{tyson/etal:1984}, as will be briefly discussed below.

Apart from the close connection to measurements, the advantage of $n$-point functions is that, upon restricting all scales $r_i$ to be greater
than a minimum scale $r_{\rm min}$, or, in Fourier space, 
all wavectors $k_i$ to be less than some prescribed $k_{\rm max}$, 
one has complete control over nonlinear and higher-derivative corrections.  
Moreover, the measurements can be made even more robust by marginalizing over the leading nonlinear corrections which have a known functional form, as we
will see in \refsec{npt1loop}.  

On the other hand, the main practical disadvantage of this method is that
large observed or simulated volumes are necessary to obtain a 
high signal-to-noise measurement of $n$-point statistics on scales
above $r_{\rm min}$ and/or below $k_{\rm max}$.  This is because we need a 
significant number of independent modes in order to reduce sample variance.  
In addition, measuring the bias parameters at cubic or higher order becomes 
increasingly difficult, since measurements of higher-order statistics become necessary, for instance the trispectrum in the case of cubic-order bias parameters.
The implementation and the required computational resources for higher-order
statistics become increasingly demanding.  

Since this is a substantial subsection, we provide a brief outline here.  
We begin with the leading two- and three-point functions in Eulerian space,
both in the Fourier- and real-space representations (\refsec{npttree}).
We then briefly discuss the corresponding results in Lagrangian space
(\refsec{nptLagr}), which are relevant for estimating bias parameters from 
halos identified in N-body simulations.  
\refsec{npttree_fisher} then provides a quantitative, albeit simplified and 
idealized, forecast of the ability of current and future galaxy surveys
to measure the bias parameters and amplitude of the matter power spectrum
using the results of \refsec{npttree}.  Next, we
derive the next-to-leading correction to the galaxy two-point function
(1-loop power spectrum) in \refsec{npt1loop}, illustrating how the
predictions of \refsec{npttree} can be taken to higher order and what 
scalings the higher-order terms obey.

\subsubsection{Two- and three-point functions at leading order}
\label{sec:npttree}

We begin with the leading-order (LO), or tree-level, predictions for the 
power spectrum and bispectrum of halos, that is, the two- and 
three-point correlation functions in Fourier space. The leading-order
calculation of the halo power spectrum and bispectrum requires, respectively,
linear- and second-order perturbation theory (see \refapp{SPT}). These leading-order predictions are accurate on sufficiently large 
scales, roughly at the level of 10\% for $k \lesssim 0.03\iMpch$ in Fourier space at $z=0$ (a more precise calculation is the subject of \refsec{npt1loop}); the range 
increases at higher redshifts \cite{jeong/komatsu:2006}.  We will present the 
corresponding real-space results, the correlation functions, at the end
of this section.  

The halo auto-power spectrum and halo-matter cross-power spectrum are 
given by
\ba
P_{hh}^\LO(k) \equiv \< \d_h(\vk) \d_h(\vk')\>'_\LO =\:& b_1^2 \Plin(k) + \Peps \vs
P_{hm}^\LO(k) \equiv \< \d_h(\vk) \d_m(\vk') \>'_\LO =\:& b_1 \Plin(k) \,,
\label{eq:Phh}
\ea
where, here and throughout, a prime on an expectation value denotes that
the momentum-conserving Dirac delta, $(2\pi)^3 \d_D(\vk+\vk')$ in case of 
\refeq{Phh}, is to be dropped (see \reftab{math}).  As mentioned in the introduction, we drop the time argument throughout this section for clarity. Again, we would
obtain the same relation for galaxies if we were able to measure their
proper rest-frame density at the true physical position, that is, without 
redshift-space distortions and other projection effects.  
$\Peps = \lim_{k\to 0} \< \eps(\vk) \eps(\vk') \>'$ is the scale-independent large-scale 
stochastic contribution [see \refeq{Pggstoch} in \refsec{stoch}].  
Note that this is a renormalized stochastic term which absorbs 
scale-independent terms from higher loop integrals (see \refsec{npt1loop}).  
We will discuss $\Peps$ in more detail in \refsec{meas:stoch}.  
The next-to-leading-order corrections to $P_{hh}(k)$ as well as $P_{hm}(k)$ from nonlinear evolution 
of both matter and bias, and from higher-derivative biases, will be 
described in \refsec{npt1loop}.  

Since the halo stochasticity contributes to the halo auto-power spectrum 
$P_{hh}(k)$ but not to the halo-matter cross-power spectrum $P_{hm}(k)$, 
the latter offers the simplest and cleanest measurement of the linear
bias parameter $b_1$ for halos
(see e.g. \cite{jing/suto/mo:2007,gao/white:2007,smith/etal:2007}).  
This technique can also be applied to galaxies, by measuring the 
matter distribution through weak gravitational lensing,     
specifically, the cross-correlation (``galaxy-galaxy lensing'') of 
the projected galaxy density with the tangential shear measured from source galaxies at 
higher redshifts \cite{brainerd/blandford/smail:1996,schneider:1998,vanwaerbeke:1998,sheldon/etal:2004,mandelbaum/etal:2013,DES1yr:gglensing} (see \cite{munshi/etal:2008} for a recent review).    
Briefly, for lens galaxies at a known comoving distance $\chi_L$ and source galaxies following a normalized redshift distribution $p(z)$, the stacked tangential shear around galaxies in angular multipole space corresponds to a projection of the real-space galaxy-matter power spectrum, $P_{gm} = b_1 P_{mm}$ at leading order, given in the Limber approximation \cite{limber:1954} by
\be
C_{g\gamma}(l) = 
\frac32 \Omega_{m0} H_0^2 
\left[\int\! dz\, p(z) 
\frac{\chi(z)-\chi_L}{\chi(z)}  
\right]
\frac{1+z(\chi_L)}{\chi_L}
P_{gm}\left(k = \frac{\sqrt{l(l+1)}}{\chi_L}, z(\chi_L)\right) \,.
\label{eq:Cgm}
\ee
By itself, this observable suffers from a degeneracy between $b_1$ and 
the matter power spectrum normalization.  This degeneracy can be broken by 
including the projected auto-correlation of galaxies $C_{gg}(l)$, and/or the cosmic shear power spectrum $C_{\gamma\gamma}(l)$, as recently applied in \cite{DES1yr:combined,DES1yr:cosmology}.

Alternatively, the degeneracy between bias and amplitude of fluctuations can be 
broken by measuring the three-point correlation function (in real 
space; \cite{fry:1994,frieman/gaztanaga:1994}) or bispectrum (in Fourier space; \cite{matarrese/verde/heavens:1997}).
Statistical homogeneity and isotropy dictate that
the bispectrum depends on three parameters which describe the shape and
scale of a triangle. The bispectrum thus encodes much more information than the two-point 
function, which is a function of one scale only.  
Moreover, the leading nonlinear (second-order) effects of bias 
become apparent in the bispectrum.  Finally, the bispectrum contains 
interesting cosmological information in its own right 
\cite{sefusatti/etal:2006}.  The leading-order (tree-level) expressions for 
the matter-matter-halo, matter-halo-halo, and halo-halo-halo bispectra
are respectively given by  
\ba
B_{mmh}^\LO(k_1,k_2,k_3) \equiv\:&  \< \d(\vk_1) \d(\vk_2) \d_h(\vk_3) \>'_\LO \vs
=\:& b_1 B_{mmm}^\LO(k_1,k_2,k_3)
+ \left[ b_2 
+  2 b_{K^2} \left( \left[ \hat\vk_1\cdot\hat\vk_2 \right]^2 - \frac13\right) \right]
\Plin(k_1) \Plin(k_2) \vs
B_{mhh}^\LO(k_1,k_2,k_3) \equiv\:&  \< \d(\vk_1) \d_h(\vk_2) \d_h(\vk_3) \>'_\LO \vs
=\:& b_1^2 B_{mmm}^\LO(k_1,k_2,k_3) 
+ 2 \Plin(k_1) \Pepsepsd 
\vs
& + \left\{
b_1 \left[ b_2 
+  2 b_{K^2} \left( \left[ \hat\vk_1\cdot\hat\vk_2 \right]^2 - \frac13\right) \right]
\Plin(k_1) \Plin(k_2) 
+
\left(\vk_2\leftrightarrow\vk_3\right)
\right\}
\vs
B_{hhh}^\LO(k_1,k_2,k_3) \equiv\:& \< \d_h(\vk_1) \d_h(\vk_2) \d_h(\vk_3) \>'_\LO \vs
=\:& b_1^3 B_{mmm}^\LO(k_1,k_2,k_3) + \Beps \label{eq:Bhhh}\\[3pt]
& + \left\{ b_1^2 \left[ b_2 
+  2 b_{K^2} \left( \left[ \hat\vk_1\cdot\hat\vk_2 \right]^2 - \frac13\right) \right] \Plin(k_1) \Plin(k_2)
+ 2\, b_1 \Pepsepsd \Plin(k_2) \right\}
 + \perm{2}
\,.\nonumber
\ea
In the first two lines, we choose $k_3$, and $k_2,\,k_3$, respectively,
to refer to halo overdensities.  
Note that the bispectra $B_{mmh}$ and $B_{mhh}$ are symmetric under
interchange of $k_1,\,k_2$ and $k_2,\,k_3$, respectively, while $B_{hhh}$
is symmetric under interchange of any of the $k_i$.  
The leading-order matter bispectrum entering \refeq{Bhhh} is given by
\ba
B_{mmm}^\LO(k_1,k_2,k_3) =\:& 2 F_2(\vk_1,\vk_2) \Plin(k_1) \Plin(k_2) + 
\perm{2}\,,
\label{eq:Bmmm}
\ea
where the $F_2$ kernel is defined in \refeq{F2}.  
\refeq{Bhhh} contains the leading stochastic terms discussed in \refsec{stoch} [cf. \refeq{Bgggstoch}], where
\ba
\Pepsepsd\equiv \lim_{k\to 0} \left<\eps(\vk)\eps_\delta(\vk)\right>'
\quad\mbox{and}\quad
\Beps \equiv \lim_{k_1,k_2\to 0} \< \eps(\vk_1)\eps(\vk_2)\eps(\vk_3) \>'
\label{eq:PBeps}
\ea
are the cross-power spectrum of the leading ($\eps$) and next-to-leading ($\eps_\d$) stochastic fields,
and the bispectrum of $\eps$, respectively;  both of these are scale-independent in the large-scale limit.  
The significance of $\eps_\delta$ was discussed in \refsec{evol1}, and we
will return to the stochastic contributions in \refsec{meas:stoch} below.
In analogy with $P_{hm}$, there are no stochastic contributions to
$B_{mmh}$, making this the cleanest statistic to measure $b_1,\,b_2,\,$ and
$b_{K^2}$ for halos. For galaxies, if lensing data are available, 
the projected, real-space matter-matter-galaxy bispectrum $B_{mmg}$ can be measured
by constructing the galaxy-shear-shear cross-correlation.

\begin{figure}[t!]
\centering
\includegraphics[width=0.85\textwidth]{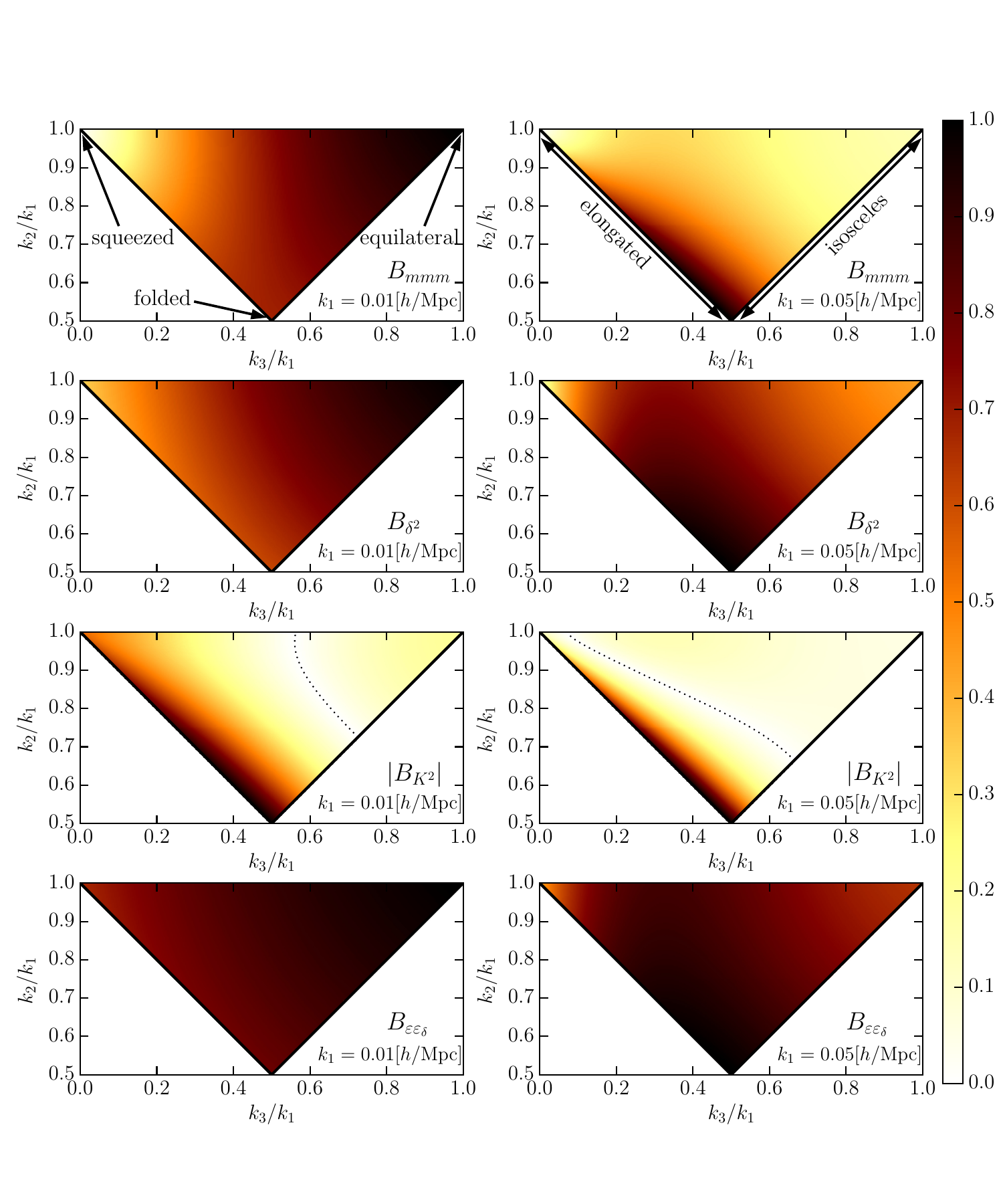}
\caption{Illustration of the dependence of the different contributions to $B_{hhh}^\LO$ 
[\refeq{Bhhh}] on triangle configuration, where $k_1 \ge k_2 \ge k_3$.  
The color scale shows the magnitude of each contribution at fixed 
$k_1 = 0.01 \iMpch$ (left column) and $k_1 = 0.05 \iMpch$ (right column).
We divide each amplitude by the maximum
value attained.
Commonly used designations for certain triangle configurations are indicated in the first row.
Shown are, from top to bottom:
$(i)$ $B_{mmm}^\LO$ [\refeq{Bmmm}],
$(ii)$ $B_{\d^2}(k_1,k_2,k_3) \equiv \Plin(k_1)\Plin(k_2) + \perm{2}$,
$(iii)$ $B_{K^2}(k_1,k_2,k_3) \equiv ([\hat{\vk}_1\cdot\hat{\vk}_2]^2-1/3)\Plin(k_1)\Plin(k_2) + \perm{2}$, and
$(iv)$ $B_{\eps\eps_\d} \equiv \Pepsepsd [\Plin(k_2) + \perm{2}]$.
The prefactor $\Pepsepsd$ is irrelevant in this representation.  
The dotted lines in the third row indicate the zero-crossing of $B_{K^2}$; 
$B_{K^2}$ is positive 
(negative) on the left- (right)-hand side of the dotted lines, respectively.  
\label{fig:Bk_shape}}
\end{figure}

The scale and configuration dependence of the halo bispectrum allows
for the degeneracy between the bias and amplitude of fluctuations to be broken.    
\reffig{Bk_shape} shows the dependence of the 
different contributions to $B_{hhh}^{\LO}$ on $(k_1,\,k_2,\,k_3)$.  The filled region in
each panel is defined by the triangle inequality ($k_1\le k_2+k_3$) with 
the condition $k_1\ge k_2\ge k_3$ that we impose without loss of generality.  
We also show names that are commonly used to refer to special triangle configurations;  
these will again play a role when discussing primordial non-Gaussianity in \refsec{NG}.  
From top to bottom we plot the contributions to \refeq{Bhhh} that are 
proportional to, respectively, 
$b_1^3$, $b_1^2 b_{2}$, $b_1^2 b_{K^2}$ and $b_1 \Pepsepsd$ 
for two different scales: the left-hand side plots 
are for fixed $k_1=0.01\iMpch$, corresponding to a scale larger than the 
matter-radiation equality turn-over of the power spectrum, while
the right-hand side plots are for fixed $k_1=0.05\iMpch$, showing 
the shape dependence on smaller scales ($k_1>k_{\rm eq}$).  
In all panels, we normalize the bispectrum to have a maximum value of 1, 
in order to highlight the shape dependence.
One can find a detailed explanation for the 
shape dependence of $B_{mmm}^\LO$ in \cite{jeong/komatsu:2009b}.
The different shape dependences are clearly visible by eye. Thus, a precise 
measurement of $B_{hhh}^\LO$ in principle yields 
clean measurements of $b_1$, $b_2$, $b_{K^2}$ and $\Pepsepsd$ that are 
independent of the power spectrum normalization.  We will quantify these 
statements in \refsec{npttree_fisher}.  

An alternative approach to measuring and fitting \refeqs{Phh}{Bhhh} was 
proposed in \cite{schmittfull/etal} and applied in \cite{modi/castorina/seljak:2016}, who considered the cross-power spectrum of 
$\d_h(\vk)$ with the Fourier transform of squared smoothed fields 
$(\d_{R_\ell}^{2})(\vk), [(K^{ij}_{R_\ell})^2](\vk)$, where $R_\ell$ is a
smoothing scale that is sufficiently large 
to suppress the higher-order nonlinear contributions.  
This method is an extension of efficient higher-order correlation function 
estimators used in the analysis of CMB temperature and polarization fields \cite{hu:2001,cooray:2001}.  
An advantage of this approach is that one can circumvent a direct measurement 
of all triangle configurations of the bispectrum,\footnote{This can be a time-consuming process when done in a brute-force way.  See \cite{scoccimarro/etal:1998,sefusatti:2005,jeong:2010} for efficient bispectrum estimators.} and that it can 
be extended to extract information in higher-order statistics
such as the trispectrum.  
The downside of this approach is that nonlinear and scale-dependent corrections
are less under control than in the full $n$-point approach, as the nonlinear
operation (squaring) performed in real space corresponds to a convolution 
in Fourier space, which leads to additional high-$k$ contributions, 
although they are suppressed by the smoothing.\\

Finally, the bias parameters can equivalently be measured from real-space
statistics.  The leading-order
halo two-point auto-correlation function and the halo-matter cross-correlation 
function are given by the Fourier transform of \refeq{Phh}, or directly from the bias expansion \refeq{deltag2}:
\be
\xi_{hh}^\LO(r) = b_1^2\xi_{\rm L}(r),
\quad\xi_{hm}^\LO(r) = b_1 \xi_{\rm L}(r),
\ee
where the linear matter two-point correlation function is given by
\be
\xi_{\rm L}(r) 
= \int_{\bfk} \Plin(k) e^{i\bfk\cdot\bfr}
= \frac1{2\pi^2}\int k^2 dk \Plin(k) j_0(kr).
\label{eq:xilin}
\ee
Here and throughout, we use the shorthand 
$\int_{\vk} \equiv \int d^3\vk/(2\pi)^3$ (see \reftab{notation}).   
Note that a constant stochastic contribution $\Peps$ in the power spectrum \refeq{Phh} contributes a term $\Peps \d_D(\vr)$ to the two-point correlation function, and we have not written it here as it only contributes at vanishing, or very small, separation (see \refsec{stoch} for a detailed discussion).  Nevertheless, this contribution is important
when evaluating the covariance of $\xi_{hh}$, which enters the likelihood
that needs to be evaluated to obtain the best-fitting $b_1$ and its measurement uncertainty.  
It also enters the moments of halos (\refsec{bmom}).  
The three-point functions are similarly obtained from the bias expansion \refeq{deltag2}, or alternatively from the Fourier transform of 
\refeq{Bhhh}, yielding \cite{gaztanaga/bernardeau:1998,barriga/gaztanaga:2002,takada/jain:2003,bel/hoffmann/gaztanaga:15}
\ba
\xi^{(3),\LO}_{mmh}(r_1,r_2,r_3) \equiv\:& \< \d(\vx) \d(\vx+\v{r}_1) \d_h(\vx+\v{r}_2) \>_\LO \vs
=\:&
b_1 \xi^{(3),\LO}_m(r_1,r_2,r_3)
+ 
b_2
\xi_{\rm L}(r_{2})\xi_{\rm L}(r_{3}) 
+
2b_{K^2}
\xi^{(0)}_2(r_{2})\xi^{(0)}_2(r_{3})\left[\mu_{23}^2-\frac13\right]
\vs
\xi^{(3),\LO}_{mhh}(r_1,r_2,r_3) \equiv\:& \< \d(\vx) \d_h(\vx+\v{r}_1) \d_h(\vx+\v{r}_2) \>_\LO \vs
=\:&
b_1^2 \xi^{(3),\LO}_m(r_1,r_2,r_3) + 
b_1 b_2
\Big\{ \xi_{\rm L}(r_{1})\xi_{\rm L}(r_{3}) 
+ \xi_{\rm L}(r_{2})\xi_{\rm L}(r_{3})
\Big\}
\vs
& + 2 b_1 b_{K^2} \left\{
\xi^{(0)}_2(r_{1})\xi^{(0)}_2(r_{3})\left[\mu_{13}^2-\frac13\right]
+ \xi^{(0)}_2(r_{2})\xi^{(0)}_2(r_{3})\left[\mu_{23}^2-\frac13\right]
\right\}
\vs
\xi^{(3),\LO}_{hhh}(r_1,r_2,r_3) \equiv\:& \< \d_h(\vx) \d_h(\vx+\v{r}_1) \d_h(\vx+\v{r}_2) \>_\LO \vs
=\:&
b_1^3 \xi^{(3),\LO}_m(r_{1},r_{2},r_{3}) + b_1^2b_2 \xi^{(3)}_{\d^2}(r_{1},r_{2},r_{3}) + 2 b_1^2b_{K^2} \xi^{(3)}_{K^2}(r_{1},r_{2},r_{3}) \,, \vs
\mbox{where} \quad  r_1 =\: & |\v{r}_1|\,,\  r_2 = |\v{r}_2|\,,\
r_3 = |\v{r}_1-\v{r}_2|\,,\  \mbox{and}\vs
\xi^{(3)}_{\d^2}&(r_{1},r_{2},r_{3}) = \xi_{\rm L}(r_{1})\xi_{\rm L}(r_{2}) 
+ \perm{2} \vs
\xi^{(3)}_{K^2}&(r_{1},r_{2},r_{3}) = 
\xi^{(0)}_2(r_{1})\xi^{(0)}_2(r_{2})\left[\mu_{12}^2-\frac13\right]
+ \perm{2} \,,
\label{eq:3pcfhhh}
\ea
where $\mu_{ij} \equiv \hat{\bfr}_{i}\cdot\hat{\bfr}_{j}$ is the cosine between the two vectors $\v{r}_i$ and $\v{r}_j$, $\xi_2^{(0)}(r)\equiv\int_{\bfk}\Plin(k)j_2(kr)$,
which is is a special case of \refeq{xielln} without smoothing ($R\to 0$), and 
\ba
\xi^{(3),\LO}_m(r_1,r_2,r_3)
=
\bigg\{ &
\frac{34}{21}\xi_{\rm L}(r_{1})\xi_{\rm L}(r_{2})
+ \frac47 \left[\mu_{12}^2 - \frac13 \right] \xi^{(0)}_2(r_{1}) \xi^{(0)}_2(r_{2})
\vs
&
-
\frac13\mu_{12}
\left[
\left(\xi_{\rm L}(r_{1})+\xi^{(0)}_2(r_{1})\right) r_1\frac{d\xi_{\rm L}(r_{2})}{dr_2}
+
(1\leftrightarrow2)
\right]
\bigg\} + \perm{2}
\label{eq:xim3}
\ea
is the three-point correlation function of the matter density field at leading order (see, e.g., App.~A of \cite{bel/hoffmann/gaztanaga:15}; we have used that $j_1(x)/x = [j_0(x)+j_2(x)]/3$).    
Note that, similar to the case of the bispectrum, statistical homogeneity and isotropy demand that the three-point 
correlation functions only depend on the three separations $r_{i}$.  
Again, we did not include the stochastic terms that are non-zero only
if at least one of the $r_{i}$ are very small (i.e., non-perturbative). 

\begin{figure}[t]
\centering
\includegraphics[width=0.89\textwidth]{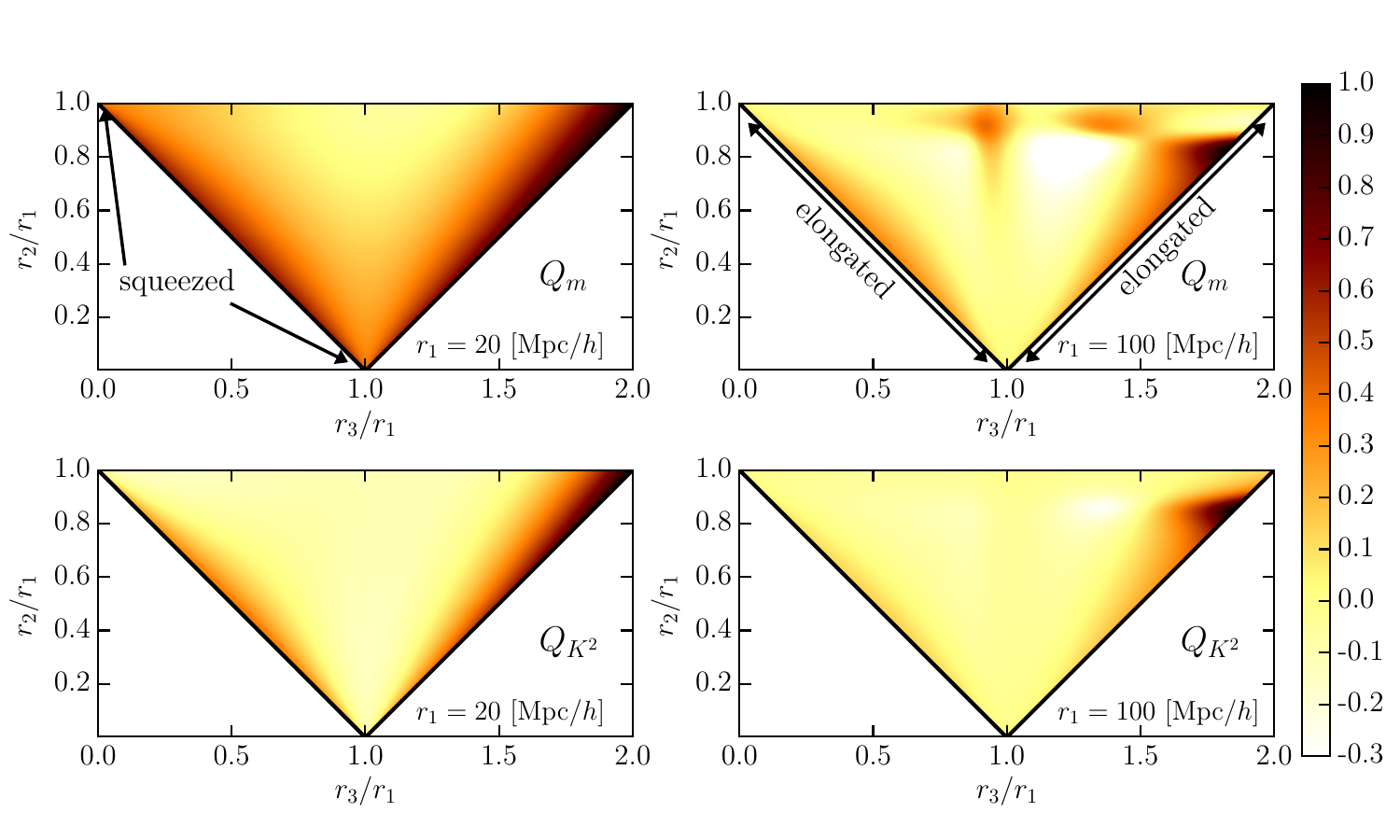}
\caption{Shape dependence of two contributions to the reduced 
halo three-point correlation function: $Q_{m}$ [\refeq{Qm}] (top panels),
and  
$Q_{K^2} = \xi^{(3)}_{K^2}(r_{1},r_{2},r_{3})/\xi^{(3)}_{\d^2}(r_{1},r_{2},r_{3})$ (bottom panels),
for two different scales: $r_{1}=20\,[{\rm Mpc}/h]$ 
(left panels) and $r_{1}=100\,[{\rm Mpc}/h]$ (right panels).  
Here, we show all possible triangle configurations satisfying 
$r_{2}\le r_{1}$
for a given $r_{1}$. In order to highlight the shape dependence, we divide each
panel by the maximum value.
A horizontal cut $r_{2}/r_{1}\equiv {\rm const.}$ in the 
$\xi_m^{(3),\LO}$ plot with $r_{1}=20\,[{\rm Mpc}/h]$ 
reveals the well-known ``U-shaped'' curves of the three-point function.
The features seen in the upper right panel (with $r_{1}=100\,[{\rm Mpc}/h]$) 
are due to the BAO feature in the linear correlation function.
\label{fig:Zeta_shape}}
\end{figure}

In order to illustrate its shape dependence, we define the reduced
halo three-point function $Q_{hhh}(r_1,r_2,r_3)\equiv \xi_{hhh}(r_1,r_2,r_3)/\xi^{(3)}_{\d^2}$, and similarly for the individual contributions defined in \refeqs{3pcfhhh}{xim3}. Further, 
\be
Q_m(r_1,r_2,r_3)\equiv \frac{\xi_m^{(3,\LO)}(r_1,r_2,r_3)}{\xi_{\rm L}(r_{1})\xi_{\rm L}(r_{2}) + \perm{2}} = \frac{\xi_m^{(3,\LO)}(r_1,r_2,r_3)}{\xi^{(3)}_{\d^2}(r_1,r_2,r_3)}\,.
\label{eq:Qm}
\ee
The nontrivial contributions are shown in \reffig{Zeta_shape}. 
Here, we impose the condition $r_{2}\le r_{1}$, and 
show all regions that satisfy the triangle conditions.  
The configuration dependence of each term is clearly distinct, similar to the 
case in Fourier space. The interpretation of \reffig{Zeta_shape} is, however, 
complicated by the fact that the covariance of the three-point function is not 
diagonal, so that different triangles are significantly correlated.  On the 
other hand, the different configurations of the bispectrum shown in 
\reffig{Bk_shape} are independent at leading order.

In this context, it is worth reiterating that the expressions for the
power spectrum and bispectrum \refeqs{Phh}{Bhhh} are only strictly correct 
on large scales.  It is not consistent to extend these to smaller scales by 
inserting the nonlinear matter power spectrum and bispectrum calibrated with 
N-body simulations, as done in some references
 (e.g., \cite{nishizawa/takada/nishimichi:2013,pollack/smith/porciani:2014}).  
This is because many higher-order and higher-derivative bias parameters are 
implicitly set to zero in this procedure.  While this might 
provide a good empirical match to halos in simulations, it is not guaranteed 
to be a good description of the statistics of galaxies.  Moreover, this
procedure will in general lead to inconsistent bias parameters when these
are measured from different statistics (e.g., power spectrum and bispectrum),
making it impossible to use the synergy between different statistics.

Finally, while we have restricted to the two- and three-point functions
here, the four-point function (trispectrum) at leading order similarly allows for 
measurements of the cubic bias parameters written in \refeq{deltag2}.  
Unfortunately, the trispectrum has a small signal-to-noise ratio on large 
scales once taking into account the covariance with the two-point function
as well as the survey geometry \cite{sefusatti/scoccimarro:2005}, and no full
measurement for either galaxies or halos has been published so far.
Hence we do not consider it further here.  

\subsubsection{Two- and three-point functions in Lagrangian space}
\label{sec:nptLagr}

In case of dark matter halos identified in N-body simulations, we can
also measure $n$-point functions in Lagrangian space, by tracing the halos
back to the initial conditions.  Then, the
contributions due to nonlinear evolution of the density field disappear
in \refeqs{Phh}{Bhhh}, 
and the expressions for the $n$-point functions simplify.  In particular,
the leading-order two- and three-point functions become
\ba
P_{hh}^{L,\LO}(k) \equiv\:& \< \d_h^L(\vk) \d_h^L(\vk') \>'_\LO = (b_1^L)^2 \Plin(k) + \PepsL \vs
P_{hm}^{L,\LO}(k) \equiv\:& \< \d_h^L(\vk) \d^{(1)}(\vk') \>'_\LO = b_1^L \Plin(k) \vs
B_{mmh}^{L,\LO} (k_1,k_2,k_3) =\:& 
\left[ b_2^L +  2 b_{K^2}^L \left( \left[ \hat\vk_1\cdot\hat\vk_2 \right]^2 - \frac13\right) \right]
\Plin(k_1) \Plin(k_2)
\label{eq:PhhL}\\
B_{mhh}^{L,\LO} (k_1,k_2,k_3) =\:& 
b_1^L \left[ b_2^L +  2 b_{K^2}^L \left( \left[ \hat\vk_1\cdot\hat\vk_2 \right]^2 - \frac13\right) \right]
\Plin(k_1) \Plin(k_2)\, 
+ 
\PepsepsdL 
\Plin(k_1) 
+ (\vk_2\leftrightarrow \vk_3)  \nonumber \\
B_{hhh}^{L,\LO} (k_1,k_2,k_3) =\:& 
 (b_1^L)^2 \left[ b_2^L +  2 b_{K^2}^L \left( \left[ \hat\vk_1\cdot\hat\vk_2 \right]^2 - \frac13\right) \right]
\Plin(k_1) \Plin(k_2) \nonumber \\
&\qquad + 
2 b_1^L \PepsepsdL \Plin(k_1) 
 + \perm{2} +  \BepsL \nonumber \;,
\ea
where Lagrangian statistics are denoted with a superscript $L$, and $\PepsL$, $\PepsepsdL$ and $\BepsL$ are the Lagrangian counterparts of the stochastic amplitudes defined in \refeq{PBeps}; again, they are constant on large scales. Moreover, for conserved tracers such as protohalos, they are the same as their Eulerian counterparts $\Peps,\,\Pepsepsd,\,\Beps$.  In that case, 
\refeq{PhhL} and \refeqs{Phh}{Bhhh} also directly map onto each other
through the relations between Eulerian and Lagrangian bias parameters 
given in \refsec{evol1}.  Thus, at each order in perturbation theory (but
excluding higher-derivative terms), a knowledge
of one set of bias parameters (either Eulerian or Lagrangian) determines
the other, and describes both late time Eulerian statistics as well as
Lagrangian statistics of proto-halos in the initial conditions;  this
includes the stochastic terms, as mentioned above.    
It is also worth noticing that the only corrections to the halo-matter 
cross-power spectrum $P_{hm}^{L,\LO}(k)$ and halo-matter-matter bispectrum
$B_{hmm}^{L,\LO}(k)$ in Lagrangian space are due to higher-derivative 
terms \cite{bardeen/etal:1986,desjacques/sheth:2010,frusciante/sheth:2012}.  
This makes $P_{hm}^L(k)$ and $B_{hmm}^L(k)$ a convenient tool for measuring the Lagrangian higher-derivative biases of proto-halos 
(see e.g. \cite{elia/ludlow/porciani:2012,baldauf/desjacques/seljak:2015}).

\subsubsection{A worked example: bias constraints from the leading-order power spectrum and bispectrum}
\label{sec:npttree_fisher}

As we have discussed in the previous section, the dependence on 
scale and triangle configuration of the halo bispectrum [\refeq{Bhhh} and \reffig{Bk_shape}]
provides ample information on the bias parameters that can be extracted from 
the observed halo sample itself.  The bispectrum thus breaks the
degeneracy between linear bias $b_1$ and the normalization of $\Plin(k)$
that is present in the leading-order halo power spectrum. The same is expected
to hold for galaxies, once projection effects are included. 
We now investigate this information gain quantitatively, though in an idealized setting, by
using the Fisher information matrix formalism \cite{tegmark:1997} and
applying it to upcoming galaxy surveys.  Since we neglect redshift-space distortions (RSD, \refsec{RSD}), 
this is clearly not a realistic, complete forecast.  We will discuss this further below.  

At leading order, i.e. at second order in perturbation theory, 
the galaxy power spectrum and bispectrum are statistically independent.
Moreover, different wavenumbers are uncorrelated; that is, the covariance matrices of power spectrum and bispectrum are diagonal. 
Then, the full Fisher information matrix is given by
\be
F_{ij} = F_{ij}^{(P)} + F_{ij}^{(B)},
\label{eq:Fisherij}
\ee
where the power spectrum and bispectrum Fisher matrices are respectively given by 
\ba
F_{ij}^{(P)} =\:& \sum_{k}
\frac{1}{\Varof{P_{gg}(k)}}
\frac{\partial P_{gg}(k)}{\partial \theta_i}
\frac{\partial P_{gg}(k)}{\partial \theta_j}
\label{eq:FisherPij} \\
F_{ij}^{(B)} =\:& \sum_{(k_1,k_2,k_3)}
\frac{1}{\Varof{B_{ggg}(k_1,k_2,k_3)}}
\frac{\partial B_{ggg}(k_1,k_2,k_3)}{\partial \theta_i}
\frac{\partial B_{ggg}(k_1,k_2,k_3)}{\partial \theta_j}\,.
\label{eq:FisherBij}
\ea
Here, the sums run over wavenumber bins specified below, and $\Varof{P_{gg}(k)}$ and $\Varof{B_{ggg}(k_1,k_2,k_3)}$ denote the 
variances of the binned power spectrum and bispectrum estimators,
respectively.  For this example, our parameter vector $\vec{\theta}$ contains all first- and second-order
bias parameters as well as stochastic amplitudes, in addition to the amplitude of
the primordial power spectrum $\mathcal{A}_s$ (\reftab{ref_cosmology}):
\be
\vec{\theta} = \left\{b_1,\, b_2,\, b_{K^2}, \, \ln \mathcal{A}_s,\, \Peps,\, 
\Pepsepsd, \, \Beps \right\}\,.
\ee 
For the biases $b_O$, we assume fiducial values as indicated in \reftab{Fisher_bias}, while for the stochastic amplitudes we assume as fiducial the values predicted by Poisson statistics, $\Peps = (\avng)^{-1}$, $\Pepsepsd = b_1/(2\avng)$, and $\Beps = (\avng)^{-2}$, as derived in \refsec{meas:stoch}.

Let us briefly discuss the effect of higher-order nonlinear 
corrections and RSD.  These can lead to a degradation of the idealized bias 
constraints, since both effects introduce additional free parameters. 
On the other hand, RSD lead to anisotropies in the clustering statistics 
that can break parameter degeneracies. Furthermore, including nonlinear 
corrections allows us to include smaller scales, and hence many additional 
modes, in the analysis. These effects in turn can lead to an improvement 
over the idealized constraints presented here.  

The leading-order variance of the binned galaxy power spectrum estimator
$\hat P_{gg}(k)$ is given by \cite{FKP:1994}
\be
\Varof{P_{gg}(k)}
=
\frac{1}{N_k} \left[ P_{gg}(k) \right]^2\,,
\label{eq:DeltaPk}
\ee
where $N_k$ is the number of independent Fourier modes in the wavenumber bin used to estimate the power spectrum $P_{gg}(k)$.  Note that in our notation, $P_{gg}(k)$ includes the stochastic (shot-noise) contribution [\refeq{Phh}]. We do not 
take into account the effect of survey geometry and assume that $N_k$ is the 
same as that of a cubic box with the survey volume $V_{\rm survey}$.  In that 
case, $N_k$ is given by
\be
N_k 
= \frac12 \frac{4\pi k^2\Delta k}{(k_F)^3}
= \frac{V_{\rm survey}}{4\pi^2} k^2 \Delta k\,,
\ee
where $k_F \equiv 2\pi/V_{\rm survey}^{1/3}$ is the fundamental wavenumber of the survey.  
Note that the constraint that $\d_g(\vx)$ is a real field reduces the number
of independent Fourier modes so that only one half of the total Fourier volume
is counted in the number $N_k$.

The leading-order variance of the estimated galaxy bispectrum 
$\hat B_{ggg}(k_1,k_2,k_3)$ is given by 
\cite{sefusatti/etal:2006,sefusatti/komatsu:2007,jeong:2010}
\be
\Varof{B_{ggg}(k_1,k_2,k_3)}
= s_B \frac{V_{\rm survey}}{N_t} 
P_{gg}(k_1)
P_{gg}(k_2)
P_{gg}(k_3) \,,
\label{eq:DeltaBk}
\ee
where $s_B$ is a symmetry factor (6 for equilateral triangles, 2 for isosceles 
triangles, 1 for other triangles), and $N_{t}$
is the number of triangle configurations in the bin considered;
for simplicity, we use the expression in the continuum limit, valid
if $N_t$ is large:
\ba
N_t \simeq\:&
\frac{1}{k_F^6}
\int_{|\vk_1-\vp_1|<\Delta k_1/2} d^3 p_1
\int_{|\vk_2-\vp_2|<\Delta k_2/2} d^3 p_2
\int_{|\vk_3-\vp_3|<\Delta k_3/2} d^3 p_3\:
\delta_D(\vp_1+\vp_2+\vp_3)
\vs
=\:& 
\left(
\prod_{i=1}^3
\frac{k_i \Delta k_i}{k_{F}^2}
\right) \times
\left\{
\begin{array}{ll}
4\pi^2\,, & {k_i=k_j+k_k} \\
8\pi^2\,, & {\rm otherwise}
\end{array}
\right.
\,,
\ea
where $i,j,k \in {\rm cyclic}(1,2,3)$. 
For this calculation, we use $\Delta k_i = k_F$ as bin width. 
We consider the following six galaxy surveys: 
HETDEX~\cite{HETDEX}, eBOSS~\cite{eBOSS}, DESI~\cite{DESI}, PFS~\cite{PFS}, 
Euclid~\cite{amendola/etal:2013} and WFIRST~\cite{WFIRST}. Their mean redshift $\bar z$ as well as volume and mean galaxy density assumed for this forecast are given in \reftab{Fisher_bias}. 

\begin{table}[t!]
\centering
\caption{Projected uncertainties on the deterministic
bias parameters, $b_1$, $b_2$, $b_{K^2}$, and the amplitude of the linear power 
spectrum, $\ln\mathcal{A}_s$, from current and upcoming galaxy surveys with listed specifications in an idealized setting.  
In all cases, we assume $b_1=1.5$, and calculate $b_2$ and $b_{K^2}$ from the relation given in \reftab{fittingbiases} ($b_2 \simeq -0.69$), and Lagrangian LIMD ($b_{K^2}\simeq -0.14$), respectively.  
The fiducial values of the stochastic amplitudes are given by Poisson sampling, $\Peps = 1/\avng,\,\Pepsepsd = b_1/(2\avng),\,\Beps = 1/\avng^{2}$.  
}\vspace*{3pt}
\label{tab:Fisher_bias}
\begin{tabular}{l||ccc||cccc|cccc}
\hline\hline
\multirow{3}{*}{survey} & 
\multirow{3}{*}{\begin{tabular}[c]{@{}c@{}}redshift\\{$\bar z$}\end{tabular}} & 
\multirow{3}{*}{\begin{tabular}[c]{@{}c@{}}$V_{\rm survey}$\\ {\small $[h^{-3} {\rm Gpc}^3]$}\end{tabular}} & 
\multirow{3}{*}{\begin{tabular}[c]{@{}c@{}}$10^4\,\avng$\\ {\small $[h^{-3}{\rm Mpc}^3]$} \end{tabular}} &
\multicolumn{4}{c|}{$k_{\rm max}=0.1\,h/{\rm Mpc}$} & 
\multicolumn{4}{c}{$k_{\rm max}=0.2\,h/{\rm Mpc}$} 
\\ \cline{5-12} 
              &      &      &         &
\multicolumn{3}{c}{$100\, \sigma(b_O)$} &
\multicolumn{1}{|c|}{\multirow{2}{*}{$\sigma_{\ln \mathcal{A}_s}$}} & 
\multicolumn{3}{c}{$100\, \sigma(b_O)$} & 
\multicolumn{1}{|c}{\multirow{2}{*}{$\sigma_{\ln \mathcal{A}_s}$}} 
\\ \cline{5-7} \cline{9-11}
              &      &      &         &
              $b_1$ & $b_2$ & $b_{K^2}$ & \multicolumn{1}{|c|}{}
&              $b_1$ & $b_2$ & $b_{K^2}$ & \multicolumn{1}{|c}{}
\\[2pt] \hline
eBOSS (LRG)   & 0.8  & 6.1  & 4.4     & 32     & 45      & 30   & 0.43  & 7.0     & 4.5      & 5.9  & 0.093    \\
eBOSS (QSO)   & 1.4  & 39   & 1.5       & 38     & 51      & 36     & 0.51      & 11     & 6.5  & 9.2 & 0.15     \\ 
HETDEX        & 2.7  & 2.7  & 3.6       & 190      & 260     & 180    & 2.6      & 59      & 35   & 49   & 0.79    \\ 
PFS           & 1.5  & 8.7  & 4.6       & 47      & 66      & 44     & 0.62     & 11     & 6.7  & 8.9  & 0.14      \\ 
DESI          & 1.1  & 40   & 3.3      & 18     & 25      & 17     & 0.25     & 4.4     & 2.7  & 3.7  & 0.059      \\ 
WFIRST        & 1.9  & 13   & 12      & 35     & 49      & 32     & 0.46     & 6.8     & 4.4  & 5.6 & 0.091      \\ 
Euclid        & 1.4  & 63   & 5.2      & 15    & 20      & 14     & 0.20     & 3.3     & 2.1  & 2.7  & 0.044      \\
\hline \hline
\end{tabular}
\end{table}

The results of the projected 1-sigma (68\% confidence level) uncertainties on
the bias parameters $b_1$, $b_2$, $b_{K^2}$, and the amplitude of the linear 
power spectrum, $\ln \mathcal{A}_s$, after marginalizing over the stochastic 
parameters $\Peps$, $\Pepsepsd$ and $\Beps$, are given in 
\reftab{Fisher_bias}.
First of all, as we have noted earlier, the bispectrum breaks the notorious 
degeneracy between the two quantities $b_1$ and $\ln \mathcal{A}_s$, and we can measure 
both parameters simultaneously by combining the power spectrum and bispectrum.  
Consequently, future galaxy surveys can constrain $b_1$, $b_2$ and $b_{K^2}$ at 
the level of tens of percent when including Fourier modes with $k<k_{\rm max}=0.1\,h$/Mpc.  
However, we stress that the constraints depend sensitively on the range of
wavenumbers included: going from $k_{\rm max}=0.1\iMpch$ to $0.2\iMpch$
improves constraints by a factor of several, and the constraints are now at 
the few-percent level.  
This highlights the importance of an accurate theoretical model
of nonlinear corrections to $P_{gg}$, $B_{ggg}$, as well as their
covariance matrices.  The precise value of $k_{\rm max}$ that leads to optimal
unbiased constraints on cosmological parameters and the $b_O$ is highly dependent on the survey considered,
in particular its redshift range and the galaxy sample being targeted, and 
we refrain from attempting to estimate survey-specific values for $k_{\rm max}$ 
here.  The constraints quoted in \reftab{Fisher_bias} should thus only be seen as illustrative figures.

\subsubsection{Next-to-leading-order corrections to the two-point functions}
\label{sec:npt1loop}

So far, we have derived the leading contributions to the two- and
three-point functions of galaxies and halos on large scales.  
In order to illustrate how higher-order corrections to the above results
can be derived, we also present the next-to-leading-order (NLO, or 1-loop) correction to the two-point function. We only discuss real-space predictions without any projection effects. Hence our expressions mostly apply to halos, but the structure of the perturbative expansion remains the same even when including projection effects. 
Deriving the nonlinear correction to the two-point functions requires a third-order calculation in perturbation theory, since, for any operator $O$, contributions of 
the type $\< O^{(1)} O^{\prime (3)}\>$ contribute at the same order as $\< O^{(2)} O^{\prime (2)}\>$ 
\cite{vishniac:1983,mcdonald/roy:2009,assassi/etal}.  Thus, the bias expansion
in \refeq{deltag2} is necessary and sufficient to derive this correction. 
Writing
\ba
P_{hm}(k) =\:& P_{hm}^\LO(k) + P_{hm}^\NLO(k) + \cdots \vs
P_{hh}(k) =\:& P_{hh}^\LO(k) + P_{hh}^\NLO(k) + \cdots\,,
\ea
and following the notation of \cite{assassi/etal}, the NLO contributions to the halo-matter and 
halo-halo power spectrum are respectively given by\footnote{We include the relevant stochastic terms which were not considered in \cite{assassi/etal}.}
\ba
P_{hm}^\NLO(k) =\:& b_1 \left[ P_{mm}^\NLO(k) - 2 C_{s,\rm eff}^2 k^2 \Plin(k)\right] + \hat{P}_{hm}^\NLO(k) \vs
\hat{P}_{hm}^\NLO(k) \equiv\:& b_{\d^2} \mathcal{I}^{[\d^{(2)},\d^2]}(k)
+ b_{K^2} \mathcal{I}^{[\d^{(2)},K^2]}(k)  
+ \left(b_{K^2} + \frac25 b_{\otd} \right) f_\NLO(k) \Plin(k)
\vs
& - b_{\lapl\d} k^2 \Plin(k) + k^2 P_{\eps \eps_m}^{\{2\}}
\label{eq:Phm1l}\\
P_{hh}^\NLO(k) =\:& (b_1)^2 \left[ P_{mm}^\NLO(k) - 2 C_{s,\rm eff}^2 k^2 \Plin(k)\right] + 2 b_1 \hat{P}_{hm}^\NLO(k) 
+ \sum_{O,O' \in \{ \d^2,K^2 \}} b_O b_{O'} \mathcal{I}^{[O,O']}(k)
+ k^2 \Plapleps\,,
\nonumber
\ea
where $P_{mm}^\NLO(k)$ is the NLO correction to the matter power spectrum [\refeq{Pm1loop}], 
$C_{s,\rm eff}^2 \equiv (2\pi) c_{s,\rm eff}^2/\knl^2$ is the scaled
effective sound speed of the \emph{matter} fluid \cite{baumann/etal:2012,carrasco/etal:2012} [cf.~\refeq{Pmmcs}] and $\knl$ 
is defined in \refeq{knldef}, while $\eps_m$ is the effective stochastic contribution to the matter density (\refapp{EFT}). Further,
\ba
f_\NLO(k) =\:& 4 \int_{\vp} \left[\frac{[\vp\cdot(\vk-\vp)]^2}{p^2 |\vk-\vp|^2}-1\right] F_2(\vk,-\vp) \Plin(p)
\vs
\mathcal{I}^{[O, O']}(k) =\:& 2 \Bigg[ \int_{\vp} S_O(\vk-\vp,\vp) S_{O'}(\vk-\vp,\vp) 
\Plin(p) \Plin(|\vk-\vp|) \vs
& \quad - \int_{\vp} S_O(-\vp,\vp) S_{O'}(-\vp,\vp) \Plin(p) \Plin(p) \Bigg]\,,
\label{eq:IOO} \\
\mbox{where}\quad
S_O(\vk_1,\vk_2) =\:& \left\{
\begin{array}{ll}
F_2(\vk_1,\vk_2), & O = \d^{(2)} \\
1, & O = \d^2 \\
(\hat{\vk}_1\cdot\hat{\vk}_2)^2-1/3, & O = K^2
\end{array}\right.
\,.\label{eq:1ldefs}
\ea
We see that, in addition to\footnote{We use the bias parameter $b_{\d^2}$ here to make the notation in \refeqs{Phm1l}{1ldefs} more compact.} $b_{\d^2} = b_2/2$ and $b_{K^2}$, which also enter the leading-order halo three-point function,
one new local bias term appears in the NLO halo 
power spectra, namely $O^{(3)}_{\otd}$ defined in \refeq{Otddef}, with associated bias
parameter $b_{\otd}$.  Thus, only one of four cubic-order local bias parameters contributes to
the next-to-leading-order halo power spectra.  
We have also included the leading higher-derivative bias $b_{\lapl\d}\propto R_*^2$ in \refeq{Phm1l}, where $R_*$ is the nonlocality scale of the halos or galaxies considered.

Correspondingly, we have also included the scale-dependent stochastic contributions to $P_{hm}^\NLO$ and $P_{hh}^\NLO(k)$. The latter is expanded following \refeq{epsOgeneral}, and is expected to scale as $|\Plapleps| \sim R_*^2 \Peps$
\cite{smith/etal:2007,hamaus/seljak/etal:2010,baldauf/seljak/etal:2013,chan/etal:2014}.  We will return to this in \refsec{meas:stoch}.  
It is often assumed that there is no stochastic contribution to the halo-matter cross-power spectrum. However, this is only true at lowest order. The nonlinear small-scale modes of the density field are responsible for both the halo stochasticity $\eps$ and the stochastic contribution to the matter density field $\eps_m$, which, as discussed in \refapp{EFT}, is due to the effective pressure of the nonlinear matter fluctuations and scales as $k^2$ in the low-$k$ limit.  Hence, one has to allow for a correlation between the two stochastic fields, leading to the term $k^2 P_{\eps \eps_m}^{\{2\}}$ in $P_{hm}^\NLO$, which is comparable to the other NLO contributions. Note that it could be either positive or negative.

\begin{figure}[!t]
\centering
\includegraphics[width=\textwidth]{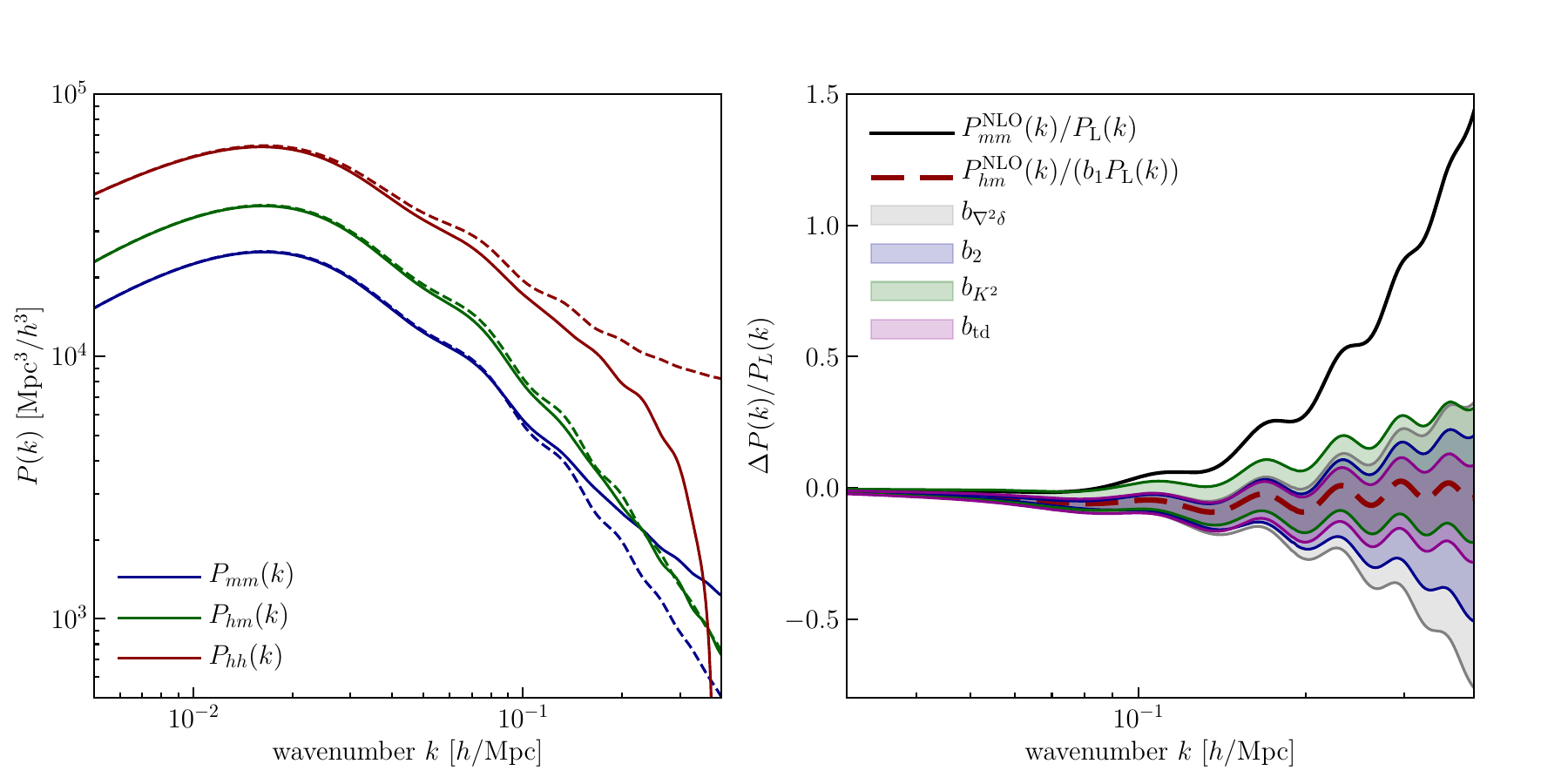}
\caption{
\textit{Left panel:} illustration of halo auto- (red, top line)
and cross-power spectra (green, middle line), and the matter power spectrum (blue, bottom line) at $z=0$.  
The solid lines show the total LO plus NLO result, while the dashed curves show the LO (linear) prediction only.  
The bias parameters used here are
$b_1=1.50$, $b_2 = -0.69$, and $b_{K^2} = -0.14$, as in \reftab{Fisher_bias}, while $b_{\lapl\d} = R_*^2$ with $R_* = 2.61 \Mpch$. 
$b_{\otd}= 23/42 (b_1-1)$ is taken from the
Lagrangian \LIMD prediction (\refsec{evol2}).
The stochastic amplitudes are taken from the Poisson expectation,
$\Peps = 1/\avnh$ and $\Plapleps = - R_*^2/\avnh$, with $\avnh = 1.41\cdot 10^{-4} (\nsMpch)^{-3}$.
We have set $P_{\eps \eps_m}^{\{2\}}=0$ in $P_{hm}^\NLO(k)$.
\textit{Right panel:} fractional size of the NLO contributions to the matter and halo-matter cross-power
spectrum at $z=0$.  The red dashed line shows the result
for $P_{hm}(k)$ for the fiducial bias parameters given above.  
The different shaded areas around $P_{hm}^\NLO$
show the effect of rescaling the various bias parameters by a factor
in the range $[0.5, 2]$.  Clearly, the contributions from different
bias parameters exhibit similar dependencies on $k$, and are in general difficult to disentangle using only the power spectrum.
The perturbative description is expected to fail for $k \gtrsim 0.25 \iMpch$, where $P_{mm}^\NLO(k)$ becomes as large as the LO prediction $\Plin(k)$. 
\label{fig:Pkg_1loop}}
\end{figure}

The magnitude and scale dependence of the NLO corrections to the halo and matter power spectra is shown in \reffig{Pkg_1loop}.  As expected, we see that the corrections become increasingly important towards smaller scales (higher $k$). We see a particularly steep suppression of $P_{hh}(k)$, which, for our fiducial parameters, is dominated by the higher-derivative stochastic contribution $k^2 \Plapleps$. 
The right panel of \reffig{Pkg_1loop} shows the fractional size of the NLO correction to $P_{mm}(k)$ and $P_{hm}(k)$. Depending on the value of the various bias and stochastic parameters, the NLO correction could be either positive or negative (shaded regions), and cancellations between the different NLO contributions can occur. In any case, as soon as the fractional size of the NLO correction approaches order unity, we expect that higher-order loop contributions which we have not included become comparable to $P_{hm}^\NLO(k)$ as well, and hence the perturbative expansion ceases to converge.

The NLO halo-matter power spectrum adds five additional free parameters to the ones present at leading order ($b_1$, $\Peps$).  These can, in principle, be disentangled due to the different scale dependence of each term.  
However, as illustrated in \reffig{Pkg_1loop}, these scale dependences are sufficiently similar that it is difficult
to disentangle the various higher-order bias parameters in practice;  
note that there is only a limited range in wavenumbers that can be used for the parameter estimation, due to the presence of higher loop and derivative corrections (see below).  Nevertheless,
the leading-order bispectrum can be used to determine $b_2$ and $b_{K^2}$, leaving only $b_{\otd}$, $b_{\lapl\d}$, and $P_{\eps \eps_m}^{\{2\}}$ to be constrained from the NLO correction to the halo-matter cross-power spectrum.

In order to gain a more detailed understanding of
the magnitude of the corrections in \refeq{Phm1l}, let us approximate
the matter power spectrum by a power law,
\be
\Plin(k) \approx \frac{2\pi^2}{\knl^3} \left(\frac{k}{\knl}\right)^n\,,
\label{eq:knldef}
\ee
where $\knl$ is the \emph{nonlinear scale} at which the dimensionless matter
power spectrum $\Delta^2(k)=k^3 \Plin(k)/(2\pi^2)$ becomes unity.  This yields, for example,
\ba
\frac{\mathcal{I}^{[\d^2,\d^2]}}{\Plin(k)}
=\:& 2 \left(\frac{k}{\knl}\right)^{3+n}
\int_{-1}^1 \frac{d\mu}2 \int_0^\infty x^2 dx
\left[\left(x \sqrt{1 + x^2 - 2 x\mu}\right)^n - x^{2n}\right]\,.
\ea
While other NLO loop-integral terms have different angular integrands, the 
scaling $\propto (k/\knl)^{3+n}$ is common to all (see also \reffig{Pkg_1loop}).  Note that, depending
on the value of $n$, the integral over $x$ might need to be regularized
in the UV (ultraviolet, small-scale, or large-$x$, limit of the integral),
while the integral is safe from divergence in the IR
(infrared, large-scale, or small-$x$, limit of the integral), because of the 
term subtracted in \refeq{IOO}; 
in any case, this does not affect the scaling with $k/\knl$.  
This scaling allows us to estimate the importance of higher-order terms.  
For example, 2-loop corrections correspondingly scale as
$(k/\knl)^{2(3+n)}$ for a scale-free power spectrum \cite{pajer/zaldarriaga:2013}.  For our reference $\Lambda$CDM cosmology, we have 
approximately\footnote{This was obtained by fitting a power law to $\Plin(k)$ over the range $k \in [0.1,0.25] \iMpch$.} $\knl(z=0) = 0.25 \iMpch$ and $n = d\ln\Plin/d\ln k|_{\knl} = -1.7$, so that the one-loop terms scale approximately as $(k/\knl)^{1.3}$.  Of course, this is only a rough approximation as $\Plin(k)$
cannot be approximated as a power law over the entire relevant range of
scales.  In particular, since $n$ becomes positive for $k \lesssim 0.02\iMpch$,
the NLO terms eventually scale as $k^2$ for sufficiently small values of $k$.  
Nevertheless, such estimates are important as they allow us to marginalize
over higher-loop corrections and rigorously take into account the uncertainty
in the prediction of \refeq{Phm1l} \cite{baldauf/etal:2016}.  

The higher-derivative term $\propto b_{\lapl\d}$ obeys a scaling with $k$ ($\propto k^2$) that is in general different from that of the 
NLO corrections ($\propto k^{3+n}$). Further, the former involves an additional scale, $R_*$. Thus, we have
\emph{two independent expansion parameters},
\be
\epsilon_\text{loop} \equiv \left(\frac{k}{\knl}\right)^{3+n}
\approx \left(\frac{k}{0.25\iMpch}\right)^{1.3} \,,
\quad\mbox{and}\quad
\epsilon_\text{deriv.} \equiv  k^2 R_*^2\,.
\ee
Thus, depending on the halo or galaxy sample, the leading higher-derivative term could be negligible
compared to the NLO corrections on the scales of interest, e.g. 
$0.01 \lesssim k\: [\iMpch] \lesssim 0.2$, or could be significantly larger.  
If $\epsilon_\text{deriv.}$ is comparable to $\epsilon_\text{loop}$
on the scales considered, then both NLO and leading higher-derivative
corrections should be included.  This is what we have assumed in 
\refeq{Phm1l}.  More generally, when going to higher orders, 
one would then include terms that involve the same powers of 
$\epsilon_\text{loop}$ and $\epsilon_\text{deriv.}$.  For example, at 2-loop
order, these are the terms of order $\epsilon_\text{loop}^2$, 
$\epsilon_\text{loop}\epsilon_\text{deriv.}$, and $\epsilon_\text{deriv.}^2$.  
On the other hand, if the two expansion parameters are substantially different,
then it is necessary to retain terms that are higher order in the larger
parameter.  
For example, if $\epsilon_\text{deriv.} \gg \epsilon_\text{loop}$, one 
should allow for additional higher-derivative terms, which
leads to contributions $\propto \{k^4 R_*^4, \, k^6 R_*^6, \cdots\}\,\Plin(k)$ 
in \refeq{Phm1l} \cite{mcdonald/roy:2009,fujita/etal:2016,nadler/perko/senatore:2017}.  The cutoff of the perturbative approach then is at $k\approx 1/R_*$. 
All of this applies analogously to the bispectrum and higher $n$-point functions.  

Finally, the higher-derivative stochastic contributions, which scale as $k^2$ (as opposed to $k^2\Plin(k)$ as the higher-derivative bias contribution), are higher order in terms of their $k$ scaling, but the amplitude could be large for very rare high-mass halos.  
Note that, in the definition of $\mathcal{I}^{[O,O']}$ [\refeq{IOO}], we have subtracted any possible 
constant term in the $k\to 0$ limit, since such a constant term is absorbed by the renormalized stochastic term $\Peps$;  this corresponds to the term $P_0\equiv b_2^2\int_{\vq}|\Plin(q)|^2$ in \cite{mcdonald:2006}, and evaluates to a finite but significant value for a $\Lambda$CDM power spectrum.  
In principle, one should also subtract the term $k^2 \partial^2\mathcal{I}^{[O,O']}/\partial k^2 |_{k=0}$ (and corresponding higher-order terms), since it is absorbed by the higher-derivative stochastic term. However, as long as the amplitude of $\Plapleps$ is allowed to be free, it can absorb the contribution from the local bias loop integrals (note that new terms with the same scaling appear at each higher loop order).

The prediction for $P_{hm}^\NLO,\,P_{hh}^\NLO$ in \refeq{Phm1l} is 
unambiguous; however, the various contributions can be broken down in a variety of
different ways \cite{mcdonald/roy:2009,saito/etal:14,angulo/etal:2015}, of which
\refeq{Phm1l} is only one option.  Also, when using \refeq{Phm1l} in conjunction
with \refeq{Phh} to fit the data, one should, strictly speaking, include the leading
connected 4-point function (trispectrum) in the covariance of $P_{hh}$ 
and $P_{hm}$, as this is also a third-order contribution in perturbation
theory.  In practice, this leads to fairly small corrections to the 
uncertainties of inferred 
bias parameters \cite{mohammed/seljak/vlah:2016}.

Finally, the standard NLO result given in \refeq{Phm1l} does not provide a very good description of the power spectrum and correlation function around the scale of the BAO feature ($0.05 \iMpch \lesssim k \lesssim 0.2\iMpch$). This is because large-scale modes introduce displacements which lead to a significant smoothing of the BAO feature. This deficiency can be improved by resumming these displacement terms \cite{crocce/scoccimarro:2008,senatore/zaldarriaga:2015,baldauf/etal:2015BAO,blas/etal:2016}. We briefly describe this resummation, which applies in the same way to biased tracers as to matter, in \refapp{IRresum}.

Previous derivations have
obtained a subset of the terms in \refeq{Phm1l};  for example,
\cite{matsubara:2008} derived the analogous result in LPT
for \LIMD Lagrangian bias, where $b_{K^2}$ and $b_{\otd}$ are set to zero at the initial time 
(see \refsec{evol2}).  
Ref.~\cite{mcdonald:2006,jeong/komatsu:2009a} assumed 
\LIMD Eulerian bias, 
setting $b_{K^2}$ and $b_{\otd}$ to zero at the final time.  
Ref.~\cite{nishizawa/takada/nishimichi:2013} used renormalized bias parameters, but also set 
$b_{K^2}=0=b_{\otd}$, and replaced 
$b_1(\Plin+P_{mm}^\NLO)$ with $b_1$ times the 
matter power spectrum given by the closure theory prescription \cite{taruya/etal:2009}, finding 
fairly good agreement with simulation measurements of $P_{hm}(k)$ up to $k = 0.3\iMpch$ at $z=0$.  Note however that this does not prove that $b_{K^2}$ and $b_{\otd}$ vanish; as emphasized above, using the correct values for all bias parameters is essential in order to obtain a consistent prediction for all halo statistics (including auto- and cross-, two- and three-point functions).  
Ref.~\cite{matsubara:2014} derived the one-loop power spectrum in 
LPT with general Lagrangian bias functions $c^L_N(\vk_1,\cdots,\vk_N)$.  
This approach is somewhat different from the renormalized, perturbative bias expansion discussed in 
\refsec{evolution} which leads to \refeq{Phm1l}, in that one needs a ``microscopic'' model for the free functions 
$c_N^L(\vk_1,\cdots,\vk_N)$.  
The example of inserting a bias parameter multiplied by $N$ powers of a filtering kernel $W_R(k)$ 
corresponds to introducing a finite physical cutoff scale $\Lambda \sim R^{-1}$ in the loop integrals.  
This is again in contrast to \refeq{Phm1l}, which is explicitly independent of any cutoff (see \refsec{pkevol} for a more detailed comparison).  
More generally, it is worth noting that predictions from Eulerian SPT and Lagrangian LPT, while 
they agree by definition on large scales, begin to diverge when extrapolated beyond their regime of validity.

\subsection{Moments}
\label{sec:bmom}

Historically, the first studies of bias considered the 
moments of the one-point distribution, i.e. number counts, of biased tracers
within volumes of a fixed size $R_\ell$, also referred to as counts-in-cells.  
Specifically, the moments are defined as $\< [\d_{g,\ell}(\vx)]^n [\d_\ell(\vx)]^m \>$, where $n,m = 0, 1, \cdots$ and $\d_{g,\ell},\,\d_\ell$ denote the galaxy and matter density fields, respectively, both smoothed on the scale $R_\ell$.  
These moments have been estimated on low-redshift galaxy catalogs as early
as the 1970s \cite{peebles:1975,fall/geller/etal:1976,fry/peebles:1978,white:1979,fry:1986b,
balian/schaeffer:1989,gaztanaga/frieman:1994,szapudi/colombi:1996}.  
The moments can be conveniently rescaled to define the \emph{hierarchical amplitudes} in terms of the skewness, kurtosis, and higher reduced 
moments of the filtered galaxy density field $\d_{g,\ell}$:
\be
\label{eq:hierarcham}
\frac{\<\d_{g,\ell}^3\>_c}{\<\d_{g,\ell}^2\>^2}\;,\qquad
\frac{\<\d_{g,\ell}^4\>_c}{\<\d_{g,\ell}^2\>^3} \;, 
\ee
and so on, where a subscript $c$ denotes the connected part of the moment. 
The skewness captures the asymmetry in the volume fraction of underdense and overdense regions of the galaxy density field.   
The original motivation for the definitions in \refeq{hierarcham} was the fact that, in ``hierarchical models'' of clustering, the amplitudes $\< \d_{g,\ell}^N \>_c/\< \d_{g,\ell}^2\>^{N-1}$ are 
independent of time and scale $R_\ell$ \citep{groth/peebles:1977,fry:1984,juszkiewicz/bouchet/colombi:1993}, 
as confirmed by gravity-only simulations with power-law initial conditions 
\cite{bouchet/schaeffer/davis:1991,fry/melott/shandarin:1993,bernardeau:1994b,
baugh/gaztanaga/efstathiou:1995,colombi/bouchet/hernquist:1996}.  
For realistic CDM power spectra, however, there are deviations from the hierarchical clustering prediction, which result 
in significant $R_\ell$-dependent corrections to $\< \d_{g,\ell}^N \>_c/\< \d_{g,\ell}^2\>^{N-1}$
\cite{bouchet/etal:1992,bernardeau:1994b,buchalter/kamionkowski:1999}.  
A generalization of the moments in \refeq{hierarcham}, the two-point moments
\cite{bernardeau:1996,gaztanaga/etal:2002} employs the joint expectation value
of powers of $\d_{g,\ell},\,\d_\ell$ at two different locations.  
Ref.~\cite{uhlemann/etal:2016,codis/etal:2016} considered rare excursions of the smoothed
density field, which are related to moments and amenable to analytical statistical techniques.  

For a \LIMD bias expansion and assuming hierarchical clustering, the second and third moments can be combined to yield estimators 
of $b_1$ and $b_2$ \cite{szapudi:1998}. 
Note that, since both numerator and denominator have some non-vanishing noise, care must be taken in constructing unbiased 
estimators of the hierarchical amplitudes \cite{hui/gaztanaga:1999}.  
Early applications of this method to simulations were presented in 
\cite{kauffmann/nusser/steinmetz:1997,mo/jing/white:1997,blanton/etal:1999,cen/ostriker:2000}.  
More recent analyses can be found in \cite{bel/marinoni:2012}.  
Ref.~\cite{pujol/etal:2016} present a method to use the moments of weak lensing convergence and projected galaxy 
density to estimate the linear bias, which is the analog in moments to the galaxy-galaxy lensing angular power spectrum 
\refeq{Cgm}.  

The advantage of moments-based methods for measuring bias is that they are simple to implement on galaxy catalogs or simulation outputs. 
Namely, we merely have to throw spheres randomly in the simulated volume, or divide the simulation volume into a grid, and compute 
fractional densities.  
The disadvantages are that large smoothing scales need to be chosen to ensure that nonlinear corrections are under 
control ($\gtrsim 30 \Mpch$ \cite{hoffmann/etal:2015,bel/hoffmann/gaztanaga:15}, see also below), and that more complicated moments than those in \refeq{hierarcham} need to be included in the analysis to 
disentangle the different bias parameters at a given order, for example $b_2$ and $b_{K^2}$.
The tidal bias $b_{K^2}$ and stochastic contributions such as $\eps_\d$, which  appear in the halo bispectrum \refeq{Bhhh} and, thus, contribute to the
skewness, were neglected in the references listed above. 
Finally, application to observational data sets is further complicated by the usually non-trivial survey geometry and mask.

In the following, we again neglect redshift-space distortions and other projection effects to discuss moments of the smoothed 
halo density field $\d_{h,\ell}$ as extracted from N-body simulations.  
We will derive the leading-order prediction for the moments in the general perturbative bias expansion, and show how the bias parameters estimated using this method are related to those obtained from halo $n$-point functions on large scales.  For this, we introduce a slightly different bias expansion to \refeq{deltag2} by writing
\ba
\d_{h,\ell}(\vx) =\:& b_1^{\rm m}(R_\ell) \d_\ell(\vx) + \frac12 b_2^{\rm m}(R_\ell) \left([\d_\ell(\vx)]^2 - \s^2(R_\ell) \right) + b_{K^2}^{\rm m}(R_\ell) \left( K_\ell^2(\vx) - \< K_\ell^2\> \right) \vs
& + \eps^{\rm m}_\ell(\vx) + \eps_{\d,\ell}^{\rm m}(\vx) \d_\ell(\vx) + \cdots\,,
\label{eq:deltahmom}
\ea
where $O_\ell$ denotes an operator $O$ smoothed on the scale $R_\ell$, and
$K_\ell^2 \equiv K_{ij,\ell} K_\ell^{ij}$. 
Note the difference to the expression obtained if one were to smooth
\refeq{deltag2}: there, one would obtain $[\d^2]_\ell$ while here we have
$[\delta_\ell]^2$.  Correspondingly, we have introduced a different set of bias parameters here, the \emph{moments biases} $b_O^{\rm m}(R_\ell)$ which depend on the scale $R_\ell$.  The relation between the $b_O^{\rm m}$, the moments, and the $b_O$ of \refeq{deltag2} will become clear momentarily.  

We begin with the variance: 
\ba
\s_{hm}^2(R_\ell) \equiv\:& \< \d_\ell(\vx) \d_{h,\ell}(\vx) \> &&\stackrel{\LO}{=} b_1^{\rm m}(R_\ell) \s^2(R_\ell)
&&\stackrel{\LO}{=} b_1 \s^2(R_\ell) \vs
\s_{hh}^2(R_\ell) \equiv\:& \< \d_{h,\ell}(\vx) \d_{h,\ell}(\vx) \> &&\stackrel{\LO}{=} [b_1^{\rm m}(R_\ell)]^2 \s^2(R_\ell) + \< (\eps_\ell^{\rm m})^2\>
&&\stackrel{\LO}{=} b_1^2 \s^2(R_\ell) + \< \eps_\ell^2\>\,,
\label{eq:shh}
\ea
where, for each line, the first relation in the $\LO$ limit is that obtained using the moments-bias expansion
\refeq{deltahmom}, while the second is what is obtained in the large-scale bias expansion \refeq{deltag2}. Further,
\be
\< (\eps_\ell^{\rm m})^2\> = \< \eps_\ell^2\> = \Peps V_\ell^{-1}\,,
\ee
where 
$V_\ell^{-1} = [\int d^3\vx\, W_\ell^2(x)]$ is the inverse of the volume of the filtering kernel $W_\ell \equiv W_{R_\ell}$.  
As emphasized in \refeq{shh}, these results hold only for sufficiently large
smoothing scales so that nonlinear and higher-derivative corrections become unimportant; specifically, these scale as $\s^4(R_\ell)$ and $R_*^2 \s_1^2(R_\ell)$, respectively, where $\s_n(R_\ell)$ is defined in \reftab{notation}.    

\refeq{shh} shows that we have $b_1^{\rm m}(R_\ell) = b_1$ at this order, and that the variance of the stochasticity agrees as well. Note that
these relations also receives corrections of order $R_*^2 \s_1^2(R_\ell)$, and
$R_*^2 \int_{\vk} k^2 W_\ell^2(k)$, respectively. These corrections not only depend on the filter scale $R_\ell$, but the shape of the filter as well, in particular its behavior at high $k$.

We now consider third-order moments.  Specifically, we focus on halo-matter-matter moments as these do not 
receive stochastic contributions at leading order and are thus convenient to single out the 
deterministic bias parameters; more generally, this holds for all moments involving only one power of the halo field.   We will see that the 
differences between the (renormalized) bias 
parameters appearing in the $n$-point functions of \refsec{bnpt} and those 
defined through \refeq{deltahmom} now become apparent.  
Consider the halo-matter-matter moment, which via \refeq{deltahmom} becomes
at leading order
\ba
\<\d_\ell^2(\vx) \d_{h,\ell}(\vx)\> \stackrel{\LO}{=}\:& 
b_1^{\rm m}(R_\ell) \< \d_\ell^3 \> 
+ \frac12 b_2^{\rm m}(R_\ell) 
\Big\< \d_\ell^2\,\left[\d_\ell^2 -\s^2(R_\ell)\right] \Big \>
+ b_{K^2}^{\rm m}(R_\ell) \Big\< \d_\ell^2 
\,\left[K_\ell^2 - \< K_\ell^2\>\right] \Big\>  
\vs
=\:& b_1^{\rm m}(R_\ell) \< \d_\ell^3 \> + b_2^{\rm m}(R_\ell) \s^4(R_\ell)
\,. \label{eq:Shmm2}
\ea
Note that in our choice of bias expansion \refeq{deltahmom}, $b_{K^2}^{\rm m}(R_\ell)$ does not
contribute to $\<\d_\ell^2 \d_{h,\ell}\>$.  This changes for example if one replaces the operator $K^2$ with 
$\mathcal{G}_2 \equiv K^2 - (2/3) \d^2$ \cite{bel/hoffmann/gaztanaga:15}. 
In order to measure both $b_2^{\rm m}$ and $b_{K^2}^{\rm m}$, we also need to
measure an independent third moment, in particular one that includes the smoothed tidal field \cite{chan/scoccimarro/sheth:2012,modi/castorina/seljak:2016},
\ba
\< K_\ell^2(\vx) \d_{h,\ell}(\vx) \>  \stackrel{\LO}{=}\:& b_1^{\rm m}(R_\ell) \< K_\ell^2 \d_\ell \> 
+ \frac12 b_2^{\rm m}(R_\ell) \Big\< \left[\d_\ell^2 - \s^2(R_\ell)\right] \,K_\ell^2 \Big \>
+ b_{K^2}^{\rm m}(R_\ell) \Big\< \left[ K_\ell^2 - \< K_\ell^2 \>\right] \, K_\ell^2\Big\>  \vs
=\:& b_1^{\rm m}(R_\ell) \< K_\ell^2 \d_\ell \>  
+ \frac8{45} b_{K^2}^{\rm m}(R_\ell) \s^4(R_\ell) \,,
\label{eq:Shmm2b}
\ea
which now yields an estimate of $b_{K^2}^{\rm m}(R_\ell)$.  Here, we have used that $\<K_\ell^2\> = (2/3) \s^2(R_\ell)$.  
Similarly to the $n$-point functions, where higher-order terms in perturbation theory lead to corrections of order
$(k/\knl)^{3+n}$, higher-order corrections to the moments enter as
higher powers of $\s^2(R_\ell)$, in particular $\s^6(R_\ell)$ in the case of \refeqs{Shmm2}{Shmm2b}.  
We will now show that there is a well-defined procedure for mapping the $b_2^{\rm m}(R_\ell),\  b_{K^2}^{\rm m}(R_\ell)$ to parameters $b_2,\  b_{K^2}$ measured from $n$-point
functions in the large-scale limit; however, they are \emph{not} equal, 
a fact which has been ignored in nearly all of the literature on moments of biased tracers 
(see e.g. \cite{manera/etal:2010,bel/marinoni:2012,bel/hoffmann/gaztanaga:15}, with the notable exception of \cite{modi/castorina/seljak:2016}). 
The precise relation depends on the matter power spectrum shape as well as the scale $R_\ell$ and type of smoothing filter used.

Let us first briefly make a technical note.  
The moments of biased tracers on some physical scale $R_\ell$ are an observable
just like the $n$-point functions discussed in \refsec{bnpt}.  Thus,
the moments biases appearing in \refeqs{Shmm2}{Shmm2b} are also physical
renormalized bias parameters, but defined using different
\emph{renormalization conditions} than those employed for the
biases $b_O$ defined through $n$-point correlation functions, as described in \refsec{renormalization}.  
That is, \refeq{deltahmom} should strictly
be written in terms of renormalized operators $[\d^2]^{\rm m}(R_\ell),\,[K^2]^{\rm m}(R_\ell)$, and so on, which are defined through the renormalization conditions
given in \refeq{renormcondM}, rather than \refeq{renormcond} which defines the
operators $[\d^2], [K^2]$ that are employed in \refsec{bnpt}.  
This means that the two sets of bias parameters can be mapped onto each
other, via a transformation that is calculable at a given order in PT. 

To obtain the relation between the $b^{\rm m}_O(R_\ell)$ and the $b_O$,
we evaluate \refeqs{Shmm2}{Shmm2b} as integrals over the leading-order
halo-matter-matter bispectrum $B_{mmh}^\LO$,
\ba
\< \d_\ell^2(\vx) \d_{h,\ell}(\vx) \> 
=\:& b_1 \< \d_\ell^3 \> + \frac12 b_2 \< \d_\ell^2\, [\d^2]_\ell \>
+ b_{K^2} \<  \d_\ell^2 \, [K^2]_\ell \> + \cdots\vs
\stackrel{\LO}{=}& \int_{\vk_1} \int_{\vk_2} W_\ell(k_1) W_\ell(k_2) W_\ell(k_{12}) 
\label{eq:Shmm}\\
 &\qquad \times
 \left\{ b_1 B_{mmm}^\LO(k_1,k_2,k_{12})
+ \left[ b_2 
+  2 b_{K^2} \left( \left[ \hat\vk_1\cdot\hat\vk_2 \right]^2 - \frac13\right) \right]
\Plin(k_1) \Plin(k_2) \right\}\,.
\nonumber
\ea
and similarly for $\< K_\ell^2(\vx) \d_{h,\ell}(\vx) \>$.  
We now see the source of the differences between the $n$-point biases $b_O$
and the moments biases $b_O^{\rm m}(R_\ell)$:  the former 
are defined without 
making reference to any scale $R_\ell$, which only enters when smoothing the
halo density field $\d_h \to \d_{h,\ell}$ \emph{after} the bias expansion;  hence, the correlators contain $[O]_\ell$.    
On the other hand, the operators $[O]^{\rm m}(R_\ell)$ corresponding to the $b_O^{\rm m}(R_\ell)$ 
make reference to the \emph{smoothed linear density and tidal field}
[\refeqs{Shmm2}{Shmm2b}], and thus depend explicitly on the scale $R_\ell$.  
Note that an exactly analogous result, with $b_O\to b_O^L$, is obtained when considering moments
of the proto-halo density field in Lagrangian space.  In that case,
the skewness of $\delta_\ell$ vanishes.  

\begin{figure}[t]
\centering
\includegraphics[width=0.7\textwidth]{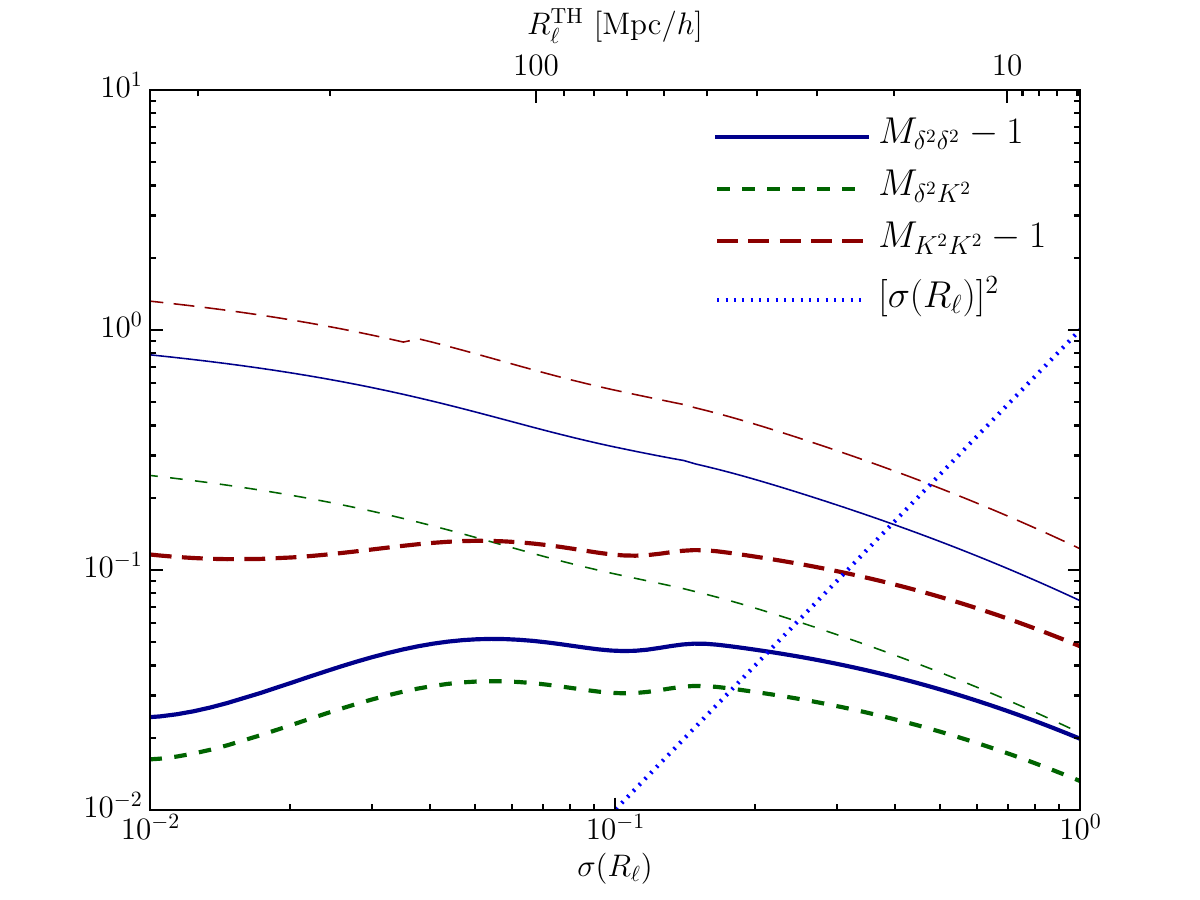} 
\caption{Elements of the bias transformation matrix $M_{OO'}$ [\refeq{bmapping}] as a function of $\s(R_\ell)$, for two filters: real-space tophat (lower, thick set of
lines) and Gaussian (upper, thin set of lines).  Specifically, we show
the departure of the linear map from the identity.  
The top axis shows the corresponding smoothing scale $R_\ell^\text{TH}$ for the tophat filter.  
The dotted line shows the order of magnitude of higher-order corrections 
$\propto \s^2(R_\ell)$. 
}
\label{fig:scaled_mom}
\end{figure}
The leading-order relation in PT between the different bias
definitions can now be read off by comparing \refeq{Shmm2} and \refeq{Shmm}.  
For the linear bias parameter, we obtain simply
\ba
b_1^{\rm m}(R_\ell) \stackrel{\LO}{=}\:& b_1
\,,
\ea
in agreement with \refeq{shh}.  Thus, as long as NLO corrections of order 
$\s^2(R_\ell)$ and $R_*^2 \s_1^2(R_\ell)$ can be neglected, the linear moments bias is exactly equal to
the linear bias inferred from the large-scale two- or three-point function.  
For the second-order biases, there is a nontrivial mapping already at leading order:
\ba
b_O^{\rm m}(R_\ell) \stackrel{\LO}{=}\:& \sum_{O'=\d^2,K^2} M_{OO'}(R_\ell) b_{O'}
\,, \quad
\mbox{where}\quad O, O' \in \{ \d^2,\,K^2 \} \quad\mbox{and}\vs
M_{OO'}(R_\ell) =\:& \s^{-4}(R_\ell) \left(\begin{array}{cc}
\frac12 \< [\d^2]_\ell \d_\ell^2 \> & 
\frac12\< [K^2]_\ell \d_\ell^2 \> \\
\frac{45}8 \< [\d^2]_\ell \,K_\ell^2 \> & 
\frac{45}8 \< [K^2]_\ell \, K_\ell^2\>
\end{array}\right)_\LO\,.
\label{eq:bmapping}
\ea
Note that $b_{\d^2} = b_2/2$, and $\<[\d^2]_\ell \,K_\ell^2\> = \< [K^2]_\ell \d_\ell^2 \>$ at leading order.  
\reffig{scaled_mom} shows the three independent elements of $M_{OO'}(R_\ell)$ as function of $\s(R_\ell)$ and for two different filters; the fourth element $M_{K^2 \d^2}$ is simply proportional to $M_{\d^2 K^2}$.  More specifically, we show the departure from
the identity matrix.  Clearly, the biases inferred from 
moments through \refeq{Shmm2} in general differ significantly from the biases inferred
from the three-point function in a strongly $R_\ell$-dependent way.  
The $R_\ell$-dependence of $M_{OO'}(R_\ell)$ that is evident from \reffig{scaled_mom} 
is due to the fact that $\Plin(k)$ is not a pure power law (for which the
coefficients would be $R_\ell$-independent), but instead the effective power law 
index depends on the smoothing scale $R_\ell$.  Again, these results hold equally for proto-halo moments in Lagrangian space.\\

Nevertheless, there is an unambiguous
transformation between the two bias definitions.  Note that it is essential that all bias
parameters that are relevant at a given order are included (only 2 at second order, but, for
example, 4 at third order) in order to obtain the correct transformed biases.  
Moreover, if $\sigma^2(R_\ell)$ is not sufficiently 
small, there are significant nonlinear corrections to the relation \refeq{bmapping}.  
The order of magnitude of higher-order contributions, i.e. $\s^2(R_\ell)$, is shown by the dotted line in \reffig{scaled_mom};  if this quantity is not small, then the $b_O^{\rm m}(R_\ell)$ cannot be related to the large-scale biases $b_O$ within the framework of perturbation theory.   

For the real-space tophat filter, the bias transformation matrix
$M_{OO'}$ is numerically close to the identity (within 3--15\% depending on $R_\ell$).  Still, for precision
bias measurements the difference between say $b_2$ and $b_2^{\rm m}(R_\ell)$
is not negligible.  For the Gaussian filter, the differences between
$b_2$ and $b_2^{\rm m}(R_\ell)$ are significant and cannot be neglected for any
interesting filter scales.  
It might seem surprising that $b_{\d^2}^{\rm m}(R_\ell)$ is fairly close to $b_{\d^2}$ 
(assuming $\s^2(R_\ell)$ is negligible) for the tophat filter, given that they are 
related to different moments.  To see why, let us write
\be
W_\ell(k_{12}) = W_\ell(k_1) W_\ell(k_2) f_\ell(k_1, k_2, \hat\vk_1\cdot\hat\vk_2)\,.
\ee
If $f_\ell=1$, then the moment \refeq{Shmm} reduces to the hierarchical result,
and we recover $b_{\d^2}^{\rm m}(R_\ell) = b_{\d^2} + \O(\s^2(R_\ell))$.  
For a Gaussian filter, we simply have $f_\ell = \exp(-k_1 k_2 \mu R_\ell^2)$.  A real-space 
tophat filter on the other hand satisfies $W_\ell^2(\vx) = W_\ell(\vx)/V_\ell$, where $V_\ell = 4\pi R_\ell^3/3$
is the volume of the filter.  This implies that
\be
\int_{\vk_1} |W(k_1)|^2 f(k_1,k,\mu) = \int_{\vk_1} |W(k_1)|^2 = 1\,,
\ee
so that $f(k_1,k,\mu)$ is close to unity.  This explains why, for the tophat
filter, $\< [\d^2]_\ell \d_\ell^2\> \approx \s^4(R_\ell)$ while $\< [K^2]_\ell \d_\ell^2\> \ll \s^4(R_\ell)$, 
so that the moments biases for this filter are close to the large-scale renormalized biases.  

In order to determine the leading corrections of order $\s^2(R_\ell)$ in
\refeq{bmapping}, one has to evaluate the NLO corrections to $P_{hm}$
and $B_{hmm}$, respectively, which include higher-order and higher-derivative
biases.  Specifically, the leading correction to $b_1^{\rm m}(R_\ell)$ [\refeq{shh}] is
given by the integral over \refeq{Phm1l}.  The contribution 
proportional to $b_1$ captures the nonlinear evolution of matter.  However, there are
additional contributions at the same order which involve four additional
bias parameters.  The number of additional biases entering $b_2^{\rm m}(R_\ell)$
due to the NLO contribution to the halo-matter-matter bispectrum
is even larger.  Thus, in order to obtain rigorous, predictive measurements
for the biases from moments, one has to restrict to the regime $\s(R_\ell) \ll 1$.  
At low redshifts, significant nonlinear corrections already appear for 
cell sizes $R_\ell\lesssim 50\hmpc$ \cite{angulo/baugh/lacey:2008}.\\  

A related, but subtly different method to measure \LIMD bias parameters was proposed in 
\cite{paranjape/sheth/desjacques:2013,paranjape/sefusatti/etal:2013}, where halos identified at a given 
redshift were traced back to the initial conditions, yielding the Lagrangian density field 
of halos $\d_h^L(\vq)$.  The Lagrangian bias parameters---including those induced by higher-derivative 
operators---can be obtained by cross-correlating the Lagrangian halo field $\d_h^L$ with nonlinear transformations of the matter density field \cite{musso/paranjape/sheth:2012,biagetti/chan/etal:2014}.
The Hermite polynomials $H_N$ provide a suitable basis to measure the Lagrangian \LIMD bias parameters.
Namely, one measures the joint moments 
\be
M_N^{\rm LCC}(R_\ell) = \left\< [1 + \d_h^L(\vq)] H_N\left[ \frac{\d^{(1)}_\ell(\vq)}{\s(R_\ell)} \right] \right\> \,,
\label{eq:bNLCC}
\ee
where LCC stands for ``Lagrangian cross-correlation.''  Crucially, unlike in the moments method discussed above, the halo density field itself is \emph{not smoothed.}  
This method can be implemented simply by summing $H_N[\d^{(1)}_\ell(\vq_i)/\sigma(R_\ell)]$ evaluated
at the Lagrangian halo positions $\vq_i$.  Since the Hermite polynomials 
are orthogonal with respect to a Gaussian weight, they have the useful property that 
$\< H_N(\nu) H_{M}(\nu) \> = \d_{NM} \<\nu^2\>^{N/2}$ 
for a zero-mean Gaussian variate $\nu$.  
At leading order, the first two bias parameters can be related directly to the large-scale renormalized biases $b_N$:\footnote{$R_\ell$ corresponds to $R_0$ in \cite{paranjape/sefusatti/etal:2013}.  Note that they use $\< \d_{R(M)}^{(1)} \d_\ell^{(1)}\>$ instead of $\<\d^{(1)}\d_\ell^{(1)}\>$ to convert their results to the large-scale bias, where $R(M)$ is the Lagrangian radius of halos.  This however has negligible numerical impact for the scales $R_0 \geq 50 \Mpch$ used there (see \refsec{bNmL} for a discussion).  Further, $\< \d_{R(M)}^{(1)} \d_\ell^{(1)}\> /\s^2(R_\ell) = S_\times/S_0$ in the notation of \cite{paranjape/sefusatti/etal:2013}.} 
\ba
M_1^{\rm LCC}(R_\ell) =\:& b_1^L \frac{\< \d^{(1)} \d^{(1)}_\ell \>}{\s(R_\ell)} + \O\left(R_*^2 \frac{ \< \lapl\d^{(1)} \d_\ell^{(1)}\>}{\s(R_\ell)}\right) \vs
M_2^{\rm LCC}(R_\ell) =\:& b_2^L \left(\frac{\< \d^{(1)} \d^{(1)}_\ell \>}{\s(R_\ell)}\right)^2 + \O\left(R_*^2\frac{\< \lapl\d^{(1)} \d^{(1)}_\ell \>}{\s(R_\ell)}\right) \,.
\label{eq:b1LCC}
\ea
Since these are Lagrangian moments, corrections to these expressions arise only from higher-derivative bias terms, starting with $b_{\lapl\d}^L$, just as is the case for $P_{hm}^L$, \refeq{PhhL}
\cite{desjacques:2008,musso/paranjape/sheth:2012,PBSpaper,paranjape/sheth/desjacques:2013,paranjape/sefusatti/etal:2013}.  
Thus, in the large-$R_\ell$ limit where $R_*^2 \< \lapl \d^{(1)} \d^{(1)}_\ell\> \ll \s^2(R_\ell)$, the $b_N^{\rm LCC}(R_\ell)$ are simply related to the large-scale renormalized Lagrangian biases.  This is a significant difference from the moments biases discussed above, and stems from the fact that the halo density field is not smoothed in this method.    For reference, $\< \d^{(1)} \d_\ell^{(1)}\> /\s^2(R_\ell)$ evaluates to $\sim1.1-1.2$ for a tophat filter, depending on $R_\ell$.

For smaller smoothing scales, corrections induced by higher-derivative terms $\propto\lapl\delta$, and so on, become important.  From the point of view of the general bias expansion, the perturbative expansion breaks down unless $R_\ell$ 
is much larger than the nonlocality scale $R_* \sim R(M)$ for the halos considered.  On the other hand, in the context of the peak approach (\refsec{peaks}), all higher-derivative contributions to \refeq{b1LCC} are controlled by the peak constraint and the filtering kernel, and measurements for any value of $R_\ell$ can be interpreted in the context of this model \citep[see, e.g.,][]{paranjape/sefusatti/etal:2013,biagetti/chan/etal:2014}.    
This method can be generalized to measure several different Lagrangian bias parameters, as was shown in \cite{biagetti/chan/etal:2014} who measured the bias coefficient $b_{(\nabla\d)^2}$.
\refsec{bNmL} describes this approach in more detail in the context of the peak formalism.

\subsection{Scatter-plot method}
\label{sec:bscatter}

\begin{figure}
\subfloat{\includegraphics[width = 0.5\textwidth]{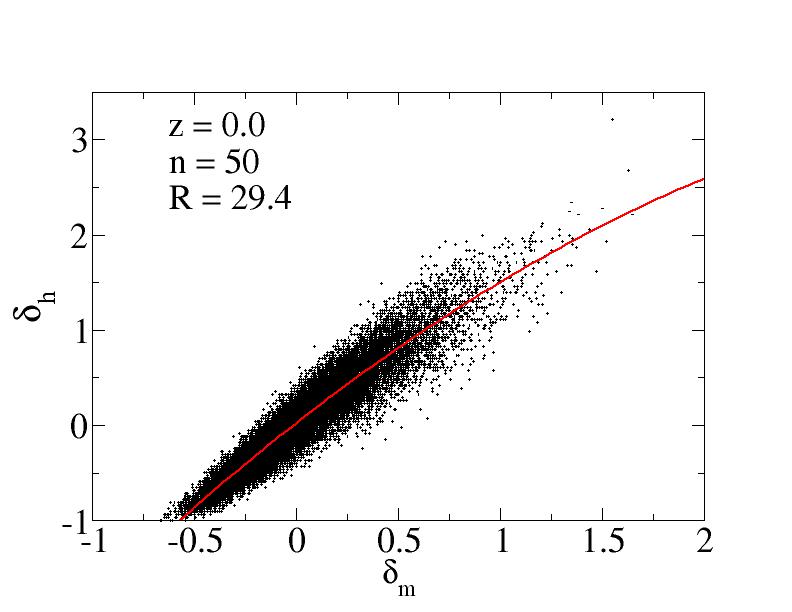}} 
\subfloat{\includegraphics[width = 0.5\textwidth]{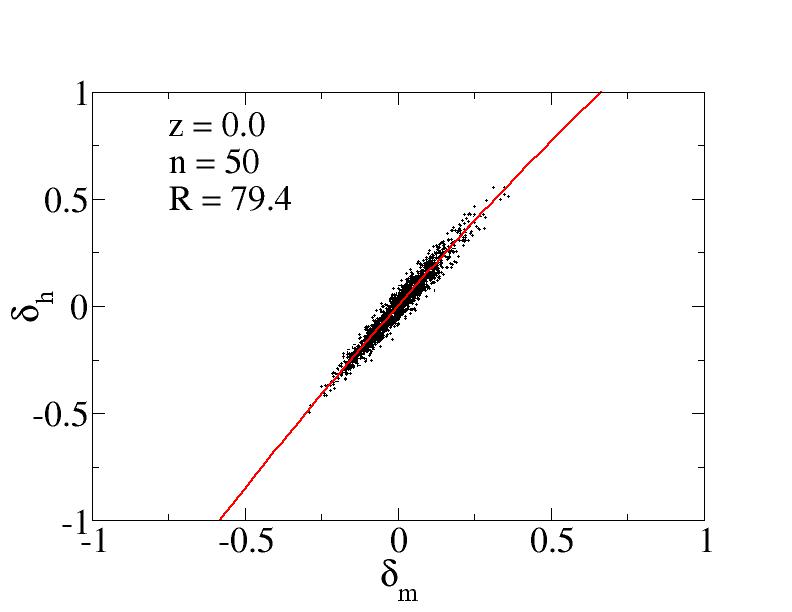}} \\  
\subfloat{\includegraphics[width = 0.5\textwidth]{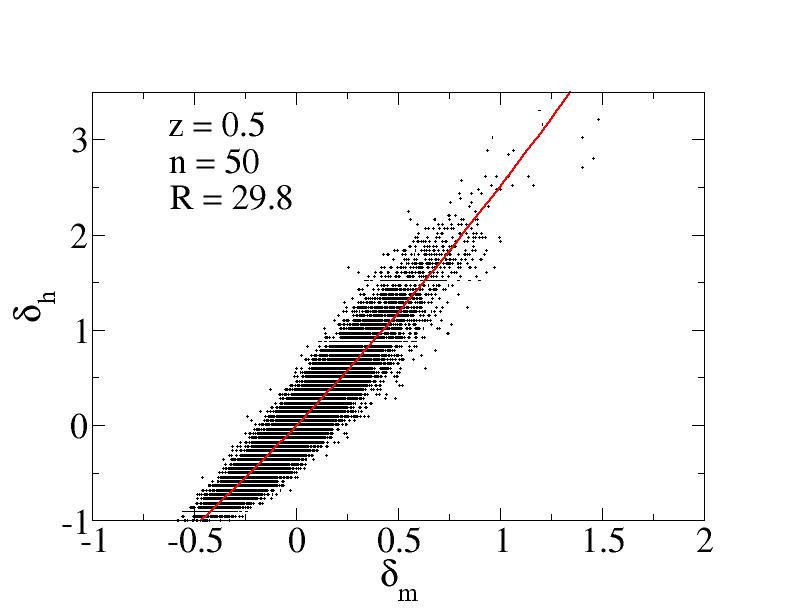}}
\subfloat{\includegraphics[width = 0.5\textwidth]{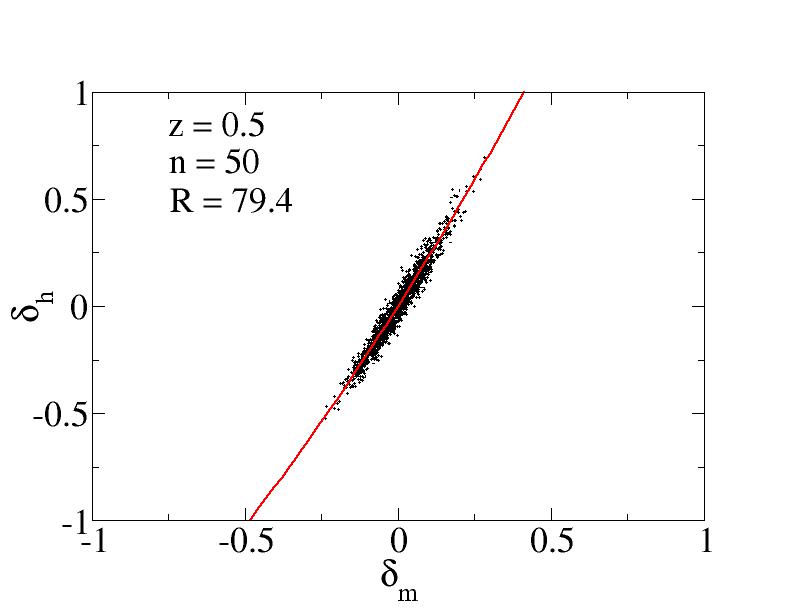}}
\caption{Scatter plots of the halo density contrast $\delta_{h,\ell}$ vs. matter density contrast $\delta_\ell$ in spherical tophat cells of different radii and redshifts, as indicated in the caption.  Only halos with $n=50$ or more particles are included; this corresponds to a minimum halo mass of $1.17\times 10^{13}\hmsun$.  The solid lines represent the least-squares fits of a quadratic polynomial yielding estimates of the first two \LIMD moments bias parameters.  
The variance of the matter density field in each case, clockwise
from the top left, is 
$\s^2(R_\ell) = 0.085,\  0.010,\  0.006,\  0.051$.  
\figsource{manera/gaztanaga:2011}  
}
\label{fig:scatterplots}
\end{figure}

We now turn to a frequently employed method to measure bias which, as we
will see, is an application of the moments discussed in \refsec{bmom}.  
Consider the conditional probability distribution of the smoothed
halo density field $\d_{h,\ell}(\vx)$ at an arbitrary location, given
a fixed value of the smoothed matter density $\d_\ell(\vx)$ at that
location:
\be
P(\d_{h,\ell}|\d_\ell) = \frac{P(\d_{h,\ell}, \d_\ell)}{P(\d_\ell)}\,.
\label{eq:Pdhd}
\ee
Note that we are not choosing any particular location,
and statistical homogeneity then allows us to drop the argument $\vx$
on the PDFs.  In the following, we will drop the subscript $\ell$ for clarity,
as all moments in the following will be defined for a fixed scale $R_\ell$, and correspondingly denote $\s \equiv \sigma(R_\ell)$.  
The conditional PDF \refeq{Pdhd} has been studied extensively
in the literature \cite{taruya/soda:1999,manera/gaztanaga:2011,roth/porciani:2011,chan/scoccimarro:2012}.  The biases can be inferred by applying 
\refeq{Pdhd} to simulations as follows \cite{fry/gaztanaga:1993,angulo/baugh/lacey:2008,manera/gaztanaga:2011,hoffmann/bel/gaztanaga:2016}.  
First, one measures the halo and matter fractional density perturbations,
$\delta_{h,\ell}$ and $\delta_\ell$, for example by counting halos in randomly
thrown spheres of radius $R_\ell$ in the simulation volume.  
One then obtains the mean relation between $\d_{h,\ell}$ and $\d_\ell$ by creating a 
``scatter plot'' (see \reffig{scatterplots}), and taking the mean of $\d_{h,\ell}$ in slices of $\d_\ell$.  Mathematically, in the limit of infinite statistics, 
this corresponds to measuring
\be
\< \d_h|\d \> = \int d\d_h\, \d_h P(\d_h|\d) \,.
\label{eq:edhd}
\ee
Then, one fits a quadratic or higher-order polynomial to $\<\d_h|\d\>$ as 
function of $\d$.  The coefficients of the polynomial then yield the estimates of the scatter-plot bias parameters. The order of the polynomial that is necessary depends
on $R_\ell$: at large smoothing radii $R_\ell$, the matter fluctuations are sufficiently small that a
truncation at third order is appropriate.  
In the following, we thus assume that the smoothing scale $R_\ell$ is sufficiently large 
so that a perturbative treatment is applicable, just as we have done for the moments derived in the previous section.  

Let us begin with linear theory, where $\d$ follows a Gaussian
distribution with variance $\s^2 \equiv \s^2(R_\ell)$, and the halo overdensity is given
by $\d_h \equiv \d_{h,\ell} = b_1 \d + \eps$ where $\eps$ is a Gaussian stochastic variable.  
The corresponding variance of $\d_h$ is given in \refeq{shh}, and hence
we have
\be
P(\d_h, \d)\Big|_{\rm linear} \propto \exp\left(-\frac12 (\d_h, \d) \bm{C}^{-1} (\d_h, \d)^\trans \right), \quad\mbox{where}\quad
\bm{C} = \left(\begin{array}{cc}
b_1^2 \s^2 + \<\eps^2\> & b_1 \s^2 \\
b_1 \s^2 & \s^2
\end{array}\right)\,,
\ee
and $\<\eps^2\> \equiv \< (\eps_\ell^{\rm m})^2\> = \Peps V_\ell^{-1}$ at this order. 
This yields 
\be 
P(\d_{h}|\d)\Big|_{\rm linear} \propto \exp\left[-\frac{\big(\d_h-b_1 \d\big)^2}{2\<\eps^2\>}\right] \;.
\label{eq:PdhdL}
\ee
The expectation value of $\d_h$ conditioned on $\d$ is then simply obtained as
\be
\< \d_h|\d \>\Big|_{\rm linear} = b_1 \d\,.
\ee
Thus, the derivative of $\<\d_h|\d\>$ with respect to $\d$, or equivalently the coefficient of the linear term of a polynomial fit to $\<\d_h|\d\>$, yields the (renormalized) bias parameter $b_1 = b_1^{\rm m}(R_\ell)$.   
\refeq{PdhdL} clearly shows that the scatter
in $\d_h$ at fixed $\d$ is related to the stochasticity $\eps$ (see \refeq{shh} and \cite{taruya/soda:1999}).  
However, both of these statements strictly hold in the limit of $R_\ell \to \infty$; for a finite scale $R_\ell$, \refeq{PdhdL} receives corrections of order $\s^2 = \s^2(R_\ell)$ from NLO terms
involving the quadratic and cubic bias parameters, just as was discussed for the quadratic moments in \refsec{bmom}.  
Note that, in case of scatter-plot bias estimates, there are further corrections of the same order from the non-Gaussianity of $P(\d_h,\d)$.  

We now turn to the leading non-Gaussian, i.e. second-order, contributions to the conditional PDF.  
At this order, the PDF acquires a skewness and, using the Edgeworth
expansion, can be written as
\be
P(\d_h, \d)\Big|_{\rm 2nd} = \left\{ 1 - \frac16 \left[
\< \d^3 \> \frac{\partial^3}{\partial\d^3}
+ 3 \< \d^2 \d_h \> \frac{\partial^2}{\partial\d^2}
\frac{\partial}{\partial\d_h}
+ 3 \< \d \d_h^2 \> \frac{\partial}{\partial\d}
\frac{\partial^2}{\partial\d_h^2}
+ \< \d_h^3 \> \frac{\partial^3}{\partial\d_h^3}
\right]\right\}
P(\d_h, \d)\Big|_{\rm linear}\,.
\label{eq:Pdhd2}
\ee
Upon inserting this into \refeq{edhd}, the second derivative with respect to $\d$ of $\< \d_h | \d\>$ evaluated at $\d=0$ yields
\be
\frac{\partial^2}{\partial\d^2} \Big\<\d_{h}\Big|\d\Big\>\Big|_{{\rm 2nd},\,\d=0} = 
\frac1{\s^4}\left[\< \d^2 \d_h\> - b_1 \< \d^3 \>\right] + \O(\s^2)
= b_2^{\rm m}(R_\ell) + \O(\s^2)\,,
\label{eq:d2Pdhd}
\ee
where we have used \refeq{Shmm2}.  This is precisely ($1/2$ of) the 
quadratic coefficient of a polynomial fit to $\<\d_h|\d\>$.  
Thus, in the large-scale limit, the second-order scatter plot bias 
measurement yields exactly the second-order moments bias discussed
in \refsec{bmom} for the same filter $W_{R_\ell}$.  
The two differ by NLO corrections that scale
as $\s^2$, but which are in principle calculable.  
Similar results hold if one were to construct the conditional
PDF with respect to the tidal field smoothed on scale $R_\ell$, $P(\d_h|K^2)$.  
In order to connect to the biases measured from $n$-point functions,
one needs to apply the transformation in \refeq{bmapping}.  
For measurements which use $R_\ell \lesssim 40 \Mpch$ and tophat filters,
the higher-order contributions which scale as $\s^2(R_\ell)$ are however still
larger than the effects of the transformation \refeq{bmapping}, as shown 
in \reffig{scaled_mom}.  

\reffig{scatterplots} shows typical scatter plots of $\delta_{h,\ell}$ vs. 
$\delta_\ell$ for different cell sizes and redshifts, and polynomial fits
to the mean relation \cite{manera/gaztanaga:2011}; 
again, the coefficients of the polynomial fits correspond to the bias parameters inferred from this method. 
The variance of the matter density in cells 
$\s^2$ for each case is indicated in the caption.  
For the larger cells shown, $R_\ell \simeq 80\Mpch$, the mean relation is well fitted by a second-order polynomial.  
The curvature of this fit corresponds to the quadratic moments bias $b_2^{\rm m}(R_\ell)$; specifically, we see a negative ($z=0$) or 
positive ($z=0.5$) value of $b_2^{\rm m}(R_\ell)$.  
In the excursion set and peaks models (\refsecs{exset}{peaks}), this is explained by the different
peak significance of halos with fixed mass $M \gtrsim 10^{13} \Msunh$ at 
$z=0$ and $z=0.5$, respectively.  

For the smaller cell result at $z=0.5$ (bottom-left panel), discreteness effects, which arise from the fact 
that there can only be an integer number of halos per cell, are apparent.  
This implies that the cell size is not much larger than the mean 
inter-halo separation, which is of order $R(M)$.  For cells this small,
non-perturbative effects such as exclusion become relevant.  This
points to the fact that scatter plot results 
for small $R_\ell$ cannot be rigorously connected with perturbative predictions on large scales.

The stochasticity between $\delta_{h,\ell}$ and $\delta_\ell$, which is responsible for the 
scatter around the mean relation in \reffig{scatterplots}, is taken into account in 
the perturbative approach by the stochastic fields $\eps,\,\eps_\d,\cdots$. 
This is appropriate as long as the scatter is small 
[note that both $\<\eps^3\>$ and $\< \eps \eps_\d\>$ contribute to \refeq{Pdhd2}
at leading order, although they do not enter the second derivative at $\d=0$
in \refeq{d2Pdhd}].  
This observations has led some authors to propose that, in the non-perturbative regime, the deterministic bias expansion be replaced by a more 
general bias distribution function \cite{dekel/lahav:1999,taruya/soda:1999}.  
Another approach is to construct the scatter plot in terms of the log-transformed fields $\ln(1+\d)$ and $\ln(1+\d_h)$ \cite{jee/etal:2012}.  
However, these small-scale measurements strictly apply to counts-in-cells only. Namely, their results 
cannot be applied to the computation of the large-scale $n$-point correlations estimated using direct pair counting.

\subsection{Response approach}
\label{sec:bsepuni}

\begin{figure}[t!]
\centering
\includegraphics[width=0.6\textwidth]{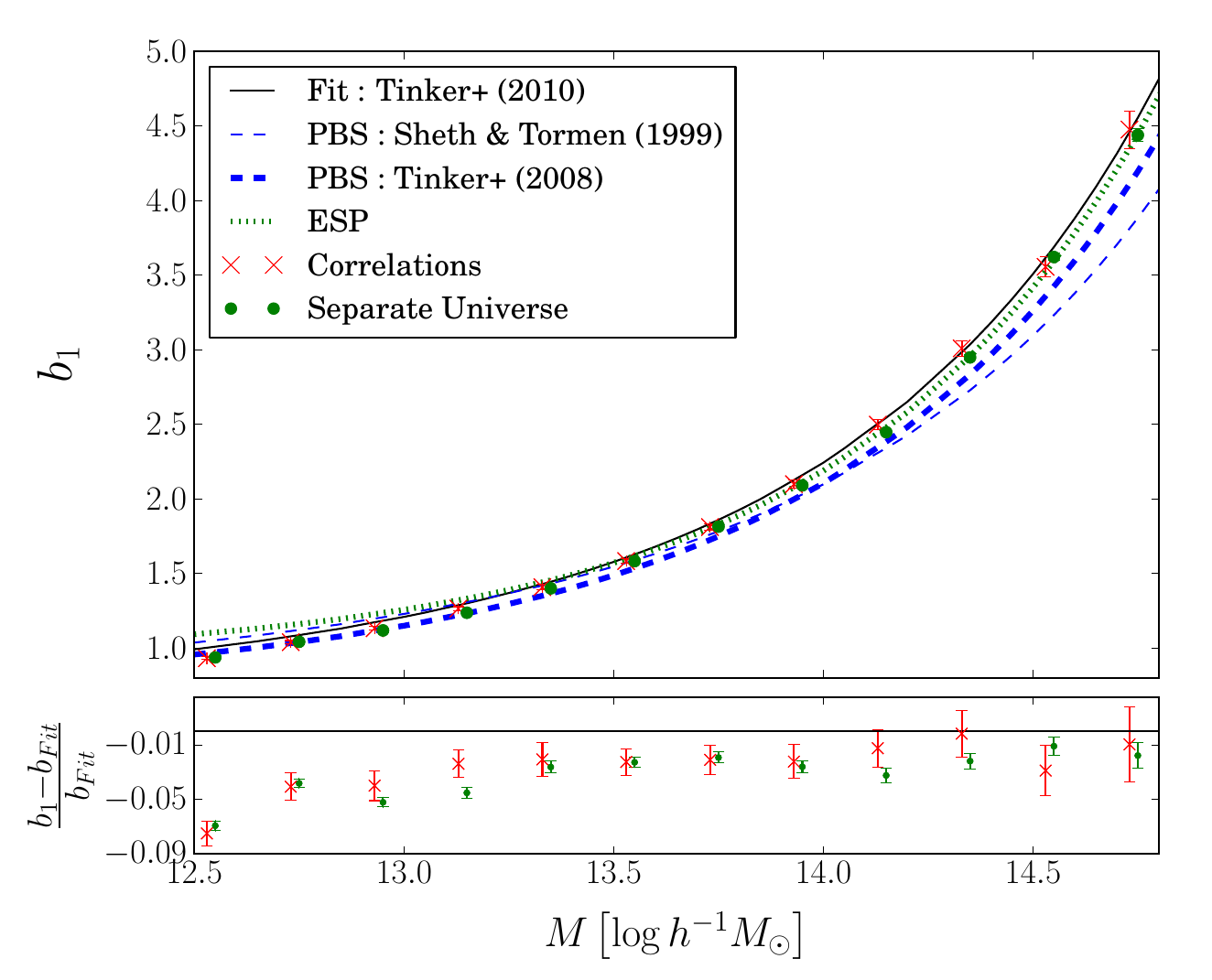} 
\caption{\textit{Top panel:} comparison between the linear halo bias from separate-universe simulations (green dots), and from clustering (red crosses; displaced slightly horizontally for clarity). Error bars that are not visible are within the marker size.  The solid black curve is the Tinker et al. (2010) \cite{tinker/etal:2010} best fit curve for $b_1$ given in \reftab{fittingbiases}, while the dot-dashed green curve is the excursion set peak (ESP) prediction [derived from \refeq{espmf} via the PBS approach, \refsec{buniv}].  Also shown are the results obtained by applying \refeq{bPBSuniv} for a constant barrier $B=\dc$ to the Tinker et al (2008) \cite{tinker/etal:2008} 
and Sheth-Tormen \cite{sheth/tormen:1999} mass functions (blue dashed curves; see also \reffig{shethtormen}). \textit{Bottom panel:} relative difference between the measurements and the fitting formula from Tinker et al. (2010).  
\figsource{lazeyras/etal}
}
\label{fig:b1SU}
\end{figure}
We now turn to an alternative approach to measuring bias which is motivated by
 the rigorous definition of the peak-background split derived in \refsec{sepuni}.  
Specifically, the \LIMD bias parameters of halos are measured through the response of the 
halo abundance to an infinite-wavelength density perturbation [\refeq{bN}], implemented by 
performing a series of N-body simulations with different amounts of spatial curvature 
following the separate-universe approach.  Moreover, it is possible to generalize 
this idea to other bias parameters, for example the tidal bias or higher-derivative 
biases, by measuring the response of the halo abundance to modified initial conditions 
that contain certain configurations of amplified long-wavelength modes.  
We use the term ``response bias'' to encompass all of these methods.  Here, we will
focus on the \LIMD bias parameters $b_N$ of dark matter halos.  

\begin{figure}[t!]
\centering
\includegraphics[width=0.49\textwidth]{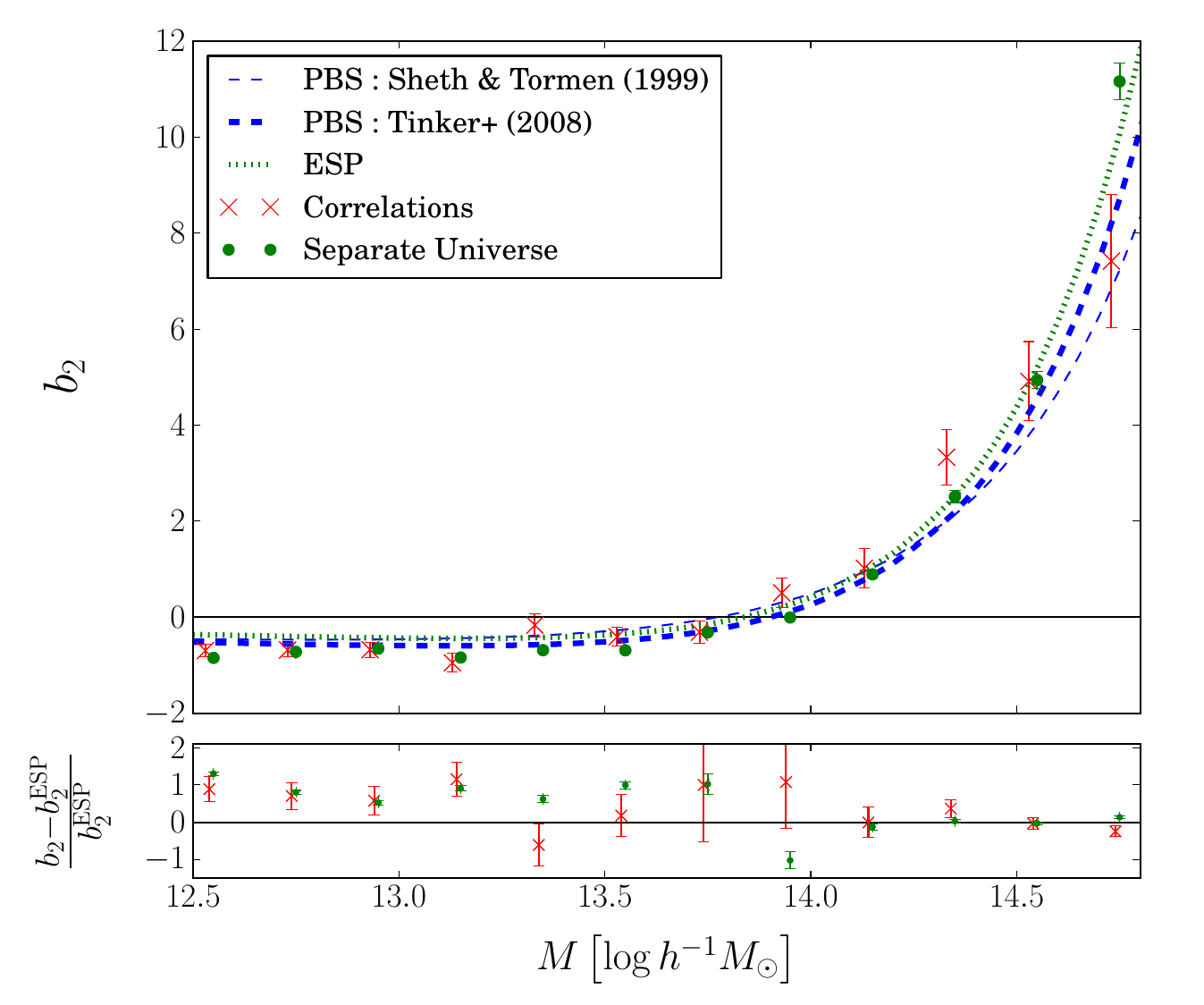} 
\includegraphics[width=0.49\textwidth]{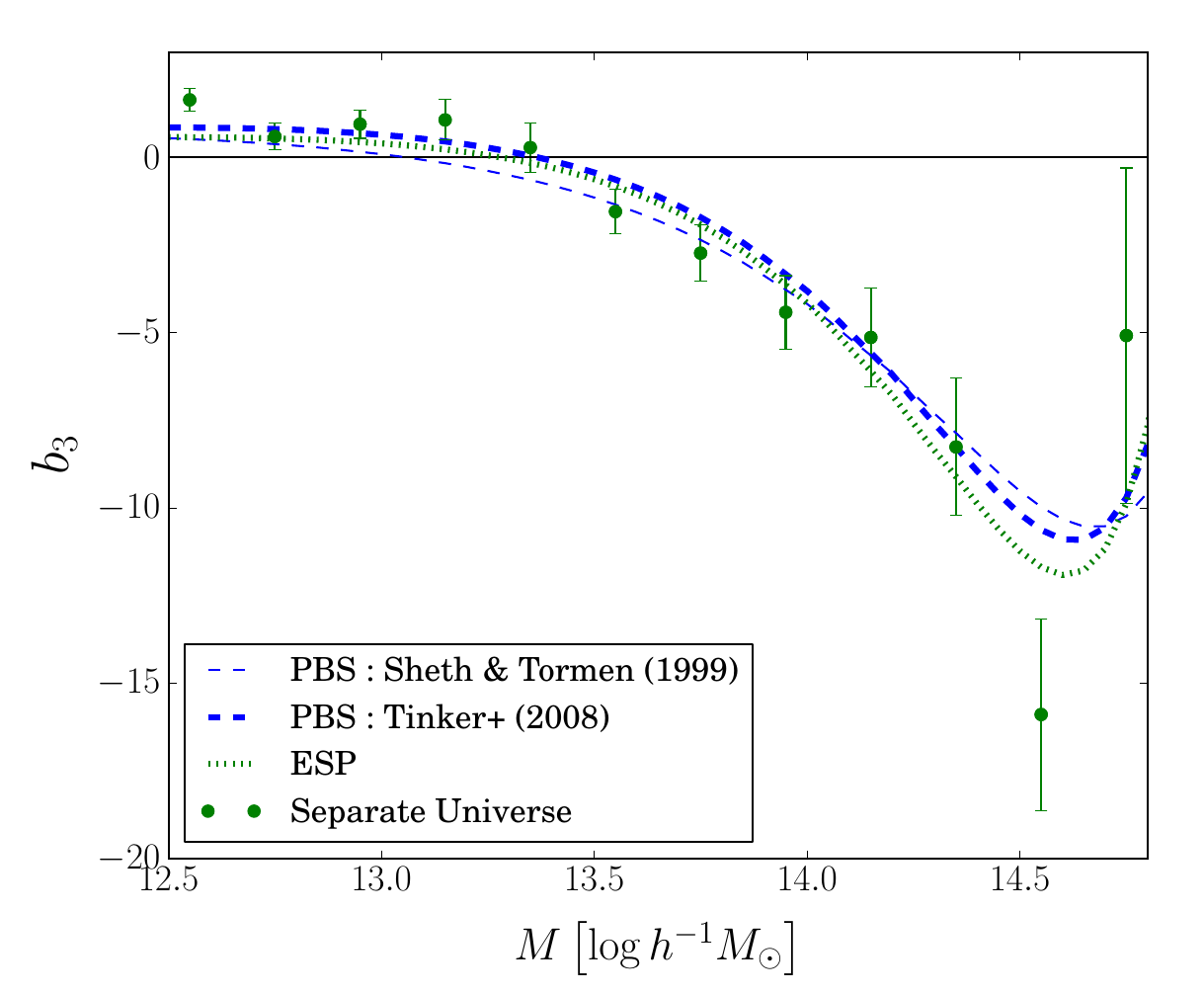} 
\caption{\textit{Left figure, top panel:} same as \reffig{b1SU}, but for the quadratic Eulerian bias $b_2$.  \textit{Bottom panel:} relative difference between measurements and the theoretical prediction of the excursion set peaks (ESP).  In each panel, the clustering points have been displaced horizontally as in \reffig{b1SU}.
\textit{Right figure:} as left, but for the cubic Eulerian bias $b_3$. 
\figsource{lazeyras/etal}
}
\label{fig:b2b3}
\end{figure}

Clearly, this method is restricted to simulated tracers, and cannot be applied to 
actual observed galaxies.  However, for simulations it offers several advantages.  
First, this approach directly isolates the desired bias parameters without needing 
to fit several parameters at the same time.  
Second, as we are dealing with an infinite-wavelength perturbation, we are not limited by the theoretical uncertainty of the 
perturbation theory predictions for finite-wavelength modes (\refsec{npt1loop}).  
Finally, by using the same initial seeds for simulations with different 
long-wavelength modes, we can cancel cosmic variance to a large extent.  
In principle, this method is able to achieve a higher statistical precision on bias parameters for a given simulation volume than $n$-point functions.  Whether this is attained in practice depends on
the precise estimator used in measuring the response (for example, binned 
halo counts in \cite{lazeyras/etal,baldauf/etal:2015}, and the cumulative mass
function in \cite{li/hu/takada:2016}).  Generally, one expects the statistical advantage
of the response approach to be more important for higher-order biases, as
can be seen when comparing the error bars in the lower panels of \reffig{b1SU} and \reffig{b2b3} (left panel).
  
\begin{figure}
\centering
\includegraphics[width=0.49\textwidth]{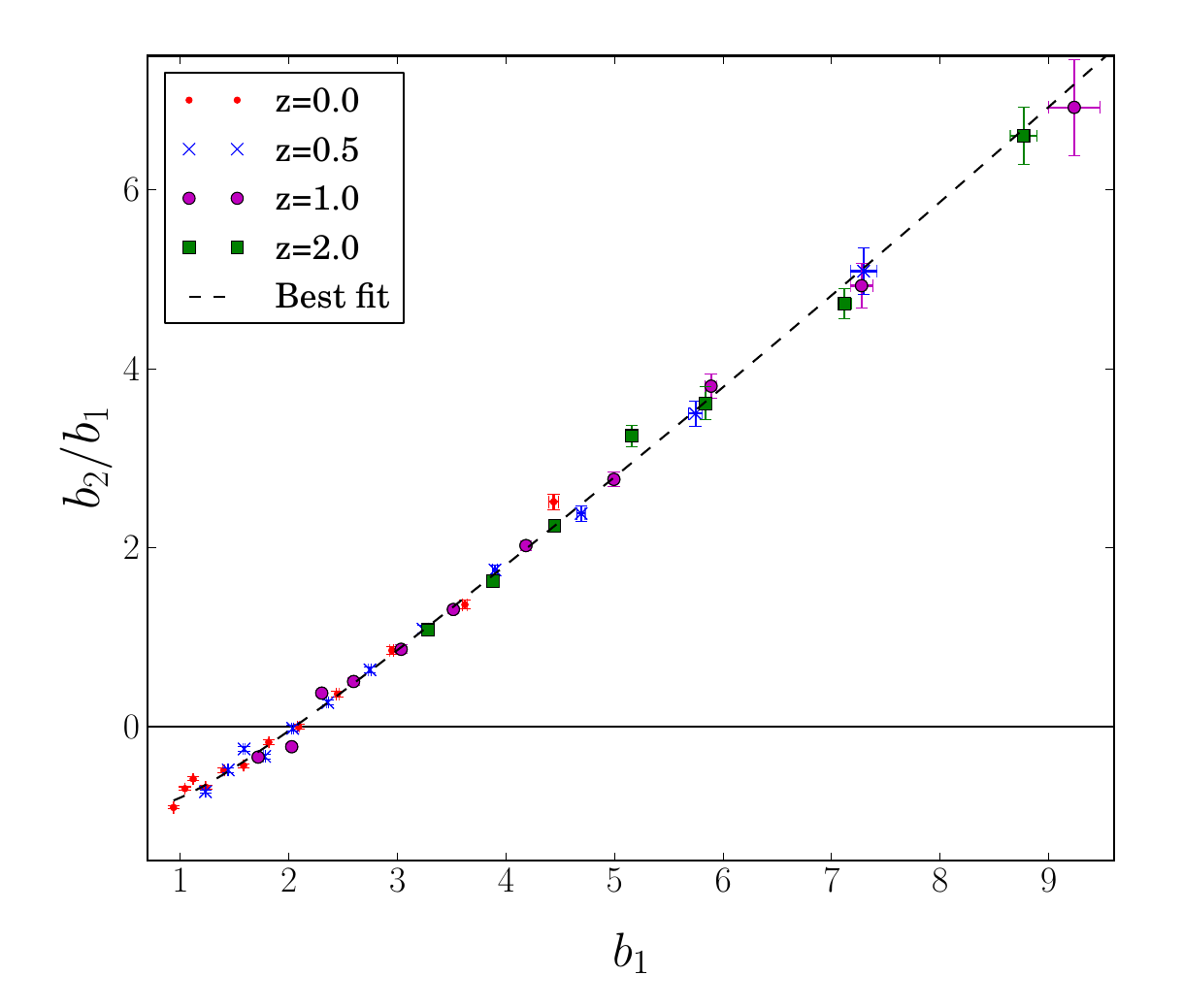} 
\includegraphics[width=0.49\textwidth]{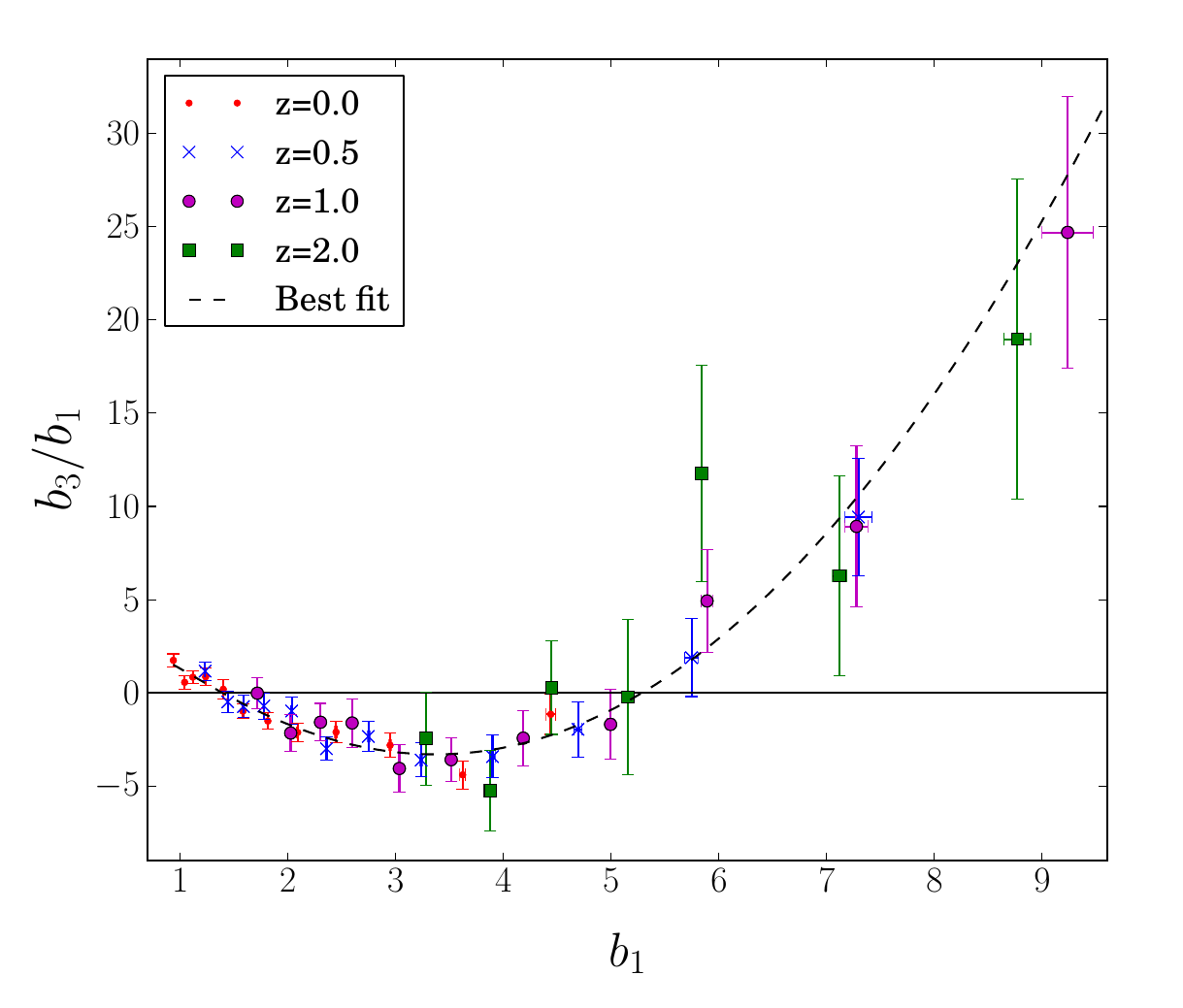}
\caption{Eulerian bias parameters $b_2$ \textit{(left panel)} and $b_3$ \textit{(right panel)}, each divided by $b_1$, as a function of $b_1$ obtained from separate-universe simulations and for different redshifts. The dashed curves show the third-order best-fit polynomials which are summarized in \reftab{fittingbiases}.  
\figsource{lazeyras/etal}
}
\label{fig:b2b3(b1)}
\end{figure}

This novel method has been applied in three recent papers 
\cite{lazeyras/etal,baldauf/etal:2015,li/hu/takada:2016}, where \cite{lazeyras/etal} 
went to nonlinear order while \cite{baldauf/etal:2015,li/hu/takada:2016} only measured 
the linear bias $b_1$.  
All three references found consistency between the response approach and the 
$n$-point function measurements of $b_1$, 
as well as $b_2$ in the case of \cite{lazeyras/etal}.  
\reffig{b1SU} shows a comparison of the linear bias of dark matter halos measured from 
the halo-matter cross-power spectrum on large scales, and that measured from the separate-universe response \cite{lazeyras/etal}.  
\reffig{b2b3} shows the analogous result for the quadratic and cubic biases $b_2,\,b_3$.  
This shows explicitly that, if carefully measured, both techniques unambiguously measure the same bias parameters (the same is true for the moments-based approaches, provided 
sufficiently large smoothing scales are chosen and the transformation \refeq{bmapping} 
is applied).  
Ref.~\cite{lazeyras/etal} also provided fitting formulas for $b_2,\,b_3$ in terms of 
$b_1$ (\reffig{b2b3(b1)}), combining results at various redshifts since the relation was found to be redshift-independent.  
The fitting formulas are given in \reftab{fittingbiases}.  
The fitting formula for $b_2(b_1)$ given in \cite{hoffmann/bel/gaztanaga:2016}, obtained 
from fits to the two- and three-point correlation functions in real space, agrees 
with the one given in \reftab{fittingbiases}, with differences much smaller than the error 
bars on the measurements.

\reffigs{b1SU}{b2b3} also compare the simulation results to predictions based on the 
peak-background split applied to universal mass functions (\refsec{buniv}).  Once a specific barrier $B$ 
is assumed, the PBS allows for a derivation of all \LIMD bias parameters $b_N$ 
from a 
given halo mass function.  
The simplest and most common choice is $B=\dc$.  
\reffigs{b1SU}{b2b3} show the result of applying this to the Sheth-Tormen  
(\cite{sheth/tormen}, see \refeq{bST} for the explicit expressions) and Tinker et al (2008) \cite{tinker/etal:2008} mass function prescriptions.  
Even though the latter provides a very accurate mass function, the linear bias derived via 
the PBS and simple collapse threshold is only accurate at the $\sim 10$\% level, as shown 
previously in \cite{manera/etal:2010,tinker/etal:2010,hoffmann/bel/gaztanaga:2015}.  
The agreement is even worse for $b_2$, with up to 50\% discrepancy at low mass, although the 
absolute difference between the PBS predictions and the measurements is similar to that in $b_1$.  

Finally, \reffigs{b1SU}{b2b3} also show the biases predicted by the excursion set peak (ESP)
approach, described in detail in \refsec{esp}, which includes a stochastic moving barrier 
motivated by simulation results.  At high mass, this performs much better, at least for $b_1$, 
showing that, in the context of the PBS applied to universal mass functions, the choice of barrier is a key ingredient in deriving accurate bias parameters.  
In this context, it is important to note that previous results on the inaccuracy of PBS bias 
parameters \cite{manera/etal:2010,hoffmann/bel/gaztanaga:2015} relied on the simple constant 
threshold $B=\dc$.  
This shows that the cause of theses inaccuracies is not the peak-background split itself.  
Interestingly, the results of \reffig{b2b3(b1)} show that there is still some universal scaling 
behavior in the higher-order bias parameters of halos: the relations $b_2(b_1)$ and $b_3(b_1)$ 
are found to be independent of redshift to within the uncertainties.
For this reason, the fitting formulas of these relations (\reftab{fittingbiases}) are expected 
to depend only weakly on cosmology.  

We conclude this discussion by reiterating that the ``inaccuracy of the peak-background split'' 
depends on what one defines PBS to mean, and summarize it as follows:
\begin{itemize}
\item As shown in \refsec{PBS}, local bias parameters derived using the PBS implemented via the 
separate-universe response approach are \emph{physically the same parameters} as those obtained 
from large-scale $n$-point correlations.  Implementations of both approaches in N-body simulations have clearly confirmed this (\reffigs{b1SU}{b2b3}).  
\item PBS biases derived using universal mass function based on a simulation-calibrated stochastic moving barrier 
\cite{robertson/etal:2009,paranjape/sheth/desjacques:2013} are accurate to a few percent for $b_1$, 
at least at high masses \cite{lazeyras/etal}.
\item The PBS using the constant spherical collapse barrier is no better than 10\% for $b_1$, 
and worse for $b_2$ \cite{manera/etal:2010,hoffmann/bel/gaztanaga:2015,lazeyras/etal}.
\end{itemize}

\subsection{An overview of bias measurements}
\label{sec:meas:meas}

We now briefly review actual measurements of bias parameters for dark matter halos in simulations and observed galaxies. We then review measurements and physical models of halo stochasticity, i.e. the stochastic amplitudes $P_\eps, B_\eps,\cdots$ which enter halo $n$-point statistics.

\subsubsection{Halo bias}
\label{sec:meas:meas:halos}

Early measurements of the large-scale halo bias using two-point functions are 
presented in, e.g., 
\cite{kauffmann/nusser/steinmetz:1997,sheth/tormen:1999,kravtsov/klypin:1999,jing:1999,seljak/warren:2004}.  
Some of the most accurate recent measurements of the linear halo bias can be found in \cite{tinker/etal:2010}, who used $P_{hh}(k)$ on large scales to infer $b_1$ for 
spherical overdensity (SO) halos in a large suite of N-body simulations (see \refapp{halofinder} for a brief description of halo finding algorithms).  
For this, they assumed that $\Peps = 1/\avnh$.  
They also cross-checked the results with the linear bias parameter measured from the halo-matter 
cross-power spectrum, and find that, within the uncertainty of the measurements, the deviation from perfect Poisson
stochasticity (\refsec{meas:stoch}) does not significantly affect the bias measurement.  
The resulting fitting function, which is accurate to 5--10\% for a wide range of SO halo definitions and $\Lambda$CDM cosmologies, is summarized in \reftab{fittingbiases}. Note that the universality of $b_1$, i.e. that its mass and cosmology dependence are completely described by the parameter $\nu_c = \dc/\s(M)$, is not perfect (see Fig.~1 in \cite{tinker/etal:2010}).
As discussed in detail in \refsec{bsepuni}, Ref.~\cite{lazeyras/etal} showed that the bias fitting function agrees very well with the response bias measurements (see \reffig{b1SU}).  
It is well established that the bias $b_1$ of halos, at fixed halo mass, also depends on other halo properties.  
This is known as \emph{assembly bias} and will be discussed in \refsec{assembly}.  

Robust measurement of the large-scale nonlinear halo bias parameters from $n$-point functions 
have only been published fairly recently. 
Refs.~\cite{guo/jing:2009,pollack/smith/porciani:2012} combined measurements of halo two- and three-point functions to infer $b_1,\,b_2$ assuming 
Eulerian \LIMD bias, i.e. setting $b_{K^2}=0$ in \refeqs{Phh}{Bhhh}.  
Ref. \cite{pollack/smith/porciani:2012} did not find complete consistency between two- and three-point function 
measurements (see also \cite{roth/porciani:2011}), which could be an indication that $b_{K^2}$ is 
an important ingredient \cite{chan/scoccimarro/sheth:2012,baldauf/etal:2012} (see below).  
Refs.~\cite{saito/etal:14,angulo/etal:2015} included the NLO corrections to $P_{hm}$ in addition to the 
leading-order bispectrum $B_{hmm}^\LO(k_1,k_2,k_3)$ of friends-of-friends (FoF) halos, to measure $b_1$, $b_2$, $b_{K^2}$, 
and $b_{\otd}$.  
Note that they used different parametrizations for the operators $K^2$ and $O_{\otd}$ 
(see \refapp{biastrans} for the relations between various different parametrizations).  
Ref.~\cite{saito/etal:14} set $b_{\lapl\d}$ to zero, and performed a fit including modes with $k \lesssim 0.1\iMpch$ at $z=0$, while 
Ref.~\cite{angulo/etal:2015} fit up to scales of $0.15\iMpch$ in the bispectrum and $0.3 \iMpch$ in the power spectrum.    
The linear and quadratic bias parameters obtained from the two-and three-point functions were all found to be 
consistent.  

Regarding real-space measurements, 
Ref. \cite{hoffmann/etal:2015} analyzed N-body simulations using the \LIMD bias assumption and found a 10\% 
deviation of the linear growth factor estimated by combining the two-point and three-point correlation functions. 
Later, the same group \cite{bel/hoffmann/gaztanaga:15} 
presented measurements of $b_1$, $b_2$ and $b_{K^2}$ from the two-point halo-halo and halo-matter correlation functions, 
as well as the halo-matter-matter three-point function in real space. 
They found consistency in the measured value of $b_1$ from the two- and three-point functions, and, furthermore, 
confirmed the necessity of including $b_{K^2}$ to correctly fit the latter. 

\begin{figure}
\centering
\includegraphics[width=0.6\textwidth]{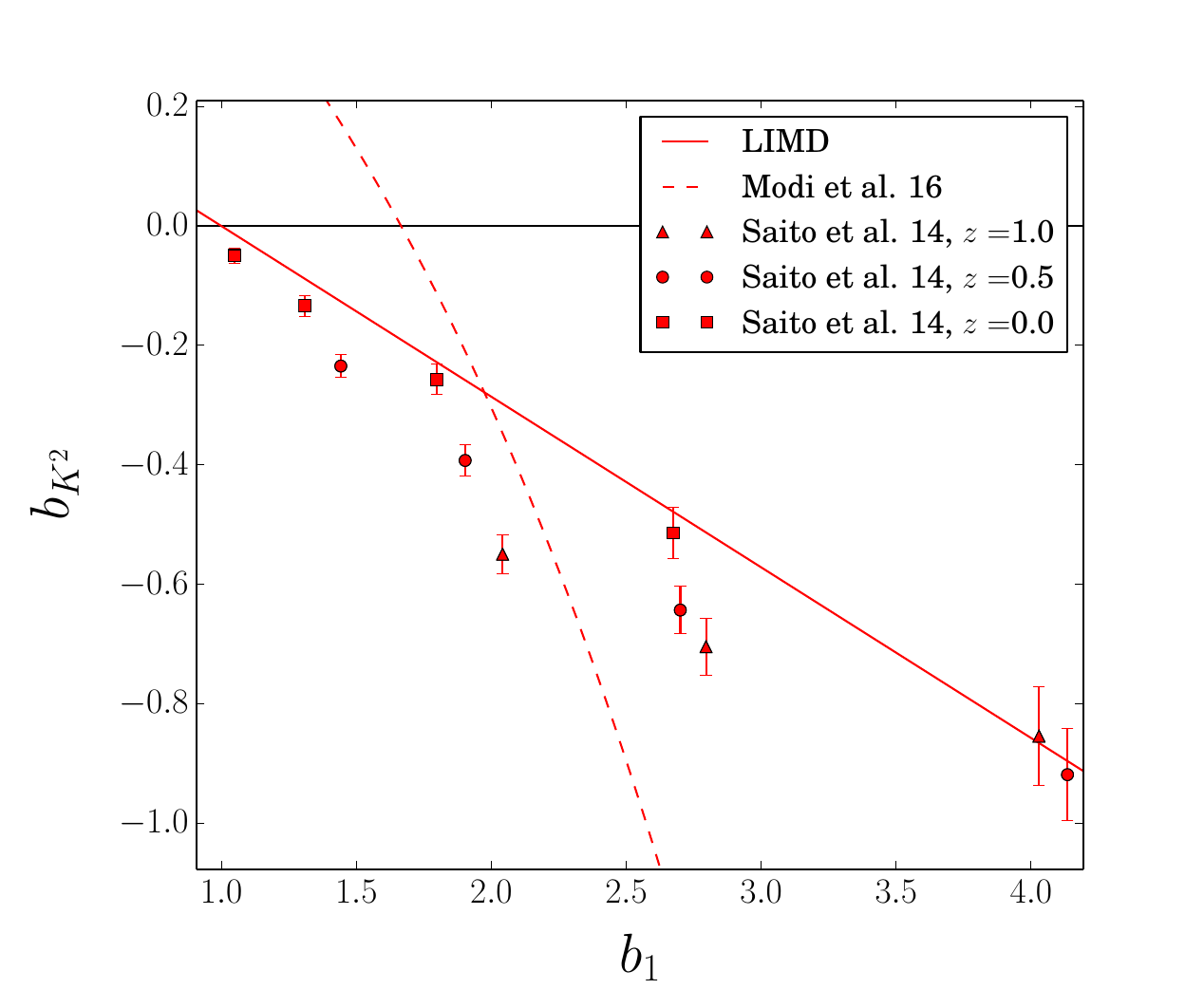} 
\caption{Eulerian tidal bias parameter $b_{K^2}$, measured for FOF-identified dark matter halos, as a function of their linear Eulerian bias $b_1$. The points show the measurements from \cite{saito/etal:14} at the redshifts indicated, while the dashed line shows the fitting formula derived by \cite{modi/castorina/seljak:2016} from their measurements (at $z=0$ and $z=1$) using halo moments in Lagrangian space, and converted to the large-scale Eulerian bias parameter. The solid line shows the prediction of the Lagrangian LIMD ansatz, $b_{K^2} = (-2/7)(b_1-1)$ [\refsec{evol1}].
}
\label{fig:bK2(b1)}
\end{figure}

Let us now briefly discuss the significance of the numerical results obtained for $b_{K^2}$ and $b_{\otd}$.
Early studies of $b_{K^2}$ found consistency with the Lagrangian \LIMD prediction of $b_{K^2} = -2/7 (b_1-1)$.  
This includes Refs. \cite{baldauf/etal:2012}, who used Eulerian measurements of $B_{mmh}$, the
same observable used by \cite{chan/scoccimarro/sheth:2012}.  
The latter, however, found evidence for a departure from the Lagrangian 
\LIMD prediction at low halo mass, with $b_{K^2} > -2/7(b_1-1)$.  
Similar deviations were also reported for low mass halos in \cite{bel/hoffmann/gaztanaga:15} from an analysis of the 
three-point correlation function. Importantly however, these authors obtained $b_{K^2} < -2/7(b_1-1)$.  
By contrast, the results of \cite{sheth/chan/scoccimarro:2012}, obtained using the 
Lagrangian bispectrum [\refeq{PhhL}], indicate a departure from Lagrangian \LIMD 
at high masses, whereas \LIMD\ provides a good fit at low masses.
Subsequently, Ref.~\cite{saito/etal:14} presented combined constraints from the cross-power spectrum $P_{hm}$ and bispectrum $B_{mmh}$, which are shown in \reffig{bK2(b1)}. While they follow the general trend of the Lagrangian LIMD prediction, there are some indications for a deviation at lower values of $b_1$, corresponding to a Lagrangian tidal bias $b_{K^2}^L < 0$ for low-mass halos.
These results were very recently confirmed by two studies \cite{lazeyras/schmidt:2018,abidi/baldauf:2018} who used the cross-correlation of halos with quadratic operators constructed from the smoothed density field, first proposed by \cite{schmittfull/etal}.
Another recent study \cite{castorina/paranjape/etal:2016} measured $b_{K^2}^L$ directly using the method described in
\refsec{bmom} (see \refsec{bNmL} for the theoretical background), and found results consistent with Lagrangian \LIMD,
albeit with large error bars.    
Finally, Ref.~\cite{modi/castorina/seljak:2016} presented different estimates of the moments bias $b_{K^2}^{\text{m},L}$ using halos in Lagrangian space. Their ``PBS estimator'' corresponds to the scatter-plot method described in \refsec{bscatter} applied in Lagrangian space using a cell size $R_\ell \approx 150 \Mpch$, which strongly suppresses higher-order corrections. Note that this yields the moments biases  in our notation, which they correct to obtain the large-scale $b_{K^2}^L$ using the results of \refsec{bmom}.
They found strong evidence for $b_{K^2}^L\ne 0$, with $b_{K^2}^L$
positive (negative) for low (high) mass halos. The fitting function for $b_{K^2}(b_1)$ they obtain from their measurements is also shown in \reffig{bK2(b1)}. There is significant disagreement with the results from the halo bispectrum by \cite{saito/etal:14}. This disagreement, which could possibly be due to the nontrivial conversion from the measured halo moments to the large-scale bias $b_{K^2}$, clearly warrants further investigation.

Only few measurements of the cubic-order bias parameter $b_{\otd}$ have been reported
\cite{saito/etal:14,angulo/etal:2015}, which were based on the NLO halo power 
spectrum.  
Ref.~\cite{saito/etal:14} found broad consistency with the Lagrangian \LIMD prediction 
for the combination [see \refeq{b3nl}]
\be
b_{3\rm nl} \equiv -\frac{32}{21} \left( b_{K^2} + \frac25 b_{\otd} \right) 
\stackrel{\text{Lagr. \LIMD}}{=} \frac{32}{315} b_1^L\,,
\ee
which is proportional to the combination of bias parameters that multiplies
$f_\NLO(k) \Plin(k)$ in the NLO halo power spectrum [\refeq{Phm1l}].  
The second equality follows from the Lagrangian \LIMD prediction for $b_{K^2}$ and 
$b_\otd$. Note, however, that the constraints on the combination $b_{3\rm nl}$ from \cite{saito/etal:14} are based on its contribution to the NLO halo power spectrum, and are thus strongly degenerate with the contribution from the higher-derivative bias $b_{\lapl\d}$ (see \reffig{Pkg_1loop}), for which Ref.~\cite{saito/etal:14} assume $b_{\lapl\d}=0$. Thus, further measurements are needed to break the degeneracy between $b_{3\rm nl}$ (or $b_{\otd}$) and $b_{\lapl\d}$.

\begin{table*}
\centering
\begin{tabular}{cccclc}
\hline
Bias & & Fitting formula & Mass range & Halo finder & Reference \\
\hline
$b_1$ & $=$ & $ 1-A \nu_c^\alpha [\nu_c^\alpha+\dc^\alpha]^{-1} + B\nu_c^\beta + C\nu_c^\gamma$ & $-0.4 \lesssim \nu_c \lesssim 0.6$ & SO
& \cite{tinker/etal:2010} \\
& & $A = 1+ 0.24 \,y\, e^{-(4/y)^4},\,\,\alpha = 0.44y-0.88$ & \\
& & $B = 0.183,\,\,\beta = 1.5$ & \\
& & $C = 0.019 + 0.107 y + 0.19 e^{-(4/y)^4},\,\,\gamma=2.4$ & \\[2pt]
\hline
$b_2$ & $=$ & $0.412-2.143\,b_1+0.929\,(b_1)^2+0.008\,(b_1)^3$ & $1 \lesssim b_1 \lesssim 9$ & SO & \cite{lazeyras/etal} \\[2pt]
\hline
$b_3$ & $=$ & $-1.028+7.646\,b_1-6.227\,(b_1)^2+0.912\,(b_1)^3$ & $1 \lesssim b_1 \lesssim 9$ & SO & \cite{lazeyras/etal} \\[2pt]
\hline
\end{tabular}
\caption{
Published fitting formulas for bias parameters of dark matter halos in $\Lambda$CDM N-body simulations. These phenomenological fitting functions should only be trusted within the mass range over which they were calibrated on simulations; the respective approximate range, in terms of $\nu_c$ or $b_1$, is indicated. 
The third column indicates how halos were identified (\refapp{halofinder}): using spherical overdensity (SO) or friends-of-friends (FOF) algorithms.
In \cite{tinker/etal:2010}, halos are identified using a SO
criterion $\Delta_\text{SO}$ defined with respect to the background matter 
density, where $y=\log_{10}\Delta_\text{SO}$, $\nu_c = \dc/\sigma(M)$, and 
$\dc = 1.686$ is fixed independently of the background cosmology; Ref.~\cite{lazeyras/etal} use an SO finder with $\Delta_\text{SO}=200$. 
Note that the fitting formulas for $b_2(b_1),\  b_3(b_1)$ provided in Refs.~\cite{hoffmann/bel/gaztanaga:2015,hoffmann/bel/gaztanaga:2016}
agree with those of \cite{lazeyras/etal} to well within the error bars. We choose the latter here as it was calibrated over a larger range of $b_1$.
\label{tab:fittingbiases}}
\end{table*}

Turning to the higher-derivative bias $b_{\lapl\d}$, the first constraints
have been placed by studies testing the scale dependence of bias on large scales 
\cite{scherrer/weinberg:1998,mann/peacock/heavens:1998}.  
More recent measurements include those of \cite{fujita/etal:2016}.  
For their mass bin ``2'', which follows the definition of
Table~1 in \cite{okumura/seljak/desjacques:2012}, 
they obtain a value of $b_{\lapl\d}$ that
is of order $3[R(M)]^2$, although $b_{\lapl\d}=0$ is only ruled out at the
$\sim 1.3\sigma$ level, and no constraint is found for the other mass bins.  
On the other hand, Ref.~\cite{angulo/etal:2015} quote values for $b_{\lapl\d}$
that are much smaller than $[R(M)]^2$ when using either of the two
different definitions of $b_{c_s}$ given there [Eq.~(18) and Eq.~(31)].  
Thus, there is still large uncertainty in the magnitude of higher-derivative
biases for halos. Note that there is a strong degeneracy in shape between the contribution from $b_{\lapl\d}$ and the NLO term controlled by $b_{K^2} +(2/5) b_{\otd}$ (e.g., \cite{assassi/etal}; see \reffig{Pkg_1loop}). Thus, without an independent constraint on (or assumption about) the third-order bias parameter $b_{\otd}$, the higher-derivative bias is very difficult to extract from the halo power spectrum.

Higher-derivative biases can be measured more easily in Lagrangian space,
using either the halo-matter power spectrum 
\cite{elia/ludlow/porciani:2012,baldauf/desjacques/seljak:2015} or the projection
method of \cite{paranjape/sefusatti/etal:2013,biagetti/chan/etal:2014} discussed near the end of \refsec{bmom}.  
In particular, \cite{elia/ludlow/porciani:2012,baldauf/desjacques/seljak:2015} 
measured the so-called peak bias $b_{01}$, which contributes to $-b_{\lapl\d}^L$
along with the leading contribution from the filtering kernel [\refeq{bhderiv_thr} in \refsec{higherderiv}], and obtained a contribution $b_{01} \approx 2 [R(M)]^2$ for halos with mass $M\geq 8\times 10^{12}\hmsun$, with only a weak 
departure from the simple $[R(M)]^2$ scaling with mass.  
The negative sign of this contribution to $b_{\lapl\d}^L$ is the opposite of what simple filtering yields, but is expected if halos collapse from Lagrangian 
patches near initial density peaks \cite{desjacques:2013} (\refsec{PBSpeaks}).  
However, due to the impact of halo velocity bias on the evolution of higher-derivative biases, it is 
not possible to relate Lagrangian higher-derivative biases to their Eulerian counterpart without using a model for the amplitude and time evolution of velocity bias \cite{desjacques/crocce/etal:2010,baldauf/desjacques/seljak:2015,chan:2015}
(see the discussion in \refsec{velbias}).  Further, 
Ref.~\cite{biagetti/chan/etal:2014} measured $b_{(\vn\d)^2}^L\sim -R(M)^{3.5}$ (which corresponds to $\chi_1$ in their notation).  

We summarize the most precise recently published fitting functions for the leading halo bias parameters in \reftab{fittingbiases}.  

\subsubsection{Galaxy bias}

As we have discussed in \refsec{history}, the linear bias parameter $b_1$
has been measured many times in galaxy surveys over the past 50 years.  
Here, we briefly review the measurements of nonlinear bias parameters from higher-order 
correlation functions.  
Ref. \cite{scoccimarro/etal:2000} and \cite{verde/etal:2001} have measured 
the linear and second-order \LIMD bias parameters from galaxies
in, respectively, IRAS (Infrared Astronomical Satellite) and 2dFGRS (2dF Galaxy 
Redshift Survey), using the bispectrum (they set $b_{K^2}=0$).  
The results for $b_2$, a marginally detected
value of $b_2 \approx -0.3$ for IRAS and a result consistent with
zero for 2dFGRS, are
broadly consistent with the expectation for dark matter halos of the
same linear bias $b_1$ ($b_1 \approx 0.76$ and $1.0$, respectively).  
Later, Ref.~\cite{nishimichi/etal:2007} showed that the relation
between $b_1$ and $b_2$ predicted by the PBS relation \refeq{bPBSuniv} 
for halo mass functions can 
explain the amplitude of the galaxy bispectrum for equilateral configurations 
extracted from SDSS data.  The three-point correlation function of the SDSS luminous red galaxy (LRG) sample was measured by \cite{marin:2011}. 
Ref.~\cite{marin/etal:2013} measured $b_2$ from 
the three-point correlation function of the WiggleZ survey.  
For the recent SDSS-III CMASS sample, Ref.~\cite{gil-marin/etal:2015} have constrained $b_2$ 
by combining the power spectrum and bispectrum, while Ref.~\cite{slepian/etal:2017} used the three-point function.  Further, Ref.~\cite{chiang/etal:2015} employed the position-dependent correlation function 
\cite{chiang/etal:2014}, which corresponds to an integral over the three-point
function in the squeezed limit, to constrain $b_2$ for the same galaxy sample.  
All three references emphasize that, unfortunately, these estimates are still limited 
by residual systematic uncertainties (in particular due to redshift-space distortions) in the model that is used to fit the bias parameters.  
A related method to estimate the bispectrum in the squeezed configuration has been applied to the cross-correlation of the Lyman-$\alpha$ 
forest power spectrum and CMB lensing \cite{doux/etal:2016} to constrain the bias 
parameters $b_1$ and $b_2$ of the Lyman-$\alpha$ forest.

\subsubsection{Stochasticity}
\label{sec:meas:stoch}

The deterministic bias contributions discussed so far in this section
are only a subset of the complete bias expansion, \refeq{deltag2} up to third order, which also contains stochastic terms.
We now discuss these stochastic contributions, restricting ourselves to the results obtained for halos in simulations.  
The qualitative picture also applies to galaxies;  we will return to
additional stochastic effects which appear for galaxies in \refsec{HOD}.  
We begin with the most frequently studied contribution $P_\eps(k) = \<\eps(\vk)\eps(\vk')\>'$.  
If halos are a perfect Poisson sample with a mean comoving density (per
logarithmic mass interval) $\avnh$, then $P_\eps(k) = \Peps = 1/\avnh$.  
Moreover, under this 
assumption the shot noise of halos of different mass is uncorrelated.  
One can succinctly summarize this prediction as \cite{schmidt:2016a}
\be
P_\eps(k;M,M') \equiv \<\eps(\vk,M)\eps(\vk',M')\>' \stackrel{\text{Poisson}}{=} \frac{\d_D(\ln M-\ln M')}{\avnh(M)}\,.
\label{eq:Pepspoisson}
\ee
However, this is only approximately correct for actual halos
\cite{CasasMiranda:2002on,smith/etal:2007,hamaus/seljak/etal:2010,gil-marin/etal:2010,manera/gaztanaga:2011}.  
On scales much smaller than the Lagrangian radii $R(M),\,R(M')$ of the halos, 
one expects \refeq{Pepspoisson} to be accurate, since the likelihood of a halo center being located within 
a given cell is then a true Poisson process.  
Thus, \refeq{Pepspoisson} is expected to hold for $k\gg [R(M)+R(M')]^{-1}$ \cite{schmidt:2016a}.  
On the other hand, on large scales $k \ll [R(M)+R(M')]$ the halo model paradigm (see \cite{cooray/sheth} for a comprehensive overview) provides 
a different constraint.  
If we make the assumption that all matter is contained within virialized halos, then mass and momentum 
conservation of matter imply that the matrix $\Peps(M,M')=\lim_{k\to 0} P_\eps(k;M,M')$, which quantifies the halo stochasticity in the large-scale limit, has a zero eigenvalue, whose eigenvector 
corresponds to weighting each halo by mass 
\cite{hamaus/seljak/etal:2010,schmidt:2016a}.  This is because any stochastic contribution to the nonlinear matter power spectrum has to scale as $k^4$ on large scales \cite{abolhasani/mirbabayi/pajer:2016}.  
Thus, within the halo model framework, the mass-weighted power spectrum of all halos enjoys zero stochasticity \cite{seljak/hamaus/desjacques:2009}.  
In the presence of a diffuse matter component that is not associated with halos of any mass, this constraint does not need to hold exactly.  
No further robust constraints on $P_\eps(k;M,M')$ are known beyond these two limiting cases, although, as mentioned above, the transition between the two limits is expected to happen at the Lagrangian radius $k \sim R^{-1}(M)$ of halos, i.e. the same scale that determines the amplitude of higher-derivative biases.   
Hence, one expects that 
\be
\left|\Plapleps \right| \sim [R(M)]^2 \Peps\,,\quad\mbox{where}\quad
\Plapleps \equiv \left.\frac{\partial P_\eps(k)}{\partial k^2}\right|_{k=0} \,,
\ee
keeping in mind the caveat about possible contributions that are controlled by $\knl$ instead of $R(M)$ (\refsec{stoch}). 
Importantly, beyond the large-scale limit, measurements of stochasticity depend on the precise definition 
of the ``deterministic'' part of halo clustering, for example whether NLO corrections which involve 
higher-order biases are included following \refsec{npt1loop}. The reason is that higher-order terms which are not accounted for in the deterministic model can contribute to the inferred stochastic contribution 
(e.g. \cite{heavens/matarrese/verde:1998,smith/etal:2007,baldauf/schaan/zaldarriaga:2016}).  
Finally, an upper bound can be placed on the leading stochastic contribution to the halo-matter cross-power spectrum [\refeq{Phm1l}] by assuming that the fields $\eps$ (halos) and $\eps_m$ (matter) are perfectly correlated. This leads to
\be
\big|P_{\eps\eps_m}^{\{2\}}\big| \leq \sqrt{P_{\eps}^{\{0\}} P_{\eps_m}^{\{4\}}}\,,
\ee
where $P_{\eps_m}^{\{4\}}$ is the leading stochastic contribution to the matter power spectrum in the EFT approach (\refapp{EFT}).

Ref.~\cite{hamaus/seljak/etal:2010} presents a detailed simulation study of the 
halo stochasticity covariance, which they estimated as
\be
\hat P_\eps(k;M,M') \equiv \Big\< \left[\d_h(\vk;M) - b_1(M) \d_m(\vk)\right]
\left[\d_h(\vk';M') - b_1(M') \d_m(\vk')\right]\Big\>'\,,
\label{eq:Pepshamaus}
\ee
by dividing the halo mass range ($\sim 10^{13} - 3\cdot 10^{15}\Msunh$) into 10 bins of equal number density.  
The linear bias was determined using $P_{hm}(k)$ for $k < 0.024 \iMpch$.  
Interestingly, they do find an eigenvalue of $\hat P_\eps(k;M,M')$ that is significantly lower than $1/\avnh$, 
whose eigenvector is close to mass-weighting, as expected from the halo model argument made above.  
This is also reflected in the components of $\hat P_\eps(k;M,M')$ at high masses, which are lower than the Poisson 
expectation.  Moreover, there is one eigenvalue that is significantly larger than $1/\avnh$.  
The components of the associated eigenvector appear to be close to the quadratic bias $b_2(M)$.  
The other eigenvalues were found to be consistent with $1/\avnh$.  All these results were found to be 
roughly scale independent for $k \lesssim 0.2\iMpch$.  The structure of the
halo stochasticity matrix is relevant for methods that use multiple, weighted tracers within the same volume to reduce the sample variance and shot noise in constraints of scale-dependent bias from primordial non-Gaussianity (see \refsec{NG:mt}).  

All these features can be qualitatively understood through a toy model of the clustering of proto-halos in Lagrangian space, combining exclusion at small-scale $r<R(M)$, and nonlinear bias at larger scales $r>R(M)$ \cite{baldauf/seljak/etal:2013}.  
In this ansatz, including quadratic \LIMD bias, the stochastic contribution to the proto-halo power spectrum in Lagrangian space takes the form
\be
\label{eq:stochtoymodel}
P_\eps(k, M)
\stackrel{k\to 0}{=} \frac{1}{\avnh} + \frac12 (b_2^L)^2 \int_{R(M)}^\infty\! d^3r \, [\xi_{\rm L}(r)]^2
- (b_1^L)^2 \int_0^{R(M)}\!d^3r\, \xi_{\rm L}(r)  - V_\text{excl} \;,
\ee
where the second-term on the right-hand side is the $k\to 0$ limit that is
subtracted in $\mathcal{I}^{[\d^2,\d^2]}$, and the exclusion
volume is $V_\text{excl}=4\pi R^3(M)/3$.  Note that while \refeq{stochtoymodel}
is given in Lagrangian space, the constant stochastic contribution to $P_{hh}(k)$ in the large-scale limit trivially transforms to Eulerian space, since
$\d_h^E = \d_h^L + \d$ at linear order.  
\refeq{stochtoymodel} clearly shows that exclusion lowers the stochasticity
(whence the low eigenvalue $<1/\avnh$), whereas nonlinear bias enhances it (whence the large eigenvalue $>1/\avnh$).
Further, the scale dependence of the stochastic contribution is controlled by the scale $R(M)$.   
Quantitative agreement is difficult to obtain with such perturbative toy models because stochastic contributions 
to the low-$k$ power spectrum are genuinely non-perturbative as discussed in \refsec{stoch} (see \cite{baldauf/codis/etal:2015} for a 
non-perturbative, albeit one-dimensional approach).  

We finally briefly discuss theoretical expectations for the other two 
stochastic amplitudes $\Pepsepsd$ and $\Beps$ that have appeared in 
our discussion of halo $n$-point functions.  First, for Poisson shot noise,
the skewness is given by $\Beps = (\avnh)^{-2}$; more specifically, if
the stochasticity in halo counts of different mass is independent, we have \cite{schmidt:2016a}
\be
B_\eps(k;M,M',M'') \equiv \<\eps(\vk,M)\eps(\vk',M')\eps(\vk'',M'')\>' 
\stackrel{\text{Poisson}}{=} \frac{\d_D(\ln M - \ln M') \d_D(\ln M-\ln M'')}{[\avnh(M)]^2}\,.
\label{eq:Bepspoisson}
\ee
Corrections to the Poisson bispectrum from exclusion and nonlinear bias are of similar magnitude as in the case of $P_\eps$
\cite{ginzburg/desjacques/chan:2016}.
Under the same assumptions as made for \refeq{Pepspoisson} and \refeq{Bepspoisson}, the cross-correlation of $\eps$ and $\eps_\d$ for halos is given by
\cite{pollack/smith/porciani:2012,kayo/takada/jain:2013,schmidt:2016a}
\be
P_{\eps\eps_\d}(k; M,M') \equiv \< \eps(\vk,M) \eps_\d(\vk',M') \>' \stackrel{\text{Poisson}}{=} \frac12 b_1(M) \frac{\d_D(\ln M - \ln M')}{\avnh(M)}\,.
\ee
The interpretation of this result is quite simple:  a large-scale density perturbation $\d$ modulates the local halo abundance, with $\d_h = b_1 \d$ at linear order, and correspondingly modulates the shot noise amplitude.\\

At this point, it is worth making the connection to some 
previous references \cite{tegmark/peebles:1998,tegmark/bromley:1999,taruya/soda:1999} which, instead of the relation \refeq{Phh}, included the stochasticity 
by defining a scale-dependent bias $\check{b}_1(k)$ and correlation coefficient $\check{r}_{gm}(k)$,
\be
P_{hh}(k) = [\check{b}_1(k)]^2 P_{mm}(k); \quad P_{hm} = \check{b}_1(k) \check{r}_{gm}(k) P_{mm}(k)\,,
\label{eq:brTP}
\ee
where in some references the Poisson expectation is subtracted from the halo auto-power spectrum $P_{hh} \to P_{hh}^\text{p.sub.}(k) \equiv P_{hh}(k) - \avnh^{-1}$.  
The real-space version of \refeq{brTP} is adopted in \cite{taruya/etal:2001,sato/matsubara:2013}.  
These can be related to the quantities defined in \refeq{Phh} via \cite{cresswell/percival:2009}
\be
\check{b}_1(k) = b_1 \sqrt{1 + \Peps/[b_1^2 P_{mm}(k)]}\: ; \quad
\check{r}_{gm}(k) = \frac{b_1}{\check{b}_1(k)}\,.
\ee
In the Poisson-subtracted case, $\Peps$ should be replaced with $\Peps-\avnh^{-1}$ here.  
Ref.~\cite{taruya:2000} includes one-loop nonlinear evolution in this parametrization.  
Ref.~\cite{tegmark/bromley:1999} estimated $\check{b}_1(k)$ and $\check{r}_{gm}(k)$
from SDSS galaxies on large scales, finding significant departures from the Poisson expectation.  
However, Ref.~\cite{blanton:2000} pointed out that this could be caused by errors in the selection 
function.  The author estimated the correlation coefficient to be larger than 0.9; a similarly high 
value $> 0.95$ was found by \cite{wild/etal:2005} for 2dF galaxies using counts-in cells.  

The parametrization \refeq{brTP} does not cleanly distinguish between the shape of the matter 
power spectrum on large scales and the small-scale physics of halo formation, which are included 
in \refeq{Phh} through the \emph{constants} $b_1,\,\Peps$.  Equivalently, the deterministic and 
stochastic (noise) contributions are not cleanly separated.  Hence, \refeq{Phh} is preferred in 
the modern literature.

\subsection{Halo assembly bias}
\label{sec:assembly}

So far, we have only considered the dependence of the halo bias parameters on mass and redshift.  However, it is now well established that halo bias depends on various other properties of halos at fixed mass and redshift.  These dependencies have come to be known summarily as \emph{assembly bias}.  This section provides a brief overview of these results.    

\begin{figure}[t!]
\centering
\includegraphics[width = 0.6\textwidth]{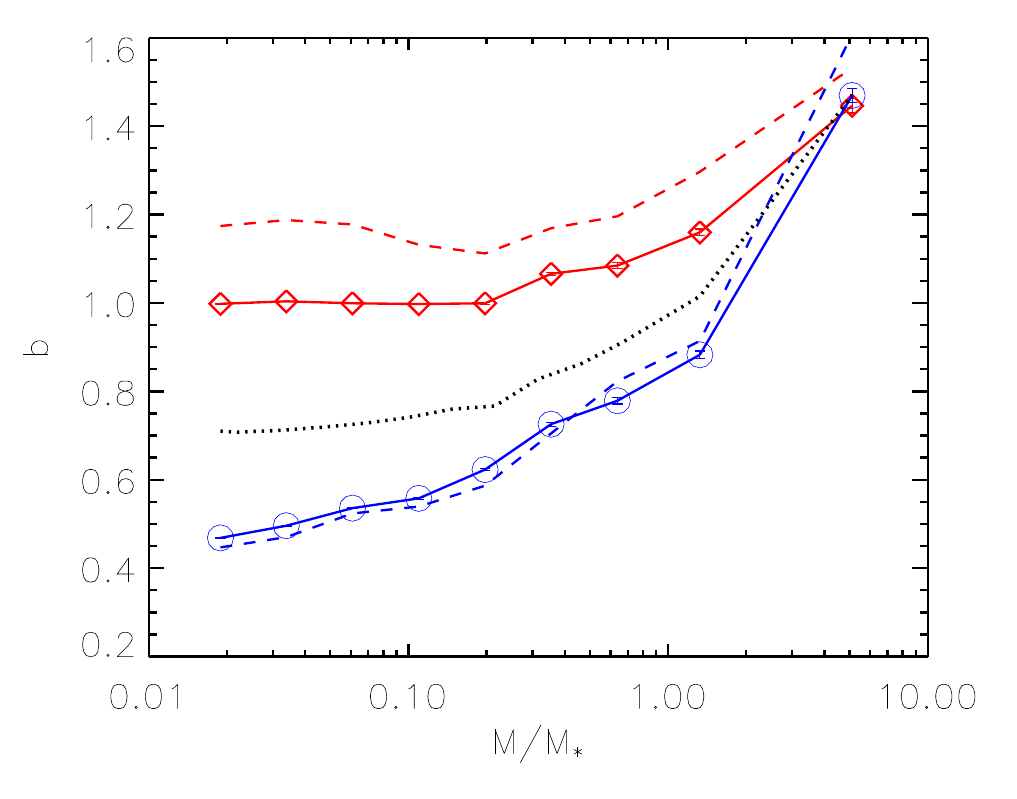}
\caption{Bias of halos identified at $z=0$, as a function of mass and formation redshift, i.e. the redshift when the most massive progenitor halo reaches 1/2 of the final halo mass.  The dotted black curve
represents the mean bias in a given mass bin, whereas the solid red and blue curve show results
for the 20 percent oldest and youngest halos, respectively. Similarly, the red and blue dashed
curves refer to the upper and lower 10th percentile. Note that the halo mass is in units of $M_\star$ [\refeq{Mstardef}].  
\figsource{gao/springel/white:2005}}
\label{fig:assembly_bias}
\end{figure}

While the first numerical studies did not provide any conclusive evidence for a dependence of 
halo clustering on additional properties \cite{lemson/kauffmann:1999,percival/scott/etal:2003},
Refs.~\cite{sheth/tormen:2004,avila-reese/colin/etal:2005,harker/cole/etal:2006} showed that, at fixed halo mass and redshift, halos in dense regions 
collapse at a slightly higher redshift than in underdense environments, 
where the local environment density was defined as the matter density in spheres of radius $R_\intscale \sim$~few$\hmpc$ centered on the halo. The local environment density is expected to correlate with the halo formation time, and indeed Ref.~\citep{gao/springel/white:2005} found a dependence of halo bias, at fixed mass, on the halo formation time, as shown in \reffig{assembly_bias}. As the halo profile shape or concentration also correlates with the halo assembly history \citep{wechsler/zentner/etal:2006,ludlow/etal:2014}, assembly bias has also been found in terms of concentration. Assembly bias has since been found in several other halo properties, and most likely cannot be explained completely by the halo formation history \cite{mao/zentner/wechsler:2017}. We now provide an overview of the trends found in the literature to date, before discussing implications for models of halo biasing:

\begin{itemize}
\item \emph{Concentration:} this parameter quantifies the shape of the spherically averaged halo profile. It is determined either by fitting a profile of the NFW form \cite{NFW} to the measured halo profile, or by matching the measured maximum circular velocity of the halo to the NFW profile prediction. The latter estimate is sensitive to substructure, and less reliably related to the halo profile shape.
\item \emph{Formation redshift:} this is commonly defined as the epoch, phrased in terms of cosmological redshift, at which the main progenitor of the halo assembled 50\% of the final halo mass. A large formation redshift thus corresponds to early-formed, or old, halos.
\item \emph{Lagrangian slope:} defined as $d\< \d^{(1)}\>_\text{proto-halo}/dM$, this is measured in \cite{dalal/white/etal:2008} by estimating the overdensity of the Lagrangian proto-halo patch, and taking its derivative with respect to mass using the measured halo assembly history (i.e., removing particles first that have joined the halo last). 
\item \emph{Spin parameter:} a measure of the angular momentum of halos, this is defined as $\lambda = |\v{J}|/(\sqrt{2} M_{\Delta} V_{\Delta} R_{\Delta})$, where $R_\Delta,\,M_\Delta$ denote the radius and mass within which the mean density reaches $\Delta$ times the background matter density $\rhob$, and $V_{\Delta}= (G M_\Delta/R_\Delta)^{1/2}$ is the circular velocity at $R_{\Delta}$.
\item \emph{Halo sphericity:} this parameter is defined as the square-root of the ratio of smallest to largest elements of the diagonalized inertia tensor within $R_{\Delta}$.
\item \emph{Velocity anisotropy:} the parameter $\beta$ is defined as $\beta = 1-\s_{v,t}^2/(2\s_{v,r}^2)$, where $\s_{v,t}^2,\,\s_{v,r}^2$ are the mean tangential and radial velocity dispersions within the halo, respectively. A value of $\beta=0$ corresponds to an isotropic velocity dispersion; positive values correspond to preferentially radial orbits, while negative values imply more circular orbits.
\item \emph{Subhalo mass fraction:} this is defined as the fraction of the total mass within $R_\Delta$ which is in the form of self-bound substructures. Similar trends hold for the mass fraction in the main subhalo \cite{gao/white:2007}. Other subhalo-related properties are the number of subhalos and their mean separation from the center-of-mass of the main halos \cite{mao/zentner/wechsler:2017}.
\item \emph{Late-time accretion rate:} this is defined as $d\ln M_{\Delta}/d\ln a = -d\ln M_{\Delta}/dz$ by following the halo's main progenitor. 
\end{itemize}

\begin{table}[t]
\centering
\begin{threeparttable}[b]
\begin{tabular}{l|c|c}
\hline
\hline
Halo property & Low-mass halos & High-mass halos \\
\hline
Concentration\tnote{1}~~\cite{wechsler/zentner/etal:2006,wetzel/cohn/etal:2007,gao/white:2007,jing/suto/mo:2007,angulo/baugh/lacey:2008,faltenbacher/white:2009,paranjape/padmanabhan:2016,lazeyras/musso/schmidt:2016,mao/zentner/wechsler:2017}
 & $\aup$ & $\adown$ \\
Formation redshift \cite{gao/springel/white:2005,wechsler/zentner/etal:2006,wetzel/cohn/etal:2007,gao/white:2007,jing/suto/mo:2007,mao/zentner/wechsler:2017}
 & $\aup$ & $\anone$ \\
Lagrangian slope \cite{dalal/white/etal:2008}
 & & $\aup$ \\
Spin parameter \cite{gao/white:2007,faltenbacher/white:2009,lazeyras/musso/schmidt:2016,mao/zentner/wechsler:2017}
 & $\aup$ & $\aup$ \\
Sphericity\tnote{2}~~\cite{faltenbacher/white:2009,lazeyras/musso/schmidt:2016}
 & $\aup$ & $\aup$ \\
Subhalo mass fraction\tnote{3}~~\cite{gao/white:2007,mao/zentner/wechsler:2017}
 & $\adown$ & $\aup$ \\
Velocity anisotropy $\beta$ \cite{faltenbacher/white:2009} 
 & $\adown$ & $\adown$ \\
Late-time accretion rate\tnote{4}~~\cite{lazeyras/musso/schmidt:2016,mao/zentner/wechsler:2017}
 & & $\adown$ \\
\hline
\hline
\end{tabular}
{\footnotesize
\begin{tablenotes}
\item[1] Note that different definitions of concentration are used in the literature. Refs.~\cite{wechsler/zentner/etal:2006,wetzel/cohn/etal:2007,jing/suto/mo:2007,paranjape/padmanabhan:2016,mao/zentner/wechsler:2017} use that obtained from a fit to the density profile, while Refs.~\cite{gao/white:2007,angulo/baugh/lacey:2008,faltenbacher/white:2009,lazeyras/musso/schmidt:2016} use that obtained from the maximum circular velocity.
\item[2] Similar trends are found for the sphericity of the velocity field within halos \cite{faltenbacher/white:2009}.
\item[3] Ref.~\cite{mao/zentner/wechsler:2017} use the number and average separation of subhalos for high-mass halos, and find the same trend.
\item[4] Ref.~\cite{lazeyras/musso/schmidt:2016} finds trends that are not strictly monotonic.
\end{tablenotes}
}
\end{threeparttable}
\caption{Overview of assembly bias trends that have been reported in the literature. An upward pointing arrow means that bias increases with increasing value of the given property, while a downward pointing arrow denotes a decreasing bias. A horizontal arrow indicates absent or weak trends, while no symbol indicates no known constraints. ``Low-'' and ``high-mass'' halos refer to halos with $M \ll M_\star$ and $M \gg M_\star$, respectively.}
\label{tab:assembly}
\end{table}

\reftab{assembly} presents a summary of the trends with the properties listed above which have been reported in the literature. We split the trends into those for low-mass and high-mass halos, distinguished through the characteristic mass $M_\star(z)$ of halos collapsing at redshift $z$, defined through
\be
\s(M_\star, z) = \dc\,.
\label{eq:Mstardef}
\ee
We focus here on the trends in the linear bias $b_1$; only Refs.~\cite{angulo/baugh/lacey:2008,paranjape/padmanabhan:2016,lazeyras/musso/schmidt:2016} have measured assembly bias in higher-order \LIMD bias parameters. They found the same qualitative trends also in the higher-order biases. Most results presented in \reftab{assembly} are for spherical-overdensity identified halos with $\Delta=200$; those of \cite{faltenbacher/white:2009} are for FoF halos. Recently, Ref.~\cite{villarreal/etal:2017} explored the dependence of assembly bias on the halo definition employed, and found that, while assembly bias can be mitigated by a judicious choice, no single halo definition can absorb all assembly bias effects.
Clearly, the trends shown in \reftab{assembly} are only a subset of all possible correlations of bias with halo properties. For example, Ref.~\cite{salcedo/etal:2017} very recently showed a significant trend with the distance to the nearest more massive halo (``neighbor bias'').

Despite the fact that the reported results in the literature are in good agreement overall, an important caveat to the comparison of different results is that they do not strictly have to agree, since the bias has been estimated using different approaches and different scales. Refs.~\cite{gao/springel/white:2005, wechsler/zentner/etal:2006, gao/white:2007, dalal/white/etal:2008, faltenbacher/white:2009,mao/zentner/wechsler:2017} used the halo correlation function on scales in the range $10-60 \Mpch$, while \cite{angulo/baugh/lacey:2008} used moments. None of these references have converted their results to the bias parameters strictly in the large-scale limit. 
Ref.~\cite{jing/suto/mo:2007} employed the halo-matter cross-power spectrum on scales $k \lesssim 0.1 \iMpch$, which, following the discussion in \refsec{bnpt}, should fairly accurately measure the large-scale linear bias. Finally,
Refs.~\cite{paranjape/padmanabhan:2016,lazeyras/musso/schmidt:2016} used the response approach based on separate-universe simulations. These strictly recover the large-scale bias parameters. Note that Ref.~\cite{sunayama/etal:2016} recently reported strongly scale-dependent features of assembly bias in the halo correlation function, although these are restricted to very small scales $r \lesssim 10 \Mpch$. 

For low-mass halos, late-forming (younger) halos have higher concentration, but lower bias than the early-forming (older), more concentrated halos.  This trend for assembly bias with respect to concentration reverses for high-mass halos. This reversal has so far not been conclusively detected for the assembly bias with respect to formation time. The trend with concentration at high masses 
can be explained within the excursion set approach, as discussed in \refsec{assembly_exset}.  
Note that this requires going beyond the canonical formulation of the excursion set theory based 
on sharp $k$-space filters \citep{bond/etal:1991,lacey/cole:93}, in which the fluctuations that 
collapse to form halos are statistically independent of the larger scales, i.e. their environment.

Another interesting trend is the dependence on the slope of the Lagrangian density profile of halos. Ref.~\cite{dalal/white/etal:2008} found that high-mass halos with steeper (more negative) slopes are less biased than those with shallower slopes. If shallower slopes are associated with lower final concentrations, then this trend matches that found for concentrations. On the other hand, if one relates the Lagrangian slope to the late-time accretion rate of halos, where shallower (less negative) slopes correspond to higher accretion rates \cite{lacey/cole:93}, the halo assembly bias found with respect to Lagrangian slope has the \emph{opposite} sign of that found with respect to direct measurements of the accretion rate \cite{lazeyras/musso/schmidt:2016}. 

At low mass ($M\lesssim M_\star$), the trend with concentration reverses sign, so that the more concentrated halos are clustered more strongly. In this regime, one also observes a strong trend with formation time, with early-forming halos being more clustered.
This low-mass trend is illustrated in \reffig{assembly_bias} where halos of a given mass are 
split according to their formation redshift. The effect becomes very significant for $M\lesssim M_\star$.  
Refs.~\citep{wang/mo/jing:2007,diemand/kuhlen/madau:2007,more/diemer/kravtsov:2015} proposed that the assembly bias seen at 
low mass originates from tidal interactions with a larger neighbor. 
The mass accretion onto these ``harassed'' or ``backsplash'' low-mass halos is suppressed at late time by the tidal field of massive neighboring halos. This tidal stripping significantly affects the formation history of many field halos. The detailed zoomed simulation study of \cite{borzysz/etal:2017}, as well as the correlation of halo bias with the tidal field (estimated on a scale of a few halo virial radii) found by \cite{paranjape/hahn/sheth:2017}, support this picture.

Observationally, constraining halo assembly bias is a difficult measurement, as one needs to identify at least two observable proxies, one for halo mass as well one that correlates with halo formation time or profile shape at fixed mass. The mean halo mass of a given sample can be constrained by gravitational lensing. Various detections have been claimed in the literature \cite{yang/etal:2006,wang/etal:2013,miyatake/etal:2016, more/etal:2016}, although all of them have been disputed \cite{lin/etal:2016,zu/etal:2016}, due to different projection and selection effects which mimic the assembly bias signature. The upper limits placed by \cite{lin/etal:2016,zu/etal:2016} are still completely consistent with the level of assembly bias expected in the standard $\Lambda$CDM cosmology.

Finally, it is worth noting that assembly bias is an important effect in phenomenological approaches to galaxy clustering that populate halos with galaxies; the most well-known approaches, which we will review in \refsec{HOD},  are halo occupation distributions and sub-halo abundance matching. This is because certain types of galaxies might preferentially reside in, say, early-forming halos within a given mass range. Then, their clustering will not follow those of all halos of this mass, but only a special subset which, due to assembly bias, has different clustering properties \cite{croton/gao/white:2007,zentner/etal:2014,hearin:2015}. We will return to this point in \refsec{HOD}, but emphasize already here that this issue does not affect the general perturbative bias expansion: since the bias parameters directly describe the statistics of galaxies, without the intermediate step of the clustering of halos at fixed mass, this approach is entirely insensitive to assembly bias.

%% file: exset.tex
\clearpage
\section{The excursion-set approach to the distribution of dark matter halos}
\label{sec:exset}

\secttoc

In this and the next section, we review the predictions for the bias of dark matter halos in two well-developed physical models of halo formation: 
the excursion-set formalism (this section) and the peak approach (\refsec{peaks}). To set the frame and illustrate the basic physical assumptions of these models, we begin with a discussion of recent simulation studies of the connection between virialized halos and their progenitors in the initial density field 
(\refsec{exset:general}).

Next, we  introduce in detail the spherical collapse solution (\refsec{sph_collapse}), which provides an idealized model of halo formation. Moreover, it yields a natural, typical value of the mean overdensity of the Lagrangian regions (proto-halos) that form halos at late times. This fractional overdensity $\dc$ can then be used, for example, in the simple Lagrangian thresholding model introduced in \refsec{localbias}. If na\"{i}vely applied, however, as is done by 
Press-Schechter theory (\refsec{PS}), the thresholding model leads to a  mass function (number density of halos per unit mass) which is not 
properly normalized because of the so-called cloud-in-cloud problem (\refsec{CICproblem}); that is, the matter density contained in halos, when integrating over all halo masses, is a factor of 2 smaller than the actually available matter density. 
The excursion-set formalism (\refsec{exset_begin}) solves the cloud-in-cloud problem by adding the 
\emph{first-crossing} condition.
In the subsections that follow, we discuss the predictions for the halo mass function and bias from the 
excursion-set formalism with a constant threshold $\dc$ (\refsecs{survival_pdf}{exset_summary}) as well 
as a general threshold (moving, fuzzy barrier in \refsec{gen_barrier}). 
Finally, in \refsec{assembly_exset}, we revisit the phenomenon of assembly bias (\refsec{assembly}) in the context of the excursion-set formalism.

In \refsec{exset} and \refsec{peaks}, the analysis will always be done in Lagrangian space, and in terms 
of the linear density contrast $\d_R^{(1)}$ smoothed at a radius $R$.  In these sections, we 
therefore simplify the notation with respect to that used in previous sections, to conform more closely 
with that followed in the literature.  
For convenience, we list frequently used symbols in \reftab{symbol56} (p.~\pageref{tab:symbol56}).

\bigskip

Before delving into the details of the standard excursion-set picture which we will describe in \refsecs{exset:intro}{excursion_HeavensPeacock}, let us summarize the qualitative predictions for halo bias in this picture.  Since the only quantity appearing in the 
calculation is the smoothed linear density field $\d_R^{(1)}(\vq)$, the Lagrangian halo density field can be 
written as a local function of $\d_R^{(1)}$ and its derivatives with respect to $R$.  Specifically,
\ba
\d_h^L(\vq)\Big|_\text{exc.~set} =\:& F\left[ \d_R^{(1)}(\vq),\, \partial\d_R^{(1)}(\vq)/\partial R,\,\cdots \right] \vs
=\:& b_1^L \d^{(1)}(\vq) + \frac12 b_2^L [\d^{(1)}]^2(\vq) + \cdots \vs
& + b_{\lapl\d}^L \lapl \d^{(1)}(\vq) + b_{\d \lapl\d}^L \d^{(1)}(\vq) \lapl \d^{(1)}(\vq) + \cdots \vs
& + b_{\laplsq\d}^L \laplsq \d^{(1)}(\vq) + \cdots\,,
\label{eq:dhL_exset}
\ea
where here and throughout this section, we drop the time arguments, as all quantities are linearly extrapolated to the time $\tau_0$ of identification of halos. 
We have used that the higher-derivative terms introduced by smoothing
$\d^{(1)} \to \d_R^{(1)}$ can all be written as $\nabla^{2n} \d^{(1)}$ (see \refsec{higherderiv}), which also holds for $\partial \d_R^{(1)}/\partial R$, leading 
to a Lagrangian \LIMD bias expansion with a subset of higher-derivative terms.  
Note that no stochastic contributions are predicted by the standard excursion set, though a shot-noise 
term $\eps$ is often added empirically.  
We will generalize this version of the excursion-set picture in \refsec{gen_barrier}, where we also take into 
account the tidal field in Lagrangian space, and discuss stochasticity in the threshold.  These lead to additional contributions to the bias expansion beyond \refeq{dhL_exset}.

\subsection{General considerations about the formation of dark matter halos}
\label{sec:exset:general}

Since numerical simulations give us access to the entire formation history of dark matter halos, they enable 
us to ascertain the validity of the approximations made in Lagrangian approaches such as excursion sets or 
peaks. The correspondence between halos identified at low redshift and their progenitors in the
initial conditions---the so-called {\it proto-halos}---can be studied on an object-by-object basis. 
Here, proto-halos are the Lagrangian regions obtained by tracing the dark matter particles belonging to virialized halos back to their initial positions. The typical comoving size of proto-halos of mass $M$ thus is the Lagrangian radius $R(M)$ which encloses $M$ at the mean comoving density $\rhob$.

In the excursion-set formalism, proto-halos are modeled as regions where the smoothed linear matter density on the scale $R(M)$ is above a threshold $\dc$. Moreover, any regions which are included in a region above threshold identified at a larger smoothing scale are excluded. On the other hand, the peak approach models proto-halos as local maxima of the linear density smoothed on the scale $R(M)$.

Ref.~\cite{frenk/etal:1988} were the first to investigate the robustness of the correspondence between peaks in the initial density field and proto-halos. On analyzing $32^3$-particle N-body simulations of fairly small box size ($L\sim 14\hmpc$), 
they concluded that the resolved dark matter halos form preferentially around high peaks of the initial 
density field. 
However, Ref.~\cite{katz/quinn/gelb:1993} used simulations evolving $144^3$ particles in a much larger box 
($L=50\hmpc$) and found a rather weak association between initial density peaks and collapsing halos. 
Their conclusions were confirmed partially by \cite{porciani/dekel/hoffman:2002}, who studied the properties 
of proto-halos in N-body simulations of $256^3$ particles in a box of size $L\simeq 85\hmpc$. 
The latter concluded that nearly half of the galactic-sized proto-halos do not contain a linear density peak 
within their Lagrangian volume.

\begin{figure}
\centering
\includegraphics[trim=1.8cm 1.7cm 1.8cm 7cm,clip,width=\textwidth]{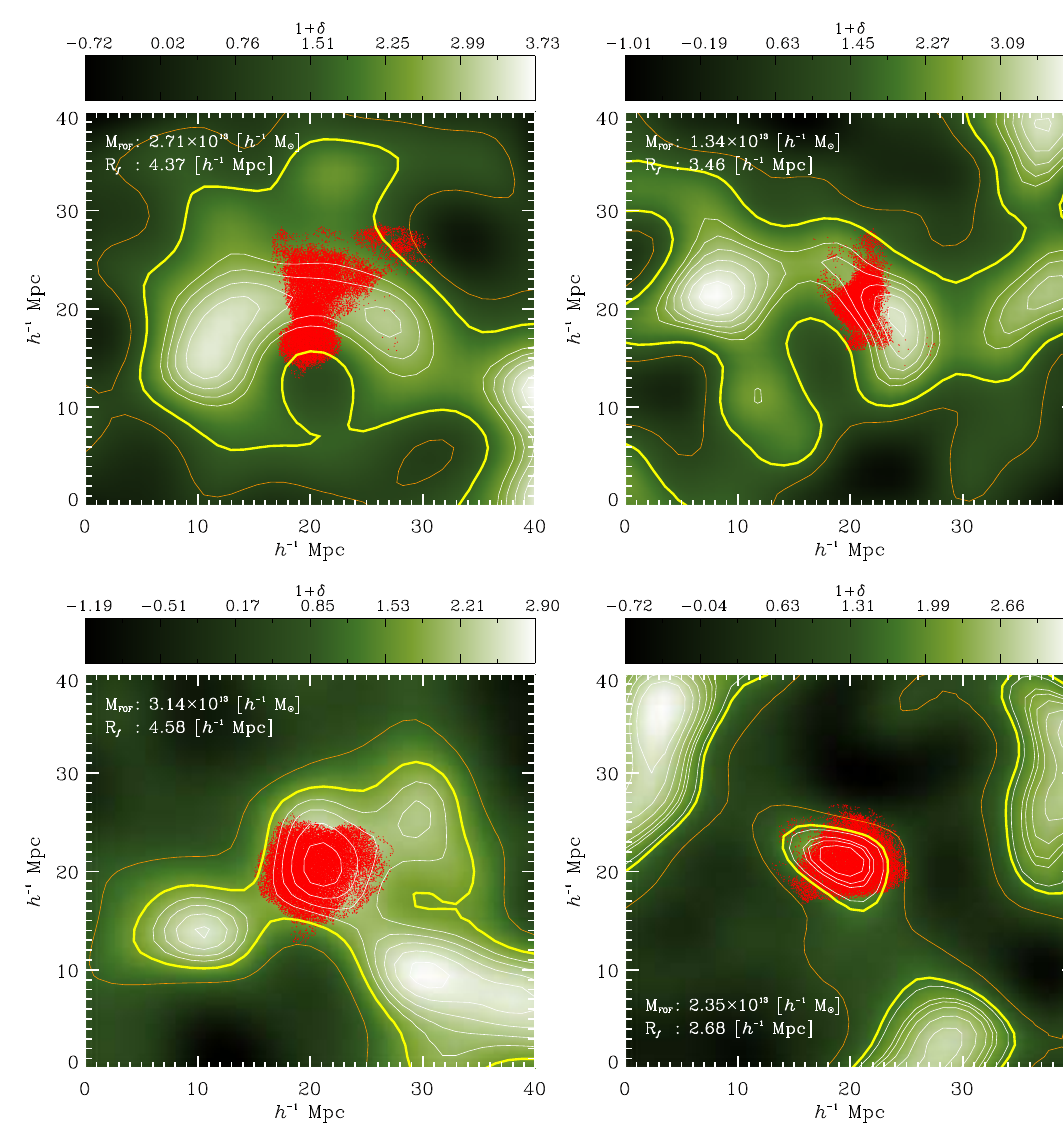}
\caption{
Examples of the overdensity field in the vicinity of four proto-halos.  
The top panels display results for halos with $M_{\rm pk}<M_\text{halo}/4$, 
where $M_{\rm pk}$ is the mass enclosed within the filter radius $R$ at which a 
peak was found. The bottom panels show results for two similar mass halos, yet with 
$M_{\rm pk} \approx M_\text{halo}$. Linearly extrapolated density fields have been 
smoothed with a tophat filter with $R = R(M_\text{halo})$.  Particles that 
belong to the FoF halos at $z=0$ are shown as red dots. In all panels, contours 
highlight the density gradients in the neighborhood of the halo. A density 
contrast $\d_R=1$ is shown as an orange curve, while the threshold for 
spherical collapse, $\d_R=\dc=1.686$, is shown as a thick yellow line.
\figsource{ludlow/porciani:2011} 
}
\label{fig:ludlow_porciani}
\end{figure}

Recently, Ref.~\cite{ludlow/porciani:2011} revisited this issue using two high-resolution $1024^3$ N-body simulations 
in boxes of size $1200$ and $150\hmpc$, respectively. 
They found that essentially all the proto-halos trace Lagrangian patches with a linearly extrapolated density 
contrast $\delta_R^{(1)}\gtrsim \mathcal{O}(1)$.
In addition, as many as $\sim$70\% of all halos identified with an FoF algorithm (with $N_{\rm FoF}>100$) in both 
of their simulations can be properly identified with a peak in the linear density field (i.e. the peak resides in 
the Lagrangian halo patch) when smoothed on mass scales $0.7M\leq M_\text{pk}\leq 1.3M$ (\reffig{ludlow_porciani}). 
Here, $M$ is the mass of the actual halo, given by the number of bound dark matter particles, while $M_\text{pk}$ 
is the \emph{peak mass}, i.e. the mass enclosed in the filter.  
The fraction of peak associations increases to 85\% upon extending the search to the mass range 
$0.5 M \leq M_\text{pk}\leq 2 M$. Furthermore, the fraction depends strongly on halo mass, with as many as $\sim$91\% of halos 
with $M> 5\times 10^{14} \hmsun$ forming in the vicinity of peaks of the expected characteristic mass.

However, a non-negligible portion of halos show a considerable disparity between the predicted and measured masses. 
For example, $\sim$20\% of halos with $>100$ particles have $M_\text{pk} < M$, $15$\% of which do not contain any 
peaks on any mass scale within a factor of four of the true halo mass scale. 
The authors refer to these as ``peakless'' halos (see the upper panels in \reffig{ludlow_porciani}).  
The increasing fraction of ``peakless'' halos with decreasing mass reflects the fact that the approximation of an 
isolated spherical collapse of a perturbation centered around a density peak becomes increasingly inaccurate as the 
halo mass decreases.

These numerical findings help us in understanding the validity of the various approximations made in Lagrangian bias 
models. The standard Press-Schechter and excursion-set theory discussed in 
\refsec{exset:intro} -- \refsec{exset_summary}
rely on two key assumptions:
\begin{itemize}
\item The collapse of dark matter halos is approximately spherical.
\item Proto-halo patches are characterized by a linearly extrapolated overdensity $\delta_R^{(1)}$ 
equal to the spherical collapse threshold $\dc$, \refeq{deltacdef}.
\end{itemize}
In light of the numerical studies reviewed above, these assumptions are expected to be valid at high mass $M\gg M_\star$ only. 
Two important additional ingredients can be taken into account in order to improve the accuracy of Lagrangian bias models:
\begin{itemize}
\item The shear field, which is expected to be partly responsible for both the ``peakless'' halos and the mass dependence of the
critical density threshold. In a first approximation, the effect of shear can be modeled as a moving barrier as discussed in \refsec{gen_barrier}.
\item The peak constraint, i.e. the fact that halos collapse around local maxima in the initial density field. This extension is the focus of \refsec{peaks}.
\end{itemize}
\begin{figure}[t!]
\centering
\includegraphics[width=0.49\textwidth]{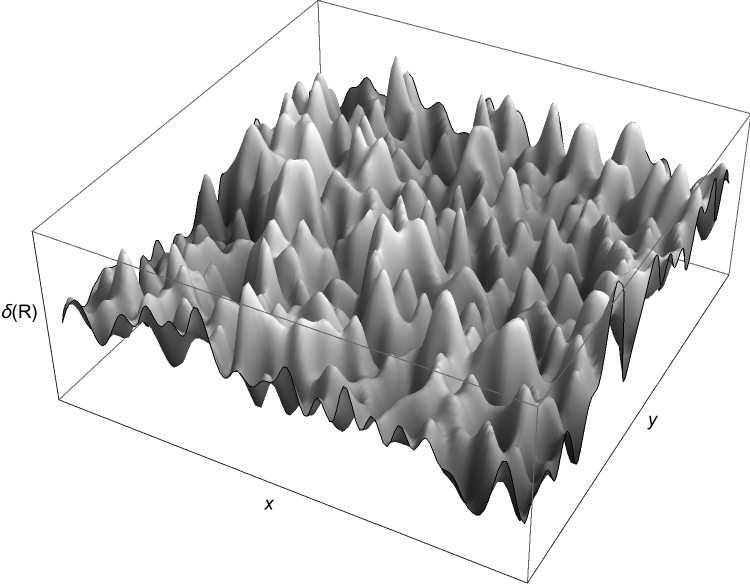}
\includegraphics[width=0.49\textwidth]{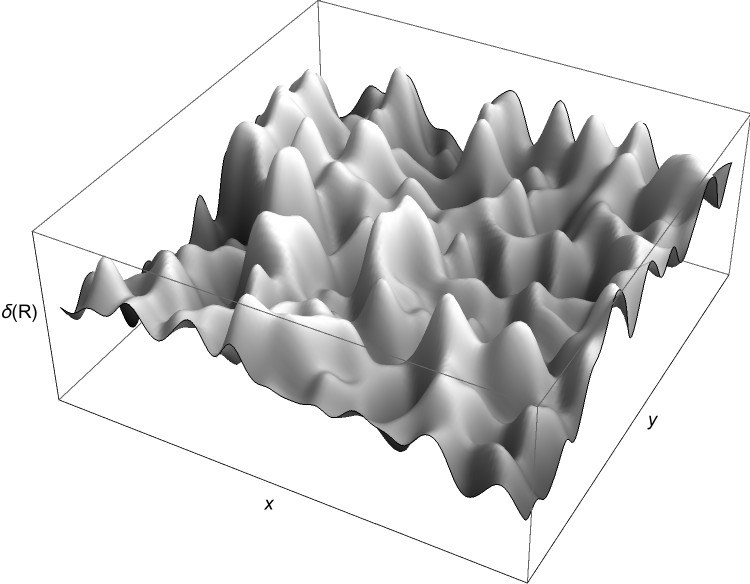}
\includegraphics[width=0.49\textwidth]{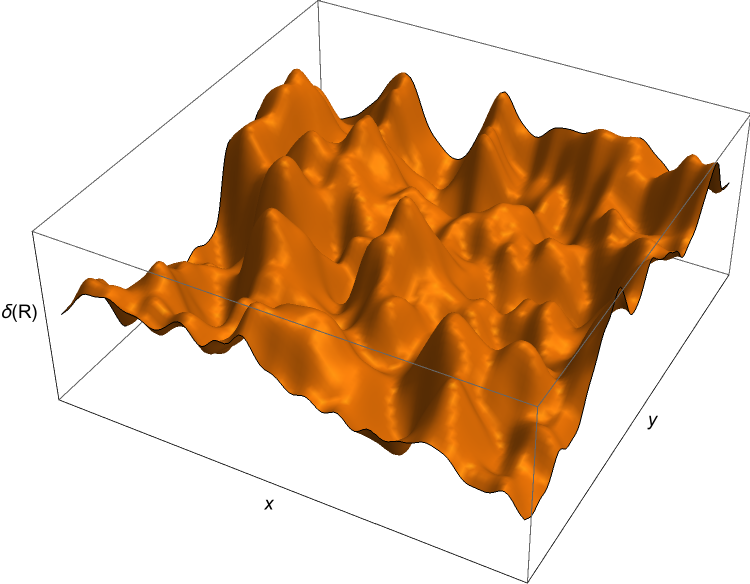}
\includegraphics[width=0.49\textwidth]{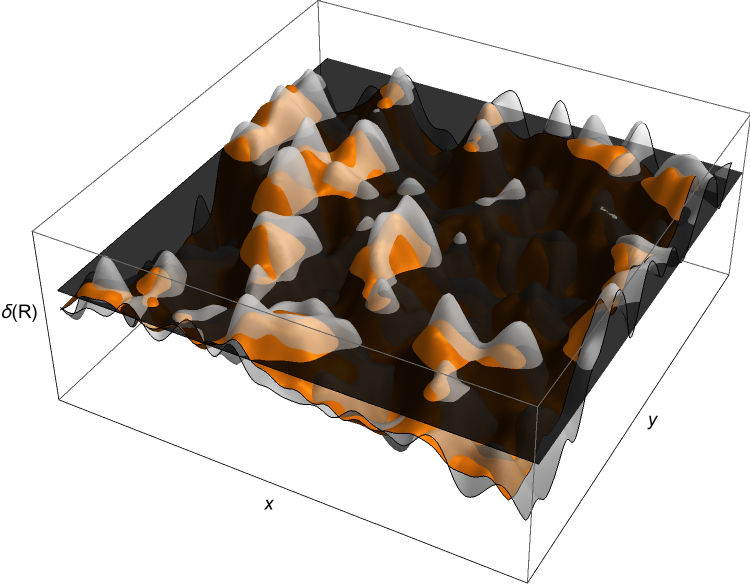}
\caption{
Illustration of a two-dimensional Gaussian density field $\delta(R)$ smoothed on different scales $R$: $0.4\Mpch$ (\textit{top left}),  $1\Mpch$ (\textit{top right}), $2\Mpch$ (\textit{bottom left}). The side length is $50\Mpch$. 
In the {\it bottom-right} panel, we superimpose the fields with $R= 1\Mpch$ and $R=2\Mpch$, along with the spherical collapse threshold ($\dc=1.686$, black plane). 
In the excursion-set approach, the mass fraction of halos is set equal to the fraction of area above the threshold at the smoothing scale $R(M)$. 
Note that there are regions which are above threshold at the larger smoothing scale $R=2\Mpch$ but below the threshold at $R=1\Mpch$. This illustrates the so-called {\it cloud-in-cloud} problem that is discussed in \refsec{CICproblem}.
}
\label{fig:illustrate_LdeltaR}
\end{figure}

The peak constraint resolves a fundamental limitation of the excursion-set formalism, which stems from the assumption 
that \emph{the total mass fraction enclosed in halos with mass greater than $M$ is the same as the total volume fraction of above-threshold smoothed Lagrangian regions,} with filter size $R = R(M)$. 
For this assumption to hold, the \emph{smoothed} density of all points inside halos must exceed the threshold.  
This is not guaranteed for actual halos. 
In particular, depending on the density profile, the outer part of halos can have values of the smoothed density that are significantly below the threshold (see \reffig{illustrate_LdeltaR} for an illustration).  
Indeed, the comparison between the halo mass function calculated from the excursion-set formalism with constant threshold 
and N-body simulations shows that the theory underpredicts the abundance of massive halos 
\cite{jenkins/frenk/etal:2001,robertson/etal:2009,paranjape/sheth:2012}.  
Moreover, this assumption is the very foundation of the excursion-set formalism.  
The two adjustable components in the excursion-set formalism, the scale dependence and fuzziness of the collapse threshold, 
are not directly related to the above-mentioned deficiency.  
This problem is addressed in the peak formalism (\refsec{peaks}) where we define the {\it center} of halos as peaks of the 
density field.  
Hence, we do not need to rely on the volume fraction in order to calculate the halo abundance. Recently, the virtues of both approaches have been combined in the \emph{excursion-set peaks} ansatz, which we will discuss in \refsec{esp}.

\subsection{From Press-Schechter to excursion sets}
\label{sec:exset:intro}
\subsubsection{The spherical collapse model of halo formation}
\label{sec:sph_collapse}

Let us consider a spherical region with uniform density of initial radius $R_i$ and 
initial average density contrast $\Delta_i$ in an expanding Universe 
\cite{gunn/gott:1972,gunn:1977} with mean comoving matter density at the initial time 
$\overline{\rho}_{m,i} \equiv \rhob(t_i)$. That is, the total mass of the overdensity is 
$M=(4\pi/3)R_i^3\overline{\rho}_{m,i}(1+\Delta_i)$.
For simplicity, we only present the case of an Einstein de-Sitter (EdS) Universe, although 
the effects of cosmological constant as well as curvature of the Universe 
can easily be incorporated \cite{lahav/etal:1991,lacey/cole:93}.

The interior of a spherically symmetric perturbation embedded in an FRW spacetime evolves 
independently of the surrounding spacetime and is only determined by the total mass 
enclosed \cite{einstein/straus,balbinot/etal:1988,carrera/giulini:2010}.   
The equation of motion of the radius $R$ of the spherical region is given by 
\be
\frac{\dd^2 R}{\dd t^2} 
= -\frac{GM(<R)}{R^2}
= -\frac{4\pi G}{3} \overline{\rho}_{m,i}(1+\Delta_i)\frac{R_i^3}{R^2}
= -\frac{H_i^2R_i^3}{2R^2} (1+\Delta_i),
\label{eq:eom_STH}
\ee
where $H_i=2/(3t_i)$ is the Hubble rate at the initial time $t_i$. It is worth noting that \refeq{eom_STH} holds for arbitrarily large-scale spherical perturbations; i.e. we do not need to assume $R \ll H^{-1}$. 
Integrating once over time, we obtain the energy conservation equation: 
\be
\frac12 \dot{R}^2
-\frac{H_i^2 R_i^3}{2R}(1+\Delta_i)
= {\rm constant} \equiv E \;.
\label{eq:energy_STH}
\ee
The spherical region is unbound if $E>0$, i.e. if the initial expanding velocity exceeds
$
\dot{R}_i \ge H_iR_i\sqrt{1+\Delta_i}
$,
but it will eventually turn around and 
re-collapse if $E<0$.  If $\Delta_i \ll 1$, then the initial velocity
can be obtained by matching the linear adiabatic growing mode,
$\dot{R_i} = H_i R_i (1-\Delta_i/3)$.  The total energy then becomes 
\be
E = -\frac53\frac{(H_iR_i)^2}{2}\Delta_i,
\ee
and the turn around, or maximum, radius is obtained by setting $\dot R=0$ and solving \refeq{energy_STH} for $R$:
\be
R_{\rm ta} = \frac35\left(\frac{1+\Delta_i}{\Delta_i}\right) R_i\,.
\ee
The exact solution of \refeq{eom_STH} is given by a cycloid, which can be
parameterized with a variable $\theta$:
\ba
R = \frac{R_{\rm ta}}{2}\left(1-\cos\theta\right),~
t = \frac{t_{\rm ta}}{\pi}(\theta-\sin\theta),
\ea
where $t_{\rm ta}$ is the time at turn-around ($\theta=\pi$).
Noting the similarity between the case at hand and the Keplerian 
radial orbit (with semi-major axis $R=R_{\rm ta}/2$ and 
period $T=2t_{\rm ta}$), we use Kepler's third law to relate 
$R_{\rm ta}$ and $t_{\rm ta}$ as $8GMt_{\rm ta}^2=\pi^2 R_{\rm ta}^3$.

The parametric solution yields a relation between $\theta$ and the overdensity $\Delta$,
\ba
1+\Delta 
=\:& 
\frac{\overline{\rho}_{m,i}(1+\Delta_i)R_i^3}{\rhob R^3}
= (1+\Delta_i)\left(\frac{t}{t_i}\right)^2\left(\frac{R_i}{R}\right)^3
=
\frac{9}{2}\frac{(\theta-\sin\theta)^2}{(1-\cos\theta)^3}.
\ea
At early times ($\theta\simeq\theta_i$), we reproduce the linear growing
modes $\Delta(t)\simeq 3\theta^2/20\simeq\Delta_i (t/t_i)^{2/3}\propto a(t)$
in the EdS Universe.  Moreover, the coefficients of the 
higher-order terms in the expansion of $\Delta(t)$ in powers of $a(t)$ are
precisely the coefficients of the operators $\d^n$ in the $n$-th order contribution to the density field in perturbation theory (e.g., $17/21$ at second order, $341/567$ at third order, and so on).  
Then, the spherical region evolves nonlinearly as time proceeds, 
reaching $1+\Delta_{\rm ta}\simeq 9\pi^2/16\simeq 5.55$ at turn-around time 
($\theta=\pi$). 
The spherical collapse model predicts the collapse to a singularity $R=0$ at $\theta=2\pi$. This is of course unphysical. Instead, gradient instabilities, which occur as soon as mass shells cross in physical space, will break spherical symmetry and lead to a complex bound structure. We can estimate the final radius and density of this structure (halo) assuming perfect virialization. The virial theorem then states that $E=T+W=W/2$, where $T$ is the kinetic energy while $W$ is the potential energy. Since $W=E$ holds at turn-around, we obtain for the radius of the virialized halo, $R_{\rm vir} = R_{\rm ta}/2$. Further assuming that virialization is completed precisely at $\theta=2\pi$, the formal collapse epoch, we obtain for the density of the virialized halo
\be
1+\Delta_{\rm vir} = 
(1+\Delta_{\rm ta})
\left(
\frac{\overline{\rho}_{m,\rm ta}}{\overline{\rho}_{m,\rm vir}}
\right)
\left(
\frac{R_{\rm ta}}{R_{\rm vir}}
\right)^3
=
(1+\Delta_{\rm ta})
\left(
\frac{t_{\rm ta}}{t_{\rm vir}}
\right)^2
\left(
\frac{R_{\rm ta}}{R_{\rm vir}}
\right)^3
=
18\pi^2\simeq 178.
\ee
The corresponding linear density contrast at the time of virialization (collapse), which
defines the critical density $\dc$, is 
\be
\dc
\equiv
\Delta^{(1)}
(t_{\rm vir}) 
= \Delta_i \left(\frac{t_{\rm vir}}{t_i}\right)^{2/3}
= \frac{3}{5}\left(\frac{3}{4}\right)^{2/3}
\left(\theta_{\rm vir} - \sin\theta_{\rm vir}\right)^{2/3} 
= \frac{3}{5}\left(\frac{3\pi}{2}\right)^{2/3} \simeq 1.686 \;.
\label{eq:deltacdef}
\ee
There is some arbitrariness in the definition of the virialization epoch.
Analytic arguments \cite{bertschinger:1985} and N-body simulations \cite{boryz/ludlow/porciani:2014} show 
that the virialization process is not instantaneous. 
Furthermore, halos are often identified with threshold values $\Delta_\text{vir}$ different from 178
\cite{tinker/etal:2008}.
Therefore, the corresponding linear density contrast $\dc$ also depends on the halo identification 
procedure. 
Nevertheless, in the high-mass limit where the spherical collapse assumption works best, Ref.~\cite{valageas:2009} 
argues that a threshold $\dc={\cal F}^{-1}(\Delta_\text{vir})$, where ${\cal F}$ is the spherical collapse 
mapping from linear to nonlinear densities, should be used for halos identified with a nonlinear threshold 
$\Delta_\text{vir}$. For $\Delta_\text{vir}=200$, this rare-event limit yields $\dc\simeq 1.59$. While this
prescription is somewhat supported by N-body simulations \cite{valageas:2011}, we will assume \refeq{deltacdef} as
default value here. 

As the linear growth factor is a monotonically increasing function of time, 
one can use it as a clock to measure the virialization time; thus,
a spherical region of radius $R$ will be virialized and form a halo
when the linearly extrapolated density contrast exceeds $\dc$.
This motivates the Press-Schechter theory that estimates the abundance of 
halos by interpreting the fraction of thresholded ($\d^{(1)}_R>\dc$)
Lagrangian volume, following \refsec{localbias}, as the fraction of mass contained in 
halos with mass greater than $M=(4\pi/3)\rhob R^3$. 

Clearly, the spherical collapse is only a very rough toy model for how gravitational collapse to bound structures
actually happens.
In particular, tidal fields, angular momentum, as well as the effect of small-scale perturbations are entirely neglected.  
Models in which halos form through ellipsoidal collapse provide a better description of the formation times of dark matter
halos \cite{giocoli/moreno/etal:2007,boryz/ludlow/porciani:2014}.
Therefore, $\dc$ should be regarded as an idealized reference value, which we will later generalize to a scale-dependent
stochastic barrier $B(\sigma)$ (see \refsec{gen_barrier}). In practice, since there is some freedom in the definition of a
virialized dark matter halos (see \refapp{halofinder}), the value of $\dc$ effectively depends on the halo identification
algorithm and, thus, can deviate from 1.686. This effective threshold can be measured by extrapolating the linear overdensity
of the Lagrangian regions that collapse to form halos to the limit $M\to\infty$ (see \reffig{barrier} on p.~\pageref{fig:barrier}).

Note that a similar calculation can be done for the formation of cosmic 
voids \cite{sheth/vandeWeygaert:2004}. In particular, for very underdense 
spherical regions with a linearly extrapolated fractional underdensity 
$\d_\ell^{(1)} \lesssim -1$, 
Ref.~\cite{bernardeau:1994a} has found that 
$\d_\ell(\tau) = (1-\d_\ell^{(1)}(\tau)/1.5)^{1.5}$ 
is a good approximation to the interior density, which might be useful to model 
the bias relation in cosmic voids along with the arguments in 
\refsec{localbias} \cite{neyrinck:2013}.

\subsubsection{Press-Schechter formalism}\label{sec:PS}

\begin{table*}[t]
\centering
\begin{tabular}{l|l}
\hline
\hline
Quantity & Symbol\\
\hline
Smoothed linear density contrast & $\delta(R)\equiv \delta_{R}^{(1)}(\vq)$ \\
Variance of the smoothed density field & $S \equiv \left<[\delta(R)]^2\right>$ \\
Cumulative collapsed mass fraction  & $F(>M)$ \\
Differential volume fraction & $f(M) = -dF(>M)/dM$\\
Multiplicity function & $f(\nu_c)= f(M)\left|dM/d\nu_c\right|$\\
Differential survival probabilities of random walks: & \\
\quad from $(\infty, 0)$ to $(R, \delta)$ & $\Pi(\delta;R)$ \\
\quad from $(R_\ell, \d_\ell)$ to $(R, \delta)$ & $\Pi[(\d;R),(\d_\ell;R_\ell)]$ \\
Conditional survival probability for a walk to arrive at $(R,\delta)$, & \\
\quad given constraint that $\d(R_\ell)=\d_\ell$ & $\Pi(\d;R|\d_\ell;R_\ell)$ \\
\hline
\end{tabular}
\caption{List of symbols used throughout \refsec{exset}.
}
\label{tab:symbol56}
\end{table*}
We now recap the original derivation of the Press-Schechter mass function 
\cite{press/schechter:1974} (see also \refsec{localbias}).  
The key assumptions are 
\begin{itemize}
\item The linear matter density contrast $\d^{(1)}$, whence the smoothed density contrast $\d_R^{(1)}$, 
follows Gaussian statistics. One must relax this assumption in the presence of primordial non-Gaussianity (see \refsec{NGexset}). 
\item A Lagrangian region of volume $V=(4\pi/3) R^3$ collapses to form a halo of mass 
$M = \rhob V$ when its density contrast $\d^{(1)}_R$ (linearly extrapolated to 
the epoch of interest) exceeds the spherical collapse threshold $\dc$.
\end{itemize}
Given the second assumption, the fraction of Lagrangian volume belonging to halos of mass {\it greater than} $M$ is [cf. \refeq{p1thr}]
\be
p_G(\d_R^{(1)}>\dc)
= \frac{1}{\sqrt{2\pi\sigma^2(R)}}\int_{\dc}^\infty 
d\delta\; \exp\left[-\frac12 \frac{\delta^2}{\sigma^2(R)}\right]
=
\frac12 \mathrm{erfc}\left[\frac{\dc}{\sqrt{2}\sigma(R)}\right] \;.
\label{eq:FM_PS}
\ee
Here, $p_G$ stands for the Gaussian cumulative PDF, and 
$\erfc$ denotes the complementary error function defined in \reftab{math} on p.~\pageref{tab:math}.  
$R$ is the Lagrangian radius corresponding to the halo mass $M$,
\be
M
=
3.1389\times 10^{11} h^{-1} M_\odot
\left(\frac{\Omega_m}{0.27}\right)
\left(\frac{R}{h^{-1} \mathrm{Mpc}}\right)^3
 \;,
\label{eq:MofR}
\ee
and $\sigma(R)$ is the root-mean-square  of the linear smoothed density perturbations on the scale $R$,
\be
S \equiv \sigma^2(R) 
\equiv \left<\left(\d_R^{(1)}\right)^2\right>
= \int_{\vk} \Plin(k) 
W_R^2(k)\;,
\label{eq:def_sigmaRsq}
\ee
where $W_R(k)$ denotes the filtering kernel in Fourier space. Note that, in \refeq{FM_PS}, the smoothing 
scale $R$ is held fixed although the volume fraction is an integral over contributions from halos with
Lagrangian radius greater than $R$.  This will be dealt with by taking a derivative with respect to mass.  

Given the fundamental assumption of the excursion set stated in \refsec{exset:general}, the Lagrangian volume fraction $F(>M)$ of the total mass enclosed in halos of mass greater than $M$ follows immediately from \refeq{FM_PS}:
\be
F(>M)
=
\frac{1}{\rhob} \int_{M}^{\infty} \dd \ln M' \, M' \avnh(M')
=
p_G(\d_R^{(1)}>\dc)\,.
\label{eq:PS_PGM}
\ee
Therefore, the differential volume fraction in halos of mass $M$ is
\be
f(M) \equiv -\frac{dF(>M)}{dM}\,,
\ee
which formally leads to the halo mass function, i.e. the comoving number density of halos per logarithmic mass interval,
\be
\avnh(M) \equiv \frac{d^2 \overline{N}_h}{dVd\ln M} 
= \rhob f(M) = - \rhob \frac{dF(>M)}{dM} \;.
\label{eq:dndM_PS_bare}
\ee
Unfortunately, this mass function does not appear to be properly normalized, as an
integration over all the mass included in halos only recovers half of the total
mass:
\be
\int_0^{\infty} d\ln M \, M \,\avnh(M)
=
-\rhob
\int_0^{\infty} dM \frac{d F}{dM}
=
-\rhob
\Big[ p_G(R=\infty) - p_G(R=0)\Big]
=
\frac12 \rhob \;.
\label{eq:PS_normal}
\ee
On the other hand, the basic premise of this argument states that, when including arbitrarily small halo masses, the mass in halos should add to the total matter density, since
\be
\lim_{R\to0}\sigma(R) = \infty \quad\mbox{and}\quad \lim_{R\to \infty} \sigma(R) = 0\,,
\ee
so that there are no regions that do not collapse to halos of some (arbitrarily small) mass.  Note that we have assumed a hierarchical density field without any cutoff here.  
To resolve this issue, Ref.~\cite{press/schechter:1974} introduced an \textit{ad hoc}, 
``fudge'' factor of two which leads to the \textit{Press-Schechter} (PS) mass function:  
\be
\overline{n}_{h,\rm PS}(M) 
= \rhob\left(-2\frac{dF}{dM}\right)
=
\frac{\rhob}{M}\sqrt{\frac{2}{\pi}} 
\nu_c e^{-\nu_c^2/2}
\left|\frac{d\ln \sigma(R)}{d\ln M}\right|,
\label{eq:PS_massfn}
\ee
where $\nu_c \equiv \dc/\sigma(R)$ is the significance of the critical density in terms of the standard deviation of matter fluctuations on the scale $R$.   Massive halos, for which $\nu_c\gg1$, are rare, while low-mass halos such that $\nu_c \ll 1$ are common.  
It is convenient to parametrize the halo mass function with a {\it multiplicity function} 
$\nu_c f(\nu_c)$ defined as
\be
\label{eq:vfv}
\nu_c f(\nu_c)\frac{d\nu_c}{\nu_c} = \frac{M}{\rhob} \avnh(M) \frac{dM}{M} \;.
\ee
The notation $d\nu_c$ implicitly signifies that $\sigma$ is varied while $\dc$ is kept fixed.
As we shall see in \refsec{exset_begin}, this multiplicity function corresponds to the 
first-crossing distribution of random walks in the excursion-set theory. 
Hence, the halo mass function can be recast into the generic form
\be
\avnh(M) 
=
\frac{\rhob}{M} 
\nu_c f(\nu_c)
\left|\frac{d\ln \sigma(R)}{d\ln M}\right| \;.
\label{eq:univ_massfn}
\ee
Note that the Jacobian often appears as $d\ln\nu_c/d\ln M$ in the literature. Still, one should keep in mind that $\sigma$ is varied while $\dc$ is held fixed.  
In Press-Schechter theory, the multiplicity function is
\be
\nu_c f_{\rm PS}(\nu_c) = 
\sqrt{\frac{2}{\pi}}\nu_c e^{-\nu_c^2/2} \;.
\ee
\refeq{univ_massfn} is frequently referred to as \emph{universal mass function} (we have already encountered this form in \refsec{buniv}), 
since it only depends on redshift, cosmological parameters, and the power spectrum 
of initial fluctuations through $\nu_c = \dc/\sigma(M)$ \cite{sheth/tormen:1999,jenkins/frenk/etal:2001}.  As discussed in \refsec{buniv}, in an EdS universe
with exact power-law initial conditions, the mass function can always be written in this form.  In a more general cosmology, for example a realistic $\Lambda$CDM model, the multiplicity function will also depend on 
variables other than the peak significance $\nu_c$. For example, the mass 
function of density peaks (see \refsec{peaks}) cannot be written in the form 
\refeq{univ_massfn}.  Still, the cosmology dependence of the peak multiplicity function is captured by a small set of moments $\sigma_n(R)$. 

\subsubsection{Cloud-in-cloud problem and its resolutions*}
\label{sec:CICproblem}
\technote{* This section is of a more technical nature and 
a slight digress.  Readers interested in the excursion set itself may go directly to \refsec{exset_begin}.}

Despite an unsatisfactory treatment of the normalization, the Press-Schechter formalism captures the qualitative features of the halo mass function measured in N-body simulations (a power-law function with an exponential cutoff; see \reffig{shethtormen}). 
As such, it provides a baseline for more detailed models of the halo mass function.
In this section, we review the origin of the factor of two difference in 
the normalization of the Press-Schechter mass function, and the 
attempts to resolve the issue. 

Refs.~\cite{peacock/heavens:1990,bond/etal:1991} explained the underlying reason for the ``fudge factor''  in the Press-Schechter formalism.  
Namely, the Press-Schechter approach does not take into account the possibility that an underdense region identified on 
a scale $R$ may be embedded in a halo on a bigger scale $R' > R$ (see the derivation of 
\refeq{dndM_PS_bare} above). 
This is the so-called {\it cloud-in-cloud} problem first identified in \cite{epstein:1983} and named in \cite{peacock/heavens:1990}. 

To properly address this problem, we have to take into account the underdensities 
$\d_R^{(1)}<\dc$ that belong to a collapsed region with larger radius
$R'>R$. That is, the fraction of Lagrangian volume 
belonging to collapsed objects with mass greater than $M$, $F(>M)$, should be written as
\be
F(>M)
=
p_G(\d_R^{(1)} >\dc) 
+
\int_{-\infty}^{\dc}
d \d
\frac{d p_G}{d\d}
p_{up}\left(\d(r_{>R}) > \dc;\d\right),
\label{eq:FM}
\ee
instead of \refeq{FM_PS}, which includes only the first term.
Here, $\d(r_{>R})$ denotes the linear density contrast smoothed over some radius $r$ 
greater than $R$, whereas the first-crossing probability $p_{up}$ (subscript {\it up} because the first crossing always happens upward)
in the second term of
\refeq{FM} is the probability that regions with $\d_R^{(1)}<\dc$ belong to a collapsed 
object with larger radius. In other words, it is the probability that there exists $R'>R$ 
such that $\d_{R'}^{(1)}>\dc$.
Importantly, \refeq{FM} is, by construction, properly normalized as $p_{up}\to 1$ in 
the small-scale limit $R\to 0$ (where $\sigma(R)\to \infty$). This means that, for a 
sufficiently small length scale, virtually all underdense regions are parts of collapsed 
objects of bigger size.
Note, however, that it is still not obvious why the second term on the right-hand side of \refeq{FM} 
(the probability that an underdense region is part of a bigger collapsed object) is 
exactly the same as the first term (the probability that the density contrast of a given 
region exceeds critical value $\dc$).

Alternatively, Ref.~\cite{jedamzik:1995} formulates the cloud-in-cloud problem as a lack 
of distinction between sub-halos, i.e.~collapsed regions (with size $R$) enclosed in a  bigger 
collapsed region (with size $R'$), and {\it isolated} collapsed regions.
In their formulation, the cloud-in-cloud problem can be resolved upon introducing a 
conditional probability function $p(M,M')$, which quantifies the probability that an 
overdense region of mass $M$ belongs to a larger isolated region of mass $M'(\ge M)$. This
can also be interpreted as a probability distribution of substructure of mass $M$ inside a 
larger, isolated virialized object of mass $M'$.
Thus, \refeq{PS_PGM} becomes
\be
F(>M)
=
\frac{1}{\rhob}
\int_{M}^\infty
d M' \avnh(M') p(M,M').
\label{eq:PG_correct}
\ee
Clearly, the function $p(M,M')$ satisfies $p(M,M')=0$ when $M>M'$, and 
$\lim_{M' \to M^+}p(M,M')=1$ by construction. In addition, 
the $M\to0$ limit of \refeq{PG_correct} yields [see \refeq{PS_PGM}
and \refeq{PS_normal}]:
\be
F(>0) = 
\frac{1}{\rhob}
\int_{0}^\infty
d M' \avnh(M') p(M\to0,M')
=
\frac12 \;.
\ee
This implies $p(M\to0,M')\to1/2$ in order to ensure that the halo mass function be correctly 
normalized.
That is, only half of the small scale ($M\ll M'$) regions in the collapsed halos of mass $M'$ 
are dense enough to form a sub-halo. 
This follows from the fact that, on extremely small scales ($M\to 0$), the threshold density 
contrast $\dc$ can be safely ignored compared to the standard deviation of matter perturbations:
$\dc\ll\sqrt{\sigma_M^2-\sigma_{M'}^2}$.
Comparing \refeq{PG_correct} to \refeq{PS_PGM}, however, we find that the Press-Schechter case corresponds 
to setting $p(M,M')=1$ for all $M'>M$, which means that the Press-Schechter formalism implicitly assumes that 
every collapsed region is part of a bigger isolated region.
This fallacious assumption also leads to an error in the normalization, as
$p(M,M')$ has the wrong asymptote in the limit $M\ll M'$.
Finally, in analogy with the step following \refeq{PS_PGM}, we take the mass derivative 
of \refeq{PG_correct} which leads to an integro-differential equation,
\be
\frac{dF(>M)}{dM}
=
-
\frac{1}{\rhob} \avnh(M) 
+
\frac{1}{\rhob}
\int_{M}^\infty
d M' \avnh(M')\frac{d p(M,M')}{d M},
\ee
which allows us to solve for $\avnh(M)$ for a given function $dp(M,M')/dM$.

In both formulations of the cloud-in-cloud problem, new functions 
$p_{up}(\delta(r_{>R})>\dc;\delta)$ and $p(M,M')$ are introduced to describe the
statistics of small-scale regions (either over-dense or under-dense) in conjunction with the 
large-scale collapsed regions.
The excursion-set formalism provides a systematic way of calculating those functions given the 
statistical properties of the underlying matter density field.
In fact, as we shall see shortly, the correct normalization of the Press-Schechter mass function can be derived within 
an excursion-set approach based on the sharp-$k$ filter, in which fluctuations on different 
smoothing scales are completely uncorrelated.

\subsection{Excursion-set formalism: setting up the scene}
\label{sec:exset_begin}

The basic building block of the excursion-set formalism 
is the smoothed, linearly evolved initial density field at a point $\vq$:
\be
\d(R)\equiv
\delta_R^{(1)}(\vq)
=
\int d^3 x\, W_R(|\vx|)\delta^{(1)}(\vq+\vx) \;.
\label{eq:deltaLxR}
\ee
Since we always center the filter at the same Lagrangian position $\vq$, 
regardless of the value of $R$, and we always deal with the linear density 
field, we will omit the explicit dependence on $\vq$ and the 
superscript $(1)$ to simply write $\d(R) \equiv \delta^{(1)}_R(\vq)$ in this 
section (see \reftab{symbol56}).  Frequently used filtering kernels $W_R(x)$ and
their Fourier transform can be found in \refapp{stat:Fourier}. 
Note that the shape of the filter in real space reflects the Lagrangian density
profile of halos. In this regard, the sharp-$k$ filter is not realistic, 
as the corresponding real-space filtering kernel $W_R(x)$ is not always positive.

\begin{figure*}[t!]
\centering
\includegraphics[width=0.8\textwidth]{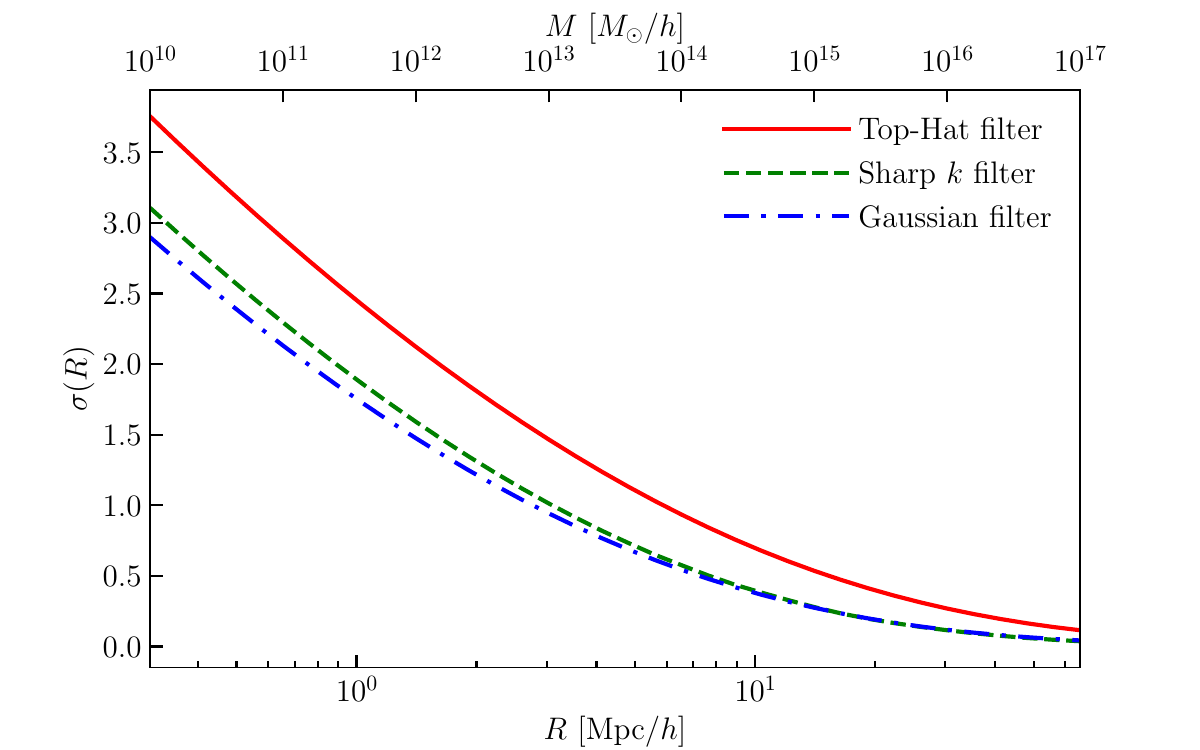}
\caption{
Relation between the halo mass ($M$), Lagrangian radius ($R$) and the 
r.m.s. density fluctuation smoothed on the scale $R$, $\sigma(R)$, 
at $z=0$
for three different filters: tophat filter (red solid line), 
Sharp-$k$ filter (green dashed line), and Gaussian filter 
(blue dot-dashed line).
Note that $\sigma(R)$ for the tophat filter is always larger than the 
others because the ringing of the spherical Bessel function captures 
power from higher-$k$ modes than the other two filters.
Throughout this section, we use $R$, $M$, and $S$ interchangeably.
\label{fig:MRsigma}
}
\end{figure*}

The statistical properties of the smoothed density field $\d(R)$ are entirely 
specified by its correlation functions. In particular, when the initial density contrast 
follows Gaussian statistics, the linear matter power spectrum extrapolated to the present 
epoch $\Plin(k)$ encodes the entire statistics (\refapp{stat:Gauss}).  

We are now ready to formulate the excursion-set formalism for halo statistics.
The key assumption is highlighted in the following box:
\begin{framed}
\noindent At a given redshift $z$, a Lagrangian point $\vq$ 
belongs to a halo of size $R$ {\it if} $R$ is 
the \textit{maximum smoothing scale} at which the smoothed 
linear density contrast $\d(R)$ exceeds the 
critical density contrast $\dc(z)$.
\end{framed}
\noindent
The spherical collapse model (\refsec{sph_collapse}) provides the connection 
between the extrapolated density field $\delta(R)$ and dark matter halos. 
Appealing to the assumption of statistical homogeneity of the Universe and the 
ergodicity of the cosmic density field, the probability that a random Lagrangian 
point $\vq$ belongs to a halo of mass between $M$ and $M+d M$ is proportional to 
the fractional number density of halos in the same mass range. This allows us to
calculate the halo mass function $\avnh(M)$.

Note, however, that the excursion set with a constant critical density threshold
generically underestimates the collapsed fraction 
\cite{jenkins/frenk/etal:2001,robertson/etal:2009,paranjape/sheth:2012}. 
As we have mentioned earlier, this is because any region below the 
threshold that is associated with a collapsed object is not counted 
as a part of the collapsed region in the excursion set.  N-body simulations show that the 
underestimation is most severe for massive halos.  
A possible remedy to this problem is to reduce the threshold in a 
mass-dependent way, although this does not address the underlying physical 
deficiency of the approach.  
An alternative is to explicitly (albeit approximately) incorporate 
the associated regions below the threshold as is done in the peak model 
(\refsec{peaks}), where all the mass associated with a peak in the smoothed
Lagrangian density field is counted towards the final halo, and one does not have to calculate the collapsed volume fraction explicitly.  

When implementing the excursion-set formalism, we start from an infinitely 
large smoothing radius $R$ around any given point in the Universe.
Since the density averaged over this region is the cosmic mean density, the 
density contrast must vanish as $R\to\infty$. 
We then gradually decrease the smoothing length $R$, and trace the evolution of 
the density contrast $\delta(R)$ smoothed at scale $R$.
We proceed further until the smoothed density contrast {\it first} exceeds the 
critical value $\dc(z)$, which signifies that the point belongs to a 
halo with mass given by the smoothing radius. 
The density contrast at any given smoothing 
scale fluctuates around zero with corresponding variance $\sigma^2(R)$. 
Unlike what is done in the Press-Schechter formalism however, excursion-set theory records the ``trajectory''  
$\delta(R)$ starting from $R=\infty$ in order to find the \textit{first-crossing} 
fraction.  In this way, the cloud-in-cloud problem is resolved by counting only {\it isolated} 
virialized regions and keeping track of underdensities within larger virialized
objects.

Trajectories can be thought of as random walk with $R$ as time variable (\reffig{randomwalk}).
Alternatively, they can also be parametrized with the variance
$S(R)\equiv \sigma^2(R)$ of density fluctuations at smoothing scale $R$
[\refeq{def_sigmaRsq}].  For standard cosmological models, $S(R)$ is a monotonically decreasing function 
of the filtering scale $R$ and, therefore, is in one-to-one correspondence with 
$R$ and $M$. 
Hence, we shall hereafter use $M$, $R$ and $S$ interchangeably when we 
describe the halo scale. 
The relation between them is shown in \reffig{MRsigma} for our fiducial 
cosmology.

In order to find the first-crossing probability as a function of either $R$, $S$ or $M$, given the matter power spectrum $\Plin(k)$, one can use Monte-Carlo techniques
\cite{bond/etal:1991,robertson/etal:2009}.  
Analytical solutions of the excursion set are available in some specific cases. 
In particular, two {\it exact} solutions for the excursion set 
formalism are known in two extreme cases, where the statistics of 
density contrast at different smoothing scales are (i) completely independent 
(for example, when employing sharp-$k$ filtering \cite{bond/etal:1991}), 
or (ii) completely correlated.  
Expansions around these exact solutions are often employed in the literature
\cite{maggiore/riotto:2010a,musso/sheth:2012} in order to solve excursion-set 
problems for more realistic filter shapes, such as a real-space tophat.

\subsection{Survival probability, halo mass function, and bias}\label{sec:survival_pdf}
One of the key quantities computed in the excursion-set formalism is the 
{\it differential survival probability} 
$\Pi(\delta;R_0)$ that the density field 
$\delta(R)$ does not exceed the critical value $\dc$ for all smoothing radii 
$R$ greater than $R_0$, and reaches some value $\delta$ at smoothing radius $R_0$: 
$\delta(R_0)= \delta$. 

The knowledge of $\Pi(\delta;R_0)$ leads to the halo mass function as follows.
Firstly, we calculate the probability that the smoothed density contrast never 
exceeded $\dc$ for all radii greater than $R_0$ by integrating the 
differential survival probability $\Pi(\delta;R_0)$ from $-\infty$ to 
$\dc$. Its complement is the probability that  $\delta(R)$ exceeded 
$\dc$ {\it at least once} at some radius $R>R_0$, i.e. it is the fraction 
$F(>M)$ of random field points belonging to halos of mass greater than $M(R_0)$:
\be
F(>M) = 1 - \int_{-\infty}^{\dc} d \d\,\Pi(\d;R_0) \;.
\ee
The differential volume fraction 
\ba
f(M) dM
=\:&
-\frac{dF(>M)}{d M} d M
= 
\frac{d R}{d M}\frac{\partial}{\partial R} 
\left[
\int_{-\infty}^{\dc}d \d \, \Pi(\d;R) 
\right]
d M \;,
\ea
must now be interpreted as a {\it first-crossing distribution}, i.e. the probability
that the smoothed density contrast exceeds the threshold $\dc$ within a mass
range $[M,M+dM]$. The halo mass function follows from \refeq{dndM_PS_bare},
\ba
\avnh(M)
=\:&
\rhob f(M)
=
\rhob
\left[ \frac{dR}{dM}\frac{\partial}{\partial R} 
\int_{-\infty}^{\dc}d\d \, \Pi(\d;R) 
\right] \;.
\label{eq:excursion_nm}
\ea
To calculate the halo bias in the excursion-set formalism, we consider a {\it conditional} 
survival probability in analogy with the peak-background split discussed in \refsec{PBS}.
Namely, the \LIMD bias is given by the response of the number density of collapsed objects 
to a long-wavelength density perturbation $\delta_\ell\equiv\delta(R_\ell)$, where $R_\ell \gg R$ 
is some arbitrary large scale, in the large-scale limit $R_\ell \to \infty$. Therefore, we must compute the fraction $F(>M|\delta_\ell)$ of
overdense regions conditioned on the presence of a long-wavelength density perturbation. 

In line with the aforementioned argument, the number density of collapsed objects with size $R$ 
in the large-scale region with density contrast $\delta_\ell$ is given by the 
{\it conditional survival probability} $\Pi(\d;R|\delta_\ell;R_\ell)$:
\be
\avnh(M|\delta_\ell; R_\ell)
=
\rhob
\left[ \frac{dR}{dM}\frac{\partial}{\partial R} 
\int_{-\infty}^{\dc}d \d \, \Pi(\d;R| \delta_\ell;R_\ell) 
\right] \;.
\label{eq:excursion_nm_deltal}
\ee
The density contrast in Lagrangian space then reads (cf. the discussion in \refsec{localbias}) 
\ba
1+\delta_h^L &\equiv
\frac{\avnh(M|\delta_\ell,R_\ell)}{\avnh(M)} \nonumber \\
&= \frac{f(M|\dell;R_\ell)}{f(M)} =
\left[ \frac{\partial}{\partial R} 
\int_{-\infty}^{\dc}\dd \d \, \Pi(\d;R) 
\right]^{-1}
\left[ \frac{\partial}{\partial R} 
\int_{-\infty}^{\dc}\dd \d \, \Pi(\d;R | \delta_\ell;R_\ell) 
\right] \;.
\label{eq:excursion_bias}
\ea
Here, the conditional probability can be evaluated as
\be
\Pi(\delta;R|\delta_\ell;R_\ell)
=
\frac{\Pi[(\delta;R), (\delta_\ell;R_\ell)]}{\Pi(\delta_\ell;R_\ell)}\,,
\ee
where $\Pi[(\delta;R),(\delta_\ell;R_\ell)]$ is the survival probability that 
the smoothed density contrast is $\delta_\ell$ at the larger smoothing scale
$R_\ell$ ({\it background} scale), and $\delta$ at the smaller smoothing 
$R$ ({\it peak} scale).  
This can be pictorially represented as random walks starting from 
$(\delta,R) = (\d_\ell,R_\ell)$ instead of the origin 
$(\delta,R_\ell) = (0,\infty)$ (see \reffig{exset_vals}).  

Similarly, with the variance $S$ as time variable, the fraction of random walks 
$f(S)$ that first cross the barrier in the range $[S,S+dS]$ is related to the 
halo mass function through
\be
\label{eq:sfs}
f(S) dS = \frac{1}{\rhob} \avnh(M) dM \;.
\ee
Therefore, we have the relation
\be
f(M) dM = f(S) dS = f(\nu_c) d\nu_c \;.
\ee
Since $\nu_c=\dc/\sqrt{S}$, this yields also
\be
\nu_c f(\nu_c) = 2 S f(S) \;.
\ee
Furthermore, the bias coefficients are computed by considering
\begin{equation}
1+\delta_h^L = \frac{f(S|\dell;S_\ell)}{f(S)}\;.
\end{equation}
Note again that the \LIMD bias parameters, which describe the response to an 
infinite-wavelength density perturbation, are obtained in the limit $S_\ell \to 0$.  
The leading correction for a finite $S_\ell$ can be mapped onto a higher-derivative 
bias $b_{\lapl\d}$ \cite{desjacques/crocce/etal:2010,musso/paranjape/sheth:2012}.

\subsection{Numerical solution of excursion sets: Langevin equation}
\label{sec:exset_numerics}

One approach to derive the first-crossing distribution is to simulate a random walk of the density at a fixed location, with a 
smoothing radius decreasing from $R=\infty$ or, equivalently, a variance increasing from $S=0$.  
In this case, the excursion-set formalism reduces to a diffusion problem that 
is described by the Langevin equation where the rate of change of the smoothed 
linear density field $\d(R)$ as a function of smoothing length is given by a 
{\it stochastic force} $\cQF(R)$ 
\cite{bond/etal:1991}
(this force also depends on the position $\vq$, which we suppress throughout since it is arbitrary, but fixed):
\be
\frac{\dd\d(R)}{\dd R} = \cQF(R) \;.
\label{eq:Langevin}
\ee
The stochastic force has a vanishing expectation value $\<\cQF(R)\>=0$, but nonzero variance.  
For any isotropic filtering kernel, the covariance, or two-point function, of the stochastic force is given by
\be
\left<\cQF(R_1)\cQF(R_2)\right>
= \int_{\vk} \Plin(k) \frac{d W_{R_1}(k)}{dR_1} \frac{d W_{R_2}^*(k)}{dR_2} 
 \,.
\label{eq:cQF1cQF2_general_one_pt}
\ee

\subsubsection{Monte-Carlo solution with sharp-$k$ filter}\label{sec:sharpk_numeric}

Solving the Langevin equation is particularly simple for the case of the sharp-$k$ 
filter, because stochastic forces at each step are independent owing to the 
Gaussian nature of the linear density field, and the statistical homogeneity of the 
Universe. For the sharp-$k$ filter, the derivative is 
$
dW_R(k)/dR
= 
\dd\Theta_H(1-kR)/\dd R
= 
- k/R \;\delta_D\left(k-1/R\right)$, where $\Theta_H$ is the Heaviside step function.  
Consequently, \refeq{cQF1cQF2_general_one_pt} becomes 
\be
\left<\cQF(R_1)\cQF(R_2)\right>
=
-\left.
\frac{k^2 \Plin(k)}{2\pi^2}
\right|_{k=\frac{1}{R_1}}
\frac{1}{R_1^2}
\delta_D\left(R_1- R_2\right) \;.
\label{eq:cQF1cQF2_sharpk}
\ee
This suggests rewriting the Langevin equation in terms of $\ln k \equiv -\ln R$:
\be
\frac{\dd\d(R=1/k)}{\dd\ln k}
= \cQF(\ln k) \;,
\label{eq:Langevin_sharpk}
\ee
where
\be
\left<
\cQF(\ln k_1)
\cQF(\ln k_2)
\right>
= \frac{k_1^3\Plin(k_1)}{2\pi^2}\delta_D(\ln k_1-\ln k_2) \;.
\label{eq:QF2_sharpk}
\ee
\refeq{QF2_sharpk} allows us to generate realizations of the stochastic force 
$\cQF(\ln k)$ from a given linear matter power spectrum $\Plin(k)$. 
We can calculate the density contrast 
$\d(R=1/k)$ from the formal solution of \refeq{Langevin_sharpk}, which may be written as
\be
\d(R) = \int_{-\infty}^{- \ln R} \dd(\ln k') \cQF(\ln k') \;,
\label{eq:Langevin_sharpk_solution}
\ee
and from there calculate the survival probability $\Pi(\d;S)$ as the fraction of random 
walks that first cross the barrier $\dc(z)$ between $S$ and $S+dS$.

Note that both the Langevin equation for sharp-$k$ filtering, \refeq{Langevin_sharpk}, 
as well as the stochastic force, \refeq{QF2_sharpk}, do not explicitly depend on the 
step $R$ or the density contrast $\delta_L$. 
Therefore, changing the {\it initial time} $R_\text{init}$ and the density contrast 
$\d(R_\text{init})$ simply re-defines the integration boundaries of 
\refeq{Langevin_sharpk_solution}.   
The usual choice for evaluating the halo mass function is to set the initial value as 
$(R_\text{init},\d(R_\text{init}))=(\infty,0)$, because the density contrast averaged 
over the entire Universe vanishes.

\begin{figure}
\centering
\includegraphics[width=0.9\textwidth]{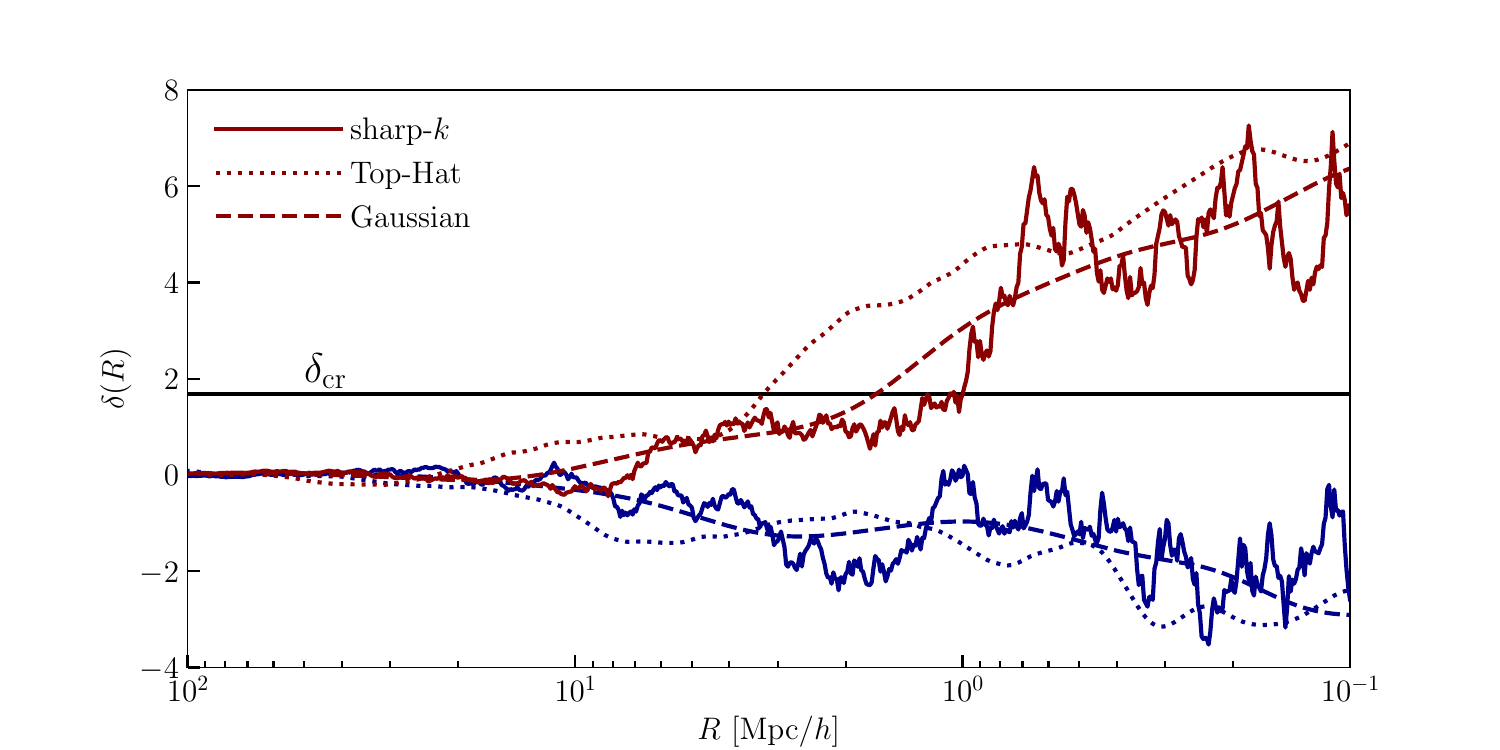}
\caption{
Two sets of three example excursion-set random walks of the Lagrangian density contrast 
$\delta(R)$ as a function of the filter radius $R$. 
Lines with the same color show the same realization of 
excursion-set random walk but with three different filter functions:
sharp-$k$ filter ({\it solid lines}),
Gaussian filter ({\it dashed lines}),
tophat filter ({\it dotted lines}).
While the blue (bottom) realization has not formed a halo for scales 
$R>0.1~[{\rm Mpc}/h]$, the red (upper) realization forms a halo at 
$R = 1 \sim 3~[{\rm Mpc}/h]$ depending on the filter function.
For this calculation, we use 4000 equal steps in logarithmic space 
between $k=10^{-4} h$/Mpc and $10^2 h$/Mpc.
}
\label{fig:randomwalk}
\end{figure}

In practice, the numerical implementation can be done as follows.
Upon discretizing the steps in Fourier space, $\ln k_i = (\Delta \ln k) i$, 
the stochastic force is drawn
from a Gaussian distribution centered around zero and with variance
\be
\sigma_{\cQF}^2 = 
\frac{k_i^3\Plin(k_i)}{2\pi^2 \Delta\ln k} \;.
\ee
Then, a random walk of the density contrast can be calculated by discretizing
the integration as
\be
\delta(R)
= \sum_{k_i< 1/R} 
r_i
\sqrt{\frac{k_i^3P(k_i)}{2\pi^2}\Delta\ln k} 
\qquad\qquad \mbox{(for the sharp-$k$ filter)}\;,
\label{eq:dLxR_sharpk}
\ee
where $r_i$ is a random number drawn from the normal distribution ${\cal N}(0,1)$. 
For illustration, we show in \reffig{randomwalk} two excursion-set random walks 
with the sharp-$k$ filter as solid lines.

\subsubsection{Monte-Carlo solution with general filters}

Dor general filter functions, such as the 
real-space tophat or Gaussian filters, the derivative $dW_R(k)/dR$ is not a Dirac delta in Fourier space, and the stochastic forces 
at different smoothing lengths $\cQF(R_i)$, $\cQF(R_j)$ are no longer 
independent from each other. That is, the stochastic force at each step
$\cQF(R_i)$ depends on the entire set of previous steps
$\{\cQF(R_0),\cQF(R_1),\cdots \cQF(R_{i-1})\}$, so that that the random walk
is {\it non-Markovian}. In principle, for a given filter function and linear
matter power spectrum, one can draw a random Gaussian realization of 
$\cQF(R_i)$ from the full covariance matrix 
$\left<\cQF(R_i)\cQF(R_j)\right>$ given in \refeq{cQF1cQF2_general_one_pt}.

Alternatively, we can exploit the fact that the linear density contrast 
$\delta(\vk)$ is Gaussian, which in particular implies that Fourier amplitudes of different wavenumber are independent. To do so, we first re-write the smoothed linear density 
field in \refeq{deltaLxR} in Fourier space as
\be
\d(R) = \int_{\vk} \d(\vk) W_R(k),
\ee
where we set the Lagrangian coordinate of interest $\vq=0$ without loss of 
generality. Comparing this to the case for the sharp-$k$ filter in \refeq{Langevin_sharpk_solution}, 
it is clear that the stochastic force $\cQF^{\rm sk}(\vk)$ in the sharp-$k$ case
precisely corresponds to the linear density contrast integrated over a spherical shell in Fourier space with fixed logarithmic width $d\ln k$.
Using this, we can write the solution of the Langevin equation in terms of the
filter function $W_R(k)$ and sharp-$k$ stochastic force $\cQF^{\rm sk}(\vk)$
as
\be
\delta(R) =
\int_0^\infty
\dd(\ln k') \cQF^{\rm sk}(\ln k') W_R(\ln k).
\label{eq:Langevin_solution_general}
\ee
Note that the integration runs from $0$ to $\infty$. 
By taking a derivative of \refeq{Langevin_solution_general} with respect to 
$R$, we find the stochastic force $\cQF(R)$ for a general filter function
which satisfies the required covariance matrix given in 
\refeq{cQF1cQF2_general_one_pt}.

For a given set of stochastic forces $\cQF^{\rm sk}(k_i)$, or a given set of random 
variables $r_i$ in \refeq{dLxR_sharpk}, the corresponding 
excursion-set random walk with filter $W_R(k)$ can be calculated
from the discretized version of \refeq{Langevin_solution_general},
\be
\delta(R)
=
\sum_{i} 
r_i
W_R(k_i)
\sqrt{\frac{k_i^3P(k_i)}{2\pi^2}\Delta\ln k}
\qquad\qquad \mbox{(for a generic filter)} \;.
\label{eq:dLxR_Wk}
\ee
Here, $i$ labels the steps in log wavenumber that are sampled.
For comparison, we overlay in \reffig{randomwalk} two random walks obtained
with a Gaussian (dashed lines) and tophat (dotted lines) filter. They were
generated with the same sequence of random numbers $r_i$ used for the sharp-$k$
random walk.
As the random walk at each radius depends on all Fourier modes (the dependence
is determined by the filter function), the random walk with real-space tophat
and Gaussian filters are correlated and, consequently appear much smoother than that 
constructed with the sharp-$k$ filter.

\subsection{Analytical approaches I: excursion set with uncorrelated steps}
\label{sec:excursion_exact}
Although the survival probability, mass function and halo bias can be 
calculated numerically as explained in \refsec{survival_pdf}, there are 
a number of analytical studies of the excursion-set formalism in the 
literature. The analytical approaches help us better understand the 
problem, in addition to providing closed form expressions for the halo 
bias. We shall now review some of the recent developments along this 
direction. Throughout, $S \equiv \s^2(R)$, while $s$ is a dummy time variable.

\subsubsection{Completely independent (Markovian) steps with sharp-$k$ filter}
\label{sec:sharpk_analytic}
Consider first the excursion-set random walks with sharp-$k$ filter.  The real-space kernel corresponding to
this choice is oscillatory and thus very different
from physical Lagrangian density profiles of halos.  
Nevertheless, this case is the most popular in the literature because 
an exact solution to the first-crossing problem can easily be derived.

As we have shown in \refsec{sharpk_numeric}, the shark-$k$ filtering guarantees (for 
Gaussian initial conditions) that the steps of the random walk are fully independent 
or Markovian.  
That is, the density contrast $\delta(s+\Delta s)$ is only determined by the 
density contrast at the previous step $\delta(s)$, and the variance $\Delta s$: 
the steps which the random walk has gone through to reach $\delta(s)$ are irrelevant.  
When performing the step from $s$ to $s+\Delta s$ (i.e. the filter decreases 
from $R$ to $R-\Delta R$), the difference $\delta(s+\Delta s)-\delta(s)$ in
the density contrast is a Gaussian random variate with zero mean and variance  
$\Delta s$.  Only Fourier modes in the range $k \in [1/R,\,1/(R-\Delta R)]$ contribute 
to this random variate, proving that the step is independent from all the others.

Since the sum of independent Gaussian random variables is also a Gaussian with a variance 
given by the sum of individual variances, the probability distribution of trajectories 
reaching $\delta(S) = \delta$ at time $S$ is
\be
p_G(\delta)d\delta 
=
\frac{1}{\sqrt{2\pi S}}e^{-\delta^2/2S}\dd\d \;.
\label{eq:P_delta_at_S}
\ee
Then, the differential survival probability is given by \cite{chandrasekhar:1943}
\be
\Pi(\delta;S)
=\left\lbrace\begin{array}{cc}
\frac{1}{\sqrt{2\pi S}}
\left[
e^{-\delta^2/2S}
-
e^{-(2\dc-\delta)^2/2S}
\right] & \mbox{$(\d<\dc)$} \\
0 & \mbox{$(\d\geq\dc)$}
\end{array}\right.
\label{eq:Pi_delta_at_S}.
\ee
This can be understood as follows. The total fraction of trajectories attaining 
$(S,\delta)$ with $\d<\dc$ is given in \refeq{P_delta_at_S}, which is the first 
term in \refeq{Pi_delta_at_S}. 
However, not all trajectories passing through $(S,\delta)$ always remained below 
the threshold  $\dc$ at earlier time.
Therefore, we must subtract all paths which crossed the threshold 
$\dc$ for some $S'<S$. To estimate this fraction, consider a path reaching 
$(\delta,S)$ after crossing $\dc$ at an earlier time $S'<S$. 
We can construct another path which arrives at $(S,\dc+(\dc-\delta))$ 
and which, at time $S'<s<S$, is the exact reflection of the original path off the 
``mirror'' $\d=\dc$. Clearly, this reflection establishes a one-to-one 
correspondence between trajectories ending up at $(S,2\dc-\delta)$ and the
paths that reach $(S,\delta)$ after crossing the threshold at time $S'<S$.
Therefore, we must subtract $p_G(2\dc-\d)$ from $p_G(\d)$, which is nothing but
\refeq{Pi_delta_at_S}.

The survival probability is thus given by
\be
\int_{-\infty}^{\dc} \dd\d\,\Pi(\d;S) = 
\int_{-\infty}^{\dc}
\frac{1}{\sqrt{2\pi S}}
\left[
e^{-\delta^2/2S}
-
e^{-(2\dc-\delta)^2/2S}
\right]
\dd\d
=
\erf\left[
\frac{\dc}{\sqrt{2S}}
\right]\,,
\ee
from which we calculate the collapsed fraction as
\be
F(>M)
=
1-
\int_{-\infty}^{\dc} \dd\d\,\Pi(\d;S) 
= 
\erfc\left[
\frac{\dc}{\sqrt{2S}}
\right].
\label{eq:Fcoll_sharpk}
\ee
This yields the halo mass function, via \refeq{excursion_nm},
\be
\overline{n}_{h,\rm sk}(M)
=
\frac{\rhob}{M} \nu_c f_{\rm sk}(\nu_c)
\left|\frac{d\ln \sigma(R)}{d\ln M}\right| \;,
\label{eq:nm_sharpk}
\ee
where
\be
\nu_c f_{\rm sk}(\nu_c) = \sqrt{\frac{2}{\pi}} \nu_c e^{-\nu_c^2/2} \;.
\label{eq:nufnu_sharpk}
\ee
This shows that the Markovian (sharp-$k$) excursion-set formalism recovers the 
Press-Schechter mass function [\refeq{PS_massfn}] with the correct normalization 
factor.

\subsubsection{Halo bias with sharp-$k$ filter}
\label{sec:sharpk_bias}
In order to calculate the Lagrangian bias coefficients, we need to evaluate the conditional
survival probability
\be
\Pi(\delta;S|\delta_\ell;S_\ell)
=
\frac{\Pi\left[(\delta;S),(\delta_\ell;S_\ell)\right]}{\Pi(\delta_\ell;S_\ell)} \;.
\ee
For the case of a sharp-$k$ filter, each step of the random walk is independent so that the 
trajectory is invariant under translation---that is, it does not matter where the walk has 
started. Consequently, the differential survival probability is given by
\be
\Pi\left[(\delta;S),(\delta_\ell;S_\ell)\right]
=
\Pi(\delta-\delta_\ell;S-S_\ell)
\Pi(\delta_\ell;S_\ell).
\ee
We can then calculate the conditional collapsed fraction from a simple rescaling of 
\refeq{Fcoll_sharpk},
\be
F(>M|\dell;S_\ell)
=
\erfc\left[
\frac{\dc-\dell}{\sqrt{2(S-S_\ell)}}
\right] \;.
\ee
The conditional mass function immediately follows,
\be
\overline{n}_{h,\rm sk}(M|\dell)
=\frac{\rhob}{M}\,2 S f_{\rm sk}(S|\dell;S_\ell)\,
\left|\frac{\dd\ln\sigma(R)}{\dd\ln M}\right| \;,
\ee
with a conditional first-crossing distribution $f_{\rm sk}(S|\dell;S_\ell)$ given by
\be
f_{\rm sk}(S|\dell;S_\ell) = \frac{1}{\sqrt{2\pi}}
\frac{\dc-\dell}{(S-S_\ell)^{3/2}}
\exp\left[-\frac{(\dc-\dell)^2}{2(S-S_\ell)}\right] \;.
\ee
Therefore, the halo density contrast in a Lagrangian region of size $R_{\ell}$ and 
overdensity $\d_\ell$ is
\be
1+\delta_h^{L}(\dell)
\equiv
\frac{f_{\rm sk}(S|\dell,S_\ell)}{f_{\rm sk}(S)}
=
\left(1-\frac{\dell}{\dc}\right)
\left(\frac{S}{S-S_{\ell}}\right)^{3/2}
\exp
\left[
- \frac{(\dc-\dell)^2}{2(S-S_\ell)}
+
\frac{\dc^2}{2S}
\right],
\ee
which, in the limit where $R_\ell\to\infty$ and $S_\ell\to 0$, reduces to the Lagrangian \LIMD bias expansion:
\be
1+\delta_h^L(\dell)
=
\left(1-\frac{\dell}{\dc}\right)
\exp
\left[
\frac{(2\dc-\dell)\dell}{2S}
\right]
=
1 
+ \frac{\dell}{\sigma(R)}\left(\nu_c - \frac{1}{\nu_c}\right)
+ \frac{1}{2}
\frac{\dell^2}{\sigma^2(R)}
\left(
\nu_c^2 - 3
\right)
+ \mathcal{O}(\dell^3),
\label{eq:sharpk_bias}
\ee
The linear- and second-order coefficients of $\dell$ in 
\refeq{sharpk_bias} are, respectively, the linear- and the second-order 
Lagrangian \LIMD bias parameters.

Let us compare this with the bias parameters predicted by the peak-background
split following \refsec{buniv}, which are given by
\be
b_n^L = (-1)^n \frac{1}{\overline{n}_{h,\rm sk}(M)}\frac{\partial^n \overline{n}_{h,\rm sk}(M)}{\partial\dc^n}\,.
\ee
It is easy to see, via \refeqs{nm_sharpk}{nufnu_sharpk}, that these are exactly the same as \refeq{sharpk_bias},
the biases derived in the excursion set from the conditional first crossing distribution in the large-scale limit $S_\ell \to 0$.  Explicitly, the first
two bias parameters are the same as \refeq{bLPS},
\ba
\label{eq:b1Lsk}
(b_1^L)_\text{sk} = (b_1^L)_\text{PS}
=\:& - \frac{1}{n_{h,\rm sk}(M)}\frac{\partial \overline{n}_{h,\rm sk}(M)}{\partial\dc}
= \frac{1}{\sigma(R)}
\left(
\nu_c - \frac{1}{\nu_c} 
\right)
\\
\label{eq:b2Lsk}
(b_2^L)_\text{sk} = (b_2^L)_\text{PS}
=\:& \frac{1}{n_{h,\rm sk}(M)}\frac{\partial^2 \overline{n}_{h,\rm sk}(M)}{\partial^2\dc}
= 
\frac{1}{\sigma^2(R)}
\left(
\nu_c^2-3
\right)\,.
\ea 
Note that, in the limit $\sigma(R)\to\infty$, the linear bias $b_1^L$ tends towards $-1/\dc$.  That is, unlike the simple thresholding procedure 
described in \refsec{localbias}, low mass objects are predicted to be anti-biased. This can be traced to the first-crossing constraint, which results in 
low-mass {\it isolated} virialized objects being preferentially found in underdense regions.

\subsubsection{Expanding around the Markovian (sharp-$k$) solution*}
\label{sec:exsetMR}
\begin{figure}
\centering
\includegraphics[width=\textwidth]{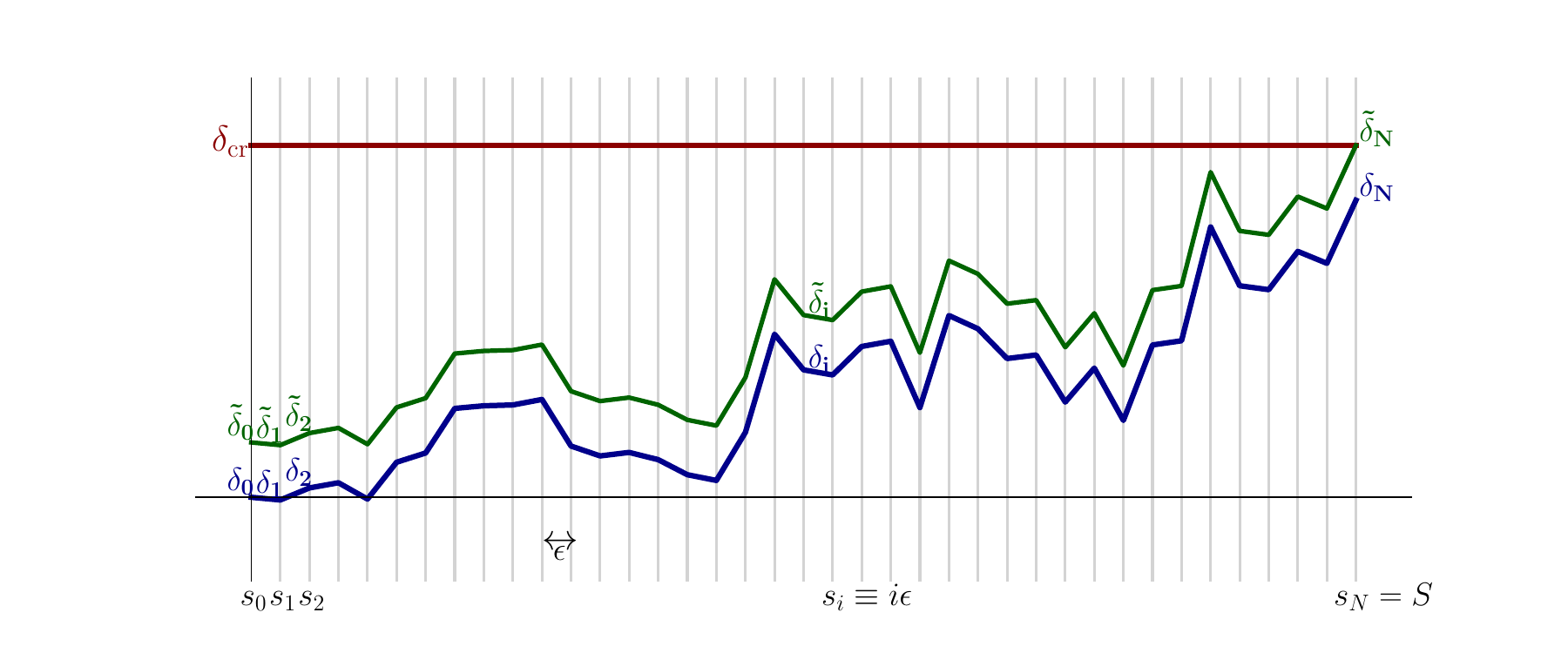}
\caption{
Illustration of the variables used in \refsec{exsetMR}.  
We show two excursion-set paths with exactly the same realization of 
the local density field, but with different large-scale environment.
The blue path starts from $\delta(S_0=0)=0$, corresponding to an environment at the cosmic mean density, 
while the green path starts from $\delta(S_0=0)=\tilde\delta_0> 0$, corresponding to an overdense environment.
} 
\label{fig:exset_vals}
\end{figure}

\technote{* This section is of a more technical nature and is not essential for the remainder of the review.  The main result of this section are 
\refeqs{MRnM}{MRb2L}.}

Above we have obtained an analytical solution for the excursion set 
in the Markovian case, i.e. with a sharp-$k$ filter. For other filters, 
the covariance matrix of the stochastic force $\cQF$ [\refeq{Langevin}] 
is non-diagonal, which means that all steps at different filter radii 
are correlated. 
As a result, the random walk is generally non-Markovian.  
We now describe generalizations of the excursion-set procedure that are 
able to deal with these non-Markovian walks, as long as the cross-correlation between steps of the random walk remains small.

An approximation due to \cite{maggiore/riotto:2010a,desimone/maggiore/riotto:2011,ma/etal:2011} 
describes the correlated random walks in terms of a path-integral of discretized trajectories
 (with step size $\epsilon\equiv s_{i+1}-s_i$, that is, $s_i = i\epsilon$) starting 
from $\d_0\equiv \d(s=0)$ and reaching $s_N\equiv S$.
They first write down a joint probability distribution function 
${\cal W}(\d_0;\,\d_1,\cdots \d_N)$ in the $N$-dimensional space of random walk trajectories  
$\delta_i \equiv \delta(s_i)$.  This is given by
(see \reffig{exset_vals} for an illustration and definitions of symbols)
\ba
{\cal W}(\delta_0;\delta_1,\cdots,\delta_N)
=\:&
\left<
\delta_D\!\big(\delta(s_1)-\delta_1\big)
\cdots
\delta_D\!\big(\delta(s_N)-\delta_N\big)
\right>
\vs
=\:&
\int_{-\infty}^{\infty}\frac{d\lambda_1}{2\pi}
\cdots
\int_{-\infty}^{\infty}\frac{d\lambda_N}{2\pi}
e^{
i\sum_{i}\lambda_i\delta_i}
\left<
e^{-i\sum_{i}\lambda_i\delta(s_i)}\right>\;,
\label{eq:def_Wdelta}
\ea
where we have used the Fourier representation of the Dirac delta distribution $\delta_D(x)$.  
The differential survival probability $\Pi[(\delta_0;s_0=0),(\d_N;s_N)]$ 
immediately follows 
from ${\cal W}(\d_0;\d_1,\cdots,\d_N)$ since it is the probability that the path 
$(\d_1,\cdots,\d_{N-1})$ never exceeded the threshold density contrast $\dc$ 
along the path:
\be
\Pi_\epsilon(\d_0;\d_N;s_N)
=
\int_{-\infty}^{\dc} \dd\d_1
\cdots
\int_{-\infty}^{\dc} \dd\d_{N-1}
{\cal W}(\d_0;\d_1,\cdots,\d_N) \;.
\label{eq:MR_Pi_}
\ee
Here, we define the shorthand notation of 
$\Pi_\epsilon(\d_0;\d_N;s_N)\equiv \Pi_\epsilon[(\d_0;s_0),(\d_N;s_N)]$
where the subscript $\epsilon$ in 
\refeq{MR_Pi_} reminds us that we are dealing with a discrete random
walk, and we have to eventually take the limit $\epsilon\to0$ at the end of the
calculation.

Because $\delta(s_i)$ follow Gaussian statistics, the random variate
$X \equiv i\sum_i\lambda_i\delta(s_i)$ is also Gaussian distributed.
Therefore, we can simplify further the expectation value of the 
exponential\footnote{
For a Gaussian random variable $X$, 
$\left<e^X\right> = \exp\left[\frac12\left<X^2\right>\right]$.
One way to prove this relation is to Taylor-expand the left-hand side and 
use Wick's theorem: 
$\left<X^{2n}\right> = (2n-1)!!\left[\left<X^2\right>\right]^n$.
} 
in \refeq{def_Wdelta} to
\be
\left<
\exp\left[{-i\sum_{i}\lambda_i\delta(s_i)}\right]\right>
=
\exp
\left[\frac12
\left<\left(i\sum_i\lambda_i\delta(s_i)\right)^2
\right>
\right]
=
\exp
\left[-\frac12
\sum_{ij}\lambda_i\lambda_j
\left<
\delta(s_i)
\delta(s_j)
\right>
\right].
\label{eq:expX}
\ee
The differential survival probability then reads, by combining
\refeq{def_Wdelta} and \refeq{expX},
\begin{align}
\Pi_\epsilon(\d_0;\d_N;s_N)
&=
\int_{-\infty}^{\dc} \dd\d_1
\cdots
\int_{-\infty}^{\dc} \dd\d_{N-1}
\int_{-\infty}^{\infty}\frac{d\lambda_1}{2\pi}
\cdots
\int_{-\infty}^{\infty}\frac{d\lambda_N}{2\pi}
\exp
\left[
i\sum_{i=1}^N \lambda_i\d_i
-
\frac12\sum_{i,j=1}^N \lambda_i\lambda_j 
\left<\d(s_i)\d(s_j)\right>
\right]  \;.
\label{eq:MR_Pi_epsilon0}
\end{align}
Note that, once the differential survival probability $\Pi_\epsilon(\d_0;\d_N;s_N)$ 
is known, the survival probability, collapsed fraction $F(>M)$ and 
mass function $f(M)$ follow from the steps described in \refsec{survival_pdf}.
Also note that, for the sharp-$k$ filter, each increment in 
$\delta(s_i)$ is independent. Therefore, the two-point correlator is given by
\be
\left<
\delta(s_i)
\delta(s_j)
\right>
=
{\rm min}\left(s_i,s_j\right)
=
\epsilon\,{\rm min}(i,j),
\ee
and the PDF can be calculated analytically: 
\be
\Pi_{\epsilon}^{\rm sk}(\d_0;\delta_N;s_N)
=
\int_{-\infty}^{\dc}
\,\dd \d_{N-1}
\frac{1}{\sqrt{2\pi\epsilon}}
\exp\left[-\frac{(\delta_N-\delta_{N-1})^2}{2\epsilon}\right]
\Pi_{\epsilon}^{\rm sk}(\d_0;\delta_{N-1};s_{N-1}).
\ee
This equation is indeed consistent with the fact that the sharp-$k$ filtering 
yields a Markovian process and, therefore, yields the differential survival probability 
\refeq{Pi_delta_at_S}. 

One intermediate result for the sharp-$k$ filter case shown in 
\cite{maggiore/riotto:2010a} is the finite-$\epsilon$ correction.
The $\epsilon\to0$ limit solution in \refeq{Pi_delta_at_S} vanishes 
when the random walk ends at the critical density $\delta=\dc$, 
$\Pi_\epsilon^{\rm sk}(\d_0;\dc;S)= 0$, or begins from the critical density
$\Pi_\epsilon^{\rm sk}(\dc;\d;S)=0$. With the finite step size $\epsilon$, 
Ref.~\cite{maggiore/riotto:2010a} has shown that 
the leading corrections for these cases are of order $\sqrt{\epsilon}$. 
Namely,
\ba
\Pi_\epsilon^{\rm sk}
(\delta_0;\dc;S) =\:& 
\sqrt{\frac{\epsilon}{\pi S^3}}
\left(
\dc-\d_0
\right)
\exp\left[-\frac{(\dc-\d_0)^2}{2S}\right] + \mathcal{O}(\epsilon)
\label{eq:finite_epsilon_d0dc}
\\
\Pi_\epsilon^{\rm sk}
(\dc;\delta;S) =\:&
\sqrt{\frac{\epsilon}{\pi S^3}}
\left(
\dc-\d
\right)
\exp\left[-\frac{(\dc-\d)^2}{2S}\right] + \mathcal{O}(\epsilon)
\label{eq:finite_epsilon_dcd}
\ea
if $\d_0$ and $\d$ are both smaller than $\dc$.
The finite-$\epsilon$ corrections are necessary to calculate the 
non-Markovian correction because the same parameter $\epsilon$ appears when
converting a finite sum, for example, in \refeq{MR_Pi_epsilon}, to the 
continuous limit as $\sum\to1/\epsilon\int ds$.

For the general filters other than the sharp-$k$ filter, different time 
steps are correlated, and the two-point correlators may be written as 
(for $0\le s_i\le s_j$)
\be
\left<
\delta(s_i)
\delta(s_j)
\right>
=
s_i % {\rm min}\left(s_i,s_j\right) % FS: since by assumption s_i < s_j in this relation
+
\Delta(s_i,s_j)
\simeq
s_i % {\rm min}\left(s_i,s_j\right)
+
\kappa\frac{s_i(s_j-s_i)}{s_j},
\label{eq:dLdL_nm}
\ee
where $\Delta(s_i,s_j)\equiv\Delta_{ij}$ is the cross-correlation 
between different smoothing scales ($\Delta_{ij}=0$ for the sharp-$k$ case). 
Therefore, $\kappa$ parametrizes the deviation from the sharp-$k$ case for
which we have an analytical solution.
For the spherical tophat filter (which \cite{maggiore/riotto:2010a} refers
to as {\it sharp-x} filter), 
$\kappa\simeq 0.4592-0.0031\left(R/[1{\rm Mpc}/h]\right)$ whereas, for the 
Gaussian filter, $\kappa\simeq 0.35$ for a $\Lambda$CDM linear power spectrum.
Next, Ref.~\cite{maggiore/riotto:2010a} integrated
\refeq{MR_Pi_epsilon0} with \refeq{dLdL_nm} by treating 
$\Delta_{ij}$ (and therefore $\kappa$) as a small perturbation.  
To linear order in $\kappa$, the differential survival probability is 
\ba
\Pi_\epsilon(\d_0;\d_N;s_N)
=\;
\Pi_{\epsilon}^{(0)}(\d_0;\d_N;s_N)
+
\frac12\sum_{i,j=1}^N 
\Delta_{ij}
\int_{-\infty}^{\dc} \dd\d_1
\cdots
\int_{-\infty}^{\dc} \dd\d_{N-1}
\frac{\partial}{\partial\delta_i}
\frac{\partial}{\partial\delta_j}
{\cal W}^{\rm sk}(\d_0;\d_1,\cdots,\d_N;s_N)
\label{eq:MR_Pi_epsilon},
\ea
where the leading-order term $\Pi_{\epsilon}^{(0)}$ is the same expression 
as the sharp-$k$ filtering case, which reduces to \refeq{Pi_delta_at_S} 
in $\epsilon\to0$ limit.  
$\mathcal{W}^{\rm sk}$ is the probability density function of the
sharp-$k$ case, that is, a multivariate Gaussian with diagonal covariance
matrix, where the variance $s(R_i)$ is calculated with the general filter of 
interest. Note that $s(R_i) = \sigma^2(R_i)$ depends on the shape of filter chosen (\reffig{MRsigma}).  

The perturbative expansion is justified as the maximum of
$\Delta_{ij}/s_i\simeq \kappa(1-s_i/s_j)<\kappa$ is less than unity; 
at the same time, however, truncation at the leading order may not lead to an
accurate approximation as $\kappa\simeq 0.3-0.4$.
The linear-order result is, in the $\epsilon\to0$ limit, given by
\ba
\Delta\Pi(\d;S)
=\:&
\kappa\frac{\partial}{\partial \delta}
\left[
\frac{\dc(\dc-\d)}{S}
\erfc\left(\frac{2\dc-\d}{\sqrt{2S}}\right)
\right]
\vs
&+
\frac{\kappa \dc}{\sqrt{2\pi S}}
\frac{\partial}{\partial \delta}
\left[
e^{-(\dc-\d)^2/(2S)}\int_0^S\frac{ds}{s}
e^{-\dc/(2s)}
\erfc\left((\dc-\d)\sqrt{\frac{s}{2S(S-s)}}\right)
\right],
\ea
which, on using \refeq{excursion_nm}, yields the halo mass function
\be
\avnh(M) = 
\sqrt{\frac{2}{\pi}}
\frac{\rhob}{M}
\left[
(1-\kappa)
\nu_c e^{-\nu_c^2/2}
+
\frac{\kappa\nu_c}{2}\Gamma\left(0,\frac{\nu_c^2}{2}\right)
\right]
\left|\frac{d\ln \sigma}{d\ln M}\right|,
\label{eq:MRnM}
\ee
where the incomplete Gamma function is 
$\Gamma(0,x)=\int_x^\infty e^{-t} \dd\ln t$.  
This mass function has been found to be in good agreement with simulations
\cite{corasaniti/achitouv:2011,achitouv/rasera/etal:2013}.
The linear and quadratic Lagrangian bias parameters can be calculated from 
the peak-background split approach in the same way as in \refeq{b1Lsk},
leading to\footnote{This expression corrects a typo in \cite{ma/etal:2011}.}
\ba
b_1^L(\nu_c)
=\:&
\frac{1}{\dc\left[1-\kappa+\frac{\kappa}{2}e^{\nu_c^2/2}\Gamma(0,\nu_c^2/2)\right]}
\left\{
(1-\kappa)(\nu_c^2 - 1) 
+ \frac{\kappa}{2}\left[
2- e^{\nu_c^2/2}\Gamma\left(
0,\frac{\nu_c^2}{2}
\right)
\right]
\right\},
\label{eq:MRb1L}\\
b_2^L(\nu_c) =\:&
\frac{1}
{\dc^2\left[1-\kappa+\frac{\kappa}{2}e^{\nu_c^2/2}\Gamma(0,\nu_c^2/2)\right]}
\left\{(1-\kappa)\nu_c^4 +(4\kappa-3)\nu_c^2-\kappa\right\}\,,
\label{eq:MRb2L}
\ea
where the appropriate value for $\kappa$ should be inserted, depending on the filter used.

\subsection{Analytical approaches II: excursion set with correlated steps}
\label{sec:excursion_HeavensPeacock}

Another method to computing the first-crossing probability as a function of 
the filtering scale $R$ for general (non sharp-$k$) filters 
is to expand around the completely-correlated solution
(the original Press-Schechter result)
\cite{paranjape/sheth:2012,paranjape/lam/sheth:2012,musso/sheth:2012,
musso/paranjape/sheth:2012,musso/sheth:2014}.
This method originated from the observation of \cite{peacock/heavens:1990} 
that the trajectories $\delta(R)$ can, to a first 
approximation, be decomposed into a series of independent steps 
(in $\ln R$) of finite size $\Delta$. 
Here, the critical step size $\Delta$ is a correlation length, i.e. 
$\delta(R)$ 
at two different smoothing scales $R_1$ and $R_2$ are strongly correlated if
$|\ln R_1-\ln R_2|\lesssim \Delta$, and weakly correlated if
$|\ln R_1-\ln R_2|\gtrsim \Delta$. In the limit $\Delta\to 0$, we recover the
Markovian walks (sharp-$k$ filtering case, \refsec{sharpk_analytic})
whereas, in the opposite limit $\Delta\to\infty$, the walks are referred 
to as completely correlated. 

\subsubsection{First-crossing with completely correlated steps}
\label{sec:completely_correlated}
As noted in \cite{paranjape/lam/sheth:2012}, an example of completely
correlated walks is the set of straight lines passing through the origin and 
a point $(\delta,S)$.
These curves are fully characterized by the slope $\nu=\delta/\sqrt{S}$. The first-crossing distribution associated with this 
ensemble of walks will depend on the assumed distribution of $\nu$. If this 
distribution is Gaussian, $p(\nu)=\exp(-\nu^2/2)/\sqrt{2\pi}$, then the 
corresponding survival probability for the constant barrier of height $\dc$ 
is given by
\begin{equation}
\int_{-\infty}^{\dc} \dd\d\,\Pi(\d;S) = 
\int_{-\infty}^{\dc/\sqrt{S}}\dd\nu\,p(\nu)=
\frac{1}{2}\Big(1+\erf\big(\dc/\sqrt{2S}\big)\Big) \;.
\end{equation}
Consequently, the fraction of walks first crossing the barrier in the range 
$(S,S+dS)$ 
is one half of the solution for completely uncorrelated steps,
\begin{equation}
S f(S) 
= \frac{1}{2}S f_{\rm sk}(S)
= \frac{1}{2} \frac{\dc}{\sqrt{2\pi S}}e^{-\dc^2/2S} \;,
\end{equation}
i.e. the original Press-Schechter solution \cite{press/schechter:1974} 
(see \refsec{PS}) with a halo mass function of
\be
\avnh(M) = 
\frac{\rhob}{M}
\frac{\dc}{\sqrt{2\pi S}}
e^{-\dc^2/2S}
\left|\frac{d\ln\sigma}{d\ln M}\right|.
\label{eq:fxPS}
\ee 
This expression characterizes very well the first-crossing distribution at 
small $S$ (that is, for high-mass halos), even when the steps are not fully 
correlated \cite{musso/paranjape:2012}. 
Comparison with Monte-Carlo realizations of random walks show that, at small $S$, 
the first-crossing distribution also asymptotes to $1/2$ times the corresponding
distribution for sharp-$k$ walks when the shape of the underlying power spectrum
is varied, or primordial non-Gaussianity (PNG) is added \cite{musso/paranjape:2012}.

PNG generates a skewness in the initial density field which affects the first-crossing distribution 
and, thus, the halo mass function.  This effect is largest at high mass because the tails of the 
density PDF $p(\d_R)$ are very sensitive to any non-zero skewness in the initial conditions
\citep{lucchin/matarrese:1988,colafrancesco/lucchin/matarrese:1989,coles/barrow:1987}.
In practice, non-Gaussian corrections to the initial density PDF can be computed
using e.g. saddle-point techniques and Edgeworth expansions
\citep{matarrese/verde/jimenez:2000,loverde/miller/etal:2008,maggiore/riotto:2010b}.  
We will not discuss this issue any further here, and refer the reader to 
\citep{desjacques/seljak:2010b} for additional details on the halo mass function in the presence of PNG.  
The effect of PNG on the clustering of biased tracers in the context of the excursion set is the topic 
of \refsec{NGexset}.  

\subsubsection{Up-crossing probability distribution function*}
\label{sec:completely_correlated_derivation}
\technote{* This section is of a more technical nature and is not essential 
for the remainder of the review. Readers can go directly to 
\refsec{completely_correlated_bias} where the halo bias obtained from this method is discussed.}

In order to calculate the survival probability, one must in 
principle enforce the constraint $\delta(s)<\dc$ for all scales $s<S$.
When the steps are strongly (yet not perfectly) correlated, however,
Refs.~\cite{musso/sheth:2012,paranjape/sheth:2012} (see also 
\cite{appel/jones:1990}) have suggested to replace this condition with the 
milder requirement that {\it the first crossing must happen from below}:
$\delta(S-\Delta S)<\dc$ and $\delta(S)>\dc$.
In the limit of small increment $\Delta S$, the condition of first crossing
happening from below is equivalent to
\begin{equation}
\label{eq:dscross}
\dc < \delta(S) < \dc + \delta' \Delta S\;,
\end{equation}
where a prime denotes a derivative with respect to $S$.
For this condition to make sense, we must have $\delta'\geq 0$. 

Therefore, when identifying the fraction of walks which {\it up-cross} 
$\dc$ in the range $(S,S+\Delta S)$ as the fraction of Lagrangian volume
enclosed in halos of the corresponding mass $M$, the multiplicity function is given by
\begin{align}
f(S)\Delta S = \int_0^\infty\!\!d\delta'\int_{\dc}^{\dc+\delta'\Delta S}\!\!d\delta\,
P(\delta,\delta') 
\approx \Delta S\, \int_0^\infty\!\!d\delta' \, \delta'\,P(\delta=\dc,\delta') 
\label{eq:fxstrongcorr1}\;.
\end{align}
Here, the joint probability distribution of $\d$ and $\d'$ is given by 
\be
P(\d,\d') = \frac{\gamma}{\pi\sqrt{1-\gamma^2}}
\exp\left[
-\frac{2\gamma^2}{1-\gamma^2}\left\{
\frac{\delta^2}{4S\gamma^2} - \delta\delta' + S \delta'^2
\right\}
\right],
\ee
where 
\begin{equation}
\gamma^2
\equiv 
\frac{\left[\la\delta(S)\delta'(S)\ra\right]^2}{\la [\delta(S)]^2 \ra\, \la[\delta'(S)]^{2}\ra}
 \;,
\label{eq:gamma}
\end{equation}
and we have $\left<\delta(S)\delta'(S)\right>=1/2$ since $\delta'(S)$ is
the derivative of $\d(S)$ with respect to $S=\< \d^2(R[S])\>$. Then,
the conditional probability $P(\delta'|\dc)$ has mean
$\la\delta'\big\lvert\dc\big\ra 
= \dc/2S$ 
and variance $\Var(\delta'|\dc)=
1/(4S\Gamma^2)$, where $\Gamma^2 \equiv \gamma^2/(1-\gamma^2)$.  
The integral in \refeq{fxstrongcorr1} can, then, be easily performed so
that the first-crossing distribution reads
\begin{align}
S f(S) 
&= \frac{e^{-\dc^2/2S}}{2\sqrt{2\pi S}}\dc
\bigg[\frac{1+\erf\big(\Gamma\dc/\sqrt{2S}\big)}{2}
+\frac{e^{-\Gamma^2\dc^2/2S}}{\sqrt{2\pi/S}\,\Gamma\dc}\bigg]\;.
\label{eq:fxstrongcorr2}
\end{align}

The bias parameters can be correspondingly computed by considering the 
first crossing distribution with the additional constraint 
that the excursion-set random walks went through $\delta_\ell\equiv\d(S_\ell)$ 
on the large scale $S_\ell\ll S$, analogously to the Markovian case 
(\refsec{survival_pdf}). The conditional first crossing can be calculated 
similarly to \refeq{fxstrongcorr1} except that we now need to integrate
the conditional probability function $P(\delta,\delta'|\delta_\ell)$
\ba
f(S|\dell;S_\ell)\Delta S 
=
\int_0^{\infty} d\d'
\int_{\dc}^{\dc+\d'+\Delta S}
d\d
P(\delta,\delta'|\delta_\ell)
\approx 
\Delta S \,
P(\dc|\d_\ell)
\int_0^\infty d\d'\,
\d'
P(\delta'|\dc,\delta_\ell).
\ea
Upon taking the limit $\Delta S\to 0$, we obtain
\begin{equation}
f(S|\dell;S_\ell) = P(\dc|\dell) \int_0^\infty\!\!d\delta'\,\delta'\,
P(\delta'|\dc,\dell)
\end{equation}
for the constant deterministic barrier that we are considering here. 
In order to obtain the conditional probability, we use that the joint probability of 
$(\d,\d',\dell)$ is a trivariate Gaussian 
\cite{musso/paranjape/sheth:2012} specified by the parameters that we have 
defined earlier and 
the variance of the long-wavelength fluctuations
$S_\ell\equiv\left<\dell^2\right> = \left<\delta^2(R_\ell)\right>$,
the covariance of long- and short-wavelength fluctuations,
$S_\times\equiv\left<\d(S)\d(S_\ell)\right>$,
and the cross-correlation coefficient between $\d'(S)$ and $\d(S_\ell)$,
$\epsilon_\times=2S\left<\d'(S)\d(S_\ell)\right>/S_\times$.  
Then, the mean and variance of the conditional Gaussian distribution function 
$P(\d'|\dc,\dell) = P(\d',\dc|\dell)/P(\dc|\dell)$ are
\ba
\bar\d'(R)\equiv
\left<\d'(R)|\dc,\dell\right>
=\:&
\frac{1}{2SQ}
\left[
\d_{c\times}
+
\epsilon_\times
\frac{S_\times}{S_\ell}
\left(\dell - \dc \frac{S_\times}{S_\ell}\frac{S_\ell}{S}\right)
\right]
\;,
\ea
and
\ba
\bar\sigma^2(R) \equiv
\Var\left(\d'(R)|\dc,\dell\right)
=\:&
\frac{1}{4S\Gamma^2}
\left[
1 - \frac{\Gamma^2 S_\ell}{QS}\frac{S_\times^2(1-\epsilon_\times)^2}{S_\ell^2} 
\right]
\;,
\ea
with
\be
\d_{c\times}\equiv\dc - \frac{S_\times}{S_\ell}\dell\;, \qquad
Q \equiv 1-\frac{S_\times^2}{SS_\ell} 
= 1-\left(\frac{S_\times}{S_\ell}\right)^2\frac{S_\ell}{S}\;.
\ee
Finally, integrating over Gaussian conditional distribution function 
$P(\d'|\dc,\dell)$ yields the conditional first-crossing probability 
distribution function as
\citep{musso/paranjape/sheth:2012}
\ba
Sf(S|\dell;S_\ell)
=\:&
SP(\dc|\dell)
\int_0^\infty d\d' \d' P(\d'|\dc,\dell)
\vs
=\:& S\frac{\bar\delta' e^{-\delta_{c\times}^2/2SQ}}{\sqrt{2\pi SQ}}
\bigg[\frac{1+ \erf\big(\bar\delta'(S)/\sqrt{2}\bar\sigma(S)\big)}{2}
+\frac{e^{-\bar\delta^{'2}(S)/2\bar\sigma^2(S)}}{\sqrt{2\pi}(\bar\delta'(S)/\bar\sigma(S))}\bigg]\;.
\label{eq:correlated_cond_SfS}
\ea
Note that explicit expressions for $S_\times$ and $\epsilon_\times$ can be derived for e.g. power-law spectra with
Gaussian filter. In the case of the sharp $k$-space filter, $S_\times/S_\ell=1$ and $\epsilon_\times=0$.
Although $f(S|\dell;S_\ell)$ in \refeq{correlated_cond_SfS} remains positive 
even when $\dell>\dc$, we need to impose the condition $\dell\leq\dc$ in order 
to satisfy the ``first crossing'' criterion.

\subsubsection{Halo mass function and bias}
\label{sec:completely_correlated_bias}
The first-crossing PDF 
in \refeq{fxstrongcorr2} can be directly translated to the halo mass function
through
\begin{align}
\avnh(M) =
\frac{\rhob}{M}
2S f(S) 
\left|
\frac{d\ln \sigma}{d\ln M}
\right|
= 
\frac{\rhob}{M}
\frac{e^{-\dc^2/2S}}{\sqrt{2\pi S}}\dc
\bigg[\frac{1+\erf\big(\Gamma\dc/\sqrt{2S}\big)}{2}
+\frac{e^{-\Gamma^2\dc^2/2S}}{\sqrt{2\pi/S}\,\Gamma\dc}\bigg]
\left|
\frac{d\ln \sigma}{d\ln M}
\right|
\;.
\label{eq:nM_completelycorrelated}
\end{align}
Here, $\Gamma^2=\gamma^2/(1-\gamma^2)$ and $\gamma$ is the cross-correlation
coefficient between $\delta(S)$ and $\delta'(S)$ defined in \refeq{gamma}.
This mass function is in general not normalized to unity owing to the imperfect
correlation between $\delta$ and $\delta'$.
Comparison with \refeq{fxPS} shows that the square bracket in 
\refeq{fxstrongcorr2} is the leading-order correction to the fully correlated 
solution, which is attained in the limit $\Gamma \to \infty$ 
i.e. when $\delta$ and $\delta'$ are completely correlated ($\gamma=1$).
Note that this limit is hardly achieved in reality; for a Gaussian filter and a power-law power spectrum 
$\Plin(k)\propto k^n$, $\Gamma^2=(3+n)/2$ for example, and the limit can only be 
achieved when $n\to\infty$.  
Therefore, the first-crossing distribution of partially correlated walks is 
fundamentally different than that of either entirely correlated or fully 
Markovian walks.  
Surprisingly, Monte-Carlo realizations of the first-crossing distributions indicate that the approximation
\refeq{nM_completelycorrelated} works very well over a large range 
of scales for a range of choices of power spectrum and filters 
(including tophat-filtered $\Lambda$CDM spectra) \cite{musso/sheth:2012}.
Note that, for ``moving barriers'' $\dc \to B(S)$, the halo mass function is 
not simply given by \refeq{nM_completelycorrelated} with $\dc$ replaced by 
$B(S)$ because the derivative with respect to $S$ in
\refeq{dscross} also brings a factor of $B'(S)$ which changes the lower 
limit of the integral in \refeq{fxstrongcorr1}.  
We will discuss this in \refsec{gen_barrier}.

We then calculate the halo bias parameters from the conditional first-crossing
probability distribution function in \refeq{correlated_cond_SfS}:
\be
\avnh(M|\delta_\ell;S_\ell)
= 
\frac{\rhob}{M}\sqrt{\frac{2S}{\pi}}
\frac{\bar\delta' e^{-\delta_{c\times}^2/2SQ}}{\sqrt{Q}}
\bigg[\frac{1+ \erf\big(\bar\delta'(S)/\sqrt{2}\bar\sigma(S)\big)}{2}
+\frac{e^{-\bar\delta^{'2}(S)/2\bar\sigma^2(S)}}{\sqrt{2\pi}(\bar\delta'(S)/\bar\sigma(S))}\bigg]
\left|
\frac{d\ln \sigma}{d\ln M}
\right|
\;,
\label{eq:nh_completely_correlated}
\ee
where we refer to \refsec{completely_correlated_derivation} for the definition of the symbols.  
Because in this case the first-crossing probability depends on two 
distinct variables $\delta$ and $\delta' \equiv d\delta/dS$, the halo bias is bivariate and we 
shall denote the corresponding (Lagrangian) bias parameters as $b_{ij}^L$.  
Note that $\delta'$ is a higher-derivative operator, and proportional to $R^2 \lapl\d$ at leading order
in derivatives (see \refsec{higherderiv}).  
Thus, the Lagrangian \LIMD biases are given by $b^L_{n0}$, while the
$b^L_{ni}$ with $i>0$ are higher-derivative biases.  We see that the Lagrangian
halo density $\d_h^L$ predicted by this ansatz can be expanded as written in \refeq{dhL_exset}.  

When the peak-background split is implemented with conditional mass functions, one considers the limit
$R_\ell\gg R$ or, equivalently, $S_\ell\to 0$ at fixed value of $\dell$ \citep{mo/white:1996,sheth/tormen:1999}. 
Particular attention must be paid to the ratio $S_\times/S_\ell$,
which tends towards ${\cal O}(1)$ in
this limit (the exact value depends on the shape of the filter). 
Hence, on taking the limit $R_\ell\to\infty$, all the corrections to $Q=1$, $\bar\sigma^2=(4S\Gamma^2)^{-1}$ 
vanish, except for $\delta_{c\times}= \dc-\dell (S_\times/S_\ell)\approx \dc-\dell$, which implies
$\bar\delta'\to(\dc-\dell S_\times/S_\ell)/(2SQ)$. 
In other words, the result is the same as differentiating w.r.t. $\dc$, and one recovers the linear Lagrangian
\LIMD bias parameter from the peak-background split, \refeq{bPBSuniv}.
Explicitly, for the constant barrier considered here, this yields
\begin{align}
\label{eq:b10Lx}
b_{10}^L &\equiv -\big[\nu_c f(\nu_c)\big]^{-1}\frac{\partial\big[\nu_c f(\nu_c)\big]}{\partial\dc} 
=\frac{\nu_c^2-1}{\dc}+\frac{1}{\dc}\left[1+\frac{1+\erf(\Gamma\nu_c/\sqrt{2})}{2} 
\frac{\sqrt{2\pi}\Gamma\nu_c}{e^{-\Gamma^2\nu_c^2/2}}\right]^{-1} \\
b_{20}^L &\equiv \big[\nu_cf(\nu_c)\big]^{-1}\frac{\partial^2\big[ \nu_c f(\nu_c)\big]}{\partial\dc^2} 
=\frac{\nu_c^4-3\nu_c^2}{\dc^2}+\frac{2+\gamma^2}{s}\left[1+\frac{1+\erf(\Gamma\nu_c/\sqrt{2})}{2} 
\frac{\sqrt{2\pi}\Gamma\nu_c}{e^{-\Gamma^2\nu_c^2/2}}\right]^{-1} \;,
\label{eq:b20Lx}
\end{align}
where we have inserted \refeq{fxstrongcorr2} for $\nu_c f(\nu_c)=2Sf(S)$.  
The Lagrangian bias for completely correlated walks (which equals the Markovian result \refeq{b1Lsk} since the 
corresponding multiplicity functions differ only by a constant factor of 2) is recovered in the high-peak limit $\nu_c\gg 1$.

Ref.~\cite{desjacques/crocce/etal:2010} showed that, on simultaneously taking the limit $S_\times/\sqrt{SS_\ell}\to 0$ but 
retaining the dependence on $\epsilon_\times$, 
one can obtain higher-derivative biases such as $b_{\lapl\d}$ from 
\refeq{nh_completely_correlated}, generalizing the PBS approach to 
higher-derivative biases.
This corresponds to the dependence of $\d_h^L$ on other variables apart from
$\d$, such as $\delta'$ or the curvature $\lapl\delta$. 
In this particular limit, we still find $Q\to 1$, $\bar\sigma^2\to (4S\Gamma^2)^{-1}$ and
$\delta_{c\times}\to \dc-\dell (S_\times/S_\ell)$. However, we now have
\begin{equation}
\bar\delta'\to  \frac{1}{2s} \bigg[\dc-\frac{S_\times}{S_\ell}\big(1-\epsilon_\times\big)\dell\bigg] \;.
\end{equation}
Taking the ratio of the conditional to unconditional first-crossing distributions and expanding in powers of 
$\dell$, we eventually obtain
\begin{equation}
\label{eq:danddprime}
1+\delta_h^{L}(\dell) = \frac{f(S|\dell;S_\ell)}{f(S)}
= 1 + \frac{S_\times}{S_\ell}\Big(b_{10}^L+ \epsilon_\times b_{01}^L\Big)\dell +
\frac{1}{2}\left(\frac{S_\times}{S_\ell}\right)^2\Big(b_{20}^L+2\epsilon_\times b_{11}^L
+ \epsilon_\times^2 b_{02}^L\Big) \dell^2 + \dots
\end{equation}
which differs from the Lagrangian \LIMD expansion owing to the presence of higher-derivative bias contributions 
induced by $\delta'$. These arise with non-zero powers of $\epsilon_\times$.
The appearance of $\epsilon_\times$ in \refeq{danddprime} indicates that the bias in Fourier space is $\propto k^2$.
Namely, we have at linear order $\delta_h^L(\vk)=c_1^L(k) \delta^{(1)}(\vk)$ with
\cite{musso/paranjape/sheth:2012}
\be
c_1^L(\vk) =  \left(b_{10}^L +\frac{k^2 S}{\la[\delta'(S)]^2\ra}b_{01}^L\right) W_R(k)
\ee
This shows that excursion-set theory generically predicts a $k$-dependence in Fourier space. The peak theory discussed
in \refsec{peaks} and, more generally, any Lagrangian bias scheme more sophisticated than \LIMD (e.g. \cite{matsubara:2011})
leads to $k$-dependent Lagrangian bias functions $c_n^L(\vk_1,\dots,\vk_n)$.  
Crucially however, the $k$-dependence can be mapped exactly onto the general set of higher-derivative terms described in
\refsec{higherderiv} in all cases.  

We can read off both the \LIMD bias parameters $b_{i0}^L$ and the higher-derivative bias parameters $b_{ij}^L$ ($j>0$) 
from \refeq{danddprime}. In particular, we have 
\be
\label{eq:b01Lx}
b_{01}^L = \frac{1}{\dc} \bigg\{1-\left[1+\frac{1+\erf(\Gamma\nu_c/\sqrt{2})}{2} 
\frac{\sqrt{2\pi}\Gamma\nu_c}{e^{-\Gamma^2\nu_c^2/2}}\right]^{-1}\bigg\} \;.
\ee
Note that the $b_{0j}^L$, which represent the $j$th-order response of the first-crossing distribution to a long-wavelength
perturbation $\delta_\ell'$, are dimensionless because the derivative is taken with respect to the variance $S$.  
Adding \refeq{b10Lx} and \refeq{b01Lx}, the first-order Lagrangian biases are linearly related through
\be
\label{eq:10b01}
b_{10}^L + b_{01}^L = \frac{\dc}{S} \;.
\ee
Similar relations hold at any order, as well as for more sophisticated Lagrangian prescriptions such as the
peak constraint \citep{musso/paranjape/sheth:2012,desjacques:2013}. They can be used to infer the value of
the higher-derivative, or ``scale-dependent'', bias parameters once the \LIMD, or ``scale-independent'' biases 
are known. These relations follow from the possibility of expanding the joint distribution of $\d$ and $\d'$  
in Hermite polynomials. In fact, the $b_{ij}^L$ can be written as an ensemble average over Hermite polynomials 
\cite{szalay:1988,musso/paranjape/sheth:2012,desjacques/gong/riotto:2013}. We will discuss polynomial expansions
in more detail in \refsec{PBSpeaks}, and show that this can be generalized to variables which do not follow Gaussian distributions.  

\begin{figure}[!t]
\centering
\includegraphics[width=0.9\textwidth]{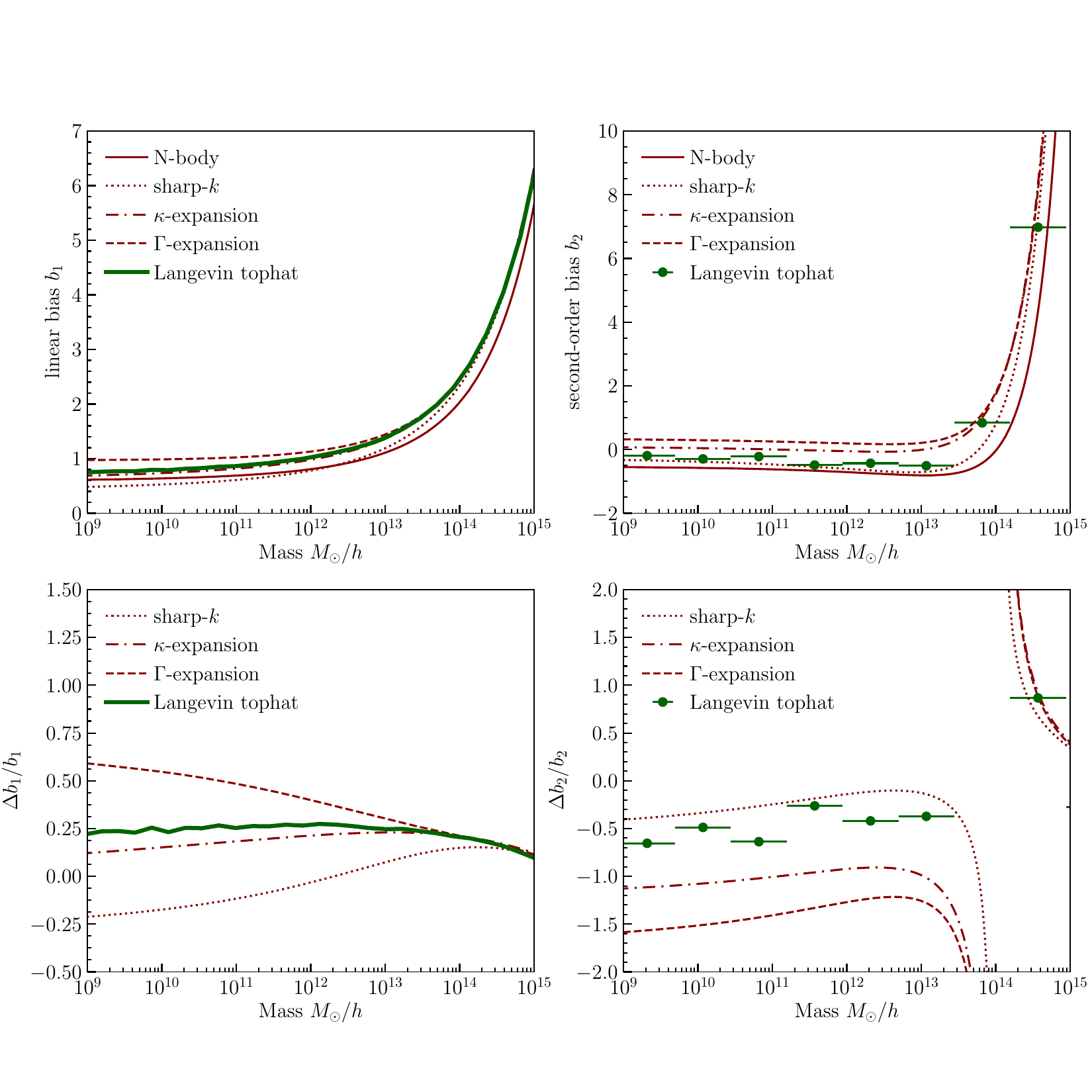}
\caption{
Excursion-set predictions for the Eulerian bias parameters $b_1$ (upper left panel) and 
$b_2$ (upper right panel) for halos at redshift $z=0$ (red lines).
Also shown is the result from N-body simulations, parametrized using the fitting functions given in \reftab{fittingbiases} in \refsec{measurements} 
(using $\Delta_\text{SO} = 200$).  The lower panels show the corresponding fractional deviation of the excursion-set predictions from the N-body results. 
We show four excursion-set predictions 
(all with $\dc = 1.686$ 
and using $b_1 = 1+b_1^L$ and $b_2 = b_2^L + 8/21b_1^L$):
(1) Markovian, or Press-Schechter case [\refeqs{b1Lsk}{b2Lsk}, dotted lines],
(2) first order in the $\kappa$-expansion around the Markovian case [\refeqs{MRb1L}{MRb2L}, dot-dashed lines], (3) first order in the $\Gamma$ expansion around the fully correlated case [\refeqs{b10Lx}{b20Lx}, dashed lines], and (4) the numerical integration result of the exact Langevin equation with a tophat filter [\refeq{dLxR_Wk}; thick solid line]. 
We have integrated the Langevin equation 
to obtain 5 billion independent random-walk chains, each with 2000 steps from 
$R=10^{3.5}\Mpch$ to $10^{-1.5}\Mpch$. An accurate determination of 
$b_2$ demands even more random-walk chains, however.
Note that, even with the numerical solution of the Langevin equation, 
the excursion-set predictions do not agree well with the bias 
parameters measured from N-body simulations.
}
\label{fig:exset_bias}
\end{figure}

\subsection{Summary: bias of halos in the standard excursion-set formalism}
\label{sec:exset_summary}

So far in this section, we have reviewed the theoretical predictions for the 
linear and second-order \LIMD bias parameters from the excursion-set formalism 
with a constant density threshold $\dc$.  
This approximation is motivated by the spherical collapse model (\refsec{sph_collapse}).  
Therefore, these results are built on a strongly simplified scenario of halo formation.

When employing this assumption, however, exact analytical solutions 
are known for two special cases of the excursion set: when each step of the 
random walk is 
(i) completely independent from the previous steps (Markovian case obtained by employing the sharp-$k$ filter, \refsec{sharpk_analytic}), or is
(ii) completely correlated with the previous steps (\refsec{completely_correlated}).  
Both cases are far from being realistic descriptions of Lagrangian halo density profiles.  
For more general filters, approximate solutions can be obtained by expanding around these analytic solutions, as shown in \refsec{exsetMR} and 
\refsec{completely_correlated_bias}.  

Specifically, the predictions for the linear and 
second-order bias parameters are given in:
\begin{itemize}
\item \refeqs{b1Lsk}{b2Lsk} for the Markovian treatment (sharp-$k$ filter, or Press-Schechter)
\item \refeqs{MRb1L}{MRb2L} for the expansion around the Markovian case (i.e. $0 < \kappa < 1$)
\item \refeqs{b10Lx}{b20Lx} for the expansion around the fully correlated case
($\Gamma\gg 1$).  
\end{itemize}
For both expansions around analytical solutions, we recover a multiplicity function that is proportional to that of the Markovian case in certain
limits, namely $\kappa\to 0$ and $\Gamma\to\infty$, respectively.  Note however that they describe expansions around two completely different solutions.  
We compare these three predictions in \reffig{exset_bias}, assuming 
a spherical tophat filter for $\s(M)$, and a constant density threshold $\dc=1.686$. Further, we choose $\kappa=0.44$ in \refeqs{MRb1L}{MRb2L}.
For \refeqs{b10Lx}{b20Lx}, we calculate $\Gamma$ from our reference cosmology 
(\reftab{ref_cosmology}) which gives $\Gamma = [0.35...0.57]$ 
in the mass range plotted in \reffig{exset_bias}.
We also show the result of a numerical integration of the Langevin equation 
\refeq{dLxR_Wk} for the tophat filter. 
For both $b_1$ and $b_2$, the $\kappa$-expansion follows the exact numerical 
result more closely than the $\Gamma$-expansion.
For both $b_1$ and $b_2$, all four predictions agree for high-mass halos, but diverge for lower-mass halos.  
Note, however, that the excursion-set predictions do not agree well with the bias parameters 
measured from N-body simulations, which are represented in \reffig{exset_bias} using the empirical fitting formulas given in \reftab{fittingbiases}
(\refsec{meas:meas}).
This also holds for the numerical integration of the Langevin equation, which corresponds to the exact solution of the excursion set. 
This result illustrates the need for a more realistic description of the collapse, to which
the spherical model is a crude approximation.  We turn to extensions 
such as the ellipsoidal collapse model and stochastic barriers next.

\subsection{Beyond the spherical collapse model}
\label{sec:gen_barrier}

For an initially spherical perturbation, the spherical collapse solution is 
a good approximation until the first orbit crossing.  However, 
actual overdense regions in a Gaussian density field are not spherical, but triaxial 
\citep{doroshkevich:1970a,bardeen/etal:1986,jing/suto:2002}.  
Moreover, the traceless part of $\partial_i\partial_j\Phi$, i.e. the tidal shear $K_{ij}$, has been
shown to play a crucial role in the formation of nonlinear structures 
\citep{hoffman:1986,peebles:1990,bond/myers:1996,delpopolo/gambera:1998}.  
N-body simulations have established that 
the principal frame of the proto-halo strongly correlates with the local tidal field
\cite{lee/pen:2000,porciani/dekel/hoffman:2002,lee/hahn/porciani:2009,ludlow/porciani:2011,
despali/tormen/sheth:2013,ludlow/boryz/porciani:2014}.
The collapse of an isolated, homogeneous ellipsoid has been studied extensively 
\citep{lynden-bell:1964,lin/mestel/shu:1965,icke:1973,white/silk:1979,hui/bertschinger:1996}, 
although these are still only approximations to actual Lagrangian proto-halo patches.    
In the formulation of \cite{bond/myers:1996}, the initial conditions and external tides are 
chosen to recover the Zel'dovich approximation in the linear regime. 
For a given fixed Lagrangian overdensity, the final density can attain a range of values
depending on the local tidal shear, as estimated through ellipsoidal collapse calculations in
\cite{lam/sheth:2008,lam/sheth:2008b}.  

In a first approximation, the dynamics of ellipsoidal collapse can be incorporated through a mass-dependent, or ``moving'' barrier $B(S)$ in the excursion-set approach. 
The tidal shear effectively slows down the collapse of low-mass objects and, therefore, yields a barrier $B(S)$ which grows monotonically with $S$.  
As a result, the relative abundance of high-mass objects increases \citep{monaco:1995}, such that the 
multiplicity function $f(\nu_c)$ furnishes a better fit to the halo mass function measured in N-body simulations
\citep{sheth/tormen:2002}.  

In practice, whereas the spherical evolution leads to a condition on the trace 
$\delta$ of the deformation tensor only, the critical density for 
non-spherical collapse will also depend on the other two invariants of 
$\partial_i\partial_j\Phi$ (equivalently, $\tr[(K_{ij})^2]$ and $\tr[(K_{ij})^3]$).  
Following \cite{ohta/kayo/taruya:2004,sheth/chan/scoccimarro:2012,lazeyras/musso/desjacques:2015},
the three invariants of the second-rank tensor $\partial_i\partial_j\Phi$ can be conveniently parametrized as 
(see \refsec{dynamics})
\begin{equation}
\delta \;, \qquad K_2 \equiv \frac{3}{2}\tr\big(K_{ij}^2\big)\;, \qquad
K_3\equiv \frac{9}{2}\tr\big(K_{ij}^3\big) \;.
\label{eq:K2K3def}
\end{equation}
In $K_2$, a multiplicative factor of $3/2$ is added such that $\big\la K_2\big\ra=S$, which implies that, for Gaussian initial conditions, $5K_2$ is 
$\chi^2$-distributed with 5 d.o.f.. 
Similarly, the factor of $9/2$ in the definition of $K_3$ ensures that $|K_3|\leq K_2^{3/2}$.  Note that $K_2=3e^2+p^2$, where $e$ and $p$ are 
the shear ellipticity and prolateness \cite{sheth/mo/tormen:2001}. 
Ellipsoidal collapse will then give rise to a dependence of $\delta_h^L$ on $K_2$, $K_3$ of the form 
(see also \refsec{evol2}) 
\begin{equation}
\delta_h^L(\vq) = 
b_1^L \delta + \frac{1}{2} b_2^L \delta^2 + b_{K_2}^L K_{2} 
+ \frac{1}{6} b_3^L \delta^3 + \frac{5}{\sqrt{7}}b_{K_3}^L K_{3} + b_{\d K_2}^L \d K_{2} + \dots \;.
\label{eq:deltah5.9}
\end{equation}
This generalizes the Lagrangian \LIMD bias obtained in the standard excursion set, as summarized in \refeq{dhL_exset},  
to include the local tidal terms induced by non-spherical collapse.
Specifically, the bias parameters introduced in \refeq{deltah5.9} are related to those defined in \refsec{evol2} through
\be
b_{K_2}^L = \frac23 b_{K^2}^L\,; \quad
b_{\d K_2}^L = \frac23 b_{\d K^2}^L\,; \quad
b_{K_3}^L = \frac{2\sqrt{7}}{45} b_{K^3}^L\,.
\ee

In order to calculate the Lagrangian tidal bias parameters $b_{K_2}^L$, $b_{K_3}^L$, and so on, we need a model for the 
barrier $B$, which now characterizes the first-crossing of multidimensional random walks $(\delta,K_2,\dots)$
\citep{chiueh/lee:2001,sheth/tormen:2002,castorina/sheth:2013,musso/sheth:2014b,castorina/paranjape/etal:2016}. 
That is, a complete theory of Lagrangian bias must take into account the dependence of the galaxy number 
density on $K_{2\ell}$ and $K_{3\ell}$ separately, but this has not been done yet.
As a rule of thumb, any additional variable adds a dimension to the first-crossing problem.
For illustration, Ref.~\cite{sheth/chan/scoccimarro:2012} considered the simple model
\begin{equation}
B(S,K_2) = \dc \left(1+\sqrt{K_2/K_c}\right) \;,
\end{equation}
which is motivated by setting $p=0$ in the moving barrier of \cite{sheth/mo/tormen:2001}. Here, $K_c$ is a 
characteristic scale for the effect of tidal shear. The barrier depends on $S$, but only through $K_2\propto S$.
Furthermore, we always have $B\geq \dc$ since $K_2\geq 0$. In analogy with our previous calculation of the
Lagrangian \LIMD bias $b_n^L$, the tidal bias $b_{K_2}^L$ can be obtained by considering the response of the 
unconditional first-crossing distribution to a long-wavelength perturbation $K_{2\ell}$. 
The conditional distribution $f(S|\dell,K_{2\ell};S_\ell)$ now describes the first-crossing distribution of
2-dimensional walks that start from some non-zero $\dell$ and $K_{2\ell}$.
Of course, if the shape of the overdensity also affects the collapse, then the barrier would also depend
on the misalignment between the tidal shear and shape tensor \cite{castorina/paranjape/etal:2016}.

In practice, a simpler approach, in which $K_2$ is replaced by its mean value $\langle K_2\rangle=S$, provides
a good approximation to the exact result \cite{sheth/tormen:2002}. 
In this approximation, the barrier becomes $B(S,K_2)\to B(S)\approx \dc (1+\sqrt{S/K_c})$, such that
$f(S|\dell,K_{2\ell};S_\ell)$ is well represented by one-dimensional walks crossing the barrier
\begin{equation}
\dc \left[1+\sqrt{(S+K_{2\ell})/K_c}\right] - \dell \;.
\end{equation}
Expanding this barrier in a series of powers of $\dell$ and $K_{2\ell}$ and dividing by the unconditional 
first-crossing distribution, we arrive at
\begin{equation}
\label{eq:bK2L}
\dc^2 b_{K_2}^L \approx -\nu_c^3 \frac{\dc}{K_c}\left(1+\frac{\dc}{\nu_c K_c}\right) \,,
\end{equation}
in the limit $\nu_c\gg 1$.  
Unsurprisingly, $b_{K_2}^L < 0$ since a large shear increases the barrier height and, therefore, impedes the 
collapse of halos. 
Furthermore, $b_{K_2}^L\propto -\frac{\sigma_0}{K_c} b_2^L$ in the high-peak limit, so that the spherical
collapse approximation---within which only the LIMD bias parameters are non-zero in Lagrangian space---is recovered.
Higher-order local bias parameters involving $K_2$ can be computed analogously.
Evidence for a non-zero $b_{K^2}^L$ has recently been presented in the literature (see the discussion in \refsec{meas:meas:halos}).

\begin{figure*}[t]
\centering
\includegraphics[width=0.6\textwidth]{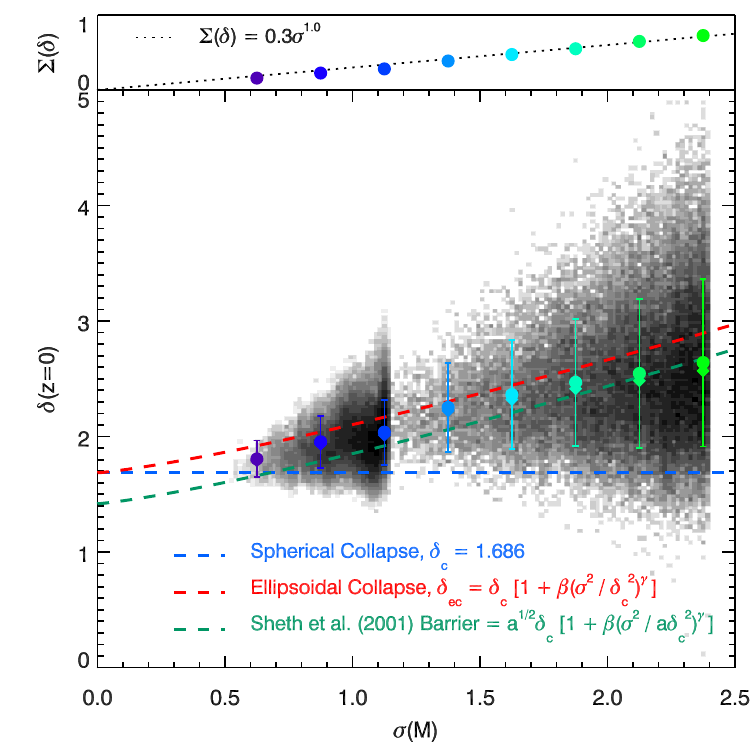}
\caption{
Smoothed linear overdensity, extrapolated to $z=0$ for the 
Lagrangian regions that collapse and form halos by $z=0$, as a function of halo mass parametrized through $\sigma(M)$ . Circles correspond to the mean overdensities,
whereas diamonds show the median overdensities. The errorbars indicate the halo-to-halo scatter.  Shown for comparison are the spherical 
collapse barrier (blue dashed line), the ellipsoidal collapse barrier of \cite{sheth/mo/tormen:2001} (red 
dashed line), and the collapse barrier associated with the Sheth-Tormen mass function 
\citep{sheth/tormen:1999} (green dashed line). The upper panel shows the scatter $\Sigma$ in the barrier 
height as a function of $\sigma(M)$.  
\figsource{robertson/etal:2009}  
\label{fig:barrier}
}
\end{figure*}

Numerical implementations of the ellipsoidal collapse \citep[e.g.,][]{sheth/mo/tormen:2001,desjacques/ellips,lithwick/dalal:2011}
and numerical studies of Lagrangian halos \citep[e.g.,][]{robertson/etal:2009,ludlow/porciani:2011} have shown 
that the collapse barrier is ``fuzzy'', i.e. there is a range of values of $\dc$ at fixed $S$. 
This is apparent in \reffig{barrier}, which shows the distribution of smoothed linear overdensities
associated with Lagrangian regions collapsing into halos \citep{robertson/etal:2009}. 
Namely, for each halo identified in N-body simulations, one can trace its constituent dark matter 
particles back to their initial position, and compute the initial overdensity in a tophat sphere centered 
at the initial center-of-mass. These overdensities, once linearly extrapolated to $z=0$, furnish a snapshot
of the collapse barrier. As seen in \reffig{barrier}, the mean as well as median barriers increase with decreasing
halo mass, in broad agreement with the ellipsoidal collapse prediction. The scatter $\Sigma$ around the mean 
barrier is generated by variables other than the density such as the local shear, coupling with the 
large-scale environment. Note that $\Sigma$ is proportional to $\sigma(M)$, which is consistent with a lognormal 
distribution of the barrier.

Thus, an alternative to explicitly including the tidal shear in the barrier is to use a ``fuzzy'' moving barrier,
which includes scatter around the mean, such as the square-root barrier \cite{paranjape/lam/sheth:2012}
\begin{equation}
\label{eq:squarerootbarrier}
B(S)=\dc\Big(1+\beta\sqrt{S}/\dc\Big) \,,
\end{equation}
where $\beta$ can be calibrated with N-body simulations using the method of \cite{robertson/etal:2009} discussed above.
For the purpose of predicting the halo mass function $\avnh(M)$ and \LIMD bias parameters $b_N$, this approximation is more convenient than introducing an explicit dependence on the shear $K_{ij}$ and other fields. 
The results in \refsecs{excursion_exact}{excursion_HeavensPeacock} can indeed be extended to this barrier. Moreover, this ansatz is adopted in the excursion-set peak approach described in \refsec{sub:esp}. 
Alternatively, the scatter has also been modeled as a ``diffusive barrier'' in the context of the path integral
approach described in \refsec{exsetMR} \cite{maggiore/riotto:2010c}.
The resulting mass functions have been found to be in good agreement with simulations \cite{corasaniti/achitouv:2011,achitouv/rasera/etal:2013}.

One should bear in mind that the scatter in the barrier reflects the existence of hidden variables
and, therefore, is a consequence of projecting the actual, multi-dimensional collapse barrier onto a simpler
subspace in which $S$ is the sole variable.  
A moving barrier $B(S)$ cannot fully capture the dependence of halo collapse on the tidal shear (or deformation) tensor $K_{ij}$ unless
the latter is explicitly accounted for.
As we shall see in \refsec{NGLagBias}, a microscopic description of the barrier $B$ in terms of $K_{ij}$ is, in fact, essential to obtain a
physically consistent prediction for the non-Gaussian bias in Lagrangian bias models.

\subsection{Halo assembly bias in the excursion-set formalism}
\label{sec:assembly_exset}

As discussed in \refsec{assembly}, studies of halo clustering in numerical simulations have established that, at fixed halo mass, the halo bias depends on other properties such as formation time, concentration, or environment density, a phenomenon known as assembly bias. In the high-mass limit and for halo properties that can be related to Lagrangian quantities, the excursion set provides a physical picture for the origin of assembly bias, as we will now discuss. This holds in particular for the environment density and formation time.

Assembly bias cannot be explained in the standard Markovian
formulation of excursion-set theory (see \refsec{sharpk_analytic})
since there is no 
correlation between the largest scales which characterize the clustering, and the small scales which characterize
the accretion history \citep{white:1996}. Of course, correlations will arise as soon as the Markovian 
assumption is relaxed.

\begin{figure*}[t]
\centering
\includegraphics[trim=0cm 5.3cm 0cm 3.4cm,clip,width=0.55\textwidth]{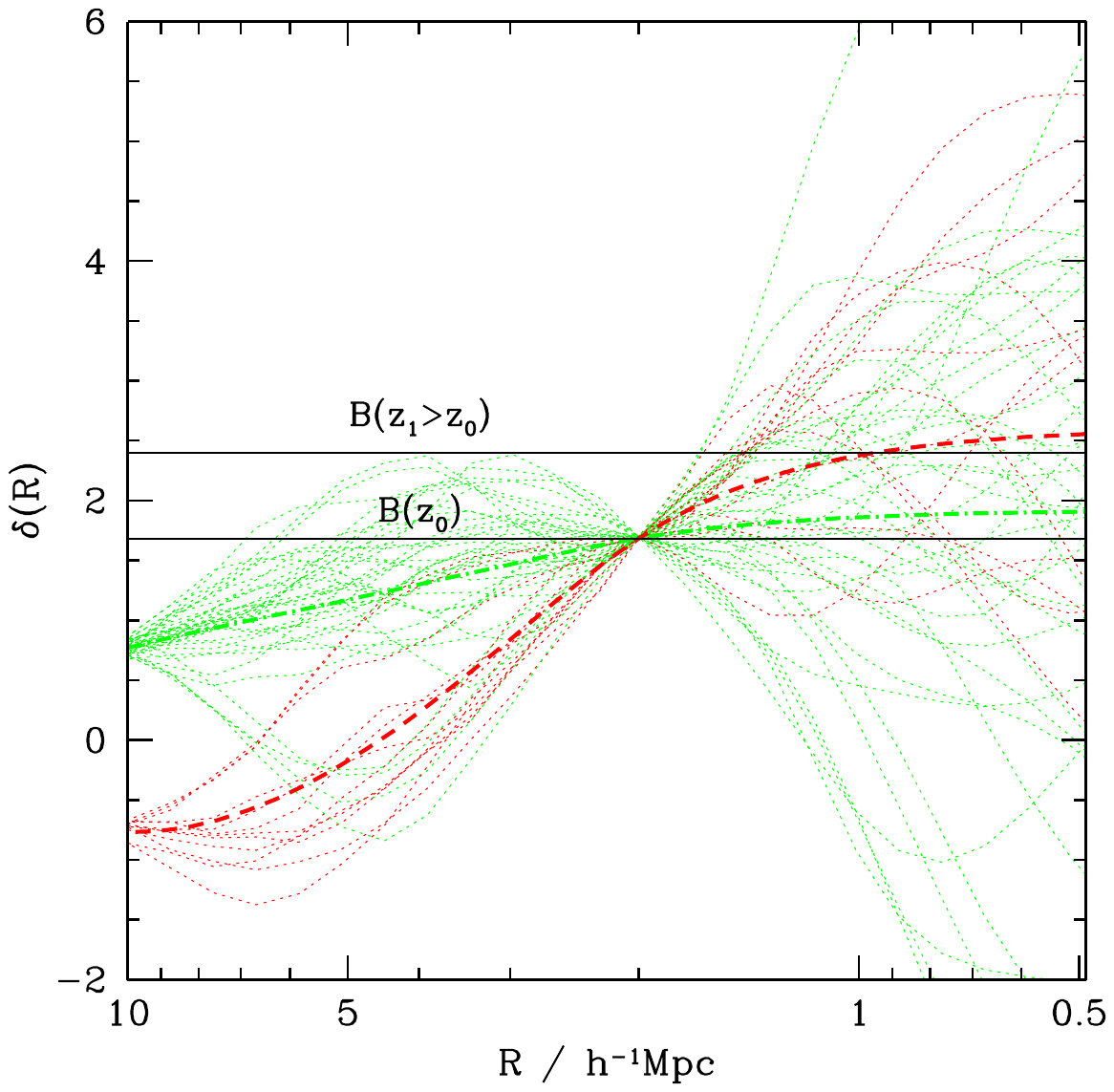}
\caption{
Trajectories $\delta(R)$ as a function of the tophat smoothing radius $R$, extracted from random realizations of the linear density field.  
All the trajectories obey the constraint $\delta(R)=\dc$ at the halo scale. 
The green and red curves show the realizations which have $\delta(R_\intscale)=+2\sigma_\intscale$ and 
$-2\sigma_\intscale$, respectively. 
The thick dashed and dot-dashed curve represent the average trajectory $\bar\delta(R|\delta_\intscale)$.  Here, we have used $R_\intscale = 5\Mpch$.  
\label{fig:walk}
}
\end{figure*}

As shown in \cite{zentner:2007} and \cite{dalal/white/etal:2008}, the trend at high halo mass can be explained
with the statistics of correlated random walks $\delta(s)$.
To see this, we follow common practice and adopt the redshift $z_{1/2}$ at which one half of the mass has been 
accreted onto the halo as a proxy for the halo formation redshift.
Simple considerations suggest a correlation between the large-scale overdensity $\d_\intscale$ and the slope 
$\delta'$ \cite{zentner:2007,desjacques/ellips}.  
A peak constraint as invoked in \cite{dalal/white/etal:2008} leads to similar conclusions 
since the peak curvature $J_1$  (see \refsec{invariants})) strongly correlates with $\delta'$ 
($J_1$ and $\delta'$ are, in fact, completely correlated for a Gaussian filter, see \refsec{esp}).
Namely, when the value of $\delta$ is held fixed to $\delta=\dc$ at the halo mass scale, we find:
\begin{align}
\label{eq:zform}
\d_\intscale > 0 &\quad \Longleftrightarrow \quad \mbox{smaller} ~\delta' \quad \Longleftrightarrow \quad \mbox{low}~~
z_{1/2} \\ 
\d_\intscale < 0 &\quad \Longleftrightarrow \quad \mbox{larger} ~~\delta' \quad \Longleftrightarrow \quad \mbox{high}~
z_{1/2} \nonumber \;.
\end{align}
To illustrate this point, we have extracted trajectories $\delta(R)$ smoothed with a tophat filter, as shown in
\reffig{walk}.
All the trajectories are subject to the two constraints $\delta(R)=\dc$ on the halo scale, and 
$\delta(R_\intscale)=\pm 2\sigma_\intscale$ [$R=1.5\hmpc$ and $R_\intscale=5\hmpc$ for illustration].  
The solid and dashed thick curves show the expected ``mean'' field $\bar\delta(R|\delta_\intscale)$ subject to the
large-scale constraint. The horizontal lines represent the collapse barrier $B(z) = \dc D(z_0)/D(z)$ at redshifts $z_0$ and $z_1>z_0$,
where $z_0$ is the redshift at which the linear density field $\delta(R)$ is extrapolated.
Clearly, the steeper the slope $|\delta'(R)|$, the higher $z_{1/2}$ is, since, at constant final mass $M$ (i.e. $R$), 
the collapse barrier $B(z)$ is reached relatively earlier. 
Furthermore, these early-forming halos reside in more isolated environment (underdense regions with $\dell<0$) 
with less late-time accretion, which leads to steeper halo profiles or higher concentration. 
The converse holds for the late-forming halos which preferentially collapse in regions with $\dell>0$.

Now, $\d_\intscale>0$ corresponds to halos with a larger bias. An easy way to see this is to re-express $b_1^L\equiv b_{10}^L$ defined in \refeq{b10Lx} as
\begin{equation}
b_1^L = \frac{\nu^2}{\dc(1-\gamma^2)} - 2 \Gamma^2 \bar{\delta'}\;,
\end{equation}
where $\bar{\delta'}$ is the average of the slope {\it given the threshold and first-crossing constraints}
$\delta\equiv\dc$, $\delta'\geq 0$. When the large scale overdensity $\d_\intscale$ is fixed, then $\bar{\delta'}$
is replaced by some $\delta'<\bar{\delta'}$ if $\d_\intscale>0$, and $\delta'>\bar{\delta'}$ if $\d_\intscale<0$. 
Consequently, $b_1^L$ is larger than average if $\d_\intscale>0$, and smaller if $\d_\intscale<0$.  
We thus reach the conclusion that ``young'' halos (i.e., those with low $z_{1/2})$ have lower concentration and are more 
strongly clustered than ``old'' halos (i.e., those with high $z_{1/2}$).  
This prediction is consistent with the N-body results for high-mass halos.
At low mass ($M\lesssim M_\star$), however, the assembly bias seen in N-body simulations is reversed, such that low-concentration halos are more biased (see \refsec{assembly}).

The excursion-set formalism can also be adapted to investigate the impact of the anisotropic cosmic web on assembly bias. 
For instance, Ref.~\cite{musso/etal:2017} investigated the effect of tides generated by filaments---modeled as one-dimensional saddle points---on the properties of neighboring halos. In the excursion-set language, the presence of a saddle point adds a contribution of the form $\rhat_i \rhat_j K_{ij}$, where $\vrhat$ is the unit separation vector between the halo and the saddle, to the mean
and covariance of $\delta$ and $\delta'$. This quadrupolar term, which is negative along the filament axis and positive perpendicular to it, can be used to quantify the response of halo counts, accretion rate and formation times to the presence of a nearby filament.

%% file: peaks.tex
\clearpage
\section{The statistics and evolution of Lagrangian density peaks}
\label{sec:peaks}

\secttoc

The peaks approach to the clustering of dark matter halos is built upon the 
assumption that halos form from peaks in the initial Lagrangian density field 
\cite{doroshkevich:1970a,doroshkevich:1970b,kaiser:1984,peacock/heavens:1985,bardeen/etal:1986}.
This is clearly an oversimplification of the complicated process of halo formation.  
Nevertheless, peaks provide an interesting non-perturbative toy model of discrete tracers which 
exhibits several interesting features of the bias expansion of general tracers (\refsec{evolution}), 
including nonlinear and higher-derivative bias as well as stochasticity.  
The fundamental quantity in peak theory is the set of local maxima of the smoothed density field; therefore, 
peaks define a point process. 
Since the evolved density field is highly nonlinear, the peak constraint is generally applied to 
the initial (Lagrangian) Gaussian density field, with the assumption that the most prominent peaks 
should be in one-to-one correspondence with luminous galaxies or massive halos in the Universe. 

The formalism became popular after \cite{kaiser:1984} demonstrated that 
the bias inherent to overdense regions of the Universe could explain the 
higher clustering amplitude of Abell clusters 
(see the discussion in \refsec{localbias}). 
The first numerical and analytic investigations of density peaks of the initial, 
Gaussian density field have  been performed in \cite{peacock/heavens:1985,bardeen/etal:1986}. 
Since then, their abundance, clustering and internal properties have been studied in detail. 
In particular, their profiles have been investigated in an attempt 
to predict internal properties of virialized halos such as 
their profiles and angular momentum 
\cite{hoffman/shaham:1985,hoffman:1986,heavens/peacock:1988,ryden:1988,vandeWeygaert/babul:1994,
catelan/theuns:1996,bond/myers:1996b,bond/myers:1996c,ma/bertschinger:2004,schaefer/merkel:2012}.  
As this review is focused on large-scale bias, however, we shall not discuss this aspect here.    
Some of these results have been used to constrain the power spectrum of matter density perturbations 
\cite{croft/gaztanaga:1998,de/croft:2010} (using the fact that the number density of peaks 
in a Gaussian random field is sensitive to the shape of the power spectrum of density perturbations); 
to model the column density distribution of the Lyman-$\alpha$ forest
\cite{hui/gnedin/zhang:1997}, the cosmic web \cite{pogosyan/etal:2009} or the alignment of 
galactic spins \cite{codis/pichon/pogosyan:2015}; 
to study velocity bias and redshift-space distortions
\citep{peacock/lumsden/heavens:1987,percival/schaefer:2008,desjacques/sheth:2010}
as well as assembly bias \citep{dalal/white/etal:2008}; and to generate constrained realizations 
of the large-scale structure \citep{vandeWeygaert/bertschinger:1996,porciani:2016}.    
Furthermore, the peak formalism has been applied to study the importance of higher-derivative bias 
on the BAO scale \citep{desjacques:2008,desjacques/crocce/etal:2010}, and to understand the interplay between 
bias and gravitational evolution \citep{desjacques/crocce/etal:2010,baldauf/desjacques/seljak:2015}.   
More recently, \cite{paranjape/sheth:2012} have proposed an approach which combines peak theory with 
the excursion-set approach described in the previous section, while \cite{lazeyras/musso/desjacques:2015} 
have shown how the peak clustering statistics can be derived from a perturbative bias expansion.  

The empirical evidence for the association of massive halos with Lagrangian density peaks is reviewed in
\refsec{exset:general}.
Briefly, while massive halos with $M\gg M_\star$ are, to a good approximation, in one-to-one correspondence 
with prominent ($\nu_c\gg 1$) initial density peaks---simply because they are very rare and their proto-halo 
patch is approximately spherical \cite{bardeen/etal:1986}---the association weakens as one goes to lower halo mass.
In fact, overdense regions of a Gaussian density fields are inherently triaxial 
\cite{doroshkevich:1970a,bardeen/etal:1986,jing/suto:2002}. 
Furthermore, as discussed in \refsec{gen_barrier}, the initial shear field, i.e. the tidal field $K^{ij}_R$ 
smoothed on the peak scale, certainly plays an important role in the  formation of nonlinear structures 
\cite{hoffman:1986,peebles:1990,dubinski:1992,bertschinger/jain:1994}.
Taking into account these complications would very likely extend the predictability of peak theory to lower halo mass. 
The peak-patch picture of \cite{bond/myers:1996, bond/myers:1996b,bond/myers:1996c} has taken steps in this 
direction by incorporating the ellipsoidal collapse, solving the cloud-in-cloud problem and relying on an 
exclusion algorithm to avoid overlap of peak-patches.
However, this approach does not allow for analytical expressions of the large-scale clustering of peak patches,
although it is guaranteed to fit into the general bias expansion of \refsec{evolution} on scales where perturbation
theory applies.
Moreover, we expect the same qualitative features to appear within the simple thresholded peaks approach.
Therefore, we shall hereafter restrict ourselves to the spherical approximation, and ignore the finite extent of
the peak-patches (except for the filtering kernel). 
We will discuss each of these issues in detail in the context of the analytical peak model.  
As we will see, many insights can be gained even with the simplifications made here.

The outline of this section is as follows:
\begin{itemize}
\item \refsec{spectralmoments} to \refsec{pknum} review the basic formalism, introduce the 
peak constraint and present a simple derivation of the peak number density (1-point function).
\item \refsec{pkcorr} reviews calculations of peak correlation functions, while 
\refsec{peakbias} delineates the peak bias expansion and its connection with the peak-background 
split and renormalization.  
\refsec{bNmL} illustrates how the \LIMD as well as higher-derivative peak bias parameters can be measured from N-body 
simulations using cross-correlations at two scales (see \refsec{bmom}).  
\item \refsec{esp} presents the excursion-set peaks (ESP) approach, which merges 
ideas from excursion-set theory (see \refsec{exset}) with the peak formalism.  
\item \refsec{sub:pkgrav} focuses on the gravitational evolution of Lagrangian peaks, with a particular 
emphasis on the peak velocity bias.
\end{itemize}
Throughout this section, the peak number densities are defined as comoving Lagrangian densities.  
The definition of symbols used in this section can be found in \reftab{symbol6}.

\subsection{Spectral moments and characteristic scales}
\label{sec:spectralmoments}

The fundamental ingredient in all calculations within the peak approach 
is the smoothed initial density field $\d_R^{(1)}$, extrapolated to the 
collapse epoch $\tau_0$
using linear theory, where $\tau_0$ does not necessarily correspond
to today's epoch.    
In this section, we will denote the smoothed 
linear density field as $\d_R(\vq)\equiv \d_R^{(1)}(\vq)$, where $\vq$ is the Lagrangian position. 
This is the same field as used in the thresholding example
considered in \refsec{localbias}.  However, owing to the peak constraint, the 
clustering properties of density peaks also depend on the statistics of the 
first and second derivatives $\partial_i\delta_R(\vq)$ and 
$\partial_i\partial_j\delta_R(\vq)$ of the smoothed density field.
Throughout, we assume $\delta_R$ to be a homogeneous Gaussian 
random field of zero mean.

It is convenient to introduce the normalized dimensionless variables 
$\nu\equiv\delta_R/\sigma_{0}(R)$ [to be distinguished from the peak significance $\nu_c \equiv \dc/\sigma_{0}(R)$], 
$\eta_i\equiv\partial_i\delta_R/\sigma_{1}(R)$ 
and $\zeta_{ij}\equiv \partial_i\partial_j\delta_R/\sigma_{2}(R)$ (\reftab{symbol6}), 
where $\sigma_{n}(R)$ are the spectral moments of the matter power spectrum smoothed 
on the scale $R$, 
\be
\sigma_n^2(R) \equiv \int_{\vk} 
k^{2n}\,  W_R^2(k)\,\Plin(k)\;,
 \label{eq:mspec}
\ee
where $W_R(k)$ is a spherically symmetric filtering kernel.  Note that
this includes $\sigma(R) \equiv \sigma_0(R)$ as a special case.  
Here and in what follows, we omit the dependence on conformal time, which should 
always be taken to be the time at which halos are identified. 
Further, we shall often drop the argument $R$ when no confusion is possible.
The window function $W_R(k)$ reflects the fact that the formation of a given, 
isolated dark matter halo is insensitive to small-scale (UV) matter
density fluctuations with $k \gg 1/R$. 
The filter shape $W_R(k)$ and its corresponding filtering scale 
$R(M)\propto M^{1/3}$ can be measured upon tracing halos back to the initial conditions 
\citep[e.g.][]{baldauf/desjacques/seljak:2015,chan/sheth/scoccimarro:2015}. 
Numerical simulations indicate that $W_R(k)$ is more extended than a tophat, but more compact than a Gaussian 
\citep{dalal/white/etal:2008,chan/sheth/scoccimarro:2015}.
In order for the peak constraint to be meaningful, the actual filtering kernel $W_R(k)$ 
used must be such that the convergence of the spectral moments up to $\sigma_2(R)$ is 
ensured.  
While this is not the case for the tophat filter, this is certainly true for the Gaussian 
filter and, apparently, also for the actual filter $W_R(k)$, which can be measured upon tracing back halos to the initial conditions
\cite{dalal/white/etal:2008,baldauf/desjacques/seljak:2015,chan/sheth/scoccimarro:2015}.  
In what follows, we will assume a Gaussian filter for simplicity, 
until we reach the excursion-set peaks, \refsec{esp}, where a tophat filter 
is used to smooth the density and a Gaussian is used to smooth its derivatives.

\begin{table*}[t]
\centering
\begin{tabular}{l|l}
\hline
\hline
Quantity & Symbol\\
\hline
Zero-lag spectral moments & $\sigma_n(R)$ \quad [\refeq{mspec}] \\
Correlation functions (non-zero lag spectral moments) & $\xi_\ell^{(n)}(R,r)$ \quad[\refeq{xielln}] \\
Spectral shape parameters & $\gamma_n(R)$ \quad [\refeq{gammas}] \\
Normalized smoothed density & $\nu(\vq) \equiv \frac{1}{\sigma_0}\delta_R(\vq)$ \\
Normalized smoothed density gradient & $\eta_i(\vq) \equiv \frac{1}{\sigma_1}\partial_i\delta_R(\vq)$ \\
Normalized smoothed density Hessian & $\zeta_{ij}(\vq) \equiv \frac{1}{\sigma_2}\partial_i\partial_j\delta_R(\vq)$ \\
Eigenvalues of $-\zeta_{ij}$ & $\lambda_1,\,\lambda_2,\,\lambda_3$ \\
Rotational invariants of $\zeta_{ij}$ & $J_1,\,J_2,\,J_3$ \\
Peak curvature & $J_1 \equiv - \tr[\zeta_{ij}]$ \\
Scaled, uniformly distributed invariant & $x_3\equiv J_3/(J_2)^{3/2}$ \\
Lagrangian number density of BBKS peaks & $\npk(\vq)$ \\
Mean comoving BBKS peak number density & $\bnpk$ \\
Derivative of $\d_R$ with respect to smoothing scale & $\mu_R=-\frac{d\d}{dR}$ \\
Lagrangian number density of ESP peaks & $n_\text{ESP}(\vq)$ \\
Mean comoving ESP peak number density & $\overline{n}_\text{ESP}$ \\
Lagrangian 2-point correlation function of peaks & $\xpk(r)$ \\
Orthogonal polynomials and their duals & $O_{\bf n}(\vw)$, $O_{\bf n}^\star(\vw)$ \\
Polynomials corresponding to $J_2$ and $J_3$ & $F_{lm}(5J_2,J_3)$ \\
$n$-th order Lagrangian bias & $c_n^L(\vk_1,\dots,\vk_n)$ \\
$n$-th order Eulerian bias & $c_n^E(\vk_1,\dots,\vk_n,\tau)$ \\
Linear peak displacement & $\vlpk(\vk,\tau)$ \\
Linear peak velocity bias & $c_{v,\text{pk}}(k)$ \\
Linear peak velocity dispersion & $\sigma_{v,\text{pk}}(\tau)$ \\
\hline
\end{tabular}
\caption{List of symbols used throughout \refsec{peaks}. Note that $J_1$ is often denoted as $u$ in the literature.}
\label{tab:symbol6}
\end{table*}

Following Ref.~\cite{bardeen/etal:1986} (BBKS), characteristic scales can be defined 
by taking ratios of spectral moments. In particular,
\be
R_n(R)\equiv \sqrt{3}\frac{\sigma_{n}(R)}{\sigma_{n+1}(R)}
\ee
defines an ordered sequence of characteristic lengths $R_0\geq R_1\geq R_2\geq \dots$
for typical values of interest of $R$, due to the shape of the linear matter power spectrum $\Plin(k)$.
The first two scales are the typical separation $R_0$  between zero-crossings of the 
density field, and the mean distance $R_1$ between stationary point 
\cite{bardeen/etal:1986,pogosyan/etal:2009}.
These are the only scales involved in the calculation of the peak 
correlation functions. For subsequent use, we also introduce the 
dimensionless parameters
\be
 \gamma_n(R)=\frac{\sigma_{n}^2(R)}{\sigma_{n-1}(R)\sigma_{n+1}(R)}
 \label{eq:gammas}
\ee
which quantifies the range over which $R^{-3} (kR)^{2(n-1)}P_{\rm L}(k)W_R^2(k)$ is significant, 
i.e. whether it is sharply peaked or broad.

The analogous quantities to $\sigma_n^2$ at non-zero separation are 
defined as follows:
\begin{equation}
 \xi_\ell^{(n)}(R,r)= 
\int_{\vk}
  k^{2(n+1)} W_R^2(k)\,\Plin(k)\, j_\ell(kr)\;,
 \label{eq:xielln}
\end{equation}
where $j_\ell(x)$ are spherical Bessel functions. As $\ell$ is increased at fixed $r$, these 
harmonic transforms become increasingly sensitive to the small scale power.  

The auto- and cross-correlations of the fields $\eta_i(\vq)$, 
$\nu(\vq)$ and $\zeta_{ij}(\vq)$ can generally be decomposed  
into components with definite transformation properties under 
three-dimensional rotations. Explicit expressions can be found in 
\cite{regos/szalay:1995,desjacques:2008,desjacques/crocce/etal:2010}.

\subsection{The Kac-Rice formula}
\label{sec:kacrice}

Let $\{\vq_1,\vq_2,\cdots,\vq_p,\cdots\}$ be the Lagrangian positions of point particles 
such as centers of halos in some (finite or infinite) volume. 
The comoving Lagrangian density $n_g(\vq)$ of these point particles (for a brief introduction to point processes in a 
cosmological context, see \cite{martinez/saar:2002}) is formally written as a 
sum of Dirac distributions
\begin{equation}
n_g(\vq)=\sum_p\delta_D\!\left(\vq-\vq_p\right)\;.
\label{eq:npk1}
\end{equation}
As shown in \cite{kac:1943,rice:1945,bardeen/etal:1986}, in the case of stationary points (maxima, minima and saddle 
points) of a random field, the density $n_\text{sp}(\vq)$ can be entirely expressed in terms of $\nu$ and its derivatives.  
To see why this is the case, note that the density of stationary points,
defined as points $\vq_p$ where the gradient of the density field satisfies $\veta_i(\vq_p)=0$, can be written as 
\be
n_\text{sp}(\vq) = \d_D[\veta(\vq)] \left|\frac{\partial\veta(\vq_p)}{\partial\vq_p}\right|\,,
\label{eq:npk2}
\ee
where the Jacobian ensures that one obtains a proper density following \refeq{npk1}.  
Thus, we need to evaluate the derivative of the gradient $\veta$ with respect to $\vq$.  
Now, in the neighborhood of a stationary point located at $\vq_p$, the gradient is given by
\begin{equation}
\eta_i(\vq) = \frac{\sqrt{3}}{R_1}\,\zeta_{ij}(\vq_p)
\left(q-q_p\right)^j + \O\Big((\vq-\vq_p)^2\Big)\;.
\end{equation}
Then, \refeq{npk2} can be rephrased as
\begin{equation}
n_\text{sp}(\vq)=\sum_p \delta_D\!\left(\vq-\vq_p\right)=
\frac{3^{3/2}}{R_1^3}\,|\det\zeta_{ij}(\vq)|
\delta_D\!\left[\v{\eta}(\vq)\right]\;,
\label{eq:next}
\end{equation}
provided that the Hessian $\zeta_{ij}$ is invertible. 
The Dirac delta $\delta_D[\v{\eta}(\vq)]$  ensures that all the 
extrema are included.
This expression, known as the Kac-Rice formula~\cite{kac:1943,rice:1945}
holds for arbitrary smooth random fields.

Since we are only interested in counting the density maxima, we further 
require that $\zeta_{ij}(\vq_p)$ be negative definite at the position $\vq_p$ of
the stationary point. Finally, one usually restricts the set to those maxima 
with a certain threshold height $\nu_c$ following 
\cite{kaiser:1984,bardeen/etal:1986}, as in the case of thresholding (\refsec{localbias}).  
The {\it localized} number density of ``BBKS peaks'' of height $\nu_c$ then reads
\begin{equation}
\npk(\vq)=\frac{3^{3/2}}{R_1^3}\,
|\det\zeta_{ij}(\vq)|\delta_D\!\left[\v{\eta}(\vq)\right]
\Theta_H(\lambda_3)\delta_D\!\left[\nu(\vq)-\nu_c\right]\;, 
\label{eq:kacrice}
\end{equation}
where $\Theta_H$ is the Heaviside step function and $\lambda_3$ is the 
smallest eigenvalue of $-\zeta_{ij}$.  
The determinant somewhat complicates calculations. 
To circumvent this difficulty, one could weight the peaks by the inverse of 
$|{\rm det}\zeta_{ij}|$ \cite{verde/jimenez/etal:2014}.
However, it is unclear how such a weighted field exactly relates to dark matter
halos.

\subsection{Rotational invariants and their distribution}
\label{sec:invariants}

Since clustering statistics of peaks are invariant under three-dimensional
rotations, they depend only on the scalar functions that can be constructed
from the independent variables $\nu$, $\eta_i$, $\zeta_{ij}$ (each of which 
contains, respectively, one, three and six degrees of freedom). 
Furthermore, since the
peak constraint does not induce a dependence on $\eta_i\zeta_{ij}\eta_j$, 
only five scalar functions are relevant: 
$\nu(\vq)$, $J_1(\vq)=-\tr\big(\zeta_{ij}(\vq)\big)$, the essentially $\chi^2$-distributed 
quantity $\eta^2(\vq)=\sum_i \eta_i^2(\vq)$, and the two rotational invariants
\begin{equation}
J_2(\vq) = \frac{3}{2}\tr\big(\bar{\zeta}_{ij}^2(\vq)\big) \qquad
J_3(\vq) = \frac{9}{2}\tr\big(\bar{\zeta}_{ij}^3(\vq)\big) \;,
\end{equation}
where $\bar\zeta_{ij}=\zeta_{ij}+(1/3)J_1 \delta_{ij}$ are the components 
of the traceless part of the Hessian ($\zeta_{ij}$). We will group all these 
rotational invariants into the vector of variables
\begin{equation}
\label{eq:w}
  \vw \equiv
  \big\{\nu(\vq), J_1(\vq), 3\eta^2(\vq), 5J_2(\vq),J_3(\vq) \big\} \;,
\end{equation}
and denote their PDF as $p(\vw)$. The reason for using $3\eta^2$ and $5J_2$ will become 
apparent shortly. The remaining $10-5=5$ angular degrees of freedom, which we group into the 
vector $\vomg$, characterize the direction of the gradient $\grad\d_R$ (two angles) and 
the orientation of the principal axis frame of $\zeta_{ij}$ (three ``Euler'' angles). The
corresponding PDF is $p(\vomg)$. These angles will factorize out of the following 
calculations, so we can simply ignore them.

For Gaussian initial conditions, the probability density $p(\vw)$ simplifies to
\begin{equation}
\label{eq:PG}
p(\vw) = \frac{5\sqrt{5}}{8\pi^2\sqrt{3}} \frac{\sqrt{\eta^2}}{\sqrt{1-\gamma_1^2}}
\exp\left[-\frac{\nu^2+J_1^2-2\gamma_1\nu J_1}{2(1-\gamma_1^2)}-\frac{3}{2}\eta^2-\frac{5}{2}J_2\right]\;.
\end{equation}
Importantly, we always have $J_2>0$ and $J_3^2\leq (J_2)^3$ since $\zeta_{ij}$ is symmetric. Therefore, 
$p(\vw)$ is normalized such that
\begin{equation}
1 =
\int_{-\infty}^{+\infty}\!\!d\nu\int_{-\infty}^{+\infty}\!\!dJ_1\int_0^\infty\!\!d(3\eta^2) 
\int_0^\infty\!\!d(5J_2)
\int_{-J_2^{3/2}}^{+J_2^{3/2}}\!\!dJ_3\, p(\nu,J_1,3\eta^2,5J_2,J_3) \;,
\end{equation}
even though it does not explicitly depend on $J_3$. Clearly, the integral over $J_3$ is trivial 
and results in $5J_2$ being $\chi^2$-distributed with 5 degrees of freedom. These 5 degrees of freedom 
correspond to the 5 independent components of $\bar\zeta_{ij}$. This is the reason for writing 
$5J_2$. Similarly, one can easily show that $3\eta^2$ is $\chi^2$-distributed with 3 degrees of freedom.  
Note also that $\int\! d\vomg\, p(\vomg)=1$.

The variable $J_3$ does not explicitly appear in \refeq{PG} because it is, in fact, uniformly distributed.  
In order to emphasize the point, we introduce $x_3\equiv J_3/(J_2)^{3/2}$. 
Then, $p(\vw)d\vw$ can be written as
\begin{equation}
\label{eq:newfactorization}
  p(\vw) d\vw
  = {\cal N}(\nu,J_1) d\nu dJ_1
  \times\chi^2_3(3\eta^2) d(3\eta^2)
  \times\chi^2_5(5J_2) d(5J_2)
  \times\frac{1}{2}\Theta_H\!(1-x_3^2)dx_3 \;.
\end{equation}
In the above expression, ${\cal N}(x,y)$ is a bivariate Normal distribution with unit variances, 
\begin{equation}
  {\cal N}(\nu,J_1) = \frac{1}{2\pi\sqrt{1-\gamma_1^2}}
\exp\left[-\frac{\nu^2+J_1^2-2\gamma_1\nu J_1}{2(1-\gamma_1^2)}\right] \;,
\label{eq:NnuJ1}
\end{equation}
whereas $\chi^2_k (x)$ is the $\chi^2$ distribution with $k$ degrees of freedom,
\begin{equation}
\chi^2_k (x) = \frac{1}{2^{k/2} \Gamma(k/2)} x^{k/2 -1} e^{-x/2} \;.
\label{eq:chi}
\end{equation}
Clearly, $x_3$ has uniform probability density in the range 
$-1\leq x_3\leq +1$; thus, we can think of 
it as the cosine of an angle. Note, however, that the ``angle'' $x_3$ has nothing to do with spatial 
rotations. In fact, as the cubic root of the characteristic polynomial det$(\zeta-\lambda \vii)$, where 
$\vii$ is the $3\times 3$ identity matrix, $x_3$ itself is a rotational invariant.  
Below, we will see that, unlike $\vomg$ which does not contribute to the clustering of peaks, $J_3(\vq)$ actually does 
contribute, and, therefore, $x_3$ does not describe a rotation in space.

\subsection{Average peak number density*}
\label{sec:pknum}
\technote{* This section is of a more technical nature and is not essential for the remainder of this section.  The main result of this section is \refeq{bnpk}.}

Owing to the peak constraint, the calculation of the $N$-point correlation 
functions of peaks of a 3-dimensional random field requires the evaluation of high-dimensional integrals 
over a joint probability distribution in 5$N$ variables, after the 5$N$ angular variables have been integrated out.  
Therefore, even the evaluation of the 1-point correlation function or average density of peaks $\bnpk$ is 
not completely trivial. 

The average number density of peaks of a 3-dimensional Gaussian random field,  
\begin{equation}
\label{eq:pointaverage}
\bnpk(\nu_c) \equiv \big\langle\npk(\vq)\big\rangle = \int\!\!d^5\vw\,\npk(\vw)\,p(\vw) \;,
\end{equation}
was first calculated in BBKS \cite{bardeen/etal:1986}. Here however, we shall not proceed along the lines of 
\cite{bardeen/etal:1986}, who explicitly wrote the volume element $d^6\zeta_{ij}$ in terms of the 
three eigenvalues $\lambda_i$ of $-\zeta_{ij}$. 
Rather, we will follow the calculation of \cite{lazeyras/musso/desjacques:2015}, which exploit the invariance 
under rotations and, hence, is far simpler.

Namely, the peak constraint implies that all three eigenvalues of the Hessian $\zeta_{ij}$ be negative. 
In terms of the rotational invariants, the restriction to local maxima of the density field translates 
into the conditions $J_1>0$, $J_2<J_1^2$ and $J_3<(J_1/2)(J_1^2-3J_2)$. 
Taking into account the symmetry of $\zeta_{ij}$, the last condition implies that $x_3$ must satisfy
\begin{equation}
  -1<x_3<\mbox{min}\big[1,(y/2)(y^2-3)\big] \;,
\label{eq:detconstraint}
\end{equation}
where $y\equiv J_1/\sqrt{J_2}$. This splits the parameter space into two different regions depending on 
whether the inequality $(y/2)(y^2-3)<1$ holds. For $0<J_2<J_1^2/4$, one finds $-1<x_3<+1$ whereas, for 
$J_1^2/4<J_2<J_1^2$, the more stringent constraint $-1<x_3<(y/2)(y^2-3)$ applies. 
Therefore, the multiplicative factor of $\Theta_H(\lambda_3)$ in the localized peak number density 
\refeq{kacrice} can also be written as 
\begin{align}
\Theta_H(\lambda_3) = \Theta_H(J_1)
\bigg\{ \Theta_H\big(J_1^2/4-J_2\big) +\, & \Theta_H\big(J_2-J_1^2/4\big)\,
\Theta_H\big(J_1^2-J_2\big)
\nonumber \\
& \times \Theta_H\big(y^3/2-3y/2-x_3\big)\bigg\} \;.
\end{align}

We are now in the position to derive the well-known BBKS formula for the 
average peak number density $\bnpk$ in a very simple way. 
To compute $\bnpk$, one usually expresses the measure $d(5J_2)$ in terms of the ellipticity $v$, the 
prolateness $w$ and three Euler angles so that, in these new variables, $J_2=3v^2+w^2$. 
The calculation would then proceed along the lines of BBKS.
However, this change of variable is, in fact, unnecessary as the calculation can be explicitly carried out 
in the variables $J_1$, $5J_2$ and $x_3$ on imposing the aforementioned conditions.
To illustrate this point, we begin by rewriting the determinant $|\det(\zeta_{ij})|$ in \refeq{kacrice}
in terms of the $J_i$. Introducing $s\equiv 5J_2$, we obtain
\begin{align}
|\det(\zeta_{ij})|
&= \frac{1}{27}\left(J_1^3-\frac{3}{5}s J_1 - \frac{2}{5^{3/2}}s^{3/2}x_3\right) \;,
\end{align}
since $\det(\zeta_{ij})$ is always negative for density maxima. The integral over the variables $5J_2$ and 
$x_3$ becomes 
\begin{align}
\int\!\!ds\, \chi_5^2(s) & \int\!\!dx_3\,\frac{1}{2}\Theta_H(1-x_3^2)\, 
\big\lvert\mbox{det}(\zeta_{ij})\big\lvert\,\Theta_H(\lambda_3) \\
&= \frac{1}{2^{7/2}3^3\Gamma(5/2)}\biggl\{\int_0^{5J_1^2/4}\!\!ds\int_{-1}^{+1}\!\!dx_3+
\int_{5J_1^2/4}^{5J_1^2}\!\!ds\int_{-1}^{\sqrt{5/s}(J_1/2)(5J_1^2/s-3)}\!\!dx_3\biggr\} \nonumber \\
&\qquad \times \left(J_1^3-\frac{3}{5}s J_1 - \frac{2}{5^{3/2}}s^{3/2}x_3\right)  s^{3/2} e^{-s/2} \;, \nonumber
\end{align}
and can be computed straightforwardly.
Taking into account two additional multiplicative
factors of $3^{3/2}$, one arising from $\npk(\vy)$ and the other from the integral over 
$\chi_3^2(3\eta^2)\delta_D(\veta)$, we find
\begin{align}
\label{eq:fJ1}
3^3\int\!\!ds\, \chi_5^2(s) & \int\!\!dx_3\,\frac{1}{2}\Theta_H(1-x_3^2)\, 
\big\lvert\mbox{det}(\zeta_{ij})\big\lvert\,\Theta_H(\lambda_3) \\
&= \sqrt{\frac{2}{5\pi}}\left[\left(\frac{J_1^2}{2}-\frac{8}{5}\right)e^{-5J_1^2/2}
+\left(\frac{31}{4}J_1^2+\frac{8}{5}\right) e^{-5J_1^2/8}\right]+\frac{1}{2}\left(J_1^3-3J_1\right) \nonumber \\
&\qquad \times 
\left[\mbox{Erf}\!\left(\sqrt{\frac{5}{2}}J_1\right)+\mbox{Erf}\!\left(\sqrt{\frac{5}{2}}\frac{J_1}{2}\right)\right]
\equiv f(J_1) \nonumber \;,
\end{align}
which is precisely the function $f(J_1)$ defined in \cite{bardeen/etal:1986}. 
The rest of the calculation is trivial, and we immediately recover their well-known expression for the average 
peak abundance $\bnpk(\nu_c)$,
\begin{equation}
\label{eq:bnpk}
  \bnpk(\nu_c) = \frac{1}{(2\pi)^2 R_\star^3} 
  G_0(\gamma_1,\gamma_1\nu_c) e^{-\nu_c^2/2}\;,
\end{equation} 
where the function $G_0$ is a special case of
\begin{align}
\label{eq:Gn}
G_n(\gamma_1,\omega)=\int_0^\infty\!\!du\,u^nf(u)\,
\frac{e^{-(u-\omega)^2/2(1-\gamma_1^2)}}{\sqrt{2\pi(1-\gamma_1^2)}}\;,
\end{align}
and $f(u)$ is the function defined in \refeq{fJ1}.  
The integration over $u$ must generally be performed numerically.
Note that, within the peak formalism, $\bnpk$ is then related to the mean halo abundance through
$\bnpk(\nu_c) d\nu_c = \avnh(\nu_c) d\ln M$.

It is worth noticing that, while the exponential 
$\exp[-(u-\omega)^2/2(1-\gamma_1^2)]$ 
decays rapidly to zero, $u^n f(u)$ are rapidly rising.
As a result, the integrands are sharply peaked, and
the functions $G_n(\gamma_1,w)$ receive most of their contribution around the 
maximum of the peak.
For large values of $\omega$, we find that $G_0$ and $G_1$ asymptote to
\begin{align}
G_0(\gamma_1,\omega) &\approx \omega^3-3\gamma_1^2\omega+B_0(\gamma_1)\,
\omega^2\, e^{-A(\gamma_1)\omega^2} \\
G_1(\gamma_1,\omega) &\approx \omega^4+3\omega^2\left(1-2\gamma_1^2\right)
+B_1(\gamma_1)\,\omega^3\, e^{-A(\gamma_1)\omega^2} \;.
\end{align}
The coefficients $A(\gamma_1)$, $B_0(\gamma_1)$ and $B_1(\gamma_1)$ are 
obtained from the asymptotic expansion of the Error function that
appears in \refeq{fJ1}. We have explicitly
\begin{equation}
A=\frac{5/2}{\left(9-5\gamma_1^2\right)},\qquad
B_0=\frac{432}{\sqrt{10\pi}\left(9-5\gamma_1^2\right)^{5/2}},\qquad
B_1=\frac{4 B_0}{\left(9-5\gamma_1^2\right)} \;.
\end{equation}
The value of $\nu_c$ at which $\npk(\nu_c)$ attains its maximum depends sensitively on the value of $\gamma_1$. 
For small $\gamma_1\lesssim 0.3$, the peak is at $\nu_c\simeq 0$, 
reflecting the fact that there is significant power on all scales. For large 
$\gamma_1\gtrsim 0.7$, most of the power is in a narrow range of wavenumber, 
so that density maxima are much more likely to reach heights well above typical, 1$\sigma$ fluctuations of the 
density field.

\subsection{Two-point correlation functions of peaks}
\label{sec:pkcorr}

While the average number density of peaks has been worked out analytically in \cite{bardeen/etal:1986}, 
progress with the computation of the peak correlation functions, 
which is difficult to perform rigorously, has been relatively slow.
Even though the peak correlation functions do not generally have explicit analytical expressions, 
they provide much information on ideas such as the peak-background split and its range of validity. 
Furthermore, in low dimensions they can be solved numerically and, therefore, furnish insights 
into non-perturbative effects such as small-scale halo exclusion.
For illustrative purposes, we will focus on BBKS peaks defined through \refeq{kacrice}, 
but the following considerations can be straightforwardly extended to the excursion-set peaks discussed in \refsec{sub:esp}.  

\subsubsection{The peak-density cross-correlation function}
\label{sec:pkcross}

Let us begin with the cross-correlation function $\xi_{\text{pk},\delta}(r)$ between peaks of height $\nu_c$ and the 
Lagrangian density field.  
This is the Fourier transform of $P_{hm}^L(k)$ considered 
in \refsec{nptLagr}, evaluated for peaks.  Further, it corresponds to the average Lagrangian density 
profile $\la \delta(\v{r})|{\rm peak~at~}\v{0}\ra$ around peaks.  
Unlike the peak auto-correlation function, the cross-correlation function can be derived in closed form \cite{bardeen/etal:1986}, 
and is given by
\begin{equation}
\xpd(r)=b_{10}^L\xi_0^{(0)}\!(R,r)+ b_{01}^L \xi_0^{(1)}\!(R, r)
=\left[b_{10}^L - b_{01}^L \lapl\right] \xi_0^{(0)}\!(R,r)
\label{eq:crosspkdensity}
\end{equation}
in the notation adopted here.  We have used the relation $\lapl\xi_0^{(0)}=-\xi_0^{(1)}$ in the second equality.  
In the notation of \cite{desjacques/sheth:2010}, 
$b_{10}^L= b_\nu$ and $b_{01}^L= b_\zeta$, while in the notation of \refsec{evolution} and 
\refsec{measurements}, $b_{10}^L = b_1^L$ and $-b_{01}^L$ contributes to the leading Lagrangian higher-derivative bias $b_{\lapl\d}^L$, 
along with the contribution from the filtering kernel in $\xi_0^{(0)}(R,r)$ derived in \refeq{bhderiv_thr} (\refsec{higherderiv}).  
The linear BBKS peak bias parameters are given by
\begin{equation}
\label{eq:b10}
b_{10}^L = \frac{1}{\sigma_0}\left(\frac{\nu_c-\gamma_1 \bar{J}_1}{1-\gamma_1^2}\right)\;, \qquad
b_{01}^L = \frac{1}{\sigma_2}\left(\frac{\bar{J}_1-\gamma_1\nu_c}{1-\gamma_1^2}\right) \;,
\end{equation}
where $\bar{J}_1\equiv G_1(\gamma_1,\gamma_1\nu_c)/G_0(\gamma_1,\gamma_1\nu_c)$ is the average 
curvature of peaks of height $\nu_c$.  We will see in \refsec{PBSpeaks} that these
bias parameters can be obtained directly from the peak abundance \refeq{kacrice} using 
a peak-background split argument.  
Note that $b_{10}^L$ and $b_{01}^L$ satisfy the relation
\begin{equation}
b_{10}^L + \left(\frac{\sigma_1^2}{\sigma_0^2}\right) b_{01}^L = \frac{\nu_c}{\sigma_0}\,,
\end{equation}
analogous to \refeq{10b01};  the additional factor of $\sigma_1^2/\sigma_0^2$ arises from the
fact that, unlike in \refeq{10b01}, $b^L_{01}$ defined here has dimensions of length$^2$.  
Let us also mention that $\xi_0^{(0)}(r)$ in Ref.~\cite{desjacques/sheth:2010} corresponds 
to Eq.~(7.10) in Ref.~\cite{bardeen/etal:1986}.
Although the latter equation appears to have an additional factor of 1/3 that multiplies 
the factors of $\lapl\xi_0^{(0)}$, this is only because 
Ref.~\cite{bardeen/etal:1986} measures $r$ in units of $R_1$.

\subsubsection{The peak auto-correlation function}
\label{sec:pkauto}

Unlike the peak-matter cross-correlation function, the auto-correlation function of peaks
is a significantly more involved calculation.  However, it exhibits interesting
features beyond those present in \refeq{crosspkdensity}, in particular
nonlinear bias and stochasticity.  To begin, we define the fractional
peak overdensity through
\begin{equation}
\dpk(\vq) = \frac{\npk(\vq)}{\bnpk}-1 \;,
\end{equation}
where $\bnpk$ is the average peak number density, \refeq{bnpk}.  
The peak auto-correlation function is then given by the expectation value of 
the product $\dpk(\vq_1)\dpk(\vq_2)$, which reads 
\begin{align}
\left\la\dpk(\vq_1)\dpk(\vq_2)\right\ra &= \frac{1}{\bnpk^2}
\Bigl\la\sum_\alpha \sum_\beta\delta_D\!\left(\vq_1-\vq_\alpha\right)
\delta_D\!\left(\vq_2-\vq_\beta\right)\Bigr\ra -1 \\ 
&= \frac{1}{\bnpk^2}\delta_D\!\left(\vq_1-\vq_2\right)
\Bigl\la\sum_\alpha\delta_D\!\left(\vq_2-\vq_\alpha\right)\Bigr\ra
+\frac{1}{\bnpk^2}\Bigl\la\sum_{\alpha\ne\beta}\delta_D\!\left(\vq_1-\vq_\alpha\right)
\delta_D\!\left(\vq_2-\vq_\beta\right)\Bigr\ra -1 \nonumber \\
&= \frac{1}{\bnpk} \delta_D\!\left(\vq_1-\vq_2\right)
+ \frac{1}{\bnpk^2}\sum_{\alpha\ne\beta}\Bigl\la\delta_D\!\left(\vq_1-\vq_\alpha\right)
\delta_D\!\left(\vq_2-\vq_\beta\right)\Bigr\ra -1 \nonumber \\
&= \frac{1}{\bnpk}\delta_D\!\left(\vq_2-\vq_1\right) + \xpk(r)  \;,
\label{eq:pk2pt}
\end{align}
with $r=|\vq_2-\vq_1|$.  
The first term on the right-hand side comes from self-pairs, which are usually
ignored in the calculation of the 2-point correlation function in real space,
but are relevant in Fourier space;   
the second term is the usual correlation function $\xpk(r)$ of peaks excluding
self-pairs.  

Before turning to the calculation of $\xpk(r)$, let us consider the issue
of stochasticity in a bit more detail.  The contribution $\bnpk^{-1} \d_D(\v{r})$ 
corresponds to Poisson shot noise, and yields an exactly constant
contribution $\bnpk^{-1}$ to the Fourier transform $\ppk(k)$ of \refeq{pk2pt}.
The actual stochastic contribution to the peak power spectrum on large scales can be obtained
by using the fact that the deterministic contribution scales as $\Plin(k) \propto k^{n_s-1}$
in the large-scale limit, as it does for any physical biased tracer.
The stochastic contribution can then be isolated as the constant that remains when sending $k$ to exactly zero:
\begin{equation}
P^{\{0\}}_{\eps,\rm pk} \equiv \lim_{k\to 0} \ppk(k) = \frac{1}{\bnpk} +\int\!\!d^3\vr\,\xpk(r) \;.
\label{eq:Ppklowk}
\end{equation}
If $\int d^3\vr\,\xpk(r)$ does not vanish, then peaks show super-Poisson 
or sub-Poisson noise in the low-$k$ limit of the power spectrum.  
In fact, simple arguments suggest that peak-peak exclusion should contribute a negative 
white-noise term, so that $P^{\{0\}}_{\eps, \rm pk} < 1/\bnpk$ 
(see \cite{smith/etal:2007,hamaus/seljak/etal:2010,baldauf/seljak/etal:2013} for a discussion of halo exclusion). 
Thus, peaks provide a nontrivial example of stochasticity beyond the
usual Poissonian counts-in-cells prediction \cite{peebles:1980,smith:2009}.  

We now turn to the proper 2-point correlation function $\xpk(r)$ of peaks of height $\nu_c$.  
Let $p_2$ be the joint probability for the vector of independent
components $\vy\equiv \{\nu,\eta_i,\zeta_{ij}\}$ at positions $\vq_1$ and $\vq_2$.  
Note that, since $\v{r}$ provides a preferred direction, it is no longer 
sufficient to write the peak abundance solely in terms of the five invariants $\v{w}$.  
Then, $\xpk(r)$ formally reads
\begin{equation}
1+\xpk(r)=\frac{1}{[\bnpk(\nu_c)]^2} \int\!\!d^{10}\vy_1 \int\!\!d^{10}\vy_2\,
\npk(\vy_1) \npk(\vy_2) \, p_2(\vy_1,\vy_2) 
\label{eq:2pk2}
\end{equation}
where, for shorthand convenience, subscripts denote quantities evaluated at different Lagrangian 
positions $\vq_1,\,\vq_2$, where $r = |\vq_2-\vq_1|$, and $\npk(\vy)$ is given by \refeq{kacrice}.  
It is now clear why the evaluation of \refeq{2pk2} is tedious: one must evaluate the joint probability 
distribution $p_2(\vy_1,\vy_2)$ for the $2\times 10$-dimensional vector of variables, which involves a
20-dimensional covariance matrix.

Several numerical and analytical investigations have been performed to evaluate \refeq{2pk2}.  
Early work \citep{peacock/heavens:1985,coles:1989,lumsden/etal:1989} focused on 
Monte-Carlo realizations of 1-dimensional Gaussian (and non-Gaussian) random fields to estimate 
the peak 2-point correlation function down to small separations where peak-peak exclusion becomes important.  
Ref.~\cite{bardeen/etal:1986} worked out analytically the peak 2-point correlation function
in the large-scale limit, neglecting spatial derivatives of the matter correlation function.  
This essentially corresponds to performing a \LIMD bias expansion as in \refsec{localbias}.    
Refs. \cite{otto/etal:1986,cline/etal:1987,matsubara:1995} obtained asymptotic expansions appropriate 
for high peak threshold $\nu_c\gg 1$.   
Leading-order derivations are presented in 
\cite{regos/szalay:1995,desjacques:2008,desjacques/sheth:2010}, whereas
\cite{desjacques/crocce/etal:2010,desjacques:2013} obtained the full NLO (1-loop) expression.

\begin{figure}[t!]
\centering
\includegraphics[trim=0cm 5cm 0cm 3.5cm,clip,width=0.49\textwidth]{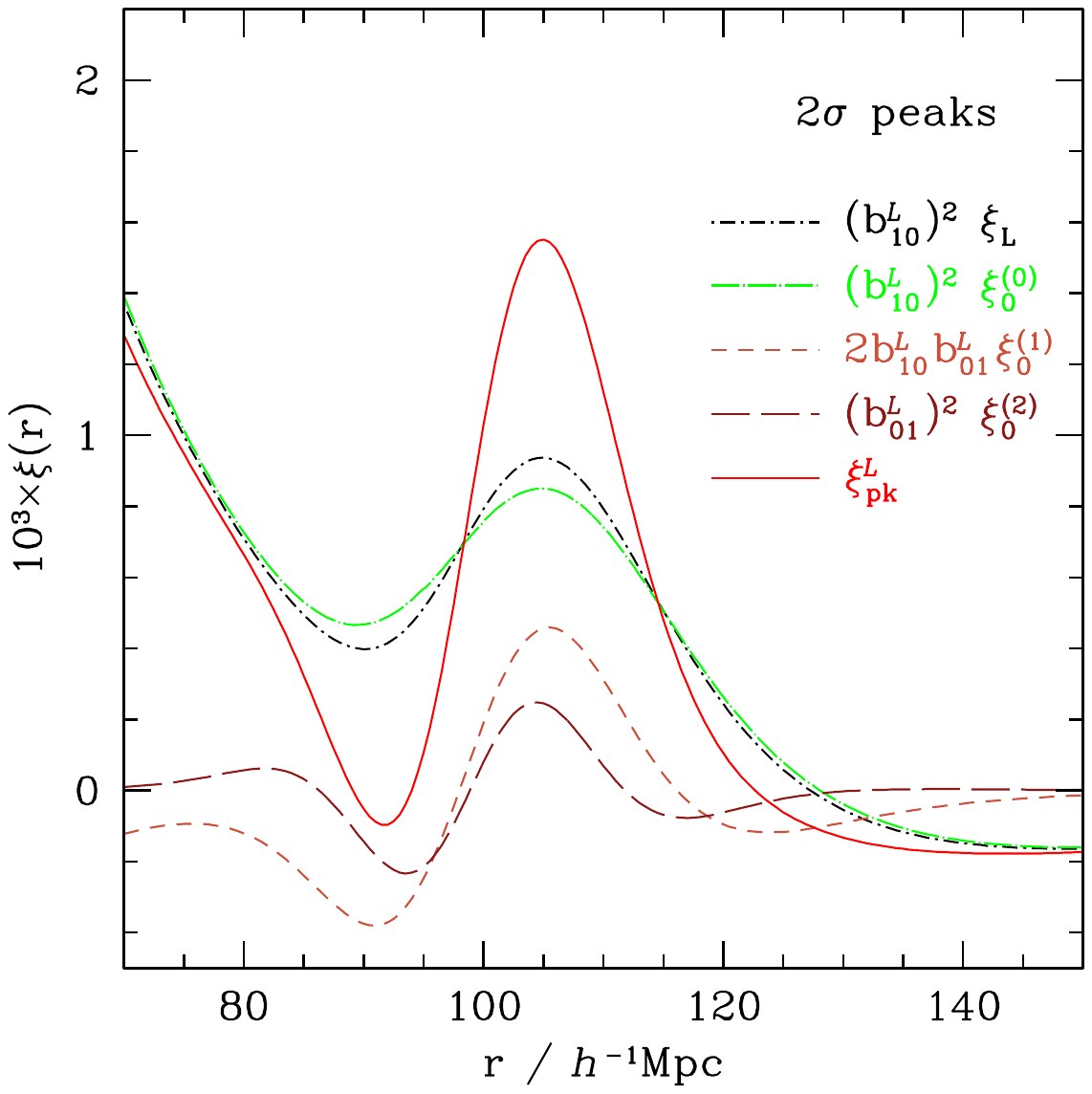}
\caption{A comparison between the {\it un-smoothed} Lagrangian matter correlation function $(b_{10}^L)^2\xi_{\rm L}(r)$ 
(black, dot-dashed)  and the Lagrangian correlation function $\xpk(r)$ of $\nu_c=2$ peaks (red, solid) around 
the BAO feature. In order to obtain the peak correlation function, the density field was smoothed with a Gaussian filter on mass 
scale $M=3\times 10^{13}\Msunh$. The dot-long dashed, short-dashed and long-dashed curves represent the 
individual contributions 
$(b_{10}^L)^2\xi_0^{(0)} = (b_1^L)^2\xi_{{\rm L},R}$, 
$2 b_{10}^Lb_{01}^L\xi_0^{(1)} = -2 b_1^L b_{\lapl\d}^L\lapl\xi_{{\rm L},R}$ 
and 
$(b_{01}^L)^2\xi_0^{(2)} = (b_{\lapl\d}^L)^2\laplsq\xi_{{\rm L},R}$ 
to the linear peak correlation function. A nonzero $b^L_{01}$ restores, and even amplifies the acoustic peak otherwise 
smeared out by the filtering of the matter density field. Note that the second- and higher-order 
contributions to $\xpk(r)$ are so small on these scales that they do not lead to visible differences.  
\figsource{desjacques/crocce/etal:2010}  
}
\label{fig:bao}
\end{figure}

In the following section, we will describe in detail a rigorous perturbative approach to deriving $\xpk(r)$ 
on large scales where the matter correlation function $\xi_{{\rm L},R}(r)$ is small.  
Before we discuss this approach, let us make a few general comments:
\begin{itemize}
\item
The peak auto-correlation function generally reaches the value $\xpk=-1$ in the limit $r \to 0$, because the
probability of finding two maxima arbitrarily close together is suppressed for the smoothed density field.
This small-scale peak exclusion is clearly seen in the 
one-dimensional calculation of \cite{lumsden/etal:1989,baldauf/codis/etal:2015}. 
As discussed above, this is related to the fact that $\int d^3\vr\,\xpk(r)$ does 
not generally vanish for density peaks. As a consequence, $P^{\{0\}}_{\eps, \rm pk} \ne 1/\bnpk$
so that the noise is not Poissonian. 
The one-dimensional calculation of \cite{baldauf/seljak/etal:2013} has shown that the 
small-scale peak exclusion can qualitatively reproduce the shot-noise matrix measurements 
of \cite{hamaus/seljak/etal:2010}.
\item
As pointed out in \cite{desjacques:2008}, the higher-derivative peak bias $b^L_{01}$ can significantly amplify
the contrast of the BAO feature in the clustering of initial density peaks if $b_{01}^L>0$, as illustrated in
\reffig{bao}. This arises from the fact that $\xi_0^{(1)}(R, r)$ is equal to 
$-\lapl \xi_0^{(0)}(R, r)$, and the BAO feature is enhanced through the derivative of the correlation 
function (see also the discussion in \refsec{higherderiv}).  
The analysis of \cite{desjacques/crocce/etal:2010} further includes the gravitational evolution to 
ascertain the extent to which the initial BAO sharpening survives at late time. A crucial component 
of this analysis is velocity bias.  
We will return to these issues in \refsec{sub:pkgrav}.  
\end{itemize}

\subsection{Perturbative peak bias expansion}
\label{sec:peakbias}

While \cite{desjacques/crocce/etal:2010} showed that the first- and some of the second-order 
contributions to $\xpk(r)$ 
(those depending on $\nu\propto\d_R$ and $J_1\propto -\nabla^2\d_R$) could be 
obtained from a peak-background split formulated in terms of conditional mass functions, 
validating thereby the peak-background split calculation of \cite{mo/jing/white:1997} of the \LIMD biases $b_{i0}^L$, 
they could not determine the physical origin of the other second-order contributions. 
This was clarified in \cite{desjacques:2013,lazeyras/musso/desjacques:2015}.

Let us go back to the definition of $\xpk(r)$ through \refeq{pk2pt}.  
Since it describes the 2-point correlation of a scalar variable $\d_{\rm pk}(\vq)$, which, 
moreover, only depends on the variables $\nu$, $\eta_i$ and $\zeta_{ij}$, $\xpk(r)$ can be 
expanded in two-point correlations of all independent scalars constructed from 
$\nu,\,\eta_i,\,\zeta_{ij}$;  that is,
\emph{exactly the five quantities in the vector} $\v{w}$ [\refeq{w}].  
Note that we have also relied on the absence of coupling between $\eta_i$ and $\zeta_{ij}$
in the BBKS peak prescription.  That is, the peak correlation function can be obtained 
perturbatively by writing the peak density perturbation $\dpk(\vq)$ as
\ba
\dpk(\vq) =\:& F\left[ \nu(\vq),\, J_1(\vq),\, 3\eta^2(\vq),\, 5J_2(\vq),\, J_3(\vq) \right] \vs
=\:& c_{\nu} \nu(\vq) + c_{\nu^2} \nu^2(\vq) + c_{J_1} J_1(\vq) + c_{\nu J_1} \nu(\vq) J_1(\vq) + \cdots\,,
\label{eq:dpkbare}
\ea
where in the second line we have written a few example terms up to second order, multiplied by bias coefficients.  
This corresponds to a specific Lagrangian bias expansion of the type discussed in \refsec{evol1} and \ref{sec:evol2}, 
see \reftab{translation} (p.~\pageref{tab:translation}).  Specifically, all terms involving the tidal field $K_{ij}$ are
absent, and the higher-derivative terms only consist of powers of $(\vn\d)^2$ and all scalar combinations of $\partial_i\partial_j\d$, 
in addition to the terms obtained by expanding the filtering kernel $W_R(k)$ (\refsec{higherderiv});  as discussed 
there, this is only a subset of all higher-derivative operators that could be present. 
The physical reason is, again, the simplification made by assuming the density peak constraint and spherical collapse approximation.

Now we are faced with two issues: first, in order to obtain a prediction for the peak correlation function, we need 
to calculate the values of the bias parameters;  second, when calculating the two-point function using \refeq{dpkbare} 
directly, we obtain a large number of zero-lag terms which spoil the perturbative expansion on large scales as described 
in \refsec{renorm:intro}.  
The second problem can be solved by replacing the fields in the second line of \refeq{dpkbare} with renormalized 
operators, multiplied by renormalized bias parameters:
\ba
\dpk(\vq) =\:& \sigma_0 b_{10}^L [\nu](\vq) + \frac12 \sigma_0^2 b_{20}^L [\nu^2](\vq) + \sigma_2 b_{01}^L [J_1](\vq)
+ \sigma_0\sigma_2 b_{11}^L [\nu J_1](\vq) + \cdots\,.
\label{eq:dpkrenorm}
\ea
Powers of $\sigma_i$ have been introduced to ensure that the renormalized bias parameters are defined relative to the physical
(unnormalized) fields.
As shown in \cite{desjacques:2013,lazeyras/musso/desjacques:2015}, the renormalized operators as well as their 
coefficients can be derived at all orders, given the Gaussian nature of the density field, by making use of orthogonal 
polynomials $O_{\bm{n}}^\star\big(\vw(\vq)\big)$: 
\begin{equation}
\label{eq:biasexpansion}
\dpk(\vq) = \sum_{\bm{n}\ne \{0\}} \sigma^{[\bm{n}]} b_{\bm{n}}^L O_{\bm{n}}^\star\big(\vw(\vq)\big) \;.
\end{equation}
Here, $\bm{n} = \{ i_1,\,i_2,\,i_3,\,i_4,\,i_5 \}$ is a list of indices that denote the highest power of each of the 5 variables
$\vw = \{\nu, J_1, \eta^2, J_2, J_3\}$ in the operator $O^\star_{\bm{n}}$, while $\sigma^{[\bm{n}]}$ is a shorthand for the product
$\sigma_0^{i_1}\sigma_1^{2i_3}\sigma_2^{i_2+2i_4+3i_5}$.
Thus, $\{ O_{\bm{n}}^\star \}_{\bm{n}}$ corresponds to the entire set of renormalized operators that can be constructed from the 5
invariants contained in $\vw$. Correspondingly, the coefficients $b^L_{\bm{n}}$ are physical, renormalized bias parameters which are measurable
for example through large-scale correlations, and are given by the \emph{1-point ensemble averages}
\begin{equation}
\label{eq:avbias}
\sigma^{[\bm{n}]} b_{\bm{n}}^L\equiv \frac{1}{\bnpk}\Big\langle \npk(\vq)\, O_{\bm{n}}[\vw(\vq)]\Big\rangle\,.
\end{equation}
Here, $O_{\bm{n}}^\star$ and $O_{\bm{n}}$ are dual polynomials satisfying (see \refapp{stat:Gauss})
\begin{equation}
\<O^\star_{\bm{n}}(\vq) O_{\bm{n'}}(\vq)\> = \delta_{\bm{n}\bm{n'}}\,,
\end{equation}
and $\delta_{\bm{n}\bm{n'}} = \prod_{k=1}^5 \d_{i_k i'_k}$. 
Thus, by making use of a perturbative bias expansion, we have reduced the very complicated problem of calculating the
peak two-point function \refeq{2pk2} to the \emph{much} simpler problem of calculating the auto- and cross-correlations of 
the scalars $\nu,\,\eta^2,\,J_i$, and the bias parameters \refeq{avbias} which are one-point expectation values.  
Moreover, as we will see in \refsec{pertpkcorr}, we can straightforwardly derive higher-order peak correlation functions in 
this approach as well, while their calculation based on the joint probabilities $p_3,\,p_4,\,\dots$ would be prohibitive.  
Note, however, that it is not possible to obtain the non-perturbative stochastic contributions in this way.  

In the following subsection, we will derive the specific form of the polynomials, and relate their associated bias parameters to 
the peak-background split. Note that the polynomials are defined up to multiplicative constants which can be re-absorbed in the 
bias parameters $b_{\bm{n}}^L$. In what follows, like in \refsec{evolution}, we will adopt a normalization such that 
the peak \LIMD bias parameters $b_{i0}^L$ correspond to the parameters $b_i^L$ used throughout (see \reftab{translation} for the 
correspondence between the peak bias parameters used here and those used in \refsecs{evolution}{measurements}).

\begin{table*}
\centering
\begin{threeparttable}[b]
\begin{tabular}{l|l}
\hline
\hline
Peak model (\refsec{peaks}) & General expansion (\refsecs{evolution}{measurements}) \\
\hline
$b_{ij}^L$ & $(-1)^j i!j!\, b^L_{\delta^i (\lapl\delta)^j}$~\tnote{1} \\
$\chi_i^L$ & $b^L_{(\vn\delta)^{2i}}$ \\
$\omega_{l m}^L$ & higher-derivative bias involving $J_2,\,J_3$ constructed from $(\partial_i\partial_j\d)$ \\
\hline
\end{tabular}
{\footnotesize
\begin{tablenotes}
\item[1] For $j>0$, there are further contributions to $b^L_{\delta^i (\lapl\d)^j}$ from the expansion of the filtering kernel (\refsec{higherderiv}).
\end{tablenotes}
}
\caption{Conversion between peak biases (left) and those appearing in the general bias expansion (right).  
\label{tab:translation}}
\end{threeparttable}
\end{table*}

\subsubsection{Polynomials, bias parameters and the peak-background split}
\label{sec:PBSpeaks}

The relevant polynomials are those associated with the distribution of the invariants $\vw$.
Namely, for the variables $\nu$ and $J_1$ distributed according to the bivariate normal distribution ${\cal N}(\nu,J_1)$ [\refeq{NnuJ1}], the orthogonal polynomials are 
the bivariate Hermite $H_{ij}(\nu,J_1)$.
For the $\chi^2$ variate $3\eta^2$ with 3 degrees of freedom, these are the generalized Laguerre polynomials $L_q^{(1/2)}\!(3\eta^2/2)$, which are orthogonal over $[0,\infty)$ with respect to the $\chi^2$-distribution $\chi^2_3(3\eta^2)$ with $k=3$ degrees 
of freedom.
Finally, for the jointly distributed variables $(J_2,J_3)$, the polynomials are \cite{lazeyras/musso/desjacques:2015}
\begin{equation}
F_{l m}(5J_2,J_3) 
= (-1)^l \sqrt{\frac{\Gamma(5/2)}{2^{3m}\Gamma(3m+5/2)}}\,L_l^{(3m+3/2)}\!(5J_2/2)\,
\mathcal{L}_m(x_3)\;,
\end{equation}
where $\mathcal{L}_m(x)$ are Legendre polynomials, and the factor of $(-1)^l$ ensures that the term with highest power of $J_2$
always has positive sign. The Legendre polynomials $\mathcal{L}_m(x_3)$ appear because they are the orthonormal polynomials 
associated with the uniform distribution on the interval $[-1,1]$, i.e. $\Theta_H(1-x_3^2)/2$ in \refeq{newfactorization}.
Note that both $L_q$ and $F_{lm}$ are self-dual, i.e. $L_q^{(1/2)\star}= L_q^{(1/2)}$ and $F_{l m}^\star = F_{l m}$ 
owing to the factorization of the PDF of $(5J_2,x_3)$. Finally, we slightly modified the notation of \cite{gay/pichon/pogosyan:2012}
so that their $F_{l0}$ agree with the above definition.
The physical origin of the appearance of these orthogonal polynomials can be found in the peak-background split:
{\it 
Long-wavelength background perturbations locally modulate the mean of the distributions ${\cal N}(\nu,u)$, $\chi_3^2(3\eta^2)$ 
and $\chi_5^2(5J_2)\Theta_H(1-x_3^2)$. 
The resulting non-central distributions can then be expanded in the appropriate set of orthogonal polynomials.}

The bias coefficients of the BBKS peaks are given by the ensemble average \refeq{avbias}.
For the variables $(\nu,J_1)$, these are 
\begin{align}
\sigma^i_0 \sigma^j_2  b_{ij}^L
&= \frac{1}{\bnpk}\Big\langle\npk\, H_{ij}(\nu,J_1)\Big\rangle \\
&= \frac{1}{\bnpk} \int\! d^5\vw\,\npk(\vw)\, H_{ij}(\nu,J_1)\, p(\vw) \nonumber \;,
\end{align}
where the peak constraint in $\npk(\vw)$ selects spatial locations corresponding to density peaks. 
Factors of $1/\sigma_0$ and $1/\sigma_2$ are introduced because bias factors are ordinarily 
defined relative to the physical field $\delta_R^{(1)}(\vq)$ and 
its derivatives, e.g. $\lapl\delta_R^{(1)}(\vq)$, rather than the 
normalized variables $\nu,\,J_1$.  In the particular cases $ij=(10)$ and $(01)$, we recover \refeq{b10}.  
Explicit expressions for the density bias parameters $b_{i0}^L$ can be found in \cite{mo/jing/white:1997}, 
whereas $b_{ij}^L$ with $j\geq 1$ can be found in \cite{desjacques/crocce/etal:2010,desjacques:2013}.

Since the bivariate Hermite polynomials can be generated through the shift 
${\cal N}(\nu,J_1)\to{\cal N}(\nu+\nu_\ell,u\to J_1+J_{1\ell})$, where $\nu_\ell$ and $J_{1\ell}$ are 
long-wavelength background perturbations uncorrelated with the (small-scale) fields $\nu(\vq)$ and $J_1(\vq)$, 
$b_{ij}^L$ are peak-background split biases as in \cite{kaiser:1984}. Thus, we find again that the renormalized, physical bias 
parameters are exactly those obtained from the peak-background split as discussed for the \LIMD biases $b_{i0}^L$ in \refsec{PBS}.   
Note also that, owing to the fact that $\nu$ and $J_1$ are correlated variables, the $b_{ij}^L$ can also be derived 
by considering only the long-wavelength perturbation $\nu_\ell$ as done in \cite{desjacques/crocce/etal:2010}.
Furthermore, since a shift in the mean curvature is equivalent to the replacement $\omega\to\omega - J_{1\ell}$ 
in \refeq{Gn}, the $b_{ij}^L$ can also be obtained as derivatives of the mean peak abundance with respect to $\omega$ as explained 
in \cite{PBSpaper}. This is in precise analogy to the peak-background split definition of the \LIMD bias parameters $b_i^L$ as 
responses of the mean abundance (\refsec{PBS}).  
Finally, the same reasoning holds if one exchanges $J_1$ for the slope 
$\delta_R'=d\d_R/dR$ considered in \refsec{completely_correlated}, 
because $\d'$ and $J_1$ are completely correlated for a Gaussian smoothing kernel.
Hence, the $b_{ij}^L$ which arise from $(\delta,\delta')$ can also be written down as ensemble average of bivariate Hermite polynomials
\cite{musso/paranjape/sheth:2012,desjacques/gong/riotto:2013}.

Similarly, the bias parameters associated with the invariant $3\eta^2$ are derived from the generalized Laguerre polynomials, i.e. 
\begin{align}
\sigma_1^{2q} \chi_q^L
&= \frac{1}{\bnpk}\Big\langle\npk\, L_q^{(1/2)}(3\eta^2)\Big\rangle \\
&= \frac{(-1)^q}{\bnpk} \int\! d^5\vw\,\npk(\vw)\, L_q^{(1/2)}\!(3\eta^2/2)\, p(\vw) 
\nonumber \;.
\end{align}
The factor of $(-1)^q$ ensures that the term with the highest power of $J_2$ always has a positive sign.
The first generalized Laguerre polynomials are $L_0^{(\alpha)}(x)=1$ and 
$L_1^{(\alpha)}(x) = -x+\alpha + 1$. 
Taking into account the peak constraint, the lowest-order bias parameter thus is \cite{desjacques:2013}
\begin{align}
\label{eq:varrho1}
\chi_1^L &= \frac{1}{\sigma_1^2\bnpk}
\int\!\!d^5\vw\,\npk(\vw) \left(\frac{3}{2}\eta^2-\frac{3}{2}\right) p(\vw) \\
&= -\frac{3}{2\sigma_1^2} \nonumber \;.
\end{align}
Note that $3\eta^2/2-3/2\geq -3/2$. Therefore, since $\int d\vw\,p(\vw)=\bnpk$, we always have 
$\chi_1^L\geq (-3/2)\sigma_1^2$ 
regardless of the peak properties (height, curvature, etc).
The effect of a background (long-wavelength) perturbation $\veta_\ell$ on the 
components of the density gradient is $\eta_i\to\eta_i+\eta_{i,\ell}$. 
The components $\eta_{i,\ell}$ need not be the same for distinct $i$.
However, owing to invariance under rotations, only the magnitude of the vector 
$\sum_{i=1}^3(\eta_{i,\ell})^2\equiv\eta_\ell^2$ matters. 
As a result, the background fluctuation effectively perturbs 
$\chi_3^2(3\eta^2)$ into a non-central $\chi^2$-distribution, with 
non-centrality parameter $\lambda=\sqrt{\eta_\ell^2}$. 
Equivalently, the $q$-th order bias parameters $\chi^L_q$ can also 
be written as 
the derivatives $\sim \langle\partial^q\npk/\partial(\eta^2)^q\rangle$,
again paralleling the case of \LIMD biases described in \refsec{bphys}. 

Finally, the bias parameters $\omega_{l m}^L$, which correspond to the variables $J_2$ and $J_3$, 
are given by
\begin{align}
\sigma_2^{2l+3m}\omega_{lm}^L
&= \frac{1}{\bnpk}\Big\langle\npk\, F_{l m}(5J_2,J_3)\Big\rangle \\
&= \frac{1}{\bnpk} \int\! d^5\vw\,\npk(\vw)\, F_{l m}(5J_2,J_3) p(\vw) \nonumber \;. 
\end{align}
Here again, the bias coefficients $\omega_{l m}^L$ can be derived through a peak-background 
split upon considering long-wavelength perturbations $J_{2l}$ and $J_{3l}$ to the second- and 
third-order invariant traces of $\bar{\zeta}_{ij}$. 
The computation of $\omega_{10}^L$ is straightforward, and one obtains \cite{desjacques:2013}
\begin{equation}
\label{eq:bzz}
\omega_{10}^L=-\frac{5}{2\sigma_2^2} \left(1+\frac{2}{5}
\partial_\alpha\ln G_0^{(\alpha)}\!(\gamma_1,\gamma_1\nu)\Bigr\rvert_{\alpha=1}\right)\;,
\end{equation}
where $G_0^\alpha$ is defined in \refeq{Gn}. 
Note that $J_2(\vq)$ quantifies the ellipticity
of the peak density profile. In the high-peak limit, $\omega_{10}^L\to 0$ reflecting the fact that 
the most prominent peaks are nearly spherical. 

The previous expressions correspond to restricted expansions in each of the
five invariants.  The fully general bias parameters $b_{\v{n}}^L$ of BBKS peaks 
take the generic form \cite{lazeyras/musso/desjacques:2015}
\begin{align}
\sigma_0^i\sigma_1^{2q}\sigma_2^{j+2l+3m} b_{ijqlm}^L 
&= \frac{1}{\bnpk}\Big\langle\npk\, O_{ijqlm}(\vw)\Big\rangle \vs
&\equiv \frac{1}{\bnpk}
\Big\langle\npk\, H_{ij}(\nu,J_1) (-1)^q L_q^{(1/2)}\!(3\eta^2/2) F_{lm}(5J_2,J_3)\Big\rangle
 \;.
 \label{eq:Oijqlm}
\end{align}
This defines the orthogonal polynomials $O_{ijqlm}(\vw)$.  Here we have again taken out 
factors of $\sigma_i$, so that the BBKS peak bias factors $b_{ijqlm}^L$ are defined relative to the 
unnormalized fields $\d_R(\vq)$, $\grad\d_R$ and so on, as is conventional in large-scale structure.  
Note that, in general, the $b_{\v{n}}^L$ do not factorize into a product 
of $b_{ij}^L$, $\chi_q^L$ and $\omega_{l m}^L$, 
which correspond to the subsets $b_{ij}^L = b_{ij000}^L $, $\chi_q^L = b_{00q00}^L$, and $\omega_{lm}^L = b_{000lm}^L$ 
because of the cross-correlation between the variables.

\begin{figure}
\centering
\includegraphics[trim=0cm 5cm 0cm 3.2cm,clip,width=0.49\textwidth]{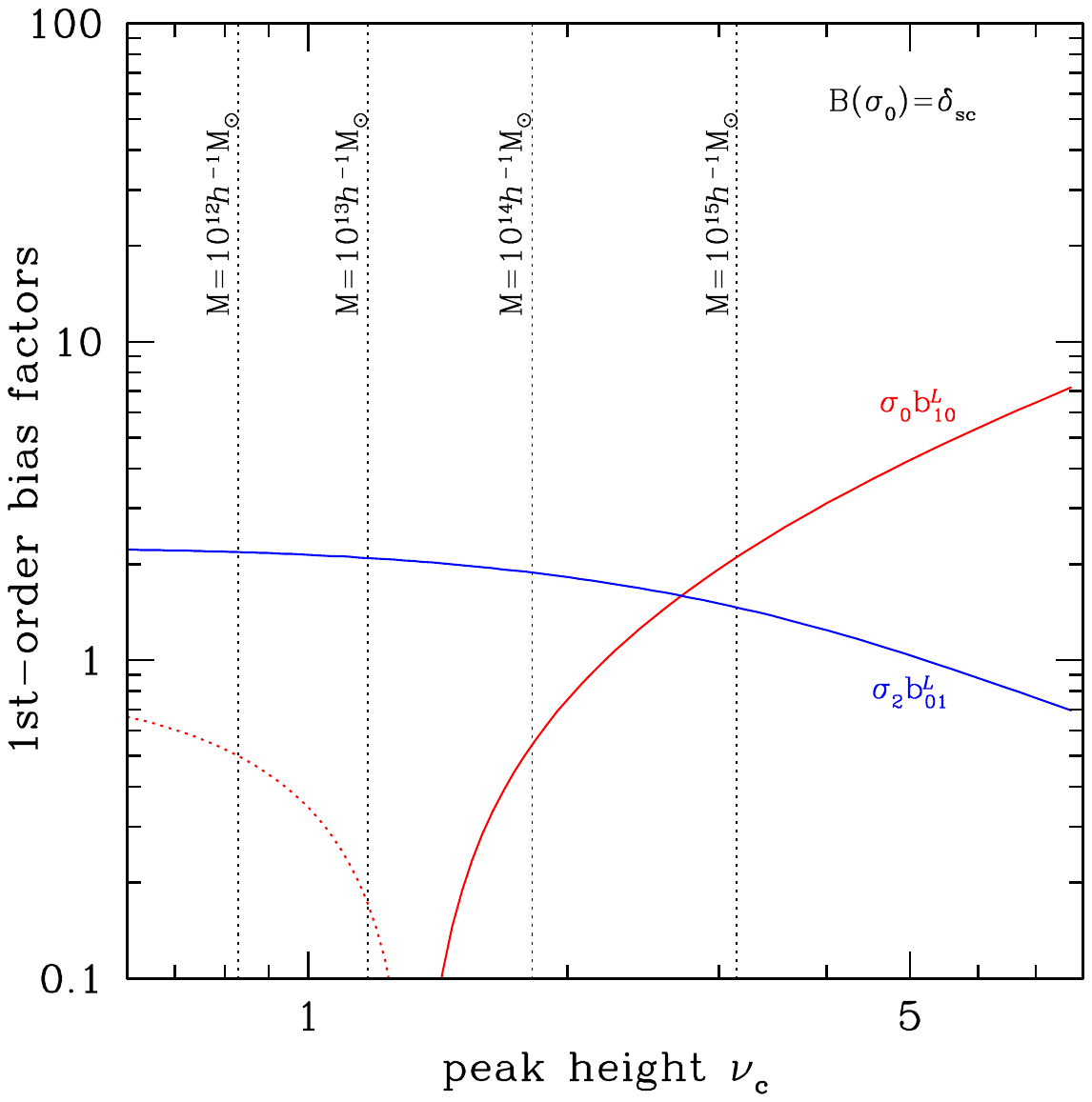}
\includegraphics[trim=0cm 5cm 0cm 3.2cm,clip,width=0.49\textwidth]{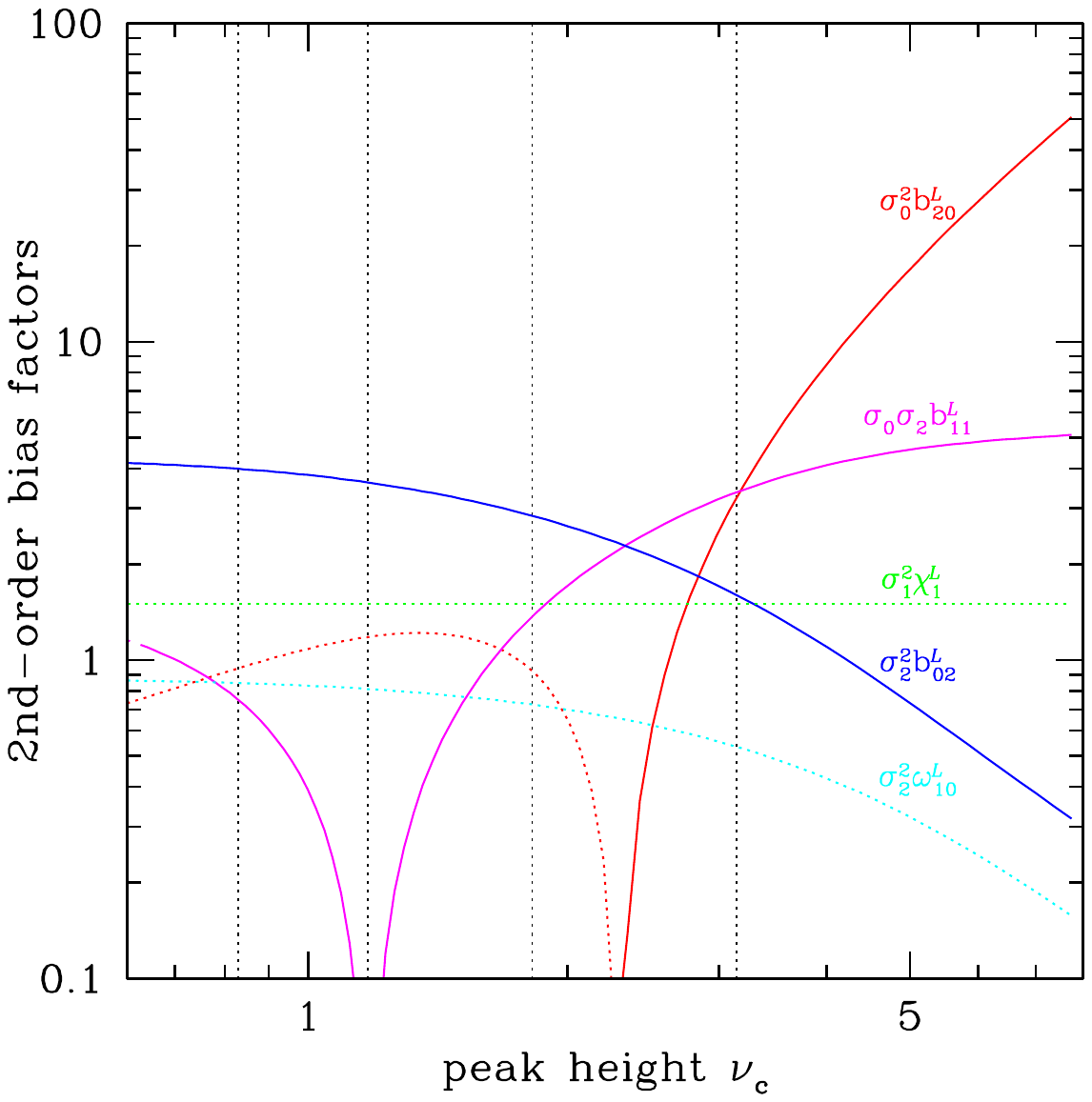}
\caption{
First-order \textit{(left panel)} and second-order \textit{(right panel)} Lagrangian peak bias parameters 
assuming a constant barrier $B(\sigma_0)=\dc$. The bias parameters have been multiplied by the
appropriate factors of $\sigma_i$, so that they all are dimensionless. 
The second-order bias factors $\sigma_0^2 b_{20}^L$, $\sigma_0\sigma_2 b_{11}^L$ and $\sigma_2^2 b_{02}^L$ 
are associated with the normalized variables $\nu(\vq)$ and $J_1(\vq)$,
whereas $\sigma_1^2 \chi_1^L$ and $\sigma_2^2 \omega_{10}^L$ multiply the contributions from  
$\eta^2(\vq)$ and $J_2(\vq)$ (see text), respectively. Vertical lines mark the peak significance 
at which the halo mass is $M=10^{12}$, $10^{13}$, $10^{14}$ and $10^{15}\hmsun$ (from left to
 right). Dashed curves indicate negative values. Results are shown at $z=0$ for the $\Lambda$CDM 
cosmology defined in \refsec{notation}. 
\figsource{desjacques:2013}  
}
\label{fig:pkbias}
\end{figure}

\reffig{pkbias} illustrates the behavior of the BBKS peak bias parameters $b_{ij}^L$, $\chi_q^L$ and $\omega_{lm}^L$ 
up to quadratic order. The bias parameter $b_{01}^L$ (which contributes to $-b^L_{\lapl\delta}$) is always positive whereas 
$\sigma_1^2\chi_1^L$ is negative and equal to $-3/2$ at all peak heights [see \refeq{varrho1}].

\subsubsection{Renormalization and peak correlation functions}
\label{sec:pertpkcorr}

Having obtained the renormalized bias parameters $b^L_{\bm n}$ in the previous section, we 
can now use \refeq{biasexpansion} to derive the $n$-point functions of peaks on large 
scales, by perturbatively expanding in the correlation functions of $\nu,\,J_1,\cdots$.  
For this, we need the orthogonal polynomials $O_{ijqlm}^\star$ to those
that appeared in \refsec{PBSpeaks}.   Their properties and construction
are described in \refapp{stat:Gauss}.  Let us give one example, 
\be
\big[\nu^2 J_1\big] \equiv H^\star_{21}(\nu,J_1) = \nu^2 J_1 - J_1 - 2\gamma_1\nu \;.
\ee
As in the case of the standard univariate Hermite polynomials, 
lower-order terms appear in addition to the leading-order term $\nu^2 J_1$
which exactly cancel the zero-lag terms or self-correlators that appear in
correlators of $\nu^2 J_1$ with other operators.  Consider for example
its auto-correlation function, 
\begin{align}
\Big\la\big[\nu^2(\vq_1) J_1(\vq_1)\big]\big[\nu^2(\vq_2) J_1(\vq_2)\big]\Big\ra 
&= 2 \la\nu(\vq_1)\nu(\vq_2)\ra^2\: \la J_1(\vq_1)J_1(\vq_2)\ra \\ 
&\quad + 4 \la\nu(\vq_1)\nu(\vq_2)\ra \: \la\nu(\vq_1)J_1(\vq_2)\ra\:\la\nu(\vq_2)J_1(\vq_1)\ra 
\nonumber \;.
\end{align}
We see that the terms of order $<3$ in $H^\star_{21}(\nu,J_1)$ exactly cancel all the 
zero-lag contributions.  Only the products of 2-point correlators at finite separation 
(which are the only nontrivial irreducible correlation functions for Gaussian random fields, 
see \ref{app:stat}) survive. 
Thus, our expansion in orthogonal polynomials is, in fact, equivalent to the renormalization 
procedure described in \refsec{renorm:intro} (see also \cite{Collins:1984xc}).  
We have correspondingly denoted the operator $H^\star_{21}(\nu,J_1)$ with $[\nu^2 J_1]$.  
Note that this property of course relies on the Gaussianity of the linear density field.  

Similar considerations apply to the other variables considered here. 
For a single density variable, the polynomials are univariate Hermite polynomials as was already pointed 
out by \cite{szalay:1988}.
Note that a diagrammatic reasoning along the lines of \cite{matsubara:2014} also explains 
why the self-correlators disappear. In fact, the Integrated Perturbation Theory (iPT) diagrammatic approach is 
fully equivalent to an expansion in orthogonal polynomials \cite{matsubara/desjacques:2016}.

We now have all the necessary ingredients to write down the perturbative bias expansion of 
BBKS peaks according to \refeq{biasexpansion}. Up to second order, we have
\begin{align}
\label{eq:dpkeff}
\dpk(\vq) &= \bigg. b^L_{10}\delta_R(\vq) - b^L_{01} \lapl\delta_R(\vq) \\
& \quad + \bigg.\frac{b^L_{20}}{2}\Big(\delta_R^2(\vq)-\sigma_0^2\Big) 
- b^L_{11} \Big(\delta_R(\vq) \lapl\delta_R(\vq)+\sigma_1^2\Big) \nonumber \\
& \quad + \bigg. \frac{b^L_{02}}{2}\Big[\big(\lapl\delta_R\big)^2(\vq)-\sigma_2^2\Big]
+ \chi^L_1 \Big[\big(\grad\delta_R\big)^2\!(\vq)-\sigma_1^2\Big] \nonumber \\
& \quad+ \omega^L_{10}\bigg[\frac{3}{2}
\bigg(\partial_i\partial_j\delta_R-\frac{1}{3}\delta_{ij}\lapl\delta_R\bigg)^2\!(\vq)-\sigma_2^2\bigg]  
+\dots\;. \nonumber
\end{align}
The absence of Lagrangian tidal shear terms 
($\propto K_2(\vq)$, $K_3(\vq)$; see \refsec{gen_barrier}) follows from the spherical collapse assumption.
This series can be used to compute all the $N$-point correlation functions of 
BBKS peaks in Lagrangian space.  
Furthermore, it is also valid in the presence of weak primordial 
non-Gaussianity (see \refsec{NGpeaks}).  
When expanded in this form,  $\dpk(\vq)$ is generally {\it not} a count-in-cell (measurable) 
density, but should be understood as a {\it mean peak field}
given a certain realization of the density field and its derivatives,   
after averaging over small-scale perturbations.  
This is because we have not included the stochastic contributions
induced by the small-scale perturbations in a given realization, as described in \refsec{stoch}.  
Since the stochastic contributions have a finite variance when averaged over a finite volume, \refeq{dpkeff} 
is equivalent to \refeq{kacrice} only in the limit of infinite ``survey'' volume, 
in which case ergodicity ensures that the series coefficients are the average bias parameters given above. 
As any perturbative expansion, this effective density contrast formally satisfies $\dpk(\vq)\geq -1$ only 
if all the terms in the infinite series expansion are included.  

Expressing the right-hand side of \refeq{dpkeff} as a Fourier transform and collecting all possible combinations
of rotational invariants involving exactly $n$ powers of the linear 
density contrast field $\delta^{(1)}$,  the perturbative peak bias expansion takes the form
\begin{equation}
\label{eq:Fdpkeff}
\dpk(\vq) = \sum_{n=1}^\infty \frac{1}{n!}
\int_{\vk_1}\dots \int_{\vk_n}\,
c_n^L(\vk_1,\dots,\vk_n)\, \Big[\delta^{(1)}(\vk_1)\dots\delta^{(1)}(\vk_n)+\cdots\Big] e^{i \vk_{1\cdots n}\cdot\vq} \;,
\end{equation}
where the Lagrangian peak bias functions $c_n^L(\vk_1,\dots,\vk_n)$ are defined at the collapse time $\tau_0$,
and the ellipsis in the square bracket stands for terms of order $n-2$, $n-4$, and so on.
These terms are present because each integrand corresponds to a sum of $n$-th order renormalized operators written in Fourier space,
which contain lower-order operators as counter-terms.
For $n=2$ for instance, the term in square brackets is
\begin{equation}
\Big[\delta^{(1)}(\vk_1)\delta^{(1)}(\vk_2)-\big\la \delta^{(1)}(\vk_1)\delta^{(1)}(\vk_2) \big\ra\Big]
\end{equation}
so that, in the calculation of $\la\dpk(\vq_1)\dpk(\vq_2)\ra$,
we obtain using Wick's theorem (see \refapp{stat:Gauss})
\begin{multline}
\bigg\la\Big[\delta^{(1)}(\vk_1)\delta^{(1)}(\vk_2)-\big\la \delta^{(1)}(\vk_1)\delta^{(1)}(\vk_2) \big\ra\Big]
\Big[\delta^{(1)}(\vk_3)\delta^{(1)}(\vk_4)-\big\la \delta^{(1)}(\vk_3)\delta^{(1)}(\vk_4) \big\ra\Big]\bigg\ra 
e^{i\vk_{12}\cdot\vq_1+i\vk_{34}\cdot\vq_2}\\
=\bigg[\big\la\delta^{(1)}(\vk_1)\delta^{(1)}(\vk_3) \big\ra \big\la\delta^{(1)}(\vk_2)\delta^{(1)}(\vk_4) \big\ra
+ \big\la\delta^{(1)}(\vk_1)\delta^{(1)}(\vk_4) \big\ra \big\la\delta^{(1)}(\vk_2)\delta^{(1)}(\vk_3) \big\ra
\bigg] e^{i\vk_{12}\cdot\vq_1+i\vk_{34}\cdot\vq_2} \;.
\end{multline}
Thus, only those correlators that involve fields at different Lagrangian positions remain. 
Note that, since $\la\delta^{(1)}(\vk_1)\delta^{(1)}(\vk_2)\ra$ is non-zero only if $\vk_1+\vk_2=0$, this term is multiplied by
a factor of $c_2^L(\vk_1,-\vk_1)$ which precisely corresponds to the second-order zero-lag terms in \refeq{dpkeff}.  
At order $n\geq 3$, terms of the form $\delta^{(1)}(\vk_1)\la\delta^{(1)}(\vk_2)\delta^{(1)}(\vk_3)\ra$, and so on, 
are correspondingly subtracted in \refeq{Fdpkeff}. 
Furthermore, we emphasize that $c_n^L(\vk_1,\cdots,\vk_n)$ are \emph{not free functions}, but a convenient way to collect all 
renormalized Lagrangian bias terms at a given order. \refeq{Fdpkeff} with the peak constraint is a particular case of a 
Lagrangian bias expansion in the iPT framework \citep{matsubara:2008,matsubara:2011,matsubara:2014}.

Armed with this result, it is straightforward to write down the $n$-point functions of peaks.  For 
the cross-correlation functions between peaks and matter in Fourier space, we have
(cf. \refsec{nptLagr})
\begin{align}
\label{eq:PBmpkL}
P_{\text{pk},m}^L(k)= c_1^L(k) \Plin(k) \;, \qquad
B_{\text{pk},mm}^L = c_2^L(\vk_1,\vk_2) \Plin(k_1) \Plin(k_2) \;,
\end{align}
where the renormalized Lagrangian peak ``bias functions'' are
\ba
\label{eq:c1pk}
c_1^L(k) \equiv\:& \left(b^L_{10} + b^L_{01} k^2\right) W_R(k) \\
\label{eq:c2pk}
c_2^L(\vk_1,\vk_2) \equiv\:& \biggl\{b^L_{20} + b^L_{11} \left(k_1^2+k_2^2\right) 
+ b^L_{02} k_1^2 k_2^2-2 \chi^L_1 \left(\vk_1\cdot\vk_2\right) \\ 
& \qquad  +\omega^L_{10}\biggl[3\left(\vk_1\cdot\vk_2\right)^2 -k_1^2 k_2^2\biggr]\biggr\}\, 
W_R(k_1) W_R(k_2) \nonumber \;.
\ea
These definitions agree with the renormalized Lagrangian bias functions introduced by \cite{matsubara:2011} 
in the context of the integrated perturbation theory (iPT) (see also \cite{matsubara:2012,matsubara:2015}).
Specifically, they are the ensemble averages of functional derivatives \cite{matsubara:2011},
\begin{align}
\label{eq:iPTcnL}
c_n^L(\vk_1,\dots,\vk_n) &= (2\pi)^{3n} \int_{\vk'}\left\langle\frac{\mathcal{D}^n\delta_X^L(\vk')}
{\mathcal{D}\delta^{(1)}(\vk_1)\dots\mathcal{D}\delta^{(1)}(\vk_n)}\right\rangle \\
&= \sum_{\alpha_1,\dots,\alpha_n=1}^D
\frac{1}{\overline{n}_X}\left\langle \frac{\mathcal{D}^n n_X}{\mathcal{D} y_{\alpha_1}\dots \mathcal{D} y_{\alpha_n}}\right\rangle
{\cal U}_{\alpha_1}(\vk_1)\dots {\cal U}_{\alpha_n}(\vk_n) \nonumber \;,
\end{align}
where $\delta_X^L\equiv n_X / \overline{n}_X -1$ is the fractional Lagrangian overdensity of a generic biased tracer,
$\vy=\{y_1,\dots,y_D\}$ 
are the linear operators from which the Lagrangian bias relation is constructed,
and the kernels ${\cal U}_\alpha$ are defined as
\begin{equation}
y_\alpha(\vk) = {\cal U}_\alpha(\vk)\,\delta^{(1)}(\vk)\;.
\end{equation}
For the BBKS peak constraint, for instance, $\vy=\{\nu,\eta_k,\zeta_{ij}\}$ and $D=10$, with corresponding kernels given by
\be
\big\{ U_\alpha(\vk) \big\} =\left\{\frac{1}{\sigma_0}W_R(k),\  \frac{i}{\sigma_1} k_k W_R(k),\  -\frac{1}{\sigma_2}k_i k_j W_R(k) \right\}\;.
\ee
For $n=1$, \refeq{iPTcnL} then gives
\begin{equation}
c_1^L(\vk) = \frac{1}{\bnpk}\left[\frac{1}{\sigma_0}\left\langle\frac{\mathcal{D}\npk}{\mathcal{D}\nu}\right\rangle
+ \frac{i}{\sigma_1}\sum_i k_i \left\langle\frac{\mathcal{D}\npk}{\mathcal{D}\eta_i}\right\rangle - \frac{1}{\sigma_2}
\sum_{i,j} k_i k_j \left\langle\frac{\mathcal{D}\npk}{\mathcal{D}\zeta_{ij}}\right\rangle\right] W_R(k)\; ,
\end{equation}
which, upon taking into account invariance under rotations and symmetrizing the arguments, returns 
\refeq{c1pk} \cite{matsubara/desjacques:2016}. Higher-order Lagrangian bias functions can be computed 
analogously.

\refeq{PBmpkL} has exactly recovered the result for the two-point function in \refsec{pkcross}.  
However, we have also easily obtained the corresponding three-point function.  
This is equivalent to the result derived for the general bias expansion in \refeq{PhhL}, 
with additional higher-derivative terms whose bias 
coefficients are unambiguously predicted by the peak formalism.  
Furthermore, the peak auto-power spectrum can formally be written as
\ba
\ppk(k) = [c_1^L(k)]^2 \Plin(k) + \sum_{n\geq 2}\frac1{n!}
\int_{\vp_1}\cdots\int_{\vp_n} \Big[c_n^L(\vp_1,\cdots,\vp_n)\Big]^2 \Plin(p_1) \cdots \Plin(p_n) 
(2\pi)^3\delta_D(\vp_{1...n}-\vk) + \frac1{\bnpk} \;,
\label{eq:Ppk1l}
\ea
in agreement with the iPT approach \cite{matsubara:2008}.  
It is instructive to compare this result, up to order $n=2$, to the NLO Eulerian halo power spectrum in the general
perturbative bias expansion derived in \refsec{npt1loop}.  In order
to compare at the same order in perturbations, we should only keep the
terms $\propto [c_1^L(k)]^2$ and $[c_2^L(\vk_1,\vk_2)]^2$ in \refeq{Ppk1l}.
First, notice that, since the renormalized bias expansion \refeq{Fdpkeff} is written in terms of the initial density field,
the Lagrangian-space power spectrum of peaks does not contain cross-correlations between linear and cubic contributions,
as these are completely absorbed in the leading-order contribution via renormalization (see also \refsec{renorm:intro}).
Second, by construction, the BBKS peak number density does not depend on the tidal field, leading to the absence of
$b_{K_2}^L$ in \refeq{Ppk1l}.
Third, there are no contributions from gravitational evolution, and tidal operators such as $K^2$ and $O_\otd$ can be
consistently set to zero.
Fourth, the peak power spectrum contains linear and second-order higher-derivative terms encoded in $c_1^L(k)$ and
$c_2^L(\vk_1,\vk_2)$; in \refeq{Phm1l}, we have only included the linear higher-derivative term.  
Finally, the loop integrals in \refeq{Ppk1l} are explicitly regularized through the 
filtering kernels $W_R(k_i)$, while no such kernel is present in \refeq{Phm1l}.  
On large scales, the difference amounts to a $k$-independent constant, which in 
the general perturbative bias expansion is absorbed by the renormalized stochastic amplitude $\Peps$.

Peaks are not the only example of biased tracers that yield higher-derivative bias terms.  
Excursion-set approaches can produce such terms as soon as the window is not a sharp $k$-space 
filter \cite{musso/paranjape/sheth:2012}, as we have seen in \refsec{gen_barrier}.  
In all cases, the form of the higher-derivative terms is dictated by the fact that the tracer 
density is a scalar quantity \cite{matsubara:2011} (see \refsec{higherderiv}).

\subsection{Bias parameters from cross-correlations at two smoothing scales*}
\label{sec:bNmL}

\technote{* This section is of a more technical nature and is not essential for the remainder of this section.}

The peak-background split also works with perturbations of long, but finite wavelength, provided one takes into account the correlation between the long 
 and short modes.
Therefore, it should be possible to measure the bias parameters using 
cross-correlations at two different scales $R$ and $R_\ell$, where $R$ 
corresponds to the halo scale and $R_\ell > R$ 
(the smoothing scale $R_\ell$ can take any value as long as it is distinct 
from the halo smoothing scale).  
This leads to the ``Lagrangian cross-correlation'' (LCC) moments discussed at the end of \refsec{bmom}.  
Some of this idea can be traced backed to \cite{desjacques/crocce/etal:2010},
where the authors showed how to recover the higher-derivative peak bias factors from
a conditional mass function. 
The point was first made clear in \cite{musso/paranjape/sheth:2012}, who demonstrated that 
the bias factors $b^L_{ij}$ can be computed from one-point measurements rather than 
computationally more expensive $n$-point correlation functions.

This approach was implemented by 
\cite{paranjape/sheth/desjacques:2013,paranjape/sefusatti/etal:2013}
to halos extracted from N-body simulations in order to test the predictions of peak theory. 
Namely, halos were traced back to their proto-halo patch in the initial conditions 
(since one is interested in measuring Lagrangian biases); the 
linear density field was smoothed on a scale $R_\ell \gtrsim R$, and the quantity 
$H_n(\nu_\ell=\delta_\ell/\sigma_{0l})$ was computed (for $n=1,2$) at the location of each proto-halo. In the peak approach, the overdensity of proto-halos is given by
$1+\delta_h^L=\npk/\bnpk$. Therefore, the average of $H_n(\nu_\ell)$ over all proto-halos 
reads
\begin{align}
\label{eq:ensHn}
M_n^{\rm LCC}(R_\ell) \equiv \frac{1}{N_h}\sum_{p=1}^{N_h} H_n[\nu_\ell(\vq_p)] &=
\int_{-\infty}^{+\infty}\!\!d\nu_\ell\,{\cal N}(\nu_\ell) 
\bigl\langle 1 + \delta_h^L\bigl\lvert \nu_\ell\bigr\rangle H_n(\nu_\ell) \\
&= \frac{1}{\bnpk}\int\!\!d^5\vw\,\npk(\vw)\,\left(-\epsilon_\nu\right)^n 
\left(\frac{\partial}{\partial\nu}+\frac{\epsilon_{J_1}}{\epsilon_\nu}\frac{\partial}{\partial J_1}\right)^n p(\vw) 
\nonumber \\
&= 
\sum_{i+j=n} 
\frac{n!}{i!j!}
(\sigma_0\epsilon_\nu)^i(\sigma_2\epsilon_{J_1})^j b^L_{ij} \nonumber \;,
\end{align}
where $\bigl\langle 1 + \delta_h^L\bigl\lvert \nu_\ell\bigr\rangle$ is the average
number density of proto-halos (relative to the mean) in the presence of a long-wavelength density perturbation,
$p\in \{1,\cdots,N_h\}$ is the index labeling the proto-halos, and
$\epsilon_X=\<\nu_\ell(\vx) X(\vx)\>$ denotes the cross-correlation 
between $\nu_\ell$ and the variables $X=(\nu,J_1)$ defined at the halo smoothing scale. 
Relations between bias factors of a given order 
(which arise owing to their close connection with Hermite polynomials, see e.g.
\citep{musso/paranjape/sheth:2012}) can then be used to extract a measurement of each $b^L_{ij}$
(see \citep{paranjape/sheth/desjacques:2013,paranjape/sefusatti/etal:2013} 
and \refsec{meas:meas} for an overview of the results).

Note that the expression \refeq{ensHn} is slightly different from that obtained in the context of the general perturbative bias expansion, \refeq{b1LCC} in \refsec{bmom}: instead of the moment $\sigma_0\epsilon_\nu = \<\nu_\ell \d_R\> = \< \delta_\ell \delta_R \> / \s(R_\ell)$, the general bias expansion yields a proportionality factor of $\< \d_\ell \d\>/\s(R_\ell)$. The difference is explained by the fact that the peak approach works with the density field smoothed on the peak scale $R(M)$, while the general bias expansion does not perform such a smoothing. The difference between the two is, as expected, absorbed by higher-derivative bias terms. Specifically, noting that \refeq{b1LCC} assumes that $R_\ell \gg R(M)$ in order for the perturbative approach to be valid, we have
\be
\< \delta_\ell \delta_R \> = \int_{\vk} W_{R_\ell}(k) W_R(k) \Plin(k)
\approx \int_{\vk} W_{R_\ell}(k) \left[1 - c_W R^2 k^2 + \cdots \right] \Plin(k)
= \< \d_\ell \d \> + c_W R^2 \< \d_\ell \lapl\d \> + \cdots\,,
\ee
where $c_W$ is a constant depending on the filter employed (for example, $c_W=1/10$ for a real-space tophat filter). The assumption that $R_\ell \gg R(M)$ implies that the smoothing kernel $W_R(k)$ can be expanded in this integral, since the contribution of modes of order $k\gtrsim 1/R$ is highly suppressed by $W_{R_\ell}(k)$. The last term is precisely a higher-derivative term of the order indicated in \refeq{b1LCC}. Thus, once the smoothing kernel $W_R(k)$ is expanded, the result from the general bias expansion agrees with the peak approach. The only difference is that higher-derivative biases contain the contributions from the smoothing kernel in the general bias expansion, while the latter remain implicit in the peak expressions.

As shown by \cite{biagetti/chan/etal:2014}, this approach can be generalized to 
measure the higher-derivative peak bias factors. The main difference is the 
appearance of other polynomials and distributions.  
Consider the bias factors $\chi^L_k$ associated with $\eta^2(\vx)$ for illustration.
In analogy with \refeq{ensHn}, the ensemble average of $L_n^{(1/2)}(3\eta_\ell^2)$ 
at the proto-halo positions is
\begin{equation}
\label{eq:ln3}
\frac{1}{N_h}\sum_{p=1}^{N_h} 
L_n^{(1/2)}\!\left(\frac{3\eta_\ell^2(\vq_p)}{2}\right) =
\int_0^\infty\!\!d(3\eta_\ell^2)\, \chi_3^2(3\eta_\ell^2)\,
\bigl\langle 1+\delta_h \bigl\lvert 3\eta_\ell^2\bigr\rangle
L_n^{(1/2)}\!\!\left(\frac{3\eta_\ell^2}{2}\right) \;.
\end{equation}
The calculation is somewhat more intricate than for Hermite polynomials because the 
bivariate Normal distributions (which arise from the conditioning on scale $R_\ell$) are
replaced by bivariate $\chi^2$-distributions. For $n=1$ for instance, a little algebra 
shows that the average of $L_1^{(1/2)}(3\eta_\ell^2/2)$ at the location of proto-halos is
\begin{equation}
\label{eq:ensL1}
\frac{1}{N_h}\sum_{i=1}^{N_h} 
L_1^{(1/2)}\!\left(\frac{3\eta_\ell^2(\vq_p)}{2}\right)
= - \epsilon^2 \sigma_1^2 \chi^L_1 \;,
\end{equation}
where the cross-correlation coefficient $\epsilon$ is
\begin{equation}
\epsilon^2 \equiv \frac{\bigl\langle\eta^2\eta_\ell^2\bigr\rangle-
\bigl\langle\eta^2\bigr\rangle\bigl\langle\eta_\ell^2\bigr\rangle}
{\sqrt{
\left(\bigl\langle\eta^4\bigr\rangle-\bigl\langle\eta^2\bigr\rangle^2\right)
\left(\bigl\langle\eta_\ell^4\bigr\rangle-\bigl\langle\eta_\ell^2\bigr\rangle^2\right)
}}
=\left(\frac{\sigma_{1\times}^2}{\sigma_{1s}\sigma_{1\ell}}\right)^2 \;.
\end{equation} 
Here, $\eta_\ell^2=(\nabla\delta_\ell)^2/\sigma_{1\ell}^2$ is the long-wavelength perturbation and
the subscript ``$\times$'' denotes the splitting of filtering scales, that is,
one filter is on scale $R$ while the second is on scale $R_\ell$.
Note that, unlike the average of $H_n(\nu_\ell)$, the right-hand side of \refeq{ensL1}
involves only one bias factor because the variable $\eta^2$ does not correlate with the others. 

This approach has been applied to measure the Lagrangian \LIMD biases 
\cite{paranjape/sheth/desjacques:2013,paranjape/sefusatti/etal:2013}, 
the higher-derivative Lagrangian bias  $\chi^L_1$, or equivalently $b_{(\vn\delta)^2}^L$ \cite{biagetti/chan/etal:2014}, and
the tidal-shear bias $b_{K_2}^L$ \cite{castorina/paranjape/etal:2016}, by stacking the tidal invariant $K_2$ defined in
\refeq{K2K3def} at the positions of proto-halos. 
For a summary of these measurements, see \refsec{meas:meas:halos}.

\subsection{Excursion-set peaks}
\label{sec:sub:esp}\label{sec:esp}

In contrast to the excursion-set approach, in which the basic ingredient is
random walks at all points in space (see \refsec{exset}), the peak 
formalism focuses on the subset 
of points that correspond to initial density maxima, i.e. peaks. In fact, these two 
approaches are not irreconcilable as was shown in \cite{paranjape/sheth:2012}.
Similar ideas can already be found in the early work of \cite{bond:1989}.
However, before we discuss the relation between peak theory and excursion sets, 
we will first address the equivalent of the {\it cloud-in-cloud} 
problem in the context of peak theory.  The cloud-in-cloud problem in the 
excursion-set formalism is discussed in detail in \refsec{CICproblem}.

So far, in order to make the connection with the abundance of dark matter halos, 
we have implicitly assumed that
there is an one-to-one correspondence between halos of mass $M$ collapsing at 
redshift $z$ and maxima of the initial density field smoothed on the 
scale $R(M)$ and of height $\dc(z)$. This is a good approximation at least for 
relatively massive objects (see \refsec{exset:general}). 
Therefore, the halo mass function (the number density of halos per 
logarithmic mass bin $d\ln M$) is given by
\cite{efstathiou/rees:1988,bond:1989,peacock/heavens:1990}
\begin{equation}
\label{eq:peakmf}
\avnh(M) d\ln M = \bnpk(\nu_c)d\nu_c\equiv \frac{\rhob}{M} \fpk(\nu_c) d\nu_c\;,
\end{equation}
where $\fpk(\nu_c)d\nu_c$ is the analog of the excursion-set multiplicity function, 
that is, the fraction of peaks (instead of random walks) at the smoothing scale $R(M)$ with height in the
range $[\nu_c,\nu_c+d\nu_c]$.
In the case of BBKS peaks (cf. \refsec{pknum}), 
\begin{equation}
\label{eq:fpk}
\fpk(\nu_c)= \frac{M}{\rhob} \bnpk(\nu_c)
=\left(\frac{V}{V_\star}\right)G_0(\gamma_1,\gamma_1\nu_c)\, 
\frac{e^{-\nu_c^2/2}}{\sqrt{2\pi}}\;,
\end{equation}
plays the role of the multiplicity function. Here, $V=M/\rhob$ is the Lagrangian volume 
occupied by a halo of mass $M$, while $V_\star=(2\pi)^{3/2}R_1^3$ is the characteristic volume 
occupied by a peak on the smoothing scale $R$. 
Note that $\fpk= \fpk(\dc,{\sigma_i})$ truly is a function of $\dc$
and the spectral moments $\sigma_i$. 
Unfortunately, the prescription \refeq{peakmf} does not ensure that a peak of height $\nu_c$ 
identified at the filtering scale $R$ is not embedded in a bigger collapsed object.

To remedy this problem, one considers trajectories $\delta(R)$ as a function of $R$ at the 
location of initial density maxima in analogy with the excursion-set formalism 
\cite{bond:1989,appel/jones:1990,paranjape/sheth:2012}.
To ensure that peaks-in-peaks are not counted, 
we must enforce the constraint that a peak identified in the density field 
smoothed on scale $R$ is not included in a peak identified on a larger smoothing scale. 
This {\it peak-in-peak} effect is difficult to handle because of the correlated 
nature of the walks $\delta(R)$ for realistic filters such as tophat or Gaussian.
Namely, one should in principle consider an infinite number of constraints for 
each filtering scale between $R$ and infinity. 
As discussed in \refsec{excursion_HeavensPeacock}, this condition can be 
approximated by the milder requirement that the density contrast of the maxima
satisfies
\begin{equation}
\delta(R) > \dc,\qquad \delta(R+dR) < \dc \;,
\end{equation}
as proposed by \cite{appel/jones:1990}.
These two inequalities enforce the condition that  maxima of height $\dc$
are just at the edge of disappearing. 
As in \refsec{excursion_HeavensPeacock}, they can be combined into the constraint
\begin{equation}
\label{eq:pkfirstcross}
\dc < \sigma_0\nu < \dc - \frac{d\delta}{dR} dR \equiv \dc - \delta'(R)dR
\;,
\end{equation}
where throughout this section a prime denotes a derivative with respect to the filtering scale.  
For the condition \refeq{pkfirstcross} to make sense, we must have $\delta'(R)\leq 0$ or, 
equivalently, the random walk $\delta(R)$ must up-cross the threshold $\dc$. 
The analysis of \cite{musso/paranjape:2012} shows that this up-crossing 
is, to a good approximation,
a first-crossing down to peak height of order unity. 
The combination of the BBKS peak constraint with this up-crossing condition has been dubbed 
{\it excursion-set peaks} or ESP by \cite{paranjape/sheth:2012}.   

The ESP approach still suffers from some of the limitations inherent to the 
excursion-set approach: 
in particular, as it is solely based on the quantities at the position $\vxpk$ of the 
host halo, information about the halo merger history 
(the presence of local density maxima within the Lagrangian patch, for example) 
is incomplete.
To overcome this, the peak-patch approach of Ref.~\cite{bond/myers:1996} (see \refsec{exset:general})
relies on  a more sophisticated description of the collapse which considers the behavior of $\delta$ 
(and the other relevant variables) in a neighborhood of $\vq_\text{pk}$ and on various filtering scales.
This issue becomes important when considering lower-mass halos, where the 
dynamics increasingly deviates from spherical collapse. 
Another issue is that any Lagrangian-based prescription 
of halo abundance and clustering requires spectral moments of the linear 
power spectrum [see \refeq{mspec}], which are not always 
finite for the real-space tophat filter.  
To alleviate this problem,
Refs.~\citep{paranjape/sheth/desjacques:2013,paranjape/sefusatti/etal:2013,biagetti/chan/etal:2014} have used a Gaussian filter 
whenever a calculation using tophat filter is not convergent. 
In practice, the Gaussian filter radius $R_G$ was determined from the tophat filter
radius $R_T$ through the requirement
\be
\la\delta_G|\delta_T\ra = \frac{\<\d_G(\vq) \d_T(\vq) \>}{\sqrt{\< \d_G^2 \> \< \d_T^2 \>}} = \dc\,,
\ee
where $\delta_G$ and $\delta_T$ are the density field smoothed with a Gaussian and real-space tophat filter, respectively.  
This ensures that a peak with $\delta_T=\dc$ also satisfies $\la\delta_G\ra=\dc$ on
average.
Clearly, it would be desirable to have a less ad hoc treatment along the lines of \cite{chan/sheth/scoccimarro:2015} 
for instance.

The number of virialized objects $\bnesp(R)$ per smoothing interval $dR$ is then equal to the number of 
trajectories that both satisfy the peak constraint and up-cross the threshold $\dc$ on scale $R$. 
For convenience, we introduce the normalized variable $\mu\equiv -\delta'/\sigma_{\d'}$ with 
$\sigma_{\d'}=\sqrt{\la\delta^{'2}\ra}$, so that the up-crossing condition becomes $\mu\geq 0$.  We then have
\begin{align}
\bnesp(R)\, \Delta R &= 
\int_0^\infty\!\!d\mu \int_{\nu_c}^{\nu_c+(\sigma_{\d'}/\sigma_0)\mu\Delta R}\!\!d\nu
\int_{-\infty}^{+\infty}\!\!dJ_1 \dots \int_{-J_2^{3/2}}^{+J_2^{3/2}}\!\!dJ_3 
\frac{3^{3/2}}{R_1^3} \big\lvert\mbox{det}(\zeta_{ij})\big\lvert \delta_D\!\big[\v{\eta}(\vq)\big]
\Theta_H\!(\lambda_3)\,p(\vw) \\
&\approx \frac{3^{3/2}}{R_1^3}
\int_0^\infty\!\!d\mu \int_{-\infty}^{+\infty}\!\!dJ_1 \dots \int_{-J_2^{3/2}}^{+J_2^{3/2}}\!\!dJ_3 
\frac{\sigma_{\d'}}{\sigma_0}\mu\,
\big\lvert\mbox{det}(\zeta_{ij})\big\lvert \delta_D\!\big[\v{\eta}(\vq)\big]\Theta_H\!(\lambda_3)
\delta_D\!\big[\nu(\vq)-\nu_c\big]\, p(\vw)\, \Delta R \nonumber \;,
\end{align}
where the number density of ESP peaks is expressed as a function of $R$ and, in the second line, we 
have assumed that $\Delta R$ is infinitesimal.  
Further, in the ESP approach the rotational invariants in \refeq{w} are 
supplemented by $\mu$, i.e. 
$\vw \equiv\big\{\nu(\vq), J_1(\vq), \mu(\vq), 3\eta^2(\vq), 5J_2(\vq),J_3(\vq) \big\}$.  
Using $d\nu_c/dR=-\nu_c\sigma_0'/\sigma_0$ and \refeq{pointaverage}, we can read off the number density of 
ESP peaks as a function of $\nu_c$ from the variable transformation $\bnesp(\nu_c)=\bnesp(R) dR/d\nu_c$.  
Therefore, we can define the ESP equivalent of the \emph{local} number density \refeq{kacrice} of BBKS peaks of height $\nu_c$ as
\begin{equation}
\label{eq:kacriceESP}
\nesp(\vq)= \left(\frac{1}{\nu_c\gamma_{\nu\mu}}\right)\mu\,
\Theta_H\!\big[\mu(\vq)\big]\times \npk(\vq)\;,
\end{equation}
upon taking advantage of the definition of the cross-correlation between $\nu$ and $\mu$: 
$\gamma_{\nu\mu}=-\sigma_0'/\sigma_{\d'}$.
For a Gaussian filter, $\mu$ is precisely equal to $J_1\, R\sigma_2/\sigma_{\d'}$, in which case both 
$\delta'$ and the peak curvature $J_1$ are perfectly correlated (see also the discussion on assembly bias 
in \refsec{assembly_exset}). 
Hence, in this specific case, the condition $\lambda_3\geq 0$ automatically ensures $\mu\geq 0$.  
This implies that ESP and BBKS peaks are the same for a Gaussian filter.

\refeq{kacriceESP} can be generalized to moving, deterministic or stochastic collapse barriers, $\dc \to B$, upon 
making the replacements $\delta_D(\nu-\nu_c)\to\delta_D(\nu-\tilde B)$ and
$\Theta_H(\mu)\to\Theta_H(\mu+\tilde B')$, where $\tilde B\equiv B/\sigma_0$. The second condition 
constrains the slope of the (correlated) walk to be steeper than that of the barrier at up-crossing.
When the barrier is stochastic however, the following two physical pictures are possible, as was pointed 
out by \cite{biagetti/chan/etal:2014}: either each walk ``sees'' a moving barrier whose shape changes
from peak to peak, or each walk ``sees'' a constant flat barrier whose height varies as a function of
$R$. In the second case, the condition $\mu\geq -B'$ must be replaced by $\mu\geq 0$. 
Here, we will follow \cite{biagetti/chan/etal:2014} and adopt the second 
interpretation. This somewhat simplifies the calculation, while yielding only 
a percent-level difference.

As discussed in \refsec{gen_barrier}, each variable 
which the collapse threshold $B$ depends on
beyond the density adds a dimension to the 
first-crossing problem, and introduces scatter in the $(\sigma, \delta)$-plane. 
To account for the scatter in the collapse threshold, 
Ref.~\cite{paranjape/sheth/desjacques:2013,biagetti/chan/etal:2014} considered the square-root barrier
\refeq{squarerootbarrier}, $\tilde B=\nu_c+\beta$, where the stochastic variable $\beta$ follows
a distribution $p(\beta)$ which is represented by a lognormal. This furnishes a good description 
of the actual collapse thresholds as a function of halo mass \cite{robertson/etal:2009}.
Using the definition of $f(J_1)$, \refeq{fJ1}, the multiplicity function of ESP peaks reads\footnote{
This expression can be simplified with Bayes' theorem,
${\cal N}(\nu,J_1,\mu)={\cal N}(\nu,J_1){\cal N}(\mu|\nu,J_1)$, so that the integral $\int_0^\infty d\mu\,\mu\,{\cal N}(\mu|\nu,J_1)$ 
analogous to \refeq{fxstrongcorr1} takes a compact form similar to \refeq{fxstrongcorr2}.}
\begin{equation}
\fesp(\nu_c)=V\bnesp(\nu_c)=\left(\frac{V}{V_\star}\right)\frac{1}{\nu_c\gamma_{\nu\mu}}
\int\!\!d\beta\,p(\beta)\int_0^\infty\!\!d\mu\,\mu\int_0^\infty\!\!dJ_1\,f(J_1)\,
{\cal N}(\nu_c+\beta,J_1,\mu) \;,
\end{equation}
where ${\cal N}(\nu,J_1,\mu)$ denotes a trivariate normal distribution with vanishing mean and unit variances.
In addition to the terms appearing in the bias expansion \refeq{dpkeff} of BBKS peaks discussed earlier in this section,
the expansion for ESP peaks will include terms depending on $\delta'$.

However, a dependence on tidal fields, $K_2(\vq)$, $K_3(\vq)$ as defined in \refsec{gen_barrier}, will arise only 
if deviations from the spherical collapse are modeled explicitly. Such extensions of the peak constraint (along 
the line of \cite{sheth/chan/scoccimarro:2012} and, recently, \cite{castorina/paranjape/sheth:2016}) are
desirable since they would provide a physical description as well as 
predictions for the scatter $\beta$. As we shall see in \refsec{NGLagBias}, 
a first-principle description of the scatter in Lagrangian bias models is
in fact essential for the non-Gaussian bias consistency relation.

\begin{figure}[t]
\centering
\includegraphics[trim=0cm 5.5cm 0cm 3.2cm,clip,width=0.6\textwidth]{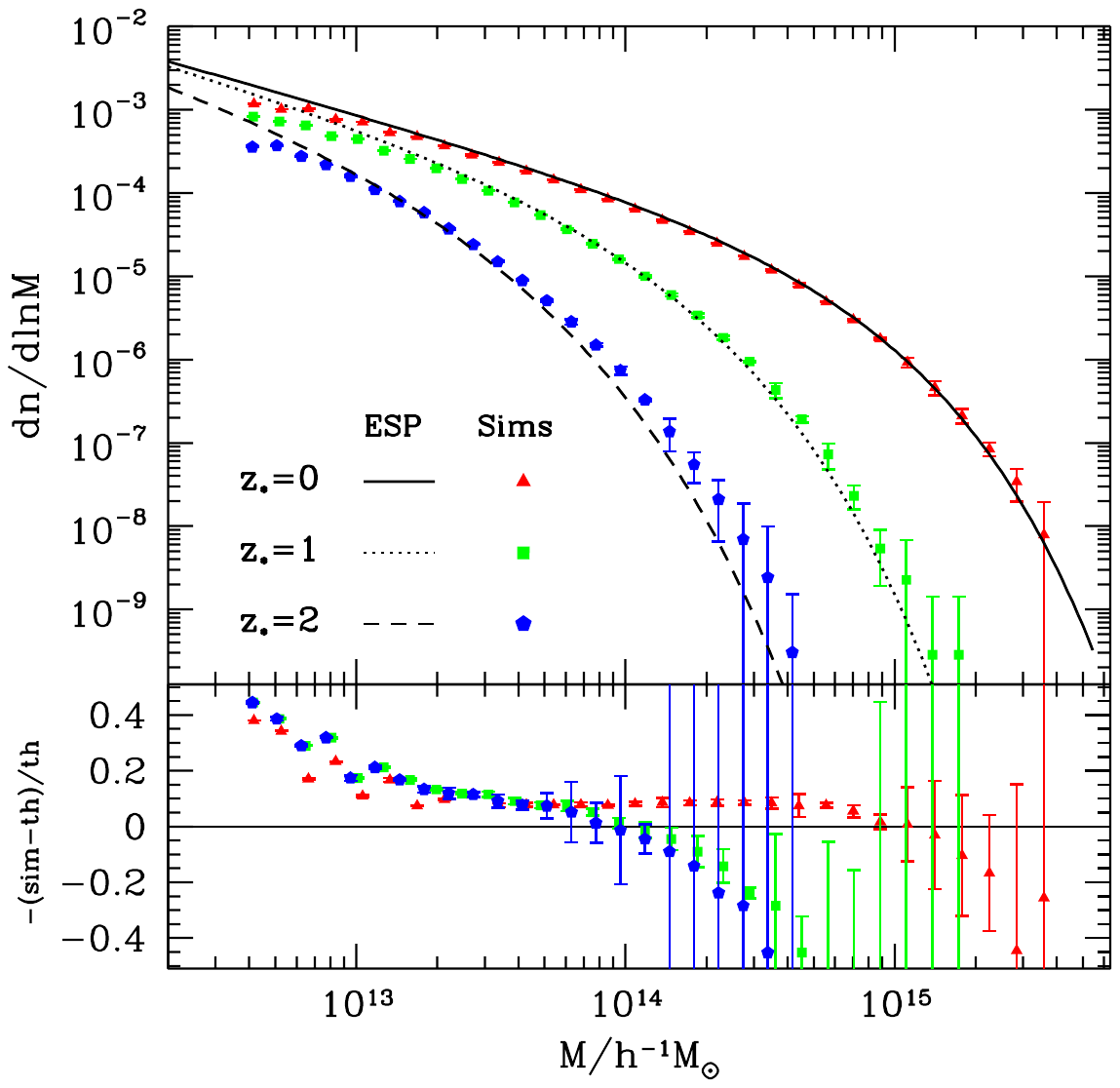}
\caption{{\it Top panel}: Logarithmic mass function of SO halos (with $\D_\text{SO}=200$) extracted from N-body simulations.
Different symbols refer to different redshifts as indicated in the figure. The solid, dotted and 
dashed curves represent the ESP prediction at $z=0$, 1 and 2. {\it Bottom panel}: 
Fractional deviation of the simulations from the ESP prediction. In both panels, error bars denote the scatter among realizations. 
\figsource{Dizgah:2015kqi}
}
\label{fig:ESPmf}
\end{figure}

Using \refeq{peakmf}, the halo mass function predicted by the ESP approach eventually is
\begin{equation}
\avnh(M) 
= \rhob \,f_\text{ESP}(\nu_c)\,\frac{d\nu_c}{dM}
= -\frac{1}{3} M R \left(\frac{\sigma_{\d'}\gamma_{\nu\mu}\nu_c}{\sigma_0}\right) V^{-1}
f_\text{ESP}(\nu_c) \;.
\label{eq:espmf}
\end{equation}
One should bear in mind that, when more than one filters are employed to 
calculate the rotational invariants (to ensure the convergence), then the 
multiplicative factor of $R$ in the above expression is the smoothing 
radius of the normalized density field $\nu$.
For instance, in the ESP implementation of 
\cite{paranjape/sheth/desjacques:2013,biagetti/chan/etal:2014}, a
tophat filter is used to define $\nu$ and $\mu$, whereas a Gaussian filter is
used for the variables $J_1$, $\eta^2$, $J_2$ and $J_3$ in order to ensure 
the convergence of $\sigma_1$ and $\sigma_2$. In this 
case, $R$ would be the tophat radius.

Predictions for the halo mass function based on \refeq{espmf} were presented 
in \cite{manrique/salvadorsole:1995}, while 
\cite{manrique/salvadorsole:1996,manrique/etal:1998} discussed the implications 
for halo mass accretion histories. 
Ref.~\cite{paranjape/sheth/desjacques:2013} included a halo mass definition which uses 
a tophat filter in real space together with the mean dependence and scatter of the critical 
collapse threshold on the halo mass. 
Comparison with N-body simulations shows that the ESP model provides a good fit to the mass 
function of spherical-overdensity (SO) halos 
(identified with a constant overdensity threshold $\Delta_\text{SO}=200$) if 
one assumes a simple square-root barrier $B(s)=\dc+\beta\sqrt{s}$
with lognormal scatter 
$p(\beta)$ \cite{paranjape/sheth/desjacques:2013,biagetti/chan/etal:2014}.
\reffig{ESPmf} indeed shows that the ESP prediction fares reasonably well at redshift
zero, which was used to calibrate the free parameters describing the square-root barrier.
Notwithstanding, it underestimates the abundance of massive halos at higher redshift. 
Note, however, that the figure is generated with the same mean $\big\langle\beta\big\rangle$ 
and variance Var$(\beta)$ 
at all redshifts \cite{Dizgah:2015kqi}, even though these were inferred from halos 
which virialized at $z=0$ only. It is likely that the mean and variance in the linear 
collapse threshold depend on redshift. Including this dependence may improve the agreement 
with simulations at higher redshift.

Finally, the results described in \refsec{bNmL} on the estimation of Lagrangian bias parameters directly apply to the ESP model,
provided $X$ is replaced by $X=(\nu,J_1,\mu)$ and $\npk$ by $\nesp$.

\subsection{Gravitational evolution of Lagrangian density peaks}
\label{sec:sub:pkgrav}

In this section, we show how the peak statistics in Lagrangian space can be related to the Eulerian 
(late-time) statistics by approximately modeling gravitational evolution. 
Specifically, we will assume that the halo centers are test particles that do not interact 
with each other and that locally flow with the dark matter.  
We will begin with a derivation of the velocity bias of peaks, and use that 
in conjunction with the continuity equation to derive the
time evolution of the linear \LIMD and higher-derivative peak bias from initial Lagrangian space to 
Eulerian space ($b_{01}^L \to b_{01}^E$). 
Next, we will discuss how a phase-space treatment can 
predict the time evolution of the Lagrangian peak correlation function at any order.  
We will employ the Zel'dovich approximation for simplicity. Because of this, we do not recover 
the correct gravitational evolution discussed in 
\refsec{evol1}--\ref{sec:evol2} at second and higher order
\cite{scoccimarro:1998,crocce/pueblas/scoccimarro:2006,mccullagh/etal:2016}.
Nevertheless, it illustrates the possibilities and challenges for 
including gravitational evolution in the peak formalism. 

\subsubsection{Velocity bias}
\label{sec:velocitybias}

The peak model is an explicit realization of biased tracers that have biased velocities. The velocity bias 
arises from the correlation between linear velocities and density gradient. 
In particular, the linear-order, 3-dimensional velocity dispersion of Lagrangian density peaks is \cite{bardeen/etal:1986}
\begin{equation}
\label{eq:sigvpk}
\sigma_{v,\text{pk}}^2(\tau) \equiv \< \v{v}_\text{pk}^2(\vq,\tau) \>  = (\cH f)^2 \sigma_{-1}^2 \left(1-\gamma_0^2\right) \;,
\end{equation}
where $\gamma_0$ is defined in \refeq{gammas}. Thus, the velocity dispersion
of peaks is smaller than that of the dark matter filtered on the same scale, 
$\< \v{v}_R^2 \> = (\cH f)^2 \s_{-1}^2$. It is because large-scale 
flows are more likely to be directed toward peaks than to be oriented randomly 
\cite{bardeen/etal:1986,szalay/jensen:1987,peacock/lumsden/heavens:1987,percival/schaefer:2008}. 
An important consequence is that the linear peak velocities are biased according to the relation
\begin{equation}
\label{eq:vpkL}
\vvpk(\vq,\tau) = \v{v}_R(\vq,\tau) - (\cH f) \frac{\s_0^2}{\s_1^2} \vn\delta_R(\vq,\tau) \;,
\end{equation}
which can be inferred from a calculation of peak pairwise velocity statistics or the redshift-space linear 
power spectrum in the distant-observer limit \cite{desjacques:2008,desjacques/sheth:2010}.
Furthermore, this also implies that the linear peak displacement $\vlpk$ is biased
\cite{desjacques/crocce/etal:2010,baldauf/desjacques/seljak:2015}. 
Namely, $\vlpk$ is given in Fourier space by
\begin{equation}
\label{eq:LpkL}
\vlpk(\vk,\tau) = \left(1-\frac{\sigma_0^2}{\sigma_1^2}k^2\right) W_R(k) \v{s}_{(1)}(\vk,\tau)\;,
\end{equation}
where $\v{s}_{(1)}$ is the linear displacement field [see \refeq{lindisp}].  
This defines the time-independent linear peak velocity bias $c_{v,\text{pk}}(k)$ \cite{desjacques:2008,desjacques/sheth:2010}
\begin{equation}
\label{eq:bvel}
c_{v,\text{pk}}(k) = \left(1 - \frac{\sigma_0^2}{\sigma_1^2} k^2\right) W_R(k) \;,
\end{equation}
such that $\vlpk(\vk,\tau)=c_{v,\text{pk}}(k)\v{s}_{(1)}(\vk,\tau)$.
In the notation of \refsec{velbias}, $\beta_{\lapl\v{v}} = \sigma_0^2/\sigma_1^2$ plus a contribution from the expansion of the filtering kernel $W_R(k)$.
Note that $c_{v,\text{pk}}$ depends on $k$ and on the smoothing scale $R$, but not on the peak height $\nu$ nor on
time; equivalently, $\beta_{\lapl\v{v}}$ only depends on the radius $R$. Since $\sigma_0/\sigma_1\propto R$, which is the only scale in the problem,  
we expect this velocity bias to increase with halo mass.  
As we can see, $\vlpk$ differs from the matter displacement $\v{s}_{(1)}$ by the filtering kernel, which arises from 
the fact that halo internal motions do not affect the motion of the host halo, and by the factor of 
$(1-k^2\sigma_0^2/\sigma_1^2)$. 

It is important to distinguish between {\it local} and {\it statistical} velocity bias.  
The former means that the velocities of biased tracers are locally different than that of the dark matter;  that is, 
a local observer would measure a relative velocity between the tracer and the dark matter.    
The latter implies that the tracers locally flow with the dark matter, but their velocities are statistically biased.  
This can occur essentially as a selection effect, if the tracers reside in special locations where the velocity is 
smaller or larger than at randomly chosen locations.   
In the peak approach, both effects are present in the form of the window function $W_R(k)$ (local part) and the 
factor $1-(\sigma_0/\sigma_1)^2 k^2$ (statistical part). 

Let us now discuss in more detail the relation between \refeq{sigvpk} and \refeq{vpkL}. Clearly, taking the 
ensemble average of the square of \refeq{vpkL} yields \refeq{sigvpk}, as it should. 
However, Ref.~\cite{desjacques/sheth:2010} also pointed out that, for \refeq{bvel}, 
\ba
(\cH f)^{-2} \sigma^2_{v,\text{pk}} 
  =\:& \frac{1}{2\pi^2}\int_0^\infty\!\!dk\,\Plin(k)\, 
    c_{v,\text{pk}}^2(k) \vs
 =\:& (\cH f)^{-2} \< \v{v}_R \cdot \v{v}_\text{pk} \> = 
   \frac{1}{2\pi^2}\int_0^\infty\!\!dk\,\Plin(k)\, 
    c_{v,\text{pk}}(k) W_R(k) \;.
\label{eq:intbvpk}
\ea
That is, the variance of peak velocities, involving an integral over 
$c_{v,\text{pk}}^2$, is equal to 
the peak--smoothed dark matter velocity variance, which involves
$c_{v,\text{pk}}$ only.  
This identity indicates that, at the peak position, the velocities 
of the peak and the filtered matter distribution are the same. 
This follows from our assumption that the peaks locally flow with the smoothed dark matter.

A further requirement on the peak velocity bias is that, in real space, it
has to be a higher-derivative operator which maps a
vector (velocity) field onto another vector field.  
Homogeneity and isotropy then require that it is built from powers of the Laplacian $\lapl$, or equivalently, 
powers of $k^2$ in Fourier space.  
Therefore, we generically expect the lowest-order $k$-dependence to scale as  
$c_{v,\text{pk}}(k)= (1- R_v^2 k^2)W_R(k)$, for some constant  $R_v$.  
This second requirement, together with \refeq{intbvpk}, then uniquely
yields the velocity bias \refeq{bvel}.  
Note that, for instance, the choice $c_{v,\text{pk}}(k) = [1 - (\sigma_{-1/2}/\sigma_0)^2\, k]W_R(k)$ also satisfies 
\refeq{intbvpk}, but does not arise from a local operator in real space.

\subsubsection{Linear evolution: continuity equation}

Having discussed the origin of the peak velocity bias, we proceed forward and compute the time evolution of the 
linear peak bias using the continuity argument invoked in \cite{desjacques/crocce/etal:2010}.  This generalizes 
the discussion of \cite{fry:1996} to a linearly biased displacement/velocity field, and parallels the discussion 
in \refsec{velbias}.  
Integrating the linear continuity equation $\partial\delta_\text{pk}/\partial\tau=-\vn\cdot\vvpk$, and evaluating the 
result in  Fourier space, the first-order Eulerian peak bias function reads
\begin{align}
\label{eq:cpk1E}
c_1^E(k,\tau) 
 &=  c_{v,\text{pk}}(k) + \frac{D(\tau_0)}{D(\tau)} c_1^L(k,\tau_0) \;,
\end{align}
where the first-order Lagrangian peak bias function is defined in \refeq{c1pk}
and we have momentarily restored the time-dependence of $c_1^L$ for clarity.
This relation is a particular case of \refeq{velbG}.  
Note however that, unlike in \refsec{evol1} and \refsec{velbias}, $\tau$ parametrizes the trajectory of the peak-patch 
from the Lagrangian ($\tau=0$ initial conditions) to Eulerian (collapse epoch at $\tau=\tau_0$) space
(though the peak trajectory can formally be extended beyond $\tau_0$).
Therefore,
\begin{equation}
c_1^E(k,\tau) \delta_R(\tau) = c_1^E(k,\tau) \frac{D(\tau)}{D(\tau_0)}\delta_R(\tau_0) 
\stackrel{\tau\to 0}{=} c_1^L(k,\tau_0) \delta_R(\tau_0) \;.
\end{equation}
That is, we recover the Lagrangian bias \refeq{c1pk} in the limit $\tau\to 0$ when peaks are defined relative to the density field 
linearly extrapolated to $\tau_0$, which is our convention throughout this section.
The Eulerian peak bias $c_1^E$ is scale-independent in the limit 
$k R\ll 1$, as expected from ``local bias theorems'' \cite{coles:1993,scherrer/weinberg:1998}.  
\refeq{cpk1E} represents the fact that peaks stream towards (or move apart from) each other in high (low) 
density environments.  This effect is higher order in derivatives, as any velocity bias (statistical or local) has to be.

On splitting \refeq{cpk1E} into its higher-derivative ($\propto k^2$) and local ($k$-independent) terms, we obtain
\begin{gather}
\label{eq:bnIE1}
b_{10}^E(\tau)\equiv 1+\frac{D(\tau_0)}{D(\tau)}\,b^L_{10}(\tau_0),\qquad
b_{01}^E(\tau)\equiv -\frac{\sigma_0^2}{\sigma_1^2} + \frac{D(\tau_0)}{D(\tau)}\,b^L_{01}(\tau_0)\;.
\end{gather}
The first relation is the usual relation \refeq{b1E} for the Eulerian, linear bias \citep{mo/white:1996}. 
The second relation implies that, in the idealized limit $\tau\to\infty$,
the higher-derivative bias parameter approaches the negative, $R$-dependent constant 
$-\sigma_0^2/\sigma_1^2$ as structure grows and $D^{-1}(\tau)$ shrinks.
The higher-derivative contribution to the linear peak bias function $c_1^E(k,\tau)$ thus persists at late time
if the linear velocities are statistically biased \cite{desjacques/crocce/etal:2010}.  
Note that since $b^L_{01}$ is always positive (see \reffig{pkbias}), the contribution from the linear velocity bias
suppresses the amplitude of $b_{01}^E(\tau_0)$.
In fact, $b_{01}^E(\tau_0)$ even becomes negative for peaks corresponding to $M\lesssim 5\times 10^{13}\hmsun$.  

A temporally constant velocity bias might seem to be at odds with the prediction of a two-fluid calculation 
(dark matter and halos) which yields a decaying velocity bias \cite{chan/scoccimarro/sheth:2012}.  
However, as we discuss in \refsec{velbias}, these two results simply correspond to different assumptions about 
the time dependence of the force difference between matter and halos;  \refeq{bnIE1} assumes a relative force 
that is constant in time, while the calculation of \cite{chan/scoccimarro/sheth:2012} assumed an instantaneous 
force in the initial conditions.  
Recent work has shown conclusive evidence for a statistical bias of the form \refeq{bvel} in two-point statistics 
of the initial halo velocities \cite{elia/ludlow/porciani:2012,baldauf/desjacques/seljak:2015}.
Since, in the linear regime, the acceleration is parallel to the initial velocity, this implies that the linear 
velocity bias $c_{v,\text{pk}}$
does not decay, but remains constant throughout time. In other words, the gravitational force acting on initial 
density peaks is biased at the linear level \cite{baldauf/desjacques/seljak:2015,biagetti/desjacques/etal:2015}
(see the discussion in \refsec{velbias}).
The recent measurements of \cite{baldauf/desjacques/seljak:2015}, which are not volume weighted and thus less
plagued by numerical artifacts (see \cite{zheng/zhang/jing:2015} for a discussion), appear consistent with this 
interpretation. 

Finally, we have thus far assumed that the peak velocity is equal to that of the smoothed dark matter component
evaluated at the peak position. In practice however, the halo velocity is commonly computed as the center-of-mass 
velocity of all the particles belonging to the halo. Assuming that effects related to the finite extent of dark 
matter halos introduce a {\it time dependence} in the window, i.e. $W_R\to W(\vk, \tau)$, the halo velocity bias at
any time $\tau \geq \tau_i$ reads \citep{chan:2015}
\begin{equation}
c_v(k,\tau) = W(k,\tau) + \big(c_v(k,\tau_i)-W(k,\tau_i)\big) \left(\frac{D(\tau_i)}{D(\tau)}\right)^{3/2}
-D^{-3/2}(\tau)\int_{\tau_i}^\tau\!\!d\tau'\,W'(k, \tau') D^{3/2}(\tau') \;,
\end{equation}
where $W'=\partial W/\partial\tau$.
There are thus two contributions to $c_v(k,\tau)$: one from $W$ [i.e. the smoothing $W_R(k)$ 
in \refeq{bvel}] and one from $W'$. The second contribution proportional to $W'$ dominates for $\tau\simeq \tau_i$, 
but is negligible at late time $\tau\gg \tau_i$. Therefore, any measurement of a higher-derivative term, apart 
from that induced by the window function $W_R$, in the velocities of proto-halos should be attributed to a 
statistical velocity bias. 
This furnishes a clean way of disentangling statistical biases from effects induced by the time dependence of the 
halo profile.

\subsubsection{Evolution at higher order}
\label{sec:pkevol}

We now compute the evolution of the peak two-point statistics 
using the phase-space distribution of peaks, that is, the joint distribution 
of peak velocities.  
Assuming that each halo center is in one-to-one correspondence with a peak, the Eulerian comoving position and 
proper velocity of a halo can in general be expressed through the mapping
\begin{equation}
\vxpk(\tau)=\vqpk+\vpsi(\vqpk,\tau)\;,\qquad 
\vvpk(\tau)=a(\tau)\,\frac{\partial}{\partial\tau}\vpsi(\vqpk,\tau)\;,
\label{eq:map} 
\end{equation}
where $\vqpk$ is the initial position of the halo center, and $\vpsi(\vq,\tau)$ is the displacement. 
For simplicity, we will work within the Zel'dovich approximation \cite{zeldovich:1970,shandarin/zeldovich:1989}, 
but the results can be extended to include higher-order Lagrangian displacements.
In this approximation, the peaks displacement is given by \refeq{LpkL},
\begin{align}
\vlpk(\vq,\tau) &=
-\frac{D(\tau)}{D(\tau_0)} \left(\frac{\vn}{\lapl}\d_R + \frac{\sigma_0^2}{\sigma_1^2}\vn\d_R\right)\!(\vq,\tau_0) \\
\frac{\partial}{\partial\tau} \vlpk(\vq,\tau) &=
- f \cH \frac{D(\tau)}{D(\tau_0)}\, 
\left(\frac{\vn}{\lapl} \d_R + \frac{\sigma_0^2}{\sigma_1^2}\vn\delta_R\right)\!(\vq,\tau_0)\;,
\label{eq:za}
\end{align} 
where we have explicitly written the dependence on $\tau_0$ to emphasize the fact that the linear fields are 
normalized relative to the collapse epoch.

Consider now an ensemble of realizations of some Lagrangian, biased point process. The correlation 
function $\xpk(r,\tau)$ is related to the zeroth moment of the joint probability 
$p_2(\vv_1,\vv_2;\vr,\tau|{\rm pk})$ to have a pair of peaks  separated by a distance $\vr$ and 
with normalized velocities $\vv_1$  and $\vv_2$ \cite{bharadwaj:1996,desjacques/crocce/etal:2010}, 
\begin{equation}
\bnpk^2\left[1+\xpk(r,\tau)\right]=\int\!\!d^3\vv_1d^3\vv_2\, p_2(\vv_1,\vv_2;\vr,\tau|{\rm pk})\;,
\end{equation}
where we have introduced the scaled velocity $\vv = \v{v}/\cH f\sigma_v$,
where $\sigma_v\equiv\sigma_{-1}(z)$ is proportional to the 3-dimensional variance of the matter velocity field, 
such that $\<\vv^2\> = 1$.  
When the peak motions are governed by \refeq{za}, $p_2(\vv_1,\vv_2;r,\tau|{\rm pk})$ can be 
easily related to the joint probability distribution at the initial time $\tau_i\ll \tau_0$,
\begin{equation}
\label{eq:pk2if}
p_2(\vv_1,\vv_2;\vr,\tau|{\rm pk})
=\int d^3\vq\: \,
\delta_D\!\left(\vq+\sigma_v[\vv_2-\vv_1]-\vr\right)
p_2(\vv_1,\vv_2;\vq,\tau_i|{\rm pk})\;.
\end{equation}
Here, $\vq$ is the Lagrangian separation vector.  Note that the 2-point correlation function depends only on $r$, even 
though the probability $p_2(\vv_1,\vv_2;\vr,\tau|{\rm pk})$ also depends on $\vv_i\cdot\vr$.  

We have implicitly assumed that each observed tracer corresponds to a unique Lagrangian patch. Like
in the peak-patch picture of \cite{bond/myers:1996} (see \refsec{exset:general}), the merging history 
depends on the details of the small-scale matter distribution inside the patch.  
Here, however, we are not following the merging of sub-clumps, but rather the motion and collapse of 
the patch as a whole. Given that the number of patches is conserved by definition, \refeq{pk2if} is 
equivalent to Liouville's theorem, which states that the phase space density of conserved tracers is conserved.

\refeq{pk2if} is especially useful when one knows how to calculate distributions in the initial
conditions. For a bias prescription that can be specified through a finite number of constraints,
e.g. for the BBKS peaks considered here,
\begin{equation}
p_2(\vv_1,\vv_2;\vq,\tau_i|{\rm pk})=
\int\!\!d^{13}\vu_1 \int\!\! d^{13}\vu_2\,\npk(\vu_1)\npk(\vu_2)\,
p_2(\vu_1,\vu_2;\vq,\tau_i)\;,
\end{equation}
where $\vu$ is a vector containing the 13 variables that describe both the BBKS peak constraint and 
the initial relative velocity (i.e., $\vu=\{\nu,\eta_k, \zeta_{ij}, \tilde{v}_l\}$), and $p_2$ is the joint-probability 
of $\vu_1=\vu(\vx_1)$ and $\vu_2=\vu(\vx_2)$. Note that the number density $\npk$ of BBKS peaks,
\refeq{kacrice}, does not depend on $\tilde{v}_i$ as this is prohibited by the equivalence principle (\refsec{renorm:boost}).  
For Gaussian initial conditions, $p_2$ is a multivariate Gaussian. 
In this case, expressing the Dirac delta in \refeq{pk2if} as the Fourier transform of a uniform distribution and 
integrating out the velocities,
the 2-point correlation function of discrete tracers can eventually be written as
\begin{align}
\label{eq:xpkz}
\bnpk^2\left[1+\xpkE(r,\tau)\right] = &
\int_{\vk} \int d^3\vq\: \,e^{i\vk\cdot(\vr-\vq)}
\int\!\!d^{10}\vy_1 \int\!\! d^{10}\vy_2\,\npk(\vy_1)\npk(\vy_2) \,
p_2(\vy_1,\vy_2;\vq,\tau_i) \\
& \quad \times 
\exp\left(-\frac{1}{2} \sigma_v^2\, \vk^\top \vcc\,\vk + i \sigma_v \vk\cdot\Delta\vpsi\right)
\nonumber \;,
\end{align}
where 
$p_2(\vy_1,\vy_2;\vq,z_i)$ is the joint PDF for 
$\vy=\{\nu,\eta_k,\zeta_{ij}\}$ at Lagrangian position $\vq_1$ and $\vq_2$,
$\Delta s_i=\big\la (s_1-s_2)_i\big\lvert{\rm pk}\big\ra$ is the relative peak displacement, and 
$\vcc_{ij}\equiv \big\la (s_1-s_2)_i(s_1-s_2)_j\big\lvert{\rm pk}\big\ra-\Delta s_i\Delta s_j$ is its covariance matrix.  
Note that the Gaussian integral over velocities introduces the displacement covariance matrix, rather than its inverse.  

\refeq{xpkz} is the exact result within the Zel'dovich approximation. Therefore, it is important to 
realize that, even though momentum conservation is ensured at leading order only, our approximation
is invariant under generalized Galilei transformations  \cite{rosen:1972,weinberg:2003,creminelli/etal:2013} 
--- that is, uniform, but time-dependent boosts, see \refsec{renorm:boost} for the explicit definition 
--- since relative displacements are unchanged. This ensures that the effect of very-long-wavelength 
perturbations vanishes in the equal-time, 2-point peak correlation function \refeq{xpkz}
\cite{scoccimarro/frieman:1996,kehagias/riotto:2013,peloso/pietroni:2013,baldauf/etal:2015BAO}.

\begin{figure}[t]
\centering
\includegraphics[trim=0cm 5cm 0cm 3.3cm,clip,width=0.6\textwidth]{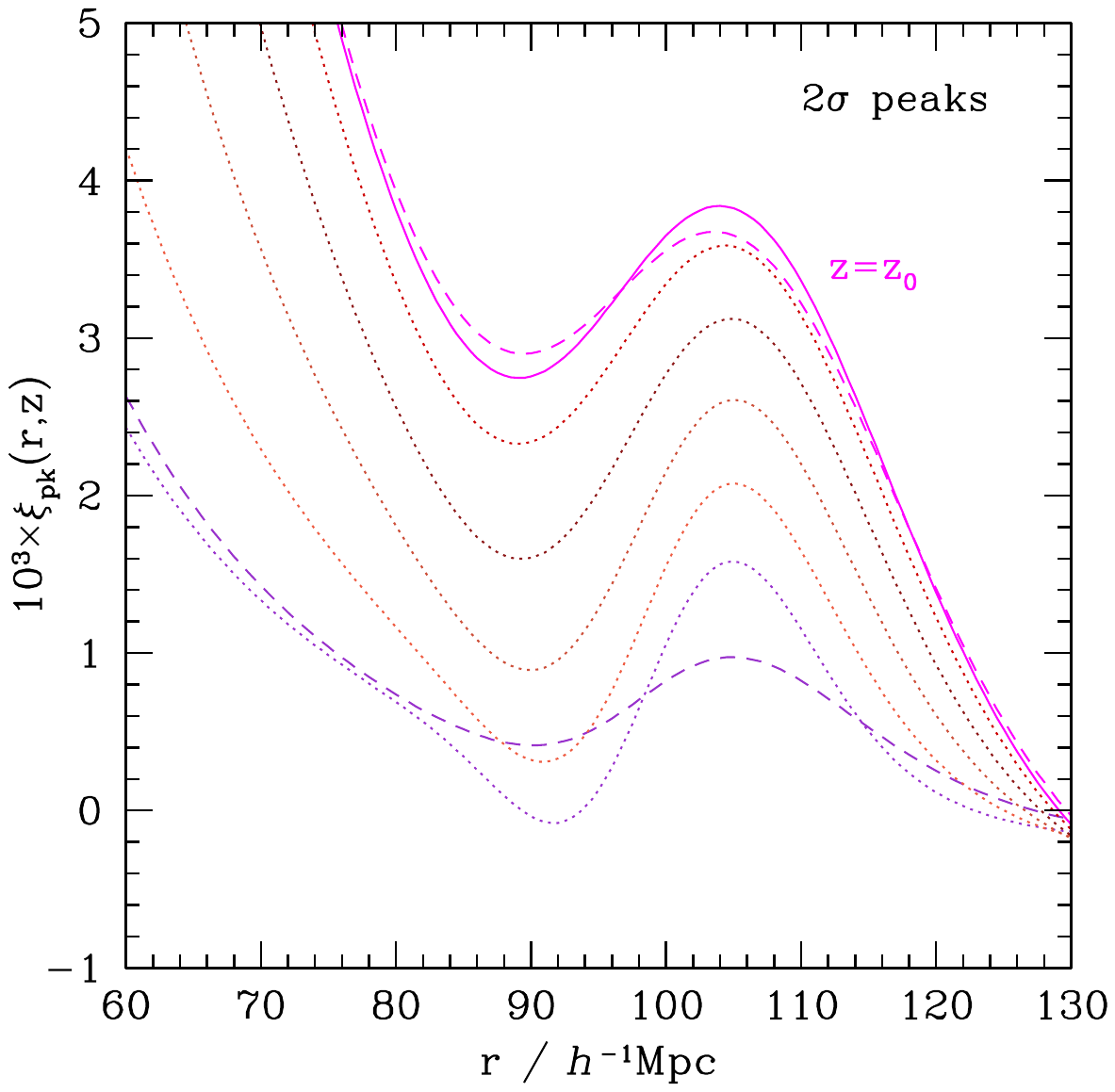}
\caption{Redshift evolution of the correlation function of $\nu_c=2$ ($2\sigma$) peaks collapsing at $z_0=0.3$ as 
predicted by \refeq{xpkz1}. The curves from bottom to top represent $\xpkE(r,z)$ at redshift $z=\infty$, 5, 2, 1, 0.5 
(dotted curves) and $z= z_0$ (solid curve). 
Only the correlation function at the collapse epoch ($z=z_0$) can be measured 
in real data. For comparison, the dashed curves show the correlation function
at $z=\infty$ and $z=z_0$ in a \LIMD bias approximation.  
\figsource{desjacques/crocce/etal:2010}
}
\label{fig:baoz}
\end{figure}

Following \cite{desjacques/crocce/etal:2010}, the integrand of \refeq{xpkz} can be expanded in powers of
the linear power spectrum $\Plin(k)$ such that, at NLO in the Zel'dovich approximation, we have 
\begin{equation}
\label{eq:xpkz1}
\xpkE(r,\tau) = \left(\frac{D(\tau)}{D(\tau_0)}\right)^2 \int_{\vk}\,e^{-\frac{1}{3}k^2\svpk^2(z)} \,
\big[c_1^E(k,\tau)\big]^2\,\Plin(k)\, e^{i\vk\cdot\vr} + \xi_\text{MC}(r,\tau) + \mathcal{O}(3) \;. 
\end{equation}
Here, $c_1^E$ is defined in \refeq{cpk1E}, and $\mathcal{O}(3)$ denotes the NNLO (next-to-next-leading order, or two- and higher-loop) contributions, 
the linear power spectrum $\Plin$ is evaluated at the collapse time $\tau_0$, and the NLO mode-coupling term $\xi_\text{MC}(r,\tau)$ is given by
\ba
\xi_\text{MC}(r,\tau) \equiv\:& \int_{\vk} P_\text{MC}(k,\tau) e^{i\vk\cdot\vr} \vs
\label{eq:xpkz2}
P_\text{MC}(k,\tau) \equiv\:&
\frac{1}{2}
\left(\frac{D(\tau)}{D(\tau_0)}\right)^4\int_{\vk_1}\int_{\vk_2}\,
\big[c_2^E(\vk_1,\vk_2,\tau)\big]^2 \Plin(k_1) \Plin(k_2) \, 
(2\pi)^3 
\delta_D(\vk-\vk_{12}) \;.
\ea
The quadratic (Eulerian) peak bias function $c_2^E$ is given by 
\begin{align}
\label{eq:cpk2E}
c_2^E(\vk_1,\vk_2;\tau) &= 
\frac{1}{2}\left[1+\left(\frac{k_1}{k_2}+\frac{k_2}{k_1}\right)\mu+\mu^2\right]
c_{v,\text{pk}}(k_1) 
c_{v,\text{pk}}(k_2) 
+ \frac{D(\tau_0)}{D(\tau)} c_1^L(k_1) c_{v,\text{pk}}(k_2) \left(1+\frac{k_1}{k_2}\mu\right) \\
&\quad + \frac{D(\tau_0)}{D(\tau)} c_1^L(k_2) c_{v,\text{pk}}(k_1)\left(1+\frac{k_2}{k_1}\mu\right)
+ \left(\frac{D(\tau_0)}{D(\tau)}\right)^2 c_2^L(\vk_1,\vk_2) \nonumber \;.
\end{align}
Therefore, the peak constraint yields a 2-point correlation function (\reffig{baoz}) consistent with that inferred 
in Lagrangian PT (e.g. \cite{bharadwaj:1996,catelan/etal:1998,matsubara:2008a,matsubara:2008})
with the additional feature that the Lagrangian bias functions contain higher-derivative terms, and the displacement
to the Eulerian position is modified owing to a velocity bias.
Note also that, although \cite{desjacques/crocce/etal:2010} did not consider Lagrangian tidal shear bias, it is
straightforward to include this contribution into $c_2^L$ following \cite{desjacques/jeong/schmidt:2017}.

It should be noted that, unlike semi-analytic methods which resum part of the perturbative expansion
(such as e.g. RPT \cite{crocce/scoccimarro:2006}), our perturbative solution is the series expansion
of the exact, Zel'dovich approximated correlation function.
Therefore, even though our truncated perturbative expansion formally violates Galilei-invariance at any order (see 
\cite{peloso/pietroni:2013} for a related discussion), it will automatically satisfy Galilei-invariance on the scales
where the expansion has converged towards the full Zel'dovich result \cite{baldauf/desjacques:2017}.
At NLO, the convergence is achieved at 
the 1\% level across the BAO. Of course, the mode-coupling is not given correctly due to the Zel'dovich 
approximation. To remedy this problem, one has to consider higher-order displacements in LPT 
\cite{bouchet/etal:1995,scoccimarro:1998,matsubara:2008a}.

We now discuss the relation of this result to the NLO halo power spectrum in the general bias expansion 
derived in \refsec{npt1loop}. First, the bias functions $c_1^E$ and $c_2^E$ contain a subset of the local bias 
terms obtained in the general bias expansion, as discussed in \refsec{pertpkcorr}.  
However, they contain additional higher-derivative terms, which are higher order following the perturbative 
counting in \refsec{npt1loop} (see the caveats of the counting mentioned there).
On the other hand, due to the Zel'dovich treatment of nonlinear evolution and the absence of Lagrangian tidal bias,
\refeqs{xpkz1}{xpkz2} do not contain contributions from the coupling of cubic and linear terms. Further, unlike the results 
in \refsec{npt1loop}, where the loop momentum attains arbitrarily high values, in \refeqs{xpkz1}{xpkz2} the 
integrals are cut off on the scale $k \sim R^{-1}$ by the filtering kernel $W_R(k)$.  
Finally, unlike the contributions $\mathcal{I}^{[O,O']}(k)$ appearing in \refsec{npt1loop}, whose low-$k$ limits 
are subtracted, here the mode-coupling term contributes a non-vanishing white noise in the limit $k\to 0$ 
which adds to the Poisson noise $1/\bnpk$.  
There are infinitely many such white-noise contributions induced by higher-order terms in the 
bias expansion. All these non-vanishing contributions in the limit $k\to 0$ renormalize $1/\bnpk$ into an 
effective white noise term, which can generally be super- or sub-Poissonian depending on the halo mass considered. 
This effect physically arises from small-scale exclusion and, therefore, cannot be modeled perturbatively. 
Note however that, in the peak approach, the ``renormalized'' white noise amplitude in the limit $k\to 0$ 
can be obtained from a numerical evaluation of $\int d^3r \xpk(r)$ through \refeq{Ppklowk}
(see the discussion in \refsec{meas:stoch} and \cite{baldauf/seljak/etal:2013,baldauf/codis/etal:2015}).  

Furthermore, the contribution proportional to $c_1^Lc_2^L\partial^{-1}\delta\cdot\partial\delta$, where $c_n^L$ 
are Lagrangian bias functions, shifts the position of the BAO peak, as was first pointed 
out in \cite{crocce/scoccimarro:2008,smith/scoccimarro/sheth:2008};  this term is of course also present in the NLO 
contribution to the two-point function derived in the general bias expansion (\refsec{npt1loop}).  
The higher-derivative correction to the second-order 
bias function does not change the strength of the effect significantly, which is dominated by the quadratic \LIMD bias 
$b^L_{20}$. Importantly, this nonlinear shift can be either accounted for in a forward analysis 
\cite{crocce/scoccimarro:2008,padmanabhan/white:2009}, or reversed using a suitable reconstruction of the linear displacement field 
\cite{eisenstein/seo/etal:2007}. 
Alternatively, the ``linear point'', which lies midway between the dip at $r\sim 90\hmpc$ and the BAO peak at
$r\sim 105\hmpc$, is weakly affected by nonlinearities and could also be used for distance measurements
\cite{anselmi/starkman/sheth:2016,anselmi/starkman/etal:2017,anselmi/corasaniti/etal:2017}.

Before concluding, let us illustrate how gravitational motions from the initial to final time wash out most 
of the Lagrangian higher-derivative bias induced by the peak constraint, by considering the peak correlation function 
$\xpkE(r,z)$ around the BAO scale (see \refsec{pkauto}).  
\reffig{baoz} shows the redshift evolution of $\xpkE(r,z)$, as predicted by \refeq{xpkz1}, for $\nu_c=2$ BBKS peaks 
from the initial conditions $\tau_i=0 $ ($z=\infty$) (bottom dotted curve) until halo collapse at $\tau=\tau_0$.  
The latter is assumed to take place at $z_0\equiv z(\tau_0) =0.3$ (top solid curve). 
For illustration purposes, $\xpkE(r)$ is also shown at the intermediate redshift values 
$z=5$, 2, 1 and 0.5.
For comparison, the bottom and top dashed curves represent the initial and final correlation function in the \LIMD approximation, 
in which all the peak bias parameters are set to zero except for $b^L_{N0}$, the velocity bias 
has been turned off so that $\svpk= \sigma_{v,\text{dm}}$, and filtering is absent  (cf. \reffig{xi_thr} in \refsec{localbias}).  
As the redshift decreases, gravitational instability generates coherent motions which amplify the large-scale amplitude of the peak correlation function, together with random motions which increasingly smear out the 
initial BAO feature. Although the randomness in the large-scale flows is less important for the peaks than 
for the \LIMD-biased tracers (owing to the velocity bias), the final correlation function of peaks is noticeably more 
similar to that of the \LIMD-restricted tracer than it was initially. Still, mild differences remain at $z=z_0$ between the peak 
and \LIMD predictions, especially around the BAO feature. This residual higher-derivative bias, 
which strongly depends on the halo mass, is a subtle combination of the higher-derivative terms in the 
Lagrangian bias functions $c_n^L(\vk_1,\dots,\vk_n)$ and of the velocity bias.

%% file: NG.tex
\clearpage
\section{Bias and primordial non-Gaussianity}
\label{sec:NG}

\secttoc

So far, we have made the assumption of adiabatic Gaussian initial perturbations, which forms a key prediction 
of the inflationary paradigm \cite{baumann/mcallister:2015}.  Moreover, the cosmic microwave background has 
placed stringent upper limits on deviations from Gaussianity in the initial conditions, known under the term 
\emph{primordial non-Gaussianity (PNG)}.  
Nevertheless, there is still significant unconstrained parameter space for PNG which offers rich insights into the physics 
of inflation (see \refsec{inflation}, and \cite{bartolo/etal:2004,liguori/etal:2010,chen:2010} for reviews).  
This provides strong motivation to study the impact of PNG on large-scale structure.

The connection between the clustering of galaxies on large scales and the statistics of the initial conditions 
has been studied since the 1980s, including the seminal papers 
\cite{matarrese/etal:1986,grinstein/wise:1986,lucchin/matarrese:1988,scherrer/bertschinger:1991,weinberg/cole:1992,fry/scherrer:1994,scoccimarro:2000} 
(see also \cite{coles:2001} for an overview).  
In the 2000s, it was realized that non-Gaussianity in the initial conditions is likely to be only a small 
correction to the non-Gaussianity induced by nonlinear gravitational evolution \cite{verde/etal:2000}.  
Refs.~\cite{scoccimarro/etal:2004,sefusatti/komatsu:2007} then developed techniques to disentangle primordial 
and late-time non-Gaussianities in the bispectrum of the large-scale structure.  
Finally, this field experienced another breakthrough with the discovery of the strong effect of PNG of the 
local type (\emph{local PNG}) on the clustering of galaxies on large scales \cite{dalal/etal:2008} (although this effect was 
already implicitly contained in the results of \cite{matarrese/etal:1986,grinstein/wise:1986}).  
We now know that galaxy clustering can provide independent constraints on the magnitude of local PNG that are 
competitive with those from the CMB and, in the long run, may even give the best constraints (\refsec{NGforecast}).  
Moreover, constraints from galaxy bias are based on small-scale modes 
$k \gtrsim 0.3 \iMpch$ (of order of the formation scale $R_*^{-1}$ of tracers), while the CMB probes 
modes with $k\lesssim 0.1 \iMpch$.    
Given the possibility of a scale dependence in the PNG, as 
predicted from several models of the early Universe,
the two should therefore be seen as complementary probes of PNG.     
The effect of PNG on large-scale structure has recently been reviewed in 
\cite{desjacques/seljak:2010b,liguori/etal:2010,verde:2010}.  
Here, we significantly expand their discussion of bias in the presence of PNG, 
and include results derived in the past several years.

We begin in \refsec{NGevolution} by describing the effect of PNG in the 
framework of the general bias expansion described 
in \refsec{evolution}, starting with the simplest type, local PNG, in \refsecs{localNG}{bNG}.  For clarity, we refer to the additional terms introduced
in the bias expansion in the presence of PNG as \emph{non-Gaussian bias}.  
Local PNG exemplifies the essential physics of the effect of PNG on galaxy clustering, without the complications of
more general non-Gaussian initial conditions, which are then considered in \refsecs{nonloc}{stochNG}.  
The discussion of additional terms appearing on smaller scales (\refsec{bbeyond}) concludes the treatment of PNG in this context.  
\refsec{NGsummary} provides a brief summary of the contributions to the general bias expansion at 
leading order in the amplitude of PNG, complementing the summary in \refsec{evol:summary}, and
also gives the contributions to the two- and three-point functions of galaxies
in their rest frame, extending the results given in \refsec{npttree}. 
We briefly discuss the implications for early-Universe physics of detecting 
and constraining the various signatures of PNG in the clustering of galaxies in \refsec{inflation}.  
In order to obtain quantitative constraints on PNG however, we need predictions for the new bias coefficients that appear in the 
presence of PNG.  The PBS argument of \refsec{PBS} can be used to derive these, as described in \refsec{PBSbiasNG}.  
We then show in \refsec{NGLagBias} how, for any given 
``microscopic'' Lagrangian model of tracers,
the effect of PNG on the $n$-point functions in Lagrangian space can be fully predicted. This includes the thresholding (\refsec{localbias}), excursion set (\refsec{exset}) 
and peak models (\refsec{peaks}), which are considered in turn in \refsecs{NGthresh}{NGpeaks}.
 
Finally, the comparison of the theoretical predictions of halo bias in the presence of PNG with N-body simulations is discussed 
in \refsec{NGsim}.  
We present current and forecasted observational constraints on local PNG 
using galaxy clustering in \refsec{NGforecast}, generalizing the forecast in \refsec{npttree_fisher}.  

\subsection{Primordial non-Gaussianity in the general bias expansion}
\label{sec:NGevolution}

In this section, we describe in detail how the general bias expansion 
described in \refsecs{general}{stoch} can be extended to include 
non-Gaussian initial conditions.  For this, we will restrict to the
leading contributions in the limit of weak PNG, which are described by the
three- and four-point functions of the initial conditions.  As we will see,
these are phenomenologically by far the most important contributions given current constraints on large-scale PNG.  

\subsubsection{Primordial non-Gaussianity of the local type}
\label{sec:localNG}

PNG deals with the statistical properties of the initial conditions for structure formation. 
These are usually phrased in terms of the primordial Bardeen potential $\phi$ \cite{bardeen:1980}, which is, in turn, 
directly related to the curvature perturbation in comoving gauge 
$\mathcal{R}$ by $\phi = (3/5) \mathcal{R}$
for modes that enter the horizon during the matter-dominated 
epoch \cite{mukhanov/etal:1992}. 
The Newtonian potential $\Phi(\vk)$ is
related to $\phi(\vk)$ during matter domination by the transfer function $T(k)$ which satisfies $T(k)=1$ for $k\ll k_{\rm eq}$, where $k_{\rm eq} \simeq 0.02 \iMpch$ is the wavenumber that entered the horizon at matter-radiation equality.  Thus,
$\phi$ is related to the linear density field by
\ba
\delta^{(1)}(\vk,\tau) = \Mm(k,\tau)\phi(\vk) \quad\mbox{where}\quad
\Mm(k,\tau) = \frac{2}{3} \frac{k^2 T(k) D_\text{md}(\tau)}{\Omega_{m0} H_0^2}\,.
\label{eq:deltaphi}
\ea
Here, $D_\text{md}(\tau)$ is the linear growth factor normalized to $a(\tau)$ during the matter-dominated epoch.  
To be precise, following our discussion in \refsec{GR}, $\d^{(1)}$ defined in \refeq{deltaphi} is the matter density contrast in synchronous-comoving gauge.  
In the following, we will drop the explicit time dependence on $\d$ and $\Mm$, as well as the bias parameters,
but include it again in \refsec{NGsummary}.  

In all previous sections, we have assumed that $\phi=\phi_G$ is a Gaussian random 
field, which is completely described by its power spectrum $P_\phi(k)$.  
The simplest way to generate non-Gaussian initial conditions is to
perform a local, nonlinear transformation of such a Gaussian field $\phi_G$, 
$\phi(\vx) = f(\phi_G(\vx))$.  Since the root-mean-squared value of $\phi$ is less than 
$10^{-4}$, a Taylor expansion converges rapidly, and so we only keep the leading nonlinear term.  This leads to
\be
\phi(\vx) = \phi_G(\vx) + \fnl \left[\phi_G^2(\vx) - \< \phi_G^2 \> \right] + \O(\phi_G^3)\,.
\label{eq:phiNGlocal}
\ee
This is the definition of \emph{local quadratic PNG}, parametrized by the parameter $\fnl$ as first introduced 
in \cite{komatsu/spergel:2001}.  The term $-\fnl \<\phi_G^2\>$ ensures that $\<\phi\>=0$.  

At leading order in $\fnl$, the only poly-spectrum beyond the power spectrum
is the bispectrum,
\ba
B_\phi(\vk_1,\vk_2,\vk_3) =\:& 2\fnl [P_\phi(k_1)P_\phi(k_2) + \perm{2}]\,.
\label{eq:Bphi}
\ea
Note that if we want to derive the $N$-point functions of $\phi$ at order
$\fnl^2$, we also have to include the next term $\gnl \phi^3$ appearing in the expansion \refeq{phiNGlocal}, as both are of the same perturbative order.    
From \refeq{deltaphi}, we immediately obtain the leading \emph{primordial contribution} to the matter bispectrum,
\ba
B^{(1)}(\vk_1,\vk_2,\vk_3) \equiv\:& \Mm(k_1)\Mm(k_2)\Mm(k_3) B_\phi(\vk_1,\vk_2,\vk_3)\,.
\label{eq:Bmloc}
\ea
That is, this is the synchronous-comoving-gauge matter bispectrum at early 
times $\tau\to 0$. A particularly important regime for the effects of
PNG on galaxy clustering is the \emph{squeezed limit}, where two
wavenumbers are much larger than the third. In this limit, the matter bispectrum \refeq{Bmloc} becomes
\ba
B^{(1)}(\vk_1,\vk_2,\vk_3) \stackrel{k_1\simeq k_2\gg k_3}{=}\:& 4 \fnl \Mm(k_3) P_\phi(k_3) \Plin(k_S)
+ \O\left(\frac{k_3^2}{k_S^2}\right) \vs
=\:& 4 \fnl \Mm^{-1}(k_3) \Plin (k_3) \Plin(k_S)
+ \O\left(\frac{k_3^2}{k_S^2}\right)
\,, \quad\mbox{where}\quad
\vk_S \equiv \vk_1 + \frac12 \vk_3\,.
\label{eq:Bmlocsq}
\ea
In the second line, we have used the relation between $\phi(\vk)$ and 
$\d^{(1)}(\vk)$, \refeq{deltaphi}.

\subsubsection{General bias expansion with local PNG}
\label{sec:bNG}

Let us go back to the expression for the galaxy-matter 
cross-power spectrum in 
the large-scale limit ($k\to 0$), \refeq{Phh} or \refeq{renorm2pt} in \refsec{renorm}:
\be
\< \d_g(\vk) \d(\vk') \> \Big|_{\LO} \stackrel{\text{Gauss}}{=}
b_1 \< \d^{(1)}(\vk) \d^{(1)}(\vk') \>\, .
\label{eq:corrG}
\ee
In the case of Gaussian initial conditions, the NLO correction
to this expression (or 1-loop power spectrum) derived in 
\refsec{npt1loop} involves the second- and third-order density field induced 
by gravitational evolution as well as nonlinear bias contributions.
The NLO contribution is suppressed on large scales by $(k/\knl)^{3+n}$, where
$n$ is the effective power spectrum index and $\knl(\tau) \gtrsim 0.2 \iMpch$ is the nonlinear scale [\refeq{knldef} in \refsec{npt1loop};
higher-derivative corrections are similarly suppressed by $(k R_*)^2$].  

Now let us consider the leading correction to the galaxy-matter cross-power spectrum induced by local-type PNG,
which corresponds to the following two contributions:
\be
\< \d_g(\vk) \d(\vk') \>\Big|_\text{PNG} =
 \frac12 b_2 \left\< \left(\d^{(1)}\right)^2(\vk) \d^{(1)}(\vk') \right\>
+ b_{K^2} \left\< \left(K_{ij}^{(1)}\right)^2(\vk) \d^{(1)}(\vk') \right\>
\,,
\label{eq:2ptNG0}
\ee
since now the bispectrum of the linear (initial) density field
$\d^{(1)}$ no longer vanishes.
The notation $(\d^{(1)})^2(\vk)$  denotes the Fourier 
transform of the square of the density field, which is given
by a convolution in Fourier space, and analogously for the tidal field $(K_{ij}^{(1)})^2(\vk)$. Now, for reasons that will become clear shortly, we re-introduce the artificial smoothing of the density field appearing in the bias operators on the scale $R=\L^{-1}$, which was employed in \refsec{renorm} to make the distinction between large (perturbative, or background) and small (non-perturbative, or peak) scales rigorous. In the end, all observables should be independent of $\L$, and we can take the limit $\L^{-1}\to 0$. Correspondingly, we denote the ``bare'' bias coefficients with $c_O$. \refeq{2ptNG0} becomes
\be
\< \d_g(\vk) \d(\vk') \>\Big|_\text{PNG} =
\frac12 c_{2,\L} \left\< \left(\d_\L^{(1)}\right)^2\!\!(\vk)\, \d^{(1)}(\vk') \right\>
+ c_{K^2,\L} \left\< \left(K_{ij,\L}^{(1)}\right)^2\!\!(\vk)\, \d^{(1)}(\vk') \right\>
\,,
\label{eq:2ptNGbare}
\ee
We now evaluate the first contribution using \refeq{Bmloc}, yielding \cite{matarrese/verde:2008,PBSpaper}
\ba
\left\< \left(\d_\L^{(1)}\right)^2\!\!(\vk)\, \d^{(1)}(\vk') \right\>'
=\:& \int_{\vp} B^{(1)}(\vp,\vk-\vp,\vk') W_\L(p) W_\L(|\vk-\vp|) \vs
\approx\:& 4 \fnl \int_{\vp} |W_\L(p)|^2\Plin(p)\:\Mm^{-1}(k) \Plin(k)
= 4 \fnl \s^2(\L)\: \< \phi(\vk) \d^{(1)}(\vk') \>'\,.
\label{eq:corrNGsq1}
\ea
In the second approximate equality, we have assumed that $k \ll p \sim \Lambda$, where the integral peaks, so that $B^{(1)}$ is evaluated in the squeezed limit \refeq{Bmlocsq}. Further, $\s^2(\L)$ is the variance of the density field smoothed on the cutoff scale $R = \L^{-1}$.  
We will discuss the sub-leading terms in $k/\L$ in \refsec{bbeyond}, which are similar to the higher-derivative bias contributions discussed in \refsec{higherderiv};  physically, the squeezed limit is 
a good assumption as long as the scales on which galaxy statistics are measured are much larger than 
the scales over which galaxies form. The second term in \refeq{2ptNGbare},
$\< (K_{ij,\L}^{(1)})^2(\vk) \d^{(1)}(\vk') \>$, precisely yields \refeq{corrNGsq1} multiplied by a factor of $2/3$. 

Perhaps surprisingly, we see that $(\d_\L)^2$ and $(K_{ij,\L})^2$ need to be renormalized at \emph{leading order} in perturbation 
theory in the presence of local PNG, since their correlation with $\d^{(1)}$ on large scales is strongly cutoff-dependent, as it is proportional to $\s^2(\L)$.
Moreover, the contribution in \refeq{corrNGsq1} is not small on large scales. Rather, it becomes even larger than the Gaussian leading-order contribution in \refeq{corrG} for sufficiently small $k$; note that, 
on large scales $k < k_{\rm eq} \simeq 0.02 \iMpch$, $\Mm^{-1}(k) \propto (k/H_0)^{-2}$.  
 This result is clearly not satisfactory, but can be cured by adding counter-terms to $\d^2,K^2$ which absorb this unphysical contribution and yield renormalized operators:
\ba
[\d^2](\vk) =\:& (\d_\L^2)(\vk) - 4 \fnl \s^2(\L) \phi(\vk) + [\mbox{Gaussian counter-terms}]\\
[ K^2 ](\vk) =\:& (K_\L^2)(\vk) - \frac83 \fnl \s^2(\L) \phi(\vk) + [\mbox{Gaussian counter-terms}]\,.
\ea
What is crucial about these counter-terms induced by PNG is that
they both involve the Bardeen potential $\phi$ directly, without any
derivatives.  Such a term cannot be generated by gravitational evolution,
as we have seen in \refsec{general}, since the gravitational potential
is not locally observable; only density and tidal effects proportional to 
second spatial derivatives of $\phi$ are.  

In the presence of local PNG, therefore, we need to add $\fnl\phi$ to the list of operators that appear in the bias
expansion \cite{mcdonald:2008,giannantonio/porciani:2010},
\be
\d_g\Big|_\text{PNG} \supset  b_\phi\,\fnl [\phi] \,.
\ee
While $[\phi]$ is strictly a renormalized operator, it does not receive
any counter-terms at leading order in $\fnl$.    
Adding this new \emph{linear-order} operator to the bias expansion leads to a leading-order large-scale galaxy-matter cross-power spectrum given by
\ba
P_{gm}^\LO(k) \equiv \< \d_g(\vk) \d(\vk') \>_\LO' =\:&
b_1 \< \d^{(1)}(\vk) \d^{(1)}(\vk') \>' 
+  b_\phi \fnl\< \phi(\vk)  \d^{(1)}(\vk') \>' \vs
 =\:& \left[b_1 + b_\phi  \fnl \Mm^{-1}(k) \right] \Plin(k)
\,,
\label{eq:corrNGsq2}
\ea
which now includes the leading terms on large scales in a proper cutoff-independent way.  
Similarly, we find for the leading-order galaxy auto-power spectrum in the presence of 
local PNG
\ba
P_{gg}^\LO(k) =\:&
\left[b_1 +  b_\phi \fnl \Mm^{-1}(k) \right]^2 \Plin(k) + \Peps\,,
\label{eq:corrNGsq2b}
\ea
where $\Peps \equiv \lim_{k\to 0} \< \eps(\vk) \eps(\vk')\>'$ is the leading-order stochastic contribution which is scale-independent on large scales and thus of the same form as in the Gaussian case.  
This shows that PNG of the local type leads to a \emph{scale-dependent bias $\propto k^{-2}$}
in the large-scale two-point function of tracers \cite{dalal/etal:2008,matarrese/verde:2008} 
(see \reffig{bNG_vs_k} on p.~\pageref{fig:bNG_vs_k}).  
This is a unique smoking-gun signal for PNG, as neither any local process nor nonlinear gravitational evolution can generate such a scale dependence, as we have explained above. 
The term proportional to $(c_{2,\L}/2) \<(\d_\L^2)(\vk) \d_\L(\vk')\>$ in \refeq{2ptNGbare} on the other hand must 
be seen as an artifact of the bare bias expansion which is absorbed in the renormalized operator $[\d^2]$. Correspondingly, the new bias parameter $b_{\phi}$ has nothing to do with $b_2$ a priori. 
We will show how $b_\phi$ can be derived using a generalized PBS argument 
in \refsec{PBSbiasNG}.  

The physical interpretation of \refeqs{corrNGsq2}{corrNGsq2b} is the following.  If the
initial potential fluctuations have a bispectrum given by \refeq{Bphi},
then long-wavelength potential perturbations modulate the initial amplitude
of small-scale perturbations.  Consider a long-wavelength mode
$\phi(\vk_\ell)$, and a patch much smaller than the wavelength of this
mode.  If we measure the local power spectrum of small-scale density perturbations within such
a patch at position $\vx$, denoted as $\Plin(\vk_S|\vx)$, then
\refeq{Bmlocsq} states that this is modulated by
\be
\Plin(\vk_S|\vx) = \left[1 + 4\fnl \phi(\vk_\ell) e^{i\vk_\ell\cdot\vx} \right]
\Plin(k_S)\,,
\label{eq:Plinmod}
\ee
where $\Plin(k)$ is the matter power spectrum averaged over the entire initial
conditions.  To see this, simply multiply this relation by
$\phi(\vk_\ell') e^{i\vk_\ell'\cdot\vx}$, take the expectation value, and integrate
over $d^3\vx$.  This recovers the initial matter bispectrum in the squeezed limit, \refeq{Bmlocsq}.  
We thus see that the amplitude of small-scale perturbations in the
initial conditions is modulated by long-wavelength potential
perturbations via $4\fnl\phi(\vk_\ell)$.  
Since the abundance of halos and
galaxies is sensitive to the amplitude of initial fluctuations,
their abundance is correspondingly modulated by $4\fnl\phi(\vk_\ell)$ 
\cite{slosar/etal:2008}, leading to the scale-dependent bias in \refeq{corrNGsq2}.  
In \refsec{general}, we were able to absorb the amplitude of small-scale 
perturbations completely through stochastic contributions such as $\Peps$ since, in the Gaussian case, 
the small-scale initial perturbations are statistically the same everywhere.  
This changes in the non-Gaussian case, and the relevant quantity that determines the effect
on bias is the squeezed-limit bispectrum, as we have seen.  

The term $b_\phi \phi(\vk)$ is the leading PNG contribution
to bias, but it is not the only one.  The non-Gaussianity
present in the initial conditions couples to gravitational
evolution, and modifies the growth of matter perturbations as well
(see \reffig{CFCsketch} on p.~\pageref{fig:CFCsketch}).   Changing the statistics of the 
small-scale initial conditions at $\vq=\xfl(0)$
will modify the subsequent evolution along the fluid trajectory $\xfl(\tau)$
\cite{angulo/etal:2015,assassi/baumann/schmidt}.  At lowest order in derivatives and linear
order in $\fnl$, the bias operators we need to include 
consist of all combinations of $\fnl \phi$ with the Gaussian
operators we have listed in \refsec{general}.  A crucial point
here is that $\phi = \phi(\xfl(0)) = \phi(\vq)$ is to be evaluated at the \emph{Lagrangian}
position $\vq$ that corresponds to the Eulerian
position $(\vx,\tau)$ \cite{giannantonio/porciani:2010,baldauf/seljak/senatore:2011},  
\be
\phi(\vq) = \phi(\vx) - s^i(\vx,\tau) \partial_i \phi(\vx) + \cdots\,,
\label{eq:phiq}
\ee
where we have expanded to second order in perturbations.  
This is because the coupling is present in the
initial conditions, and not induced by evolution.  In fact, this holds
for any imprints present in the initial conditions, including isocurvature
modes between baryons and CDM (see \refsec{baryons}).  In summary,
working in Eulerian frame, the basis of operators introduced in \refsec{basisE}, \refeq{EulBasis},
needs to be augmented by the following additional terms up to third order in case 
of local-type PNG: 
\bea
{\rm 1^{st}} \ && \  \fnl \phi(\vq)  \label{eq:EulBasisNG} \\[3pt] 
{\rm 2^{nd}} \ && \ \fnl {\rm Tr}[\Pi^{[1]}(\vx)] \phi(\vq)  \nonumber  \\[3pt]
{\rm 3^{rd}} \ && \ \fnl {\rm Tr}[(\Pi^{[1]}(\vx))^2] \phi(\vq)\,,\  \fnl ({\rm Tr}[\Pi^{[1]}(\vx)])^2 \phi(\vq) \,, \nonumber
\eea
where $\Pi_{ij}^{[1]} = K_{ij} + (\d_{ij}/3)\d \propto \partial_i\partial_j\Phi$.  
We have emphasized the distinction between Eulerian and Lagrangian
arguments.  The continuation to higher orders in perturbation theory is now obvious.  When deriving the statistics of galaxies at a given order in
perturbation theory, one also expands $\phi(\vq)$ following \refeq{phiq}
as we will see in \refsec{NGsummary}.  This however does not lead to additional
bias parameters, as the amplitude of the displacement terms is controlled
by the corresponding bias parameter (for example, $b_\phi$).  
The fully Lagrangian basis can be analogously constructed out
of $\phi(\vq)$ and \refeq{LagrBasis}.   

\refeq{EulBasisNG} only
gives the leading non-Gaussian contribution $\O(\fnl)$.  At
$\O(\fnl^2)$, one needs to add $\phi^2(\vq)$ to the list, and, following our
discussion after \refeq{Bphi}, also needs to keep cubic non-Gaussian terms
$\propto \gnl$.  However, all these terms have an extremely small observable effect on 
LSS statistics \cite{assassi/baumann/schmidt}.  For example, one would have to measure the bispectrum of tracers with all three $k$ modes of order $a H$, which
is unlikely to yield a significant detection in the foreseeable future, unless the cubic non-Gaussianity amplitude is very large, $\gnl \gtrsim 10^4$.  
For this reason, we restrict to linear order in $\fnl$ in the
bias expansion throughout.  Of course, even this will generate
terms of order $\fnl^2$ and $\fnl^3$ in the tracer power spectrum and 
bispectrum, respectively.  These terms should be kept as they become important
if a \emph{single} $k$ mode becomes of order $aH$.   

The bias terms listed in \refeq{EulBasisNG} only apply to the case of PNG of the local type.  However,
the generalization to other forms of the squeezed-limit bispectrum
is straightforward, and only amounts to replacing $\phi(\vq)$ with a 
nonlocal transformation of the potential.  We will describe this next.

\subsubsection{Beyond local-type PNG}
\label{sec:nonloc}

We now consider the generalization of the results of the last section to arbitrary quadratic non-Gaussianity, that is, PNG described to leading order by a three-point function or bispectrum.  
The impact of an arbitrary three-point function on the two-point function of galaxies  was first considered in \cite{verde/matarrese:2009}, who used the thresholding model described in \refsec{localbias} (based on \cite{matarrese/etal:1986}; see \refsec{NGthresh}).  
Ref.~\cite{Tseliakhovich:2010kf} explored the effects of multi-field inflation on bias (we will discuss this in \refsec{stochNG}).  
The leading scale-dependent bias induced by PNG on large scales for a generic primordial
bispectrum was derived by \cite{schmidt/kamionkowski:2010}, who generalized the physical argument of \cite{slosar/etal:2008}, while \cite{desjacques/seljak:2010c} derived the same in the high-peak limit, and \cite{matsubara:2012} adopted the iPT approach (\refsec{pertpkcorr}).  Finally,
\cite{long,scoccimarro/hui/etal:2012,sefusatti/crocce/desjacques:2012} identified a missing contribution
to the scale-dependent bias from PNG when the PNG is not of the local type; 
we will return to this issue in \refsec{PBSbiasNG}.

Our treatment here continues to follow the philosophy of the general perturbative bias expansion (\refsec{general}) and is most closely related to that in \cite{assassi/baumann/schmidt}, which in turn had precursors in \cite{mcdonald:2008,PBSpaper}.  The only assumptions made in this treatment are (i) gravity is described by General Relativity, and (ii) the initial conditions are given by weakly non-Gaussian adiabatic perturbations $\phi$.  
No assumptions are made regarding universality of the galaxy or halo mass function,
or the high-peak limit.

General quadratic PNG is often parametrized by a generalization of the 
local expansion \refeq{phiNGlocal} \cite{schmidt/kamionkowski:2010,scoccimarro/hui/etal:2012}, 
\ba
\phi(\vk) = \phi_G(\vk) + \int_{\vk_1}\int_{\vk_2} \Kernl(\vk_1,\vk_2) \phi_G(\vk_1)\phi_G(\vk_2) (2\pi)^3 \d_D(\vk-\vk_{12})\,,
\label{eq:PNGkernel}
\ea
where $\phi_G(\vk)$ is a Gaussian field, and $\Kernl(\vk_1,\vk_2)$ is a kernel.  The kernel $\Kernl$ is not uniquely determined by 
the bispectrum of $\phi$ \cite{schmidt/kamionkowski:2010,scoccimarro/hui/etal:2012},
which raises the worry that a prediction for the non-Gaussian bias based only on $B_\phi$ is ambiguous. Fortunately, this is not the case. Instead of adopting the expansion \refeq{PNGkernel} to derive the effects of PNG on galaxy bias, we will build the bias expansion upon the $n$-point statistics of $\phi$ directly.  

As discussed in \refsec{bNG}, since we are interested in the clustering of galaxies on scales much larger
than the scales ($\sim R_*$) on which galaxy formation happens, the kinematic regime of the three-point function that is relevant for bias 
is the \emph{squeezed limit} where one mode $\vk_3 = \vk_\ell$ is much smaller than the other two 
modes $\vk_1,\,\vk_2$. The bispectrum can then be written as
\ba
B_\phi(\vk_1,\vk_2,\vk_\ell) =\:& A(\vk_S,\,\vk_\ell) P_\phi(k_\ell) P_\phi(k_S)
+ \O\left(\frac{k_\ell^2}{k_S^2}\right) \vs
=\:& \sum_{J=0,2,4,...} A_J(k_S, k_\ell) \mathcal{L}_J(\hat{\vk}_S\cdot\hat{\vk}_\ell) 
P_\phi(k_\ell) P_\phi(k_S)
+ \O\left(\frac{k_\ell^2}{k_S^2}\right) \,,
\label{eq:Bphigen}
\ea
where $\vk_S \equiv \vk_1 + \vk_\ell/2=-\vk_2-\vk_\ell/2$ as above and $\mathcal{L}_J$ are the Legendre polynomials.  
In the squeezed limit, only even multipoles can contribute to the bispectrum for symmetry reasons.  
This can be seen by considering the local small-scale power spectrum of $\phi$, $P_\phi(\vk_S|\vx)$, analogously to \refeq{Plinmod}:   
since this is the power spectrum of a real scalar field, we have to have $P_\phi(\vk_S|\vx) = P_\phi(-\vk_S|\vx)$.  
This requires the absence of odd multipole contributions in \refeq{Bphigen} \cite{schmidt/hui:2013,assassi/baumann/schmidt}.  
In the following, we will assume that $B_\phi$ is scale-invariant and write
\be
A_J(k_S,k_\ell) = 4 a_J (k_\ell/k_S)^\alpha\,,
\label{eq:ALdef}
\ee
where $a_J$ and $\alpha$ are dimensionless constants.  
General scale-invariant bispectrum shapes can be approximated by considering
a sum of several contributions $\{a_{J,i}, \alpha_i\}$ \cite{smith/zaldarriaga:2006};  
note that $\alpha$ can be a real number.  
The local form \refeq{Bphi} is a special case of \refeq{Bphigen} with $J=0$, $\alpha=0$ and $a_0=\fnl$.  The factor of 4 in \refeq{ALdef} is motivated by the 
kernel expansion \refeq{PNGkernel};  the dimensionless coefficients $a_J$ quantify the quadratic coupling strength of the field $\phi$.  

Let us consider the case $J=0$ first, and look again at the galaxy-matter
cross-power spectrum before renormalization.  As in the case of local PNG, we need to deal with the term 
$(1/2) c_{2,\L} \d_\L^2(\vx)$ in the bare bias expansion, which now becomes \cite{schmidt/kamionkowski:2010}, in analogy with \refeq{corrNGsq1},
\be
\left \< \left(\d^{(1)}_\L\right)^2(\vk) \d^{(1)}(\vk') \right\> = 4 a_0 \s_{-\alpha/2}^2(\L)\, k^\alpha \<\phi_\L(\vk) \d^{(1)}(\vk') \>\,,
\label{eq:corrNGsq1gen}
\ee
where the spectral moment $\s_{-\alpha/2}(\L)$ is defined in \reftab{notation} [see also \refeq{mspec}; 
note that this definition is valid for non-integer $n$].  
Again, we find a cutoff-dependence that needs to be renormalized.  In this
case, due to the factor $k^\alpha$ in \refeq{corrNGsq1gen}, we have to add a different counter-term to obtain the renormalized operator $[\d^2]$,
\be
[\d^2](\vk) = (\d_\L^2)(\vk) - 4 a_0 \s_{-\alpha/2}^2(\L) k^\alpha \phi(\vk)
+ [\mbox{Gaussian counter-terms}]\,.
\ee
The same again holds for the renormalized operator $[K^2]$, up to a factor of $2/3$. 
The reasoning of \refsec{bNG} now goes through in exactly the same way. We have
to introduce an additional bias operator defined by 
\be
 \psi(\vq) \equiv \int_{\vk} k^\alpha \phi(\vk) e^{i\vk\cdot\vq}\,.
\label{eq:psidef}
\ee
The Eulerian basis of bias operators in general PNG with $J=0$ 
is then directly obtained from \refeq{EulBasisNG} by replacing $\fnl \phi \to a_0 \psi$, yielding
\bea
J=0\,:\qquad
{\rm 1^{st}} \ && \  a_0 \psi(\vq)  \label{eq:EulBasisNGgen} \\[3pt] 
{\rm 2^{nd}} \ && \ a_0 {\rm Tr}[\Pi^{[1]}(\vx)] \psi(\vq)  \nonumber  \\[3pt]
{\rm 3^{rd}} \ && \ a_0 {\rm Tr}[(\Pi^{[1]}(\vx))^2] \psi(\vq)\,,\  a_0 ({\rm Tr}[\Pi^{[1]}(\vx)])^2 \psi(\vq) \,. \nonumber
\eea

The physical reason behind the new operator $\psi(\vq)$ is the same as for local PNG: \refeq{Bphigen}
for $J=0$ states that the local initial matter power spectrum is rescaled by a single long-wavelength mode $\psi(\vk_\ell)$ via
\be
\Plin(\vk_S|\vx) = \left[1 + 4 a_0 \psi(\vk_\ell) e^{i\vk_\ell\cdot\vx} k_S^{-\alpha} \right]
\Plin(k_S)\,.
\label{eq:Plinmodgen}
\ee
This modulation is mediated by $\psi(\vk_\ell)$, which now appears in the bias expansion. 
From \refeq{EulBasisNGgen}, the leading-order cross- and auto-correlations of galaxies are then given by
\ba
P_{gm}^\LO(k) =\:&
\left[b_1 +  b_\psi a_0 k^\alpha \Mm^{-1}(k) \right] \Plin(k)\vs
P_{gg}^\LO(k) =\:&
\left[b_1 +  b_\psi a_0 k^\alpha \Mm^{-1}(k) \right]^2 \Plin(k) + \Peps\,.
\label{eq:corrNGsqgen}
\ea
We see that, on large scales, there is a scale-dependent bias $\propto k^{-2+\alpha}$.
For $\alpha=0$, we clearly recover the case of local-type PNG.
For $\alpha=2$, the non-Gaussian contribution to the galaxy two-point
function is scale-independent on large scales and thus degenerate with
the Gaussian term (linear bias $b_1$). This is not surprising, since for $\alpha=2$,
$\psi \propto \nabla^2 \phi$, which on large scales is simply proportional
to the matter density.  This applies to the equilateral and orthogonal shapes of 
PNG generated during single-field inflation, and raises the question of
whether these shapes can be constrained using the scale-dependent bias.  
A scale-dependence in the bias $\propto T^{-1}(k)$ does arise due to the transfer function on smaller
scales $k \gtrsim k_{\rm eq}$.  On those scales, the transfer function is no longer constant.  
However, for adiabatic perturbations, we can expand the inverse of the transfer function as $T^{-1}(k) = 1 + t_1 (k/k_{\rm eq})^2 + t_2 (k/k_{\rm eq})^4 + \cdots$, where $t_i$ are factors of order
unity.  This correspondingly leads to a scale-dependent bias of
\be
b_\psi \fnl^{\a=2} k^2 \Mm^{-1}(k) \sim \fnl^{\a=2} R_*^2 H_0^2 T^{-1}(k)
\sim {\rm const} + \fnl^{\a=2} R_*^2 H_0^2 \left( \frac{k^2}{k_{\rm eq}^2} + \frac{k^4}{k_{\rm eq}^4} + \cdots \right)\,,
\label{eq:fnla2}
\ee
where we have introduced $\fnl^{\a=2} = a_0(\alpha=2)$ and assumed that $b_\psi \sim R_*^2$ is controlled by the characteristic scale of the tracer $R_*$, the Lagrangian
radius in case of halos (this will be justified in \refsec{PBSbiasNG}).  
These terms scale in the same way with $k$, but are much smaller than the Gaussian higher-derivative bias 
$b_{\nabla^2\d} \nabla^2\d, ...$ (\refsec{higherderiv}) unless $\fnl^{\a=2} \gtrsim 10^3$ \cite{assassi/baumann/schmidt}.  However, this degeneracy can be broken if the scale dependence of the bias at $k\gtrsim k_{\rm eq}$ can be measured with sufficient precision, so that the different scales $R_*$ and $k_{\rm eq}^{-1}$ involved in the higher-derivative and PNG contributions can be disentangled. Ref.~\cite{gleyzes/deputter/etal:2016} show that constraints on $\fnl^{\alpha=2}$ from the scale-dependent bias can still be obtained; however, the partial degeneracy with the higher-derivative and NLO (1-loop) contributions to the galaxy power spectrum increases the uncertainty on $\fnl^{\alpha=2}$ by a factor of $\sim 40$ over the case where all higher-derivative and nonlinear bias contributions are absent. 

Let us now briefly consider the case $J=2$, i.e. a primordial bispectrum
that is anisotropic in the squeezed limit \cite{shiraishi/etal:2013}.  Such a form can, among other
non-standard physics, indicate the presence of a massive spin-2 (tensor) 
field during inflation (see \refsec{inflation}).  
In this case, $[\d^2]$ 
receives a somewhat different counter-term \cite{assassi/baumann/schmidt}
\be
[\d^2](\vx) = \d^2(\vx) - \frac{16}{105} a_2 \s_{-\alpha/2}^2(\L) \psi_{ij}(\vq) K^{ij}(\vx)\,,
\label{eq:d2NGnonloc}
\ee
where $\psi_{ij}(\vq)$ is a traceless tensor field derived from the Bardeen potential via
\be
\psi_{ij}(\vq) \equiv \int_{\vk} \frac32\left(\frac{k^i k^j}{k^2} -\frac13 \d^{ij}\right) k^\alpha \phi(\vk) e^{i\vk\cdot\vq}\,.
\ee
This type of PNG corresponds to an \emph{anisotropic} modulation of the local small-scale 
power spectrum by long-wavelength potential perturbations via \cite{schmidt/chisari/dvorkin}
\be
\Plin(\vk_S|\vx) = \left[1 + 4
a_2 \psi_{ij}(\vk_\ell) \frac{k_S^i k_S^j}{ k_S^{2+\alpha} } e^{i\vk_\ell\cdot\vx} \right]
\Plin(k_S)\,.
\label{eq:PlinmodgenL2}
\ee
If the small-scale matter power spectrum can be measured directly, this effect
can also be used to search for spin-2 fields by constructing an estimator
for $\psi_{ij}$ based on the local power spectrum $\Plin(\vk_S|\vx)$ \cite{jeong/kamionkowski:2012}.  

Up to cubic order, the following terms then enter the bias expansion for PNG with $J=2$: 
\bea
J=2\,:\qquad
{\rm 1^{st}} \ && \  -   \label{eq:EulBasisNGL2} \\[3pt] 
{\rm 2^{nd}} \ && \ a_2\, \Pi^{[1]}_{ij}(\vx) \psi^{ij}(\vq)  \nonumber  \\[3pt]
{\rm 3^{rd}} \ && \ a_2\,{\rm Tr}[\Pi^{[1]}(\vx)] \Pi^{[1]}_{ij}(\vx) \psi^{ij}(\vq)\,,\  \psi^{ij}(\vq) \Pi^{[2]}_{ij}(\vx) \,. \nonumber
\eea
Note that the leading non-Gaussian contribution to the bias
expansion, $\Pi^{[1]}_{ij} \psi^{ij} = \psi_{ij} K^{ij}$, is a second-order term.  For this reason,
anisotropic PNG with $J=2$ 
does not lead to a prominent large-scale signature in the galaxy 
two-point function.  It does however produce a scale dependence
$\propto k^{-2+\alpha}$ in the quadrupole of the galaxy bispectrum
\cite{assassi/baumann/schmidt} (see \refsec{NGsummary}),
which for $\alpha<2$ is a smoking-gun 
signature of inflationary physics just as the scale-dependent bias from 
local PNG is.  
Further, the same effect produces a strong scale-dependence in the two-point
function of galaxy \emph{shapes} which are expected to keep a memory
of an anisotropic local initial power spectrum of small-scale fluctuations
\cite{schmidt/chisari/dvorkin,chisari/etal:2016}.  

Finally, in the presence of the next-higher spin contribution to PNG, $J=4$, 
we need to introduce a tensor field $\psi_{ijkl}$ which is trace-free and symmetric with respect to all indices 
\cite{assassi/baumann/schmidt}.  Thus, symmetry dictates that the leading contribution (at lowest order in derivatives) is at cubic order, 
\bea
J=4\,:\qquad
{\rm 1^{st}} \ && \  -   \label{eq:EulBasisNGL4} \\[3pt] 
{\rm 2^{nd}} \ && \  - \nonumber  \\[3pt]
{\rm 3^{rd}} \ && \ a_4\, \Pi^{[1]}_{ij}(\vx) \Pi^{[1]}_{kl}(\vx) \psi^{ijkl}(\vq) \,. \nonumber
\eea
This leads to a characteristic scale-dependent signature only in the galaxy 4-point function (trispectrum).  

\subsubsection{Stochasticity from PNG}
\label{sec:stochNG}

We now turn to the contributions from PNG to the stochastic terms
in the bias expansion, discussed for Gaussian initial conditions in \refsec{stoch},
beginning with the isotropic case, $J=0$. 
By assumption, the stochastic variables $\eps,\, \eps_O$ introduced there only
depend on the statistics of the small-scale initial perturbations.  
As long as the coupling between long and short
modes is completely captured by \refeq{Plinmodgen}, all effects
 are accounted for in our non-Gaussian basis \refeq{EulBasisNGgen}.  
In this case, the list of terms in the Gaussian case, \refeq{stochBasis} in \refsec{stoch}, only needs to be augmented by terms
of the same type multiplied by $a_0\psi(\vq)$, 
\bea
J=0\,:\qquad
{\rm 1^{st}} \ && -  \label{eq:stochbasisNG} \\[3pt]
{\rm 2^{nd}} \ && \ a_0 \eps_{\psi}(\vx) \psi(\vq)  \nonumber
 \\[3pt]
{\rm 3^{rd}} \ && \ a_0\eps_{\psi\delta}(\vx) \psi(\vq) \hskip 1pt {\rm Tr}[\Pi^{[1]}(\vx)] \,. \nonumber
\eea
Here, the fields $\eps_{\psi},\,\eps_{\psi\delta}$ are characterized by
their 1-point moments just as their counterparts for Gaussian initial conditions.
 For anisotropic PNG, for example $J=2$, a stochastic term $\eps_{\v{\psi}}^{ij} \psi_{ij}$ is the leading contribution. However, since the small-scale modes do not know of any preferred direction at leading order, this contribution has to be a higher-derivative contribution: $\eps_{\v{\psi}}^{ij}(\vk) = k^i k^j \hat{\eps}_{\v{\psi}}(\vk)$, where $\hat{\eps}_{\v{\psi}}$ approaches white noise on large scales. This further suppresses its numerical importance.

One can show that \refeq{stochbasisNG} together with \refeq{stochBasis} comprise the complete set of stochastic fields as along as \emph{the initial conditions are
derived from a single statistical field, corresponding to a single set of random phases} (App.~B of \cite{assassi/baumann/schmidt}).  This is the case for the \emph{ans\"atze} \refeq{phiNGlocal} and \refeq{PNGkernel} considered above. However, if the initial Bardeen potential $\phi$ consists of a superposition of two
or more independent fields with different amounts of non-Gaussianity, for
example
\be
\phi(\vq) = \varphi_G(\vq) + \sigma_G(\vq) + \fnl^{\sigma} \sigma_G^2(\vq)\,,
\label{eq:phitaunl}
\ee
where $\varphi_G$ and $\sigma_G$ are independent Gaussian random fields, 
then \refeq{EulBasisNGgen} together with \refeq{stochbasisNG} are not sufficient.  
Such models are characterized by a large trispectrum 
$\< \phi(\vk_1)\phi(\vk_2)\phi(\vk_3)\phi(\vk_4)\>$
(Fourier transform of the four-point correlation function)
in the collapsed limit,
where the $k_i$ are comparable in magnitude but either $|\vk_{13}|$ or $|\vk_{24}|$ 
is much smaller \cite{Smith:2011if,Baumann:2012bc}; the trispectrum amplitude in this limit is customarily parametrized by a parameter $\taunl$.  A contribution of this type leads to a cutoff-dependent loop contribution in the galaxy two-point function which is not absorbed by $\psi(\vq)$ or any other term in the basis \refeq{EulBasisNGgen}.  Thus, for this type of PNG it is necessary to 
add an independent field $\hat\psi$, with associated bias parameter $b_{\hat\psi}$, to the bias expansion, where $\hat\psi$ is uncorrelated with the Gaussian part of the initial conditions $\phi$
\cite{Tseliakhovich:2010kf,Baumann:2012bc,assassi/baumann/schmidt} [in contrast with $\psi$, which is completely correlated with $\phi$, \refeq{psidef}].  
This potentially large contribution to galaxy clustering on large scales was already pointed out by \cite{allen/grinstein/wise:1987}.  

Since $\hat\psi$ is uncorrelated with long-wavelength modes of $\phi$ and only correlates with itself, it does not affect the galaxy-matter cross-correlation, but gives a non-vanishing contribution to the galaxy power spectrum
\be
P_{gg}^\LO(k)\,\supset\,  \taunl (b_{\hat\psi})^2\, P_{\hat\psi}(k)\, .
\ee
For the local-type example given in \refeq{phitaunl}, we have $P_{\hat\psi}(k) = P_{\phi}(k)$.  Thus, this term will be one of the dominant terms on 
sufficiently large scales.  
In this case, the leading-order correlation coefficient between matter and galaxies becomes
\ba
r_{gm}^\LO(k) \equiv \left.\frac{P_{gm}(k)}{\sqrt{P_{gg}(k) P_{mm}(k)}}\right|_\LO\,
\stackrel{\fnl\neq0}{=}
\, 
\frac{b_\psi P_{\d\psi}(k)}{\sqrt{\Big[ (b_{\psi})^2  P_{\psi}(k) + (\taunl/\fnl^2) (b_{\hat\psi})^2 P_{\hat\psi}(k)\Big] \Plin(k)}}\,,
\label{eq:rk}
\ea
where $P_{\psi\d}(k) \equiv \< \psi(\vk) \d^{(1)}(\vk')\>'$.  
This is equal to unity if and only if $\taunl b_{\hat\psi}=0$, otherwise the
correlation coefficient between matter and galaxies is less than one.  
Note the very different scale dependence of the correlation coefficient
compared to the Gaussian case,
\refeq{crosscorrlinear}, discussed in \refsec{evol1}.  
Hence, by measuring the correlation coefficient between galaxies and
matter on large scales, we can determine whether the collapsed
limit of the four-point function exceeds the value predicted for initial conditions sourced by a single degree of freedom.

\subsubsection{Beyond the squeezed limit}
\label{sec:bbeyond}

The leading non-Gaussian contribution to the galaxy two-point function 
for general scale-invariant quadratic PNG is given by \refeq{corrNGsqgen}, which  
was derived by evaluating the primordial bispectrum in the squeezed limit.   
This limit however is only valid if modes with wavelength of order the separation $k$ do
not contribute appreciably to the variance $[\s(R_*)]^2$ of the small-scale density field 
on the scale $R_*$ relevant for the formation of the tracer.  Empirically, one finds that the squeezed limit
is no longer a good approximation when $k$ approaches the peak of the
matter power spectrum ($k \sim 0.02 \iMpch$) \cite{schmidt/kamionkowski:2010}.  
We can deal with this issue perturbatively by expanding the initial matter bispectrum beyond the leading squeezed limit.  Assuming a scale-free bispectrum of primordial potential perturbations as given in \refeq{Bphigen}, we obtain (see \cite{schmidt:13} for details), 
\ba
B^{(1)}(\vk_1,\vk_2,\vk_\ell) 
=\:& 4 \sum_{J=0,2,4,...} \left\{ a_J^{\{0\}} \left(\frac{k_\ell}{k_S}\right)^\alpha 
+ a_J^{\{2\}} \left(\frac{k_\ell}{k_S}\right)^{\alpha+2} f_J^{\{2\}}(k_S) + \cdots \right\}
 P_\phi(k_\ell) \Plin(k_S)\,.
\label{eq:Bphibeyondsq}
\ea
The subleading term is no longer scale-free, since it involves derivatives of the transfer function $dT(k)/dk$.  However, given a primordial power spectrum and bispectrum, the subleading coefficient $a_J^{\{2\}}$ and function $f_J^{\{2\}}(k_S)$ are
uniquely determined.  Moreover, if $\alpha$ is the only (non-analytic) scaling dimension in the full bispectrum,
which holds for commonly considered shapes such as local, equilateral, orthogonal, quasi-single-field, and the folded template (see \refsec{inflation}),
then the leading correction beyond the squeezed limit is suppressed by
$(k_\ell/k_S)^2$ relative to the leading term, which is the scaling we assume here.  
The reasoning in the following does not rely on this assumption however.  
Note that even if, at leading order, only the $J=0$ term exists, terms of higher
order in $J$ will in general appear at subleading order.  

The key point to notice is that the correction beyond the squeezed limit is of the
same general form as the leading term.  Thus, specializing to a $J=0$ contribution, we merely have to introduce
another field $\tilde\psi$ in the bias expansion, where $\tilde\psi$ is 
defined as a nonlocal transformation of $\phi$, \refeq{psidef} with $\alpha \to \alpha+2$.  
Adding $\tilde\psi$ as counter-term to $[\d^2]$ in \refeq{d2NGnonloc} then removes the cutoff dependence up to order $(k/\Lambda)^4$ as desired \cite{schmidt:13}.  
At leading order, this leads to a galaxy auto-power spectrum given by
\ba
P_{gg}^\LO(k)\Big|_{J=0} =\:&
\left[b_1 + \left(a_0^{\{0\}} b_\psi k^\alpha + a_0^{\{2\}} b_{\tilde\psi} k^{2+\alpha} \right)
\Mm^{-1}(k) \right]^2 \Plin(k) + \Peps\,.
\label{eq:corrNGbeyondsq}
\ea
Correspondingly, for the $J>0$ contributions in \refeq{Bphibeyondsq} a similar reasoning applies, as described in \refsec{nonloc}.  These become
relevant for loop corrections to the two-point function as well as the 
galaxy three- and higher-point functions. We will see in \refsec{PBSbiasNG} that, if we consider halos which approximately follow a universal mass function, then $b_{\tilde\psi} \sim R_*^2 b_\psi$, where $R_*$ is the Lagrangian radius of the halos. This can serve as a rough estimate of the magnitude of the beyond-squeezed-limit contributions.

At this point, it is also worth noting that there are other terms
which contribute to the galaxy power spectrum \refeq{corrNGbeyondsq} at a similar order as the
beyond-squeezed-limit terms written in \refeq{corrNGbeyondsq}.  
First, the Gaussian higher-derivative terms combine with the leading non-Gaussian term to a contribution
\[
-2 b_\psi b_{\nabla^2\d} k^2 \< \psi(\vk) \d(\vk')\>'.
\]
Second, in the same way as higher-derivative terms, such as $b_{\nabla^2\d} \nabla^2\d$, appear in the general bias expansion for Gaussian initial conditions (\refsec{higherderiv}), 
we expect related terms to be present for the field $\psi$ in the non-Gaussian case, i.e. 
$b_{\lapl\psi} \lapl_q \psi(\vq)$ at leading order.  For example, if the
abundance of galaxies depends on the statistics of initial fluctuations within a region of scale $R_*$, then such a term is obtained after formally expanding the spatial convolution, with $b_{\lapl\psi} \propto R_*^2$.
All these terms scale as $R_*^2 k^2$ times the leading PNG contribution and are thus comparable to the beyond-squeezed-limit contribution given in \refeq{corrNGbeyondsq}. Note that this is in contrast to the scale dependence induced by the transfer function entering through $\Mm^{-1}(k)$, which is controlled by $k_{\rm eq}^{-1} \gg R_*$ [see the discussion around \refeq{fnla2}].

In \refsec{NGexset}, we will show that if we make the strong assumption that galaxy formation is an exactly
local function of the initial density field smoothed on a single scale $R$, then both the beyond-squeezed-limit 
contributions as well as terms of the type $b_{\lapl\psi} \lapl_q \psi(\vq)$ are uniquely determined by an integral over the bispectrum.

\subsubsection{Summary}
\label{sec:NGsummary}

In \refsecs{localNG}{bbeyond}, we have derived the contributions to the general 
galaxy bias expansion that appear in the presence of PNG, focusing on the leading, quadratic (in the Bardeen potential) PNG parametrized by the bispectrum of potential perturbations.  
We now summarize the complete set of bias contributions up to cubic order
in perturbations.  This complements the expansion given in \refeq{dgsummary}
for Gaussian initial conditions (\refsec{evol:summary}).  We first set $\taunl=0$, to obtain
\ba
\d_g(\vx)\Big|_{a_0,\,a_2,\,a_4} =\:& a_0 \Bigg\{ b_\psi \psi(\vq) + b_{\psi\d} \psi(\vq) \d(\vx)
+ b_{\psi\d^2} \psi(\vq) \d^2(\vx) + b_{\psi K^2} \psi(\vq) (K_{ij})^2(\vx) \vs
& \qquad + \eps_\psi(\vx) \psi(\vq) + \eps_{\psi\d}(\vx) \psi(\vq) \d(\vx)
 + b_{\lapl\psi} \lapl_q \psi(\vq) \Bigg\} \vs
& + a_2 \bigg\{  b_{\psi K} \psi^{ij}(\vq) K_{ij}(\vx) 
+ b_{\psi K \d} \psi^{ij}(\vq) K_{ij}(\vx) \d(\vx)
+ b_{\psi \Pi^{[2]}} \psi^{ij}(\vq) \Pi^{[2]}_{ij}(\vx) \vs
& \qquad\quad +  \eps_{\psi K}(\vx) \psi^{ij}(\vq) K_{ij}(\vx)
\bigg\} 
 + a_4   b_{\psi K K} \psi^{ijkl}(\vq) K_{ij}(\vx) K_{kl}(\vx) 
\,,
\label{eq:dgsummaryPNG}
\ea
where we restrict the expansion to linear order in the parameters $a_J$,
as discussed at the end of \refsec{bNG}.  
Here, we have emphasized that the fields $\psi,\,\psi_{ij},\,\psi_{ijkl}$ are
evaluated at the Lagrangian position $\vq$ corresponding to the Eulerian
position $(\vx,\tau)$.  Further, these three fields are defined as 
nonlocal transformations of the primordial Bardeen potential $\phi$,
\ba
\psi(\vq) \equiv\:& \int_{\vk} k^\alpha \phi(\vk) e^{i\vk\cdot\vq}
\vs
\psi_{ij}(\vq) \equiv\:& \int_{\vk} \frac32\left(\frac{k_i k_j}{k^2} -\frac13 \d_{ij}\right) k^\alpha \phi(\vk) e^{i\vk\cdot\vq} \vs
\psi_{ijkl}(\vq) \equiv\:& \int_{\vk} \frac{35}{8} \mathcal{P}_{ijkl}(\hat{\vk})  k^\alpha \phi(\vk) e^{i\vk\cdot\vq}
\,,
\label{eq:psidefs}
\ea
where $\mathcal{P}_{ijkl}(\hat{\vk})$ denotes the complete trace-free projection operator with respect to $\vk$ [see \refeq{Pijlm} for the explicit expression].

All operators appearing in \refeq{dgsummaryPNG} are understood to be renormalized;  for clarity,
we have omitted the square brackets.  
The first two lines, $\propto a_0$, contain the contributions for isotropic PNG.  The first line contains the generalization of the deterministic \LIMD bias expansion, while the second line gives the stochastic and higher-derivative contributions.  Here, we have made the same approximation as in \refeq{dgsummary},
and included only the leading, linear higher-derivative term.  Note that in the case of PNG, this also includes beyond-squeezed-limit contributions (\refsec{bbeyond}) which in general obey a different scaling than other higher-derivative contributions.  

The third and fourth lines, $\propto a_2$ and $a_4$, contain the deterministic and stochastic contributions for anisotropic PNG ($J=2$ and 4).  Due to symmetry, these terms start at second and third order in perturbations, respectively.  Correspondingly, we have not included a higher-derivative contribution, although the same caveats regarding beyond-squeezed-limit terms mentioned above also apply here.  

Finally, we give contributions $\propto \taunl$, which are present if the initial conditions are a superposition of several random fields, as discussed in \refsec{stochNG}.  Note that, by definition, $\hat\psi$ is not correlated with the other long-wavelength perturbations $\psi,\,\d,\,K_{ij}$, and, at this order, is completely specified by its power spectrum $P_{\hat\psi}(k)$.  We obtain, in precise analogy with the terms $\propto a_0$ in \refeq{dgsummaryPNG},
\ba
\d_g(\vx)\Big|_{\taunl} =\:& \taunl \Bigg\{ b_{\hat\psi} \hat\psi(\vq) + b_{\hat\psi\d} \hat\psi(\vq) \d(\vx)
+ b_{\hat\psi\d^2} \hat\psi(\vq) \d^2(\vx) + b_{\hat\psi K^2} \hat\psi(\vq) (K_{ij})^2(\vx) \vs
& \qquad + \eps_{\hat\psi}(\vx) \hat\psi(\vq) + \eps_{\hat\psi\d}(\vx) \hat\psi(\vq) \d(\vx)
 + b_{\lapl\hat\psi} \lapl_q \hat\psi(\vq) \Bigg\} \,.
\label{eq:dgsummarytauNL}
\ea

We can further restrict \refeq{dgsummaryPNG} to the most commonly considered case of local PNG by setting $a_0\to \fnl$, $\psi\to\phi$, and $a_2=0,\,a_4=0,\,\taunl=0$.  This yields
\ba
\d_g(\vx)\Big|_{\fnl} =\:& \fnl \Bigg\{ b_\phi \phi(\vq) + b_{\phi\d} \phi(\vq) \d(\vx)
+ b_{\phi\d^2} \phi(\vq) \d^2(\vx) + b_{\phi K^2} \phi(\vq) (K_{ij})^2(\vx) \vs
& \qquad + \eps_\phi(\vx) \phi(\vq) + \eps_{\phi\d}(\vx) \phi(\vq) \d(\vx)
 + b_{\lapl\phi} \lapl_q \phi(\vq) \Bigg\} \,.
\label{eq:dgsummaryfnl}
\ea
Ref.~\cite{assassi/baumann/schmidt} was the first to give the complete expressions \refeqs{dgsummaryPNG}{dgsummaryfnl}.  
Initial studies of bias expansions in the presence of local-type PNG only considered the first term, $b_\phi \phi$.  
Ref.~\cite{giannantonio/porciani:2010} extended the expansion by including all terms of the form $\phi^n(\vq) \d^m(\vx)$.  They did not, however, include the tidal term $\propto \phi (K_{ij})^2$, nor the stochastic and higher-derivative terms in the second line of \refeq{dgsummaryfnl}.  This also applies to \cite{baldauf/seljak/senatore:2011}.   Ref.~\cite{tellarini/etal:2016} included the Gaussian tidal term $(K_{ij})^2$, but did not write down its non-Gaussian counterpart  $\phi (K_{ij})^2$ as they restricted to a second-order bias expansion.  They also did not consider the stochastic term $\eps_\phi \phi$ which is however relevant at second order (see below).  

After having given the general bias expansion in the presence of quadratic PNG, we now turn to galaxy statistics, generalizing the results of \refsec{npttree}.  
Specifically, we provide a succinct summary of the leading-order galaxy 
power spectrum and bispectrum in the rest frame (without RSD and other projection effects).  We further make explicit the time dependence of all quantities.

First, the leading-order cross- and auto-power spectra of galaxies are given by
\ba
P_{gm}^\LO(k,\tau) =\:&
\left[b_1(\tau) + \DeltabNG(k,\tau) \right] \Plin(k,\tau) \vs
P_{gg}^\LO(k,\tau) =\:&
\left[b_1(\tau) + \DeltabNG(k,\tau) \right]^2 \Plin(k,\tau) + \Peps(\tau)
+ \taunl [b_{\hat\psi}(\tau)]^2 P_{\hat\psi}(k)\,, 
\label{eq:corrNGsqgen2}
\ea
where
\be
\DeltabNG(k,\tau) \equiv a_0 b_\psi(\tau) k^\alpha \Mm^{-1}(k,\tau)
\stackrel{\text{local PNG}}{=} \fnl b_\phi(\tau) \Mm^{-1}(k,\tau) 
\label{eq:Deltab1def}
\ee
is the \emph{scale-dependent bias} induced by PNG.  This is, of course, merely a rephrasing of the 
\emph{nonlocal bias} $b_\psi(\tau) \psi(\vq)$.  
Here, we have let $\alpha$ denote the squeezed-limit scaling index of the 
$J=0$ contribution in \refeq{Bphigen}.  For example, for local and equilateral 
type PNG \cite{babich/creminelli/zaldarriaga:2004,creminelli/etal:2006,creminelli/etal:2007}, we have $\alpha=0$, and $2$, respectively.  
Quasi-single field inflation \cite{Chen/Wang:1,Chen/Wang:2} yields
$0 \leq \alpha \leq 3/2$.

Note that we have not included the 
beyond-squeezed limit and higher-derivative corrections, which typically scale 
as $R_*^2 k^2 \DeltabNG(k) \Plin(k)$; see \refsec{bbeyond} for a discussion. Further, while the set of cubic operators listed above is sufficient to derive the NLO (1-loop) contribution from PNG to the galaxy power spectrum, these terms are highly suppressed and we thus refrain from writing the full expression here; they involve three new bias parameters: $a_0 b_{\psi\d},\,a_2b_{\psi K},\,a_2b_{\psi\Pi^{[2]}}$. In order to obtain an order-of-magnitude estimate, one can replace $\Plin(p)$ in the standard NLO terms, \refeq{Phm1l} on p.~\pageref{eq:Phm1l} in \refsec{npt1loop}, with $p^\alpha \Mm^{-1}(p) \Plin(p)$. Even for $\alpha=0$, this leads to terms which are suppressed by $\sim 10^{-4}a_J$ relative to the standard NLO terms, and would thus only be relevant if $a_J \gtrsim 100$.

Next, we give the complete expression for the rest-frame galaxy bispectrum
at leading order, including all relevant terms from \refeq{dgsummaryPNG} \cite{assassi/baumann/schmidt}.  Here, there are considerably more terms, since all quadratic combinations of $\psi$ and, for $J=2$, $\psi_{ij}$ appear.  
It is useful to decompose the separable contributions to the galaxy bispectrum in terms of 
Legendre polynomials $\mathcal{L}_l(\hat{\vk}_i\cdot\hat{\vk}_j)$ of the cosine between two wavevectors:
\begin{align}
B_{ggg}^\LO(k_1,k_2,k_3,\tau) = b_1^3 B^{(1)}(k_1,k_2 ,k_3,\tau)
+\sum_{l=0,1,2}\left[\Plin(k_1,\tau)\Plin(k_2,\tau){\cal B}^{[l]}(k_1,k_2,\tau)\,\mathcal{L}_{l}(\hat\vk_1\cdot\hat\vk_2)+\perm{2}\,\right] .
\label{eq:BgNG}
\end{align}
The first contribution here is simply the leading-order matter bispectrum 
\refeq{Bmloc}.  Note that \refeq{BgNG} is \emph{not restricted to the squeezed limit, but valid for all configurations.}  Hence, all multipoles $l$ including odd ones appear here.  As in \refsec{stoch}, we define the large-scale stochastic amplitudes
\be
P^{\{0\}}_{\eps_O\eps_{O'}}(\tau) \equiv \lim_{k\to 0} \< \eps_O(\vk,\tau) \eps_{O'}(\vk',\tau) \>'
\ee
and $\Beps(\tau)$ [\refeq{Bepsdef}].  While we restrict to $B_{ggg}$ here, the results for $B_{ggm}$ and $B_{gmm}$ can be obtained analogously.

The monopole consists of deterministic and stochastic contributions given by
\ba
{\cal B}^{[0]}(k_1,k_2,\tau) =\:& \big(b_1(\tau)+\DeltabNG(k_1,\tau)\big)\big(b_1(\tau)+\DeltabNG(k_2,\tau)\big)
\vs &\times\bigg[\frac{34}{21}b_1(\tau)+b_2(\tau) + \frac{b_{\psi\delta}(\tau)}{b_\psi(\tau)}\big(\DeltabNG(k_1,\tau)+\DeltabNG(k_2,\tau)\big)\bigg] \vs
& + \taunl b_1(\tau) b_{\hat\psi}(\tau) b_{\hat\psi\delta}(\tau) \left(\frac{P_{\hat\psi}(k_1)}{\Plin(k_1,\tau)}+\frac{P_{\hat\psi}(k_2)}{\Plin(k_2,\tau)}\right) + {\cal B}^{[0]}_{\eps}(k_1,k_2,\tau) \vs
{\cal B}^{[0]}_{\eps}(k_1,k_2,\tau) =\:& \frac{\Beps(\tau)}{3\Plin(k_{1},\tau) \Plin(k_2,\tau)} 
\vs
&+ \left[(b_1(\tau)+\DeltabNG(k_1,\tau))\left(\Pepsepsd(\tau)+\frac{P^{\{0\}}_{\eps\eps_\psi}(\tau)}{b_\psi(\tau)}\DeltabNG(k_1,\tau)\right)\frac{1}{\Plin(k_2,\tau)}+\{1\leftrightarrow2\}\right] \vs
& + \taunl b_{\hat\psi}(\tau) P^{\{0\}}_{\eps\eps_{\hat\psi}}(\tau)\frac{P_{\hat\psi}(k_1)+P_{\hat\psi}(k_2)}{\Plin(k_1,\tau)\Plin(k_2,\tau)}\, .
\label{eq:BgNG0}
\ea
The dipole term, whose only contributions at this order come from expansions of the Lagrangian position $\vq$ around the Eulerian position $\vx$ [cf. \refeq{phiq}], is given by
\ba
{\cal B}^{[1]}(k_1,k_2,\tau) =\:& \big(b_1(\tau)+\DeltabNG(k_1,\tau)\big)\big(b_1(\tau)+\DeltabNG(k_2,\tau)\big)\left[\frac{k_1}{k_2}\big(b_1(\tau)+\DeltabNG(k_1,\tau)\big)+\frac{k_2}{k_1}\big(b_1(\tau)+\DeltabNG(k_2,\tau)\big)\right] \vs
& + \taunl b_1(\tau) [b_{\hat\psi}(\tau)]^2\left(\frac{k_1}{k_2}\frac{P_{\hat\psi}(k_1)}{\Plin(k_1,\tau)}+\frac{k_2}{k_1}\frac{P_{\hat\psi}(k_2)}{\Plin(k_2,\tau)}\right)\,.
\label{eq:BgNG1}
\ea
That is, ${\cal B}^{[1]}$ is induced by the fact that the fields $\psi,\,\hat\psi$ are evaluated at the Lagrangian position.  It is completely determined by the bias parameters $b_1,\,b_\psi,\, b_{\hat\psi}$ that appear in the galaxy power spectrum, and can hence serve as a clean, independent cross-check of the scale-dependent bias measured in the power spectrum \cite{assassi/baumann/schmidt}.

Finally, the quadrupole
\ba
{\cal B}^{[2]}(k_1,k_2,\tau) =\:& \frac{4}{3}\big(b_1(\tau)+\DeltabNG(k_1,\tau)\big)\big(b_1(\tau)+\DeltabNG(k_2,\tau)\big)\left[b_{K^2}(\tau)+\frac{2}{7}b_1(\tau)\right] \label{eq:BgNG2}\\
& + a_2  b_{\psi K} (\tau)
\big(b_1(\tau)+\DeltabNG(k_1,\tau)\big)\big(b_1(\tau)+\DeltabNG(k_2,\tau)\big)
\left[ k_1^{\alpha_2} \Mm^{-1}(k_1,\tau) + k_2^{\alpha_2} \Mm^{-1}(k_2,\tau) \right]
\, ,
\nonumber
\ea
contains contributions from tidal bias [$\propto b_{K^2}$, see \refsec{npttree}] and
from anisotropic PNG with $J=2$ ($\propto a_2 b_{\psi K}$).  
Note the different prefactor in the second line, which is due to the definition of the field $\psi_{ij}$ multiplied by the bias $b_{\psi K}$ [\refeq{psidefs}].  
Here, $\alpha_2$ denotes the scaling index of the $J=2$ contribution in \refeq{Bphigen}.  The bispectrum quadrupole thus allows for $J=0$ and $J=2$ contributions to be disentangled.  As we will explain in the next section, this means that galaxy clustering can in principle probe the presence of additional fields in the early Universe as well as determine their spin. 

Finally, while we have restricted to the two- and three-point functions here,
higher $n$-point functions can also be used to constrain PNG.   For example, 
the estimator constructed in \cite{jeong/kamionkowski:2012} (out of the matter density field) essentially corresponds to measuring the collapsed trispectrum.

\subsection{Probing inflation with galaxy clustering}
\label{sec:inflation}

In the previous section, we have seen that the squeezed-limit bispectrum
and collapsed trispectrum of the initial conditions can lead to
a rich array of signatures in the clustering of galaxies on large scales.  
We now briefly summarize which aspects of the physics of inflation are
probed by these signatures.\\  

\textbf{\emph{Single-field inflation:}} In single-field inflation, one
single scalar degree of freedom drives the expansion during
the inflationary quasi-de Sitter phase, and leads to the primordial
curvature perturbations that provide the seeds of large-scale structure. 
Interestingly,
Ref.~\cite{maldacena:2003} showed that in single-field inflation and in the 
attractor regime, the squeezed-limit bispectrum is universal and
corresponds to a local PNG with $\fnl \propto n_s-1$, where 
$n_s-1 \approx 0.04$ is the deviation from the scale-invariance of the power 
spectrum of primordial curvature perturbations.  
This prediction, known as \emph{consistency relation}, would suggest that 
there is a definite nonzero value of local $\fnl$ predicted in single-field 
inflation that LSS surveys could target. 
On the other hand, multi-field inflationary models, which we discuss below, 
generically produce larger values of local $\fnl$.  While $n_s-1$ is small for the simplest
single-field models, oscillations in the primordial power spectrum
can significantly enhance this effect \cite{chen/easther/lim:2008,flauger/pajer,cyr-racine/schmidt,cabass/pajer/schmidt:2018}.  

However, the calculation of the bispectrum in \cite{maldacena:2003} was done 
in a particular gauge (comoving gauge), and evaluated during the epoch of 
inflation, raising the question of how this effect transfers to late times 
and how it manifests in actual observations of galaxy clustering.
Several recent papers 
\cite{tanaka/urakawa:2011,baldauf/etal:2011,creminelli/damico/etal:2011,CFCorig,CFCpaper2,deputter/etal:15} 
have shown that the single-field consistency relation is equivalent to the statement that there is 
\emph{no physical coupling} of long-wavelength potential perturbations $\phi_\ell$
to small-scale perturbations.  That is, \emph{in single-field inflation, there
is no scale-dependent bias in the rest frame of galaxies}; the leading effect
of a large-scale perturbation enters as $\partial_i\partial_j\phi_\ell$,
precisely as argued in \refsec{GR}.  
The apparent contradiction with Refs.~\cite{Camera/etal:14,Bartolo/etal:15}, who argue that terms in second-order relativistic perturbation theory (see \cite{villa/etal:14} for a recent review) actually lead to an effective $\fnl$ of order 1, is most likely resolved by including the mapping to local observables defined in terms of proper length and time units.  Unlike the squeezed-limit bispectrum and hence value of $\fnl$, which depend on the coordinate system in which they are calculated, the absence of long-/short-mode coupling in the rest frame of comoving observers is a physical, gauge-invariant statement.  

It is important to stress however that variants of single-field inflation that
are \emph{not} in the attractor regime, due for example to a non-Bunch-Davies
state \cite{chen/etal:2007,meerburg/etal:2009,agullo/parker:2010} or an initial kick of the inflaton \cite{namjoo/etal:2012}, do lead to a local-type
bispectrum \emph{over a finite range of values of} $k_\ell/k_S$.  That is, while they
lead to a physical scaling of $(k_\ell/k_S)^2$, as discussed above, for sufficiently small values of
$k_\ell/k_S$ (that is, in comoving coordinates the bispectrum obeys the
consistency relation), there can be an intermediate regime where $k_\ell/k_S$ is small
 and yet the bispectrum shows a scaling with $\alpha < 2$ (see \cite{chen:2010} for a discussion).  This also applies to resonant non-Gaussianity \cite{chen/easther/lim:2008,flauger/pajer}.  Thus, the simple separable
ansatz \refeq{Bphigen} is not sufficient for such models.  A full derivation
of the observable scale-dependent bias due to PNG in these models has
been performed recently in \cite{cabass/pajer/schmidt:2018}, who showed that
for models that satisfy CMB bounds as well as theoretical consistency, the
scale-dependent bias is unlikely to be observable. Previously, Ref.~\cite{cyr-racine/schmidt} had reached different conclusions, as they included the unphysical
contribution from the consistency relation.  

Note that the mapping from the galaxy rest frame to the observer's frame on 
Earth by way of tracing photon geodesics through the perturbed spacetime,
which includes gravitational redshift, Doppler effect, and lensing,
does lead to an apparent scale-dependent bias in the observed galaxy
clustering. We review these contributions in \refsec{observations}.  
However, these effects are present even for perfectly Gaussian initial
conditions, and unrelated to inflationary physics. They can easily be calculated given a fiducial cosmological 
model and the luminosity function of the sampled galaxy population.  
Thus, the observation 
of any residual scale-dependent bias $\propto k^{-2}$ in galaxy 
clustering---after all light-propagation effects have been taken into
account---would rule out single-field inflation in the attractor regime.  
Moreover, the projected constraints on $\fnl$ from planned future galaxy surveys are at the level of $\sigma(\fnl) \sim 1$ (\refsec{NGforecast}) and thus expected to improve upon current CMB constraints ($|\fnl| \lesssim 5$ \cite{planck:2013c}) significantly.  

By definition, in single-field inflation there is only a single source of random phases which precludes any large-scale stochasticity between galaxies and matter of the type described in \refsec{stochNG}. A detection of the latter would thus similarly rule out single-field inflation.  
Detecting this signature requires observations of both the matter-galaxy 
cross-power spectrum, for example from weak gravitational lensing, and the galaxy power spectrum over the same volume.  
No detailed study of the detectability of such a signature has been published yet.

Finally, single-field inflation does generally produce \emph{nonlocal} 
non-Gaussianities with $\alpha=2$ \cite{maldacena:2003,chen/etal:2007,cheung/etal:2008}.  In particular, the precise shapes are of the equilateral and
orthogonal types \cite{senatore/smith/zaldarriaga:2010}.  
While canonical single-field slow-roll inflation leads to (physical) amplitudes
of $\fnl^{\rm eql} \sim 0.01$ \cite{cabass/pajer/schmidt:2016}, these non-Gaussianities are enhanced 
in more general single-field models where the inflaton has a small sound speed $c_s$.  
In this case, one has $\fnl^{\rm eql}\simeq 1/c_s^2$ \cite{cheung/etal:2008}.  
Note that, since these types of PNG have $\alpha=2$ in the squeezed limit [\refeq{corrNGsqgen}], 
they do not lead to a large-scale scale-dependent bias.  In fact, improving upon CMB constraints 
on $\fnl^{\rm eql}$ with large-scale structure will be quite challenging \cite{baldauf/etal:2016}.  

\textbf{\emph{Multifield inflation:}}  By definition, these models involve additional light degrees 
of freedom beyond a single scalar, and generically produce local PNG with $\fnl$ of order 1 or larger  
(see \cite{byrnes/choi:2010} for a review). 
Moreover, if some of the fields have a mass that is comparable to the Hubble rate during inflation, 
then a scale-dependent bias [\refeq{corrNGsqgen}] $\propto k^{\alpha-2}$ with $0 \leq \alpha \leq 3/2$ 
is induced, and a measurement of $\alpha$ allows for a measurement of the mass $m$ of the field, 
provided $m < 3H/2$ \cite{Chen/Wang:1,Chen/Wang:2,Baumann/Green,mirbabayi/simonovic:2016}.  
For higher-mass fields, oscillatory features are present in the bispectrum
\cite{noumi/etal:2013,arkani-hamed/maldacena:2015,chen/namjoo/wang:2016}.  
Further, the large-scale stochasticity in the relation between galaxies
and mass described in \refsec{stochNG} is a unique feature of multifield models 
(see \cite{Tseliakhovich:2010kf,Baumann:2012bc} for concrete examples).  

\textbf{\emph{Higher-spin fields:}}  An anisotropic squeezed-limit bispectrum
[$J=2,4,\cdots$ in \refeq{Bphigen}] leads to further unique signatures
in galaxy clustering, as we have seen in \refsec{nonloc}.  This type of
PNG can signal the presence of higher-spin fields during inflation 
\cite{arkani-hamed/maldacena:2015,lee/etal:2016,flauger/etal:2016}; ``solid inflation'' 
\cite{endlich/etal,dimastrogiovanni/etal:2014} and ``Chromo-natural inflation'' 
\cite{adshead/wyman:2012} are models of a different type which also lead to anisotropic PNG.  
In addition, inflationary models with \emph{anisotropic} non-Bunch-Davies (i.e. non-vacuum) 
state \cite{Agullo12,ashoorioon/casadio/koivisto:2016} as well as those with large primordial magnetic fields 
\cite{ShiraishiB1,ShiraishiB2} in general also produce anisotropic squeezed-limit bispectra.\\

In summary, we see that large-scale galaxy statistics provide numerous possibilities for 
probing the statistics of the initial conditions and putting constraints on inflationary models.  
However, in order to derive constraints on $\fnl,\, a_J,\,\taunl, \cdots$, we need 
predictions for the corresponding bias parameters $b_\psi,\,b_{K \psi},\,b_{\hat\psi},\cdots$.  
The peak-background split provides a possibility for deriving such predictions from semi-analytical 
methods and simulations.

\subsection{Non-Gaussian bias parameters from the peak-background split}
\label{sec:PBSbiasNG}

In \refsec{bphys}, we were able to derive the renormalized \LIMD bias 
parameters $b_N \equiv N!\,b_{\d^N}$ as a response of the abundance of 
halos to a change in the background density via the PBS argument. 
We now present similar identifications for the 
scale-dependent bias from PNG, specifically the parameters 
$b_{\phi\d^N},\,b_{\psi\d^N}$ $(N \geq 0)$ which appear in the expansions
\refeq{EulBasisNG} and \refeq{EulBasisNGgen} for local and nonlocal PNG, respectively (we restrict to isotropic PNG with $J=0$ throughout).  
Similar to \refsec{PBS}, we begin with the general case which applies
to galaxies and other tracers, and then specialize to the case of halos
following a universal mass function.  

In order to derive $b_{\phi\d^N},\, b_{\psi\d^N}$, we need to consider the response of the galaxy number density to 
a change in the amplitude of initial density fluctuations,
which is where the scale-dependent bias physically originates,
through \refeq{Plinmod} and \refeq{Plinmodgen}.  
The simplest way to parametrize such a dependence
is to rescale initial perturbations by a factor of $1+ 2 \epsilon k^{-\alpha}$ from their fiducial
value, where $\epsilon$ is an infinitesimal parameter [corresponding to $a_0\psi$, \refeq{Plinmodgen}].  For example, for
a given realization of initial conditions of an N-body simulation, one
can obtain a realization with a different power spectrum normalization 
and shape by rescaling the initial density perturbations by $(1+2 \epsilon k^{-\alpha})$.  
We then define the Lagrangian non-Gaussian PBS bias $b_{\psi\d^N}^L$ by generalizing the expression \refeq{bN} to
\be
b_{\psi \d^N}^L \equiv \frac1{N!} \frac{1}{\avng(0)} \frac{\partial^{N+1} \avng(\Delta,\epsilon)}{\partial \Delta^N \partial \epsilon}\bigg|_{\Delta=0,\epsilon=0}\,.
\label{eq:bNM}
\ee
This includes $b_\psi$ as a special case for $N=0$, and $b_\phi$ for
$N=0,\,\alpha=0$.    Note that the distinction between Lagrangian and Eulerian
bias parameters matters only for $N \geq 1$, as the $\epsilon$-derivative is always defined with respect to the initial fluctuations.  For $N \geq 1$, 
the mapping between Lagrangian $b_{\psi \d^N}^L$ and Eulerian $b_{\psi \d^N}^E$ can be derived in analogy to the Gaussian case of \refsec{localbias};  specifically, $b_{\psi \d^N}^E$ is given by $b_{\psi \d^N}^L$ plus corrections involving the $b_{\psi \d^m}^E$ with $0 \leq m < N$.  

The bias parameters defined in \refeq{bNM} can be understood as follows.  The mean comoving galaxy number density is a function of the mean comoving matter density $\rhob$ and the initial power spectrum of density fluctuations (and higher $n$-point functions for PNG).  $b_{\psi \d^N}^L$ then denotes a specific
derivative of this function with respect to $\rhob$ and $\epsilon$
at fiducial values of $\rhob$ and the primordial power spectrum normalization.    
Thus, through a simple generalization, the PBS argument can be used to derive the leading bias parameters in the non-Gaussian case as well.  Specifically, the leading effect
of local PNG ($\alpha=0$), quantified by $b_{\phi}$, is controlled by the response
of the mean number density of galaxies to a rescaling of the amplitude $\mathcal{A}_s$ of
initial fluctuations,\footnote{Note that under the rescaling with $\alpha=0$, $\mathcal{A}_s$ and $\sigma_8$ change by a factor $(1+4\epsilon)$ and $(1+2\epsilon)$, respectively.}
\be
b_\phi = 4 \frac1{\avng}\frac{\partial \avng}{\partial\ln \mathcal{A}_s}
= 2 \frac1{\avng}\frac{\partial \avng}{\partial\ln \sigma_8}\,.
\label{eq:bphiPBS}
\ee
Here, we have used that a scale-independent rescaling of the initial conditions can be equally parametrized through $\mathcal{A}_s$ or $\sigma_8$.
This relation, and more generally \refeq{bNM}, can be evaluated for any given prescription that predicts the mean abundance of galaxies or halos.  
We discuss the application of \refeq{bphiPBS} to halos identified in N-body simulations, as performed in \cite{baldauf/etal:2015,biagetti/etal:2016}, in \refsec{NGsim}.  For now, we begin with some general considerations.

Let us assume that the number density of the given galaxy sample depends on the amplitude of initial 
fluctuations chiefly on a particular scale $R$.  Then, we can conclude that:
\begin{itemize}
\item For \emph{local} PNG, the particular scale $R$ is irrelevant for $b_{\phi \d^N}$, as all perturbations 
$\d(\vk)$ are rescaled uniformly by a long-wavelength potential perturbation.  
\item For \emph{nonlocal separable} bispectra with index $\alpha$ as in \refeq{Bphigen}, the
scale of the small-scale perturbations that govern the abundance of galaxies
does matter for the bias parameters. Note that $[b_{\psi \d^N}] = (\text{Mpc})^\alpha$. By dimensional analysis, we thus expect that $b_{\psi \d^N} \propto R^{\alpha}$.  
If the galaxy number density depends on the amplitude of initial
perturbations on several different scales, then $b_\psi$
becomes a linear combination of these different dependencies with
relative weights controlled by $\alpha$, i.e. the shape of the primordial bispectrum.  
\item \emph{Non-separable} bispectra first need to be approximated by 
a linear combination of separable shapes 
(see e.g. \cite{smith/zaldarriaga:2006}).  In that case, the 
galaxy statistics in general involve several different $b_{\psi_i}$ with different
$\alpha_i$.  
\end{itemize}
Thus, a given galaxy population responds differently
to different shapes of primordial non-Gaussianity, i.e. $b_{\psi}$ 
and $b_{\psi O}$ depend on the galaxy sample as well as the
shape of the primordial bispectrum.  

We now turn to the special case of dark matter halos.  
Specifically, we consider a slight generalization of the universal mass function discussed in \refsec{buniv}.  
We write the mean abundance of halos as 
\ba
\avnh =\:& \avnh\left(\rhob,\sigma\right)\,J\,, \quad \mbox{where}\quad
\s \equiv \s(R[M])\,,\  J \equiv \left|\frac{d\ln \s}{d\ln M}\right|\,.
\label{eq:nhsimple}
\ea
That is, $\avnh$ is given as a function of the mean density of the Universe and the variance of the density 
field smoothed on the scale $R(M)$, as well as its derivative with respect to scale.  
The Jacobian $J$ transforms from an interval in $\ln\s$ to an interval in $\ln M$, and we thus 
assume that $\avnh$ is linearly proportional to it (as is the case for the universal mass function introduced 
in \refsec{buniv}).  
Under the rescaling $\d(\vk) \to [1+2\epsilon k^{-\alpha}]\d(\vk)$, the small-scale root-mean-square density fluctuation 
$\s$ transforms to lowest order as
\be
\sigma \to \left[1 + 2\epsilon \frac{\sigma_{-\alpha/2}^2}{\s^2}\right] \sigma\,,
\ee
where $\sigma_{-\alpha/2} \equiv \sigma_{-\alpha/2}(R)$.
The Jacobian transforms as 
\ba
J \to\:& \left[ 1 +
4 \epsilon \frac{\s_{-\alpha/2}^2}{\s^2} \left( \frac{d\ln \s_{-\alpha/2}^2}{d \ln \s^2}
-  1 \right) \right] J\;,
\label{eq:Jtransf}
\ea
where we have used $d/d\ln M = 2 J \: d/d\ln \s^2$.  
Note that for local quadratic PNG, where $\alpha=0$, the local Jacobian is not 
affected by long-wavelength modes, as expected. Using \refeqs{bphiPBS}{nhsimple}, we can then derive 
the leading non-Gaussian bias through \refeq{bNM} \cite{long,PBSpaper}:
\ba
b_{\psi} =\:& \left[ b_\phi
+ 4\left( \frac{d\ln \s_{-\alpha/2}^2}{d \ln \s^2}
-  1 \right) \right]\frac{\s_{-\alpha/2}^2}{\s^2} \;.
\label{eq:bpsiuniv}
\ea
Note that we use a convention different than \cite{long} for $b_{\psi}$, whence our factor of 4.  
Here, $b_\phi=b_\psi(\alpha=0)$ is the PBS bias parameter quantifying the
effect of local PNG for halos following \refeq{nhsimple}.
Thus, for halos following a universal mass function, the bias parameters quantifying the response to general nonlocal PNG
are directly related to those for local PNG.  Refs.~\cite{long,letter} first pointed out the 
contribution by the Jacobian $J$, which is numerically important for the match to the scale-dependent bias measured 
from N-body simulations of nonlocal PNG,
as well as cubic local-type ($\gnl\phi^3$) PNG \cite{letter,scoccimarro/hui/etal:2012}. We will not discuss $\gnl$ 
in detail, as its signature is strongly degenerate with that of $\fnl$ (see, e.g., \cite{roth/porciani:2012}), but note that the
different mass-dependence could be exploited with multi-tracer techniques (as discussed in \refsec{NG:mt}) in order to isolate $\gnl$ from $\fnl$.

Note that the bias parameters associated with the additional subleading contributions from the squeezed-limit
expansion of the bispectrum (\refsec{bbeyond}) can be calculated in the same
manner.  Specifically, the bias parameter $b_{\tilde\psi}$ appearing in \refeq{corrNGbeyondsq} is defined as the 
response of the mean abundance $\avnh$ to a change in the amplitude of initial density perturbations
\be
\d^{(1)}(\vk) \to \left[1 + 2\epsilon k^{-\alpha-2} f_0^{\{2\}}(k) \right] \d^{(1)}(\vk)\,.
\ee
Thus, for a mass function of the form \refeq{nhsimple}, the subleading contribution is directly related to the 
leading scale-dependent bias $b_\psi$ via spectral moments involving $k^{-\alpha-2} f_0^{\{2\}}(k)$.  

We now specialize \refeq{nhsimple} to the well-known universal form
[\refeq{univ1}],
\be
\avnh = \frac{\rhob}{M}\:\nu_c f(\nu_c)\: J\,,\quad \nu_c\equiv \frac{\dc}{\s}\,,
\label{eq:univ}
\ee
where $f(\nu_c)$ is in general an arbitrary function of $\nu_c$.  
In addition to the relation between $b_\psi$ and $b_\phi$ [\refeq{bpsiuniv}], the 
specific form \refeq{univ} further allows us to connect $b_\phi$
to the linear Lagrangian bias $b_1^L$ derived from the PBS:
\ba
b_1^L = b_1 - 1 =\:& \frac1{\avnh} \frac{\partial \avnh}{\partial\ln \rhob}
= - \frac1{\s} \frac1{ \nu_c f(\nu_c)} \frac{d[ \nu_c f(\nu_c) ]}{d\nu_c} \vs
b_\phi =\:\frac{1}{\avnh}\frac{\partial \avnh}{\partial\epsilon}
=\:& 2 \frac1{\avnh} \frac{\partial \avnh}{\partial\ln \sigma}
= - 2 \frac{\dc}{\s} \frac1{ \nu_c f(\nu_c)} \frac{d[ \nu_c f(\nu_c) ]}{d\nu_c} = 2 \dc b_1^L\,.
\label{eq:b01univ}
\ea
This is the original relation
between the density bias parameter and the response to primordial
non-Gaussianity derived in \cite{dalal/etal:2008,slosar/etal:2008,giannantonio/porciani:2010}, resulting in a scale-dependent bias for local PNG of
\be
\DeltabNG(k,\tau) = \fnl b_\phi \Mm^{-1}(k,\tau)
= \frac{3\fnl\dc b_1^L \Omega_{m0} H_0^2}{k^2 T(k) D_\text{md}(\tau)}\,.
\label{eq:DeltabDalal}
\ee
We will discuss the accuracy of this result for actual halos in 
\refsec{NGsim} (see \reffigs{bNG_vs_k}{bNG_vs_b}).  
The relation between non-Gaussian and Gaussian PBS bias parameters can
be continued to higher order via \refeq{bNM}.  For example, one 
easily obtains \cite{giannantonio/porciani:2010}
\begin{align}
b_{\phi\d}^L &= -b_1^L + \dc b_2^L\,, \quad\mbox{and}\quad
b_{\phi\d}^E = b_{\phi\d}^L + b_\phi
\,.
\label{eq:b11univ}
\end{align}
The second relation follows immediately from the relation between Lagrangian and Eulerian halo densities (\reftab{EulLagr}
in \refsec{dynamics}).  

As in the Gaussian case, the non-Gaussian PBS biases $b_{\psi \d^N}$
are the physical, renormalized bias parameters in the large-scale limit,
which enter the $n$-point functions of galaxies and halos (\refsec{NGsummary}) 
as well as other halo statistics, such as moments, as described in \refsec{measurements}.  

\subsection{Non-Gaussian bias from Lagrangian bias models}
\label{sec:NGLagBias}

In the previous sections, we have seen how the effect of primordial non-Gaussianity, which couples long- and short-wavelength
modes, can be included in a model-independent way through an expansion of the bispectrum in the squeezed limit.  
We now discuss the non-Gaussian bias in the context of Lagrangian bias models.
The essential difference between this approach and the general bias expansion \refeq{dgsummaryPNG} is that the ``microscopic''
perturbative bias expansion in the Lagrangian models remains the same as in the Gaussian case.  That is,
no new terms are introduced in the local description of the tracer abundance.  
Throughout, we will neglect gravitational evolution and perform
the computation in Lagrangian space.  This is justified as long as we consider 
sufficiently large scales where linear theory describes the evolution of the
cosmic density field well. 
As we will see below, all the Lagrangian bias models predict the correct low-$k$ 
scaling (for example, $\DeltabNG(k)\propto k^{-2}$ for local PNG), but 
they differ in their prediction for the amplitude of the non-Gaussian scale-dependent bias.  
Note that, in all these Lagrangian approaches, the scale $R$ is a
physical scale associated with the scale of galaxy formation, unlike the 
cutoff $\L^{-1}$ employed in \refsec{bNG}.

\subsubsection{Thresholding}
\label{sec:NGthresh}

We begin with a simple but illuminating example, the simple Lagrangian \LIMD ansatz discussed in \refsec{localbias},
in which the comoving Lagrangian halo number density is given by  
\begin{equation}
n_h(\vq) = n_{\rm thr}(\vq) \propto \Theta_H\left[\delta_R(\vq)-\dc\right] \;,
\label{eq:thresholding}
\end{equation}
where $\Theta_H$ is the Heaviside function and $\delta_R$ is the density field 
smoothed on the
\emph{physical} scale $R$. The normalization constant, which we omitted to write here, cancels in the 
computation of correlation functions. The threshold $\dc$ can be identified with the spherical collapse 
threshold (\refsec{sph_collapse}), so that the expectation value of \refeq{thresholding}, after taking a 
derivative with respect to $M$, corresponds to the Press-Schechter mass function (\refsec{PS}).  
The two-point correlation function of thresholded regions in case of Gaussian initial conditions was 
derived in \refsec{localbias} [\refeq{xihthr}].  In the \emph{high-peak limit} $\nu_c\gg 1$, \refeq{xihthr} 
simplifies to \cite{kaiser:1984,politzer/wise:1984,jensen/szalay:1986}
\begin{equation}
\label{eq:xthreshG}
\xi_h^L(r)\stackrel{\nu_c\gg 1}{=}\exp\left(\frac{\nu_c^2}{\s^2(R)}\xi_{{\rm L},R}(r)\right)-1
\approx \left(\frac{\nu_c}{\s(R)}\right)^2 \xi_{{\rm L},R}(r) \equiv
(b_1^L)^2 \xi_{{\rm L},R}(r)\,,
\end{equation}
where the linear Lagrangian bias in the high-peak limit is $b_1^L\approx \nu_c/\s(R)$.
In order to calculate the non-Gaussian bias in the same high-peak regime, Ref.~\cite{matarrese/verde:2008} considered
the extension of \refeq{xthreshG} to non-Gaussian initial conditions derived in \cite{matarrese/etal:1986,grinstein/wise:1986}, 
\begin{equation}
\label{eq:xthreshNG}
\xi_h^L(r) \stackrel{\nu_c\gg 1}{=}-1+\exp\biggl\{\sum_{N=2}^\infty\sum_{j=1}^{N-1}
\frac{\nu_c^N\s^{-N}(R)}{j!(N-j)!}\xi_{{\rm L},R}^{(N)}\left[\vq_1,\cdots,\vq_1,
\vq_2,\cdots,\vq_2\right]\biggr\} \;.
\end{equation}
Here, 
\be
\xi_{{\rm L},R}^{(N)}(\vq_1,\cdots\, \vq_N) \equiv \< \d_R^{(1)}(\vq_1)\cdots \d_R^{(1)}(\vq_N) \>_{c}
\ee
are the connected $N$-point functions of the initial, linearly extrapolated density field smoothed on the scale $R$.  
Assuming that the primordial three-point function dominates all other $(N>3)$-point functions in \refeq{xthreshNG},
the non-Gaussian correction to the two-point correlation function of thresholded regions reads, at linear order in $\xi^{(3)}_{{\rm L},R}$,
\begin{equation}
\label{eq:dxiNG}
\Delta\xi_h^L(r) = \frac{\nu_c^3}{2\s^3(R)}
\left[
\xi_{{\rm L},R}^{(3)}(\vq_1,\vq_1,\vq_2)+\xi_{{\rm L},R}^{(3)}(\vq_1,\vq_2,\vq_2)
\right]\;.
\end{equation}
In Fourier space, this relation can be written as follows:
\begin{equation}
P_h^L(k)= \left[b_1 + 2 b_1 \DeltabNG(k)\right] \Plin(k)\;,
\end{equation}
where the scale-dependent bias is related to the primordial three-point function
through \cite{matarrese/verde:2008}
\ba
\DeltabNG(k) =\:& 2 \nu_c^2 \Fs{3}_R(k) \Mm_R^{-1}(k)
\vs \mbox{where}\quad 
\Fs{3}_R(k) \equiv\:& \frac{1}{4\s^2(R) P_\phi(k)}
\int_{\vk_1} \,\Mm_R(k_1) \Mm_R(p) B_\phi(k_1, |\vk-\vk_1|, k)\;.
\label{eq:DbNGthr1}
\ea
Before discussing this result which, again, is only valid in the high-peak limit, let us 
illustrate how the non-Gaussian bias can be derived from the perturbative bias expansion corresponding 
to thresholding. The latter takes the familiar \LIMD form
\begin{equation}
\label{eq:deffthr}
\delta_\text{thr}^L(\vq) = b_1^L \big[\delta_R\big](\vq) + \frac{1}{2} b_2^L \big[\delta_R^2\big](\vq) + \dots \;,
\end{equation}
where, for Gaussian initial conditions, the renormalized bias parameters $b_N^L$ and operators $[\delta_R^N(\vq)\big]$ are 
related to orthogonal polynomials as discussed in \refsec{peaks}. For the simple thresholding models, 
these are Hermite polynomials as in \refeq{dNHermite} \cite{szalay:1988}, and the $b_N^L$ are given by [see \refeq{bPBSthr}] 
\be
b_N^L = \sqrt{\frac2\pi} \left[
\erfc\left(\frac{\nu_c}{\sqrt{2}}\right)\right]^{-1}
\frac{e^{-\nu_c^2/2}}{\sigma^N(R)}
H_{N-1}(\nu_c)
\stackrel{\nu_c\gg 1}{=} \frac{\nu_c^N}{\s^N(R)}\,.
\ee
The leading-order contribution induced by a primordial three-point function is
\begin{equation}
\label{eq:DbNGthr2}
\Delta\xi_{h}^L(r) = b_1^L b_2^L \left\langle \big[\delta_R\big](\vq_1)\big[\delta_R^2\big](\vq_2)\right\rangle
= b_1^L b_2^L \left\langle \delta_R(\vq_1)\delta_R^2(\vq_2)\right\rangle \;.
\end{equation}
Transforming to Fourier space and taking the high-peak limit, we recover \refeq{DbNGthr1}.  
Let us now briefly discuss this result:
\begin{itemize}
\item One could identify $\nu_c^2$ in \refeq{DbNGthr1} with $b_1^L\dc$
and, thus, recover the prediction for a universal mass function, \refeq{b01univ}. However, \refeq{DbNGthr2} shows that 
the actual prediction of thresholding is $b_\phi\propto b_2^L$, which is clearly at odds with 
measurements from simulations. 
In fact, any Lagrangian \LIMD bias expansion yields $b_\phi\propto b_2^L$.
This point was first made by \cite{taruya/koyama/matsubara:2008,sefusatti:2009,jeong/komatsu:2009b}.  
Note that, here, this contribution 
is not absorbed by a counter-term $b_\phi \phi$ because the bias parameters are defined as coefficients of powers of the density smoothed on a \emph{physical} scale $R$, rather than the artificial cutoff $\Lambda$ used in \refsec{bNG}.  
This is a consequence of using a microscopic Lagrangian ansatz such as \refeq{thresholding}, rather than an effective large-scale expansion as in \refsec{NGevolution}.
\item \refeq{DbNGthr1} generically applies to any primordial three-point function, i.e. it is not restricted to 
the case of local PNG. However, as will be shown shortly, for nonlocal PNG it misses an important contribution 
to the scale-dependent bias.  
\item This derivation does not rely on expanding the primordial bispectrum in the squeezed limit. This is a 
consequence of the precise \LIMD ansatz \refeq{thresholding}, which does not restrict us to scales $k \ll 1/R$.  
One can think of this as resumming the beyond-squeezed-limit contributions discussed in \refsec{bbeyond}, by making an assumption about how precisely 
the Lagrangian halo number density depends on the statistics of the small-scale modes. 
The accuracy of this approach depends on whether this dependence is well approximated by \refeq{thresholding}.  Indeed, we will see below that the prediction \refeq{DbNGthr1} changes if the Lagrangian bias model includes higher-derivative operators (\refsec{NGpeaks}).  Nevertheless, \refeq{DbNGthr1} does match the results of \refsec{bbeyond} at next-to-leading order in the squeezed-limit expansion when evaluated for halos following a universal mass function.  
\end{itemize}

\subsubsection{Excursion-set approach}
\label{sec:NGexset}

Clearly, a thresholding ansatz of the type \refeq{thresholding} is not sufficient for a realistic description of 
halo bias (see \refsec{exset}). One should instead also include the first-crossing constraint, which leads to a 
correction to the scale-dependent bias.
The correction is significant for all types of PNG beyond local quadratic PNG \cite{long,letter}. 
This correction term comes about because the scale-dependent coupling between long- and short-wavelength modes 
in the case of general PNG modifies the conversion from an 
interval in $\ln\nu_c$ to an interval in $\ln M$, as shown by \refeq{Jtransf}.  
A general relation for the scale-dependent non-Gaussian bias can be obtained by considering the conditional mass
function $\avnh(M|\delta_\ell)$, where the long mode $\delta_\ell$ correlates with the small-scale density 
fluctuations owing to PNG [see \refeq{Plinmod} and \refeq{Plinmodgen}]
\cite{long,letter}.

Namely, in the standard excursion-set approach, the halo mass function conditioned on a long mode $\d_\ell$ on scale $R_\ell$ reads
\begin{equation}
\avnh(M|\delta_\ell) = -2\rhob
\frac{\partial}{\partial M}\int_{\dc}^\infty\!\!d\delta\,
p\big(\delta;R\big\lvert\delta_\ell;R_\ell\big) \;,
\end{equation}
where now the conditional probability $p\big(\delta;R\big\lvert\delta_\ell;R_\ell\big)$ for having a small-scale 
overdensity $\delta$ on scale $R$ given a large scale overdensity $\delta_\ell$ on scale $R_\ell\gg R$ is no longer
Gaussian.  Expanding $p\big(\delta;R\big\lvert\delta_\ell;R_\ell\big)$ in a series of reduced cumulants, computing 
\begin{equation}
\delta_h(\delta_\ell) = \frac{\avnh(M|\delta_\ell)}{\avnh(M)}-1
\end{equation}
as in \refsec{exset}
and identifying the coefficients of the cumulant series with the Gaussian bias parameters, we 
can eventually read off the scale-dependent bias $\DeltabNG(k)$ as \cite{long}
\begin{equation}
\DeltabNG(k) = \sum_{N=3}^\infty \frac{4}{(N-1)!}\,
\Bigg\{b_{N-2}^L\,\dc + b_{N-3}^L \left[N-3 + 
\frac{\partial\ln \Fs{N}_R(k) }{\partial\ln\s(R)}\right ]
\Bigg\} \Fs{N}_R(k) \Mm_R^{-1}(k)\;,
\label{eq:DbNGthr3}
\end{equation}
where $b_N^L$ are the Lagrangian halo bias parameters, with $b_0^L\equiv 1$.
No assumption about the shape of the smoothing filter has been made to derive this result.  
Furthermore, contributions from all connected primordial $N$-point functions can be included by defining the general shape factor
\begin{equation}
\Fs{N}_R(k) \equiv \frac{1}{4\s^2(R) P_\phi(k)}
\left\{\prod_{i=1}^{N-1}\int_{\vk_i} \Mm_R(k_i)\right\}
\left\< \phi(\vk_1) \cdots \phi(\vk_{N-1}) \phi(\vk) \right\>_c \;.
\end{equation}
In the squeezed limit, $\Fs{N}_R$ corresponds to taking 
one of the $N$ momenta of the $N$-point function on the right-hand side 
to zero and integrating over the remaining $N-1$ momenta (with the total momentum constraint enforced).  
In the case of a primordial bispectrum ($N=3$), we obtain the modified variance given in \refeq{DbNGthr1}.  
Let us emphasize two points here:
\begin{itemize}
\item In general, a primordial $n$-point function will contribute a modified $(n-1)$-cumulant \cite{jeong/komatsu:2009b}.  
Therefore, when including higher-order ($n>3$) primordial $n$-point functions, it is important to include all terms that are of 
similar order as predicted by a given inflationary model.  
For example, when including terms of order $\gnl$ ($N=4$) in higher-order local PNG, one should also include terms of order 
$\fnl^2$ which are generally comparable. 
Note that \refeq{DbNGthr2} does not include the stochastic contributions that are induced by an enhanced collapsed limit of the 
primordial trispectrum (\refsec{stochNG}).
\item As \refeq{DbNGthr2}, \refeq{DbNGthr3} exactly recovers the results of 
\cite{matarrese/verde:2008,desjacques/seljak:2010a,verde/matarrese:2009,shandera/dalal/huterer:2011} 
in the high-peak limit $\nu_c\gg 1$ where $b_{N-2}\approx (\nu_c/\s(R))^{N-2}$.
The key new term in \refeq{DbNGthr3} is the contribution from $\partial\ln \Fs{N}_R(k)/\partial\ln\s(R)$, which arises due to the 
mass ($R$-)dependence of the reduced cumulants.  We have already obtained this contribution in the PBS approach of \refsec{PBSbiasNG}, 
where it corresponds to the change in the Jacobian $J$ under a scale-dependent rescaling of the initial conditions [\refeq{Jtransf}].  
In fact, evaluating \refeq{DbNGthr3} for $N=3$ in the large-scale limit, corresponding to the bispectrum in the 
squeezed limit, one exactly recovers the result \refeq{bpsiuniv} presented in \refsec{PBSbiasNG} \cite{PBSpaper}.  
\end{itemize}
Finally, for quadratic local PNG, the first term in the curly bracket of \refeq{DbNGthr3} (including the multiplicative factor of
2 in front) recovers $b_\phi=2\dc b_1^L$, in agreement with the prediction \refeq{b01univ} for a universal mass function 
(which is the case of Press-Schechter, see \refsec{PS}).
Therefore, it appears that we have fixed the problem of \refeq{DbNGthr2}, which predicts a non-Gaussian bias amplitude $\propto b_2^L$.

To understand this better, let us now derive \refeq{DbNGthr3} from the perturbative bias expansion corresponding to the thresholding 
ansatz with first-crossing. The corresponding number density is \citep[e.g.,][]{matsubara:2012}
\begin{equation}
n_{\rm PS}(\vq) = -2 \rhob \frac{\partial}{\partial M}\Theta_H\left[\delta_R(\vq)-\dc\right] 
= 2 \rhob\mu_R(\vq) \delta_D\!\big[\delta_R(\vq)-\dc\big] \frac{dR}{dM}\,,
\label{eq:nexset}
\end{equation}
where, following \cite{desjacques/gong/riotto:2013}, we have performed the derivative with respect to $M$ in the second equality and introduced
the variable $\mu_R\equiv -d\delta_R/dR$ as in \refsec{esp}. 
Therefore, the density contrast $\delta_{\rm PS}(\vq)$ is described by a bivariate perturbative
expansion, whose renormalized operators $[\delta^i\mu_R^j](\vq)$ are, in case of Gaussian initial conditions, bivariate Hermite polynomials. Namely,
\begin{align}
\delta_{\rm PS}^L(\vq) &= b_{10}^L H_{10}(\delta_R,\mu_R)+ b_{01}^L H_{01}(\delta_R,\mu_R)  \\ 
& \quad + \frac{1}{2}b_{20}^L H_{20}(\delta_R,\mu_R) + b_{11}^L H_{11}(\delta_R,\mu_R) + \frac{1}{2} b_{02}^L H_{02}(\delta_R,\mu_R) 
+ \cdots \;. \nonumber
\end{align}
The corresponding bias parameters $b_{ij}^L$ are ensemble averages of bivariate Hermite polynomials as in \refsec{PBSpeaks}. 
For instance, we have \citep{desjacques/gong/riotto:2013}
\begin{gather}
b_{10}^L = \frac{1}{\s(R)}\left(\nu_c-\frac{1}{\nu_c}\right) \;,
\qquad b_{01}^L = -\frac{1}{\nu_c}\left(\frac{d\s(R)}{dR}\right)^{-1}\\
b_{20}^L = \frac{\nu_c^2-3}{\s^2(R)} \;,
\qquad b_{11}^L = -\frac{1}{\s(R)}\left(\frac{d\s(R)}{dR}\right)^{-1}
\;,\qquad b_{02}^L = 0 \;.
\end{gather}
Consequently, the leading-order non-Gaussian contribution to the 2-point 
correlation function of the tracers is, for a primordial bispectrum,
\begin{equation}
\label{eq:Dbexset}
\Delta \xi_{h}^L(r) = b_{10}^L \bigg[b_{20}^L \Big\langle \delta_R(\vq_1)\delta_R^2(\vq_2)\Big\rangle+
2 b_{11}^L \Big\langle \delta_R(\vq_1)\delta_R(\vq_2)\mu_R(\vq_2)\Big\rangle + b_{02}^L
\Big\langle\delta_R(\vq_1)\mu_R^2(\vq_2)\Big\rangle \bigg] \,,
\end{equation}
in the large-scale limit $r\to\infty$, so that we can ignore $b_{01}^L$, which amounts to a higher-derivative term $\propto k^2$ as discussed in \refsec{excursion_HeavensPeacock}.  
Taking the Fourier transform of \refeq{Dbexset}, we eventually find that, in the limit $k\to 0$, the non-Gaussian bias amplitude induced by local quadratic PNG is given by \citep{matsubara:2012,desjacques/gong/riotto:2013} 
\begin{equation}
\label{eq:genericLagNGbias}
b_\phi = 2 \int_{\vk} c_2^L(\vk,-\vk) \Plin(k) \;,
\end{equation}
where
\begin{equation}
c_2^L(\vk_1,\vk_2) = b_{20}^L W_R(k_1) W_R(k_2) + \frac{1}{\s^2(R)} \frac{\partial}{\partial\ln\s(R)}\Big(W_R(k_1) W_R(k_2)\Big)
\label{eq:c2Les}
\end{equation}
for the particular model \refeq{nexset} considered here. However, \refeq{genericLagNGbias} is valid for any microscopic Lagrangian bias 
model. The term proportional to $\partial/\partial\ln\s(R)$ in \refeq{c2Les} generates the new contribution, $\propto \partial\ln \Fs{N}_R(k)/\partial\ln\s(R)$ in \refeq{DbNGthr3}, for a generic primordial bispectrum.  
Substituting \refeq{c2Les} into \refeq{genericLagNGbias}, we obtain 
\begin{equation}
b_\phi = 2 \bigg(b_{20}^L \s^2(R) + 2 \bigg) = 2 \dc b_{10}^L \;,
\end{equation}
the prediction \refeq{b01univ} for a universal mass function, which is expected since the Press-Schechter mass function is universal.

Finally, one can show \cite{schmidt:13} via the approach described in \refsec{bbeyond} that the agreement between the excursion-set 
result \refeq{DbNGthr3} and the general bias expansion even holds at subleading order in the squeezed limit when adopting a universal 
mass function $\propto f(\nu_c)$, consistent with \refeq{thresholding}.   
Fundamentally, this is a consequence of the fact that in the thresholding approach, as in general for universal mass functions, there 
is only a single scale $R$ that enters all predictions of halo statistics. 

\subsubsection{Lagrangian density peaks}
\label{sec:NGpeaks}

We now turn to the peak approach, which is an example of a multivariate Lagrangian bias model.  We will focus on BBKS \cite{bardeen/etal:1986} peaks, rather 
than the excursion-set peaks (ESP) here for simplicity.  BBKS peaks are
described in detail in \refsecs{kacrice}{peakbias}, and we will adopt
the same notation here.  
Like in the simple excursion-set approach considered
in the previous section, the 
effective peak perturbative  bias expansion, 
\refeq{dpkeff} in \refsec{pertpkcorr}, in terms of the variables
$\nu \propto \d,\, J_1 \propto \lapl\d,\, \eta^2 \propto (\partial_i\d)^2$, and $J_2, J_3$ which are invariants constructed from $\partial_i\partial_j\d$,  
holds regardless of the statistical properties of the linear density field.  In complete analogy with 
\refsecs{NGthresh}{NGexset}, the leading-order contribution to the non-Gaussian 
2-point correlation of BBKS peaks induced by a primordial bispectrum reads
\begin{align}
\Delta\xi_\text{pk}^L(r) &= 2\sigma_0^3 b_{10}^L b_{20}^L \la\nu^2(\vq_1) \nu(\vq_2)\ra 
+4\sigma_0^2\sigma_2 b_{10}^L b_{11}^L\la\nu(\vq_1) J_1(\vq_1)\nu(\vq_2)\ra 
+ 2\sigma_0\sigma_2^2 b_{10}^L b_{02}^L\la J_1^2(\vq_1)\nu(\vq_2)\ra
\nonumber \\
& \quad 
+ 4\sigma_0\sigma_1^2\chi_1^L b_{10}^L\la\eta^2(\vq_1)\nu(\vq_2)\ra + 4\sigma_0\sigma_2^2
\omega_{10}^L b_{10}^L\la J_2(\vq_1)\nu(\vq_2)\ra \nonumber \\
& \quad 
+ \big(\sigma_0, b_{10}^L, \nu(\vq_2)\big)\to\big(\sigma_2, b_{01}^L, J_1(\vq_2)\big) \;,
\end{align}
where the symbols are defined in \reftab{symbol6} on p.~\pageref{tab:symbol6}, and, as in \refeq{dpkeff}, all moments are evaluated on the scale $R$.  The various Lagrangian, second-order peak bias parameters are given in \refsec{PBSpeaks} (see also \reffig{pkbias}).  
The 5 additional terms in the last line are obtained upon replacing 
each occurrence of $(\sigma_0, b_{10}^L, \nu(\vq_2))$ by $(\sigma_2, b_{01}^L, J_1(\vq_2))$.  
Restricting to local PNG, the leading-order non-Gaussian contribution to the Lagrangian peak power spectrum in the low-$k$ limit reads
\cite{desjacques/gong/riotto:2013}
\begin{equation}
\Delta P_\text{pk}^L(k) = 2 \fnl \Mm_R^{-1}(k) b_\phi \left(b_{10}^L+b_{01}^L k^2\right)W_R^2(k) \Plin(k) \;,
\end{equation}
where
\begin{equation}
b_\phi = 2 \biggl(\sigma_0^2 b_{20}^L+2\sigma_1^2 b_{11}^L+\sigma_2^2 b_{02}^L +2\sigma_1^2\chi_1^L+2\sigma_2^2\omega_{10}^L\biggr) \, .
\label{eq:DPkNGpk}
\end{equation}
Note that, with $c_2^L(\vk_1,\vk_2)$ given by \refeq{c2pk} in \refsec{pertpkcorr} for BBKS peaks, $b_\phi$ can also be computed from \refeq{genericLagNGbias}. 
Density peaks follow a universal mass function, albeit of a generalized form, since their abundance depends on several spectral moments 
$\s_i(R)$. Thus, we would not expect them to obey \refeq{b01univ}.  However, the general PBS prediction \refeq{bphiPBS} should apply. For BBKS peaks, the
halo mass function $\bnpk(M)$ is given by \refeq{peakmf}, with $\fpk$ being a function of both $\nu_c$ and the root-mean-square values $\sigma_i$. 
One can show that
\begin{equation}
\label{eq:bngBBKS}
b_\phi = \frac{\partial\ln\bnpk}{\partial\ln \eps} = 2 \sum_{i=0}^2 \frac{\partial\ln\bnpk}{\partial\ln\sigma_i}
\end{equation}
indeed yields \refeq{DPkNGpk} \cite{desjacques/gong/riotto:2013}.  
The physical interpretation is the same as in \refsec{PBSbiasNG}: in the presence of local quadratic PNG, a 
long-wavelength background perturbation of wavenumber $k_\ell$ rescales the amplitude of the power spectrum $\Plin(k_S)$ 
of the linear density field in a scale-independent manner, 
\begin{equation}
\Plin(k_S) ~\to~  \left(1+4\epsilon \right)\Plin(k_S) \;.
\label{eq:pstransform}
\end{equation}
Thus, for local quadratic PNG, all the spectral moments are rescaled proportionally, $\sigma_i\to (1+2\epsilon)\sigma_i$, 
so that the parameters $\gamma_1$ and $R_\star$ remain unchanged.  
On setting $\epsilon \equiv \fnl \Mm^{-1}(k)$, we recover the full non-Gaussian $k$-dependent correction to the linear halo bias.  
For a generic PNG, $\epsilon$ should be replaced by $a_0(k_\ell/k_S)^\alpha \Mm^{-1}(k)$, where $\alpha$ is the scaling of the 
leading contribution to the primordial bispectrum in the squeezed limit [see \refeq{Bphibeyondsq}
and \refeq{Plinmodgen}].  In this case, the different moments that $\bnpk$ depends on scale differently in \refeq{bngBBKS}, and 
the amplitude of the non-Gaussian bias of peaks differs from that obtained for a universal mass function, \refeq{bpsiuniv}.   

Note that the equivalence of $b_\phi$ computed from \refeq{bngBBKS} on the one hand, and \refeq{DPkNGpk} on the other, only holds for a 
deterministic barrier \cite{desjacques/gong/riotto:2013}. For a fuzzy moving barrier with a phenomenological 
description of the scatter (cf. \refsec{gen_barrier}) as in current excursion-set peak implementations  
(cf. \refsec{esp}), $b_\phi$ does not consistently recover $\partial\ln\overline{n}/\partial\ln \eps$ \cite{biagetti/desjacques:2015}. 
As shown in \refsec{NGsim}, this is at odds with measurements from N-body simulations. 

This problem can be solved with a ``microscopic'' model of the scatter in the collapse barrier $B$ (see \refsec{gen_barrier}), 
such as the one proposed in \cite{castorina/paranjape/etal:2016} for instance.
In this case, the renormalized Lagrangian bias functions $c_n^L$, \refeq{iPTcnL}, acquire an explicit dependence on the shear 
$K_{ij}$. It can then be shown that, since the collapse barrier $B$ now is independent of $\sigma_8$, the non-Gaussian bias 
amplitude $b_\phi$, \refeq{DPkNGpk}, recovers the peak-background split expectation $\partial\ln\overline{n}/\partial\ln\epsilon$
\cite{desjacques/jeong/schmidt:2017}. 
In practice, if all the scatter in the barrier $B$ arises from the tidal shear, then $b_\phi$ includes two additional contributions 
proportional to $b_{K_2}^L$ and $b_{Q_2}^L$, where $K_2$ is defined in \refeq{K2K3def}, and $Q_2 \propto \tr\big(K^{ij}\zeta_{ij}\big)$. 
Therefore, the consistency relation \refeq{bphiPBS} can be satisfied without violating the constraints on $b_{K_2}^L$ (see \refsec{meas:meas}),
provided that $b_{Q_2}^L$ is non-zero.\\

To summarize the results of \refsecs{NGthresh}{NGpeaks}, in the ``microscopic'' Lagrangian bias framework, the 
fundamental reason why the standard Lagrangian \LIMD model 
$\delta_h(\vx)=b_1^L[\delta](\vq)+b_2^L[\delta^2](\vq)/2+\cdots$ fails at reproducing the correct amplitude of 
the scale-dependent bias induced by PNG is the fact that, unlike the discrete density peaks and the thresholded 
regions considered here, it does not include variables other than the density, especially the filter derivative 
$\mu = -d\d_R/dR$ which ensures the first-crossing condition. 
In contrast to the approach described in \refsec{NGevolution}, the perturbative Lagrangian bias expansions 
considered here remain identical regardless of the statistical properties of the initial conditions, since the 
proto-halo number density is constructed directly from the Lagrangian matter density field using a model that is assumed valid 
on all scales.  The coefficient $b_\phi$ of the scale-dependent contribution from PNG is given by a sum of generalized
second-order bias parameters, which must ensure that the PBS scaling \refeq{bphiPBS} is recovered.  
This condition is not trivially enforced for a microscopic Lagrangian bias model such as ESP where the scatter 
in collapse threshold is only modeled in a phenomenological way, so that the model does not provide a complete description of the local relation between proto-halo density and the linear density field.  
As an alternative to ESP peaks in which the halo mass function is predicted from first principles, one could 
generalize the Press-Schechter number density \refeq{nexset} to \citep{matsubara:2012} 
\begin{equation}
\label{eq:exsetfree}
n_{\Sigma\text{PS}}(\vq) = -2\rhob\frac{\partial}{\partial M}\Sigma\big[\delta_R(\vq)-\dc\big]\;.
\end{equation}
The free function $\Sigma$ can be chosen such that the predicted halo mass function reproduces the simulated 
one \cite{matsubara/desjacques:2016}. Here again, one obtains a bivariate bias expansion $b_{ij}^L$, reflecting the 
dependence on both $\delta_R$ and $\mu_R$.  In this model, the quantity $(\partial_i\d)^2$ does not appear in the bias expansion, due to the absence of the peak constraint.  For this reason, the renormalized bias functions 
$c_n^L(\vk_1,\dots,\vk_n)$ predicted by this model depend only on the wavenumbers $k_i=|\vk_i|$. Therefore, a measurement of the scale- and configuration-dependence of the halo bispectrum 
would help distinguish between the predictions of different Lagrangian bias schemes.

\subsection{Non-Gaussian halo bias in simulations}
\label{sec:NGsim}

\begin{figure}[t]
\centering
\includegraphics[width=0.49\textwidth]{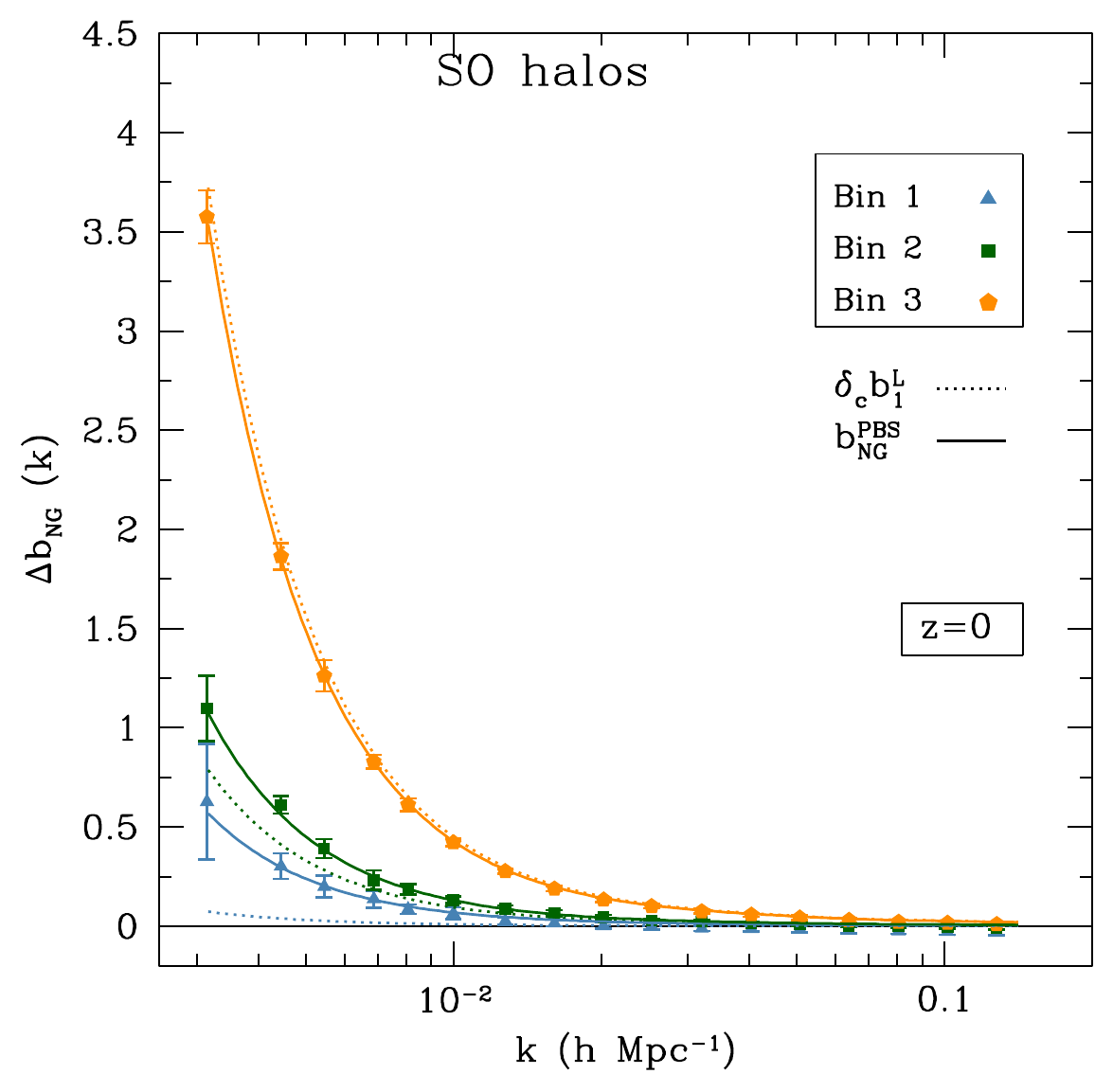}
\includegraphics[width=0.49\textwidth]{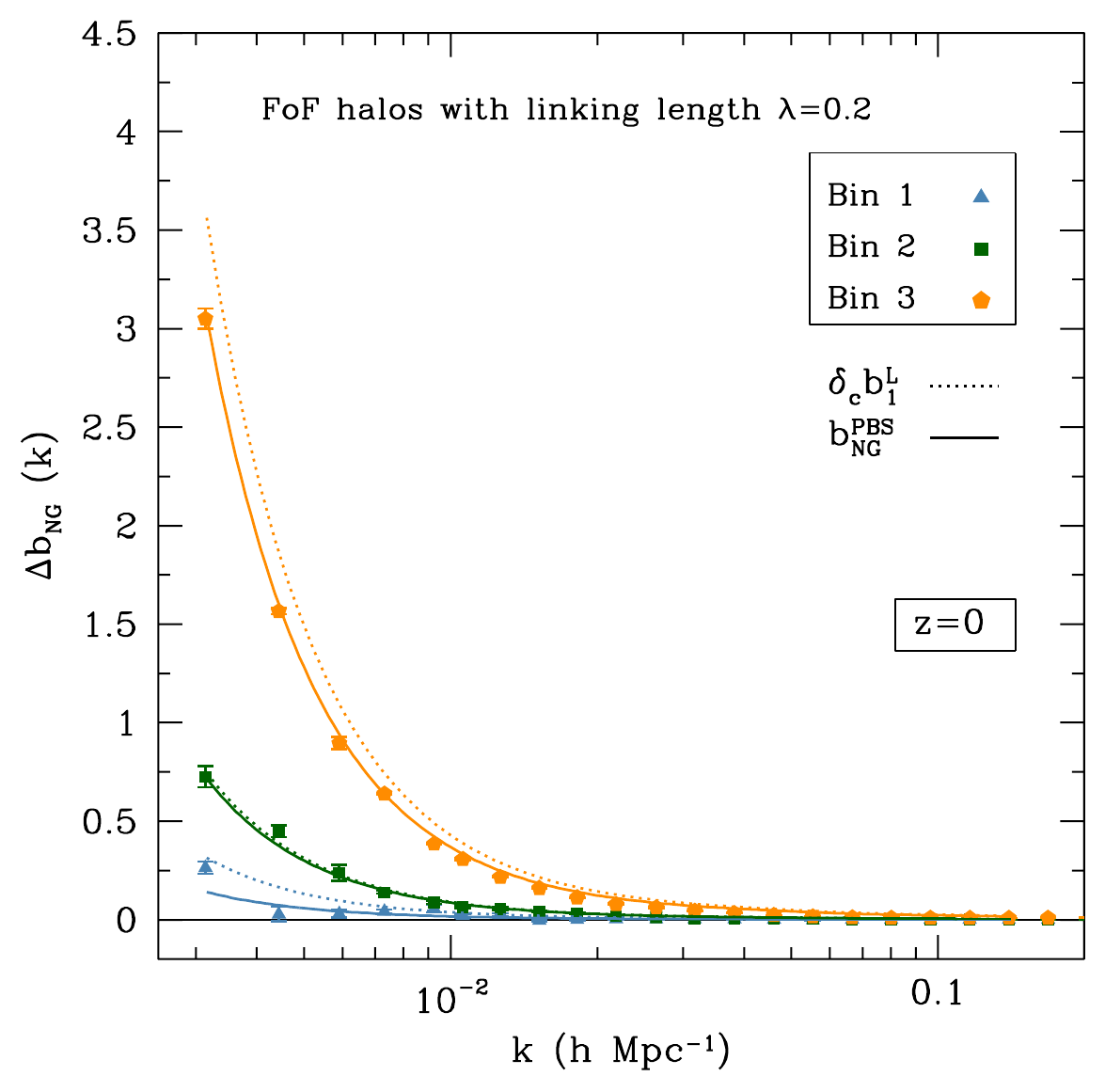}
\caption{Scale-dependent bias $\DeltabNG(k) \simeq P_{hm}(k)/P_{mm}(k) - b_1$ 
measured for dark matter halos in N-body
simulations with local-type PNG with $\fnl = \pm 250$ (points).  The \emph{left panel} 
shows the result for halos identified using the spherical overdensity method ($\Delta_\text{SO}=200$), while the \emph{right panel} 
presents measurements for friends-of-friends halos identified with a linking length $\lambda=0.2$.  
The solid lines show the general peak-background split prediction \refeq{bphiPBS}, while the dotted lines 
represent the PBS evaluated for universal mass functions, \refeq{b01univ}, using the linear bias $b_1$ 
measured for the same mass bins in simulations with $\fnl=0$.  The mass bins, in units of $10^{13} \Msunh$, are given by
Bin 1: $[0.9,1.4]$;
Bin 2: $[1.4,2.7]$;
Bin 3: $[2.7,\infty)$. 
\figsource{biagetti/etal:2016}  
\label{fig:bNG_vs_k}}
\end{figure}

A number of studies have tested the theoretical prediction for the 
scale-dependent bias induced by local quadratic PNG against the result of 
large N-body simulations, including 
\cite{dalal/etal:2008,desjacques/seljak/iliev:2009,grossi/etal:2009,pillepich/porciani/hahn:2010,giannantonio/porciani:2010,nishimichi/etal:2010,scoccimarro/hui/etal:2012,baldauf/etal:2015} (see \reffig{bNG_vs_k}).  
For the purpose of an accurate comparison between the prediction \refeqs{corrNGsqgen2}{Deltab1def} and simulations, one should also include two subtle 
corrections \citep{desjacques/seljak/iliev:2009,giannantonio/porciani:2010}.  
They arise because both the mean halo number density $\avnh(M,\tau)$ and the 
matter power spectrum $\Plin(k,\tau)$ are modified by PNG.
Firstly, the change in the mean number density (mass function) of halos in the presence of PNG induces a (scale-independent) shift $b_1 \to b_1(\fnl)$ of the linear bias, which is negative for massive halos if $\fnl > 0$, as massive halos become more common 
for $\fnl > 0$ compared to the $\fnl=0$ case, and hence less biased.  
Secondly, one should also take into account the loop contribution $\propto\fnl$ to the matter power spectrum \cite{assassi/etal:2015} if measurements extend into the mildly nonlinear regime.  

\begin{figure}[t]
\centering
\includegraphics[width=0.49\textwidth]{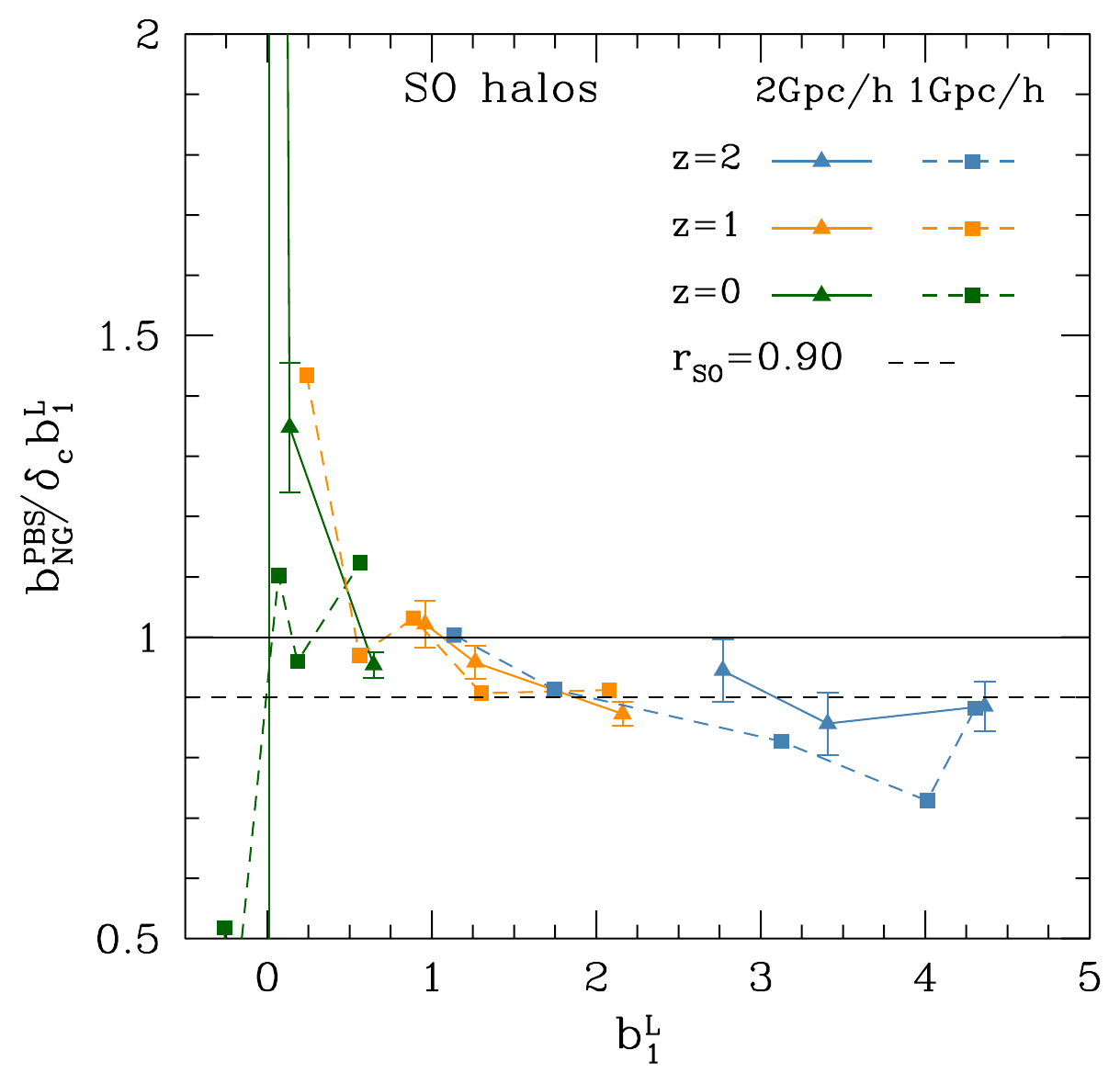}
\includegraphics[width=0.49\textwidth]{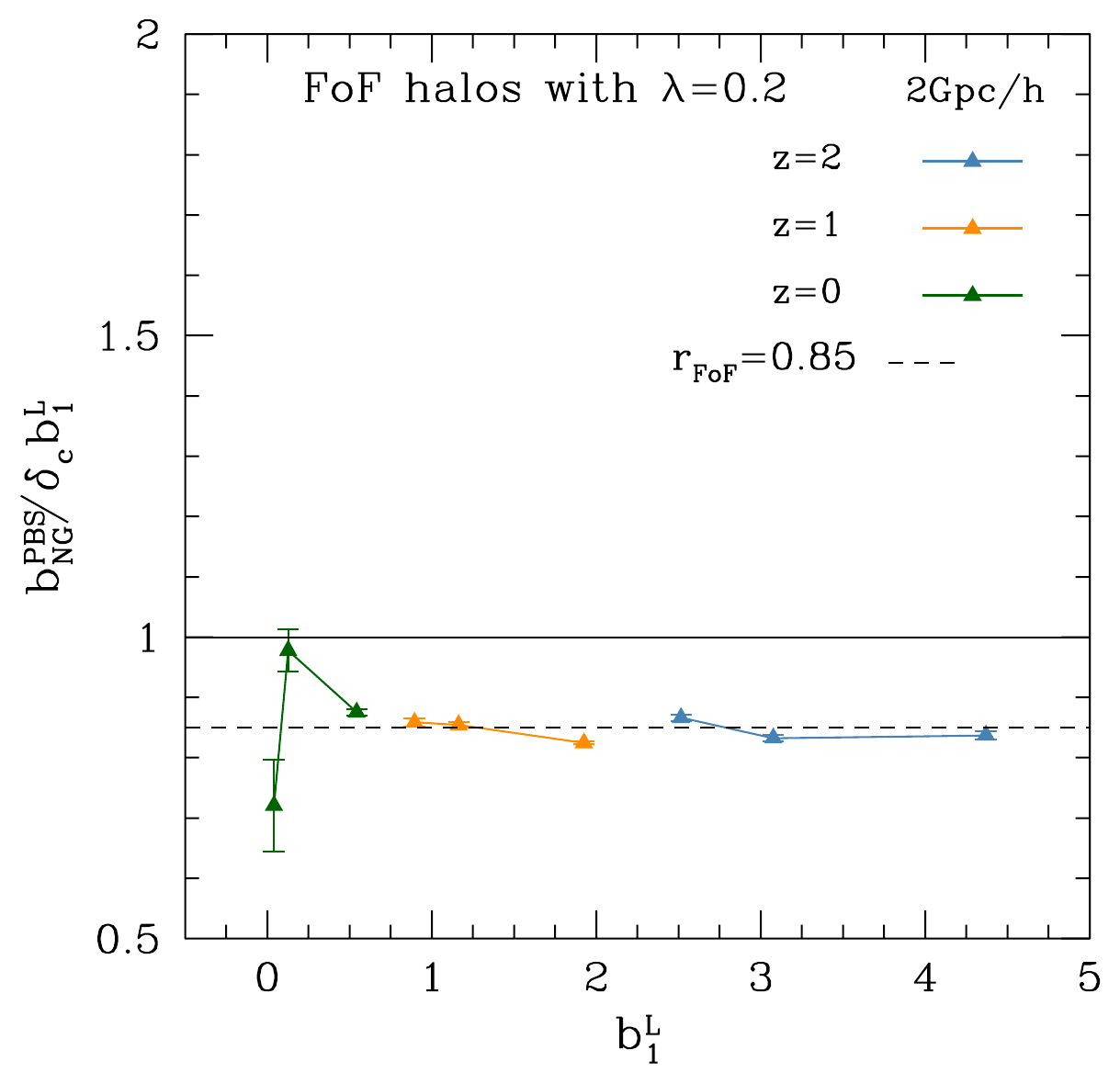}
\caption{Measured amplitude of the scale-dependent bias of halos $b_{\rm NG} \equiv b_\phi/2$ [through \refeq{bphiPBS}], 
divided by the universal mass function prediction $b_1^L \dc$ [\refeq{b01univ}], shown as function of $b_1^L$.  The 
results are from different output redshifts and different simulation box sizes, as labeled.  The \emph{left panel} 
displays the measurements for halos identified using the spherical overdensity method ($\Delta_\text{SO}=200$), while the \emph{right panel} 
shows results for friends-of-friends halos with linking length $\lambda=0.2$.  
\figsource{biagetti/etal:2016}  
\label{fig:bNG_vs_b}}
\end{figure}

The most recent results are from \cite{biagetti/etal:2016}, who measured the scale-dependent bias 
$\DeltabNG(k) = \fnl b_\phi \Mm^{-1}(k)$ for local PNG with $\fnl=250$ for different halo finders, taking into account the corrections discussed above.
\reffig{bNG_vs_k} shows their measurement for SO and FoF halos (see \refapp{halofinder}). The solid lines are the prediction of the general PBS
from \refeq{bphiPBS}, which was implemented as in \cite{baldauf/etal:2015} by performing simulations with
different initial power spectrum amplitudes $\mathcal{A}_s$, and taking the derivative
of the measured halo mass function with respect to $\ln\mathcal{A}_s$.  
The result is in excellent agreement with the data for all mass bins and halo finders. 
By contrast, the dotted curves, which assume $b_\phi = 2 \dc b_1^L$, with $\dc=1.687$, as obtained for a universal 
mass function [\refeq{b01univ}], are not in good agreement with the simulation results. 
This is more apparent in \reffig{bNG_vs_b}, which displays the relative deviation of $b_\phi$, measured using \refeq{bphiPBS}, 
from the universal mass function prediction. Clearly, the universal mass function result overpredicts $b_\phi$ for rare halos with $b_1^L \gtrsim 1$
for both halo finders. For SO halos, $2\dc b_1^L$ leads to an underestimate at lower masses, in the range $0\lesssim b_1^L\lesssim 0.5$.  
These results confirm the previous findings of 
\cite{desjacques/seljak/iliev:2009,hamaus/seljak/desjacques:2011}
who studied SO halos, and
\cite{pillepich/porciani/hahn:2010,grossi/etal:2009,scoccimarro/hui/etal:2012,baldauf/etal:2015}
who used FoF halos.  
Ref.~\cite{scoccimarro/hui/etal:2012} express the derivative of $\avnh$ with respect to $\sigma_8$ as a derivative of the mass function with respect to halo mass.  While this is motivated by the excursion-set formalism, the exact PBS prediction for $b_\phi$ strictly corresponds to a derivative with respect to the primordial amplitude of fluctuations, as we have seen.

At high mass, the discrepancy can be interpreted as due to the particular choice of $\dc=1.687$, a value motivated by the spherical collapse
approximation (see \refsec{sph_collapse}).  
For instance, there is some line of evidence that FoF halos with a linking length $\lambda=0.2$ correspond to smoothed Lagrangian overdensities of less than $1.687$ 
\citep[e.g.,][]{more/etal:2011}, which would explain why $2\dc b_1^L$ with $\dc=1.687$ overestimates $b_\phi$ at high mass.  
After all, as explained in \refsec{sph_collapse}, there is some freedom in the choice of $\dc$ because actual dark matter halos never precisely correspond to the collapse of an isolated spherical perturbation.
Therefore, in order to account for the dependence on the halo finding algorithm, it is sensible to make the replacement
\begin{equation}
\dc \to \sqrt{q} \dc \;,
\label{eq:qfudge}
\end{equation}
where the parameter $q$ encodes information about the departure between the halos considered and the spherical collapse approximation.  
Refs.~\cite{desjacques/seljak/iliev:2009,grossi/etal:2009,pillepich/porciani/hahn:2010,wagner/verde:2012} have used different halo finding prescriptions and, 
therefore, found different values of $q$, as reviewed in \cite{desjacques/seljak:2010c}.  
\reffig{bNG_vs_b} shows that, for FoF halos with linking length $\lambda=0.2$, $\sqrt{q} \equiv r_\text{FoF} \approx 0.85$ whereas, for the SO
halos, $\sqrt{q}\equiv r_\text{SO}\approx 0.9$.

There are two further important points to take away from \reffigs{bNG_vs_k}{bNG_vs_b}~:
\begin{itemize}
\item The departure from $2\dc b_1^L$ observed for SO halos at low mass 
cannot be absorbed by a change in the overdensity criterion used in the definition of SO halo masses (here, $\Delta_\text{SO}=200$ with respect to 
matter).  This is because such a change would affect the results even more strongly at high mass, where the mass function is steep.  
Therefore, the departure from universality observed here is unrelated to the effect discussed in \cite{tinker/etal:2008}, which is
induced by their particular choice of $\dc(z)$ as pointed out by \cite{diemer/more/kravtsov:2013,despali/etal:2015}.  
Another possible explanation is the failure of the spherical collapse approximation at low mass, which we have assumed to compute $2\dc b_1^L$.  
In this case, one would replace the critical threshold $\dc$ for spherical collapse via \refeq{qfudge} by, for example, the corresponding value in the 
ellipsoidal collapse \cite{afshordi/tolley:2008}.  
Clearly, this can only explain part of the deviation, since 
we see significant evidence that $b_\phi$ changes sign at a different mass than that corresponding to $b_1^L=0$, which cannot be explained by a 
change of $\dc$.
\item The non-Gaussian bias prediction of current excursion-set peak (ESP) implementations is inconsistent with the simulation data. 
In this approach (see \refsec{esp} and \refsec{NGpeaks}
for details), the amplitude of the non-Gaussian bias is a weighted sum of all the second-order
bias parameters \cite{desjacques/gong/riotto:2013}. This generally holds for 
any ``microscopic'' Lagrangian bias models \cite{matsubara:2012}, 
in contrast to models which perform a large-scale bias expansion (\refsec{NGevolution}; \cite{PBSpaper}).  
However, while, when adopting a deterministic barrier, the ESP approach predicts a value of $b_\phi$ that matches the PBS prediction \refeq{bphiPBS} 
\cite{desjacques/gong/riotto:2013}, in 
agreement with the data, the stochastic barrier of \cite{paranjape/sheth:2012} yields a value of $b_\phi$ that is greater than that obtained from 
\refeq{bphiPBS} \cite{biagetti/desjacques:2015}, which is clearly ruled out by the measurements in \reffig{bNG_vs_b}. 
To remedy this issue, the scatter around the mean barrier must be described at a ``microscopic'' level \cite{desjacques/jeong/schmidt:2017}, through the tidal shear $K_{ij}$ for 
instance, rather than through a white noise term as done in \cite{paranjape/sheth/desjacques:2013,biagetti/desjacques:2015}.
\end{itemize}

Finally, \refeq{b01univ} assumes that the clustering of halos is entirely specified by the halo mass $M$.  
This may, however, not be true for some observed tracers such as quasars whose activity may be triggered by mergers of halos. Ref.~\cite{slosar/etal:2008}
used the excursion-set formalism described in \refsec{exset} to estimate the bias correction $\Delta b_{\rm merger}$ induced by mergers, and
obtained
\begin{equation}
\label{eq:bNGmerger}
\Delta b_{\rm merger}=\dc^{-1}\;.
\end{equation}
Therefore, for a universal mass function, $\dc b_1^L$ should be replaced by $\dc b_1^L -1$. The validity of this prediction
was assessed by \cite{reid/etal:2010} using N-body simulations. On splitting the halos by the formation time identified 
using merger trees, they found a significant dependence of $b_\phi$ on the formation time in agreement with \refeq{bNGmerger}.  Such a shift (albeit of smaller magnitude) would explain the different zero-crossing of $b_\phi$ and $b_1^L$ found for SO halos (right panel of \reffig{bNG_vs_b}).  

Other numerical studies have considered the effect of PNG beyond local PNG, in 
particular the cubic coupling $\gnl\phi^3$ \citep{desjacques/seljak:2010a,smith/ferraro/loverde:2012}; the orthogonal and equilateral bispectrum 
templates \citep{wagner/verde:2012,scoccimarro/hui/etal:2012}, and multi-field inflation 
\citep{Tseliakhovich:2010kf,smith/loverde:2011}. These analyses suggest that the non-Gaussian bias of dark matter halos is 
consistent with the PBS prediction for universal mass functions, \refeq{bpsiuniv} with \refeq{b01univ}, or equivalently \refeq{DbNGthr3}.  
For these shapes however, it is crucial to take into account the second term 
on the right-hand side of \refeq{bpsiuniv} and \refeq{DbNGthr3}.  
Here, the same caveat regarding $b_\phi=\dc b_1^L$ discussed above applies to the first term in \refeq{DbNGthr3}.  

Overall, these efforts to calibrate $b_\phi$ and $b_\psi$ are important ingredients for the goal of constraining primordial non-Gaussianity using the scale-dependent bias of 
galaxies, since any systematic uncertainty on $b_\phi$ (e.g., of order 10--20\%, \reffig{bNG_vs_b}) translates into a similar uncertainty in the constraint on $\fnl$.   

\subsection{Observational prospects}
\label{sec:NGforecast}

Although the cosmic microwave background (CMB) constraints from the Planck satellite already have 
put stringent constraints on 
various PNG shapes, in particular $\fnl=0.8\pm 5$ for local 
quadratic PNG \cite{planck:2013c}, there is still room left for interesting phenomenology at the level of $|\fnl| \simeq 1$ that is unconstrained by current CMB limits.  
By making use of the scale-dependent features induced by PNG in the galaxy power spectrum and bispectrum (\refsec{NGsummary}), future surveys of the large-scale structure (LSS) of the Universe, 
such as DESI, Euclid, LSST, and others, are one of our best hopes for improving the current CMB limits on PNG.  
Most of the signal-to-noise from the scale-dependent bias in the power spectrum comes
from the largest scales accessible to a given survey.  While the theoretical
model for galaxy clustering on those scales is very robust, as linear theory
is sufficient (although including projection effects is important, as described in \refsec{GRdelta}), observational systematics can provide difficult obstacles.  
This is because the intrinsic signal is small on those scales,
while the apparent galaxy clustering induced by systematic effects 
(for example, fluctuations of the 
survey depth across the sky, stellar contamination, and photometric
calibration uncertainties) is largest on large scales.  These effects
can mimic the signature of a non-zero $\fnl$ and must be carefully
controlled for \cite{ross/percival/etal:2013,leistedt/peiris:2014,ho/agarwal/etal:2015}.  On the other hand,
the signatures of PNG in the galaxy three-point function (bispectrum) are present on
smaller scales, and the latter can thus provide important cross-checks.  

Current limits on $\fnl$ from LSS surveys, which are all based on power spectrum measurements only, are at the level of CMB pre-Planck constraints. For instance, we have (at 68\% C.L.)
\begin{align}
&-16 < \fnl < +26\,, \quad \mbox{from galaxies \cite{Giannantonio:2013uqa}, and} \\
&-39 < \fnl < +23\,, \quad \mbox{from quasars \cite{Leistedt:2014zqa}}\;.
\end{align}
Forecasts indicate that the statistical errors on $\fnl$ should decrease by $1-2$ orders of magnitude with the next generation of large redshift surveys
\cite{carbone/mena/verde:2010,Agarwal:2013qta,dePutter:2014lna,Raccanelli:2014kga,Camera:2014bwa,SPHEREx:2014,Alonso:2015sfa,Raccanelli:2015oma}.  We will discuss quantitative forecasts 
for constraints on local $\fnl$ achievable with future surveys
in the remainder of this section, focusing on the combination of 
power spectrum and bispectrum for a single tracer, and multi-tracer techniques.  
We will adopt an idealized setting as in \refsec{npttree_fisher}, neglecting
redshift-space distortions and assuming a trivial survey window function.

\subsubsection{Galaxy power spectrum and bispectrum for a single tracer}
\label{sec:fnl_fisher}

In this section, we forecast constraints on local $\fnl$ from future
surveys using the galaxy power spectrum and bispectrum.  For this, 
we use the Fisher information matrix as described in detail in \refsec{npttree_fisher} 
[see \refeqs{Fisherij}{FisherBij}].  
We use the same assumptions and survey parameters as in 
\refsec{npttree_fisher}, and calculate the projected uncertainties under the 
null hypothesis $\fnl=0$; hence, the uncertainties of power spectrum and bispectrum measurements are still given by \refeq{DeltaPk} and \refeq{DeltaBk}, respectively.  
We then expand the parameter vector considered in \refsec{npttree_fisher} 
to include $\fnl$,
\be
\vec{\theta}=\Big\{b_1,\  b_2,\  b_{K^2},\  \ln {\cal A},\  \Peps,\   
\Pepsepsd,\  \Beps,\   \fnl\  \Big\}\,,
\ee
using the same fiducial values as given in \reftab{Fisher_bias}.  We
include the effects of $\fnl$ at second order in the general bias expansion, 
as summarized in the relations of \refsec{NGsummary}, restricting to terms
that are linear in $\fnl$.  This is sufficient for a fiducial value of 
$\fnl=0$, since the Fisher matrix only involves first derivatives with respect to parameters, and
the contribution of any higher-order term to 
$\partial P_{gg}/\partial\fnl$ and $\partial B_{ggg}/\partial\fnl$ vanishes
upon evaluation at $\fnl=0$.  Then, the power spectrum is given by
\be
P_{gg}(k)
=
P_{gg}^{(G)}(k)
+2b_1\DeltabNG(k)\Plin(k)\,,
\label{eq:Pk_localfnl}
\ee
where $P_{gg}^{(G)}(k)$ is the leading galaxy power spectrum with Gaussian initial conditions [\refeq{Phh}].  Further, using the notation of \refsec{npttree}, the galaxy bispectrum is given by
\ba
B_{ggg}&(k_1,k_2,k_3) = B_{ggg}^{(G)}(k_1,k_2,k_3)
\vs
& +
2 b_1^3 \fnl\Mm(k_1)\Mm(k_2)\Mm(k_3)
\left[
P_\phi(k_1)P_\phi(k_2) 
+
\perm{2}
\right]
\vs
&+
\Bigg\{
b_1^2
\Plin(k_1)\Plin(k_2) \vs
&
\quad\quad\times
\left[
\frac{(2\dc-1) (b_1-1) + \dc b_2^L}{2\dc(b_1-1)}
\left(\DeltabNG(k_1) + \DeltabNG(k_2)\right)
+
\mu_{12}
\left\{
\frac{k_1}{k_2}
\DeltabNG(k_1)
+
\frac{k_2}{k_1}
\DeltabNG(k_2)
\right\}
\right] 
\vs
&\quad\quad+
b_1
\left[\DeltabNG(k_1) + \DeltabNG(k_2)\right]
\Plin(k_1)\Plin(k_2)
\left[
2F_2(\vk_1,\vk_2)b_1 + b_2 + 2b_{K^2}\left(\mu_{12}^2-\frac13\right)
\right] 
\vs
&\quad\quad+ 
2 \Pepsepsd
\left[\DeltabNG(k_1) \Plin(k_1) + \DeltabNG(k_2) \Plin(k_2) \right]\vs
& \quad\   + \perm{2} \Bigg\}\,,
\label{eq:Bk_localfnl}
\ea
where $B_{ggg}^{(G)}(k_1,k_2,k_3)$ is the galaxy bispectrum with Gaussian initial conditions [\refeq{Bhhh}].  In both expressions, we take $\DeltabNG$ from \refeq{DeltabDalal},
\be
\DeltabNG(k) =  2\fnl\dc(b_1-1)\Mm^{-1}(k)\,.
\ee
We have further inserted the universal mass function prediction for $b_{\phi\d}^E$ 
[\refeq{b11univ}], using the second-order Lagrangian bias given by \refeq{b2E}, $b_2^L = b_2 - 8/21(b_1-1)$, and set $P^{\{0\}}_{\eps\eps_\phi} \approx (b_\phi/b_1) \Pepsepsd$ as predicted when stochasticity is described by Poisson shot noise.  
For the purposes of these idealized forecasts, these approximations are sufficient.  
\begin{figure}[t!]
\centering
{\includegraphics[width=\textwidth]{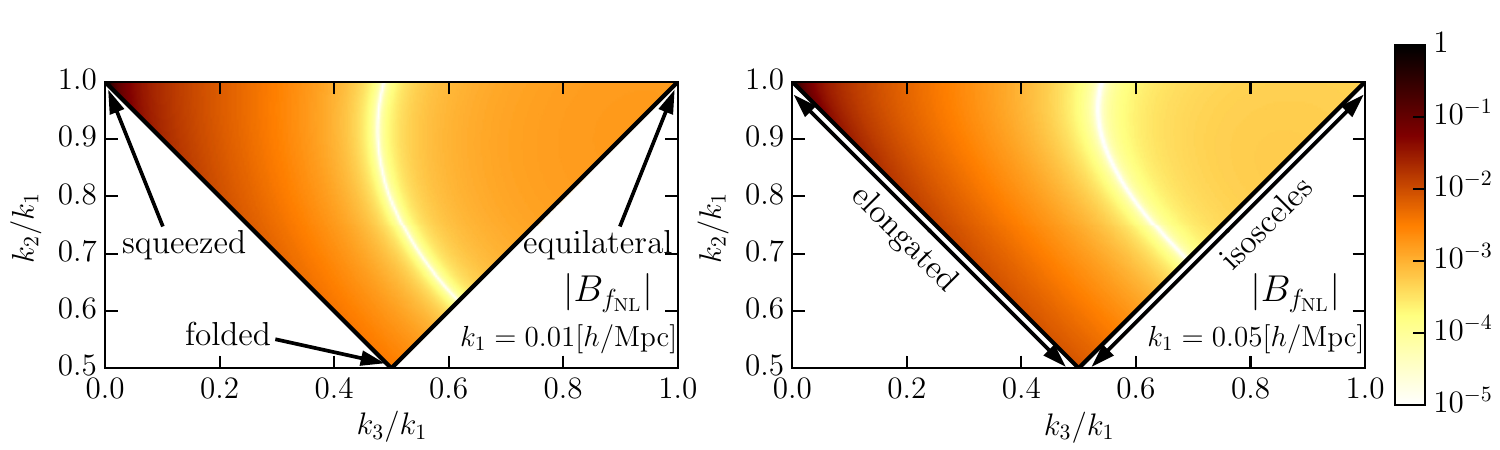}}
\caption{
Shape dependence of the contribution from local-type PNG 
to the galaxy bispectrum, i.e. all terms except for $B_{ggg}^{(G)}$ in \refeq{Bk_localfnl}.  
We calculate the shape dependence at $z=2$ with $\fnl=5$ as fiducial value.  
The remaining parameters are as listed in \reftab{fisher_fNL}.  
In order to highlight the shape dependence, we have divided the bispectrum
by the maximum value attained in each panel.  
The contribution from local-type PNG is sharply peaked 
in the squeezed limit; this is very different from the shape dependence of 
the contributions present for Gaussian initial conditions which are shown in \reffig{Bk_shape}.  
}
\label{fig:BkfNL_shape}
\end{figure}

The galaxy bispectrum in the presence of local-type PNG has been considered in several previous references \cite{jeong/komatsu:2009b,nishimichi/etal:2010,giannantonio/porciani:2010,baldauf/seljak/senatore:2011,sefusatti/crocce/desjacques:2012,tellarini/etal:2016}.  Unlike the majority of these, we have restricted to linear order 
in $\fnl$, which, as explained above, is sufficient for a Fisher forecast with fiducial value $\fnl=0$.  Further, given the constraints set by the Planck satellite, higher-order terms in $\fnl$ are unlikely to be relevant in the future.  
We have also neglected the effect of PNG on the statistics of the matter density field beyond the leading-order bispectrum.  This is consistent at the order of perturbations we are working in;  the one-loop correction to the matter power spectrum from PNG is negligible for the wavenumbers and values of $\fnl$ considered, while that to the bispectrum begins to be relevant at $k_{\rm max} \sim 0.2 \iMpch$ \cite{assassi/etal:2015}.  On the other hand, the above-mentioned references did not include the complete set of bias parameters and stochastic amplitudes that appear in the general perturbative bias expansion. Ref.~\cite{gleyzes/deputter/etal:2016} forecast the constraints obtained on $\fnl$ of the equilateral ($\alpha=2$) and quasi-single-field ($0 \leq \alpha \leq 3/2$) types from the power spectrum alone, including the NLO correction as well as the relevant higher-derivative terms. Further, Ref.~\cite{moradinezhad/dvorkin:2017} forecast the detectability of higher-spin fields in the galaxy power spectrum through the anisotropic squeezed-limit primordial bispectrum.
The latter can also be constrained using galaxy shape correlations, as forecasted in \cite{schmidt/chisari/dvorkin,chisari/etal:2016}. 

\reffig{BkfNL_shape} shows the configuration dependence of the contributions
to the galaxy bispectrum from PNG, in the same form as \reffig{Bk_shape} in 
\refsec{bnpt}.  These contributions peak for squeezed triangle configurations,
showing a distinct difference to the Gaussian contributions shown in
\reffig{Bk_shape}.  For this reason, we expect the bispectrum to help significantly
in improving constraints on $\fnl$ from galaxy clustering.  

The result of the Fisher matrix calculation is summarized in 
\reftab{fisher_fNL} for the seven galaxy surveys that we consider here:
HETDEX~\cite{HETDEX}, eBOSS~\cite{eBOSS}, DESI~\cite{DESI}, PFS~\cite{PFS}, 
Euclid~\cite{amendola/etal:2013} and WFIRST~\cite{WFIRST}.  We use
the same fiducial values for the bias and stochastic parameters as
in \reftab{Fisher_bias}, and show results for $k_{\rm max}=0.1\iMpch$
and $k_{\rm max}=0.2\iMpch$. For all surveys, combining the 
galaxy power spectrum and bispectrum gives a factor of several better constraints on $\fnl$ than the power spectrum alone.
Moreover, when using the galaxy bispectrum, future galaxy surveys are 
expected to improve on the CMB constraints on $\fnl$. 
Comparing the different choices for $k_{\rm max}$, we see that, while the 
power-spectrum-only constraints
show only incremental changes, the combined constraints improve by about a 
factor of two for the higher value of $k_{\rm max}$. This is because the signatures of PNG in the galaxy 
power spectrum are only prominent on large scales (see \reffig{bNG_vs_k}), 
while those in the galaxy bispectrum are present for triangles in the
squeezed configuration on \emph{all} scales (\reffig{BkfNL_shape}). 
Further, the forecasted error on $\fnl$ depends 
sensitively on the fiducial values of the bias parameters. This is because 
both the signal from PNG ($\Delta b_1\propto (b_1-1)$) and the 
signal-to-noise ratio per $k$-mode ($\bar{n} P_g(k)\propto b_1^2$) change
favorably for highly biased tracers (when the number density is fixed). 
For example, we find that the projected uncertainties shown in 
\reftab{fisher_fNL} are improved by a factor of $2$ to $3$ 
when assuming $b_1=2$ instead of $b_1=1.5$. In this case, 
both DESI and Euclid reach $\sigma_{\fnl}<1$ for 
$k_{\rm max}=0.2~h/{\rm Mpc}$. 
Note that previous forecasts in the literature \cite{giannantonio/etal:2012,CITA-NG:2014,dePutter/Dore:2014,raccanelli/etal:2015,tellarini/etal:2016} 
fixed the stochastic parameters, if included at all, as well as the primordial power spectrum amplitude to their respective fiducial values,
leading to more optimistic forecasts than those shown in \reftab{fisher_fNL}.  

\begin{table}[t]
\centering
\caption{Projected constraints on the amplitude of local-type 
PNG, $\fnl$, from the power spectrum and bispectrum of a single galaxy sample. The power-spectrum-only constraints are obtained after 
marginalizing over $b_1$ and $\Peps$, while the combined constraints are obtained
after marginalizing over 7 parameters: $\{$ $b_1$, $b_2$, $b_{K^2}$, 
$\Peps$, $\Pepsepsd$, $\Beps$, $\ln {\cal A}$ $\}$.  
For all cases, we use the following fiducial values:
$\fnl=0$; $b_1=1.5$; 
 $b_2(b_1)$ from the fitting formula in \reftab{fittingbiases} 
($b_2 \simeq -0.69$); and $b_{K^2}(b_1)$ by assuming Lagrangian \LIMD bias
($b_{K^2}=-2/7(b_1-1)\simeq -0.14$).  
Further, we adopt $\Pepsepsd=b_1/(2\avng)$ and 
$b_2^L = b_2 - 8/21(b_1-1) \simeq -0.88$ as fiducial values.
Note that $\sigma_{\fnl}$ depends sensitively on the fiducial value
of $b_1$ (see text).
}
\label{tab:fisher_fNL}
\begin{tabular}{l|ccc|cc|cc}
\hline\hline
\multirow{2}{*}{survey} & 
\multirow{2}{*}{\begin{tabular}[c]{@{}c@{}}redshift\\{\small $\bar z$}\end{tabular}} & 
\multirow{2}{*}{\begin{tabular}[c]{@{}c@{}}Volume\\ {\small $\left[h^{-3} {\rm Gpc}^3\right]$}\end{tabular}} & 
\multirow{2}{*}{\begin{tabular}[c]{@{}c@{}}$10^4\,\avng$\\ {\small $\left[h^{-3}{\rm Mpc}^3\right]$} \end{tabular}} & 
\multicolumn{2}{c|}{$\sigma_{\fnl}$, $k_{\rm max}=0.1\,h/{\rm Mpc}$} & 
\multicolumn{2}{c}{$\sigma_{\fnl}$, $k_{\rm max}=0.2\,h/{\rm Mpc}$} 
\\ \cline{5-8} 
              &      &      &         &
{\small $P(k)$}  & 
{\small $P(k)+B(k_1,k_2,k_3)$}  & 
{\small $P(k)$}  & 
{\small $P(k)+B(k_1,k_2,k_3)$}
\\ \hline 
eBOSS (LRG)   & 0.8  & 6.1  & 4.4     & 19     & 15      & 18    & 7.5 \\
eBOSS (QSO)   & 1.4  & 39   & 1.5     & 7.6    & 6.5     & 7.2   & 3.9 \\ 
HETDEX        & 2.7  & 2.7  & 3.6     & 24     & 22      & 22    & 14 \\
PFS           & 1.5  & 8.7  & 4.6     & 13     & 11      & 12    & 6.4 \\
DESI          & 1.1  & 40   & 3.3     & 6.1    & 5.1     & 5.7   & 2.9 \\
WFIRST        & 1.9  & 13   & 12      & 7.8    & 6.9     & 7.3   & 4.2\\
Euclid        & 1.4  & 63   & 5.2     & 4.0    & 3.4     & 3.8   & 2.0\\
\hline \hline
\end{tabular}
\end{table}

\subsubsection{Multi-tracer methods}
\label{sec:NG:mt}

In the previous section, we have derived the optimal constraints from
the galaxy power spectrum and bispectrum assuming a single tracer.  
Interestingly, a significant reduction of the statistical error on the 
scale-dependent bias, and thus the constraint on $\fnl$, from the galaxy power
spectrum alone can in principle 
be achieved with multi-tracer techniques and optimal weights 
\citep{seljak:2009,slosar:2009,mcdonald/seljak:2009,seljak/hamaus/desjacques:2009,cunha/huterer/dore:2010,gil-marin/etal:2010}.
These make use of the fact that the scale-dependent
signature in the \emph{relative} clustering of different tracers does
not suffer from cosmic variance.  Recently, Ref.~\cite{chisari/etal:2016}
applied the same technique to galaxy shape correlations, which constrain
anisotropic PNG as discussed in \refsec{nonloc}.  
Moreover, the variance due to shot noise
of tracers can be reduced by applying suitable weights.

The core of this method works as follows. First, the observed galaxy 
distribution is split into a number of subsamples with different bias 
properties. These subsamples, then, can be weighted differently to take 
advantage of the shot-noise suppression for massive halos,
which is seen in the halo clustering measured from N-body simulations
(see \refsec{meas:stoch}).  
In the limit of infinitely fine binning, and vanishing minimum halo mass,
a weight $\omega(M)=M$ proportional to the host 
halo mass minimizes the stochasticity of halos relative to the dark matter distribution 
and, thus, reduces the shot noise contribution \citep{hamaus/seljak/etal:2010}. 
Note that it is essential, though not always sufficient, that both the shot noise 
and cosmic variance suppression be considered at the same time, otherwise the 
net improvements on $\sigma_{\fnl}$, the 68\% confidence level uncertainty on $\fnl$, are small
\citep{gil-marin/etal:2010}.  
Given predictions for $b_\phi$, taken from the PBS applied to universal mass functions [\refeq{b01univ}]
and the covariance matrix of the shot noise among the different subsamples [estimated through \refeq{Pepshamaus}], 
one can construct the optimal weighting function as well as estimate the projected uncertainty on $\fnl$
\citep{hamaus/seljak/desjacques:2011}, again assuming a Gaussian halo density field which is accurate on
the large scales relevant for $\fnl$ constraints. 

\begin{figure*}[t]
\centering
\resizebox{10cm}{!}{\includegraphics{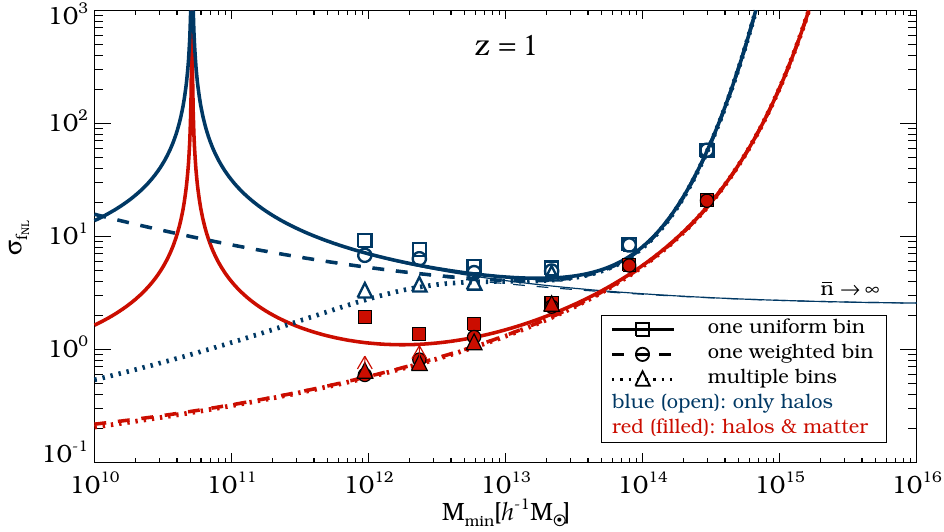}}
\caption{68\% confidence-level uncertainty on $\fnl$ for a survey centered at $z=1$ as a function 
of the minimum halo mass resolved by the survey. 
An effective survey volume $V_{\mathrm{eff}}\simeq50h^{-3}\mathrm{Gpc}^3$ (roughly comparable to Euclid) is assumed, 
taking into account all modes from $k_{\rm min}=0.0039$ to $k_{\rm max}=0.032\hmmpc$.
Note that the value of $k_{\rm max}$ is not important numerically.  
Solid and dashed lines show results
from uniform- and mass-weighted halos of a single mass bin. The N-body measurements are
overlaid as squares and circles for different minimum-mass cuts. 
Forecasts which assume knowledge of both halos and dark matter are plotted in red 
(filled symbols), whereas those based on the halo distribution only are displayed in 
blue (open symbols). The dotted lines (triangles) show the result of splitting the
halo catalog into multiple mass bins. For the single mass bin, the high-sampling
limit ($\avnh\rightarrow\infty$) is over-plotted for the uniform- (thin solid line) 
and the weighted case (thin dashed line).
The arrows represent the effect of adding a log-normal scatter of $\sigma_{\ln M}=0.5$ 
to all halo masses. They are omitted in all cases where the scatter has negligible impact.
\figsource{hamaus/seljak/desjacques:2011}
}
\label{fig:fnl_forecast}
\end{figure*}

Results are shown in \reffig{fnl_forecast} for halos at redshift $z=1$. 
When the dark matter density field is available (red lines and filled symbols),
weighting the halos by their mass (red dashed lines and filled circles) 
is always superior to the conventional uniform-weighting case 
(red solid lines and filled squares), especially when considering lower-mass 
halos. In particular, the uncertainty $\sigma_{\fnl}$ 
substantially decreases with decreasing $M_{\mathrm{min}}$ in the weighted 
case, while for uniform weighting it exhibits a spike 
at $M_{\mathrm{min}}\simeq5\times10^{10}\hmsun$.  
This is because the effective linear bias $b_1$ of the weighted halo sample is
always greater than 1 for mass-weighting, but reaches 1 at the above
mentioned value of $M_{\mathrm{min}}$ for uniform weighting.  Since this implies
$b_\phi \propto b_1-1 = 0$ assuming the PBS prediction, $\sigma_{\fnl}$ becomes
infinite.  
Simulation results are over-plotted as symbols for a few values of 
$M_{\mathrm{min}}$.  The simulations match well with the analytical 
predictions based on a parametrization of the halo stochasticity covariance (see \refsec{meas:stoch}).  
The simulations yield a minimum error of $\sigma_{\fnl}\simeq0.8$ at $M_{\mathrm{min}}\simeq10^{12}\hmsun$ in the optimally weighted case with the dark 
matter available. 

The results without the dark matter are shown as blue lines and open symbols.  
Let us focus on a single bin first.  
$\sigma_{\fnl}$ exhibits a minimum at $M_{\mathrm{min}}\sim 10^{13}\hmsun$ 
with $\sigma_{\fnl}\sim 5$ for both uniform and weighted halos.  
Thus, in the single-tracer case, it is optimal for $\fnl$ constraints to only consider the 
highest-mass halos.  However, if the halo sample is divided into multiple bins, $\sigma_{\fnl}$ does continue to improve when including lower-mass halos.  

Comparing to the results for Euclid in \reftab{fisher_fNL}, which has
a similar volume and mean redshift to what is assumed in \reffig{fnl_forecast},
and taking into account that the latter shows the theoretical optimum without
marginalizing over other parameters,
we see that the combination of power spectrum and bispectrum is expected
to outperform the power spectrum with the multi-tracer method (assuming
the matter density is not available, as assumed for \reftab{fisher_fNL}).  
However,
multi-tracer methods can also be applied to the bispectrum, combining
the advantages of both methods and potentially leading to a further 
improvement in constraints on PNG.

%% file: beyondCDM.tex
\clearpage
\section{Beyond cold dark matter, cosmological constant, and General Relativity}
\label{sec:beyondCDM}

\secttoc

So far in this review, we have focused on a Universe filled with pressureless cold
matter, and with a cosmological constant. However, the real Universe also
contains a background of free-streaming neutrinos, which we now know to have
nonzero masses.  Further, baryons are subject to non-gravitational forces.
The impact of these stress-energy
components on the galaxy bias expansion is reviewed in \refsec{neutrinos}
and \refsec{baryons}, respectively.  In addition, the cosmological
constant could in fact be dynamical, for example caused by a slowly rolling
scalar field (quintessence), or caused by modifications to General Relativity.
We consider the impact of these classes of new physics in \refsec{modgrav}.

\subsection{Massive neutrinos}
\label{sec:neutrinos}

Direct measurements of neutrino oscillations based on solar, atmospheric and
reactor neutrinos \cite{PDG} have shown that the masses of at least two of the three
mass eigenstates ($m_{\nu_1}$, $m_{\nu_2}$, $m_{\nu_3}$) of neutrinos must
be non-zero.  The experimentally constrained mass gaps are
$\Delta m_{21}^2 = m_{\nu_2}^2-m_{\nu_1}^2 \simeq (8.58\,{\rm meV})^2$ and
$|\Delta m_{32}^2| = |m_{\nu_3}^2-m_{\nu_2}^2|\simeq(49.6\,{\rm meV})^2$ \cite{PDG}.
Neutrinos constitute a small but non-negligible fraction of the cosmological
energy budget, and their mass plays an important role in the growth of
large-scale structure.  The cosmic neutrino background, thermally produced during Big Bang nucleosynthesis, has a temperature of
approximately
$T_{\nu}(z) \simeq 1.95(1+z)\,{\rm K} \simeq 0.168(1+z)\,{\rm meV}$.
Thus, at high redshifts, when $T_{\nu}(z) \gg m_i$, neutrinos are fully
relativistic and do not contribute to the growth of structure in the CDM and baryon components. 
Once neutrinos transition to a non-relativistic form of matter at a redshift
$z_{\text{nr},i}$ of roughly
\begin{equation}
1+z_{\text{nr},i}\approx 1890\, \left(\frac{m_{\nu_i}}{1\,{\rm eV}}\right) \;,
\end{equation}
perturbations in the neutrino density begin to grow appreciably on scales larger
than their free-streaming scale, effectively forming another (though not cold) dark-matter component.
The neutrino free-streaming scale, the analog of a Jeans scale for collisionless matter, is of order
the distance traveled by neutrinos per Hubble time, and correspondingly shrinks as the neutrinos
become non-relativistic.  More precisely, it is given by \cite{shoji/komatsu:2010}
\be
k_{\rm fs} (m_\nu, z) = \lambda^{-1}_{\rm fs}(m_\nu, z)
= \frac{0.068}{(1+z)^2}\left(\frac{m_\nu}{0.1\,{\rm eV}}\right)
\left[
\Omn (1+z)^3 + \OLn
\right]^{1/2}\iMpch\,.
\label{eq:kfs}
\ee
Within the free-streaming scale $\lambda_{\rm fs}$, neutrino density fluctuations are damped due to their large velocity dispersion. 
On the other hand, on scales larger than $\lambda_{\rm fs}$, neutrinos effectively behave as a pressureless fluid coupled gravitationally
to CDM and baryons.  Depending on their wavelength, modes can initially be inside the free-streaming scale and exit this scale later as
$\lambda_{\rm fs}$ shrinks.

At redshifts $z\ll \text{min}_i \{ z_{\text{nr},i} \}$, the fraction of the total, non-relativistic matter energy density in the form of neutrinos is
\be
f_\nu \equiv \frac{\Omega_{\nu 0}}{\Omega_{c+b,0}+\Omega_{\nu 0}}
\simeq \frac{\sum_i m_{\nu.i}}{93.1\Omn h^2\,{\rm eV}}\;.
\ee
where the subscripts $m$ and $c+b$ correspond to the total matter and the sum
of CDM and baryons, respectively.
We can conveniently express perturbations to the total matter density
as a weighted sum of CDM+baryon and neutrino fluctuations,
\begin{equation}
\label{eq:dmwithnu}
\delta_{m}=(1-f_\nu)\delta_{c+b}+f_\nu\delta_\nu \;,
\end{equation}
where we have assumed that CDM and baryons perfectly trace each other.
This is appropriate since we are interested in large scales
(but see \refsec{baryons} for the effect of relative density and velocity perturbations between CDM and baryons). In the following, we assume that the masses of all three active neutrino species are comparable. This can easily be generalized.
On scales larger than the free-streaming scale of the neutrino species, $k \ll k_\text{fs}(m_\nu)$, neutrino density
perturbations $\d_\nu$ follow the adiabatic prediction and are proportional
to $\d_{c+b}$, with a redshift-dependent factor (assuming there are no
primordial neutrino isocurvature perturbations).
On scales much smaller than the free-streaming scale, $k\gg k_\text{fs}(m_\nu)$, neutrino perturbations are erased,
$\delta_\nu\approx 0$ so that $\delta_m \approx (1-f_\nu)\delta_{c+b}$.
In addition, the power spectrum $P_{c+b}(k)$ of the CDM+baryon component itself is also modified by massive neutrinos due to their contribution to the Poisson
equation and the change in the expansion rate.
As a result, the growth of CDM+baryon perturbations in matter domination is slowed down according to
$\delta_{c+b}\propto a^{1-\frac{3}{5}f_\nu}$
(see \citep{lesgourgues/pastor:2006} for a detailed review of the effect of massive neutrinos in cosmology).
The following approximation \cite{hu/eisenstein:1998,hu/eisenstein/tegmark:1998},
\be
\Plin(k) \stackrel{k \gg k_\text{fs}}{\longrightarrow} (1-8f_\nu) \Plin^{(m_{\nu}=0)}(k)\,,
\ee
is a good fit to the numerical data. Numerical simulations indicate that, at redshift $z=0$, nonlinear effects
enhance the suppression, to $\Delta P_m(k)/P_m(k)\simeq -10 f_\nu$ at $k\sim 1\hmmpc$, before it diminishes at
smaller scales \cite{brandbyge/hannestad:2010,bird/viel/haehnelt:2012,wagner/verde/etal:2012}.
Thus, neutrinos leave a characteristic, scale-dependent feature in the power spectrum of total matter
as well as the non-relativistic part ($c+b$)
\cite{wong:2008,saito/takada/taruya:2009,ichiki/takada:2012}.
Current CMB- and expansion-history-based constraints on the sum of neutrino masses are $\sum m_\nu \lesssim 0.6$~eV \cite{planck:2015-parameter},
which lead to a neutrino fraction $f_\nu\lesssim 0.05$, while the minimum value allowed
by neutrino oscillation measurements is $\sum m_\nu \geq 0.06$~eV
 \cite{PDG}, corresponding to $f_\nu \gtrsim 0.005$.
For a $\sum m_\nu$ in this range, the neutrino free-streaming scale is sufficiently large
that the neutrino perturbations remain in the linear regime down to redshift $z\lesssim 10$,
as has been shown numerically
\cite{villaescusa/marulli/etal:2014,castorina/sefusatti/etal:2014,costanzi/villaescusa/etal:2013}.

The effect of massive neutrinos on the clustering of halos
has been studied both using semi-analytic methods
(e.g. \cite{ringwald/wong:2004,saito/takada/taruya:2009,ichiki/takada:2012,loverde:14a,loverde:14b})
and full N-body simulations
(e.g. \cite{singh/ma:2003,viel/haehnelt/springel:2010,marulli/carbone/etal:2011,ali-haimoud/bird:2013,villaescusa/marulli/etal:2014,castorina/sefusatti/etal:2014,castorina/carbone/etal:2015,banerjee/dalal:2016,villaescusa-navarro/etal:2017,chiang/etal:2017}).
Like in the pure CDM case, analytic approaches provide useful insights into the effect of massive
neutrinos on large-scale structure. Still, detailed N-body simulations are needed to fully
capture their impact in the nonlinear regime.

A good first ansatz for semi-analytic predictions is to assume that halos form out of the CDM and baryon components only.
That is, when the peak height is defined as $\nu_c\equiv \dc/\sigma_{c+b}$, where $\sigma_{c+b}$
is the variance of CDM+baryon fluctuations computed with the appropriate {\it scale-dependent} linear growth
factor, then the halo mass function in this approximation is close to universal as in the standard $\Lambda$CDM case
\cite{castorina/sefusatti/etal:2014}.
Correspondingly, we can perform the bias expansion of the halo density field in terms of $\delta_{c+b}$, such that the halo power spectrum reads
\begin{equation}
P_{hh}^\LO(k) = [b_1^{c+b}(\nu_c)]^2 P_{c+b,{\rm L}}(k) + \Peps\,,
\label{eq:Phhnu}
\end{equation}
where we have emphasized that the linear bias $b_1^{c+b}$ is the bias with respect to the baryon+CDM components.
\refeq{Phhnu} is fairly accurate on linear scales and, moreover, $b_1^{c+b}(\nu_c)$ is a nearly universal function
of the peak significance $\dc/\sigma_{c+b}$ \cite{ichiki/takada:2012,castorina/sefusatti/etal:2014}.

However, \refeq{Phhnu} is not completely correct, since a scale-dependent
growth generically leads to a scale-dependent bias \cite{hui/parfrey:2008,parfrey/hui/sheth:2011}
(see also \refsec{modgrav}).
In particular, if halo positions are defined in Lagrangian space and follow the
density locally there, as is the case in the peak and excursion-set pictures,
then their relation to the Eulerian matter density field will be
scale dependent.  This effect was studied in detail for the massive
neutrino case in \cite{loverde:14a}, who showed that $b_1^{c+b}$ in \refeq{Phhnu}
in fact becomes
\be
b_1^{c+b}(\nu_c; k) = 1+ \frac1{\avnh}\frac{\partial \avnh}{\partial \dc}
\frac{\partial\dc}{\partial \d_{c+b,\ell}(k)}\,,
\label{eq:b1neut}
\ee
where $\d_{c+b,\ell}(k)$ is a long-wavelength single Fourier mode perturbation
in the $c+b$ fluid.  $\partial\dc/\partial \d_{c+b,\ell}(k)$ can be calculated
by following a multi-fluid spherical collapse calculation \cite{loverde:14b},
where neutrino perturbations are treated linearly while a spherical tophat
shell of the $c+b$ fluid is followed to collapse.
Far inside the free-streaming scale
$k \gg k_{\rm fs}$, $\partial\dc/\partial \d_{c+b,\ell}(k) = -1$
as derived for the pure cold matter case in \refsec{PBS}, while it is modified on larger
scales.
This leads to a scale-dependent feature in the halo bias \refeq{b1neut} at the
few percent level, which contributes to the scale-dependent effects of neutrinos
in galaxy clustering.
Specifically, the scale dependence in $P_{hh}^\LO$ is predicted to be reduced by 20--40\% compared to that
in $P_{c+b,{\rm L}}(k)$ \cite{loverde:14a}.
The scale-dependent effects induced by neutrinos are controlled by a different scale, the free-streaming scale $\lambda_\text{fs}=1/k_\text{fs}$ [cf. \refeq{kfs}], than the higher-derivative bias contributions, which involve the scale $R_*$. The two sources of scale dependence can therefore be isolated provided that the range of wavenumbers probed is sufficiently large \cite{Biagetti:2014pha}. 
While the scale dependence in galaxy bias induced by neutrinos can also be
formally captured by the higher-derivative terms described in \refsec{higherderiv}, this expansion is only valid on scales $k\ll k_\text{fs}$, and is thus very limiting given the large value of $1/k_\text{fs}$. More explicit ways to include the effects of massive neutrinos in the general perturbative bias expansion should thus be developed (e.g., along the lines of \cite{senatore/zaldarriaga:2017}). 
Observationally, the scale dependence arising from massive neutrinos could be extracted with the multi-tracer techniques described in \refsec{NG:mt} \cite{loverde:16}.
For this, at least two tracers with different amplitudes of scale-dependent bias are necessary. It is important to stress that, if present, similar scale-dependent effects are expected
for the higher-order biases, e.g. $b_2^{c+b}$, as well.

The presence of a scale-dependent bias with respect to $\d_{c+b}$ is difficult to confirm unequivocally in N-body simulations including massive neutrinos due to its small amplitude and the degeneracy with the scale dependence of halo clustering induced by nonlinear and higher-derivative contributions. Refs.~\cite{castorina/sefusatti/etal:2014,castorina/carbone/etal:2015,villaescusa-navarro/etal:2017} reported no clear evidence for a scale dependence of $P_{h(c+b)}(k)/P_{c+b}(k)$ beyond what is expected from nonlinear contributions in the CDM+baryon sector. On the other hand, Ref.~\cite{chiang/etal:2017} used a generalization of the separate-universe technique described in \refsec{sepuni} and \refsec{bsepuni} to isolate the scale-dependent bias. They found a scale dependence consistent with \refeq{b1neut}, with an overall amplitude of roughly $6 f_\nu$ acting in the opposite direction as the matter power suppression due to neutrinos. Very recently, Ref.~\cite{chiang/etal:2018} reported a detection from halo clustering measured in a large suite of N-body simulations with neutrinos. In order to increase the signal while keeping the free-streaming scale in the linear regime, the number of neutrino species was multiplied by a factor $\sim 10$ in the simulations. The amplitude was found to be largely consistent with the separate-universe results of \cite{chiang/etal:2017}.

\subsection{Imprints of primordial baryon acoustic oscillations}
\label{sec:baryons}

Above we have seen that neutrinos leave a scale-dependent imprint
on structure formation below the free-streaming scale, despite the
fact that they constitute at most a few percent of the cosmic energy budget
today.  We now turn to the two most important matter components: cold dark
matter (CDM, $c$) and baryons ($b$, i.e. all standard model particles that
have been non-relativistic since the end of radiation domination;  this
excludes standard model neutrinos within the allowed mass range).
So far, we have assumed that on large scales, where non-gravitational
forces can be neglected, baryons and CDM comove and, therefore, can be
treated as a single fluid.
The density and velocity differences due to pressure forces and feedback
are, in the general bias expansion, taken into account through the
higher-derivative biases such as $b_{\lapl\d} \lapl\d$ (\refsec{higherderiv}).
This assumption is true for the adiabatic, growing-mode density perturbations
which dominate structure formation at low redshifts. However, before the epoch
of recombination, where protons and electrons first combined to form neutral
hydrogen, baryons were tightly coupled to photons in a
plasma and thus behaved very differently from CDM.  When setting the
initial conditions around recombination (strictly speaking, baryon-photon
decoupling), this introduces decaying
modes which can have a lasting imprint on galaxy statistics.
Note that in most cases, these effects are not taken into account when
setting up the initial conditions for N-body+hydro simulations.  That is,
typically the initial conditions only include the adiabatic growing mode
for both baryons and CDM.  Studies considering baryons and
CDM separately in perturbation theory and simulations can be found
in \cite{shoji/komatsu,somogyi/smith:2010,bernardeau/vdr/vernizzi,lewandowski/perko/senatore}.
We now study these decaying modes, focusing on their contributions
to the general galaxy bias expansion.

Consider the coupled evolution of the baryon and CDM fluids
under gravity, i.e. after baryon-photon decoupling.
At linear order, we have the following set of evolution equations:
\ba
\frac{\partial}{\partial\tau}\d_s =\:& -\theta_s \,,\quad s \in \{b,\,c\} \vs
\frac{\partial}{\partial\tau} \theta_s + \cH \theta_s =\:& - \frac32 \Om(a) \cH^2 \d_m\,,
\label{eq:deltalinearEOM1}
\ea
where $\d_s \equiv \d\rho_s/\overline{\rho}_s$, $\theta_s = \partial_j v_s^j$,
while $\d_m \equiv \d_{c+b} = f_b \d_b + (1-f_b) \d_c$ is the total matter density
perturbation and $f_b = \Omega_b/\Om$ is the baryon fraction (we neglect
the small contribution from massive neutrinos throughout this section).
Note that the CDM density perturbation $\d_c$ used in this section is not to be confused with the critical density for collapse $\dc$.

It is useful to combine these equations and to rewrite them in terms of
$\d_m$, and the relative density perturbation $\d_r \equiv \d_b-\d_c$ and relative velocity divergence
$\theta_r \equiv \vn\cdot\v{v}_r \equiv \vn\cdot(\v{v}_b-\v{v}_c)$ \cite{schmidt:2016b,ahn:2016,blazek/etal:15}: 
\ba
\frac{\partial^2}{\partial\tau^2} \d_m + \cH \frac{\partial}{\partial\tau}\d_m - \frac32 \Om(a) \cH^2 \d_m
=\:& 0 \vs
\frac{\partial^2}{\partial\tau^2} \d_r + \cH \frac{\partial}{\partial\tau}\d_r =\:& 0 \vs
\frac{\partial}{\partial\tau} \theta_r + \cH \theta_r =\:& 0
\,.
\label{eq:deltalinearEOM2}
\ea
We can now immediately obtain the general solution of these three ODE as
\ba
\d_m(\vx,\tau) =\:& A_+(\vq) D(\tau) + A_-(\vq) H(\tau) \vs
\d_r(\vx,\tau) =\:& R_+(\vq) - R_-(\vq) D_r(\tau) \hspace*{2cm}\mbox{(linear)}\vs
\theta_r(\vx,\tau) =\:& \frac{H_0}{a(\tau)} R_-(\vq)\,,
\label{eq:drgen}
\ea
where $A_\pm(\vq),\, R_\pm(\vq)$ are the initial conditions (hence evaluated at the Lagrangian position $\vq$, with $\vx=\vq$ at the order considered here), and
\be
D_r(\tau) \equiv H_0 \int_0^\tau \frac{d\tau'}{a(\tau')}
\ee
is the growth factor of the $R_-$ mode.\footnote{Note this was defined with a different sign in \cite{schmidt:2016b}.} The total matter density contrast $\d_m$ contains
the growing and decaying modes $\propto A_\pm$ of adiabatic perturbations.
We have employed the former extensively throughout all of
Secs.~\ref{sec:evolution}--\ref{sec:NG}.  A third mode, $R_+(\vq) \equiv \d_{bc}(\vq)$,
is a constant compensated density perturbation $\d\rho_c = -\d\rho_b$, corresponding to $\d_m=0$ while $\d_r \neq 0$ \cite{barkana/loeb:11,grin/dore/kamionkowski,soumagnac/etal:16}. 
This mode can be seen as modulating the local baryon-CDM ratio (see below).
The fourth mode $\propto R_-$ 
corresponds to an initial relative velocity between the two fluids, which scales as $a^{-1}(\tau)$ (just as expected for an unsourced peculiar velocity) and is commonly denoted as $\v{v}_{bc}$ in the literature. In particular, we can relate the latter to $R_-$ through
\be
\v{v}_{bc}(\vq,\tau_0) \equiv H_0 \v{R}_-(\vq) \equiv H_0 \frac{\vn_q}{\lapl_q} R_-(\vq)\,.
\label{eq:Rminus}
\ee
Correspondingly, we denote $\theta_{bc} \equiv \vn\cdot\v{v}_{bc}$, with
$\theta_{bc}(\vq,\tau_0) = H_0 R_-(\vq)$. While the notation
$\d_{bc}, \v{v}_{bc}$ is more common in the literature, the mode amplitudes $R_+,R_-$ (or $\v{R}_-$) are more convenient for deriving
the contributions to the bias expansion. However, both notations are entirely equivalent.
We refer to these two modes as \emph{baryon-CDM perturbations} in the following.

Note the distinction in notation between
the (in general) \emph{nonlinearly evolved relative density and velocity perturbations} $\d_r,\,\v{v}_r$, and the
\emph{linearly extrapolated initial amplitudes of the independent modes} of the system, $R_+,\,R_-$, or, equivalently,
$\d_{bc},\,\v{v}_{bc}$. This is analogous to the distinction between $\d^{(1)}$ and $\d\equiv\d_m$ in the adiabatic case. 
During nonlinear gravitational evolution, the decaying modes couple
with the growing mode, leading to nonlinear evolution of the baryon-CDM
perturbations (in particular the relative velocity), and making this distinction
important.
Moreover, $\d_r$ receives contributions from
both modes $R_+$ and $R_-$ [\refeq{drgen}].  As we will see however,
the contributions to the bias expansion can be written completely in terms
of the initial amplitudes of the independent modes $R_+,\,R_-$, so that we do not
need to consider $\d_r,\,\v{v}_r$ explicitly. 
\reftab{bcnotation} provides a summary of the notation used in this section.

\begin{figure}[t!]
\centering
\includegraphics[width=0.6\textwidth]{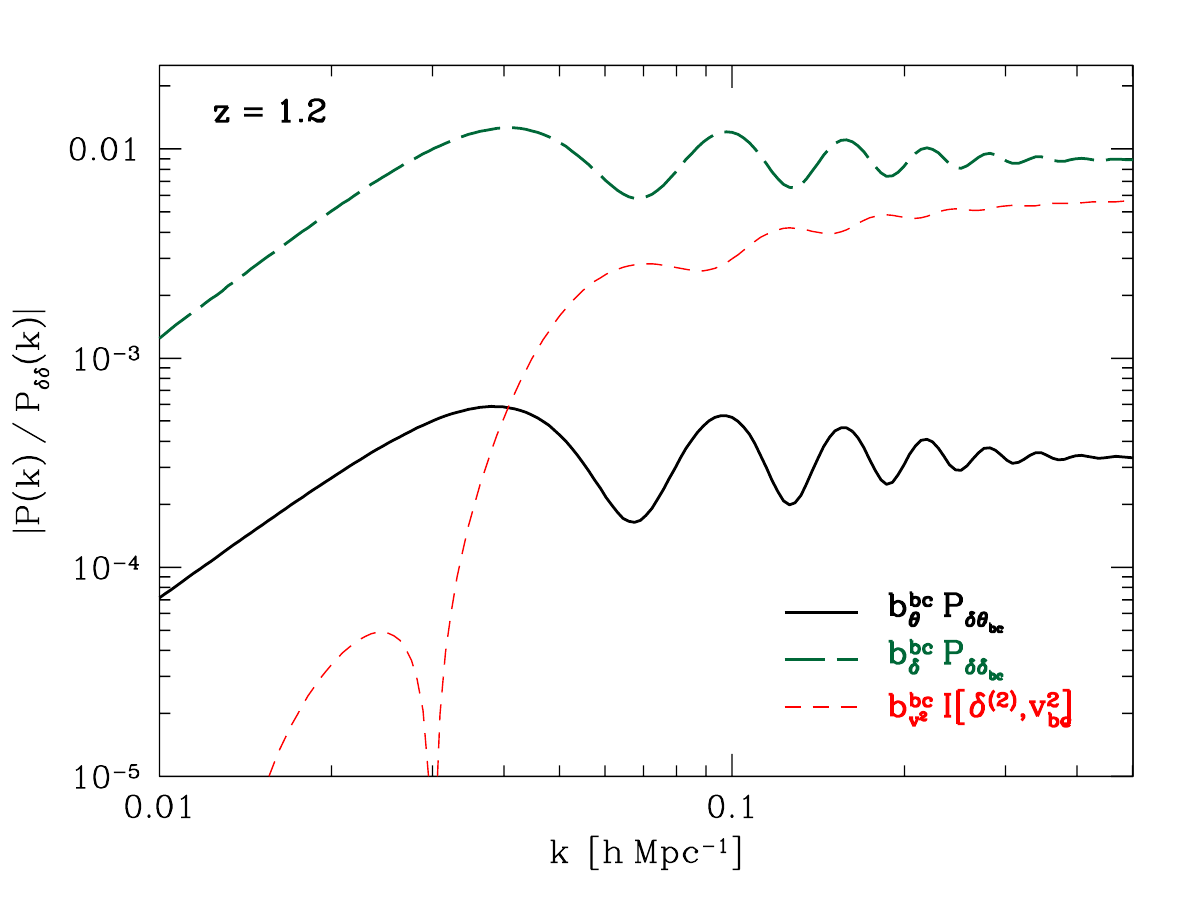}
\vspace*{-0.5cm}
\caption{Leading baryon-CDM relative density and velocity contributions to the galaxy power spectrum (at $z=1.2$), relative to the linear matter power spectrum (solid: $\theta_{bc}$; long-dashed: $\d_{bc}$).  Here, $b_\delta^{bc}=1$ and $b_{\theta}^{bc} = 6.8/[(1+z) H_0]$ [\refeq{btheta}, setting $b_1=2$] was assumed.  Also shown is the largest of the NLO contributions from the term $|\v{v}_{bc}|^2$ [\refeq{dgv2}], $b_{v^2}^{bc} \mathcal{I}^{[\d^{(2)}, v_{bc}^2]}(k)$ [\refeq{Iv2}] as short-dashed line, assuming $b_{v^2}^{bc} = 0.01 \<|\v{v}_{bc}|^2\>^{-1}$, close to the maximum value allowed by current constraints.
\figsource{schmidt:2016b}
\label{fig:Pkvcb}}
\end{figure}

Before deriving the contributions to the bias expansion, let us discuss the significance of the modes $R_+, R_-$. 
Ref.~\cite{tseliakhovich/hirata:2010} pointed out that pre-recombination plasma waves, the baryon acoustic oscillations, source $R_-$ and so lead to a supersonic
streaming velocity at the epoch of baryon-photon decoupling $\tau_{\rm dec}$.  This can leave an imprint in low-redshift structures which
assembled out of low-mass halos at high redshifts \cite{dalal/etal:2010,yoo/etal:2011,yoo/seljak,slepian/eisenstein,blazek/etal:15};  note that in many
of these references, the initial amplitude of the relative velocity is normalized to its variance, $\v{v}_{bc} \to \v{v}_{bc}/\sqrt{\< |\v{v}^2_{bc}|\>}$, which we do not do here.
Similarly, the $R_+$-mode is also sourced during recombination
\cite{barkana/loeb:11}.
Both $R_-$ and $R_+$ have significant large-scale correlations and
in particular retain large BAO features.  However, they are small numerically:  the root-mean-square relative density perturbation is $\sqrt{\<R_+^2\>} \lesssim 0.03$,
depending on smoothing scale,
while the root-mean-square relative velocity perturbation is approximately $0.03\,{\rm km/s}$ at the present epoch;  their fractional impact on the large-scale galaxy
power spectrum is expected to be at the sub-percent level (\reffig{Pkvcb}).

\begin{table}[t]
\centering
\begin{tabular}{l|l}
\hline
\hline
$\d_s$ & Density perturbation of species $s \in b,c$ \\
$\d_m = f_b \d_b + (1-f_b) \d_c$ & Total matter density perturbation \\
$\d_r = \d_b - \d_c$~~[\refeq{drgen}]  & Relative density perturbation \\
$\v{v}_r = \v{v}_b-\v{v}_c$ & Relative velocity \\
\hline
$R_+(\vx) \equiv \d_{bc}(\vx)$ & \emph{Initial amplitude} of constant relative density perturbation \\
$R_-(\vx) \equiv H_0^{-1} \partial_iv_{bc}^i(\vx,\tau_0)$ & \emph{Initial amplitude} of decaying relative velocity, extrapolated to $\tau_0$ \\
\hline
\hline
\end{tabular}
\caption{Notation for baryon-CDM perturbations used in this section.  Actual relative density and velocity perturbations are denoted with a subscript $r$, while their initial amplitudes (extrapolated to $z=0$ using linear theory), which appear in the bias expansion, are denoted with a subscript $bc$, or equivalently as $R_{\pm}$.  Note that the CDM density perturbation $\d_c$ is to be distinguished from the spherical collapse threshold $\dc$.}
\label{tab:bcnotation}
\end{table}

Since galaxy formation depends sensitively on both baryons and CDM, it is crucial to include the additional modes $R_+, R_-$ when making predictions for galaxy clustering.  We now briefly derive which operators these modes add to the bias expansion in the general case, under the same assumptions that were made in \refsec{general}.
We will ignore the adiabatic decaying mode ($A_-$) throughout, as it is numerically much smaller \cite{schmidt/beutler:2017}.  At linear order, the result is then already clear from \refeq{drgen}, as we have to allow for the remaining three modes of the baryon-CDM system to appear in the galaxy overdensity $\d_g$:
\ba
\d_g^{(1)}(\vx,\tau) =\:& b_1 \d_m^{(1)}(\vx, \tau) + b_{R+}(\tau) R_+(\vq[\vx,\tau],\tau)
+ b_{R-}(\tau) R_-(\vq[\vx,\tau],\tau) \vs
=\:& b_1 \d_m^{(1)}(\vx, \tau) + b_\d^{bc} \d_{bc}(\vq) + b_\theta^{bc} \theta_{bc}(\vq, \tau)\,,
\label{eq:dg2}
\ea
where in the second line we have used \refeq{Rminus}.  Note that $R_+,R_-$ (or $\theta_{bc},\,\d_{bc}$) have to be evaluated at the Lagrangian position $\vq[\vx,\tau]$ corresponding to $(\vx,\tau)$
\cite{blazek/etal:15}.   This is analogous to the case of primordial non-Gaussianity, where the additional operators are also evaluated at the Lagrangian position (\refsec{bNG});  in both cases, additional modes are present in the initial conditions.  Strictly speaking, in case of the relative baryon-CDM perturbations, we should use the position of the fluid trajectory at the time of recombination.  However, the distinction between these two is very small, and of similar order as other nonlinear contributions at recombination that are neglected throughout.

At second order, we expect all combinations of $R_+, R_-$ ($\d_{bc},\,\theta_{bc}$) with themselves as well as $\d_m$ to appear. However, there is a further contribution with a different structure, given by
\be
\d_g^{(2)}(\vx,\tau) \supset
b_{\v{R}_-^2}(\tau) \left[ |\v{R}_-|^2(\vq,\tau) - \< |\v{R}_-|^2 \> \right]
= b^{bc}_{v^2}(\tau) \left[|\v{v}_{bc}|^2(\vq,\tau) - \<|\v{v}_{bc}|^2\> \right]
\,.
\label{eq:dgv2}
\ee
This is the leading contribution to the galaxy density of a \emph{uniform relative velocity} between baryons and CDM;  since the velocity is a vector, it cannot contribute to the scalar galaxy density at linear order.

In order to derive all further nonlinear baryon-CDM contributions to galaxy clustering, consider the full nonlinear evolution equations of the two-fluid system.  Generalizing the results of \refsec{evol2} to two fluids, this can be written as \cite{schmidt:2016b}
\ba
\frac{D}{D\tau} \d_s +\theta_s  =\:& - \d_s \theta_s
- g_s v_{r}^i \partial_i \d_s \label{eq:deltanlEOM}\\
\left( \frac{D}{D\tau} + \cH \right) \theta_s + \frac32 \Om \cH^2 \d_m =\:&
- (\partial^i v_s^k)^2 - g_s v_{r}^i \partial_i \theta_s\,,
\nonumber
\ea
where $s \in b,c$ and
\be
\frac{D}{D\tau} \equiv \frac{\partial}{\partial\tau} + v_m^i \frac{\partial}{\partial x^i}
\ee
is the convective time derivative with respect to the total fluid velocity
$\v{v}_m = f_b \v{v}_b + (1-f_b)\v{v}_c$, and we have introduced the constants $g_b = 1-f_b$ and $g_c = -f_b$.  Note that the relative velocity perturbation is not sourced even at nonlinear order;  that is, subtracting the Euler equation for $\theta_c$ from that for $\theta_b$ only yields source terms that are proportional to $\v{v}_r$ (see also \refsec{velbias}).  Moreover, the typical linear displacement between the baryon
and CDM fluids induced by $\v{v}_{r}$ is very small:\footnote{The value given here is at $z=0$, and hence the maximal value.  The bulk of the displacement occurs at high redshifts $z \gtrsim 30$.} $\v{s}_r = \int d\tau\,\v{v}_{r} \lesssim 0.04 \Mpch$.  This is much smaller than the nonlinear scale at redshifts where we observe galaxies, as well as the characteristic scale $R_*$ expected for realistic galaxies.  We can then effectively assume that baryons and CDM still travel on the same fluid trajectories $\xfl(\tau)$, since the leading effects of the different fluid trajectories are given by higher-derivative terms of order $\v{s}_r\cdot\vn\d_s$ which are negligibly small compared to the higher-derivative terms which are present for realistic galaxies, as well as the effective fluid contributions for matter.

One can then easily verify that it is sufficient to include \cite{angulo/etal:2015,schmidt:2016b} $R_+(\vq)$ and $\v{R}_-$ defined in \refeq{Rminus} 
in the bias expansion, as well as higher spatial derivatives of these quantities and their combination
with the observables derived for the adiabatic single-fluid case (\refsec{general}).  Note that, unlike the contributions from the growing mode derived in detail in \refsec{evolution},
we do not have to include time derivatives of $R_+,\,\v{R}_-$,
since these are fixed amplitudes defined in the initial conditions.
This is analogous to the additional fields $\phi,\,\psi$ appearing in the
case of PNG (\refsec{NGevolution}).  Another way of seeing this is to allow
for a dependence of the galaxy density on $\d_r=\d_b-\d_c$, $\v{v}_r=\v{v}_b-\v{v}_c$ along the fluid
trajectory.  Up to any order in perturbation theory, and neglecting the displacements between the two fluids as described above, this relative density and velocity along the fluid trajectory can be written, using the equations of motion \refeq{deltanlEOM}, as a local function along the fluid trajectory of $R_+,\,\v{R}_-$ and the $\Pi_{ij}^{[n]}$ ($n \geq 1$) appearing in the galaxy bias expansion of \refsec{basisE}.  Thus, by allowing for $R_+,\,\v{R}_-$ to appear in all combinations with the $\Pi^{[n]}$, we allow for a general dependence of the galaxy density on the baryon-CDM relative density and velocity perturbations.
\refeq{Rminus} also shows how spatial derivatives of $R_+, \v{R}_-$ should be counted.
Specifically, $\partial^j R_-^i$ is of the same order as $R_+$.  Indeed, both relative
density perturbations ($R_+$) as well as relative velocity ($\v{R}_-$) and relative velocity shear ($\partial^i R_-^j$) between different fluids are leading local observables. 
On the
other hand, operators involving two or more derivatives of $\v{R}_-$, as well
as those with at least one derivative on $R_+$, will be suppressed by the spatial scale $R_*$ associated with the formation of the galaxies considered (e.g., the Lagrangian radius of the host halos).  Again, this holds because this scale is much larger than the absolute displacement $\v{s}_r$ due to the decaying relative velocity.

Explicitly, we make use of the single-fluid basis of operators
presented in \refsec{basisE}.
The contributions of baryon-CDM perturbations can then be succinctly
summarized in matrix notation by defining
\be
(\nabla R_-)_{ij}(\vq) \equiv \partial_q^i R_-^j(\vq) = \frac{\partial_q^i \partial_q^j}{\lapl_q} R_-(\vq)\,.
\ee
Up to third order in perturbation theory, the relative density and velocity contributions to the general Eulerian bias expansion then are \cite{schmidt:2016b}:
\bea
{\rm 1^{st}} \ && \ R_+(\vq)\,,\  {\rm tr}[(\nabla R_-)](\vq)
\label{eq:listR} \\[3pt]
{\rm 2^{nd}} \ && \ R_+ {\rm tr}[\Pi^{[1]}]\,,\  \v{R}_-^2\,,\  {\rm tr}[(\nabla R_-) \Pi^{[1]} ] \,,\
{\rm tr}[(\nabla R_-)]\, {\rm tr}[\Pi^{[1]}]
\nonumber\\[3pt]
{\rm 3^{rd}} \ && R_+\, {\rm tr}[\Pi^{[1]}\Pi^{[1]}]\,,\  R_+ ({\rm tr}[\Pi^{[1]}])^2 \,,\  \v{R}_-^2 {\rm tr}[\Pi^{[1]}]\,,\vs
&& \v{R}_- \Pi^{[1]} \v{R}_-\,,\  {\rm tr}[(\nabla R_-) \Pi^{[1]}\Pi^{[1]}]\,,\
{\rm tr}[(\nabla R_-) \Pi^{[1]}]\,{\rm tr}[\Pi^{[1]}]\,, \vs
&& {\rm tr}[(\nabla R_-)]\, {\rm tr}[\Pi^{[1]}\Pi^{[1]}]\,,\  {\rm tr}[(\nabla R_-)] ({\rm tr}[\Pi^{[1]}])^2\,,\  {\rm tr}[(\nabla R_-) \Pi^{[2]}]
\,, \nonumber
\eea
where $R_+, R_-$ are all evaluated at the Lagrangian position $\vq[\vx,\tau]$
while $\Pi^{[n]}_{ij}$ are evaluated at $(\vx,\tau)$.  The first-order
terms are precisely those included in \refeq{dg2}, while the second line
includes \refeq{dgv2}, among several others.  Here, we have only
included terms linear in $R_+,\,\nabla R_-$, while we have included
terms up to second order in $\v{R}_-$.  This is justified because the
contributions to galaxy statistics are percent-level corrections
to the standard adiabatic contributions, so that higher-order terms
are highly suppressed.  When extending \refeq{listR} to include
higher-derivative operators, one should correspondingly also include
terms of the type $(\v{R}_-\cdot\vn) \tr[\Pi^{[1]}]$,
i.e. $v_{bc}^i\partial_i\d$ (see \cite{lewandowski/perko/senatore},
who perform a resummation of this type of term).
While we have defined derivatives on $R_-^i$ to
be with respect to $\vq$, this distinction is in fact not important, since
the terms obtained when transforming $\partial/\partial q^i$ to a derivative with respect to
Eulerian coordinate $\vx$ are already included in the list \refeq{listR}.
\refeq{listR} is sufficient to derive all contributions to the leading-order
galaxy bispectrum and the one-loop galaxy power spectrum; the latter is
derived in detail in \cite{schmidt:2016b}.

In addition, as noted by \cite{blazek/etal:15}, the fact that $R_+, R_-$ are evaluated at the Lagrangian position introduces further terms.  This is again analogous to the case for primordial non-Gaussianity, \refsec{bNG}. 
Expanding the Lagrangian position $\vq$ around the Eulerian position $\vx = \vq+\v{s}$ induces displacement terms for each operator involving $R_+, R_-$. However, they are multiplied by the bias parameter of the corresponding operator in \refeq{listR}, and thus do not introduce additional free parameters. 
Finally, in addition to the deterministic operators listed in \refeq{listR}, there are also stochastic contributions of the same type as discussed in \refsec{stoch}.  That is, for each operator $O$ in the list \refeq{listR}, one introduces a stochastic field $\eps_O$ with zero mean that is, at lowest order in derivatives, fully characterized by its one-point moments.

The baryon-CDM perturbations also affect the velocities of galaxies.
Specifically, at linear order one has
\be
\v{v}_g(\vx,\tau) = \v{v}_m(\vx,\tau) + \beta_v^{bc}(\tau) \v{v}_{bc}(\vq)\,,
\label{eq:vgbc}
\ee
where $\beta_v^{bc}$ is a velocity bias parameter.  This contribution captures the fact that galaxy velocities could either follow the baryons ($\beta_v^{bc}=1-f_b$) or CDM ($\beta_v^{bc}=-f_b$), or some linear combination of the two.
Thus, one expects that $\beta_v^{bc}$ is at most of order one.  The contribution \refeq{vgbc} is relevant for galaxy surveys through the effect of redshift-space distortions (\refsec{RSD}), where, in case of the leading galaxy two-point function, it leads to a term comparable in magnitude to the linear-order terms in \refeq{dg2} \cite{schmidt:2016b}.

This completes the incorporation of baryon-CDM perturbations in the general perturbative bias expansion.
However, in order to assess the quantitative impact of these contributions,
we need estimates for the bias parameters associated with each operator
in \refeq{listR}.  For this, we divide the terms into three distinct
classes, corresponding to different physical mechanisms:

\emph{(i) operators involving $R_+ = \d_{bc}$:}  the $R_+$ mode essentially modulates the local baryon-CDM ratio through $(\Omega_b/\Omega_c)_{\vx} = (\overline{\Omega_b/\Omega_c}) [1 + R_+(\vq)]$.  That is, it can be described by a generalization of the separate-universe picture of \refsec{PBS}, where instead of varying $\Omn,\,\Omega_{K0}$ we keep them fixed and vary $\Omega_{b0},\,\Omega_{c0}$ instead.
Thus, in the presence of a long-wavelength $R_+$ perturbation, galaxies form in an environment with slightly different composition.  Refs.~\cite{barkana/loeb:11,schmidt:2016b} argue that the bias with respect to this mode is expected to be of order unity, since a modulation of $\Omega_b/\Omega_c$ corresponds to a modulation of the baryonic mass available to form stars.   Moreover, it is not unreasonable to expect an enhancement for rare massive galaxies, like in the case of the bias parameters $b_1,\,b_2,\cdots$ for halos, due to the similar exponential cutoff of the galaxy stellar mass function at high masses. Recently, Ref.~\cite{beutler/seljak/vlah:2017} constrained $|b_{R_+}| = |b_\d^{bc}| \lesssim 6$ (95\% confidence level) from the BOSS CMASS sample.

\emph{(ii) operators involving $R_-^i R_-^j \propto v_{bc}^i v_{bc}^j$:}  the second-order effect of a uniform relative velocity, first introduced by \cite{tseliakhovich/hirata:2010}, has been investigated extensively in the recent literature
\cite{dalal/etal:2010,yoo/etal:2011,yoo/seljak,slepian/eisenstein,blazek/etal:15}.  Ref.~\cite{dalal/etal:2010} argued that $\v{v}_{bc}$
increases the effective sound speed $c_s$ of the neutral gas, so that the Jeans mass $M_J$ increases by a factor $[1 + v_{bc}^2/c_s^2]^{3/2}$.
This leads to large effects on low-mass halos prior to reionization, as investigated using small-box simulations in \cite{tseliakhovich/barkana/hirata,Visbal/etal:12,popa/etal}.    These could be transferred to galaxies of much larger mass observed at low redshifts by modulations of the metal enrichment and hence star formation efficiency, for example.  To what extent this occurs quantitatively is still unclear.  $b^{bc}_{v^2}$ could be as small as $\sim 10^{-5} \<|\v{v}_{bc}|^2\>^{-1}$ \cite{blazek/etal:15,tseliakhovich/barkana/hirata}.  The fiducial value adopted in previous studies for redshifts $z \lesssim 2$
is \cite{dalal/etal:2010,yoo/etal:2011,yoo/seljak,blazek/etal:15},
\be
b^{bc}_{v^2} \sim 0.01\, \<|\v{v}_{bc}|^2\>^{-1} \approx 9\times 10^{11} \,(1+z)^{-2}\,,
\label{eq:bv2}
\ee
where we work in units where the speed of light is $c=1$.  Note that, given the non-detection in current data, $b^{bc}_{v^2}$ cannot be much larger than this \cite{yoo/seljak,slepian/etal:2016};  in particular, \cite{slepian/etal:2016} find an upper limit of $b^{bc}_{v^2} < 0.01$ from the three-point function of the SDSS BOSS DR12 CMASS sample \cite{BOSSdr12:2016}.

\emph{(iii) operators involving $\partial^i R_-^j \propto \partial^i v_{bc}^j$:}  these include the linear operator $\theta_{bc}$.  Physically, this corresponds to a uniform initial relative velocity divergence between baryons and CDM.  A very simple estimate for the associated bias parameter can be obtained by noting that from \refeq{drgen},
this induces an associated relative density perturbation of order
$\d_{r} \simeq a^{1/2} \theta_{bc}/H_0$.  Assuming that the response of the galaxy density to this relative density perturbation is of order unity, as for the constant compensated mode $\d_{bc}$, one obtains $b_{\theta}^{bc}(\tau) \sim a^{1/2}(\tau) H_0^{-1}$.
However, the physics of this mode, an initial relative velocity divergence,
is quite different from that of $\d_{bc}$, which corresponds to a changed composition.
A potentially more accurate estimate was presented in \cite{schmidt:2016b} using a spherical collapse calculation.  By following baryon and CDM shells separately, one can approximately take into account the initial relative velocity divergence in the collapse.  This in turn can be used to derive the effect on the collapse threshold $\dc$, which can be used to estimate the bias through the excursion-set argument (\refsec{sharpk_bias}).  Ref.~\cite{schmidt:2016b} obtained
\ba
b_{\theta}^{bc}(z) =\:& [(1+z)H_0]^{-1} \frac{\partial\dc(z)}{\partial(\theta_{bc,0}/H_0)} (1-b_1)
\approx 6.8 [(1+z)H_0]^{-1} (b_1-1)\,.
\label{eq:btheta}
\ea
Thus, we expect the bias parameters associated with the operators in this class to be of order several times $H_0^{-1}$.  The streaming velocity effects on very low-mass halos discussed above can induce an additional contribution to $b_\theta^{bc}$ that is proportional to $b_{v^2}^{bc}$.  Depending on the value of the latter bias parameter, this could be as large as the estimate \refeq{btheta}, but is likely to be smaller \cite{schmidt:2016b}.\footnote{Ref.~\cite{blazek/etal:15} adopt an effective $b_{\theta}^{bc}$ obtained from a loop integral that is approximately 20  times larger than \refeq{btheta} for the assumed value of $b_{v^2}^{bc}$.  Here, we have absorbed this loop integral into a renormalized bias parameter $b_\theta^{bc}$ whose value needs to be determined from the data, following renormalized perturbation theory (\refsec{renorm}).  Note that both parametrizations are equivalent as long as one allows for both $b_{v^2}^{bc}$ and $b_\theta^{bc}$ to be free parameters in the fit to galaxy statistics.}

\reffig{Pkvcb} shows the leading contributions to the galaxy power spectrum
from each of these three classes of terms.  We see that the $R_+ = \d_{bc}$
contribution dominates, while the $\theta_{bc} = \partial_i R_-^i$ contribution
is the smallest.  The contribution from $v_{bc}^2$ shown here is only one of several terms
at NLO \cite{blazek/etal:15,schmidt:2016b}, namely
\be
\mathcal{I}^{[\d^{(2)}, v_{bc}^2]}(k) = -2 \int_{\vp} F_2(\vp,\vk-\vp) \frac{\vp\cdot(\vk-\vp)}{p^2 |\vk-\vp|^2}
 P_{\d \theta_{bc}}(p) P_{\d \theta_{bc}}(|\vk-\vp|)  \,.
\label{eq:Iv2}
\ee
Note however that the bias parameter $b_{v^2}^{bc}$
is highly uncertain, and the value adopted for \reffig{Pkvcb} is near the maximum
currently allowed value.  While \reffig{Pkvcb} clearly shows that the
baryon-CDM contributions are very small at low redshifts, their prominent
BAO features, which are not exactly in phase with those in the adiabatic growing mode of matter,
make them relevant for the BAO feature as standard ruler in galaxy correlations (see the recent discussion in \cite{BOSSdr12:2016}).

Finally, Ref.~\cite{schmidt/beutler:2017} recently pointed out another physical effect which enters the large-scale galaxy bias expansion. After reionization, the free electrons in the Universe are weakly coupled to the free-streaming CMB photons via Thomson scattering. The electrons in turn are bound to the nuclei via Coulomb forces. If the baryons are at rest in the CMB frame, this scattering has no dynamical effect. On the other hand, if the gas is moving relative to the CMB, it sees a dipole in its rest frame, which leads to a drag force which is proportional to the velocity. In fact, this is the same drag that baryons experience before recombination. This drag force supplies a source term to the baryon-CDM relative velocity given by \cite{schmidt/beutler:2017}
\ba
\v{v}_r' + \cH \v{v}_r =\:& - x_e \alpha \cH\, \v{v}_b
\,,
\label{eq:euldrag}
\ea
where $x_e(\tau)$ is the electron ionization fraction, $\v{v}_b$ is the baryon fluid velocity relative to the CMB frame, and the dimensionless coefficient $\alpha(\tau)$ is given by
\ba
\alpha(\tau) =\:& a(\tau) \frac{\sigma_\text{T} u_\gamma(\tau)}{Y_e m_p c \cH(\tau)}  \propto (1+z)^4 H^{-1}(z)\,.
\label{eq:alpha}
\ea
Here, $\sigma_{\rm T}$ is the Thomson cross section, $u_\gamma(T)$ is the energy density of blackbody radiation of temperature $T$, $Y_e\approx 1.08$ is the electron molecular weight and $m_p$ is the proton mass ($\alpha$ is of order $10^{-6}$ today). The drag contribution leads to a relative velocity which, at redshifts $z \lesssim 10$, is of the same order of magnitude as the primordial contribution $\v{v}_{bc}$ discussed above. However, the scale dependence is very different, as $\v{v}_{bc}$ carries an imprint of the baryon acoustic oscillations, while the Compton drag is controlled by $\v{v}_b \approx \v{v}$, which is dominated by dark matter, and its transfer function is proportional to $T(k)/k$, where $T(k)$ is the transfer function of the adiabatic growing mode (\refsec{localNG}). This means that we have to separately account for the Compton drag contribution to the baryon-CDM relative velocity in the bias expansion. As shown in \cite{schmidt/beutler:2017}, the leading contributions are second order, and given by
\be
\d_g^{(2)}\Big|_\text{drag} = b_\text{drag} v^2 + b_\text{drag.bc} \v{v}\cdot\v{v}_{bc}\,,
\label{eq:dgbc}
\ee
where the first term is induced by Compton drag alone, while the second term corresponds to the coupling between Compton drag and primordial relative velocity. Note that, when considering gravity alone, the equivalence principle forbids the velocity itself from appearing in the bias expansion (\refsec{GR}). However, the CMB radiation provides a preferred frame, so that the local velocity with respect to the CMB is an observable; the coupling to the gas induced by Thomson scattering after reionization provides the physical process by which this observable affects galaxy formation and thus enters the bias expansion.

\subsection{Galaxy bias with dark energy and modified gravity}
\label{sec:modgrav}

Throughout this review, we have assumed a cosmological constant $\Lambda$ as
explanation of the current accelerated expansion of the Universe.  While all current cosmological observations appear consistent with this scenario, it is worth exploring other physical paradigms.  The most popular alternative to $\Lambda$, \emph{dark energy}, is usually described as a
light scalar field whose potential energy provides the source of the accelerated expansion (see \cite{Frieman:2008sn,Copeland:2006wr} for reviews).

Most models of galaxy clustering incorporate dark energy approximately by including the effects of the modified expansion history in the linear growth factor $D(\tau)$.  This approximation neglects perturbations in the dark energy density, which scale as $1+w$, where $w$ is the equation of state which is observationally constrained to be close to -1.  For this reason, perturbations in the dark energy typically have a very small effect on the growth of structure.  More precisely, this case parallels closely that of neutrinos (\refsec{neutrinos}), with the free-streaming scale being replaced by the sound horizon, or Jeans scale, $k_J^{-1} = c_s/\cH$ of the dark energy, where $c_s$ is the sound speed.  For a canonical scalar field, often referred to as \emph{quintessence}, $c_s=1$
so that the sound horizon is given by the comoving horizon $\cH^{-1}$.  Hence, structure formation happens on scales far inside the sound horizon of the dark energy, where dark energy perturbations are negligible.  In that case, the above-mentioned approximation is accurate.

In the opposite limit, $c_s=0$, realized for example by $k$-essence models, dark energy can be accurately modeled as a collisionless fluid \cite{creminelli/etal:09}. In principle, the bias expansion should now contain the dark energy and matter density separately, including the relative velocity between the two \cite{lewandowski/maleknejad/senatore:2016}. However, in the absence of initial isocurvature perturbations between dark matter and dark energy, the relative velocity vanishes, and both matter and dark energy follow the same fluid trajectories. This implies that the density perturbations in dark energy and matter are proportional to each other. In that case, no new terms are added to the bias expansion even in the presence of clustering dark energy.

If the sound horizon of the dark energy is on intermediate, observable scales, then one expects the bias parameters to become scale dependent analogously to the case of massive neutrinos.
The scale dependence of the linear bias was studied numerically in \cite{chiang/etal:2016}, by considering the two limiting cases of scales that are far inside and far outside the dark energy sound horizon. This study was based on a generalization of the separate-universe approach (\refsec{sepuni}) to include pressure perturbations \cite{hu/etal:2016}.  The fractional difference in the Lagrangian bias $b_1^L$ of dark matter halos between the sub-Jeans and super-Jeans limits was found to be at the few-percent level, for a dark energy with equation of state $w=-0.5$, and independent of halo mass (Fig.~11 in \cite{chiang/etal:2016}; note that such a strong departure from $w=-1$ is already ruled out by observations).  This implies that in the presence of dark energy, halos show a scale-dependent bias around the scale $k \sim k_J$ with an amplitude very roughly given by $\Delta b_1 \sim 0.1(1+w)\,b_1^L$.

A fundamental alternative to dark energy is to modify General Relativity (GR)
on large scales in order to yield an accelerating Universe without an exotic
stress-energy component.  We now discuss the implications of such
modifications to GR (\emph{modified gravity}), where we restrict to modified gravity models that obey the weak
equivalence principle at the level of the particle action.  This means that
there is a well-defined spacetime whose geodesics govern the motion of test particles in the absence of non-gravitational forces (universality of free fall).
Typically, modified gravity theories introduce an additional scalar degree of
freedom, so that searching for this degree of freedom is a promising avenue to
test gravity.  We refer to \cite{Clifton2012,Joyce2014} for comprehensive
and \cite{joyce/lombriser/schmidt:2016} for a concise review of modified
gravity in the context of cosmology.

As discussed in \refsec{GR}, the local gravitational observables corresponding to long-wavelength perturbations (i.e., at lowest order in derivatives) consist
of the local Hubble rate, tidal field, and spatial curvature on constant-proper-time slices ($K_F$ in \refsec{GR}).  These are supplemented by the matter density $\d$
and the velocity divergence $\theta$ and shear tensor $\partial^i v^j$.
The derivation of this fact is purely geometrical, i.e. it does not rely
on the validity of the Einstein equations.  Thus, all of this still holds in modified gravity.  Moreover, $\d$ and $\theta$ are related by the continuity equation, and $\theta,\,\partial^i v^j$ are related to the tidal field through the Euler equation, both of which are unmodified.  On the other hand, for non-relativistic tracers, the local effect of the spatial curvature is suppressed by $(v/c)^2$.  Thus, even though the relations between $\d,\,\theta,$ and $K_F$ are modified from those in GR, the reasoning in \refsec{framework} still holds, and it is sufficient to include %$\partial_i\partial_j\Phi$
the matter density, tidal field, and convective time derivatives thereof in the bias expansion.  However, the reduction of these terms to only a handful of terms at each order in perturbation theory, which holds in GR as described in \refsec{basisL}--\ref{sec:basisE}, is no longer possible in general for modified gravity.

To illustrate this, we adopt a scalar-tensor theory of
Brans-Dicke type \cite{Brans:1961sx} as toy example of modified gravity.
The well-studied $f(R)$ \cite{carroll/etal:2003,starobinsky:2007} gravity model falls into this class \cite{chiba:2003}.
This theory introduces
a scalar degree of freedom $\phi$ with potential $V(\phi)$ and a specific
coupling strength to matter.  The linear growth factor equation, \refeq{Deom} in GR, is then modified to
\be
\frac{d^2}{d\tau^2} D(k,\tau) + \cH \frac{d}{d\tau} D(k,\tau)  - \frac32 \Om(a) \cH^2 \left[1 + \alpha(\tau) \frac{k^2}{k^2 + a^2 m^2(\tau)} \right] D(k,\tau) = 0\,,
\label{eq:DeomBD}
\ee
where $\alpha(\tau)>0$ is a coupling constant, while $m^2(\tau) = d^2V(\phi)/d\phi^2|_{\bar\phi(\tau)}$ is the mass of the scalar field at its cosmological background value $\bar\phi(\tau) = \< \phi(\vx,\tau) \>$. \refeq{DeomBD} is derived by solving the usual linearized Euler-Poisson system augmented by the Klein-Gordon equation for $\phi$, where time derivatives of the latter are neglected (the so-called quasi-static approximation appropriate on subhorizon scales). 
Clearly, the growth factor becomes scale-dependent unless $m=0$.  For $k \ll a m$, corresponding to scales larger than the Compton length of the field, gravity reduces to GR, while on scales within the Compton length ($k\gg a m$) gravity is enhanced by a factor $1+\alpha$.

Now consider the evolution of a linear \LIMD bias relation.  At time $\tau=\tau_*$, we write $\d_g^* = b_1^* \d^*$. Then, linear evolution via \refeq{dgcons} immediately yields \cite{hui/parfrey:2008,parfrey/hui/sheth:2011}
\be
\d_g(\vk,\tau) = b_1^E(k,\tau) \d(\vk,\tau)\,,\quad
b_1^E(k,\tau) = 1 + (b_1^* - 1) \frac{D(k,\tau_*)}{D(k,\tau)}\,.
\ee
Thus, a \LIMD (scale-independent) initial bias becomes nonlocal (scale-dependent) at a later time, unless $D(k,\tau)$ is separable in $k$ and $\tau$.  This holds in the same way for the scale-dependent growth induced by free-streaming massive neutrinos (\refsec{neutrinos}).
From \refeq{DeomBD} we infer that, in the scalar-tensor example, the scale dependence will appear at $k\sim a m(\tau)$.  Since a general bias expansion should be able to describe the special case of a conserved, initially locally biased tracer, we clearly see that, in the case of a modified gravity scenario with scale-dependent growth, additional terms need to be included in the bias expansion already at linear order.

In full generality,
a scale-dependent growth factor $D(k,\tau)$ that is not separable
in $k$ and $\tau$ precludes us from constructing a rigorous bias expansion
in terms of a finite set of bias parameters.  Recall that this construction relied on
modes evolving at the same rate on large scales, so that time derivatives
could be reordered to be successively higher order in perturbation theory.
This no longer holds for a general $D(k,\tau)$.  However, as shown by
\refeq{DeomBD}, on scales much larger than the Compton length of the
additional degree of freedom, we can perform a perturbative expansion in
$k^2/(a m)^2$.  Effectively, higher-derivative biases then absorb the
effects of the fifth force consistently.

Viable modified gravity models typically include screening mechanisms which suppress the
additional degrees of freedom in dense regions to satisfy Solar System
constraints on gravity (see \cite{Joyce2014} for a general discussion).
These are nonlinear mechanisms and hence need
to be taken into account for nonlinear bias.  Screening mechanisms
provide motivation to use low-mass or low-density tracers such as
dwarf galaxies or voids to probe gravity.
For screening of the chameleon or symmetron types, we have to include $\phi$ itself in the bias expansion, since the screening threshold depends on the ambient field value.  Note that for $k\ll am$, $\phi(\vk) \sim k^2/(am)^2 \Phi(\vk)$ is proportional to the density, rather than the potential.  Models that invoke screening of this type have a Compton wavelength that is constrained to be less than $\sim 10 \Mpch$ in order to satisfy Solar System tests \cite{Wang:2012kj}.  Thus, this dependence on $\phi$ can be taken into account via the higher derivative terms mentioned above.
For screening of the MOND or $k$-essence type, the relevant variable is $\partial_i \phi/\mathfrak{a}_0$, where $\mathfrak{a}_0 \sim H_0$ is the MOND acceleration scale.  Thus, the lowest-order contribution from MOND-type screening to the galaxy density is  $(\partial_i\phi)^2/\mathfrak{a}_0^2$, which is of order $(v/c)^2 \lesssim 10^{-4}$ and thus expected to have a very small impact numerically.
  Finally, for models with screening of the Vainshtein type, the screening depends on $\partial_i \partial_j\phi$.  This does not lead to new terms in the bias expansion, since $\partial_i \partial_j\phi$ is already captured by including $\partial_i\partial_j\Phi$ and $\partial_i v_j$ (as well as their time derivatives) separately.

%% file: observations.tex
\clearpage
\section{Connection to observations}
\label{sec:observations}

\secttoc

So far, we have described the clustering of galaxies and halos in their
rest frame.  This was appropriate, since the focus of this review are
physical bias expansions, which naturally describe the rest-frame galaxy 
density. 
We now turn to the connection of these rest-frame statistics to observations.   
The purpose of this section is to describe briefly all effects that 
enter the \emph{theoretical prediction for the observed statistics of galaxies}, starting from
the general bias expansion in the rest-frame of galaxies.
We begin with the local (statistical) connection between 
galaxies and halos in \refsec{HOD}, commonly described via halo occupation
or abundance matching approaches.  
Although, in the context of the general bias expansion, \emph{there is no
need to describe galaxies in terms of their relation to halos,} 
as the bias expansion effectively takes into account the small-scale physics
of galaxy formation, the relation 
between galaxies and halos provides useful physical insights, and can be 
used to extend the models of halo statistics described in \refsecs{exset}{peaks} 
to galaxies.  

We then review astrophysical selection effects in \refsec{selection}.  
We show that the general bias expansion is able to describe a diverse population
of galaxies with a single set of effective bias parameters.  However,
certain selection effects can lead to additional terms in the bias expansion, 
which we describe there. 

\refsec{projection} then derives how the galaxy density is mapped
to observed redshifts and positions on the sky (\emph{projection effects});  
this includes the important complication of redshift-space distortions, as
well as so-called relativistic effects.    
Finally, \refsec{lightcone} deals with the issue that galaxy surveys
do not map out the galaxy distribution on a fixed time slice and a flat sky, 
but rather on the past light cone, which is especially relevant for 
large-scale perturbations on scales comparable to the typical radial distance to the observed galaxies in the survey.

\subsection{The connection between galaxies and halos}
\label{sec:HOD}

Galaxy formation is a complex process involving the inflow and cooling of
gas, and self-regulation via feedback effects (see \cite{mo/vandenbosch/white:2010} 
for an overview).  
Clearly, these processes go beyond the collapse of pressureless matter 
which governs the formation of dark matter halos in N-body simulations
and which forms the starting assumption of both excursion-set and peak approaches. On the other hand, realistic numerical simulations of the 
large-scale distribution of galaxies remain computationally extremely challenging. 
Since the fact that galaxies are hosted by dark matter halos has been 
verified by both simulations and observations, analytic 
approaches which rely on the halo model 
\citep{neyman/scott:1952,kauffmann/nusser/steinmetz:1997,
ma/fry:2000,seljak:2000,peacock/smith:2000,scoccimarro/etal:2001} 
(see \cite{cooray/sheth} for a review) have been developed to describe 
the clustering of galaxies.
In standard \emph{halo occupation distribution (HOD)} models
\citep{berlind/weinberg:2002,kravtsov/berlind/etal:2004,zheng/berlind/etal:2005}, 
galaxies are assumed to follow the matter distribution within their host dark matter halo, implemented for example by sampling the positions of
dark matter particles or by using the universal mean mass
profile of the NFW form \cite{NFW}. Furthermore, the host halo 
mass $M$ is usually the sole property determining the number of galaxies \cite{vale/ostriker:2004}.  We will also assume steady 
emission from galaxies;  the relation between abundance and clustering is 
modified for intermittent sources such as quasars, which can be used to 
constrain their lifetime \cite{martini/weinberg:2001}.  

Adopting the halo model ansatz, contributions to the clustering of galaxies 
can be split into those terms arising from distinct host halos (``two-halo term'', 
in case of the galaxy two-point function), and those 
involving the same host halo (``one-halo term'').  In keeping with the focus
of this review, we here derive the large-scale clustering properties
of galaxies predicted in this approach.

In most implementations of HOD approaches, galaxies are divided into ``central'' and ``satellite'' galaxies. Every halo hosting one or more galaxies has
exactly one central galaxy which resides at or close to the center-of-mass of
the halo. All other galaxies within the same halo are then denoted as satellite
galaxies. 
Let $N_c\in\{0,1\}$ and $N_s\in \{0,1, 2,\cdots \}$ be the number of central and satellite galaxies
in a given halo of mass $M$, which we consider to be random variables. Since the existence of satellite
galaxies is conditioned on the presence of a central galaxy, it is convenient to define $\mathcal{N}_s$ through $N_s=N_c \mathcal{N}_s$.
Assuming that $N_c$ and $N_s$ are independent random variables, the average number density of galaxies reads
\begin{equation}
\avng = \int\!d\ln M\,\avnh(M) \big\langle N_c\big\rangle_M
\Big[1+\big\langle \mathcal{N}_s\big\rangle_M\Big]\,,
\end{equation}
where $\big\langle N_c\big\rangle_M$ and $\big\langle \mathcal{N}_s\big\rangle_M$ 
are the expectation values of central and satellite galaxy numbers in halos of mass $M$, 
respectively.  By definition, $\< N_c \>_M \leq 1$.  The linear bias 
parameter $b_{1,g}$ is usually written as a number-weighted integral over the 
halo bias $b_1(M)$ \cite{benson/etal:2000}, 
\be
b_{1,g} = {\avng}^{-1} \int\!d\ln M\,\avnh(M) \big\langle N_c\big\rangle_M
\left[1+\big\langle \mathcal{N}_s\big\rangle_M\right] b_1(M)\,.
\label{eq:b1HOD}
\ee
According to the general peak-background split argument, which is exact (\refsec{PBS}), 
$b_{1,g}$ is the linear response of the mean galaxy density to a change in the mean density of the Universe.  We then see that \refeq{b1HOD} is derived using the assumption that $\< N_c\>_M$ and $\< \mathcal{N}_s\>_M$ are
independent of the background cosmology.  
That is, while the halo number density changes due to a long-wavelength density perturbation via $b_1(M)$, 
the occupation statistics at a given fixed halo mass are assumed to be unchanged.  
This might be a good first-order assumption in practice, although it is important to keep in mind that 
it is an approximation.  
The PBS argument can similarly be applied (with the same assumptions) to derive the higher-order \LIMD bias 
parameters $b_{N,g}$.  

One can also calculate the large-scale stochasticity of galaxies, assuming that the stochasticity of halos 
of mass $M$ is given by Poisson shot noise $1/\avnh(M)$.  Then, one obtains (e.g., \cite{abramo/etal:2015})
\begin{align}
P^{\{0\}}_{\eps,g} &= \frac1{\avng^{\,2}}\int\!d\ln M\,\avnh(M) \big\langle N_c\big\rangle_M
\left[1+2\big\langle \mathcal{N}_s\big\rangle_M + \big\langle \mathcal{N}_s^2 \big\rangle_M \right] \\
&= \frac1{\avng}+\frac1{\avng^{\,2}}\int\!d\ln M\,\avnh(M) \big\langle N_c\big\rangle_M
\left[2\big\langle \mathcal{N}_s\big\rangle_M + \left(\big\langle \mathcal{N}_s\big\rangle_M\right)^2 \right]  \;,
\end{align}
where the second line further assumes a Poisson distribution for $\mathcal{N}_s$.  
The stochastic contribution to the galaxy power spectrum thus also depends on the second moment of the distribution of satellite numbers at fixed halo mass.

Various parametrizations of $\< N_c\>_M$ and $\< \mathcal{N}_s\>_M$ exist in the literature. 
$\< N_c \>_M$ typically follows a step-like function, in agreement with what is found for the distribution of subhalos (bound substructures of halos) in N-body 
simulations \cite{kravtsov/berlind/etal:2004}, whereas $\< \mathcal{N}_s\>_M$ can be 
parametrized by a power-law with logarithmic slope $\alpha\approx 1$. 
In the model of \cite{zheng/coil/zehavi:2007} for instance, it is assumed that halos
hosting satellite galaxies in a given luminosity-limited sample also host a central
galaxy from the same sample,
\begin{align}
\big\langle N_c\big\rangle_M &=
\frac{1}{2}\left[ 1 +{\rm erf}\left(\frac{\log M-\log M_\text{min}}
{\sigma_{\log M}}\right)\right] \\
\big\langle \mathcal{N}_s\big\rangle_M &= \left\{\begin{array}{ll}
\left(\frac{M - M_0}{M_1'}\right)^\alpha\;,
& M > M_0 \\
0\;, & M \leq M_0\;.
\end{array}\right.
\label{eq:HODparam}
\end{align}
Here, $M_\text{min}$ represents the halo mass cutoff of central galaxies and $\s_{\log M}$
takes into account the scatter between galaxy luminosity and halo mass. 
For the satellite galaxies, $M_0$ is the halo mass cutoff, $M_1'$ is the characteristic
mass of halos harboring one satellite, and $\alpha\approx 1$ is the power-law slope.  
$\mathcal{N}_s$ is usually assumed to follow a Poisson distribution.  
All these model parameters depend on the galaxy sample considered, for example the galaxy luminosity.

\begin{figure}
\includegraphics[width=.5\textwidth]{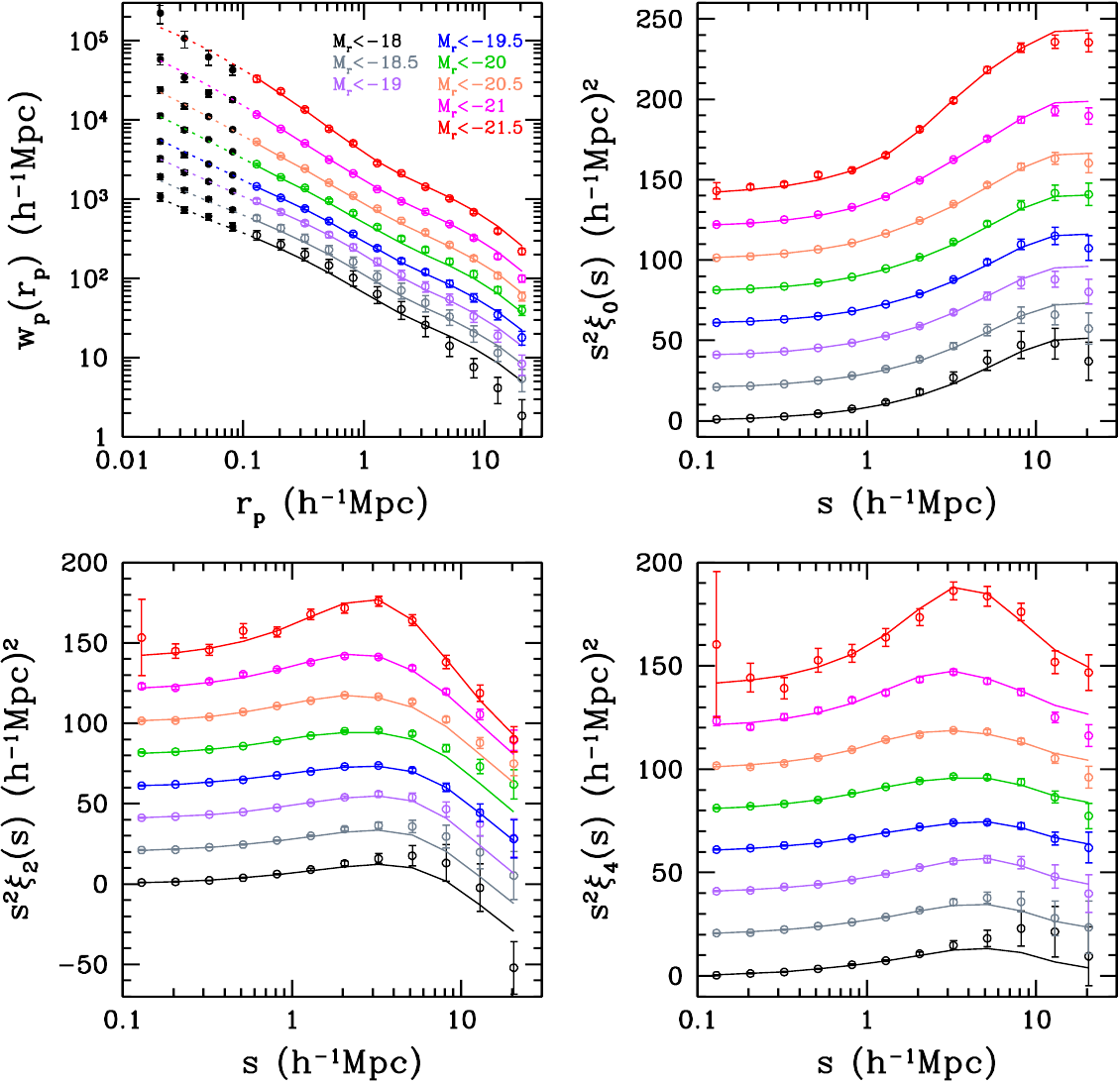}
\includegraphics[width=.5\textwidth]{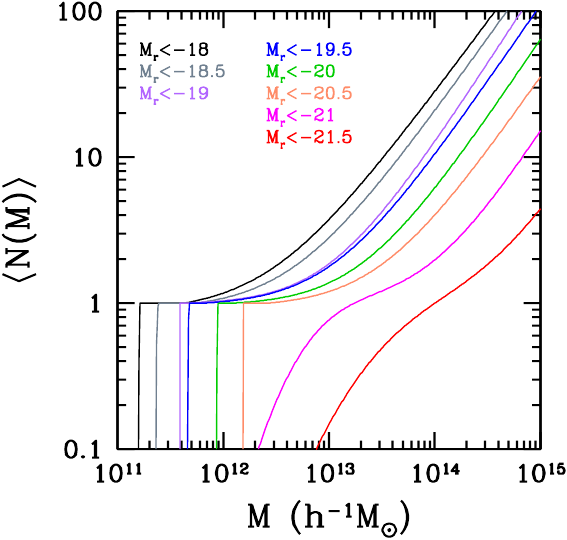}
\caption{{\it Left panel}: 2-point projected galaxy correlation function $w_p(r_p)$ measured 
from the SDSS DR7 main sample with different luminosity thresholds as indicated in the figure. 
The solid curves represent the best-fit HOD models. Only the data at projected separation
$r_p>0.1\hmpc$ was included in the fit. {\it Right panel}:
Mean halo occupation function $\< N(M) \>$ [\refeq{NMHOD}] of the best-fitting models.
A value $\sigma_{\log M}=0$ was assumed for the faint galaxy samples, for which the cutoff mass scale is not well constrained, which is visible as a sharp cut at low masses in the right panel. 
\figsource{guo/zheng/etal:2015}  
\label{fig:bestfithod}
}
\end{figure}

For illustration, the left panel of \reffig{bestfithod} shows the projected 
two-point correlation function of galaxies with different luminosity 
threshold along with the best-fit HOD models.
In order to use an N-body simulation to fit the HOD parametrization \refeq{HODparam} to 
a given galaxy sample, centrals are assigned, for each halo in the simulation, with a 
probability $\< N_c \>_M$ 
to the center-of-mass position and velocity of halos.  
Further, $\mathcal{N}_s$ is sampled from a Poisson distribution with mean $\<\mathcal{N}_s\>_M$,
and satellite galaxies 
are assigned the position and velocity of random dark matter particles within the halo.    
Then, the mean number density and the projected correlation function of this HOD sample 
is measured and compared to the data (see \cite{guo/zheng/etal:2015} for details).  Overall, the N-body-based HOD approach describes the shape of the galaxy correlation function at different luminosity thresholds quite well over a wide range of scales. 
More luminous galaxies reside on average in more massive halos and, therefore, are more biased. 
In the right panel, the mean halo occupation function
\be
\big\langle N(M)\big\rangle \equiv \< N_c\big\rangle_M \big[1+ \< \mathcal{N}_s\big\rangle_M \big]
\label{eq:NMHOD}
\ee
of the best-fit models is shown as a function of halo mass. 
The characteristic halo mass cutoff $M_\text{min}$ and $M_1$ increase with the luminosity 
threshold.  Note however, that $\sigma_{\log M}$ is weakly constrained for the faint galaxy 
samples, which leads to the value $\sigma_{\log M}\approx 0$ adopted here.  
Fitting the small-scale clustering of galaxies via the HOD approach makes a prediction 
for the large-scale bias via \refeq{b1HOD}.  Ref.~\cite{more:2011} pointed out that 
simple HOD models tend to overestimate the actual large-scale galaxy bias $b_1$ measured 
on scales $\gtrsim 60 \Mpch$.  
Note that the possible dependence of $\< N_c \>_M$ and $\<\mathcal{N}_s\>_M$ on 
large-scale density perturbations discussed after \refeq{b1HOD}, which is usually neglected, 
leads to a correction to the linear bias which could help explain this discrepancy.

Abundance matching techniques are an alternative to HOD approaches, and are based on the assumption that there is 
a monotonic relation between some galaxy property and the host halo (or subhalo) mass.  
In particular, this has been applied to the case of the stellar mass for central galaxies
\cite{kravtsov/berlind/etal:2004}, or the stellar mass in connection with the subhalo mass for 
satellite galaxies \cite{vale/ostriker:2006,conroy/wechsler/kravtsov:2006}. 
In both cases, 
the underlying assumption is that there is exactly one galaxy per dark matter halo, with
the most massive galaxies residing in the most massive (sub)halos.  
These non-parametric methods, which by construction rely heavily on simulations, predict very well the clustering 
of observed galaxies (e.g. \cite{torres/chuang/etal:2016}).  
Based on this technique, Refs.~\cite{leauthaud/etal:2012,behroozi/wechsler/conroy:2013} present
an empirical mapping between stellar mass and halo mass at redshifts $z=0-1$, and $z=0-8$, respectively. These results offer interesting insights into the efficiency of galaxy formation at different mass scales.

HOD and abundance matching approaches provide simple, physically motivated procedures to generate mock galaxy catalogs 
from N-body simulations \cite[see][for a recent review]{monaco:2016}.
However, they clearly rely on oversimplified parametrizations of galaxy formation.  
Much progress on the understanding of the physics of galaxy formation has been made recently through detailed 
hydrodynamical simulations of galaxy formation.  
Recent suites of simulations \cite{illustris,horizonAGN,MassiveBlack,CosmoOwls} reproduce 
fairly well the evolution of early- and late-type galaxies, quasars and their distribution
in the cosmic web. They suggest that feedback from supernovae and active galactic nuclei 
has a significant impact on the local galaxy abundance, and even on halo masses.  
These effects can influence the clustering of galaxies and thus need to be accounted for in order to achieve  the 
accuracy required by the analysis of forthcoming galaxy surveys.  This highlights the crucial difference  between the 
perturbative bias expansion on the one hand, and HOD and abundance matching approaches on the other.  
The former is agnostic regarding the small-scale processes that shape galaxy formation, while the latter necessarily 
relies on a specific (simplified) model.  

\subsection{Astrophysical selection effects}
\label{sec:selection}

In practice, all galaxy samples collected in sky surveys are selected
on some observable property such as luminosity and color.  The previous
section described how these complications can be modeled through an HOD 
approach.  However, employing an HOD model or other empirical method
for connecting galaxies and halos is not necessary for the 
general bias expansion of \refsec{evolution} to be valid, as we now show.  
\emph{This is the case as long as all deterministic and stochastic terms are 
included at the relevant order, with in general free bias parameters} (see \refsec{evol:summary} for the result up to third order).  We
continue to work with the galaxy density in the galaxy rest frame, and defer
redshift-space distortions and other projection effects to the next section.  

Let $\v{\alpha}$ denote a set of physical observables 
(e.g., type, luminosity, color, half-light radius, shape etc.) 
that is used to select galaxies.  Again, we ignore any projection
effects on these observables, and assume that they are true intrinsic
properties in the galaxy rest frame.  
Given a selection function $S(\v{\alpha})$, the 
observed galaxy density at the spacetime coordinate $(\vx,\tau)$ is 
\be
n_g^{\rm sel}(\vx,\tau) = \int d\v{\alpha}\,S(\v{\alpha}) n_g(\vx,\tau; \v{\alpha})
= \int d\v{\alpha}\, S(\v{\alpha}) \avng(\tau;\v{\alpha})\left[1 + \d_g(\vx,\tau;\v{\alpha})\right]\,,
\label{eq:ngobsalpha}
\ee
where we allow for both the mean galaxy density $\avng$ and density contrast $\d_g$ to depend on the set of galaxy properties $\v{\alpha}$.  
Let us apply the bias expansion to this ``conditional'' density contrast $\d_g(\vx,\tau; \v{\alpha})$:
\be
\d_g(\vx,\tau; \v{\alpha}) = \sum_O \left[ b_O(\tau,\v{\alpha}) + \eps_O(\vx,\tau;\v{\alpha})\right]O(\vx,\tau) + \eps(\vx,\tau;\v{\alpha})\,.
\label{eq:dgalpha}
\ee
Here operators and bias parameters are understood to be renormalized.  
Note that this expansion consistently takes into account that the distribution of the observables $\v{\alpha}$ at 
each point $(\vx,\tau)$ can depend on the environment via $\{O\}$ as well
(for example, a higher fraction of red galaxies in dense regions),
in addition to having a stochastic component.  Thus, inserting \refeq{dgalpha} into \refeq{ngobsalpha}, we obtain
\be
n_g^{\rm sel}(\vx,\tau) = \avng(\tau) \left[1 + \sum_O  \left[b^S_O(\tau) + \eps^S_O(\vx,\tau)\right]\,O(\vx,\tau) + \eps^S(\vx,\tau)\right]\,,
\label{eq:ng_selection}
\ee
where
\ba
\avng(\tau) =\:& \int d\v{\alpha}\, S(\v{\alpha}) \avng(\tau;\v{\alpha}) \vs
b^S_O(\tau) =\:& \int d\v{\alpha}\, S(\v{\alpha}) b_O(\tau,\v{\alpha}) \vs
\eps^S_O(\vx,\tau) =\:& \int d\v{\alpha}\, S(\v{\alpha}) \eps_O(\vx,\tau;\v{\alpha})\,.
\ea
Thus, the seemingly complicated and nonlinear sample selection effects are
consistently taken into account by effective mean bias parameters $b_O^S$ and
stochastic fields $\eps_O^S$.  While not necessary for the bias expansion to apply,
there are still good reasons to split observed galaxy samples by physical
properties (or proxies thereof) \cite{hamilton:1988,boerner/mo/zhou:1989,willmer/dacosta/pellegrini:1998,
benoist/cappi/etal:1999,norberg/2df:2002,zehavi/eisenstein/etal:2005}.  
This allows for insights into galaxy 
formation and evolution, by measuring the bias parameters of different
galaxy populations (see \reffig{bestfithod}).  Moreover, the relative bias
parameters can be measured without sampling variance, and allow for multi-tracer
techniques to be applied (\refsec{NG:mt}).  
Alternatively, ``marked correlation'' analyses, in which galaxies are weighted by some property
or ``mark,'' help in quantifying how the galaxy properties $\v{\alpha}$ correlate with their large-scale environment \cite{beisbart/kerscher:2002,sheth/tormen:2004,sheth:2005,skibba/sheth/etal:2006,skibba/sheth:2009}.

There is, however, an additional selection effect which is nontrivial.  
In the bias expansion discussed throughout this review, we have written the galaxy
density as a 3-scalar, which implies that there are no preferred directions
and all tensor indices in bias terms are contracted with each other, such as
$K_{ij} K^{ij}$.  However, in reality, galaxies are observed through
the photons emitted along the line of sight $\vnhat$ to the observer.  
Thus, there is a preferred direction involved through the observational selection function,
i.e. the probability that a given galaxy is included in the survey sample.  
Two prominent cases where this can happen are:
\begin{itemize}
\item The selection function depends on the orientation of the galaxy \cite{hirata:2009,krause/hirata:2011,fang/etal:2011}.  This is particularly relevant for disk galaxies, since a disk is typically dimmer when viewed edge-on compared to face-on due to dust obscuration.  Orientations of galaxies in turn tend to correlate with large-scale tidal fields \cite{catelan/kamionkowski/blandford:2001,hirata/etal:2007,okumura/jing:2009,singh/mandelbaum/more:2014}, leading to a dependence of the selection probability on the tidal field projected along the line of sight.
\item Galaxies are identified through emission or absorption lines, whose observed strength depends on the line of sight due to radiative transfer effects;  for example, the escape probability of a resonance line photon depends on the local velocity gradient of matter \cite{zheng/etal:2011,wyithe/dijkstra:2011} (see also \cite{behrens/etal:2017}), which again is proportional to the tidal field projected along the line of sight.  This applies in particular to the Lyman-$\alpha$ line.  
\end{itemize}
At linear order, these effects add one additional term, the line-of-sight projection of the tidal field,
\be
b_{K_\parallel} \nhat^i \nhat^j K_{ij}\,.
\label{eq:ZZbias}
\ee
In Fourier space, this term becomes $b_{K_\parallel} (\mu^2-1/3) \d(\vk)$, where 
$\mu = k_i \nhat^i/k$.  
The significance of this term is that it is degenerate with the leading
RSD contribution $f\mu^2 \d$, and can thus hamper the use of RSD to measure the growth rate $f$.
However, this degeneracy can be broken by using the galaxy bispectrum
as discussed in Refs.~\cite{krause/hirata:2011,greig/etal:2013} for 
the two cases mentioned above.  
At higher orders in perturbations and derivatives, the additional terms
introduced by the line-of-sight dependence of the selection function
multiply rapidly; see \cite{pkgspaper} for a complete list up to third
order. For this reason,
keeping such selection effects small is an important consideration in the
design of large galaxy surveys.

Finally, we consider the bias of density fields constructed through 
\emph{intensity mapping}.  In this approach, no target selection is done, 
and spectra are obtained for many lines of sight across a survey region.  
Then, the flux in each pixel of a spectrum is interpreted as a line intensity
for a given transition at the corresponding redshift.  This has or will
be applied to the 21cm hydrogen hyperfine-structure transition
\cite{furlanetto/etal:2006}, molecular lines 
such as CO \cite{visbal/loeb:2010}, as well as atomic transitions,
in particular H$\alpha$~6563, as employed by the proposed SPHEREx \cite{SPHEREx:2014} mission.  
The resulting intensity map can be seen as a biased tracer of large-scale structure
just as galaxies, and obeys an analogous bias expansion.  One simply
replaces $n_g^{\rm sel}(\vx,\tau)$ in \refeq{ng_selection} with the
observed intensity ${\cal I}_\nu^{\rm obs}(\vx,\tau)$. Due to the fact that it
measures line fluxes, intensity mapping
is often affected by the radiative transfer effects mentioned above,
and thus usually requires the bias \refeq{ZZbias} to be included in the model predictions.

\subsection{Projection effects: from proper to observed galaxy density}
\label{sec:projection}

In practice, all tracers of large-scale structure are observed via 
photon arrival directions (right ascension and declination) and redshifts,
inferred from the shift in frequency of the observed 
spectral energy distribution (SED) of the galaxy relative to the rest-frame
frequency.  That is, we do not
have access to the rest-frame galaxy density (or line intensity) directly.  
Hence, an essential ingredient in the interpretation of large-scale structure
is the mapping from rest-frame quantities to observations, which is the 
subject of this section.  
We refer to the contributions to the observed galaxy density obtained from transforming the rest-frame galaxy density to observed coordinates as \emph{projection effects}.  
Ref.~\cite{jeong/schmidt:2015} provide a concise recent review of the subject.  

Using the observed arrival direction, described by the unit vector $\vnhat$,
and observed redshift $\tilde z$ of a given galaxy, the standard practice
is to assign the galaxy a comoving position using the photon geodesics 
in a given unperturbed flat FRW spacetime described by fiducial cosmological
parameters, which, when using conformal time $\tau$,
are simply parametrized as straight lines:
\be
(x^0(\tau),\vx(\tau)) = 
\left(
\tau_0-\tau, \bar{\chi}(\tau)\vnhat
\right)\,,
\label{eq:xbar}
\ee
where $\tau_0$ is the conformal time at present (i.e., at the time of observation), and $\bar{\tau}(z)$ and $\bar{\chi}(z)$ are, respectively, 
the conformal-time-redshift relation and the comoving-distance-redshift relation
in the adopted fiducial background cosmology.  Thus, a galaxy with given 
$(\tilde z, \vnhat)$ is assigned the comoving position
\be
(\tilde\tau,\tilde{\vx}) = (\bar{\tau}(\tilde{z}),\bar{\chi}(\tilde{z})\vnhat)\,.
\label{eq:tildex}
\ee
In general, this position of course only corresponds to the true physical position of the galaxy in the absence of spacetime perturbations (and if the fiducial cosmology is the true one).  On the other hand, in the presence of perturbations as in \refeq{metriccN}, this is a convenient coordinate (gauge) choice because the variables are directly related to large-scale structure observations.  
Hereafter, quantities denoted with a tilde are directly related to observables,
while barred quantities refer to quantities evaluated for the fiducial 
background cosmology.  
Moreover, throughout we assume that the true background cosmology is used to 
assign apparent positions via \refeq{tildex}.  In practice, we have to allow
for the possibility that the true expansion history is different from the
fiducial assumption.  These deviations can effectively be taken into
account as additional coordinate rescalings, known as \emph{Alcock-Paczy\'nski} distortions \cite{APtest:1979}, which themselves contain information
on the expansion history \cite{seo/eisenstein:2007,padmanabhan/white:2008,shoji/jeong/komatsu:2009}.

\begin{figure}
\centering
\includegraphics[width=.45\textwidth]{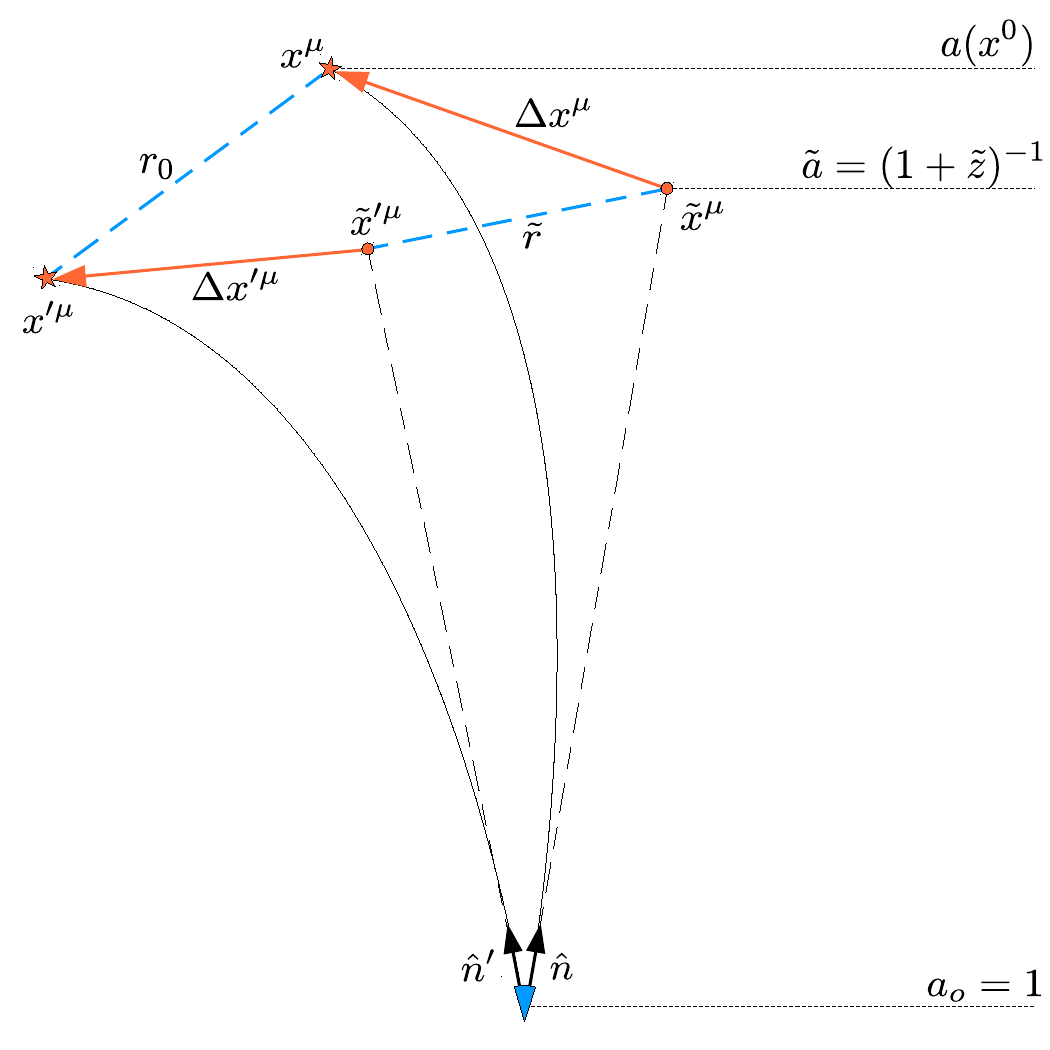}
\caption{Illustration of the projection effects, i.e. the mapping from observed photon directions and redshifts to the galaxy rest frame. The actual geodesics of photons emitted from two galaxies at $x^\mu$ and $x'^\mu$ are depicted as
solid lines.  The observer at the bottom of the plot infers radial 
distances and angular positions of the galaxies based on the observed redshift and arrival direction of photons, assigning them the spacetime positions $\tilde{x}^\mu$ and 
$\tilde{x}'^{\mu}$ respectively. The deviation four-vectors $\Delta x^\mu$ and 
$\Delta x'^\mu$ can be calculated by following the photon geodesics, taking into account the velocity four-vectors of each galaxy and 
the observer. Note that the deflection of the photon geodesics is greatly
exaggerated here for illustration purposes.
\figsource{schmidt/jeong:2012}  
\label{fig:sketch_projection}
}
\end{figure}

The calculation is illustrated in \reffig{sketch_projection}. Here, we show two
galaxies with inferred positions $\tilde{x}^\mu$ and $\tilde{x}'^\mu$ based on 
the observables $(\zt,\vnhat)$ and $(\zt',\vnhat')$; the true coordinates of 
these galaxies, in some given global coordinate system, are $x^\mu\equiv \tilde{x}^\mu +\Delta x^\mu$ 
and $x'^\mu = \tilde{x}'^\mu + \Delta x'^\mu$, respectively. We define $\Delta x^\mu$ as the 
spacetime deviation between the true spacetime coordinate and the inferred
coordinate of a galaxy.
The deviation $\Delta x^\mu$ can be calculated by integrating the geodesic 
equation from the observer's location to the source, given the photon momentum at the observer specified by $\tilde z, \vnhat$.  
General-relativistic effects such as Shapiro time delay, 
Sachs-Wolfe effect, Integrated Sachs-Wolfe effect, lensing magnification 
as well as redshift-space distortions contribute to 
$\Delta x^\mu$.  Explicit expressions for 
$\Delta x^\mu$ can be found in, for example, \cite{yoo/etal:2009,yoo:2010,jeong/schmidt:2015} and \cite{bertacca/martens/clarkson:2014}, at linear 
and second order in perturbations, respectively.  
Here, we present the linear-order results in terms of the general perturbed FRW metric, i.e. without
restricting the gauge degrees of freedom (also known as {\it gauge-ready form})
\be
\dd s^2 = a^2(\tau)\left[
-(1+2A[\vx,\tau]) \dd\tau^2 - 2B_i[\vx,\tau] \dd\tau \dd x^i 
+ \left(\d_{ij}+h_{ij}[\vx,\tau]\right) \dd x^i \dd x^j \right]\,.
\label{eq:metric}
\ee

First, we evaluate the proper time $t_F$ at which a given observed photon was
emitted, for an observer comoving with the source.  In case of an unperturbed Universe, this is simply given by
$t_F=\bar t(\tilde a)$, where $\tilde a = 1/(1+\zt)$ and $\bar t(a)$ is the time-scale factor relation in the fiducial background.  For convenience, instead of using the proper time $t_F$, we transform to $\ln a(t_F)$ using the scale factor $a(t)$ in the fiducial background.  Moreover, we phrase the departure of the actual ``log-scale factor of emission'' from the prediction in the unperturbed background, $\ln \tilde a \equiv - \ln(1+\tilde z)$, through 
\ba
{\cal T}(\tilde{\vx},\tilde\tau) \equiv
\:&
\ln\left(\frac{a(t_F[\tilde{\vx},\tilde\tau])}{\tilde{a}}\right)
=
\Delta\ln a(\tilde{\vx},\tilde\tau)
+\tilde H \int_0^{\tilde\tau} A[\vx,\tau'] a(\tau') \dd\tau' \,,
\label{eq:Tdef} 
\ea
where $\tilde H \equiv H(\tilde a)$, and 
\ba
\Delta\ln a \equiv\:&
A_o - A + v_\parallel - v_{\parallel o} 
- \int_0^{\tilde{\chi}} d\chi\,
\frac{\partial}{\partial\tau}\left[
A - B_\parallel - \frac12 h_{\parallel} \right]_{\chi\vnhat,\,\tau_0-\chi}
- H_0 \int_0^{\tau_o} A(\v{0},\tau') a(\tau')  \dd\tau'
\,,
\label{eq:Dlna}
\ea
where all quantities without arguments are evaluated at $(\tilde{\vx},\tilde\tau)$, while a subscript $o$ indicates a quantity evaluated at the observer: $v_{\parallel o} \equiv v_\parallel(\v{0},\tau_0)$.
The subscript $\parallel$ denotes quantities projected  
onto the line of sight; for example, $B_\parallel \equiv B_i\nhat^i$,
$h_{\parallel}\equiv h_{ij}\nhat^i\nhat^j$.  
Thus, ${\cal T}$ is 
the time shift, phrased in terms of $\ln a$, between a constant-observed-redshift surface (defined for a fixed central observer) and a constant-proper-time surface.
In \refeq{Tdef}, $\D\ln a$ gives the difference $\cH \delta\tau$ in coordinate  
time, while the second term  maps coordinate time to proper time.  
Note that ${\cal T}$ is an observable and gauge-invariant: if the observed source is a clock,
i.e. communicates its proper time since the Big Bang to the distant observer, then ${\cal T}$
can be measured directly.  Indeed, the observed CMB temperature fluctuation on
large scales, outside of the sound horizon, is exactly given by ${\cal T}$
\cite{cosmic_clock}.  $\D\ln a$ on the other hand is a coordinate-dependent and unobservable quantity.  Only if a gauge choice is made such that $A=0$ in \refeq{metric} does $\D\ln a = \mathcal{T}$ become an observable.  

In addition, the spatial displacement, decomposed into components parallel
($\Delta x_\parallel=\nhat^i\Delta x_i$) and 
perpendicular to the line of sight ($\Delta x_\perp^i={\cal P}^{ij}\Delta x_j$, where 
${\cal P}^{ij} \equiv \delta^{ij}-\nhat^i\nhat^j$ is the projection operator on the sky) 
can be written as 
\ba
\Delta x_\parallel =\:& -\int_0^{\tau_o} A(\v{0},\tau')  a(\tau')  \dd\tau' + \int_0^{\tilde \chi} d\chi\left[ A - B_\parallel - \frac12 h_\parallel
\right]_{\chi\vnhat,\,\tau_0-\chi}
- \frac{1+\tilde{z}}{H(\tilde{z})} \Delta \ln a\,.
\label{eq:DxparG}
\\
\Delta x_\perp^i  =\:& \left[\frac12 {\cal P}^{ij} (h_{jk})_o\, \nhat^k + B^i_{\perp o} - v^i_{\perp o}\right] \tilde{\chi} \vs
& - \int_0^{\tilde{\chi}} \dd\chi \bigg[
\frac{\tilde{\chi}}{\chi} \left(
B_\perp^i + {\cal P}^{ij} h_{jk}\nhat^k\right) 
+(\tilde{\chi}-\chi)\partial_\perp^i
\left( A - B_\parallel - \frac12 h_\parallel\right)
\bigg]_{\chi\vnhat,\,\tau_0-\chi}\,.
\label{eq:DxperpGl} 
\ea
Here, the spacetime arguments follow those in \refeq{Dlna}. 
We see that $\D\ln a$ appears in the line-of-sight displacement.  Further,
the combination of metric perturbations $A-B_\parallel - h_\parallel/2$ is
the gravitational lensing potential in a general gauge, which reduces to
$\Phi+\Psi$ in conformal-Newtonian gauge [\refeq{metriccN}].  The dominant
terms in the displacements on small scales are those which involve spatial
derivatives of the metric potentials.  This is $(v_\parallel-v_{\parallel o})$, which appears in $\Delta \ln a$ and is responsible for redshift-space distortions, and $\partial_\perp^i(A-B_\parallel - h_\parallel/2)$ in $\Delta x_\perp^i$
which is responsible for lensing magnification effects.

The displacement $\tilde x^\mu \to x^\mu = \tilde x^\mu + \Delta x^\mu$ defines a coordinate transformation from observed to true positions.  We can now derive how the galaxy density transforms under this coordinate transformation.  Specifically, the physical (rather than comoving) rest-frame galaxy density $\ngp$ is the 0-component 
of the galaxy four-momentum $\ngp u_g^\mu = \ngp u^\mu$ (since we restrict to large scales here, we neglect velocity bias, see \refsec{velbias}).  The number of galaxies within a
volume $V$ on the past light-cone of the observer, defined in terms of observed Cartesian coordinates $\tilde{\vx}$, is
then given by an integral of the three-form that is dual to $\ngp u^\mu$.  
In components, this becomes
\be
N(V) = \int_V d^3\tilde{\vx}\; 
\sqrt{-g(x)}\, \ngp(x) \varepsilon_{\mu\nu\rho\sigma}u^\mu(x)
\frac{\partial x^\nu}{\partial\tilde{x}^1}
\frac{\partial x^\rho}{\partial\tilde{x}^2}
\frac{\partial x^\sigma}{\partial\tilde{x}^3}\,,
\label{eq:N1}
\ee
where $\sqrt{-g(x)}$ is the determinant of the metric, $\ngp(x)$ is the physical number 
density of galaxies at spacetime position $x$, 
$u^\mu=dx^\mu/dt_F= a^{-1} (1-A, v^i)$
the galaxy velocity four-vector, and 
$\varepsilon_{\mu\nu\rho\sigma}$ the Levi-Civita symbol.  
\refeq{N1} is fully nonlinear, and we will evaluate it in two limits:  
first, restricting to linear order in perturbations but including all
relativistic terms;  second, working to nonlinear order in perturbations
but restricting the projection effects to those relevant on small scales, 
i.e. those terms that involve two spatial derivatives on metric perturbations.  
As discussed in \refsec{GR}, these two limits essentially cover the
parameter space where projection effects are numerically important.

\subsubsection{Observed galaxy density contrast at linear order}
\label{sec:GRdelta}

Combining the expression \refeq{N1} with \refeqs{Tdef}{DxperpGl}, we can
now obtain the fully relativistic expression for the observed galaxy density contrast 
at linear order in perturbations.  For this, we need an expression for the galaxy density 
$\ngp(x)$, which is, of course, dependent on the coordinate system
used to evaluate \refeq{N1}.  Rather than expressing $\ngp$ in terms of
the galaxy density perturbation $\d_g$ in some arbitrary gauge, which is
of course absolutely legitimate, we choose to fix coordinates to the 
\emph{constant-\textbf{o}bserved-\textbf{r}edshift} (``or'') \emph{gauge}.  
The reason is that the mean density of galaxies $\avngp$ is measured at
fixed observed redshift.  For this reason, observations of the galaxy
overdensity are naturally in the ``or'' gauge, simplifying the calculation
both practically and conceptually.  

We thus write $\ngp$ in terms of the mean \emph{comoving} number density $a^3 \avngp$ and the
perturbation $\d^{\rm or}_g$ to the comoving number density, in 
the constant-observed-redshift gauge, as
\be
a^3 \ngp(x) = \tilde{a}^3
\avngp(\tilde{z})
\left[
1 + \delta^{\rm or}_g(\vx,\zt)
\right]\,,
\label{eq:deltag1}
\ee
where $\tilde{z}$ is the observed redshift corresponding to the spacetime
location $x$, and $\tilde a = 1/(1+\tilde{z})$.  \refeq{deltag1} can be understood as the
definition of $\d^{\rm or}_g$.  At linear order in perturbations, 
we can neglect the distinction between $\vx(\tilde x)$ and
$\tilde\vx$ in the argument of $\d_g^{\rm or}$, since the latter is already 
first order.  Then, the right-hand side of \refeq{N1} becomes
\ba
N(V)
&= \int_V d^3\tilde{\vx}\:
\left(1+A+\frac{h}2\right)
\tilde{a}^3\avngp(\tilde{z})
\left[1+\delta^{\rm or}_g(\tilde{\vx},\zt) \right]
\left(
(1-A)\left|\frac{\partial x^i}{\partial \tilde{x}^j}\right|
+
v_\parallel
\right).
\label{eq:NVtrue}
\ea
The observationally inferred galaxy number density $\ngptilde$ is defined by evaluating \refeq{N1} on the unperturbed (fiducial) background, yielding
\be
N(V) = \int_V d^3 \tilde{\vx}\: \tilde{a}^3\ngptilde(\tilde{\vx},\zt)\,.
\label{eq:NVtilde}
\ee
By equating \refeq{NVtrue} and \refeq{NVtilde}, we find the observed galaxy density contrast as
\be
\tilde{\delta}_g(\tilde{\vx},\zt)
\equiv
\frac{\ngptilde(\tilde{\vx},\zt)}{\avngp(\tilde{z})}
-1
=
\delta^{\rm or}_g(\tilde{\vx},\zt) 
+\frac{h}2
+ \partial_{\parallel}\Delta x_\parallel
+
\frac{2\Delta x_\parallel}{\tilde{\chi}} - 2 \hat{\kappa}
+
v_\parallel\,.
\label{eq:deltag22}
\ee
Here, $h \equiv \d^{ij} h_{ij}$, and we have expanded the Jacobian through
\be
\left|\frac{\partial x^i}{\partial \tilde{x}^j}\right|
=
1+\frac{\partial \Delta x^i}{\partial \tilde{x}^i}
=
1 
+ \partial_{\parallel}\Delta x_\parallel
+
\frac{2\Delta x_\parallel}{\tilde{\chi}} - 2 \hat{\kappa}\,,
\ee
where the \emph{coordinate} (i.e. gauge-dependent) lensing convergence $\hat{\kappa}$ is defined as the divergence on the sky of the perpendicular displacement,
\be
\hat\kappa
=
-\frac12 {\cal P}^{ij}\partial_{i}\Delta x_{\perp j}\,.
\ee
All contributions in \refeq{deltag22} apart from $\d_g^{\rm or}$ are the 
induced, apparent modulation of the galaxy abundance due to volume distortion
effects.   

Next, we have to relate $\delta^{\rm or}_g$ in \refEq{deltag22} to the matter
density through a bias relation.  
The galaxy density contrast $\delta^{\rm or}_g$ in \refEq{deltag22} is defined in the 
constant-observed-redshift slicing.  On the other hand, as discussed in detail in \refsec{GR}, the linear bias relation between galaxy density contrast and matter density contrast is valid on \emph{constant-\textbf{p}roper-\textbf{t}ime} (``pt'') slices: $\d_g^{\rm pt} = b_1 \d^{\rm pt}$.  This is because in the
large-scale limit, galaxies only know about the local age of the Universe
and the local matter density.\footnote{This assumes Gaussian initial conditions, and that there are no additional degrees of freedom relevant on large scales, such as dark energy perturbations, or modified gravitational forces.  The generalizations are discussed in \refsec{NG} and \refsec{beyondCDM}, respectively, and can easily be included here.}  
At linear order, the transformation of the galaxy density between the two gauges is completely determined by the time shift between them.  This shift is precisely the observable $\T$ that we have derived in \refeq{Tdef}.  
Then, the relation between $\delta^{\rm or}_g(\tilde{\vx},\zt)$ and the galaxy density perturbation $\d_g^{\rm pt} = b_1 \d^{\rm pt}$ in the constant-proper-time (or synchronous) gauge is given by a standard, linear gauge transformation,
\be
\delta^{\rm or}_g(\tilde{\vx},\zt) 
=  b_1\,\delta^{\rm pt}(\tilde{\vx},\zt)  
 + b_e \T(\tilde{\vx},\zt)\, ,\qquad
b_e \equiv \frac{d\ln (a^3 \avngp)}{d\ln a}\,,
\label{eq:bias}
\ee
where we have introduced the dimensionless parameter $b_e$
quantifying the evolution of the mean comoving number density of galaxies.  
Note that this relation only involves observable quantities, so that
both $b_1$ and $b_e$ are well defined and gauge-invariant.  It also serves as the
unambiguous starting point for extending the bias relation to higher order
in perturbations, for example by adding a term $(b_2/2)(\d^{\rm pt})^2$ to
the right-hand side.  For halos following a universal mass function,
Ref.~\cite{gaugePk} derived $b_e=\dc f(b_1-1)$.  
 
Finally, $\d^{\rm pt}$ is related to the matter density
perturbation $\d$ in the chosen gauge through
\be
\delta^{\rm pt} 
= \delta + 3 \tilde H\int_0^{\tilde{\tau}} A(\vx,\tau)a(\tau) d\tau\,.
\ee
Combining the last two equations, we find the galaxy density contrast on the 
constant-observed-redshift slice in terms of the density contrast in 
an arbitrary gauge as
\be
\delta^{\rm or}_g(\tilde{\vx},\zt) 
=
b_1\left[\delta 
+ 3\tilde{H}\int_0^{\tilde{\tau}} A(\tilde{\vx},\tau)a(\tau) d\tau\right]
+ 
b_e \T.
\ee
This yields our final expression: 
\be
\tilde \d_g(\tilde{\vx},\zt) 
= b_1\left[\delta + 3\tilde{H}\int_0^{\tilde{\tau}} A(\tilde{\vx},\tau)a(\tau) d\tau\right]
+ b_e \,\T 
+ \frac12 h + \partial_{\chit} \D x_\parallel + \frac{2\D x_\parallel}{\chit} - 2\hat{\kappa} + v_\parallel\,.
\label{eq:dgtilde}
\ee
Here, the line-of-sight derivative of the longitudinal displacement is given by
\ba
\partial_{\chit} \D x_\parallel 
=\:&
A - B_\parallel - \frac12 h_\parallel
- H(\tilde{z})\left(\frac{\partial}{\partial\tilde{z}}\frac{1+\tilde{z}}{H(\tilde{z})}\right) \Delta\ln a \vs
&
- \frac{1+\tilde{z}}{H(\tilde{z})}
\left(
-\partial_\parallel A
+
\partial_\parallel v_\parallel
-\frac{\partial}{\partial\tau}v_\parallel
+ \frac12 \frac{\partial}{\partial\tau}h_\parallel + \frac{\partial}{\partial\tau}{B}_\parallel
\right)\,.
\ea

One subtlety we have neglected so far is that observational selection 
effects can modify the observed galaxy density, \refEq{dgtilde}.  
Usually, surveys observe galaxies above a certain apparent flux, or magnitude threshold.
Weak lensing magnifies/de-magnifies the flux of the source galaxies and 
therefore induces another contribution to the observed galaxy density
(\emph{magnification bias} \cite{moessner/jain:1998}).  
For a population of galaxies at fixed redshift $\tilde{z}$ with cumulative 
luminosity function $\avngp(>L_{\rm min})$, we define 
\be
\Q \equiv -\frac{d\ln \avngp(>L_{\rm min})}{d\ln L_{\rm min}} \,.
\ee
More generally, other selection effects, such as galaxy size, can contribute to $\Q$ \cite{schmidt/etal:2009b}. In that case, the coefficient should be defined as $\Q = d\ln \avngp/d\Mag$, where $\Mag$ is the gauge-invariant magnification. 
Then, magnification bias adds a contribution $\Q\,\Mag$ to $\tilde\d_g$,
where 
\be
\Mag = -2\Delta \ln a - \frac12 (h-h_\parallel) - \frac{2}{\tilde{\chi}}
\Delta x_\parallel + 2 \hat{\kappa}\,.
\ee
Note that we neglect the effect from the evolution of 
intrinsic luminosity of the galaxies which can in principle contribute 
to the magnification 
(see \cite{schmidt/jeong:2012} for the complete expression); 
this contribution is typically much smaller than $\Mag$ itself.  
We finally obtain the observed density contrast including magnification bias as
\ba
\tilde \d_g(\tilde{\vx},\zt) 
=\:& b_1\left[\delta + 3\tilde{H}\int_0^{\tilde{\tau}} A(\tilde{\vx},\tau)a(\tau) d\tau\right]
+ b_e \T
+ 2\Q\tilde{H}\int_0^{\tilde{\tau}} A(\tilde{\vx},\tau)a(\tau) d\tau
+ \frac12(1-\Q) h + \frac{\Q}{2}h_\parallel 
\vs&+ \partial_{\chit} \D x_\parallel 
+ (1-\Q)\frac{2}{\chit} \D x_\parallel
+ 2(\Q-1)\hat\kappa + v_\parallel\,.
\label{eq:dgtilde_M}
\ea
\refeq{dgtilde_M} provides the complete result for the observed overdensity
of a tracer at linear order in a general gauge.
When restricted to conformal-Newtonian gauge, this agrees with
\cite{challinor/lewis:2011,baldauf/etal:2011} (note the discussion around Eq.~(31)
of the former reference);  restricting to synchronous-comoving gauge 
yields the results derived in \cite{gaugePk}.  
Here, we have assumed a sharp source redshift, but the projection over a 
finite redshift bin is straightforward.  

\begin{figure}[t!]
\begin{center}
\rotatebox{90}{
\includegraphics[width=0.7\textwidth]{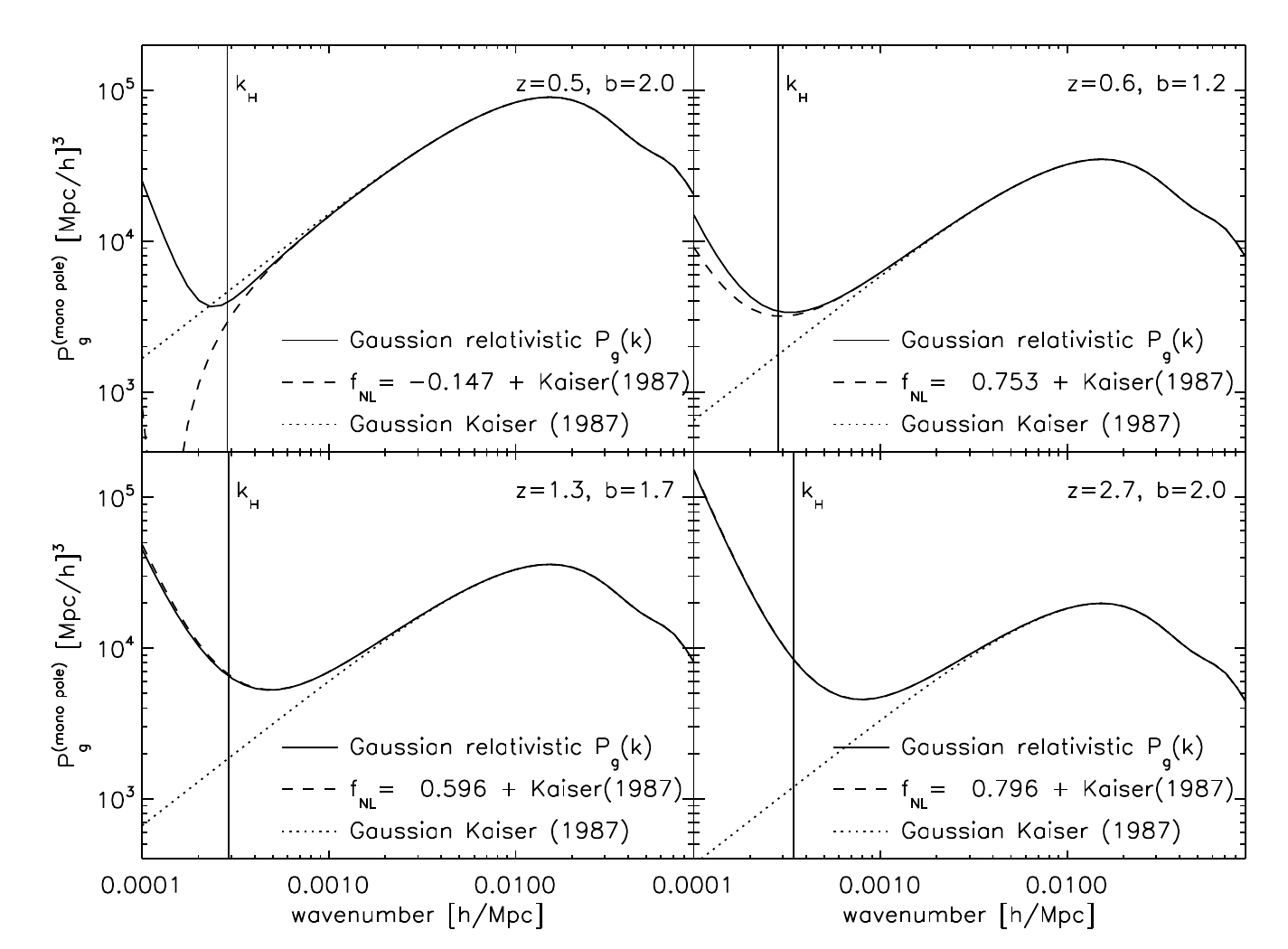}
}
\end{center}
\caption{
Linear galaxy power spectrum including relativistic projection  
effects following \refeq{dgtilde_M} (solid lines), and corresponding
prediction keeping only terms relevant on small scales, \refeq{dgleading} (dotted).  
The four panels show different values of linear bias parameters (with $b\equiv b_1$) and redshift. $b_e$ is calculated by using
the prediction for the universal mass function, $b_e=\dc f(b_1-1)$ \cite{gaugePk}. 
In each case, we perform an angle-average over $\hat{\vk}$ (power spectrum monopole), and drop contributions that are integrated along the line of sight (see discussion in the text).  
Vertical lines show the comoving horizon wavenumber ($k_H=\cH$) at each redshift.  
For comparison, we also show the galaxy power spectrum in the small-scale approximation but including the scale-dependent bias induced by local-type PNG 
via \refeq{Pk_localfnl},  
with a value of $\fnl$ adjusted in each case
to match the amplitude of the relativistic projection effects (dashed lines).  
Depending on bias and redshift, the relativistic projection effects 
are of the same order as the scale-dependent bias for PNG with $\fnl\lesssim 1$.  
\figsource{gaugePk}  
\label{fig:GRfNL}
}
\end{figure}

Assuming that the coefficients $b,\,b_e,\,\Q$ are all of order unity, the
various terms in \refeq{dgtilde_M} can be ranked in terms of relative
importance according to their scaling, relative to $\d$, with $\cH/k$ 
in Fourier space.  The largest terms, ``order 1'', are in conformal-Newtonian gauge given by
\be
\tilde\d_g^{\O(1)} = b_1\, \d + \frac{1+\zt}{\tilde H} \partial_\parallel v_\parallel + 2(\Q-1) \hat\kappa\,.
\label{eq:dgleading}
\ee
Note that in conformal-Newtonian gauge, $\d = \d^{\rm pt} + \O(\cH^2/k^2)$.  
\refEq{dgleading} is the standard small-scale result for the apparent galaxy overdensity,
including the leading redshift-space distortion \cite{kaiser:1987} and magnification bias.  
The subleading corrections in \refeq{dgtilde_M} scale as $i \cH/k$ (``velocity-type'') and $(\cH/k)^2$ (``potential-type'').  
\reffig{GRfNL} shows the angle-averaged auto-power spectrum of the galaxy density including relativistic projection effects [\refeq{dgtilde_M}, solid line], and in the small-scale approximation [\refeq{dgleading}, using $\d^{\rm pt}$, dotted line].  We have dropped terms that are integrated along the line of sight.  These cannot be represented by a three-dimensional power spectrum.  Moreover, on such large scales, employing the flat-sky approximation is not sufficient (we will generalize this in \refsec{lightcone}).  Thus, \reffig{GRfNL} is only meant as an order-of-magnitude illustration of the relativistic projection effects, which clearly become numerically relevant only for $k/k_H \equiv k/\cH \lesssim 10$.  

The power spectrum of the relativistic linear galaxy density perturbation \refeq{dgtilde_M}, again neglecting all contributions that involve integrals along the line of sight, contains three types of contributions beyond the small-scale limit \refeq{dgleading}: the auto-correlation of velocity-type terms, which scales as $(\cH/k)^2 \Plin(k)$, compared to the leading contribution $\propto \Plin(k)$;  the cross-correlation of potential-type terms with leading terms, which obeys the same scaling;  and the auto-correlation of potential-type terms, which scales as $(\cH/k)^4\Plin(k)$.  Note that the cross-correlation of velocity-type terms with either potential- or density-type terms vanishes by symmetry when considering the auto-correlation of any tracer.  
The first two contributions have the same
$k$-dependence on large scales as the leading contribution to $P_{gg}(k)$ from the scale-dependent bias $\D b(k)$ induced by local-type PNG \cite{dalal/etal:2008}, \refeq{Pk_localfnl}.  We can thus find the effective $\fnl$ that would lead to a scale-dependent term of similar magnitude.  These results are also shown in \reffig{GRfNL}.  We see that
relativistic projection effects amount to the effect of local primordial non-Gaussianity 
with $\fnl \lesssim 1$ \cite{gaugePk,bruni/etal:2012} for typical expected values of $b_1, b_e, \Q$.  
Given our forecasts in \refsec{fnl_fisher}, the projection effects
of potential-type are thus expected to be marginally detectable in future
galaxy surveys.  The velocity-type 
contributions $\propto \cH/k$, which come from terms involving $v_\parallel$ and $\partial_\parallel \Psi$
in case of conformal-Newtonian gauge, are larger and more easily detectable
\cite{hamaus/seljak/desjacques:2011,yoo/etal:2012,bonvin/hui/gaztanaga:2014}.  As mentioned above however, 
they cancel out in the auto-correlation of tracers, and necessitate the
use of two different tracers to measure the dipole of their cross-power spectrum.  
Note that projection effects will also amount to a non-zero $\fnl\lesssim 1$ in the galaxy bispectrum,
although the exact amplitude of this effect is still being debated in the literature
\cite{bertacca/martens/clarkson:2014,yoo/zaldarriaga:2014,didio/etal:2016,umeh/jolicoeur/etal:2016,
didio/perrier/etal:2016}.

Ref.~\cite{GWpaper} derived the contribution of tensor modes to the observed galaxy clustering statistics, which, at linear order, only enter through projection effects.  Unfortunately, for currently allowed amplitudes of primordial gravitational waves, these contributions are much smaller than even the relativistic projection effects from scalar perturbations shown in \reffig{GRfNL}.

\subsubsection{Nonlinear galaxy density contrast in redshift space}
\label{sec:RSD}

We now turn to the projection effects on small scales, where the
``potential-type'' terms discussed at the end of the \refsec{GRdelta} 
can be neglected.  The dominant projection contributions to the galaxy density then
come from the line-of-sight velocity $v_\parallel$ and its derivatives, and
from the lensing convergence $\hat\kappa$.  In the following, we will neglect 
the contribution from $\hat\kappa$, which, while of the same order in derivatives as the other leading terms, is suppressed by the integration over the line of sight.  However, this effect can be taken into account in a straightforward manner \cite{hui/etal:2007,hui/etal:2008,schmidt/etal:2008}.  

On the other hand, on small scales linear perturbation theory no longer applies.  Hence, we now evaluate \refeq{N1} without expanding at linear order in perturbations, but instead making the approximations 
\be
{\cal T}\to0,~\Delta x_\parallel 
\to -\frac{1+\tilde{z}}{H(\tilde{z})} v_\parallel
,~\Delta x_\perp \to 0\,.
\label{eq:small_scale}
\ee
Here, we have also dropped the observer's velocity $v_{\parallel o}$, since it
only contributes to the dipole of the galaxy density and
can thus also be neglected when focusing on small scales.  
Further, we will continue to denote the galaxy velocity as $\v{v}$ for clarity, although this can now differ from the matter velocity due to velocity bias; the matter velocity will never appear explicitly in the following, however.
With these assumptions, \refeq{N1} becomes
\be
N(V) = 
\int_V d^3\tilde{\vx} \,
a^3
\avngp(z)(1+\delta_g(\vx[\tilde{\vx}]))
\left|
\frac{\partial x^i}{\partial \tilde{x}^j}
\right|
\equiv
\int_V d^3\tilde{\vx} \,
\tilde{a}^3
\avngp(\tilde{z})(1+\tilde{\delta}_g(\tilde{\vx},\zt))\,,
\ee
where, from \refeq{small_scale} 
\be
\tilde{x}^i = x^i + u_\parallel(\vx) \hat{n}^i
\label{eq:rsd_cT}
\ee
is the (purely spatial) coordinate shift between real space ($x^i$) and 
redshift space ($\tilde{x}^i$) coordinates, using the scaled galaxy velocity
${\vu}(\vx)\equiv \v{v}(\vx)/\cH$, and $u_\parallel(\vx)\equiv \hat{\bm n}\cdot\vu(\vx)$.  
In general, the inverse function $\vx(\tilde{\vx})$ of \refeq{rsd_cT} is multi-valued.  Therefore, a simply connected volume $V$ in observed
coordinates does not in general correspond to a simply connected region in 
real space.
This multi-to-one mapping of the redshift-space distortion 
can happen for galaxies residing in massive halos which exhibit large peculiar (virial) velocities.   
In the following, we will ignore this non-perturbative effect, since our 
focus is on the perturbative description of galaxy clustering. 
Then, we can invert \refeq{rsd_cT} and write the observed galaxy density perturbation as 
\be
1+\tilde{\delta}_g(\tilde{\bfx})
=
\left[1+\delta_g(\bfx)\right]
\left|
\frac{\partial x^i}{\partial \tilde{x}^j}
\right|
=
\left[1+\delta_g(\bfx)\right]
\left|
\delta_{i}^j
+
\hat{n}^j \frac{\partial}{\partial x^i} u_\parallel(\vx)
\right|^{-1}.
\ee
Here and in the following, the galaxy density and velocity will always be evaluated at the observed redshift $\zt$,
i.e. at coordinate time $\bar\tau(\zt)$. Hence, we drop the time argument in what follows for clarity.
Sylvester's theorem allows us to evaluate the determinant via
\be
\left|
\delta_{i}^j
+
\hat{n}^j \frac{\partial}{\partial x^i} u_\parallel(\vx)
\right| = 1 + \nhat^j \partial_j u_\parallel(\vx) = 1 + \partial_\parallel u_\parallel(\vx)\,,
\ee
and we obtain
\be
\tilde{\delta}_g(\tilde{\bfx})
=
\frac{1+\delta_g(\bfx)}{1+\partial_\parallel u_\parallel(\bfx)}
-1\,.
\label{eq:delta_rsd_full}
\ee 
This result is the \emph{fully nonlinear} expression of the redshift-space galaxy 
density contrast in configuration space, as long as \refeq{rsd_cT} describes a one-to-one mapping from real to redshift space.  
Finally, we evaluate the right-hand side of the above equation using 
the quantities at the observed (redshift-space) coordinate. 
For some scalar function $f(\bfx)$ (e.g., $\d_g(\vx)$), expanding $\vx$ around $\tilde{\vx}$ via \refeq{rsd_cT} yields
\be
f(\bfx) 
=
f(\tilde{\bfx})
-
\left[
1 - 
\partial_\parallel u_\parallel
\right]
u_\parallel
\partial_\parallel f(\tilde{\vx})
+
\frac12
u_\parallel^2
\partial_\parallel^2f(\tilde{\vx})
+
\O( u_\parallel^3 )\,,
\label{eq:fxtilde}
\ee
where all fields on the right-hand side are evaluated at $\tilde{\vx}$,
and the $1-\partial_\parallel u_\parallel$ factor comes from 
expanding the argument of $u_\parallel(\vx)$.  
Note that, even though the velocity appears without a spatial derivative here,\footnote{Strictly speaking, $u_\parallel$ denotes the relative velocity between source and observer [cf.~\refeq{Dlna}], which is boost-invariant and observable.}
it always enters in combination with a spatial derivative acting on 
another perturbative quantity.  Hence, terms of order $u_\parallel \partial_\parallel$ in the expansion
\refeq{fxtilde} are of the same order in derivatives as $\partial_\parallel u_\parallel$.  

We can now expand the right-hand side of \refeq{delta_rsd_full} to obtain
the perturbative expression for the observed \emph{galaxy density in redshift space}, i.e. including the leading projection effects on small scales.  
Up to third order in perturbations, counting each power of $\d$ as well as $u_\parallel$, \refeq{delta_rsd_full} becomes
\ba
\tilde{\delta}_g(\tilde{\bfx})
=\:&
\delta_g(\bfx)
+
\left[
1+
\delta_g(\bfx)
\right]
\left\{
-
\partial_\parallel u_\parallel(\bfx)
+
\left[\partial_\parallel u_\parallel(\bfx)\right]^2
\right\}
-
\left[\partial_\parallel u_\parallel(\bfx)\right]^3
+ \O(\d^4)\,.
\ea
Applying the coordinate transformation relation in \refeq{fxtilde}, we 
obtain the third-order galaxy density perturbation in redshift space as 
\ba
\tilde{\delta}_g
=\:&
\delta_g
-
\partial_\parallel \left[ u_\parallel \left(1+\delta_g\right) \right]
+
\frac12 \partial_\parallel^2 \left[u_\parallel^2\left(1+\delta_g\right)\right]
-
\frac16 \partial_\parallel^3 \left(u_\parallel^3\right)
+ \O(\d^4)\,,
\label{eq:RSD_real}
\ea
where all quantities are evaluated at the redshift-space coordinate, 
$\tilde{\bfx}$.  This expression agrees with the Fourier-space expression 
derived in \cite{scoccimarro:2004,jeong:2010},
\be
\tilde{\delta}_g(\bfk)
=
\delta_g(\bfk)
+
\int\,d^3{\bfx}\, e^{-i\bfk\cdot\bfx}
\left(
e^{-ik_\parallel u_\parallel(\bfx) }
-1
\right)
\left[
1+\delta_g(\bfx)
\right]\,.
\ee
This in turn is formally equivalent in configuration space to a series expansion valid at all orders,
\be
\tilde{\delta}_g
=
\delta_g
+
\sum_{n=1}^\infty \frac{(-1)^n}{n!}\partial_\parallel^n
\left[
u_\parallel^n (1+\delta_g)
\right]\,.
\label{eq:RSD_real2}
\ee

The contributions in \refeq{RSD_real} that involve $u_\parallel$ are known as \emph{redshift-space distortions (RSD)}. 
We now derive the effect of RSD on observed galaxy statistics, in particular the leading-order power spectrum. This calculation simplifies considerably in the flat-sky limit, i.e. when the line of sight $\vnhat$ is taken to be a constant unit vector. We will describe approaches to galaxy clustering on the full sky in the next section. First, at linear order in perturbation theory, we have
\be
u_\parallel(\vk,\tau) = i \frac{\vk\cdot\vnhat}{k^2} f\,\d^{(1)}(\vk,\tau)\,,
\ee
where $f = d\ln D/d\ln a$ is the linear growth rate.  
The leading-order prediction for the galaxy power spectrum in redshift space in this limit (``Kaiser formula'') is given by
\be
P_{g,s}^\LO(k,\mu) = (b_1 + f\mu^2)^2 \Plin(k) + \Peps\,,
\label{eq:PkKaiser}
\ee
where $\mu \equiv \hat{\vk}\cdot\vnhat$ is the cosine of the wavevector with the line of sight. This expression was first derived by \cite{kaiser:1987}; the corresponding derivation in configuration space can be found in \cite{hamilton:1992}. Note that we use the term ``configuration space'' for the complement to Fourier space, to distinguish from ``real space,'' which is commonly used to denote the rest-frame galaxy density. These results correspond to the auto-power spectrum and correlation function, respectively, of \refeq{dgleading} in the flat-sky limit and when neglecting magnification bias. 
Note that at this order, RSD do not add any free parameters to the galaxy power spectrum. Instead, the anisotropy ($\mu$-dependence) of the observed galaxy power spectrum can be used to constrain the growth rate $f$. Thus, a measurement of RSD on large scales yields constraints on the growth of structure without the complications of bias \cite{matsubara:2004,percival/white:2009}. 

The idea of using the observed anisotropic distribution of 
galaxies in redshift space as an indicator of velocity structure was first
proposed by \cite{Jackson:1972} and further extended by 
\cite{sargent/turner:1977}, who showed that the mean anisotropy of the galaxy
two-point function is a probe of the matter density parameter $\Om$, via the growth rate $f$.  
This technique has been applied
to real data, providing constraints on modified gravity and non-canonical dark energy \cite{guzzo/etal:2008,blake/etal:2011,reid/etal:2012,beutler/etal:2012,samushia/etal:2014,satpathy/etal:2016}.
The multi-tracer method discussed in \refsec{NG:mt} can also be applied to RSD, improving constraints on the growth rate, provided that the stochasticity $\Peps$ of the galaxy samples is well understood
\cite{mcdonald/seljak:2009,gil-marin/etal:2010,bernstein/cai:2011,hamaus/seljak/desjacques:2011,abramo:2012}. This has recently been applied to the GAMA survey  sample \cite{blake/etal:2013}.
Ref.~\cite{hamilton:1998} provides a pedagogical review on 
redshift-space distortions at linear order.  

It is important to stress that \refeq{PkKaiser} can only be
considered a clean probe of velocities if the real-space galaxy density
$\d_g$ itself does not depend on $\partial_\parallel u_\parallel$.  However, since, at linear order,
$\partial_\parallel u_\parallel \propto \partial_\parallel^2\Phi \propto K_{ij} \hat n^i \hat n^j$, 
this is only true if the galaxy density does not depend on the tidal field
projected along the line of sight.  As we have seen in \refsec{selection},
such a dependence can be induced by selection or radiative-transfer effects.  
If those are present, leading-order RSD no longer provide a direct probe of velocities.  
In case of the Lyman-$\alpha$ forest, for example,
such radiative-transfer effects are important \citep[see, e.g.,][]{gunn/peterson:1965,croft/weinberg/etal:1998,croft/weinberg/etal:1999,viel/haehnelt/springel:2004,
mcdonald/seljak/etal:2005,mcdonald/seljak/etal:2006,slosar/etal:2011,seljak:2012,palanque/bosslya:2013,gontcho/etal:2014,
lee/hennawi/etal:2014}, as the optical depth of Lyman-$\alpha$ photons depends strongly on the velocity gradient along the line of sight.

Going beyond linear order, other references considered the galaxy three-point function \cite{gaztanaga/scoccimarro:2005} and bispectrum \cite{verde/etal:1998,scoccimarro/etal:1999} in redshift space.  
Ref.~\cite{mcCullagh/szalay:2014} presented the
calculation of RSD in the two-point function using the Zel'dovich approximation.  
The formulation of RSD in Lagrangian perturbation theory can be found in \cite{hivon:1995, sato/matsubara:2011}. 
For the galaxy power spectrum, the expressions derived in either \cite{scoccimarro:2004} (``Scoccimarro'' model) or \cite{taruya/nishimichi/saito:2010} (``TNS'' model) are widely applied for the interpretation of galaxy survey data beyond linear order. It is worth noting that the coupling between nonlinear bias and nonlinear RSD, as well as higher-derivative, velocity-bias and stochastic-bias contributions, are not completely accounted for in the expressions given in these references. Recently, Ref.~\cite{perko/etal:2016} (see also \cite{fonseca/etal:2018,pkgspaper}) presented complete expressions for the NLO galaxy power spectrum in redshift space (that is, the generalization of the results presented in \refsec{npt1loop} to redshift space) including the velocity bias contributions which become relevant at this order. In the context of the general bias expansion, the NLO redshift-space contributions to the galaxy power spectrum add three additional parameters, which are all related to velocity bias. Two parameters describe deterministic velocity bias (see \refsec{velbias}), while the third describes the amplitude of random, small-scale galaxy velocities (see \refsec{stoch}). It is worth emphasizing that displacement-type terms such as $u_\parallel\partial_\parallel \d_g$, which become relevant beyond the leading-order galaxy two-point function, are protected from selection and radiative-transfer effects as well as velocity bias (on large scales) through the equivalence principle. Thus, even in the presence of all these observational complications, they in principle allow for robust constraints on the growth of structure from observed galaxy statistics.

While the rigorous perturbative approach (EFT of matter coupled with the general bias expansion) is an approach whose advantages have been explained in detail in \refsec{evolution} and \refsec{measurements}, 
it is worth pointing out that alternative approaches to the modeling of small-scale RSD have been developed \cite{tinker/etal:2006,tinker:2007,reid/etal:2014}
in the context of the halo occupation distributions (HOD, see \refsec{HOD}).

Finally, redshift-space distortions also complicate significantly the interpretation of nonlinear transformations of the galaxy density field, such as void 
catalogs \cite{shoji/lee:2012,hamaus/etal:2015,chuang/etal:2016,cai/etal:2016,VIMOS_VOID} 
and  ``clipped'' \cite{simpson/etal:2011,simpson/etal:2016}
or log-transformed galaxy density fields \cite{hubble:1934,hamilton:1985,coles/jones:1991,falck/etal:2012,mccullagh/etal:2013,carron/szapudi:2013}.

\subsection{Galaxy statistics on the sky}\label{sec:lightcone}

After having described how the rest-frame galaxy density is transformed to
observed coordinates ($\tilde z, \vnhat$), we are now in a position
to describe the statistics of galaxies in these observed coordinates beyond
the flat-sky limit adopted in the previous section.
We now consider the fully general case, allowing for wide angles
[see \cite{szalay/etal:1997,matsubara:1999b,bharadwaj:1999,szapudi:2004,papai/szapudi:2008,raccanelli/etal:2010} who use \refeq{dgleading}, and \cite{yoo/desjacques:2013,raccanelli/etal:2013} for treatments including the full expression \refeq{dgtilde_M}], and for deep surveys. 
This is of particular importance when one is interested in measuring 
long-wavelength fluctuations which extend across a significant redshift range 
and/or portion of the sky, for example to probe the signature of 
primordial non-Gaussianity.  
Here, we focus on the galaxy two-point function (see \cite{scoccimarro:2000,verde/heavens/matarrese:2000,didio/etal:2016} for extensions to the 
three-point function allowing for wide angles), and neglect subtleties involved in its optimal estimation (such as discussed in \cite{hamilton:1993}).  
Moreover, we will not consider the effect of the survey window, which
further complicates the treatment.  
Since we will always deal with observed, redshift-space coordinates in this
section, we will drop the tilde, i.e. $\tilde{\vx}\to \vx$.  

The standard approach of analyzing galaxy statistics on the sky
is to decompose the observed galaxy overdensity in terms of spherical 
harmonics $Y_{lm}$, and radial window functions $W_l(p,\chi)$, labeled by
a parameter $p$:
\ba
&\tilde{\d}_g(p;\vnhat) = \sum_{l m} \delta_{l m}(p) Y_{l m}(\vnhat),
\qquad\mbox{where}~~\d_{l m}(p) = \int_0^\infty \chi^2 d\chi \int d^2\vnhat\: Y_{l m}^*(\vnhat) 
W_l(p, \chi) \d_g(\vx,\tau[\chi])\,,
\label{eq:deltalm}
\ea
and $\chi$ is the comoving radial coordinate, $W_l(p,\chi)$ is an $l$-dependent window function parametrized by a parameter $p$ (specified below), and
\be
\tilde{\d}_g(\vx,\tau) = \frac{n_g^{\rm obs}(\vx,\tau)}{\avng^{\rm obs}(\tau)} - 1
\ee
is the \emph{observed} fractional galaxy overdensity.  Throughout, we let
$\chi = \bar\chi(\tilde z)$ be a proxy for the observed redshift $\tilde{z}$, 
and similarly $\tau \equiv \bar\tau(\tilde z)$ (see \refsec{projection}). 
Correspondingly, $\vx \equiv \chi \vnhat $ parametrizes the observationally 
inferred (redshift-space) spatial location. Similarly, $\avng^{\rm obs}$ is 
defined on a constant-observed-redshift slice. 

The parameter $p$ of the window function $W_l(p,\chi)$ parametrizes the radial
selection function.  
One option is to choose $W_l(p,\chi) = \chi^{-2} \d_D(\chi - \bar\chi[p])$, 
which yields, at the two-point level, the angular two-point correlation function 
in narrow redshift bins $C_l(z,z')$ \cite{matarrese/etal:97,challinor/lewis:2011,bonvin/durrer:2011,GWpaper}.  
Alternatively, choosing $W_l(p,\chi) = j_l(p\chi)$, we recover the spherical 
Fourier-Bessel approach developed by 
\cite{binney/quinn:1991,fisher/lahav/etal:1995,heavens/taylor:1995}. 
The spherical Fourier-Bessel basis has been applied to the analysis of the IRAS 
Redshift Survey \cite{fisher/scharf/etal:1994}, and to the reconstruction 
of the velocity and the gravitational potential fields \cite{fisher/lahav/etal:1995}.  
More recently, Ref.~\cite{rassat/refregier:2012} computed the spherical power spectrum of matter density 
fluctuations on the BAO scale, Ref.~\cite{shapiro/crittenden/percival:2012} cross-correlated their spherical Fourier-Bessel analysis of RSD with 
the CMB temperature anisotropies, while 
Ref.~\cite{yoo/desjacques:2013} computed the spherical power spectrum including
all terms in \refeq{dgtilde_M}.  The advantages and disadvantages of the spherical 
Fourier-Bessel decomposition with respect to angular correlations in redshift bins $C_l(z_i,z_j)$ are reviewed in \cite{nicola/refregier/etal:2014}.  
Ref.~\cite{salazar/etal:2016} applied this method to the latest data release
of the BOSS CMASS sample from the Sloan Digital Sky Survey.  
Finally, Ref. \cite{hamilton:1996} proposed the so-called logarithmic spherical 
waves $e^{i\omega \ln r}$ which can be used to express 
``Pseudo-Karhunen-Lo\`{e}ve'' (signal-to-noise) eigenmodes.

Let us then write the redshift-space galaxy density as
\be
\tilde{\d}_g(\vx,\tau) = \sum_{O} b_{O}(\tau) O_s(\vx,\tau)\,,
\ee
where the $O_s$ are renormalized redshift-space operators.  For now, we 
neglect stochasticity, but return to this below.  
Since the effect of tensor perturbations on the galaxy density contrast is
negligibly small \cite{GWpaper}, we can restrict to scalar perturbations
here.  Then, 
any linear redshift-space operator $O_s$ can be derived from 
a real-space scalar $O_r$ [cf. \refeq{dgleading} and \refeq{RSD_real2}] as,
\be
O_s(\vx,\tau) = \vnhat^{i_1} \cdots \vnhat^{i_n} \partial_{i_1} \cdots \partial_{i_n} O_r(\vx,\tau)\,,
\ee
where $n$ can be any number, although the leading terms only involve $n=1$ and $n=2$.  Equivalently, in Fourier space
\be
O_s(\vk,\tau) = (i\vnhat\cdot\vk)^n O_r(\vk,\tau)\,.
\ee
More generally, this applies to all leading-order contributions to
redshift-space $N$-point functions.  At loop level, $\vnhat$ can also be
contracted with the loop momentum, which makes the derivations in the 
following more complicated.  However, loop terms only become important
on small scales, where the flat-sky limit applies and the treatment becomes
much simpler.  We will thus restrict to leading-order correlations, and 
consider in particular the two-point function.  Then, the operators $O_s$ are linear order in perturbation.

Using homogeneity and isotropy of the two-point function
of the real-space scalar $O_r$, the galaxy two-point function in redshift-space can be written as
\ba
\< \tilde{\d}_g(\vx,\tau) \tilde{\d}_g(\vx',\tau') \> =\:& \sum_{O_s,O_s'} b_{O_s}(\tau) b_{O_s'}(\tau')
\int_{\vk} (i\vk\cdot\vnhat)^n  (-i\vk\cdot\vnhat')^{n'} P_{O_rO_r'}(k; \tau,\tau' )
\exp\left[i \vk\cdot(\vx-\vx')\right] \vs
=\:& \sum_{O_s,O_s'} b_{O_s}(\tau) b_{O_s'}(\tau')
\int_{\vk} \left(\frac{\partial}{\partial \chi}\right)^n 
e^{ik\chi(\vkhat\cdot\vnhat)}
\left(\frac{\partial}{\partial \chi'}\right)^{n'}
e^{-ik\chi'(\vkhat\cdot\vnhat')}
P_{O_rO_r'}(k; \tau,\tau' ) \vs
=\:& \sum_{O_s,O_s'} b_{O_s}(\tau) b_{O_s'}(\tau') \frac2\pi \sum_{l,m}
Y_{lm}(\vnhat) Y_{lm}^*(\vnhat') \vs
& \qquad\qquad\times
\int k^2 dk \,k^{n+n'}\left(\frac{\partial}{\partial x}\right)^n j_l(x) 
\left(\frac{\partial}{\partial x'}\right)^{n'} j_{l'}(x') 
P_{O_rO_r'}(k; \tau,\tau' )\,,
\ea
where $x=k\chi,\,x'=k\chi'$.  Inserting this into \refeq{deltalm}, we obtain the  
two-point correlation of the $\{l,m,p\}$-decomposition of the galaxy
overdensity as
\ba
\left\< \d_{lm}(p) \d_{l'm'}^*(p')\right\> =\:& \d_{ll'}\d_{mm'} \sum_{O_s,O_s'} \frac2\pi
\int k^2 dk\, F_{O_s}^l(k;p) F_{O_s'}^l(k;p') P_{O_rO_r'}(k)\,,
\label{eq:dglm2pt}
\ea
where
\ba
F_{O_s}^l(k;p) \equiv\:& k^{n_O} \int \chi^2 d\chi\,W_l(p,\chi) \left[\left(\frac{\partial}{\partial x}\right)^{n_O} j_l(x)\right]_{x=k\chi} b_{O_s}(\tau[\chi]) D_{O_r}(\tau[\chi]) \quad\mbox{and}\vs
P_{O_rO_r'}(k; \tau,\tau' ) =\:& D_{O_r}(\tau) D_{O_r'}(\tau') \, P_{O_rO_r'}(k)\,.
\ea
Let us consider the three operators that appear at linear order in
$\tilde{\d}_g$.  In order to simplify notation, we relate them directly to the
density field $\d$ [i.e., $P_{O_rO_r'} \to \Plin$ in \refeq{dglm2pt}].  We obtain:
\ba
\mbox{(i)}\  O_s = \d:\:& \quad F^l_\d(k;p) =  \int \chi^2 d\chi\,W_l(p,\chi)  j_l(k\chi) b_1(\tau[\chi]) D(\tau[\chi]) \vs
\mbox{(ii)}\  O_s = v_\parallel = \vnhat^i \v{v}_i:\:& \quad F^l_{v_\parallel}(k;p) =  -k^{-1} \int \chi^2 d\chi\,W_l(p,\chi)  j'_l(k\chi) (\cH f D)(\tau[\chi]) \vs
\mbox{(iii)}\  O_s = \partial_{\parallel} v_\parallel = \vnhat^j \vnhat^i \partial_j\v{v}_i:\:& \quad F^l_{\partial_\parallel v_\parallel}(k;p) =  - \int \chi^2 d\chi\,W_l(p,\chi)  j''_l(k\chi) b_{\partial_\parallel v_\parallel}(\tau[\chi]) (\cH f D)(\tau[\chi]) \,.
\ea
The above relations can easily be generalized to include contributions that
are integrated along the line of sight (e.g., \cite{challinor/lewis:2011,GWpaper}). 
For terms (i) and (iii), we have allowed for bias parameters, where
$b_{\partial_\parallel v_\parallel} \propto b_{K_\parallel}$ is only induced by the specific observational selection effects discussed in \refsec{selection}.  On the other hand, the term (ii) is necessarily a pure projection effect, and hence unbiased, since the galaxy density cannot depend on the local matter velocity by way of the equivalence principle.  
Including the leading-order stochastic contribution is straightforward; it is given by
\ba
\left\< \d_{lm}(p) \d_{l'm'}^*(p')\right\>\Big|_{\rm stoch} 
=\:& \d_{ll'}\d_{mm'} 
\int \chi^2 d\chi\,W_l(p,\chi) W_l(p',\chi) \Peps(\tau[\chi])\,.
\ea
The Poisson shot-noise approximation corresponds to inserting $\Peps(\tau) = 1/\avng(\tau)$.  

Finally, one should note that actual surveys involve a window function
describing the survey footprint on the sky.  This complicates the
expression \refeq{dglm2pt}, since the orthogonality of $\d_{lm}(p)$ with
respect to $lm$ no longer holds \cite{deLaix/starkman:1998}.  

Let us now consider a survey with limited footprint on the sky, such
that the angle between different lines of sight within the survey footprint 
can be approximated as infinitesimal (i.e., wide-angle effects are negligible).
Further, we assume that the extent of the survey in the line-of-sight 
direction is small; this can of course be achieved by dividing the survey into redshift
bins.  Then, if we define $\d_{lm}(p)$ using $W_l(p,\chi) = j_l(p\chi)$,
we can construct the flat-sky version of the angular multipoles (App.~C of \cite{hu:2000}),
\ba
\d^{\rm fs}(\v{l}, p) \equiv\:& \sqrt{\frac{4\pi}{2l+1}} \sum_{m=-l}^l i^{-m} \d_{lm}(p) e^{i m \varphi_{\v{l}}}\,,
\ea
where $(l,m)$ is replaced with $\v{l} = (l_x,l_y)$, and $\varphi_{\v{l}} = \arctan(l_y/l_x)$.  The flat-sky multipoles $\d^{\rm fs}(\v{l},p)$ in turn can
be mapped onto the three-dimensional Euclidean Fourier transform $\d_g(\vk)$
used extensively throughout this review, via
\ba
\d_g(\vk) \equiv\:& \d^{\rm fs}(\bar\chi\vk_\perp, k_\parallel) \quad\mbox{where}\quad
k_\parallel \equiv \vnhat\cdot\vk;\  \vk_\perp \equiv \vk - k_\parallel \vnhat\,.
\ea
Here, $\bar\chi$ is the mean comoving distance of the redshift bin.  
Note that $\d_g(\vk)$ has dimension (length)$^3$.  
In this flat-sky-shallow-survey limit, the two-point function $\< \d_g(\vk) \d_g(\vk')\>$
then directly yields an estimate of the three-dimensional galaxy
power spectrum in redshift space at an effective redshift $\bar z$,
convolved with the three-dimensional window function of the survey.  
This is the approach commonly adopted in the analysis of current data sets
such as 2dF, SDSS, 6dFGRS, Wigglez, and BOSS
\cite{BAO/2dF,percival/etal:2007,blake/etal:2010,granett/etal:2012,beutler/etal:2016}.  
Future large-scale galaxy surveys such as 
SPHEREx \cite{SPHEREx:2014}, DESI \cite{DESI}, and Euclid 
\cite{amendola/etal:2013}, whose footprints exceed 10,000 square degrees,
must include the wide-angle effect in order to correctly interpret 
the measured galaxy clustering on large scales. This is particularly important for detecting 
primordial non-Gaussianity (\refsec{NG}) and relativistic projection 
effects (\refsec{projection}).

%% file: summary.tex
\clearpage
\section{Summary and outlook}
\label{sec:summary}

Bias describes the relation between the observed number density of galaxies and the underlying matter density and spacetime perturbations, and is an indispensable ingredient in our model of the observed Universe.  Key progress has been made on this problem in the past few years. 

We now have a general framework for galaxy bias on large scales 
in the context of perturbation theory (\refsec{evolution}).  
These bias parameters have well-defined physical interpretations, which 
become most clear by way of the generalized peak-background split argument 
(PBS, \refsec{PBS}, \refsec{PBSpeaks}, and \refsec{PBSbiasNG}).  In fact, one can rigorously define the PBS argument to 
yield \emph{exact} predictions for the bias parameters, provided one can 
accurately simulate the formation of the tracers of interest for 
different cosmologies (with different curvature in the case of the \LIMD
bias). This is certainly possible for dark matter halos.  However, the
perturbative bias expansion is much more general and applies, with caveats
described below, to any tracer of the large-scale structure, including
clusters of galaxies, voids, the Lyman-$\alpha$ forest, and 21cm line emission 
from neutral hydrogen.  

Assuming that values are given for the relevant bias parameters, the perturbation-theory framework makes definite
predictions for all observables related to galaxy clustering 
(\refsec{measurements}): 
$n$-point functions (power spectrum, bispectrum), statistics of counts-in-cells, 
and cross-correlations with the matter and among different tracers.  
Cross-correlating different tracers can yield precise measurements of
\emph{relative} bias parameters which cancel cosmic variance at leading order.  
Crucially, at a given order in perturbation theory, one set of bias parameters 
describes all observables involving a given tracer.  
These predictions are, however, valid only on sufficiently large scales (see below), and in practice some or all of these bias parameters have to be determined from the data.

The excursion set (\refsec{exset}) and the peak approach (\refsec{peaks}) have already 
furnished many insights into the scale dependence, nonlinearity and stochasticity of 
bias, as well as the validity of the peak-background split and its extension to 
variables other than the density.
Furthermore, they provide quantitative predictions for the bias parameters, which 
can be tested against N-body simulations.  
However, the models considered so far rely on a number of assumptions, such as 
the spherical collapse approximation, which we expect to become increasingly less accurate towards lower halo masses $M\lesssim M_\star$. 

A key application of bias is the incorporation of the effect of non-Gaussian initial 
conditions, which encode a rich array of signatures of early Universe physics,
and lead to additional scale dependencies that in many cases cannot be mimicked by nonlinear gravitational evolution and baryonic effects (\refsec{NG}).

Even though the bulk of this review focuses on structure forming out of collisionless matter in a universe described by General Relativity,
which is an excellent first-order approximation, 
the perturbative bias approach can be extended to tracers 
of a multi-component fluid made of CDM, baryons and massive neutrinos (\refsec{beyondCDM}), as is the case in the standard $\Lambda$CDM cosmology.  Further, it can be extended to incorporate the effects of a dynamical dark energy, or modifications to General Relativity.
The perturbative approach also allows for a consistent mapping from the local rest-frame galaxy density to the observed galaxy redshifts and positions on the sky, which we refer to as \emph{projection effects} (\refsec{observations}).  This includes relativistic effects which become relevant on the large scales targeted by forthcoming surveys. Moreover, a physical bias expansion of the galaxy velocity field (\refsec{velbias}; \refsec{velocitybias}) is a crucial ingredient for the prediction of projection effects beyond linear order.

While theoretically well-defined and rigorous, the inevitable downside of the
perturbative bias expansion is that it breaks down on small scales.  In
addition to the perturbative description of the matter density itself, which breaks
down at the nonlinear scale $\Rnl$ at which the density contrast is of order 
one, galaxy bias adds another scale, the nonlocality scale $R_*$ (\refsecs{general}{higherderiv}).  By definition,
the formation of galaxies in a given observed sample depends on the detailed distribution of matter within a region of this scale.  Thus, we cannot hope to describe galaxy clustering
perturbatively on scales that are of order $R_*$ or smaller.  Which of these two scales, $R_*$ or $\Rnl$, is more limiting, depends on the galaxy sample at hand, as well 
as its redshift.  While for halos we expect that $R_*$ is of order the Lagrangian 
radius $R(M)$, little is known about this scale for galaxies.  
The halo occupation distribution approach (\refsec{HOD}), in which the 1-halo term is calibrated with N-body simulations to describe the distribution of galaxies at small scales, could 
circumvent this limitation for the two-point function, but only if the transition region between 1- and 2-halo terms can be described accurately and robustly. \\

Beyond the state of the field reviewed here, various important questions remain open, including:
\begin{itemize}
\item What is the scale $R_*$ for galaxies?  How much cosmological information can we extract from the galaxy power spectrum, bispectrum, and possibly higher $n$-point functions, in the context of the general perturbative bias expansion?
\item Are there hierarchies and/or relations between the bias parameters of dark matter halos?  For example, are the biases in Lagrangian space that involve tidal fields, and time derivatives thereof, systematically smaller than the Lagrangian \LIMD bias parameters $b_N^L$?  Further, to what level do we need to understand galaxy formation in order to make use of any such relations for actual galaxy samples?  Does halo assembly bias play a significant role for galaxy bias?
\item Lagrangian bias models such as the peaks and excursion-set formalisms can in principle provide a bias relation in Lagrangian space that is valid on all scales, and not restricted to $r \gg R_*$.  Is there a way to derive a connection to Eulerian statistics for these models that does not break down on small scales?
\item It would be desirable to extend the validity of these Lagrangian bias models to lower-mass halos, such as those targeted by current and upcoming galaxy redshift surveys. In particular, how can we incorporate deviations from spherical collapse and other effects due to the nonlinear evolution of small-scale density perturbations into these models? 
\end{itemize}

Sophisticated models of galaxy bias will play a major role in the interpretation of 
upcoming large-scale structure surveys.  We now finally have a rigorous framework of galaxy bias, which allows for the development of a robust theoretical
description of galaxy clustering that matches the level of statistical and systematic uncertainties expected from these surveys.  
By advancing our understanding of these open issues, we will thus be able to makes the most of the data's potential to further our knowledge on galaxy formation, the history of the Universe, and fundamental physics.

\vfill
\textbf{Acknowledgments:}  
We thank the editor, Marc~Kamionkowski, for his invitation, in 2011, to write this review and his tremendous 
patience with us during the lengthy process of completion.  
We are grateful to the Aspen Center for Physics (National Science Foundation grant PHY-1066293),  
Marc~Kamionkowski and
the Department of Physics and Astronomy at Johns Hopkins University, 
Eiichiro Komatsu and the Max-Planck Institute for Astrophysics in Garching,
the Sexten Center for Astrophysics, 
and the Technion--Israel Institute of Technology,
for bringing the three of us together during various stages
of the writing process.

We would like to thank Volker Springel, Marc Manera and Enrique Gazta\~naga, and Aaron Ludlow, for providing high-resolution versions of \reffig{pie}, \reffig{scatterplots}, and \reffig{ludlow_porciani}, respectively. We further thank Titouan Lazeyras for generating \reffig{bK2(b1)}.

It is our pleasure to thank
Marcelo~Alvarez,
Valentin~Assassi,
Tobias~Baldauf,
Francis~Bernardeau,
Giovanni~Cabass,
Emmanuele~Castorina,
Kwan~Chuen~Chan,
Nico~Hamaus,
Nick~Kaiser,
Koki~Kakiichi,
Marc~Kamionkowski,
Titouan~Lazeyras,
Takahiko~Matsubara,
Nuala~McCullagh,
Marcello~Musso,
Mark~Neyrinck,
Jorge~Nore\~{n}a,
Aseem~Paranjape,
Cris~Porciani,
Dmitry~Pogosyan,
Antonio~Riotto,
Rom\'an~Scoccimarro,
Emiliano~Sefusatti,
Uro$\check{\text{s}}$~Seljak,
Ravi~Sheth,
Marko~Simonovi\'c,
Zvonimir~Vlah,
and Jaiyul~Yoo
for many helpful discussions.
Finally, we are indebted to
Diego~Blas,
Jonathan~Blazek,
Giovanni~Cabass,
Xingang~Chen,
Chi-Ting~Chiang,
Jens~Chluba,
Elisa~Chisari,
Kai~Hoffmann,
Dragan~Huterer,
Koki~Kakiichi,
Eiichiro~Komatsu,
Arthur~Kosowsky,
Titouan~Lazeyras,
Takahiko~Matsubara,
Nuala~McCullagh,
Minh~Nguyen,
Alvise~Raccanelli,
Ashley~Ross,
Ariel~S\'anchez,
Emiliano~Sefusatti,
Xun~Shi,
An$\check{\text{z}}$e~Slosar,
David~Spergel,
Michael~Strauss,
Yun~Wang,
Yvette~Welling,
and Drian~van~der Woude
for comments on the manuscript.

VD acknowledges support by the Swiss National Science Foundation, and by the Israel Science Foundation (grant no. 1395/16).
DJ acknowledges support from National Science Foundation grant AST-1517363.
FS acknowledges support from the Marie Curie Career Integration Grant  (FP7-PEOPLE-2013-CIG) ``FundPhysicsAndLSS,'' and Starting Grant (ERC-2015-STG 678652) ``GrInflaGal'' from the European Research Council.

%% file: App_statistics.tex
\section{Statistical field theory}
\label{app:stat}

In standard cosmological models, the initial conditions of the large-scale
structure of the Universe is generated from quantum mechanical vacuum 
fluctuations. Due to its quantum mechanical origin,
we cannot predict, in a deterministic sense, the precise initial conditions 
of the Universe we observe. Our theoretical treatment of the large-scale 
structure is, therefore, based on describing random fields.  
Throughout the review we draw on results
from basic statistical field theory which we briefly review here.  
More detailed introductions to this topic can be found in 
\cite{bernardeau/etal:2001,martinez/saar:2002,mo/vandenbosch/white:2010}.  

Here and throughout the review, we assume that, on a spatial slice of fixed proper time, all the cosmological random fields are statistically 
homogeneous and isotropic, in accordance with the 
\emph{cosmological principle}.  
The unperturbed FRW metric [\refeq{metriccN} with $\Phi=0=\Psi$] is a manifestation of this, since spatial 
slices ($\tau=$~const) are homogeneous and isotropic (maximally symmetric).  Furthermore, we shall 
assume the \emph{fair sample hypothesis}, which states that samples extracted from regions of the
Universe that are sufficiently distant from each other are independent realizations of the 
same physical process.
Observationally, the fair sample hypothesis has been found to be satisfied on scales above $\sim 80\hmpc$ \citep{sarkar/yadav/etal:2009,scrimgeour/davis/etal:2012}. 
Therefore, ensemble averages can be traded with spatial averages provided that the volume is large enough: this is the \emph{ergodic hypothesis}. 

A fundamental issue in the analysis of cosmic structures is to find the most appropriate observables
to retrieve information on the distribution of fluctuations (the matter density field, for example), 
their initial conditions and subsequent evolution. Here, we shall concentrate on poly-spectra and correlation functions, which include moments and encompass all observables considered in this review.

\subsection{Random fields in 3D Euclidean space}

In accordance with the cosmological principle, we consider random fields 
defined on a given spatial slice through spacetime;  in practice, one
should think of this as defined through a fixed proper time.  
We choose Euclidean coordinates $\vx$ for this slice, as written in 
\refeq{metriccN}.    
Further, we will focus on real scalar fields.  This applies
to most examples encountered in large-scale structure (temperature, density, 
pressure, gravitational potential and so on).  We will briefly consider
the generalization to vector and tensor fields at the end of this section.  

A random scalar field in 3D Euclidean space is a set of random variables 
$\rho(\vx)$, together with a collection of distribution functions 
$p_n(\rho(\vx_1)\,\dots,\rho(\vx_n))$, with $n\geq 1$.
In order to provide a complete statistical description of the random field $\rho(\vx)$ it is in general necessary (but not sufficient \cite{carron:2011}) to specify all of its correlation functions (i.e. moments).

The $n$-point correlation function is a specific expectation value given by
\begin{equation}
\left\la \rho(\vx_1)\dots \rho(\vx_n)\right\ra = \int\!\! d\rho_1\dots d\rho_n\, 
P_n(\rho_1,\dots,\rho_n)\rho_1 \dots \rho_n \;,
\end{equation}
where angle brackets denote ensemble averages and $\rho_i\equiv \rho(\vx_i)$. In particular, 
the 1-point correlation is the mean $\la \rho\ra$ of the random field. Higher-order 
correlation functions follow a particular hierarchy, which we demonstrate in the case of the
2-point function
\begin{equation}
\left\la \rho(\vx_1) \rho(\vx_2)\right\ra\;.
\end{equation}
In the event that we move one point (say $\vx_1$) far away from the other, the value of 
the field at the points become independent, so that the expectation value tends towards
\begin{equation}
\left\la \rho(\vx_1) \rho(\vx_2)\right\ra \to \la \rho(\vx_1)\ra\la\rho(\vx_2)\ra  = \la \rho \ra^2 \;,
\end{equation}
where we have used the cosmological principle, which implies translation invariance and hence 
the same expectation value of $\rho(\vx)$ at any point on the slice. Therefore, we can write
\begin{equation}
\left\la \rho(\vx_1) \rho(\vx_2)\right\ra =
\la \rho\ra^2 \left[1+\xi^{(2)}\!(\vx_1,\vx_2)\right] \;.
\end{equation}
This defines the \emph{reduced} or \emph{connected} 2-point correlation function $\xi_2(\vx_1,\vx_2)$.
The $n$-point \emph{connected correlation function} $\xi^{(n)}$ is recursively 
defined in such a way that $ \left\la \rho(\vx_1)\dots \rho(\vx_n)\right\ra$ is a sum of terms, where each term is associated with a partition of the set of 
$n$ points $\vx_1,\dots,\vx_n$.  For the first four orders for instance, we have
\begin{align}
\left\la \rho(\vx_1) \rho(\vx_2)\right\ra &= \la \rho\ra^2
\Bigl[1+\xi^{(2)}\!(\vx_1,\vx_2)\Bigr] \nonumber  \\
\left\la \rho(\vx_1) \rho(\vx_2)\rho(\vx_3)\right\ra &= 
\la \rho\ra^3 \Bigl[1+\xi^{(2)}\!(\vx_1,\vx_2)+\perm{2}+\xi^{(3)}\!(\vx_1,\vx_2,\vx_3)\Bigr] 
\nonumber \\
\left\la \rho(\vx_1) \rho(\vx_2)\rho(\vx_3)\rho(\vx_4)\right\ra &=
\la \rho\ra^4 \Bigl[1+\xi^{(2)}\!(\vx_1,\vx_2)+\perm{5} \nonumber \\ 
&\qquad\qquad + \xi^{(3)}\!(\vx_1,\vx_2,\vx_3)+\perm{3}
+\xi^{(2)}\!(\vx_1,\vx_2)\xi^{(2)}\!(\vx_3,\vx_4)+\perm{2} \nonumber \\
&\qquad\qquad + \xi^{(4)}\!(\vx_1,\vx_2,\vx_3,\vx_4)\Bigr] \;.
\end{align}
The ``$\perm{n}$'' indicates that the term immediately preceeding is repeated with all
possible cyclic permutations of the indices. In the case of a random field with zero
mean, $\la \delta(\vx)\ra = 0$ (which in this review denotes the matter density contrast), the hierarchy of correlation functions reads
\begin{align}
\left\la\delta(\vx_1)\delta(\vx_2)\right\ra &= \xi^{(2)}\!(\vx_1,\vx_2) \nonumber  \\
\left\la\delta(\vx_1)\delta(\vx_2)\delta(\vx_3)\right\ra &= \xi^{(3)}\!(\vx_1,\vx_2,\vx_3) \nonumber  \\
\left\la\delta(\vx_1)\delta(\vx_2)\delta(\vx_3)\delta(\vx_4)\right\ra &= 
\xi^{(2)}\!(\vx_1,\vx_2)\xi^{(2)}\!(\vx_3,\vx_4)+\perm{2}
+\xi^{(4)}\!(\vx_1,\vx_2,\vx_3,\vx_4) \;.
\label{eq:corrhierarchy}
\end{align}
We will also use the notation
\be
\left\la\delta(\vx_1)\delta(\vx_2)\cdots\delta(\vx_n)\right\ra_c \equiv 
\xi^{(n)}\!(\vx_1,\vx_2, \cdots,\vx_n) \,.
\ee
for the connected correlators.

The cosmological principle dictates that all expectation values and hence all correlation functions are invariant under global 
translations $\vx_i\to\vx_i+\Delta\vx$ on a fixed time slice, which is also known as \emph{statistical homogeneity}.  
For the reduced 2-point correlation, this implies
\begin{equation}
\xi^{(2)}\!(\vx_1,\vx_2) \equiv \xi^{(2)}\!(\vx_1-\vx_2) \;,
\end{equation}
so that it depends only on the separation between the two points.  Similarly, in the absence of preferred directions, expectation values are invariant under a global rotation $\v{R}$ of the coordinate system $\vx \to \v{R}\cdot\vx$, which is known as \emph{statistical isotropy}.  This implies that the 2-point correlation function depends only on the magnitude $|\vx_1-\vx_2|$ of the separation vector,
\begin{equation}
\xi^{(2)}\!(\vx_1,\vx_2)\equiv \xi^{(2)}\!(|\vx_1-\vx_2|) \;.
\end{equation}
This holds also if one cross-correlates different fields, and for higher $n$-point correlation functions, which can only depend on $x_{ij} \equiv |\vx_i-\vx_j|$.  Note that projection effects such as redshift-space distortions induce a dependence on $(\vx_i-\vx_j)\cdot \vnhat$ in the observed $n$-point functions, where $\vnhat$ is the line of sight (\refsec{projection}).  
While the cosmological principle requires statistical homogeneity, the ergodic hypothesis is valid only if the $n$-point correlation functions
decay sufficiently rapidly to zero in the limit of large separations. This is indeed the case in the standard $\Lambda$CDM cosmology. 

Note that these definitions can be extended to point processes, i.e. for distributions rather than continuous fields (in which case the correlators 
$\left\la \rho(\vx_1)\dots \rho(\vx_n)\right\ra$ are also called joint intensities), such as the peaks discussed 
in \refsec{peaks}, but there are some subtleties (such as shot noise 
corrections) owing to discreteness.

Finally, we turn to vector and tensor fields. Homogeneity and isotropy imply 
that two-point scalar-vector cross- and vector auto-correlation functions can 
be written as
  \ba
  \< \d(\v{0}) v^i(\v{r}) \> =\:& \hat r^i \xi_{\d v}(r) \vs
  \< v^i(\v{0}) v^j(\v{r}) \> =\:& \d^{ij} \xi_{v,1}(r) + \left(\hat r^i\hat r^j - \frac13 \d^{ij}\right) \xi_{v,2}(r)\,,
  \ea
  where $\xi_{\d v}(r)$, $\xi_{v,1}(r)$, and $\xi_{v,2}(r)$
are scalar functions.  
Note that the scalar-vector correlation $\xi_{\d v}(r)$ is generated only 
for longitudinal vector fields, while the vector two-point auto-correlation
functions $\xi_{v,1}(r)$ and $\xi_{v,2}(r)$ receive contributions from both 
longitudinal and transverse vectors. $\xi_{v,1}(r)$ and $\xi_{v,2}(r)$ 
are regular at $r=0$. 
The auto- and cross-correlations of symmetric tensor fields $K_{ij}$ can be similarly constructed out of $\d_{kl}$ and $(\hat r_m \hat r_n - \d_{mn}/3)$. While these relations in general become lengthy, the auto-correlation function of a trace-free tensor field at zero lag simplifies to
  \begin{gather}
\big\la K_{ij}(\vx) K_{lm}(\vx)\big\ra =
\frac{1}{15} \s^2 \left(\delta_{il}\delta_{jm}+\delta_{im}\delta_{jl}-\frac{2}{3}\delta_{ij}\delta_{lm}\right)\,,\quad
\mbox{where}\quad
\s^2 = \frac32 \< K_{ij} K^{ij} \>\,.
  \end{gather}
  Note that, if we identify $K_{ij}$ with the tidal field [\refeq{Del} on p.~\pageref{eq:Del}], then $\sigma^2$ corresponds to the variance of the density field.

\subsection{Fourier representation}
\label{app:stat:Fourier}

In a statistically homogeneous Universe, it is convenient to represent
random fields by their Fourier components.
In the following, we assume a field $\d$ with zero mean.  Adopting the Fourier
convention
\begin{equation}
\label{eq:FourierConv}
\delta(\vk) = \int\!d^3\vx\, \delta(\vx)\, e^{-i\vk\cdot\vx} \;, \qquad
\delta(\vx) = \int\! \frac{d^3\vk}{(2\pi)^3} \delta(\vk)\, e^{i\vk\cdot\vx}
\equiv \int_{\vk}\! \delta(\vk)\, e^{i\vk\cdot\vx} \;,
\end{equation}
the power spectrum $P(k)$ of the field is the expectation value 
\begin{equation}
\la \delta(\vk) \delta^*(\vk') \ra 
\equiv \la \delta(\vk) \delta^*(\vk') \ra'\: (2\pi)^3 \delta_D(\vk-\vk')
= P(\vk) \:(2\pi)^3  \delta_D(\vk-\vk')
 \;.
\end{equation}
Here, the superscript $*$ stands for the complex conjugate, which 
makes $P(k)$ positive definite.  
The Dirac delta is a consequence of translational invariance (homogeneity). Otherwise, the 
ensemble average $\la \delta(\vk) \delta(\vk')\ra$ would acquire a phase factor when $\vx\to\vx+\Delta\vx$.  
If the field $\delta(\vx)$ is real, $\delta^*(\vk) = \delta(-\vk)$, and we obtain
\begin{equation}
\la \delta(\vk) \delta(-\vk')\ra = P(\vk) (2\pi)^3 \delta_D(\vk+\vk') \;.
\end{equation}
Requesting further 
rotational invariance implies that the power spectrum depends only on $k=|\vk|$, i.e.
\begin{equation}
\la \delta(\vk) \delta^*(\vk')\ra = P(k) (2\pi)^3 \delta_D(\vk-\vk') \;.
\end{equation}
Note that, with our Fourier convention, the Dirac delta distribution is given by
\begin{equation}
\delta_D(\vk)=\frac{1}{(2\pi)^3}\int\!d^3\vr\, e^{-i\vk\cdot\vr} \;.
\end{equation}

The connected 2-point correlation function is the Fourier transform of the power spectrum. 
For a three-dimensional homogeneous and isotropic random field with zero mean, we have
\begin{equation}
\xi^{(2)}\!(r) = \int_{\vk}\!P(\vk)\, e^{i\vk\cdot\vr}
=\frac{1}{2\pi^2}\int_0^\infty\!\!dk\,k^2 P(k) j_0(kr) \;.
\end{equation}
Here, $j_0(x)=\sin(x)/x$ is a spherical Bessel function.
For sake of completeness, note that, in one and two dimensions, we have
\begin{align}
\xi(r) &= \frac{1}{\pi}\int_0^\infty\!\!dk\, P(k) \cos(kr) \qquad \mbox{(1D)}\\
\xi(r) &= \frac{1}{2\pi}\int_0^\infty\!\!dk\,k P(k) J_0(kr) \qquad \mbox{(2D)}
\end{align}
respectively, where $J_0(x)$ is a regular Bessel function.  The fact that $P(k)$ is positive definite however does not imply that $\xi^{(2)}\!(r)$ is also positive definite.  Indeed, we have
\be
\int d^3\vr \,\xi^{(2)}(r) = \lim_{k\to 0} P(k)\,,
\ee
which should vanish following the ergodic hypothesis for the zero-mean field $\d$.  This implies that
$\xi^{(2)}(r)$ has to be negative for some values of $r$.  
The variance of an isotropic random field is given by
\begin{equation}
\sigma^2 = \xi(0) = \frac{1}{2\pi^2}\int_0^\infty\!\!dk\,k^2 P(k)
= \int_{0}^{\infty}\! \frac{dk}k \Delta^2(k) \;,
\label{eq:variance}
\end{equation}
where we have introduced the dimensionless power spectrum $\Delta^2(k)$
which quantifies the variance of the density field per unit $\ln k$ and, unlike $P(k)$, is independent of the Fourier convention used. In our convention,
$\Delta^2(k) \equiv k^3 P(k)/(2\pi^2)$.

These results can be extended to higher-order correlation functions. For instance, the bispectrum
$B(\vk_1,\vk_2,\vk_3)$ is the expectation value
\begin{equation}
\la \delta(\vk_1) \delta(\vk_2) \delta(\vk_3)\ra 
= B(\vk_1,\vk_2,\vk_3) (2\pi)^3 \delta_D(\vk_1+\vk_2+\vk_3) \;.
\end{equation}
The Dirac delta ensures that the wavevectors $\vk_i$ correspond to the three sides of a
triangle, so that the three-point function is invariant under translations. Isotropy further 
implies that $B$ be a function of the wavenumbers $k_i=|\vk_i|$ solely. In analogy with the 
power spectrum, the 3-point connected correlation function is the Fourier transform of the bispectrum,
\begin{align}
\xi^{(3)}(\vx_1,\vx_2,\vx_3) &= (2\pi)^3\int_{\vk_1}\int_{\vk_2}\int_{\vk_3} 
\delta_D(\vk_1+\vk_2+\vk_3)
B(\vk_1,\vk_2,\vk_3)
e^{i\left[\vk_1\cdot\vx_1+\vk_2\cdot\vx_2+\vk_3\cdot\vx_3\right]} \;.
\end{align}
The trispectrum and higher-order poly-spectra are defined analogously.

  Finally, the Fourier-space two-point correlation functions of vector and tensor fields, such as
  \be
\< \d(\vk) v^i(\vk')\>'\,,\  
\< v^i(\vk) v^j(\vk')\>'\,,\  
\< \d(\vk) K_{ij}(\vk')\>'\,,\  
\< K_{ij}(\vk) K_{lm}(\vk')\>'\,,
  \ee
  can be decomposed into products of the vector $\hat k^i$, the  tensors $\d_{ij}$ and $(\hat k_m \hat k_n - \d_{mn}/3)$ [cf. \refeq{KijF}], and the totally trace-free projection tensor
  \ba
  \mathcal P_{ijlm} \big( \hat{\vk} \big) &= \hat k_i \hat k_j  \hat k_l \hat k_m - \frac{1}{7} \left( \d_{ij} \hat k_l \hat k_m + \perm{5} \right)
  + \frac1{35} \left( \d_{ij} \d_{lm} + \perm{2}\right)\,.
  \label{eq:Pijlm}
  \ea

We often deal with a smoothed version of the density field $\d_R$, obtained 
by convolving with a spherically symmetric filtering kernel $W_R$, where $R$ denotes the filter scale.  Throughout, the kernels are normalized through
$\int d^3\vx\,W_R(|\vx|)=1$, which implies $\lim_{k\to 0} W_R(k)=1$.  
The three most popular filtering kernels
$W_R(x)$ and their Fourier transform $W_R(k)$ are, respectively,
\begin{itemize}
\item{sharp-$k$ filter:\\
\ba
W_R(x) = \frac{3}{4\pi R^3}
\left[
3\frac{j_1(x/R)}{x/R}
\right]
, & & W_R(k) = \Theta_H(1 - kR)
\label{eq:sharp-k_filter}
\ea
}
\item{Gaussian filter:\\
\ba
W_R(x) = 
\frac{1}{\left[2\pi R^2\right]^{3/2}}
e^{-\frac{1}{2} x^2/R^2}
, & & 
W_R(k) = 
e^{-\frac{1}{2}R^2 k^2}
\label{eq:gaussian_filter}
\ea
}
\item{Tophat filter:\\
\ba
W_R(x) =
\frac{3}{4\pi R^3}
\Theta_H\left(1 - \frac{x}{R}\right)
, & & 
W_R(k) =
3\frac{j_1(kR)}{kR}\,.
\label{eq:top-hat_filter}
\ea
}
\end{itemize}
Here, $\Theta(x)$ is the Heaviside step function (\reftab{math}) 
and 
\be
j_1(x) =  \frac{\sin x - x \cos x}{x^2}
\ee
is a spherical Bessel function.
Bearing in mind that convolution in real space is a multiplication in Fourier space (and vice versa),
the variance of the random field on scale $R$ thus is
\begin{equation}
\sigma^2(R) = \< \d_R^2 \> = \int_{0}^{\infty}\!\frac{dk}k\, \Delta^2(k)\,|W_R(k)|^2 \;.
\end{equation}

\subsection{Gaussian random fields}
\label{app:stat:Gauss}

Gaussian random fields are essential in cosmology for mainly two reasons. Firstly, the
inflationary paradigm predicts that the primordial fluctuations which gave rise to the 
large-scale structure of the Universe closely followed Gaussian statistics; moreover, this is confirmed by CMB observations, which find that the CMB temperature is Gaussian to roughly one part in a thousand \cite{planck:2013c}.
Secondly, the central limit theorem, which states that the superposition of a large
number of (uncorrelated) random processes asymptotes to a Gaussian distribution, can often be applied in large-scale structure; for example, in the case of projected density fields 
\cite{jeong/schmidt/sefusatti:2011}.  

The definition of a Gaussian random field, with zero mean, is that
its distribution functions are given by Gaussian multi-variates,
\begin{equation}
P_n(\vy) = \frac{\sqrt{\det\vcc^{-1}}}{(2\pi)^{n/2}}
\exp\left(-\frac{1}{2}\vy^\trans \vcc^{-1}\vy\right)\;,
\end{equation}
where $\vy$ is the random vector $(\delta(\vx_1),\dots,\delta(\vx_n))$ and $\vcc^{-1}$ denotes the
inverse of the covariance matrix, with entries given by
\begin{equation}
\vcc_{ij} = \left\langle \delta(\vx_i) \delta(\vx_j)\right\ra \;.
\end{equation}
Note that the covariance matrix is symmetric and non-negative definite 
(which also implies $|\vcc|\ge0$).

A key property of Gaussian random fields is that, in the Fourier representation, the 
phases of the Fourier modes follow independent white-noise distributions.  
Equivalently, the real and imaginary part of the coefficients follow independent Gaussian distributions. 
Let us assume that the fluctuation field $\delta(\vx)$ is periodic over a volume $V=L^3$, 
so that the integral $\int_{\vk}$ in \refeq{FourierConv} is replaced by $(1/V)\sum_{\vk}$, 
where the sum runs over integer multiples $(n_x,n_y,n_z)k_F$ and $k_F=2\pi/L$ is the fundamental wavenumber.  
Then, $\delta(\vk)$ is dimensionless.
Furthermore, let us denote the real and imaginary parts of $\delta(\vk)$ by $\Re\delta_{\vk}$ and $\Im\delta_{\vk}$. 
Their joint probability distribution is the bivariate Gaussian
\begin{equation}
p\bigl(\Re\delta_{\vk},\Im\delta_{\vk}\bigr)d\Re\delta_{\vk} d\Im\delta_{\vk} =
\frac{1}{\pi\sigma_k^2} \exp\left[-\frac{(\Re\delta_{\vk})^2+(\Im\delta_{\vk})^2}{\sigma_k^2}\right]
d\Re\delta_{\vk} d\Im\delta_{\vk} \;,
\end{equation}
where $\sigma_k^2=(2\pi)^3 P(k)/k_F^3$ is the variance of 
$\Re\delta_{\vk}$ and $\Im\delta_{\vk}$ (which depends only on $k$ due to isotropy). 
In terms of the amplitude $|\delta_{\vk}|$ and phase $\phi_{\vk}$, we have
\begin{equation}
p(|\delta_{\vk}|,\phi_{\vk})d|\delta_{\vk}|d\phi_{\vk} =
\frac{1}{\pi\sigma_k^2} \exp\left[-\frac{|\delta_{\vk}|^2}{\sigma_k^2}\right] 
|\delta_{\vk}| d|\delta_{\vk}| d\phi_{\vk} \;,
\end{equation}
i.e. $|\delta_{\vk}|$ follows a Rayleigh distribution.  Note that for real scalar
fields $\d(\vx)$, we have the additional constraints $\Re \d_{-\vk} = \Re \d_{\vk}$,
$\Im \d_{-\vk} = -\Im \d_{\vk}$,
in which case $\delta_{\vk}$ must be generated over half of the plane (e.g. $k_z\geq0$) only.

For a homogeneous Gaussian random field, all the statistical information is 
contained in the connected 2-point correlation $\xi^{(2)}(x)$ or, equivalently, in the power 
spectrum $P(k)$. All the higher-order connected correlation functions are identically zero, 
so that their measurement from observations strictly tests Gaussianity. 
The $n$-point correlation functions $\la\delta(\vx_1)\dots\delta(\vx_n)\ra $ then simplify 
to a sum of products of irreducible 2-point functions $\xi^{(2)}$. 
This property is known as \emph{Wick's theorem}.
For instance, the 4-point correlator in \refeq{corrhierarchy} simplifies to
\be
\left\la\delta(\vx_1)\delta(\vx_2)\delta(\vx_3)\delta(\vx_4)\right\ra = 
\xi^{(2)}\!(\vx_1,\vx_2)\xi^{(2)}\!(\vx_3,\vx_4)+\perm{2}\;.
\ee

While the \emph{initial conditions} are currently found to be consistent with
perfect Gaussianity, in particular using measurements of the CMB, the observed, late-time distribution of galaxies is highly non-Gaussian due to gravitational 
instability and galaxy bias. Therefore, unlike the primary CMB anisotropies whose
distribution is almost perfectly Gaussian and hence completely described
by the power spectrum, higher-order correlation functions are necessary to specify 
the galaxy distribution.
 Note that the density contrast (or CMB temperature, for that matter) cannot be \emph{exactly} Gaussian because the field has 
to satisfy the physical constraint $\delta > -1$. Nevertheless, Gaussianity can be a good approximation
when the amplitude of fluctuations is small, i.e. when the density fluctuations are still linear.  

The fact that the likelihood of a Gaussian statistical field is so simple
allows for several important analytical results, including the 
number density of peaks in the initial density field (\refsec{kacrice}).  Exact renormalized bias operators
can be derived for Gaussian fields, as described in \refsec{peakbias} and 
\cite{PBSpaper}.  
For this, we define orthogonal polynomials \cite{withers:2000} with respect to a given weight
$p(\v{w})$, for example the joint PDF of the 5 peak invariants.  
Orthogonal polynomials satisfy
\be
\int\! d^n \v{w} \,p(\v{w}) O_{\v{n}}(\v{w}) O^\star_{\v{n}'}(\v{w}) = \d_{\v{n} \v{n}'}\,,
\ee
where $O^\star_{\v{n}}$ is the dual polynomial to $O_{\v{n}}$, $\v{n}, \v{n}'$ are sets of indices, $\d_{\v{n} \v{n}'}$ denotes the 
Kronecker symbol in the space of these sets,  
and $p(\v{w})$ is the joint 1-point distribution of the 5 peak invariants.  
Univariate orthogonal polynomials, which depend on only one element of the vector
$\v{w}$, are always their own dual. 

We now give the first few dual polynomials for the bivariate Hermite polynomials which are employed in \refsec{PBSpeaks}
\citep[see][for details]{lazeyras/musso/desjacques:2015}:  
\begin{gather}
  H^\star_{n0}(\nu,J_1) = H_n(\nu)\;,
  \qquad H^\star_{0n}(\nu,J_1)= H_n(J_1) \;,
  \nonumber \\
  {H}^\star_{11}(\nu,J_1)
  = \nu J_1 - \gamma_1 \nonumber \;,\\
  {H}^\star_{21}(\nu,J_1)
  = \nu^2 J_1 - J_1 -2\gamma_1\nu \nonumber\;, \\
  {H}^\star_{12}(\nu,J_1)
  = \nu J_1^2 - \nu -2\gamma_1J_1 \nonumber \;,\\
  {H}^\star_{31}(\nu,J_1)
  = \nu^3 J_1 - 3\nu J_1 - 3\gamma_1\nu^2 +3\gamma_1 \nonumber \;,\\
  {H}^\star_{22}(\nu,J_1)
  = \nu^2 J_1^2 - \nu^2 - J_1^2 - 4\gamma_1\nu J_1 +1 +2\gamma_1^2 \;.
\end{gather}
Note that one could work with the variable $z=(J_1-\gamma_1\nu)/\sqrt{1-\gamma_1^2}$ rather than $J_1$ as in 
\cite{gay/pichon/pogosyan:2012} and, thus, end up with univariate Hermite polynomials.

%% file: App_SPT.tex
\section{Cosmological perturbation theory}
\label{app:SPT}

In this appendix, we review the PT approach to the
description of the nonlinear gravitational evolution of dark matter density and 
velocity fields. On large scales, where the perturbative bias expansion is 
valid, the vorticity as well as baryonic pressure effects are small, and 
the nonlinear evolution of the cosmic matter density field can be well 
approximated by a self-gravitating, presureless, longitudinal-velocity 
single fluid. We will turn to the corrections induced by neglecting the 
aforementioned effects in \refapp{EFT}.  

\subsection{Standard perturbation theory}
\label{app:SPT:SPT}

In the pressureless-fluid approximation, the time evolution of the density 
$\d(\bfx,\tau)$ and velocity fields $\v{v}(\bfx,\tau)$  are 
governed by the continuity, Euler and Poisson equations:
\ba
&\frac{\partial}{\partial \tau}\delta(\bfx,\tau)
+
\vn\cdot
\left\{
\left[1+\delta(\bfx,\tau)\right]\v{v}(\bfx,\tau)
\right\} =0
\label{eq:cont}
\\
&\frac{\partial}{\partial \tau}\v{v}(\bfx,\tau)
+
\left[
\v{v}(\bfx,\tau)
\cdot
\vn\right]
\v{v}(\bfx,\tau)
+
\mathcal{H}(\tau)\v{v}(\bfx,\tau)
= -\vn\Phi(\bfx,\tau)
\label{eq:euler}
\\
& \lapl \Phi(\bfx,\tau)  
= \frac32 {\cal H}^2 \Omega_m(\tau) \delta(\bfx,\tau)\,.
\label{eq:Poisson}
\ea
These equations are strictly valid for 
$\Lambda$CDM only;  however, they are also highly accurate for canonical, 
non-clustering dark energy models where the sound horizon of the dark energy
component is of order $\cH^{-1}$ (see \refsec{modgrav} for a brief discussion).  

In Fourier space, we can eliminate
the gravitational potential $\Phi$ and express the longitudinal velocity
field in terms of the velocity divergence $\theta \equiv \vn\cdot\v{v}(\bfx,\tau)$ to obtain the following two equations:
\ba
&\frac{\partial \delta(\bfk,\tau)}{\partial \tau}
+
\theta(\bfk,\tau)
=
-\int_{\bfk_1}\int_{\bfk_2} (2\pi)^3\delta_D(\vk-\vk_{12})
\alpha(\bfk_1,\bfk_2)
\theta(\bfk_1,\tau)\delta(\bfk_2,\tau),
\label{eq:contF}
\\
&\frac{\partial \theta(\bfk,\tau)}{\partial \tau}
+
{\cal H}(\tau)\theta(\bfk,\tau)
+
\frac32{\cal H}^2(\tau)\Omega_{m}(\tau)\delta(\bfk,\tau)
=
-\int_{\bfk_1}\int_{\bfk_2} (2\pi)^3\delta_D(\vk-\vk_{12})
\beta(\bfk_1,\bfk_2)
\theta(\bfk_1,\tau)\theta(\bfk_2,\tau).
\label{eq:eulerF}
\ea
where $\bfk_{ij\cdots}\equiv\bfk_i+\bfk_j+\cdots$, and 
\be
\alpha(\bfk_1,\bfk_2)
=
\frac{\bfk_{12}\cdot\bfk_{1}}{k_1^2},
\quad
\beta(\bfk_1,\bfk_2)
=
\frac{k_{12}^2(\bfk_1\cdot\bfk_2)}{2k_1^2k_2^2}.
\ee
In SPT, \refeqs{contF}{eulerF} are solved perturbatively in terms of the linear
density contrast $\delta^{(1)}(\bfk,\tau)$:
\be
\delta(\bfk,\tau) = \sum_{n=1}^\infty \delta^{(n)}(\bfk,\tau),
\quad
\theta(\bfk,\tau) = \sum_{n=1}^\infty \theta^{(n)}(\bfk,\tau),
\label{eq:SPTexpansion}
\ee
where the superscript $(n)$ indicates that a term involves
$n$ powers of the linear density contrast; we refer to such a term as being
\emph{$n$-th order in perturbation theory}.  

On very large scales, $k\to0$, $\d^{(1)}$ is much less than 1, and the
quadratic source term on the 
right-hand side of \refeqs{contF}{eulerF} can be neglected.  
Then, all Fourier modes evolve independently, and we obtain a single
second-order ordinary differential equation for the evolution of the linear density contrast, 
$\delta^{(1)}(\bfk,\tau)$:
\be
\frac{\partial^2}{\partial \tau^2}
\delta^{(1)}(\bfk,\tau)
+
{\cal H}(\tau)
\frac{\partial}{\partial \tau}
\delta^{(1)}(\bfk,\tau)
-
\frac32 \Omega_m(\tau){\cal H}^2(\tau)\delta^{(1)}(\bfk,\tau)=0\,.
\ee
We see that the time evolution equation is scale-independent;  that is,
all Fourier modes evolve at the same rate, independent of $\vk$.  In fact,
this is the reason why time derivatives of the density field can be reordered to be successively
higher order in perturbation theory in \refsecs{basisL}{basisE}.  
We can then factor out the \emph{linear growth factor} $D(\tau)$ by writing
 $\delta^{(1)}(\bfk,\tau) = [D(\tau)/D(\tau_0)] \delta^{(1)}(\bfk,\tau_0)$, where
$\tau_0$ is a reference time, and $D(\tau)$ obeys
\be
\frac{d^2}{d\tau^2} D(\tau) + \cH \frac{d}{d\tau} D(\tau)  - \frac32 \Om(\tau) \cH^2 D(\tau) = 0\,,
\label{eq:Deom}
\ee
where initial conditions are setup so that $D(\tau)$ follows the growing mode, which corresponds to $D(\tau) \propto a(\tau)$ during matter domination. Unless otherwise indicated, all relations in this review are independent of the overall normalization of the growth factor. For reference, two popular choices for the latter are $i)$ $D(\tau) = a(\tau)$ during matter domination (for $a\ll1$), which is denoted as $D_\text{md}(\tau)$ in the text; and $ii)$ $D(\tau_0) = 1$, where $\tau_0$ corresponds to today's epoch.

The linear-theory velocity divergence field is then obtained from the linearized
continuity equation [\refeq{contF}],
\be
\theta^{(1)}(\bfk,\tau) 
= -\frac{d\ln D(\tau)}{d\tau} \delta^{(1)}(\bfk,\tau)
= - f \cH(\tau) \delta^{(1)}(\bfk,\tau)\,,
\label{eq:theta1}
\ee
where the logarithmic growth rate $f(\tau)\equiv d\ln D/d\ln a$.  Note that the identity
\be
\frac{d({\cal H}f)}{d\tau} =
{\cal H}^2 \left(\frac32\Omega_m - f - f^2\right)
\label{eq:dHfdtau}
\ee
follows from the linear equations of motion.

The equations of motion \refeqs{contF}{eulerF} motivate an ansatz of
writing the $n$-th order solution in \refeq{SPTexpansion} as
\ba
\delta^{(n)}(\bfk,\tau)
=&
\int_{\bfk_1}\cdots\int_{\bfk_{n}}(2\pi)^3\delta_D(\bfk-\bfk_{12\cdots n})
F_n(\bfk_1,\cdots,\bfk_n,\tau)\d^{(1)}(\bfk_1,\tau)\cdots\d^{(1)}(\bfk_n,\tau)
\vs
\theta^{(n)}(\bfk,\tau)
=&
-{\cal H}(\tau)f(\tau)
\int_{\bfk_1}\cdots\int_{\bfk_{n}}(2\pi)^3\delta_D(\bfk-\bfk_{12\cdots n})
G_n(\bfk_1,\cdots,\bfk_n,\tau)\d^{(1)}(\bfk_1,\tau)\cdots\d^{(1)}(\bfk_n,\tau),
\label{eq:SPTkernels}
\ea
with symmetrized density and velocity divergence kernels, respectively, 
$F_n$ and $G_n$. From the linear solutions, 
$F_1=G_1=1$ is obvious.  
Inserting the ansatz \refeq{SPTkernels} into \refeqs{contF}{eulerF} yields evolution equations for the kernels (see, e.g. \cite{jeong:2010}).  
In general, the kernels $F_n$ and $G_n$ depend on time, except when
 $\Omega_m/f^2$ is constant in time.  This is indeed the case for the EdS Universe where $\Omega_m=f=1$ and the kernels become particularly simple. The usual practice in SPT is to 
calculate the time-independent kernels in the EdS Universe and apply the same
kernels to other cosmologies.  In this case, the linear growth function 
encodes the entire cosmology dependence of the nonlinear solutions in SPT.  
For $\Lambda$CDM and quintessence cosmologies, this approach provides a 
very accurate description of the full solution \cite{takahashi:2008}.  
Specifically, the second-order kernels $F_2$ and $G_2$ in the EdS cosmology are
given by
\ba
F_2(\bfk_1,\bfk_2)
=\:& \frac57 + \frac27\frac{(\bfk_1\cdot\bfk_2)^2}{k_1^2k_2^2}
+ \frac{\bfk_1\cdot\bfk_2}{2k_1k_2}\left(\frac{k_1}{k_2}+\frac{k_2}{k_1}\right),
\vs
G_2(\bfk_1,\bfk_2)
=\:& \frac37 + \frac47\frac{(\bfk_1\cdot\bfk_2)^2}{k_1^2k_2^2}
+ \frac{\bfk_1\cdot\bfk_2}{2k_1k_2}\left(\frac{k_1}{k_2}+\frac{k_2}{k_1}\right)\,.
\label{eq:F2}
\ea
The second-order density and velocity field can also be written in real space as
\ba
\d^{(2)}(\vx, \tau) =\:&
\frac{17}{21} [\d^{(1)}]^2 + \frac27 [K_{ij}^{(1)}]^2 
- s^i_{(1)} \partial_i \d^{(1)}\,,
\label{eq:d2}
\\
-\frac{1}{{\cal H}(\tau)f(\tau)}\theta^{(2)}(\bfx,\tau)
=\:&
\frac{13}{21} [\d^{(1)}]^2 + \frac47 [K_{ij}^{(1)}]^2 
- s^i_{(1)} \partial_i \d^{(1)}
\label{eq:t2}
\,,
\ea
where on the right-hand side of \refeq{d2} all terms are evaluated at 
$\vx, \tau$. Here,
\be
\v{s}_{(1)}(\vq,\tau) = \xfl^{(1)}(\tau) -\vq = - \frac{\vn}{\lapl} \d^{(1)}(\vq,\tau)
\label{eq:lindisp}
\ee
is the first-order Lagrangian displacement, and 
\be
K_{ij}^{(1)} = \left[\frac{\partial_i\partial_j}{\lapl} - \frac13 \d_{ij} \right] \d^{(1)}
\ee
is the first-order tidal field [\refeq{Del}].  
The third-order expressions in real space are given in \refeq{delta3}
together with \refeq{Psigtotal} below.  

Finally, we give the result for the next-to-leading (NLO, or 1-loop) contribution
to the matter and velocity power spectra. We will drop the time argument for clarity. It
can be easily restored in the linear power spectra and perturbation theory
kernels, if desired. The NLO matter power spectrum is 
given by $P_{mm}(k) = \Plin(k) + P_{mm}^\NLO(k)$, where
\ba
P_{mm}^\NLO(k) =\:& P_{mm}^{(22)}(k) + 2 P_{mm}^{(13)}(k) \,,
\label{eq:Pm1loop}\\
P_{mm}^{(22)}(k) \equiv\:& \< \d^{(2)}(\vk) \d^{(2)}(\vk')\>' 
= 2\int_{\vp} \left[F_2(\vp,\vk-\vp)\right]^2\Plin(p)\Plin(|\vk-\vp|)
= \mathcal{I}^{[\d^{(2)},\d^{(2)}]}(k) 
\vs
P_{mm}^{(13)}(k) \equiv\:& \< \d^{(1)}(\vk) \d^{(3)}(\vk')\>' 
= 3\Plin(k)\int_{\vp} F_3(\vp,-\vp,\vk)\Plin(p)\,,
\ea
where $\mathcal{I}^{[O,O']}(k)$ is defined in \refeq{IOO}, and the explicit
expression for $P_{mm}^{(13)}(k)$ can be found in 
\cite{makino/sasaki/suto:1992}.
Similarly, the NLO matter-velocity cross-power spectrum and 
velocity-velocity power spectrum are given by 
$P_{m\theta}(k) = -f\cH \Plin(k) + P_{m\theta}^\NLO(k)$
and
$P_{\theta\theta}(k) = (f\cH)^2\Plin(k) + P_{\theta\theta}^\NLO(k)$,
where
\ba
P_{m\theta}^\NLO(k) =\:& P_{m\theta}^{(22)}(k) 
+ 2 P_{m\theta}^{(13)}(k) \,, \label{eq:Pt1loop}\\
P_{m\theta}^{(22)}(k) \equiv\:& 
\< \d^{(2)}(\vk) \theta^{(2)}(\vk')\>' 
= 
-2\cH f
\int_{\vp} F_2(\vp,\vk-\vp)G_2(\vp,\vk-\vp)\Plin(p)\Plin(|\vk-\vp|)
\vs
P_{m\theta}^{(13)}(k) \equiv\:& 
\frac12\left[
\< \theta^{(1)}(\vk) \d^{(3)}(\vk')\>' 
+ \< \d^{(1)}(\vk) \theta^{(3)}(\vk')\>' 
\right]
=
-\frac{1}{2}
\left[
 (\cH f) P_{mm}^{(13)}(k) 
+ (\cH f)^{-1} P_{\theta\theta}^{(13)}(k) 
\right] \vs
P_{\theta\theta}^\NLO(k) =\:& P_{\theta\theta}^{(22)}(k) 
+ 2 P_{\theta\theta}^{(13)}(k) \,,
\label{eq:Ptt1loop}\\
P_{\theta\theta}^{(22)}(k) \equiv\:& \< \theta^{(2)}(\vk) \theta^{(2)}(\vk')\>' 
= 2(\cH f)^2\int_{\vp} \left[G_2(\vp,\vk-\vp)\right]^2\Plin(p)\Plin(|\vk-\vp|)
\vs
P_{\theta\theta}^{(13)}(k) \equiv\:& \< \theta^{(1)}(\vk) \theta^{(3)}(\vk')\>' 
= 3 (\cH f)^2 \Plin(k)\int_{\vp} G_3(\vp,-\vp,\vk)\Plin(p)\,.
\ea

The accuracy of the SPT predictions for the nonlinear 
matter power spectrum and bispectrum has been studied extensively using 
simulations
\cite{juszkiewicz/bouchet/colombi:1993,catelan/moscardini:1994,gaztanaga/baugh:1995,baugh/gaztanaga/efstathiou:1995,scoccimarro/frieman:1996b,scoccimarro/etal:1998,jeong/komatsu:2006,smith/etal:2007,crocce/scoccimarro:2008,nishimichi/etal:2010,taruya/nishimichi/bernardeau:2013}.  Moreover, Ref.~\cite{jeong/komatsu:2006} showed that the accuracy improves significantly at higher redshifts ($z>1$) where the nonlinear scale is smaller (see below). 

\subsection{Feynman rules of large-scale structure}
\label{app:feynman}

The calculation of matter and galaxy statistics in perturbation theory can be performed efficiently by using a diagrammatic representation. The Feynman rules are as follows (see \cite{abolhasani/mirbabayi/pajer:2016} for a detailed description): 
\begin{enumerate}
\item An $n$-point correlation function is represented by a collection of diagrams with $n$ outgoing external legs. 
\item Interaction vertices have $m\geq 2$ ingoing lines $\vp_1,\cdots,\vp_m$ coupling to a single outgoing line $\vp$. Each such vertex is assigned a factor
  \be
   m! F_m(\vp_1, \cdots, \vp_m) (2\pi)^3 \d_D(\vp-\vp_{1\cdots m})\,.
  \ee
  One assigns a positive (negative) sign to outgoing (ingoing) momenta. Each ingoing line has to be directly connected to a propagator (linear power spectrum).\footnote{This is because diagrams that involve interaction vertices directly connected to each other are absorbed into higher-order interaction vertices.}

\item Propagators are represented as vertices with 2 outgoing lines of opposite momentum $\pm\vk$, {\includegraphicsbox[scale=0.8]{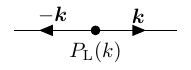}}, and they are assigned a factor $\Plin(k)$. 

\item All momenta that are not fixed in terms of momentum constraints are integrated over via
  \be
  \int_{\vp} \equiv \int \frac{d^3\vp}{(2\pi)^3}\,.
  \ee
A diagram without any loop integral is said to be a leading-order (LO), or tree-level diagram.
\item Each diagram is multiplied by the {\it symmetry factor}, which accounts for the number of all nonequivalent labelings of external lines and degenerate configurations of the diagram.
\end{enumerate}

As an example, the NLO contribution to the matter power spectrum [\refeq{Pm1loop}] can be represented as:
\ba
P_{mm}^\NLO(k) =\:& P_{mm}^{(22)}(k) + 2 P_{mm}^{(13)}(k)
=
\raisebox{-0.0cm}{\includegraphicsbox[scale=0.8]{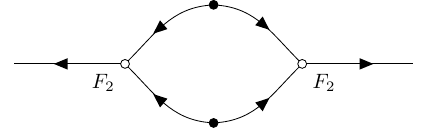}}
+
\raisebox{0.26cm}{\includegraphicsbox[scale=0.8]{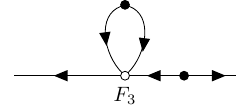}}
\,.
\label{eq:Pm1loopdiag}
\ea
  
\subsection{Effective field theory and the nonlinear scale}
\label{app:EFT}

The pressureless fluid equations \refeqs{cont}{euler} that we have
considered so far are not strictly correct, as they neither take into account
shell crossing of the dark matter, nor the presence of pressure in the baryonic component.   In reality, dark matter is governed by the
collisionless Boltzmann, or Vlasov equation, which predicts
that multi-streaming occurs on small scales.  Indeed, \refeqs{cont}{euler}
are obtained from the Vlasov equation by truncating the hierarchy of
velocity moments, and dropping the second- and higher-order moments, which contain the effective pressure
and anisotropic stress.  The pressure of the baryon fluid, 
on the other hand, cannot be neglected on small scales.
The Effective Field Theory approach to Large-Scale Structure 
(EFTofLSS \cite{baumann/etal:2012,carrasco/etal:2012})
provides a rigorous approach to take into account these beyond-pressureless-perfect-fluid contributions from small-scale perturbations.  Essentially, this
can be seen as a bias expansion for a specific tracer that obeys
stress-energy conservation.  The latter in fact ensures that \refeqs{cont}{euler}
are only corrected by higher-derivative contributions.  

The derivation of the EFT contributions proceeds by smoothing the 
density $\delta_\Lambda(\vx,\tau)$ and velocity $\v{v}_\Lambda(\vx,\tau)$ fields on the arbitrary scale $\L$,
retaining only modes $k\lesssim\Lambda$ (see \refsec{renorm}).  While this erases the small-scale
perturbations, the latter contribute stochastic terms, and moreover are 
modulated by $\delta_\Lambda$ and $\v{v}_\Lambda$, leading to additional long-wavelength contributions.  In the end, one obtains a contribution $-\partial_j \tau^{ij}/\rho_m$ on the right-hand-side of the Euler equation, where the effective stress tensor $\tau_{ij}$ 
captures the pressure and viscosity forces induced by the small-scale fluctuations. Expanding this to leading order in the large-scale fluctuations, the effective stress tensor can be written as \cite{baumann/etal:2012,carrasco/etal:2012} 
\be
\left[\tau_{ij}\right]_\Lambda
=
p_{\rm eff}(\L)\delta_{ij}
+
\rhob
\left[
c_s^2(\L)\delta_\Lambda \delta_{ij}
-
\frac{c_{bv}^2(\L)}{aH}\delta_{ij}\vn\cdot\v{v}_\Lambda
-
\frac34\frac{c_{sv}^2(\L)}{aH}
\left(
\partial_iv_{\Lambda}^j
+
\partial_jv_{\Lambda}^i
-
\frac32 \delta_{ij}\vn\cdot\v{v}_\Lambda
\right)
\right]
+ \cdots\,.
\ee 
Here, $p_{\rm eff}$, $c_s$, $c_{bv}$, $c_{sv}$
are, respectively, effective pressure, adiabatic sound speed, bulk viscosity
coefficient, and shear viscosity coefficient, which depend on $\L$. Note that $p_{\rm eff}$ leads to a stochastic contribution to the matter velocity $v^i(\vk)$ which in Fourier space is proportional to $i k^i$ (see also \refsec{stoch}). Since the quantities $c_s,\,c_{bv},\,c_{sv}$ are due to the dependence of the small-scale density and velocity fields 
on the large-scale environment, they can in principle be calculated
numerically by following a peak-background split argument similar to the
one discussed in \refsec{PBS}.  Further, the effective stress tensor $\tau_{ij}$ can be expanded up to any order in terms of a basis of counter-terms that is constructed out of the same $\Pi^{[n]}_{ij}$ that are employed in the general Eulerian bias expansion in \refsec{basisE} \cite{abolhasani/mirbabayi/pajer:2016}.  

Up to third order in perturbations, which is the highest order that is
relevant for the expressions given in this review, there is only a single
EFT contribution:
\be
\delta_{\Lambda}^{(3)}\Big|_\text{c.t.}(\bfk,\tau)
=
-\alpha(\tau,\L) D^3(\tau)\frac{k^2}{\knl^2}\delta^{(1)}(\vk,\tau)
W_{\Lambda}(\vk)\,,
\ee
where ``c.t.'' stands for counter-term, $W_{\Lambda}$ denotes the smoothing kernel used to integrate out the small-scale modes,
and $\knl$ is the nonlinear scale which we will define below.  
The free coefficient $\alpha$ cannot be predicted from perturbation theory
and needs to be determined by fitting to simulations or observations.  
Essentially, it corresponds to an effective sound speed.  
We can now choose a vanishingly small smoothing scale $\L^{-1}\to 0$,
in which case we obtain another contribution to the 
NLO matter power spectrum [\refeq{Pm1loop}],
\be
P_{mm}^\NLO\Big|_\text{c.t.}(k,\tau)
= -2  D^2(\tau) (2\pi) c_s^2 \frac{k^2}{\knl^2}\Plin(k,\tau)\,,
\label{eq:Pmmcs}
\ee
where we have traded $\lim_{\L^{-1}\to 0}\alpha(\L)$ for $c_s^2$.  This results in the contribution
given in \refeq{Phm1l}, where we have absorbed the growth factor $D^2(\tau)$ 
into $c_{s,\rm eff}^2$.  Furthermore, by allowing $c_s$ to be fit to simulations,
this correction term absorbs the dependence of the SPT contribution $P_{mm}^{(13)}$ in \refeq{Pm1loop} on fully nonlinear small-scale
modes that cannot be described by perturbation theory.  Empirically, one
finds values for $c_s$ of order 1 (e.g., \cite{carrasco/etal:2014}).  Note that
the leading stochastic EFT contribution to the matter power spectrum scales as $k^4 P_{\eps_m}^{\{4\}}(\tau)$ \cite{baldauf/schaan/zaldarriaga:2016}, where $\eps_m \propto p_{\rm eff}$ and $P_{\eps_m}^{\{4\}}(\tau)$ is its power spectrum amplitude in the low-$k$ limit which only depends on time. The scaling with $k^4$ is required by mass and momentum conservation \cite{peebles:1980,abolhasani/mirbabayi/pajer:2016}.  In terms of the perturbative counting of terms, this is a higher-order contribution; however, $\eps_m$ is relevant in the NLO halo-matter cross-power spectrum (\refsec{npt1loop}). 

We now turn to the general expectation for how large higher-order nonlinear terms
to the $n$-point functions of matter and biased tracers are at a given
scale $k$.  The most important scale is the \emph{nonlinear scale} which
we here define as the scale where the dimensionless matter
power spectrum $k^3 \Plin(k)/(2\pi^2)$ becomes unity. Note that
various different choices are possible here, since this is only a rough
estimate for the scale where higher-order corrections become of order one.  
For our reference cosmology, this corresponds to $\knl(z=0) \simeq 0.25 \iMpch$.  
At and below this scale, $\Plin(k)$ can be approximated by a power law
$\Plin(k) \propto k^n$ with $n = -1.9\cdots -1.7$ depending on the 
precise fit range chosen.  As discussed in \refsec{npt1loop}, 
higher-order SPT contributions to the $n$-point functions roughly scale
as $(k/\knl)^{n+3} \sim (k/\knl)^{1.1\cdots 1.3}$, while the leading
EFT term scales as $(k/\knl)^2$, for $c_s = \O(1)$.

\subsection{IR resummation}
\label{app:IRresum}

The perturbative expansion and counting of terms discussed so far in this Appendix
(as well as in \refsec{npt1loop}) relies on the assumption of a smooth, slowly varying linear matter power spectrum. In reality, the plasma oscillations in the baryon-photon fluid prior to recombination imprint a small-amplitude oscillatory pattern in the linear matter power spectrum, the well-known BAO feature first pointed out by \cite{eisenstein/hu}, with oscillation period $k_{\rm BAO} = 2\pi / r_s \approx 0.06 \iMpch$, where $r_s \approx 106 \Mpch$ is the sound horizon at recombination. These oscillations are damped towards higher $k$ by Silk damping.  

Now consider a perturbation with wavenumber $p \sim k_{\rm BAO}$.  This mode induces a displacement $\v{s}$ which moves small-scale perturbations with $k \gg p$. Perturbatively, this effect is captured at second order by the contribution $\v{s}^i \partial_i \d$. However, the effect of higher-order displacement terms on the oscillatory (``wiggle'') part of the power spectrum is not suppressed, since the displacement itself is modulated over scales of order $r_s$.  Fortunately, since this is a pure displacement effect, i.e., a coordinate transformation, these terms can be resummed \cite{crocce/scoccimarro:2008,senatore/zaldarriaga:2015,baldauf/etal:2015BAO,blas/etal:2016}. This resummation relies on the fact that $(i)$ the relevant modes sourcing the displacement are in the perturbative regime ($p \lesssim 0.2 \iMpch$), and $(ii)$ that the shape of the oscillatory feature is well known. Crucially, only the long-wavelength displacement effects on this oscillatory feature are resummed, as they are the ones which are enhanced.

First, we split the linear power spectrum into smooth ($s$) and oscillatory parts ($w$) via
\be
\Plin(k) = \Plin^{s}(k) + \Plin^{w}(k)\,,
\label{eq:Poscdecomp}
\ee
where
\be
\Plin^{w}(k) = \sin\left(\frac{k}{k_{\rm BAO}}\right)\, f_{\rm env}(k)
\ee
 is a strictly oscillating function whose amplitude is controlled by the smooth envelope $f_{\rm env}(k)$, while $\Plin^{s}(k)$ is defined as the remainder. Then, the IR resummation acting on the oscillatory part yields \cite{blas/etal:2016}\footnote{Eq.~(7.4) there; we have dropped the 2-loop-order displacement terms $\propto D^6$ in that expression, since they are numerically small. Note also the different Fourier convention adopted in \cite{blas/etal:2016}.} (see \cite{senatore/zaldarriaga:2015,baldauf/etal:2015BAO,vlah/etal:2016} for similar results)
\ba
P_{mm}^\text{IR-resum.}(k) =\:& \Plin^{s}(k)
+ \exp\left[-\Sigma^2(\epsilon k) k^2\right] \left(1 + \Sigma^2(\epsilon k) k^2 \right) \Plin^{w}(k) \vs
& + P_{mm}^{\NLO}\left[\Plin(k')\to \Plin^{s}(k') + \exp\left(-\Sigma^2(\epsilon k') k'^2\right) \Plin^{w}(k')\right] (k)\,.
\label{eq:PmIRresum}
\ea
Here, $P_{mm}^{\NLO}[\cdots](k)$ stands for the NLO matter power spectrum \refeq{Pm1loop} evaluated using the IR-resummed version of the linear power spectrum. Finally, 
\be
\Sigma^2(\Lambda) = \frac1{6\pi^2} \int_0^\Lambda dp\,\Plin(p) \left[1 - j_0(p r_s) + 2 j_2(p r_s) \right]
\ee
is essentially the covariance $\< s^i s^j \>$ of the linear Lagrangian displacement, after angle-averaging and smoothing with a sharp-$k$ filter at the scale $\L$. $\epsilon \ll 1$ is a parameter controlling down to what scale the IR resummation should occur; Ref.~\cite{baldauf/etal:2015BAO} choose $\epsilon=1/2$. 
The term $+\Sigma^2 k^2$ in parentheses in the first line of \refeq{PmIRresum} prevents the double counting of terms that appear in the NLO expression on the second line. As shown in \cite{senatore/zaldarriaga:2015,baldauf/etal:2015BAO,vlah/etal:2016,blas/etal:2016}, the IR-resummed matter power spectrum and its Fourier transform, the correlation function, significantly improve the match in the shape of these correlations with full N-body simulations on the scales around the BAO feature ($r \sim r_s$ in real space; $0.05 \iMpch \lesssim k \lesssim 0.3 \iMpch$ in Fourier space). 

At this, leading order in the wavelength $1/p$ of the soft modes that lead to displacements, the displacements are the same for both matter and LSS tracers, due to the equivalence principle (\refsec{velbias}; see also \cite{ivanov/sibiryakov:2018}).  This means that we can immediately make use of the IR resummation in the prediction for halo and galaxy clustering in real space, by generalizing \refeq{PmIRresum} to:
\ba
P_{hm}^\text{IR-resum.}(k) =\:& b_1 \Plin^{s}(k)
+ b_1 \exp\left[-\Sigma^2(\epsilon k) k^2\right] \left(1 + \Sigma^2(\epsilon k) k^2 \right) \Plin^{w}(k) \vs
& + P_{hm}^{\NLO}\left[\Plin(k')\to \Plin^{s}(k') + \exp\left(-\Sigma^2(\epsilon k') k'^2\right) \Plin^{w}(k')\right] (k)\,,
\label{eq:PhmIRresum}
\ea
and
\ba
P_{hh}^\text{IR-resum.}(k) =\:& b_1^2 \Plin^{s}(k)
+ b_1^2 \exp\left[-\Sigma^2(\epsilon k) k^2\right] \left(1 + \Sigma^2(\epsilon k) k^2 \right) \Plin^{w}(k) + P_\eps^{\{0\}} \vs
& + P_{hh}^{\NLO}\left[\Plin(k')\to \Plin^{s}(k') + \exp\left(-\Sigma^2(\epsilon k') k'^2\right) \Plin^{w}(k')\right] (k)\,,
\label{eq:PhhIRresum}
\ea
where the second line in each equation is obtained by replacing the linear power spectrum with its IR-resummed version in \refeq{Phm1l}. Again, this is because the IR resummation only deals with displacement terms which, at leading order in derivatives, are only modified by biasing through the trivial factor $b_1$. This holds analogously for the displacement from the galaxy rest frame into redshift space (\refsec{RSD}), whose IR contributions can similarly be resummed. Refs.~\cite{senatore/zaldarriaga:2014,perko/etal:2016,fonseca/etal:2017,ding/etal:2017,ivanov/sibiryakov:2018} provide resummed results for two-point functions in redshift space.

\subsection{Convective SPT approach and conserved evolution at third order}
\label{app:convSPT}

In this section, we provide some details on the calculation of the evolution
of a conserved biased tracer at third order in perturbation theory that
is briefly presented in \refsec{evol2}.  The ``convective SPT'' system, consisting of 
the continuity and Euler equation \refeqs{cont}{euler} for matter and
the continuity equation \refeq{contg} for the biased tracer, can be written in compact form as
\ba
\convD \v{\Psi} =\:& - \v{\sigma}\cdot\v{\Psi} + \v{S} 
\label{eq:PsigeomA}
\ea
where 
\ba
\v{\Psi}(\vx,\tau) = \left(\begin{array}{c}
\d_g(\vx,\tau) \\
\d(\vx,\tau) \\
\theta(\vx,\tau)
\end{array}
\right); \quad
\v{\sigma}(\tau) =\:& \left(\begin{array}{ccc}
0 & 0 & 1 \\
0 & 0 & 1 \\
0 & \frac32 \Om \cH^2 & \cH
\end{array}
\right); \quad
\v{S}(\vx,\tau) = \left(\begin{array}{c}
- \d_g\,\theta \\
- \d\,\theta \\
- (\partial_i v_j)^2
\end{array}
\right)\,.
\label{eq:Psigdef}
\ea
The fact that $\v{\sigma}$ is degenerate already shows that the three
equations are not really coupled, but rather the equation for $\d_g$ 
can be integrated separately as done in \refsec{evol1}.  
\refeq{PsigeomA}
allows for a convenient, compact derivation of both $\d$ and $\d_g$.  
In the following, we will solve this system up to third order.  

\subsubsection*{Source term}
\label{sec:evol2src}

The goal is to integrate \refeq{PsigeomA} along the fluid trajectory.  However, 
the source term $\v{S}$ involves some subtleties,
in particular the third component $S_3 = (\partial_i v_j)^2$,
since the derivative is with respect to Eulerian coordinate $\vx$.  The
velocity shear $\partial_i v_j$  
is nonlocally related to the degrees of freedom $\d,\,\theta$ themselves
[since $\v{v}=(\vn/\lapl) \theta$].  We can use a trick, 
namely the fact that the displacement is the integral of the peculiar
velocity along the fluid trajectory $\xfl(\vq,\tau)$, and therefore
\be
\v{v}(\xfl[\vq,\tau],\tau) = \convD \v{s}(\vq,\tau)\,.
\ee 
In the remainder of this section, we will abbreviate convective derivatives $\convDinline$ as primes.  Further, we need to transform the
derivative from $\vx$ to the fluid flow or Lagrangian coordinate $\vq$ via
\be
\partial_x^i = \left(\left[\v{1} + \v{M}\right]^{-1}\right)^i_{\  j} \partial_q^j\,,
\label{eq:derivxq}
\ee
where $M^{ij} = \partial_q^i s^j$.  At second order, we then obtain
\be
\partial_x^i v^j = \left[ \partial_q^i + ( \partial_q^i s_k ) \partial_q^k \right] s'^j + \O([\d^{(1)}]^3)\,.
\ee
Here, the left-hand side is defined at the Eulerian coordinate 
$(\vx,\tau)$ while the right-hand side quantities are defined at 
the corresponding Lagrangian coordinate $\vq$.  
With these expressions, we can construct source terms that are expressed 
purely in terms of convective time and Lagrangian spatial derivatives.  
Specifically, the third component of the
source term, $S_3$, is at first and second order given by (App.~B in \cite{MSZ})
\ba
(S^{(1)})_3 =\:& - [a'(\tau)]^2 \left[(K^{(1)})^2 + \frac13 (\d^{(1)})^2 \right]_{\tau_0} \vs
(S^{(2)})_3 =\:& 2 a\,[a'(\tau)]^2 \left[
-\frac23 \d^{(1)} \sigma^{(2)} - 2 K^{(1)}_{ij} \Del^{ij} \sigma^{(2)} + \frac19 (\d^{(1)})^3 + \d^{(1)} (K^{(1)})^2 + (K^{(1)})^3 
\right]_{\tau_0}\,,
\label{eq:S1and2}
\ea
where
\ba
\sigma^{(2)} \equiv\:& \partial_k s^{(2)k} = \frac12 \left[ -\frac27 (\d^{[1]})^2 + \frac37 (K^{[1]})^2 \right]
\label{eq:sigma2L}
\ea
is the divergence of the second-order Lagrangian displacement.

\subsubsection*{Initial conditions}

In order to integrate \refeq{PsigeomA}, we need to provide an expression for the initial conditions at some initial time $\tau_*$.  
The quantities $\d^*,\,\theta^*$ are obtained by integrating only the
matter part $[\v{\Psi}]_{i=2,3}$ from $\tau=0$ to $\tau=\tau_*$.  
Throughout, we only consider the fastest-growing-mode solutions for the density 
and velocity field. That is, the final time $\tau$ is assumed to be late enough so that all the slowly
growing modes are subdominant \cite{scoccimarro:1998,mccullagh/etal:2016}.

Under the assumptions explained in \refsec{evol2}, the galaxy density $\d_g^*$ at the ``formation time'' is given by
\refeq{dgIC}, which we repeat here for convenience:
\ba
\d_g^* \equiv \d_g(\xfl(\tau_*),\tau_*) =\:& \sum_{n=1}^3 \frac{b_n^*}{n!} [\d^*]^n 
+ b_{K^2}^* \tr \left[ (K_{ij}^*)^2 \right]
+ b_{K^3}^* \tr \left[ (K_{ij}^*)^3 \right]
+ b_{\d K^2} \d^*\, \tr \left[ (K_{ij}^*)^2 \right] \vs
& + \eps^* + \eps^*_{\d} \d^* + \eps^*_{\d^2} [\d^*]^2
+ \eps^*_{K^2} \tr \left[ (K_{ij}^*)^2 \right] 
\,,
\label{eq:dgICA}
\ea
where here and throughout a superscript $*$ indicates that a quantity is evaluated at
$\vx_* \equiv \vx_{\rm fl}(\tau_*)$ and $\tau_*$.  The stochastic fields 
$\eps^*,\,\eps^*_X$ ($X=\d$, $\d^2$, $K^2$) are
assumed to be first-order random variables.

\subsubsection*{Solution}
\label{sec:dgthird}

We define the matrix $\v{A}(\tau,\tau_i)$ as the solution to the following matrix ODE with boundary condition
\be
\partial_\tau \v{A}(\tau, \tau_i) + \v{\sigma}(\tau) \v{A}(\tau,\tau_i) = 0;\quad
\v{A}(\tau,\tau) = \v{1}\,.
\ee
In order to obtain closed analytical expressions, we restrict to an EdS Universe.  In this case, $\v{A}$ assumes a simple form constructed out 
of the growing and decaying modes which scale as $a(\tau)\propto\tau^2$ and $H(\tau) \propto \tau^{-3}$, respectively (App.~B in \cite{MSZ}).  In the final result, we then replace scale factors $a(\tau)$ with growth factors $D(\tau)$, as described after \refeq{SPTkernels}. 

Given the initial condition for \refeq{PsigeomA} from above,
$\v{\Psi}(\vx_*, \tau_*) = (\d_g^*,\:\d^*,\:\theta^*)$ with \refeq{dgICA},
the particular solution to \refeq{PsigeomA} is then given by
\be
\v{\Psi}(\vx,\tau) = 
\v{A}(\tau, \tau_*) 
\left(\begin{array}{c}
\d_g^* \\
\d^*\\
\theta^*
\end{array}\right)
+ \int_{\tau_*}^\tau d\tau'\:\v{A}(\tau,\tau')
\v{S}(\vx_{\rm fl}(\tau'),\,\tau') \,.
\label{eq:Psigsol}
\ee
The interpretation of \refeq{Psigsol} is clear:  the
density and velocity of the fluid and galaxy at position $(\vx,\tau)$ is given by
an integral over the fluid trajectory of the source term multiplied by the
Green's function $\v{A}(\tau,\tau')$.  \refeq{Psigsol} leads to
the SPT result for matter, in particular for 
the terms that are invariant under
time-dependent coordinate shifts $\vx \to \vx + \v{\xi}(\tau)$.  The non-invariant
terms, which are not captured by \refeq{Psigsol}, are precisely the 
displacement terms relating the solutions in Lagrangian and Eulerian 
coordinates.  Finally, expanding in these displacements to the same perturbative order 
then yields exactly the SPT result. Thus, the convective SPT
calculation keeps track of the displacement terms exactly (i.e., it resums them), similar to
Lagrangian perturbation theory.  

At second order, we then recover our previous result, \refeq{dg2nd_inst2} as well as \refeqs{d2}{t2}.  Specifically, the solution reads
\be
\v{\Psi}^{(2)}(\vx,\tau) 
= \left(
\begin{array}{c}
b_1^E(\tau) \d^{(2)} + b_2^E(\tau) (\d^{(1)})^2/2 + b_{K^2}^E(\tau) (K^{(1)}_{ij})^2 +  \eps^E_{\d}(\tau) \d^{(1)} 
\\
D^2(\tau) \left[\frac{17}{21} (\d^{(1)})^2 + \frac27 (K^{(1)}_{ij})^2 \right] \\
-D(\tau)\dot D(\tau) \left[ \frac{13}{21} (\d^{(1)})^2 + \frac47 (K^{(1)}_{ij})^2 \right]
\end{array}
\right)_{\vq}\,.
\label{eq:Psig2}
\ee
At third order, we obtain for the matter fields \cite{MSZ}\footnote{Note that $\theta^{(3)}$ in Eq.~(B23) of \cite{MSZ} lacks an overall minus sign.}
\ba
\d^{(3)} =\:& \frac{341}{567} \d^3 + \frac{11}{21} K^2 \d + \frac29 K^3 
+ \frac16 O_{\otd} \vs
\theta^{(3)} =\:& - f \cH \left[\frac{71}{189} \d^3 + \frac57 K^2 \d + \frac23 K^3 + \frac12 O_{\otd} \right]\,,
\label{eq:delta3}
\ea
where $K^n$ stands for $\tr[K_{ij}^n]$, and all quantities on the right-hand side are evaluated at linear order and 
at $\tau$.  $O_\otd$ is defined in \refeq{Otddef}. 
Note that, although all quantities appearing in \refeq{delta3} are defined
using derivatives with respect to $\vx$, 
the distinction between Eulerian and Lagrangian derivatives in these
terms only matters at higher order.  
Notice further that the coefficient of the $\d^3$ term in $\d^{(3)}$ is 
exactly the third-order coefficient of the perturbative expansion of 
spherical collapse in an EdS Universe [\refeq{scseries}], as expected: for
a spherically symmetric perturbation, $K_{ij}=0$, and $\d^{(3)}$ has to reduce
to the cubic term in the perturbative expansion of the spherical collapse solution.  
To the authors' knowledge, despite its fairly simple form, Ref.~\cite{MSZ} was the
first to give the complete expression for $\delta^{(3)},\,\theta^{(3)}$ in real space.  Interestingly, the expression for the nonlinear density derived by Ref.~\cite{ohta/kayo/taruya:2004} from ellipsoidal collapse \cite{white/silk:1979,bond/myers:1996} does not contain a term of the form $K^{ij} \Del_{ij} \sigma^{(2)}$, while it matches the PT result at second order.  The same is found in the local tidal approximation (LTA) of \cite{hui/bertschinger:1996}, as well as the fully relativistic local approximation derived in \cite{ip/schmidt:2016}.  
The underlying reason is that the evolution equations of an isolated ellipsoidal perturbation are only approximate, since they neglect the dynamical interaction with the large-scale environment.  When allowing for general configurations,
the nonlinear evolution of tidal fields in cosmology is nonlocal  \cite{bertschinger/hamilton:1994,kofman/pogosyan:1995,bertschinger:1995}.  

Finally, in order to obtain the density at a fixed order in standard Eulerian perturbation
theory, we need to displace $\v{\Psi}$ from a fixed Lagrangian position to the Eulerian position, by expanding in the argument.  Let us define the Eulerian solution $\v{\Psi}^E$ through
\be
\v{\Psi}^E(\vx) = \v{\Psi}(\vx - \v{s}[\vq,\tau])\,.
\ee
$\v{\Psi}$ on the right-hand side contains all invariant terms, which we have
obtained above.  We now perform a Taylor expansion in $\v{s}$
as well as a perturbative expansion of $\v{s}$, noting that $\v{s}$ is itself
a function of the Lagrangian rather than Eulerian position.  We obtain at third order
\ba
\v{\Psi}^{E,(3)} (\vx,\tau)
=\:& \v{\Psi}^{(3)} - s_{(1)}^i \partial_i \v{\Psi}^{(2)}
- \left[s_{(2)}^i - s_{(1)}^j (\partial_j s_{(1)}^i) \right] \partial_i \v{\Psi}^{(1)} 
+ \frac12 s_{(1)}^i s_{(1)}^j \partial_i \partial_j \v{\Psi}^{(1)} \,,
\label{eq:Psigtotal}
\ea
where on the right-hand side all quantities are evaluated at $\vx,\,\tau$, and
$\v{\Psi}^{(n)} = \{\d_g^{(n)},\d^{(n)},\theta^{(n)}\}$, are given
in \refeq{Psig2} for $n=2$, and \refeq{dg3f}, \refeq{delta3} for $n=3$, respectively.   
As expected, the displacement from Lagrangian to Eulerian position, being merely a coordinate shift, does not affect the bias relation at any order.

\subsection{Conserved evolution and bias expansion beyond the EdS background}
\label{app:convSPT:b}

We now allow for a more general expansion history, in particular $\Lambda$CDM
or quintessence (neglecting perturbations in the dark energy component), 
and derive how this changes the evolution equations of a conserved tracer
in the convective SPT approach.  
For this, we transform the time coordinate in the equations of motion \refeq{Psigeom} to $\ln D$, where $D$ is the linear growth factor [\refeq{Deom}].  Since $d\ln D/d\tau = \cH d\ln D/d\ln a = \cH f$ by definition of the growth rate $f$, we introduce
\be
\tilde{\v{v}} \equiv (\cH f)^{-1} \v{v} \qquad\mbox{and}\qquad
\convDD \equiv \frac{\partial}{\partial\ln D} + \tilde v^i \partial_i\,.
\ee
Note that $\tilde\theta = \partial_i \tilde{v}^i = -\d$ at linear order.  Using \refeq{dHfdtau}, we then obtain
\ba
\convDD \tilde{\v{\Psi}} =\:& - \tilde{\v{\sigma}}\cdot\tilde{\v{\Psi}} + \tilde{\v{S}} \label{eq:Psiggen}\\
\tilde{\v{\Psi}}(\vx,\tau) = \left(\begin{array}{c}
\d_g(\vx,\tau) \\
\d(\vx,\tau) \\
\tilde\theta(\vx,\tau) 
\end{array}
\right); \quad
\tilde{\v{\sigma}}(\tau) =\:& \left(\begin{array}{ccc}
0 & 0 & 1 \\
0 & 0 & 1 \\
0 & \frac32 \frac{\Om}{f^2} & \frac32 \frac{\Om}{f^2} - 1
\end{array}
\right); \quad
\tilde{\v{S}}(\vx,\tau) = \left(\begin{array}{c}
- \d_g\,\tilde\theta \\
- \d\,\tilde\theta \\
-  (\partial_i \tilde{\v{v}}_j)^2
\end{array}
\right)\,.
\label{eq:sigmagen}
\ea
We see that the equations maintain the same structure, and that the departure from EdS of the equations of motion is completely
quantified by $\Om(a)/f^2(a)-1$.  In particular, the time dependence of the second-order matter density field is, instead of $[D(\tau)]^2$, given by a second-order growth factor $D_2(\tau)$, obtained by integrating $[D(\tau)]^2$ against the Green's function corresponding to \refeq{sigmagen}.  Only starting at third order do we obtain two different time dependences in the contributions to $\d$ at the same perturbative order;  specifically, $D_{3,1}(\tau)$ and $D_{3,2}(\tau)$, obtained respectively by integrating $[D(\tau)]^3$ and $D_2(\tau) D(\tau)$ against the Green's function.  

We now go back to the construction of the set of general local bias operators in \refsecs{basisL}{basisE}.  
The first instance of a new operator being induced by the different time evolution in quintessence cosmology is at third order, where $\Pi^{[3]}_{ij}$ defined via \refeq{Pindef} in \refsec{basisE} separates into two different operators $\Pi^{[3,1]}_{ij},\,\Pi^{[3,2]}_{ij}$.  
Now, the reason for why we do not need to include convective time derivatives of $\d$ itself, or equivalently $\tr[\Pi^{[n]}]$, in the perturbative bias expansion also holds in the more general case considered here:  by including time derivatives of all other operators constructed out of $\partial_i\partial_j\Phi$, we already obtain the operators appearing in the time derivatives of $\d$, or equivalently in the perturbative contributions $\d^{(2)},\,\d^{(3)},\cdots$.  Thus, we do not need to include $\tr[\Pi^{[3,i]}]$, and the first new operator in the Eulerian basis \refeq{EulBasis} of the bias expansion appears at fourth order, where
\be
\tr[ \Pi^{[1]} \Pi^{[3]} ] \longrightarrow \tr[ \Pi^{[1]} \Pi^{[3,1]} ]\,,\  
\tr[ \Pi^{[1]} \Pi^{[3,2]} ]\,.
\label{eq:PibeyondEdS}
\ee
Conversely, in the EdS limit these two operators have to combine into
$\tr[ \Pi^{[1]} \Pi^{[3]} ]$.  This means that the bias coefficients of
the two different fourth-order operators will be approximately equal, with 
a fractional difference scaling as $D_{3,1}(\tau)/D_{3,2}(\tau) - 1 \ll 1$.  
Thus, the bases of bias operators described in \refsecs{basisL}{basisE} can
be straightforwardly extended beyond EdS to quintessence cosmologies.  
Moreover, all explicit results for galaxy and halo statistics derived in 
\refsec{measurements}, which only rely on perturbative results up to third order, are independent
of the EdS assumption.  
While the time evolution of bias parameters for conserved tracers [\refeq{biasEthirdorder}] is modified from the EdS case, this is not of practical relevance if all the bias parameters are determined from the data at a fixed time.

%% file: App_biastrans.tex
\section{Bias conventions and their relation}
\label{app:biastrans}

The general bias expansion consists of a set of operators $\{ O_i^{[n]} \}$,
$i = 1,\cdots N_n$, at each order $n$ in perturbation theory (as discussed
in \refsec{stoch}, the stochastic
terms involve the same basis of operators as well, hence we do not
need to consider them explicitly here).  A superscript $[n]$ indicates
that the lowest-order contribution to this operator is at $n$-th order
in perturbation theory. Of course, there are contributions at order $n+1, n+2$, and so on as well. Any linearly
independent combination of the $O_i^{[n]}$ at a given order $n$ leads
to an equivalent bias expansion.  For example, for $n=1$,
we only have a single operator $O^{[1]} = \d$, but $O^{[1]} = \theta/\cH$ or
$O^{[1]} = \lapl\Phi/\cH^2$ constitute equivalent choices.  Correspondingly, 
several different conventions for the first few orders of the bias
expansion have been used in the literature.  We provide a brief summary
of the relations between the most commonly used bias conventions here.

Let us denote the set of operators in a given basis at a fixed order
$n$ in perturbation theory as $\bm{O} = \{ O_i^{[n]} \}_{i=1}^{N_n}$.  The bias expansion
at $n$-th order can then be treated as an $N_n$-dimensional vector space.  
Consequently, a different basis $\tilde{\bm{O}}$ can always be written as
\be
\tilde{\bm{O}} = \bm{M}\cdot \bm{O}\,,
\ee
where $\bm{M}$ is an invertible $N_n\times N_n$ matrix.  Then, the
coefficients, i.e. the bias parameters which we write as
$\bm{b} = \{ b_{O_i^{[n]}}\}_{i=1}^{N_n}$, transform via the transpose of the inverse of $\bm{M}$: 
\ba
\d_g \supset\:& \sum_{i=1}^{N_n} b_{O_i^{[n]}} O_i^{[n]} = \sum_{i=1}^{N_n} 
\tilde b_{\tilde O_i^{[n]}} \tilde O_i^{[n]} \quad\Rightarrow\quad
\tilde{\bm{b}} = \left(\bm{M}^{-1}\right)^\trans \cdot \bm{b}\,.
\ea
Note that even if one or more of the $\tilde O_i^{[n]}$ is the same as
$O_i^{[n]}$, the corresponding bias parameter $\tilde b_{O_i^{[n]}}$ in general 
differs from $b_{O_i^{[n]}}$ due to the transformation in the $N_n$-dimensional
vector space.  We now derive these transformations for frequently used
bias conventions at second and third order.

\subsection{Second order}

At second order, the basis used in this review consists of $\bm{O} = \{ \d^2, (K_{ij})^2 \}$.
Another frequently used basis is $\tilde{\bm{O}} = \{ \d^2, \G_2 \}$, where 
\ba
\G_2 \equiv \left[ (\partial_i\partial_j\hat\Phi)^2 - (\lapl\hat\Phi)^2 \right] = (K_{ij})^2 - \frac23 \d^2\,,
\label{eq:G2def}
\ea
where $\hat\Phi \equiv 2\Phi/(3\Om\cH^2)  = \ilapl\d$.  We then obtain
\be
\tilde b_{\d^2} = b_{\d^2} + \frac23 b_{K^2} \quad\mbox{and}\quad
\tilde b_{\G_2} = b_{K^2}\,.
\label{eq:bG2}
\ee
Finally, using the tensor
\be
\Pi^{[1]}_{ij} \equiv \partial_i\partial_j\hat\Phi = K_{ij} + \frac13 \d_{ij} \d
\ee
introduced in \refsec{basisE}, one can choose
\be
\left(\tr\left[ \Pi^{[1]} \right]\right)^2 = \d^2\,,\qquad
\tr\left[ \left(\Pi^{[1]}\right)^2 \right] = K^2 + \frac13 \d^2 
\ee
as second-order basis. This yields
\be
\tilde b_{(\tr \Pi)^2} = b_{\d^2} - \frac13 b_{K^2}
\quad\mbox{and}\quad
\tilde b_{\tr[\Pi^2]} = b_{K^2}\,.
\ee

\subsection{Third order}

At third order, an operator nonlocally related to $\partial_i\partial_j\Phi$ 
appears, which can be defined in a variety of ways.  In this review, we mostly use the definition in \refeq{Otddef},
\ba
O_{\otd}^{(3)} \equiv \frac{8}{21} 
K^{(1)}_{ij} \Del^{ij} \left[(\d^{(1)})^2 
-\frac32 (K^{(1)}_{ij})^2 \right]
\,.
\ea
Note that, unlike the operators $\d^2, (K_{ij})^2$, it is not obvious how the definition of this operator is to be extended to higher order in perturbations. We will return to this shortly. 
Let us relate $O_\otd^{(3)}$ to the second-order tensor appearing in the general Eulerian bias expansion (\refsec{basisE}),
\be
\Pi^{[2]}_{ij} \equiv  (\cH f)^{-1}\convD \Pi^{[1]}_{ij} - \Pi^{[1]}_{ij}\,.
\label{eq:piij2def}
\ee
Note that \refeq{piij2def} is valid at any order in perturbation theory. 
At second order in PT, this yields 
  \ba
  \label{eq:piij2}
\Pi^{[2]}_{ij}\Big|^{(2)} =\:& \partial_i\partial_k\hat\Phi \partial^k\partial_j\hat\Phi
+ \frac{10}{21} \frac{\partial_i\partial_j}{\lapl} \left(\d^2 - \frac32 (K_{kl})^2 \right) \,.
\ea
This result can be derived in a variety of ways. \refapp{Pi2deriv} gives some details on the Fourier-space derivation which makes use of the SPT kernel expansion (\refapp{SPT:SPT}).  \refeq{piij2} immediately yields
\ba
\tr[\Pi^{[1]} \Pi^{[2]}]\Big|^{(3)} =\:& \frac54 O_{\otd}^{(3)} + (K_{ij})^3 + \frac{16}{21} \d (K_{ij})^2 + \frac{17}{63} \d^3 \,.
\label{eq:OtdPi2}
\ea
This can now be used to define $O_\otd$ at all orders in PT, although we do not need the result for any of the expressions presented in this review.

$O_\otd^{(3)}$ can also be related to the second-order Lagrangian distortion tensor $M_{ij}^{(2)}$.  Using that 
\ba
M^{(2)}_{ij} =\:& \frac{\partial_q^i \partial_q^j}{\nabla_q^2} (\vn_q\cdot\v{s}_{(2)})
= \left[f \cH \partial_\tau M_{ij} - M_{ij}\right]^{(2)} \label{eq:M2ij} \\
\mbox{where} \quad 
\vn_q\cdot\v{s}_{(2)} =\:& - \frac17 \left[ \left(\d^{(1)}\right)^2 - \frac32
 \left(K_{ij}^{(1)}\right)^2\right]
= -\frac34\left[\d^{(2)} + \frac{\theta^{(2)}}{\cH f}\right]
\ea
is the divergence of the second-order Lagrangian displacement, we obtain 
\be
O_{\otd}^{(3)} = \frac83 \left[ M^{(1)\,ij} - \frac13 \d^{ij} \tr M^{(1)}\right] 
M^{(2)}_{ij}\,.
\label{eq:OtdLagr}
\ee
In \refeq{M2ij} we have used that convective time derivatives reduce to ordinary time derivatives for operators defined in terms of Lagrangian coordinates.  

Further, relating $O_{\otd}^{(3)}$ to the velocity potential $\Phi_v \equiv -(f\cH)^{-1} \ilapl\theta$, we can also write \cite{mcdonald/roy:2009,assassi/etal}
\ba
O_{\otd}^{(3)} =\:& 
\Gamma_3 + \frac{16}{63} \d^3 - \frac8{21} \d (K_{ij})^2 \,,\vs
\mbox{where}\quad \Gamma_3 \equiv\:& (\partial_i\partial_j \hat\Phi)^2 - (\lapl \hat\Phi)^2
- \left[(\partial_i\partial_j \Phi_v)^2 - (\lapl \Phi_v)^2 \right]\,.
\label{eq:Gamma3def}
\ea
Note that, at linear order, $\Phi_v^{(1)} = \Phi^{(1)}$ so that $\Gamma_3 = \Gamma_3^{[3]}$ starts at third order in PT. Specifically, we have at this order
\be
\Gamma_3^{(3)} = 2 K^{ij} \Del_{ij} \left[ \d + (f\cH)^{-1} \theta \right]^{(2)} - \frac43 \d \left[ \d + (f\cH)^{-1} \theta \right]^{(2)}\,,
\label{eq:Gamma3third}
\ee
where
\be
\left[ \d + (f\cH)^{-1} \theta \right]^{(2)} = \frac4{21} \left[\d^2 - \frac32 K^2\right]\,.
\ee
Using that
\ba
\G_3 \equiv\:& -\frac12 \left[2 (\partial_i\partial_j\hat\Phi)^3 + \d^3 - 3 (\partial_i\partial_j\hat\Phi)^2 \d \right] \vs
=\:& - (K_{ij})^3 + \frac12 \d (K_{ij})^2 - \frac19 \d^3\,,
\ea
as defined in \cite{assassi/etal}, we then obtain the relation between the third-order operator
basis adopted in \cite{assassi/etal} and the basis used here, which is summarized in \reftab{biastrans1}.

\begin{table*}[t]
\begin{minipage}[t]{0.49\textwidth}
\centering
\begin{tabular}{|c|c|c|c|c|}
\hline
 $M_{\tilde{O} O}$ & $\d^3$ & $\d K^2$ & $K^3$ & $O_\otd^{(3)}$ \\[3pt]
\hline
\hline
 $\d^3$ & 1 & \Ctabzero & \Ctabzero & \Ctabzero \\
\hline
 $\d \G_2$ & -2/3 & 1 & \Ctabzero & \Ctabzero \\
\hline
 $\G_3$ & -1/9 & 1/2 & -1 & \Ctabzero \\
\hline
 $\Gamma_3$ & -16/63 & 8/21 & \Ctabzero & 1 \\
\hline
\end{tabular}
\end{minipage}
\hfill
\begin{minipage}[t]{0.49\textwidth}
\centering
\begin{tabular}{|c|c|c|c|c|}
\hline
 $(M^{-1})^\trans_{\  \tilde{O} O}$ & $b_{\d^3}$ & $b_{\d K^2}$ & $b_{K^3}$ & $b_\otd$ \\[3pt]
\hline
\hline
 $\tilde b_{\d^3}$ & 1 & 2/3 & 2/9 & \Ctabzero \\
\hline
 $\tilde b_{\d\G_2}$ & \Ctabzero & 1 & 1/2 & -8/21 \\
\hline
 $\tilde b_{\G_3}$ & \Ctabzero & \Ctabzero & -1 & \Ctabzero \\
\hline
 $\tilde b_{\Gamma_3}$ & \Ctabzero & \Ctabzero & \Ctabzero & 1 \\
\hline
\end{tabular}
\end{minipage}
\caption{Transformation between the third-order basis of operators
$\v{O} = \{ \d^3,\, \d K^2,\, K^3,\, O_\otd^{(3)} \}$ adopted in this review, and
$\tilde{\v{O}} = \{ \d^3,\, \d \G_2,\, \G_3,\, \Gamma_3 \}$ adopted in 
\cite{assassi/etal} (left).  The right table gives the corresponding
transformation of bias parameters. Vanishing matrix elements are indicated
as empty cells for clarity.
\label{tab:biastrans1}}
\end{table*}

\begin{table*}[t]
\begin{minipage}[t]{0.49\textwidth}

 \centering
\begin{tabular}{|c|c|c|c|c|}
\hline
 $M_{\tilde{O} O}$ & $\d^3$ & $\d K^2$ & $K^3$ & $O_\otd^{(3)}$ \\[3pt]
\hline
\hline
 $\d^3$ & 1 & \Ctabzero & \Ctabzero & \Ctabzero \\
\hline
 $\d K^2$ & \Ctabzero & 1 & \Ctabzero & \Ctabzero \\
\hline
 $st$ & \Ctabzero & \Ctabzero & \Ctabzero & -1/2 \\
\hline
 $\psi_{\rm MR}$ & 328/3969 & -4/49 & -8/63 & -8/21 \\
\hline
\end{tabular}
\end{minipage}
\hfill
\begin{minipage}[t]{0.49\textwidth}
\centering
\begin{tabular}{|c|c|c|c|c|}
\hline
 $(M^{-1})^\trans_{\  \tilde{O} O}$ & $b_{\d^3}$ & $b_{\d K^2}$ & $b_{K^3}$ & $b_\otd$ \\[3pt]
\hline
\hline
 $\tilde b_{\d^3}$ & 1 & \Ctabzero & 41/63 & \Ctabzero \\
\hline
 $\tilde b_{\d K^2}$ & \Ctabzero & 1 & -9/14 & \Ctabzero \\
\hline
 $\tilde b_{st}$ & \Ctabzero & \Ctabzero & 6  & -2 \\
\hline
 $\tilde b_{\psi_{\rm MR}}$ & \Ctabzero & \Ctabzero & -63/8 & \Ctabzero \\
\hline
\end{tabular}
\end{minipage}
\caption{Transformation between the third-order basis of operators
$\v{O} = \{ \d^3,\, \d K^2,\, K^3,\, O_\otd^{(3)} \}$ adopted in this review, and
$\tilde{\v{O}} = \{ \d^3,\, \d K^2, st, \psi_{\rm MR} \}$ adopted in  
\cite{mcdonald/roy:2009} (left).  The right table gives the corresponding
transformation of bias parameters.  Note that Ref.~\cite{mcdonald/roy:2009}
use $b_3^{\rm MR} = 6 \tilde b_{\d^3}$ and $b_{\d s^2}^{\rm MR} = 2 \tilde b_{\d K^2}$, and that $s^3 \equiv K^3$ can be re-expressed in terms of the other operators in $\tilde{\v{O}}$ and hence is redundant.  Vanishing matrix elements are indicated as empty cells for clarity.
\label{tab:biastrans2}}
\end{table*}

Finally, we can relate $O_{\otd}^{(3)}$ to the operators defined in \cite{mcdonald/roy:2009} through (note that $s_{ij}$ there equals $K_{ij}$ as defined here)
\ba
\eta \equiv\:& \lapl\Phi_v - \lapl\hat\Phi\,, 
\vs
t_{ij} \equiv\:& \left(\partial_i\partial_j  - \frac13 \d_{ij} \lapl\right) \Phi_v - K_{ij} = \Del_{ij} \eta\,.
\ea
At second order in PT, this yields
\ba
\eta^{(2)} = -\frac4{21} 
\left(\left[\d^{(1)}\right]^2 - \frac32 \left[K_{ij}^{(1)}\right]^2 \right)\,.
\ea
We then have
\ba
st \equiv\:& s_{ij} t^{ij} = K_{ij} \left(\frac{\partial_i\partial_j}{\lapl} \right) \eta = -\frac12 O_{\otd}^{(3)}\,. \label{eq:stMR}
\ea
Using \refeq{delta3}, the operator $\psi = \psi_{\rm MR}$ defined in \cite{mcdonald/roy:2009} can eventually be expressed at third order as
\ba
\psi_{\rm MR}^{(3)} \equiv\:& \left[\eta - \frac27 (K_{ij})^2 + \frac4{21} 
\d^2\right]^{(3)} \vs
=\:&  -\frac8{21} O_{\otd}^{(3)} 
+ \frac{328}{3969} \d^3
- \frac4{49} \d (K_{ij})^2
- \frac8{63} (K_{ij})^3\,.
\ea
\reftab{biastrans2} summarizes the relation between the third-order operator
basis adopted in \cite{mcdonald/roy:2009,saito/etal:14} and the basis used here. Note that $K^3$ (denoted as $s^3$ in their notation) is linearly dependent on the other operators in their basis and hence not included here.  Alternatively, one could replace either $st$ or $\psi_{\rm MR}$ with the operator $K^3$. 

Refs.~\cite{mcdonald/roy:2009,saito/etal:14} introduce a bias parameter
$b_{3\rm nl}$ which multiplies the term proportional to $f_\NLO(k) \Plin(k)$ in the NLO galaxy power spectrum, \refeq{Phm1l}, where $f_\NLO(k)$ is defined in \refeq{IOO}.  
In our notation, this parameter is given by
\be
b_{3\rm nl} = -\frac{32}{21} \left( b_{K^2} + \frac25 b_{\otd} \right) \,,
\label{eq:b3nl}
\ee
where the prefactor comes from matching $f_\NLO(k)$ and the function
$\sigma_3^2(k)$ defined in \cite{mcdonald/roy:2009}.  Inserting our results
from \refsec{evol2} for Lagrangian \LIMD bias [\refeq{b2E} and \refeq{biasEthirdorderLL}], we obtain
\be
b_{3\rm nl} = \frac{32}{315} b_1^L\,,
\ee
in agreement with \cite{saito/etal:14}.  Note, however, that $b_{3\rm nl}$
is different from the $b_O$ discussed throughout the review, since it 
does not multiply a single bias operator.  Instead, it is defined as a 
coefficient of a specific term in the NLO (1-loop) contribution to the galaxy
power spectrum (indeed, we see from \refeq{b3nl} that it involves both
second- and third-order biases).

\subsection{Derivation of $\Pi^{[2]}$}
\label{app:Pi2deriv}

Using the definition of the convective time derivative [\refeq{DDtau}], and the perturbation theory expansion of the density and velocity fields [\refeq{SPTexpansion}], $\Pi^{[2]}_{ij}$ can be formally written as a sum over perturbation-theory contributions \cite{MSZ}:
  \be
\begin{split}
\Pi^{[2]}_{ij}(\vk,\tau) =\:\sum_{n=1}^{\infty}D^n(\tau)
\Bigg[&\frac{\vk^i \vk^j}{|\vk|^2} (n-1)\delta^{(n)}(\vk)\\[8pt]
      &+\frac{1}{\cH f}\sum_{m=1}^{n-1} \int_{\vk_1}\int_{\vk_2} (2\pi)^3\d_D(\vk-\vk_{12})
\frac{\vk_1\cdot\vk_2}{k_1} \frac{k_2^ik_2^j}{k_2^2} \theta^{(m)}(\vk_1)\delta^{(n-m)}(\vk_2)\Bigg]\,.
\end{split}
\ee
At second order, we can restrict to $n=2$ and $m=1$. This yields
\ba
\Pi^{[2]}_{ij}(\vk,\tau) =\:&
\int_{\vk_1}\int_{\vk_2} (2\pi)^3\d_D(\vk-\vk_{12})
\left[
\frac{k_{12}^ik_{12}^j}{k_{12}^2} F_2(\vk_1,\vk_2)
- \frac12 \frac{\vk_1\cdot\vk_2}{k_1^2 k_2^2} \left(k_1^ik_1^j + k_2^i k_2^j\right)\right] \d(\vk_1)\d(\vk_2) \vs
=\:&
\int_{\vk_1}\int_{\vk_2} (2\pi)^3\d_D(\vk-\vk_{12})
\Bigg[
  \frac{k_{12}^ik_{12}^j}{k_{12}^2} \left( \frac{17}{21} + \frac27 \left(\mu_{12}^2-\frac13\right) + \frac12 \vk_1\cdot\vk_2 \left(\frac1{k_1^2}+\frac1{k_2^2}\right) \right) \vs
  & \hspace*{3.8cm}
- \frac12 \frac{\vk_1\cdot\vk_2}{k_1^2 k_2^2} \left(k_1^ik_1^j + k_2^i k_2^j\right)\Bigg] \d(\vk_1)\d(\vk_2)\,,
\ea
where $\mu_{12} = (\vk_1\cdot\vk_2)/k_1 k_2$ and we have used \refeq{SPTkernels}. The two displacement terms which are linear in $\vk_1\cdot\vk_2$ can be combined over a denominator $1/(k_1^2 k_2^2 |\vk_{12}|^2)$ to give
\ba
\Pi^{[2]}_{ij}(\vk,\tau) =\:&
\int_{\vk_1}\int_{\vk_2} (2\pi)^3\d_D(\vk-\vk_{12})
\left[
  \frac{k_{12}^ik_{12}^j}{k_{12}^2} \left( \frac{17}{21} + \frac27 \left(\mu_{12}^2-\frac13\right) - \mu_{12}^2 \right)
  + \frac{\vk_1\cdot\vk_2}{2k_1^2 k_2^2} \left(k_1^ik_1^j + k_2^i k_2^j\right)\right] \d(\vk_1)\d(\vk_2)\,.
\nonumber
\ea
We can now read off the real-space expression corresponding to $\Pi^{[2]}$:
\ba
\Pi^{[2]}_{ij}(\vx,\tau) =\:& \frac{10}{21} \frac{\partial_i\partial_j}{\lapl} \left[ \d^2 - \frac32 K^2 \right] + \partial_i\partial_k\hat\Phi \partial^k\partial_j\hat\Phi\,.
\ea
This is \refeq{piij2}. More generally, the following relations are useful to commute displacement terms with an inverse Laplacian:
  \ba
\frac1{\lapl} \left[\partial_i A \frac{\partial^i}{\lapl} B + (A \leftrightarrow B) \right] =\:& \int_{\vk_1}\int_{\vk_2} (2\pi)^3 \d_D(\vk-\vk_{12})
\left(- \frac1{k_{12}^2} \right) \left[\frac1{k_1^2} + \frac1{k_2^2} \right] (\vk_1\cdot\vk_2) A(\vk_1) B(\vk_2) \vs
=\:& \int_{\vk_1}\int_{\vk_2} (2\pi)^3 \d_D(\vk-\vk_{12})
\left(- \frac1{k_{12}^2 k_1^2 k_2^2} \right) \left[k_{12}^2 - 2\vk_1\cdot\vk_2  \right] (\vk_1\cdot\vk_2) A(\vk_1) B(\vk_2) \vs
=\:& \int_{\vk_1}\int_{\vk_2} (2\pi)^3 \d_D(\vk-\vk_{12})
\left[ - \frac1{k_1^2 k_2^2}  (\vk_1\cdot\vk_2)
  + \frac2{k_{12}^2} \frac{(\vk_1\cdot\vk_2)^2}{k_1^2k_2^2} \right] A(\vk_1) B(\vk_2) \vs
=\:& \left(\frac{\partial_i}{\lapl} A\right)\left(\frac{\partial^i}{\lapl} B\right) - \frac2{\lapl} \left[ \left(\frac{\partial_i\partial_j}{\lapl} A\right)\left(\frac{\partial^i\partial^j}{\lapl} B\right) \right]\,.
\ea
In particular, this yields
\ba
\frac{\partial^i \partial^j}{\nabla^2} \left[ \v{s}^k \partial_k \d \right ]
=\:& - \frac12 \partial^i \partial^j [ \v{s}^2 ] + \frac{\partial^i \partial^j}{\nabla^2} \left[ K^2 + \frac13 \d^2 \right]
\vs
\frac{\partial^i}{\nabla^2} \left[ \v{s}^k \partial_k \d \right ]
=\:& -\frac12 \partial^i \left[\v{s}^2 \right] + \frac{\partial^i}{\nabla^2} \left[ K^2 + \frac13 \d^2 \right] 
= - \v{s}^k (\partial_k \v{s}^i)  + \frac{\partial^i}{\nabla^2} \left[ K^2 + \frac13 \d^2 \right]
\,.
\label{eq:nonloctransf}
\ea
These relations can also be used to directly derive $\Pi^{[2]}_{ij}$ from the real-space expression for $\d^{(2)}$, \refeq{d2}.

%% file: App_halofinder.tex
\section{Halo finding algorithms}
\label{app:halofinder}

Collections of bound particles, or dark matter halos, identified in N-body simulations are a central tool
for testing the predictions of bias models.
It would be desirable to process halo catalogs into weighted samples with statistical properties similar to those hosting
the observed galaxies, quasars, clusters, or other tracers. However, this proves to be quite 
challenging owing to uncertainties in the relation between galaxies and dark matter halos. 
Moreover, there is freedom in defining even the most basic halo property, its mass.  

Halo identification algorithms, or ``halo finders'', can be broadly divided into two categories: 
Friends-of-Friends (FoF) finders \cite{davis/etal:1985} and spherical 
overdensity (SO) finders \cite{lacey/cole:1994}.
The choice of halo finder and mass definition is somewhat arbitrary and is often made 
to suit specific purposes. For example, the mass assignment of SO halos is more closely 
connected to the predictions of the spherical collapse model, which many of the results for 
Lagrangian models of halos presented in \refsecs{exset}{peaks} rely on, and to observable mass proxies such as the weak lensing signal or the temperature of thermal X-ray emitting gas.

Particle membership of SO halos is defined by being inside a spherical shell centered on the center-of-mass of particles which encloses a fixed interior density $\rho(<R) =\Delta_\text{SO}\,\rhob(z)$. The halo mass is given by the sum of the member particle masses. Unbound particles inside this shell are sometimes pruned. 
A common choice for the interior density threshold is $\Delta_\text{SO} = \Delta_{\rm vir}(z) \approx 200$, 
which ensures that the mean effective linear collapse threshold $\dc$ is close to the spherical
collapse prediction ($\dc=1.686$).
Larger values $\Delta_\text{SO} = 500 - 2500$ are also adopted to obtain a closer match to
observationally inferred halo masses.  On the other hand, particle membership, and hence mass, of an FoF halo is determined by being separated by less than a linking length $\lambda$ from another member particle, where, typically,
$\lambda\sim 0.15-0.2$ in units of the mean interparticle distance. 
Such linking lengths can yield a mean effective critical threshold $\dc$ which is lower than the spherical collapse prediction.

The question of how SO halo masses can be mapped onto friends-of-friends masses remains 
a matter of debate (see \cite{lukic/reed/etal:2009,more/etal:2011} for a discussion).  
The halo mass definition is clearly of relevance to Lagrangian models of halos as well as 
universal mass functions phrased in terms of the peak significance $\nu_c \equiv \dc/\s(M)$:  
changing the mass assigned to halos also changes $\nu_c$ for each halo through $\s(M)$, and, thereby, changes the multiplicity $\nu_c f(\nu_c)$.